\documentclass[prb,aps,nofootinbib,amssymb,twocolumn,superscriptaddress,10pt]{revtex4-2}
\usepackage[T1]{fontenc}
\usepackage{graphicx}
\usepackage{xcolor}
\usepackage{dcolumn}
\usepackage{comment}
\usepackage{amsmath,amssymb,bbold,bm}
\usepackage{hyperref}
\usepackage{amsfonts}
\usepackage{float,latexsym}
\usepackage{multirow}
\usepackage{color}
\usepackage[normalem]{ulem}
\usepackage{cleveref}
\usepackage{physics}
\usepackage{tabularx}
\usepackage{array}
\usepackage[caption=false]{subfig}
\usepackage{siunitx}
\usepackage{gensymb}
\usepackage{pifont}
\usepackage{placeins}
\usepackage{mathtools}
\usepackage{longtable}
\usepackage{multirow}
\usepackage{nicematrix}
\usepackage{tikz}
\usepackage{xstring}
\usepackage{makecell}
\usepackage{xr}
\usepackage{ifthen}
\usetikzlibrary{decorations.markings}
\usetikzlibrary{positioning}
\usetikzlibrary{arrows.meta}
\usepackage{tikz-feynman}
\usetikzlibrary{calc}
\tikzfeynmanset{warn luatex=false}

\tikzset{my node/.style={circle}, 
	strike through post/.append style={
		decoration={markings, mark=at position 0.75 with {
				\draw[-] ++ (0,-5pt) -- (0,5pt);}
		},postaction={decorate}}
}
\tikzset{my node/.style={circle}, 
	strike through pre/.append style={
		decoration={markings, mark=at position 0.25 with {
				\draw[-] ++ (0,-5pt) -- (0,5pt);}
		},postaction={decorate}}
}
\tikzset{
	->-/.style={decoration={
			markings,
			mark=at position #1 with {\arrow{>}}},postaction={decorate}},
	-<-/.style={decoration={
			markings,
			mark=at position #1 with {\arrow{<}}},postaction={decorate}}
}

\allowdisplaybreaks

\newcolumntype{C}{>{$} c <{$}}
\let\vec\mathbf

\DeclareMathAlphabet\mathbfcal{OMS}{cmsy}{b}{n}

\DeclarePairedDelimiterX{\setDef}[2]\{\}{%
  \, #1 \;\delimsize\vert\; #2 \,
}
\definecolor{darkgreen}{cmyk}{0.9,0,0.9,0.5} 

\makeatletter
\def\maketag@@@#1{\hbox{\m@th\normalfont\normalsize#1}}
\makeatother

\crefname{appendix}{Appendix}{Appendices}
\crefname{equation}{Eq.}{Eqs.}
\crefname{figure}{Fig.}{Figs.}
\crefname{table}{Table}{Tables}
\crefname{section}{Section}{Sections}
\crefname{enumi}{Point}{Points}
\creflabelformat{appendix}{[#2#1#3]}
\crefmultiformat{figure}{Figs.~#2#1\xdef\mycreffirstarg{#1}#3}%
 { and~#2#1#3}{, #2#1#3}{ and~#2#1#3}

\makeatletter     
\renewcommand\onecolumngrid{
\do@columngrid{one}{\@ne}%
\def\set@footnotewidth{\onecolumngrid}
\def\footnoterule{\kern-6pt\hrule width 1.5in\kern6pt}%
}

\makeatletter
\AtBeginDocument{\let\LS@rot\@undefined}
\makeatother 

\usepackage{xr}
\makeatletter
\newcommand*{\addFileDependency}[1]{
  \typeout{(#1)}
  \@addtofilelist{#1}
  \IfFileExists{#1}{}{\typeout{No file #1.}}
}
\makeatother

\crefname{appendix}{Appendix}{Appendices}
\crefname{equation}{Eq.}{Eqs.}
\crefname{figure}{Fig.}{Figs.}
\crefname{table}{Table}{Tables}
\crefname{section}{Section}{Sections}
\creflabelformat{appendix}{[#2#1#3]}

\makeatletter
\AtBeginDocument{\let\LS@rot\@undefined}
\makeatother

\makeatletter
\renewcommand\onecolumngrid{\do@columngrid{one}{\@ne}\def\set@footnotewidth{\onecolumngrid}\def\footnoterule{\kern-6pt\hrule width 1.5in\kern6pt}}

\newcommand{\siSection}{appendix}
\newcommand{\SiSection}{Appendix}
\allowdisplaybreaks

\begin{document}
\title{Obtaining the Spectral Function of Moir\'e Graphene Heavy-Fermions Using Iterative Perturbation Theory}
\author{Dumitru C\u{a}lug\u{a}ru}
	\thanks{These authors contributed equally to this work.}
	\affiliation{Department of Physics, Princeton University, Princeton, New Jersey 08544, USA}
	\affiliation{Rudolf Peierls Centre for Theoretical Physics, University of Oxford, Oxford OX1 3PU, United Kingdom}
	\author{Haoyu Hu}
	\thanks{These authors contributed equally to this work.}
	\affiliation{Donostia International Physics Center, P. Manuel de Lardizabal 4, 20018 Donostia-San Sebastián, Spain}
	\affiliation{Department of Physics, Princeton University, Princeton, New Jersey 08544, USA}
	\author{Lorenzo Crippa}
	\affiliation{Institut f{\"u}r Theoretische Physik und Astrophysik and W{\"u}rzburg-Dresden Cluster of Excellence ct.qmat, Universit{\"a}t W{\"u}rzburg, 97074 W{\"u}rzburg, Germany}
	\affiliation{I. Institute of Theoretical Physics, Universität Hamburg, Notkestraße 9-11, 22607 Hamburg, Germany}
	\author{Gautam Rai}
	\affiliation{I. Institute of Theoretical Physics, University of Hamburg, Notkestra{\ss}e 9-11, 22607 Hamburg, Germany}
	\affiliation{The Hamburg Centre for Ultrafast Imaging, Luruper Chaussee 149, 22761 Hamburg, Germany}
	\author{Nicolas Regnault}
	\affiliation{Center for Computational Quantum Physics, Flatiron Institute, 162 5th Avenue, New York, NY 10010, USA}
	\affiliation{Department of Physics, Princeton University, Princeton, New Jersey 08544, USA}
	\affiliation{Laboratoire de Physique de l'Ecole normale sup\'{e}rieure, ENS, Universit\'{e} PSL, CNRS, Sorbonne Universit\'{e}, Universit\'{e} Paris-Diderot, Sorbonne Paris Cit\'{e}, 75005 Paris, France}
	\author{Tim O.~Wehling}
	\affiliation{I. Institute of Theoretical Physics, University of Hamburg, Notkestra{\ss}e 9-11, 22607 Hamburg, Germany}
	\affiliation{The Hamburg Centre for Ultrafast Imaging, Luruper Chaussee 149, 22761 Hamburg, Germany}
	\author{Roser Valent\'\i}
	\affiliation{Institut f\"ur Theoretische Physik, Goethe Universit\"at Frankfurt, Max-von-Laue-Strasse 1, 60438 Frankfurt am Main, Germany}
	\author{Giorgio Sangiovanni}
	\affiliation{Institut f{\"u}r Theoretische Physik und Astrophysik and W{\"u}rzburg-Dresden Cluster of Excellence ct.qmat, Universit{\"a}t W{\"u}rzburg, 97074 W{\"u}rzburg, Germany}
	\author{B.~Andrei Bernevig}
	\email{bernevig@princeton.edu}
	\affiliation{Department of Physics, Princeton University, Princeton, New Jersey 08544, USA}
	\affiliation{Donostia International Physics Center, P. Manuel de Lardizabal 4, 20018 Donostia-San Sebastián, Spain}
	\affiliation{IKERBASQUE, Basque Foundation for Science, Bilbao, Spain}

\begin{abstract}
	The spectral functions of twisted bilayer graphene (TBG) in the absence of strain have recently been investigated in both the symmetric~\cite{HAU19,HU23,DAT23,RAI23a} and symmetry-broken~\cite{RAI23a} phases using dynamical mean-field theory (DMFT). The theoretically predicted Mott-Hubbard bands and gapless semimetallic state at half-filling have since been confirmed experimentally. Here, we develop several second-order perturbation theory approaches to the topological heavy-fermion (THF) model of TBG and twisted symmetric trilayer graphene (TSTG). In the symmetric phase, we adapt, implement, and benchmark an iterative perturbation theory (IPT) impurity solver within DMFT, enabling computationally efficient yet accurate spectral function calculations. We present momentum- and energy-resolved spectra over a broad range of temperatures and fillings for both symmetric and symmetry-broken states. In addition, we derive analytic expressions for the spectral function within the ``Hubbard-I'' approximation of the THF model and, as expected, find that while it provides a tractable description of Mott physics, it does not capture the low-energy Kondo peak or the finite lifetime broadening of the bands. Our methodology can be extended to include strain, lattice relaxation, and parameter variations, thereby allowing systematic predictions of TBG and TSTG spectral properties across a wide range of physical regimes. Because our perturbative approaches are far less computationally intensive than DMFT with numerically exact impurity solvers, they can be used to efficiently benchmark and scan extensive phase diagrams of the THF parameters, paving the way for full DMFT analyses of the TBG spectral function in the presence of strain and relaxation~\cite{CRI25}.
\end{abstract}
\maketitle

\section{Introduction}\label{sec:introduction}

The emergence of flat bands near charge neutrality~\cite{BIS11,SUA10,LOP07} has led to the observation of a wide range of experimentally confirmed phases, including strongly correlated states~\cite{CAO18,KER19,XIE19,SHA19,JIA19,CHO19,POL19,YAN19,LU19,STE20,SAI20,SER20,CHE20b,WON20,CHO20,NUC20,CHO21,SAI21,LIU21c,PAR21c,WU21a,CAO21,DAS21,TSC21,PIE21,STE21,CHO21a,XIE21d,DAS22,NUC23,YU23c}, superconductivity~\cite{CAO18a,YAN19,LU19,STE20,SAI20,LIU21c,CAO21,DE21a,OH21,DI22a,TIA23}, and other exotic phases~\cite{TOM19,CAO20,ZON20,LIS21,BEN21,LIA21c,ROZ21,SAI21a,LU21,HES21,DIE23,HUB22,GHA22,JAO22,PAU22,GRO22,ZHO23a,HU24a,GHO24,ZHA25a,XIA25} in twisted bilayer graphene (TBG) and twisted symmetric trilayer graphene (TSTG)~\cite{CHE19,CHE19a,PAR21,HAO21,CAO21b,LI22d,TUR22,LIU22b,KIM22,ZHA22d,ZHA22c,SHE23,KIM23,KIM25a,PAR25b}. At the same time, extensive theoretical efforts have been devoted to developing models~\cite{LOP07,SUA10,BIS11,UCH14,WIJ15,DAI16,JAI16,NAM17,EFI18,KAN18,ZOU18,PO19,LIU19a,TAR19,MOR19,LI19,FAN19,KHA19,CAR19a,CAR19,RAD19,KWA20,CAR20,TRI20,WIL20,PAR20,CAR20a,FU20,HUA20a,CAL20,WU21d,REN21,HEJ21,CAL21,BER21,BER21a,WAN21a,RAM21,SHI21,LEI21,CAO21a,SHE21,WU21c,KOS18,GUE22,DAV22,CLA22,LIN22,SAM22,KAN23b,VAF23,SHI23} that capture these systems' nontrivial single-particle topology~\cite{ZOU18,HEJ19,AHN19,PO19,SON19,HEJ19a,LIU19a,XIE20,LIA20,SON21}, correlated phases~\cite{OCH18,THO18,XU18b,KOS18,PO18a,VEN18,YUA18,DOD18,PAD18,KEN18,RAD18,HAU19,LIU19,HUA19,WU19,CLA19,KAN19,SEO19,DA19,ANG19,XIE20b,BUL20b,REP20,CEA20,ZHA20,KAN20a,BUL20a,CHI20b,CHR20,EUG20,WU20,VAF20,XIE21,KAN21,DA21,LIU21a,THO21,KWA21a,LIA21,ZHA21,VAF21,KWA21b,CHE21,POT21,XIE21b,XIE21a,CHA21,KWA21,HOF22,WAG22,CHR22,SON22,BRI22,CAL22d,HON22,ZHA23a,BLA22,XIE23a,KWA23,YU23a,WAN23c,FER20,DAT23,RAI23a,WAN24b,HER24,HU25,LED25,LED25a}, fractional insulating phases~\cite{ABO20,LED20,REP20a}, various excitations~\cite{WU20,VAF20,BER21b,XIE21,KUM21,KAN21,KWA22}, superconducting states~\cite{GUO18,YUA18,XU18,DOD18,PO18a,LIU18a,VEN18,ISO18,PEL18,KEN18,WU18,GUI18,GON19,HUA19,ROY19,WU19a,WU19,CLA19,LIA19,HU19a,JUL20,XIE20,CHI20a,LOP20,KON20,CHR20,WAN21,LEW21,FER21,QIN21,PHO21,CHO21d,LAK21,CHO21c,LI22c,FIS22,YU22,SCA22,CHR22,KWA22,GON23,CHR23,MAC23,WAG23}, and experimental responses~\cite{MOO13,LIU20b,PAD20,GAR20,CAL22d,HON22,KRU23,OCH23,WAN23b}.

Among the many effective descriptions proposed for TBG~\cite{SON22,HU23,ZHO24,LAU23,CAL23,CHO23,HU23i,LI23a,RAI23a,WAN24,CAL24,HER24a,HER24,HER24c,HU25,YOU24} and TSTG~\cite{YU23a}, the topological heavy-fermion (THF) model and its related counterparts~\cite{HAU19,CAL20,SHI22a,DAT23} stand out. Fundamentally, the THF model is a basis transformation applied to the low-energy spectrum of TBG (or TSTG). In this transformed basis, the active bands of TBG or TSTG are predominantly composed of localized, strongly correlated \emph{heavy} ($f$) electrons that hybridize with itinerant, strongly dispersive \emph{conduction} ($c$) electrons around the $\Gamma_M$ point, thereby reproducing the nontrivial topology of the system. Beyond providing a framework suitable for heavy-fermion techniques, the THF model also reconciles seemingly contradictory experimental observations. On one side, local-moment physics is manifested through quantum-dot-like features observed in scanning tunneling microscopy~\cite{BEN21,TUR22,KIM22,CAL22d,ZHO23a,KIM23,NUC23}, the Pomeranchuk effect~\cite{ROZ21,SAI21a}, the emergence of ferromagnetic ground states~\cite{SHA19,CHE20b,SAI21,TSC21,LIU22b}, and entropy measurements~\cite{ZHA25a}. On the other side, signatures of itinerant behavior -- such as metallic transport~\cite{CAO20}, Chern insulating states~\cite{SHA19,LU19,SER20,CHE20b,STE20,CHO20,NUC20,CHO21,SAI21,PAR21c,WU21a,DAS21,TSC21,PIE21,CHO21a}, Dirac-like compressibility~\cite{TOM19,ZON20}, and superconductivity~\cite{CAO18a,YAN19,CHE19a,LU19,SHE20,STE20,SAI20,HAO21,CAO21b,DE21a,OH21,TIA23,KIM22,DI22a} -- highlight the presence of itinerant electrons.

The interacting spectral functions of TBG have been theoretically predicted~\cite{HAU19,HOF22,DAT23,RAI23a,CAL25a,HU25,LED25a,LED25,ZHA25b,ZHA25c} and, more recently, directly observed~\cite{XIA25} to exhibit Mott-Hubbard flat bands along with highly dispersive states that are either gapless (at charge neutrality) or gapped, depending on the filling. At charge neutrality, the spectral function at the magic angle is characterized by gaps at the moir\'e $\mathrm{K}_M$ and $\mathrm{M}_M$ points (where the ${}_M$ subscript indicates that the high-symmetry points refer to the moir\'e Brillouin zone) but remains gapless at $\Gamma_M$. This behavior is the exact inverse of the non-interacting band structure, which is gapless at $K_M$ yet gapped at $\Gamma_M$. Away from the magic angle, experimental measurements of the spectral function align more closely with the non-interacting band picture. For non-integer fillings, recent spectroscopic studies~\cite{ZHA25a,KIM25a,PAR25b} have explored the emergence of zero-bias (Kondo) peaks and briefly examined the roles of strain and superconducting order on these theoretically predicted features~\cite{HU23,DAT23,RAI23a}.

While the overall physics of the system remains qualitatively similar across different parameter regimes ({\it e.g.}{}, varying Coulomb interaction strengths), certain quantitative features -- such as the Fermi velocity of the gapless Mott phase at charge neutrality -- are sensitive to the system parameters~\cite{HU25}. The development of the quantum twisting microscope~\cite{XIA25} provides a powerful means to probe the influence of various ``perturbations'', including strain and lattice relaxation, by directly comparing experimental data with comprehensive interacting calculations. However, unbiased dynamical mean-field theory (DMFT)~\cite{GEO96,KOT06a} combined with quantum Monte Carlo (QMC)~\cite{HAU19,HU23,DAT23,RAI23a,CAL25a} or numerical renormalization group (NRG)~\cite{ZHO24,WAN24,YOU24} impurity solvers -- methods that have been employed to study the spectral function of TBG in the presence of correlations -- remains computationally demanding. Consequently, these approaches have so far been applied only to idealized scenarios ({\it i.e.}{}, neglecting strain and often relaxation effects, while focusing on magic-angle parameters~\cite{SON22}). 

In contrast, the THF formalism can readily account for the impact of strain and relaxation~\cite{HER24,HER24a}, as well as variations in single-particle parameters ({\it e.g.}{}, the gap between the flat and remote bands, or the bandwidth of the active bands) and interaction parameters (different forms of screened Coulomb potentials)~\cite{CAL23}, all while preserving the charge density consistent with the Bistritzer-Macdonald (BM) model~\cite{BIS11}. Therefore, it is crucial to develop an approach that is far less computationally costly than DMFT with QMC or NRG solvers, yet retains much of their accuracy, allowing us to explore a broad parameter space. This becomes especially important away from integer fillings, where the spectral properties of the Kondo peak remain analytically intractable, unlike the Mott phase at integer fillings, which does permit an analytic description~\cite{HU25,LED25a,LED25}.

The goals of this paper are twofold. First, we develop and apply two complementary approaches to study the THF model beyond the Hartree-Fock approximation, both based on second-order perturbation theory applied to the THF Hamiltonian (in idealized conditions). For symmetry-broken phases, we employ self-consistent second-order perturbation theory~\cite{SCH89,SCH90,SCH91} to compute the interacting THF spectral function. For the symmetric phase, we introduce a \emph{modified} iterative perturbation theory (IPT) solver~\cite{MAR86,GEO92,YEY93,KAJ96,POT97,ANI97,LIC98,YEY99,MEY99,YEY00,SAS01,SAV01,FUJ03,LAA03,KUS06,ARS12,DAS16,WAG21,MIZ21,VAN22,CAN24,CAN25,VAN22} within the DMFT framework, and benchmark its performance against results obtained using a QMC solver. Applied to the THF model, these schemes present additional complexity due to (i) the large orbital multiplicity, (ii) novel interaction terms between light electrons and between light and heavy electrons, and (iii) the purely momentum-space nature of the topological $c$-electrons. 

Second, we implement these methods numerically to obtain momentum- and energy-resolved THF spectral functions for both TBG and TSTG across a wide range of fillings and temperatures. Our DMFT with IPT approach successfully captures both the Hubbard bands and the Kondo-like zero-energy peak originating from the heavy-electron bands. These techniques also enable us to compute carrier lifetimes, which are subsequently used in our companion transport study~\cite{CAL24}. We present spectral functions for both symmetric and symmetry-broken states and analyze the temperature dependence of the Hubbard bands and the Kondo peak. Furthermore, we derive analytic expressions for the spectral functions within the ``Hubbard-I'' approximation~\cite{HUB63,HUB64} (shown to be identical to the ``non-local moments'' picture~\cite{HU25,LED25,LED25a}) across several parameter regimes. For completeness, all definitions and detailed derivations are included in the appendices.

The main text is structured as a concise summary of the logical flow, key equations, and main results, while the detailed derivations and intermediate steps -- owing to their considerable length -- are presented in the appendices. Thanks to the modular organization of these appendices, readers who are already familiar with certain parts of the methodology can directly focus on the sections of interest. The remainder of the paper is organized as follows. In \cref{sec:thf_notation}, we review the THF model and establish the notation used throughout this work. \Cref{sec:sym_br} examines the symmetry-broken phases of the THF model at finite temperature and doping using second-order self-consistent perturbation theory. The symmetric phase of the THF model is analyzed in \cref{sec:symmetric}, where we also introduce the IPT impurity solver and benchmark it against numerically exact methods (QMC). Finally, we present our conclusions in \cref{sec:conclusions}.
 
\section{Topological heavy fermion model and notation}\label{sec:thf_notation}

We begin by outlining the model and notation used throughout this work. The details of the THF model for TBG~\cite{SON22} and TSTG~\cite{YU23a} are reviewed in \cref{app:sec:HF_review}. The parameters adopted in this work are identical to those used in Refs.~\cite{SON22,YU23a,CAL24}. 

\subsection{Model}\label{sec:thf_notation:model}

TBG consists of two graphene layers rotated relative to one another by the ``magic'' angle $\theta$~\cite{BIS11}, which gives rise to two fundamentally distinct types of fermions~\cite{HAU19,CAL20,SON22,SHI22a}. First, for each spin $s = \uparrow, \downarrow$ and moir\'e momentum $\vec{k}$ in valley $\eta = \pm$, the heavy fermions $\hat{f}^\dagger_{\vec{k},\alpha, \eta, s}$ -- a pair of $\alpha = 1,2$ ({\it i.e.}{}, $p_x \pm i p_y$) orbitals localized at the moir\'e AA-sites -- carry zero kinetic energy but dominate the interaction physics. Second, the ``light'' fermions $\hat{c}^\dagger_{\vec{k},a,\eta,s}$ -- with orbital numbers $a = 1,2$ ($a = 3,4$) representing the $\Gamma_3$ ($\Gamma_1 \oplus \Gamma_2$) irreducible (reducible) representations near the $\Gamma_M$ point of the MBZ -- form the itinerant, anomalous, and semimetallic degrees of freedom. These $c$-fermions ``carry'' the system's nontrivial topology and hybridize with the heavy fermions, resulting in the nearly flat topological bands of TBG.

The main distinction of TSTG~\cite{LI19,KHA19,CAR20,CAL21} compared to TBG lies in the presence of additional high-velocity ``Dirac'' fermions, $\hat{d}^\dagger_{\vec{p},\alpha, \eta, s}$~\cite{YU23a}. These Dirac fermions belong to the sublattice $\alpha = A,B$ of single-layer graphene and have momenta $\vec{p}$ measured from the $\mathrm{K}_M$ or $\mathrm{K}'_M$ points of the MBZ (located at $\eta \vec{q}_1$ in valley $\eta$). They hybridize exclusively with the heavy fermions and only in the presence of a perpendicular displacement field $\mathcal{E}$ (see \cref{app:sec:HF_review:single_particle:TSTG}).

For convenience, we adopt a unified notation (see also \cref{app:sec:hartree_fock})
\begin{equation}
\label{main:eqn:shorthand_gamma_not}
	\hat{\gamma}^\dagger_{\vec{k},\eta,i,s} \equiv \begin{cases}
		\hat{c}^\dagger_{\vec{k},\eta,i,s}, & \qq{for} 1 \leq i \leq 4 \\
		\hat{f}^\dagger_{\vec{k},\eta,i-4,s}, & \qq{for} 5 \leq i \leq 6
	\end{cases}.
\end{equation} 
The $c$-fermions are energetically relevant only near the $\Gamma_M$ point, but we define them across the entire first MBZ. For TSTG, we further introduce two additional fermion operators for each spin and valley flavor, denoted by $\hat{\gamma}^\dagger_{\vec{k},\eta,i,s}$ with $7 \leq i \leq 8$, which correspond to the additional high-velocity Dirac electrons. 

The single-particle THF Hamiltonian matrix $h^{\text{TBG},\eta} \left( \vec{k} \right)$ for TBG in valley $\eta$ can be written in block form as 
\begin{equation}
	h^{\text{TBG},\eta} \left( \vec{k} \right) =
	\begin{pmatrix}
		h^{cc,\eta} \left( \vec{k} \right) &  h^{cf,\eta} \left( \vec{k} \right) \\
		h^{cf,\eta,\dag} \left( \vec{k} \right) & \mathbb{0}
	\end{pmatrix}
\end{equation}
where $h^{cc,\eta}(\vec{k})$ and $h^{cf,\eta}(\vec{k})$ denote, respectively, the kinetic term of the $c$-electrons and the $f$-$c$ hybridization term. At the magic angle, and in the absence of strain and relaxation effects, the $f$-electrons have zero single-particle energy. The explicit forms of these terms are presented and discussed in \cref{app:sec:HF_review}.

As discussed in \cref{app:sec:HF_review:interaction:TBG}, the Coulomb interaction in the THF Hamiltonian for TBG -- derived under the assumption of a double-gated experimental setup with relative dielectric constant $\epsilon = 6$ and gate-to-gate distance $\xi = \SI{10}{\nano\meter}$ -- includes~\cite{SON22,SHI22a}: a dominant onsite $f$-electron repulsion ($H_{U_1}$, $U_1 = \SI{57.95}{\milli\electronvolt}$), a smaller nearest-neighbor $f$-electron repulsion ($H_{U_2}$, $U_2 = \SI{2.239}{\milli\electronvolt}$), the $c$-electron Coulomb interaction ($H_V$), which is characterized by the double-gate screened Coulomb potential $V(\vec{q})$, the $f$-$c$ density-density interaction ($H_{W}$, $W_1 = \SI{44.03}{\milli\electronvolt}$, $W_3 = \SI{50.2}{\milli\electronvolt}$), the $f$-$c$ exchange interaction ($H_{J}$, $J = \SI{16.38}{\milli\electronvolt}$), the double-hybridization interaction ($H_{\tilde{J}}$, characterized by the same energy scale as $H_J$), and the density-hybridization interaction ($H_{K}$, $K = \SI{4.887}{\milli\electronvolt}$). All interaction terms are normal ordered with respect to the charge neutrality point, and the relevant interaction strengths have been provided explicitly.

The THF interaction parameters in TSTG are \emph{approximately} larger by a geometric factor of $\sqrt{2}$, reflecting the fact that its magic angle is larger by this same factor (see \cref{app:sec:HF_review:interaction:TSTG}). In addition, compared to the THF model of TBG, there are extra Coulomb terms involving the $d$-electrons: repulsion among the $d$-electrons ($H_V^{d}$), interactions between $c$- and $d$-electrons ($H_V^{cd}$), as well as between $f$- and $d$-electrons ($H_W^{fd}$, $W_{fd} = \SI{94.71}{\milli\electronvolt}$)~\cite{YU23a}.

\subsection{Single-particle Green's function}\label{sec:thf_notation:excitations}

In this work, we investigate the single-particle excitations of the system, which are characterized by the imaginary-time Green's function
{\small\begin{equation}
	\label{eqn:matsubara_gf_THF_tau}
	-\left\langle \mathcal{T}_{\tau} \hat{\gamma}_{\vec{k},i,\eta,s} \left( \tau \right) \hat{\gamma}^\dagger_{\vec{k}',i',\eta',s'} \left( 0 \right)  \right\rangle  = \delta_{\vec{k},\vec{k}'} \mathcal{G}_{i \eta s; i' \eta' s'} \left(\tau, \vec{k} \right),
\end{equation}}where $\mathcal{T}_{\tau}$ denotes time ordering and $\left\langle \hat{\mathcal{O}} \right\rangle = \frac{1}{Z} \Tr \left[ e^{-\beta K} \hat{\mathcal{O}} \right]$ is the grand-canonical expectation value with partition function $Z \equiv \Tr \left[ e^{-\beta K} \right]$. Here, the fermion operators evolve with the grand-canonical Hamiltonian $K = H - \mu N$. By Fourier transforming \cref{eqn:matsubara_gf_THF_tau} with respect to imaginary time $\tau$, we obtain the Matsubara-frequency Green's function, which depends on $\left( i \omega_n, \vec{k} \right)$ with $\omega_n = \frac{(2n+1)\pi}{\beta}$. The \emph{non-interacting} Matsubara Green's function is obtained by replacing the full Hamiltonian $H$ with the non-interacting Hamiltonian $H_0$. In this case, the Green's function is explicitly given by $\mathcal{G}^{0} \left(i\omega_n,  \vec{k} \right) = \left[\left( i\omega_n + \mu \right) \mathbb{1} - h \left( \vec{k} \right)  \right]^{-1}$. 

The self-energy matrix $\Sigma \left( i \omega_n, \vec{k} \right)$ relates the interacting and non-interacting Green's functions through the Dyson equation~\cite{MAH00}
\begin{equation}
	\label{eqn:dyson_equation}
	\mathcal{G} \left( i \omega_n, \vec{k} \right) = \left[ \left( \mathcal{G}^{0} \left( i \omega_n, \vec{k} \right) \right)^{-1} - \Sigma \left(i \omega_n, \vec{k} \right) \right]^{-1}. 
\end{equation} 
Analytic continuation~\cite{MAH00} to real frequencies yields the retarded Green's function $\mathcal{G}_{i \eta s; i' \eta' s'} \left( \omega + i 0^{+}, \vec{k} \right)$ defined along the real $\omega$-axis and the associated spectral function
\begin{equation}
	A \left( \omega, \vec{k} \right) = \frac{-1}{2 \pi i} \left( \mathcal{G} \left(\omega + i 0^{+}, \vec{k} \right) - \mathcal{G}^\dagger  \left(\omega + i 0^{+}, \vec{k} \right)  \right).
\end{equation}
From the spectral function, the Green's function at any complex frequency $z$ above or below the real axis is obtained through
\begin{equation}
	\mathcal{G} \left( z, \vec{k} \right) = \int_{-\infty}^{\infty} \frac{\dd{\omega}}{z-\omega} A \left( \omega, \vec{k} \right).
\end{equation}
The density matrix of the system is defined as $\varrho \left(\vec{k} \right) = \int_{-\infty}^{\infty} \dd{\omega} n_{\mathrm{F}} \left( \omega \right) A\left( \omega, \vec{k} \right) - \frac{1}{2} \mathbb{1}$, where $\mathbb{1}$ is the identity matrix and $n_{\mathrm{F}} \left( \omega \right)$ is the Fermi-Dirac distribution. The fillings of the $c$- and $f$-electrons are then given by $\nu_c = \frac{1}{N_0} \sum_{i=1}^{4} \sum_{\vec{k},\eta,s} \varrho_{i \eta s; i \eta s} \left(\vec{k} \right)$ and $\nu_f = \frac{1}{N_0} \sum_{i=5}^{6} \sum_{\vec{k},\eta,s} \varrho_{i \eta s; i \eta s} \left(\vec{k} \right)$, where $N_0$ denotes the number of moir\'e unit cells. The individual fillings $\nu_c$ and $\nu_f$ are not independently fixed (except in the absence of one-body $f$-$c$ hybridization~\cite{HU23i}); however, their sum, the total electron filling $\nu = \nu_c + \nu_f$, remains fixed. A more detailed discussion, including the extension to TSTG, is provided in \cref{app:sec:hartree_fock}.

We employ the THF Hamiltonian to explore the correlation physics, focusing on dynamical self-energy effects in both the low-temperature symmetry-broken phase and the high-temperature symmetric phase. The main text presents the results for TBG, while the corresponding TSTG calculations are included in the appendices. Throughout, our analysis is restricted to states that preserve moir\'e translational symmetry.

\section{Symmetry-broken states beyond Hartree-Fock}\label{sec:sym_br}

This section examines the symmetry-broken states of the THF model. A review of the Hartree-Fock treatment of the THF model is given in \cref{app:sec:hartree_fock}. To compute the spectral function for various ground-state candidates, we employ second-order self-consistent perturbation theory in the $f$-electron interactions ($H_{U_1}$ and $H_{U_2}$). The method is summarized here, with full details provided in \cref{app:sec:se_correction_beyond_HF,app:sec:additional_mb_results}. We then present and discuss results for the finite-temperature and finite-doping symmetry-broken states of TBG, whose zero-temperature physics at integer fillings was discussed in Ref.~\cite{SON22}. The corresponding results for TSTG at small displacement field~\cite{YU23a}, along with additional results and discussion for TBG, are included in \cref{app:sec:results_corr_ins}.

\subsection{Symmetry-broken states considered}\label{sec:sym_br:states}

The nine symmetry-broken parent states of the many-body model considered here are summarized in \cref{app:tab:model_states} of \cref{app:sec:hartree_fock:ground_states:model} for integer fillings $\nu_0$. These include valley-polarized (VP) and/or Kramers-inter-valley coherent (K-IVC) states, which have been identified as the ground states at integer fillings of TBG~\cite{BRI22, BUL20a, BUL20b, CEA20, CHA21, CHE21, CHI20b, CHR20, CLA19, DA19, DA21, DOD18, EUG20, HOF22, HUA19, KAN19, KAN20a, KAN21, KEN18, KOS18, KWA21, LIA21, LIU19, LIU21a, OCH18, PO18a, REP20, SEO19, SON22, THO18, VAF20, VEN18, WAG22, WU19, WU20, XIE20b, XIE21, XIE23a, XU18b, YUA18, ZHA20, ZHA23a} and TSTG~\cite{CHR22,XIE21b,YU23a} in the absence of strain. States at negative and positive fillings are related by a many-body charge-conjugation symmetry (defined in \cref{app:sec:hartree_fock:ground_states:ph_symmetry} for the Hamiltonian without lattice relaxation effects). Consequently, and without loss of generality, we restrict our analysis to positive chemical potentials.

Within the Hartree-Fock approximation, the THF interaction Hamiltonian for a symmetry-broken state is decoupled into a quadratic form,
\begin{equation}
	H_{I,\text{MF}} = \sum_{\substack{i, \eta, s \\ i', \eta', s'}} h^{I,\text{MF}}_{i \eta s;i' \eta' s'} \left( \vec{k} \right) \hat{\gamma}^\dagger_{\vec{k},i, \eta, s} \hat{\gamma}_{\vec{k},i', \eta', s'},
\end{equation}
where the Hartree-Fock interaction matrix $h^{I,\text{MF}}(\vec{k})$ is obtained by convolving the interaction with the density matrix of the chosen state, as detailed in \cref{app:sec:hartree_fock}. The \emph{total} mean-field Hamiltonian $h^{\text{MF}}(\vec{k})$ is then the sum of the single-particle term $h(\vec{k})$ and the interaction term $h^{I,\text{MF}}(\vec{k})$. At self-consistency, the Hartree-Fock Hamiltonian and the density matrix satisfy $\varrho^{T} \left( \vec{k} \right) = \left\lbrace \exp \left[  \beta \left( h^{\text{MF}} \left( \vec{k} \right) - \mu \mathbb{1} \right) \right] + \mathbb{1} \right\rbrace ^{-1} - \frac{1}{2} \mathbb{1}$, with the chemical potential $\mu$ fixed to match the desired filling.

\subsection{Second-order self-energy}\label{sec:sym_br:so_self_en}

Before extending the above analysis beyond the Hartree-Fock level, we begin with some general remarks on the self-energy $\Sigma \left( i \omega_n, \vec{k} \right)$ of an interacting system, and then specialize to our case. In the Hartree-Fock approximation, the self-energy is static ({\it i.e.}{}, independent of $\omega$) and simply equals the Hartree-Fock interaction Hamiltonian, $\Sigma \left(i\omega_n, \vec{k} \right) = h^{I,\text{MF}} \left( \vec{k} \right)$, which determines the dispersion and energy gaps of the charge-one excitations. The \emph{lifetime} of these excitations -- relevant, for example, in transport~\cite{CAL24} -- is related to the imaginary part of the self-energy~\cite{MAH00}. This imaginary component arises at least at \emph{second order} in the interaction strength within perturbation theory and, in our analysis, is computed considering only elastic electron-electron scattering.

The complete Feynman rules for the self-energy calculation are derived and discussed in detail in \cref{app:sec:se_correction_beyond_HF}. In a \emph{self-consistent} treatment, only two distinct diagrams contribute at second order. Various aspects of applying second-order perturbation theory to the THF model are reviewed in \cref{app:sec:se_correction_beyond_HF}, which provides a fully self-contained formulation. For completeness, \cref{app:sec:additional_mb_results} also re-derives the second-order self-energy correction for the $f$-electrons directly from the action, without relying on the Feynman diagrammatic framework.

\subsubsection{Approximating the second-order self-energy of the THF model}\label{sec:sym_br:so_self_en:THF}
 
As detailed in \cref{app:sec:se_correction_beyond_HF}, the second-order self-energy is generically a complicated function of both momentum \emph{and} Matsubara frequency. It depends on the Green's function of the system, not only on the density matrix~\cite{SCH90}, and involves double summations over both momenta and Matsubara frequencies. For our purpose, we are interested in capturing the first nonzero contribution to the finite lifetime of excitations in the THF model. The strong onsite repulsion of $f$-electrons renders their self-energy more important than that of strongly dispersing $c$-electrons. We hence focus only on the second-order dynamical self-energy of the $f$-electrons. An artificial broadening of the $c$-electrons (smaller than any other energy scale of the system) is added by hand (in order to compute the total spectral function) and found not to influence the physics.
 
We assume the system to be in an ordered phase achieved at integer fillings or by doping the correlated ground state candidates of \cref{app:tab:model_states} away from integer fillings. Following Refs.~\cite{SCH89,SCH90,SCH91,MET89,GEO96,DAT23,RAI23a}, and because we are mostly interested in the leading effect of interactions on the charge-one excitation lifetimes, we also make a DMFT-like approximation and only consider the site-diagonal (momentum-independent, see \cref{app:sec:se_correction_beyond_HF:all_so}) second-order self-energy correction for the $f$-electrons and neglect correlations between the localized $f$-electrons from different lattice sites. 
Furthermore, we only consider the second-order self-energy correction arising from (most importantly) $H_{U_1}$ and $H_{U_2}$. 
A detailed justification of all approximations is given in \cref{app:sec:se_correction_beyond_HF:all_so_corrections:particularization}.

As a consequence of the above-mentioned approximations, we take the second-order self-energy of the system to be
\begin{widetext}
\begin{equation}
	\label{eqn:total_self_sym_br}
	\Sigma_{i \eta s{};i' \eta' s'} \left( \omega + i 0^{+}, \vec{k} \right) = h^{I,\text{MF}}_{i \eta s{},i' \eta' s'{}} \left( \vec{k} \right) + \begin{cases}
		\Sigma^{f,(2)}_{(i-4) \eta s;(i'-4) \eta' s'} \left( \omega + i 0^{+} \right) & \qq{if} 5 \leq i,i' \leq 6  \\
		- i \Gamma_{c} \delta_{ii'} \delta_{\eta \eta'} \delta_{ss'} & \qq{otherwise}
	\end{cases},
\end{equation}
\end{widetext}
where $\Gamma_{c}$ denotes the small artificial broadening introduced for the $c$-electrons to ensure numerical stability (since our calculations are performed directly in real frequency), and $\Sigma^{f,(2)}\left( \omega + i 0^{+} \right)$ is the site-diagonal $f$-electron self-energy computed to second order in $H_{U_1}$ and $H_{U_2}$. The latter is evaluated self-consistently in real frequency using the interacting $f$-electron Green's functions, as described in \cref{app:sec:se_correction_beyond_HF:all_so}.

It is worth emphasizing some fundamental differences between the second-order approach employed here and the DMFT treatment (such as the one using the IPT solver described in \cref{sec:symmetric}). The second-order expansion of the THF self-energy is generically both frequency- and momentum-dependent; neglecting its momentum dependence constitutes an additional approximation. In contrast, within the DMFT framework the self-energy is \emph{intrinsically} momentum-independent by construction. As we show in \cref{sec:symmetric}, the advantage of DMFT (even with the approximate IPT impurity solver) is that it can capture the Mott transition and the associated strong-coupling physics, which remain inaccessible to a second-order expansion performed directly at the lattice level of the THF model.

\subsubsection{Numerical implementation}\label{sec:sym_br:so_self_en:numerical}

We solve the THF model in the symmetry-broken phase using the self-consistent second-order perturbation method described in \cref{sec:sym_br:so_self_en:THF}. The solution is characterized by the self-consistent density matrix $\rho \left( \vec{k} \right)$ and the second-order $f$-electron self-energy $\Sigma^{f,(2)}(\omega + i 0^{+})$. From these quantities, the full Green's function is obtained via \cref{eqn:dyson_equation,eqn:total_self_sym_br}. Since both the second-order self-energy and the density matrix depend on the interacting Green's function, the problem must be solved iteratively, as described in detail in \cref{app:sec:se_correction_beyond_HF:sc_problem_and_numerics}.

For a given correlated integer-filled phase, the iteration begins from the zero-temperature density matrix of that phase, together with an initial constant-broadening ansatz for the $f$-electron self-energy, $\Sigma^{f,(2)} \left( \omega + i 0^{+} \right) = -i \Gamma_c \mathbb{1}$. The interacting Green's function is then computed, which is used to update both the self-energy and the density matrix. This process is repeated until convergence. Once a converged finite-temperature solution is obtained at integer filling, doping is introduced incrementally. At each new filling, the converged solution from the nearest filling is rigidly doped and used as the initial condition, followed by iteration to convergence.

For each symmetry-broken phase, temperature, and filling, we compute the density of states (DOS)
\begin{equation}
	\label{eqn:dos_def}
	\mathcal{A} \left( \omega \right) = \frac{1}{N_0} \sum_{\vec{k},i, \eta, s} A_{i \eta s;i \eta s} \left(\omega, \vec{k} \right)
\end{equation}
for fillings $\nu$ in the range $\abs{\nu - \nu_0} \leq 0.5$ around the integer filling $\nu_0 \in \mathbb{Z}$. We focus on cases with $\nu_0 \geq 0$, as the spectra for $\nu_0 < 0$ can be obtained via particle-hole conjugation. In addition, we calculate the momentum-resolved spectral function
\begin{equation}
	\label{eqn:k_spec_def}
	\mathcal{A} \left( \omega, \vec{k} \right) = \sum_{i, \eta, s} A_{i \eta s;i \eta s} \left(\omega, \vec{k} \right)
\end{equation}
to analyze the $\vec{k}$-dependent structure of the excitations.

\subsection{Results}\label{sec:sym_br:results}

\begin{figure*}[!t]\includegraphics[width=\textwidth]{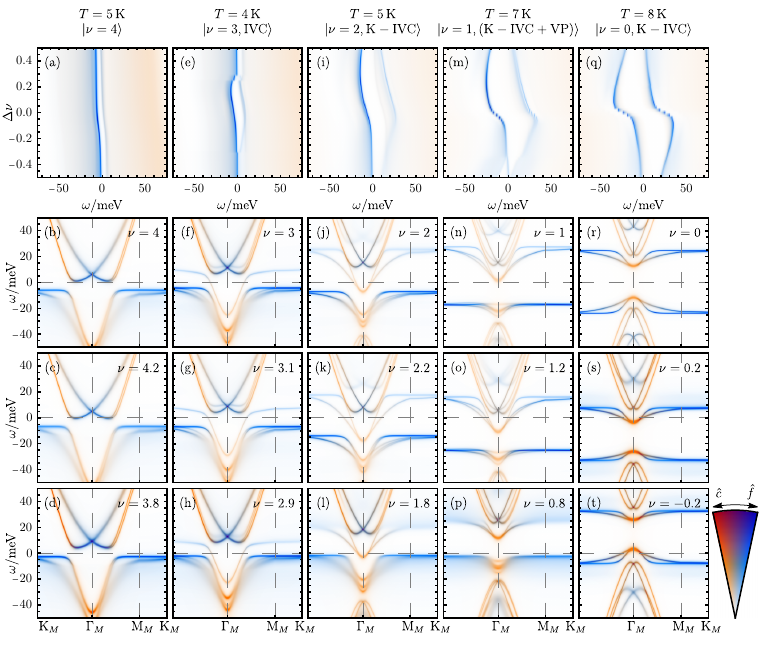}\subfloat{\label{fig:sym_br_results:a}}\subfloat{\label{fig:sym_br_results:b}}\subfloat{\label{fig:sym_br_results:c}}\subfloat{\label{fig:sym_br_results:d}}\subfloat{\label{fig:sym_br_results:e}}\subfloat{\label{fig:sym_br_results:f}}\subfloat{\label{fig:sym_br_results:g}}\subfloat{\label{fig:sym_br_results:h}}\subfloat{\label{fig:sym_br_results:i}}\subfloat{\label{fig:sym_br_results:j}}\subfloat{\label{fig:sym_br_results:k}}\subfloat{\label{fig:sym_br_results:l}}\subfloat{\label{fig:sym_br_results:m}}\subfloat{\label{fig:sym_br_results:n}}\subfloat{\label{fig:sym_br_results:o}}\subfloat{\label{fig:sym_br_results:p}}\subfloat{\label{fig:sym_br_results:q}}\subfloat{\label{fig:sym_br_results:r}}\subfloat{\label{fig:sym_br_results:s}}\subfloat{\label{fig:sym_br_results:t}}\caption{Spectral functions of TBG within the THF model in its symmetry-broken phases. Both the DOS and the momentum-resolved spectral functions are shown, as defined in \cref{eqn:dos_def,eqn:k_spec_def}. Each column corresponds to a specific symmetry-broken phase or to the band insulator, as indicated above the panels. Panels (a)-(d) illustrate the $\protect\IfStrEqCase{1}{{1}{\ket{\nu={}4} }
		{2}{\ket{\nu={}3, \mathrm{IVC}}}
		{3}{\ket{\nu={}3, \mathrm{VP}}}
		{4}{\ket{\nu={}2, \mathrm{K-IVC}}}
		{5}{\ket{\nu={}2, \mathrm{VP}}}
		{6}{\ket{\nu={}1, (\mathrm{K-IVC}+\mathrm{VP})}}
		{7}{\ket{\nu={}1, \mathrm{VP}}}
		{8}{\ket{\nu=0, \mathrm{K-IVC}}}
		{9}{\ket{\nu=0, \mathrm{VP}}}
	}
	[nada]
$ band-insulating state: (a) shows the doping- and energy-dependence of the DOS over a range $\abs{\Delta\nu} \leq 0.5$ around integer filling, while (b)-(d) display the $\vec{k}$-resolved spectral function at integer filling, electron doping, and hole doping, respectively, with the filling indicated inside each panel. The same arrangement is used for the other correlated phases. The simulation temperature is indicated above each panel. The color scale is described in the main text.}\label{fig:sym_br_results}\end{figure*}

We now present and analyze the DOS and $\vec{k}$-resolved spectral functions for four symmetry-broken states and the band insulator of TBG within the THF model. The five states considered are $\IfStrEqCase{1}{{1}{\ket{\nu={}4} }
		{2}{\ket{\nu={}3, \mathrm{IVC}}}
		{3}{\ket{\nu={}3, \mathrm{VP}}}
		{4}{\ket{\nu={}2, \mathrm{K-IVC}}}
		{5}{\ket{\nu={}2, \mathrm{VP}}}
		{6}{\ket{\nu={}1, (\mathrm{K-IVC}+\mathrm{VP})}}
		{7}{\ket{\nu={}1, \mathrm{VP}}}
		{8}{\ket{\nu=0, \mathrm{K-IVC}}}
		{9}{\ket{\nu=0, \mathrm{VP}}}
	}
	[nada]
$, $\IfStrEqCase{2}{{1}{\ket{\nu={}4} }
		{2}{\ket{\nu={}3, \mathrm{IVC}}}
		{3}{\ket{\nu={}3, \mathrm{VP}}}
		{4}{\ket{\nu={}2, \mathrm{K-IVC}}}
		{5}{\ket{\nu={}2, \mathrm{VP}}}
		{6}{\ket{\nu={}1, (\mathrm{K-IVC}+\mathrm{VP})}}
		{7}{\ket{\nu={}1, \mathrm{VP}}}
		{8}{\ket{\nu=0, \mathrm{K-IVC}}}
		{9}{\ket{\nu=0, \mathrm{VP}}}
	}
	[nada]
$, $\IfStrEqCase{4}{{1}{\ket{\nu={}4} }
		{2}{\ket{\nu={}3, \mathrm{IVC}}}
		{3}{\ket{\nu={}3, \mathrm{VP}}}
		{4}{\ket{\nu={}2, \mathrm{K-IVC}}}
		{5}{\ket{\nu={}2, \mathrm{VP}}}
		{6}{\ket{\nu={}1, (\mathrm{K-IVC}+\mathrm{VP})}}
		{7}{\ket{\nu={}1, \mathrm{VP}}}
		{8}{\ket{\nu=0, \mathrm{K-IVC}}}
		{9}{\ket{\nu=0, \mathrm{VP}}}
	}
	[nada]
$, $\IfStrEqCase{6}{{1}{\ket{\nu={}4} }
		{2}{\ket{\nu={}3, \mathrm{IVC}}}
		{3}{\ket{\nu={}3, \mathrm{VP}}}
		{4}{\ket{\nu={}2, \mathrm{K-IVC}}}
		{5}{\ket{\nu={}2, \mathrm{VP}}}
		{6}{\ket{\nu={}1, (\mathrm{K-IVC}+\mathrm{VP})}}
		{7}{\ket{\nu={}1, \mathrm{VP}}}
		{8}{\ket{\nu=0, \mathrm{K-IVC}}}
		{9}{\ket{\nu=0, \mathrm{VP}}}
	}
	[nada]
$, and $\IfStrEqCase{8}{{1}{\ket{\nu={}4} }
		{2}{\ket{\nu={}3, \mathrm{IVC}}}
		{3}{\ket{\nu={}3, \mathrm{VP}}}
		{4}{\ket{\nu={}2, \mathrm{K-IVC}}}
		{5}{\ket{\nu={}2, \mathrm{VP}}}
		{6}{\ket{\nu={}1, (\mathrm{K-IVC}+\mathrm{VP})}}
		{7}{\ket{\nu={}1, \mathrm{VP}}}
		{8}{\ket{\nu=0, \mathrm{K-IVC}}}
		{9}{\ket{\nu=0, \mathrm{VP}}}
	}
	[nada]
$, with their variational zero-temperature Slater-determinant ansatz given explicitly in \cref{app:tab:model_states}. \Cref{fig:sym_br_results} shows, for each state, the DOS of the integer-filled phase as a function of both energy and doping $\Delta\nu$. In addition, the $\vec{k}$-resolved spectral functions are displayed for the integer-filled phase and for states obtained by electron or hole doping. We focus here on low-temperature results; corresponding high-temperature data and analogous results for TSTG are provided in \cref{app:sec:results_corr_ins}. The color scheme encodes two pieces of information: the saturation reflects the spectral weight, while the hue (ranging from blue to orange) indicates the $f$- or $c$-character, determined by taking partial traces of the spectral function over the $f$- or $c$-electron indices, respectively.

Our results display several notable features. The order parameter of the $f$-electrons -- directly related to the $\mathrm{K}_M$-point gap between $f$-electron states above and below the Fermi level~\cite{SON22} -- is reduced by fluctuations compared to the Hartree-Fock predictions. For instance, at $\nu = 3$ the Hartree-Fock gap at $\mathrm{K}_M$ is $\Delta\omega \approx \SI{30}{\milli\electronvolt}$~\cite{SON22}, whereas in our simulations [see \cref{fig:sym_br_results:f}] it is reduced to $\Delta\omega \approx \SI{15}{\milli\electronvolt}$. This reduction in the single-particle gap becomes more pronounced for states further from charge neutrality; for example, the $\nu = 2$ state exhibits very small or even vanishing spectral gaps at zero strain, as illustrated in \cref{fig:sym_br_results:j}. Moreover, in the range $1 \leq \nu \leq 3$, the hole-excitation bands of the integer-filled correlated insulators are predominantly of $f$-electron character. As $\nu$ moves away from charge neutrality, these bands approach the Fermi level, making them more susceptible to charge fluctuations and thereby further suppressing the order parameter. Compared to the DMFT results for the symmetry-broken case~\cite{RAI23a}, the spectral functions obtained from second-order perturbation theory appear more coherent, as expected since perturbative approaches generally underestimate correlations. Moreover, the inclusion of the nearest-neighbor $f$-electron repulsion $H_{U_2}$ (neglected in Ref.~\cite{RAI23a}) produces a larger Hartree shift between the $f$- and $c$-electrons, thereby further enhancing the $f$-character of the hole excitations relative to the results of Ref.~\cite{RAI23a}.

As shown in \crefrange{fig:sym_br_results:r}{fig:sym_br_results:t}, for small doping ($\abs{\Delta\nu} \lesssim 0.1$) around $\nu = 0$, the band structure remains essentially the same as in the undoped case (apart from an overall energy shift). This is because the $f$-electrons lie far from the Fermi level, experience negligible fluctuations, and thus shift rigidly with doping. As the doping level increases, the $f$-electrons move closer to the Fermi energy, begin to fluctuate, and the $\mathrm{K}_M$ gap correspondingly shrinks. In this regime, the $f$-electron peaks in the spectral function become more incoherent. The DOS shown in \cref{fig:sym_br_results:p} closely matches STM measurements in ultra-low-strain devices~\cite{NUC23}, which display a gapped state without $\sqrt{3} \times \sqrt{3}$ graphene unit cell enlargement -- consistent with a $\IfStrEqCase{8}{{1}{\ket{\nu={}4} }
		{2}{\ket{\nu={}3, \mathrm{IVC}}}
		{3}{\ket{\nu={}3, \mathrm{VP}}}
		{4}{\ket{\nu={}2, \mathrm{K-IVC}}}
		{5}{\ket{\nu={}2, \mathrm{VP}}}
		{6}{\ket{\nu={}1, (\mathrm{K-IVC}+\mathrm{VP})}}
		{7}{\ket{\nu={}1, \mathrm{VP}}}
		{8}{\ket{\nu=0, \mathrm{K-IVC}}}
		{9}{\ket{\nu=0, \mathrm{VP}}}
	}
	[nada]
$ ground state~\cite{CAL22d}.

For small doping around $\nu = 1$ or $\nu = 2$, the second-order self-consistent calculation preserves the electron-light / hole-heavy dichotomy reported in Ref.~\cite{VAF20,BUL20a,KAN21}. In contrast to Hartree-Fock simulations, our results show that at $\nu = 2$ the light-fermion bands form a tiny Fermi surface near the $\Gamma_M$ point, as seen in \cref{fig:sym_br_results:j}; however, this pocket can be completely gapped by slightly adjusting the model parameters. Small electron doping of $\nu = 1$ or $\nu = 2$ fills $c$-fermions near $\Gamma_M$, leaving the $f$-electron order parameter essentially unaffected and producing only a rigid band shift. Larger electron doping, however, brings the empty $f$-electron states close to the Fermi level, where their enhanced fluctuations suppress the order parameter. 

In contrast, even small \emph{hole} doping of $\nu = 1$ or $\nu = 2$ pins the $f$-electrons close to the Fermi level, rendering their spectral weight incoherent and reducing the $\mathrm{K}_M$ gap. This behavior makes the $f$-electron peaks appear to ``move fast'' with doping when far from the Fermi level and ``move slowly'' when pinned near it. For the doped $\nu = 1$ correlated insulator, the DOS in \cref{fig:sym_br_results:m} exhibits a three-peaked structure near $\nu \approx 0.6$, signaling a phase transition in which the $f$-electron order parameter switches between distinct ordered states. 

For the $\nu = 3$ correlated state, we find no indirect gap but instead a $c$-electron Fermi surface with rigid-band-like behavior at small doping. Around $\abs{\Delta \nu} \sim 0.3$, an order-disorder transition occurs, driving the system into a gapless symmetric phase.

The $\nu = 4$ state is a band insulator and, as such, is largely unaffected by the inclusion of the second-order perturbative correction. The corresponding spectral function for the ordered state in TSTG is presented in \cref{app:sec:results_corr_ins}. We note that alternative parameter choices -- such as a smaller $U_1$ -- can, in principle, be made and yield qualitatively similar results~\cite{HU25}.

\section{The symmetric state of the THF model}\label{sec:symmetric}

At higher temperatures -- above approximately $\SIrange{5}{10}{\kelvin}$ but still below a scale set by $U_1$ -- TBG and TSTG are expected to be in a non-ordered (non-symmetry-broken) or \emph{symmetric} phase, in which interaction effects remain significant~\cite{HAU19,HOF22,HU23,DAT23,RAI23a,CAL25a,HU25,LED25a,LED25,ZHA25b,ZHA25c}. In this regime, dynamical self-energy contributions play a key role, influencing \emph{both} the lifetime \emph{and} the dispersion of charge-one excitations~\cite{RAI23a}. We now briefly examine the symmetric phase of the THF model, with a primary focus on TBG. Complete derivations, detailed formalism, and additional results (including those for TSTG) are provided in \cref{app:sec:se_symmetric,app:sec:se_symmetric_details,app:sec:results_symmetry}. 

We begin with a pedagogical review of the DMFT and Hartree-Fock framework as applied to the THF model~\cite{RAI23a}. We then introduce the IPT approximate impurity solver and its adaptation to the THF setting, and benchmark its performance against numerically exact DMFT results obtained with a QMC impurity solver. Finally, we compare the spectral functions from DMFT with IPT to those obtained within the Hubbard-I approximation~\cite{HUB63,HUB64}.

\subsection{DMFT and Hartree-Fock theory}\label{sec:symmetric:dmft_hf_theory}

To access the interacting symmetric phase of the THF model, we follow Ref.~\cite{RAI23a} and use a combination of DMFT and Hartree-Fock theory. Specifically, we treat the interactions in the THF TBG Hamiltonian between the $c$-electrons ($H_V$), between the $c$- and $f$-electrons ($H_W + H_J + H_{\tilde{J}} + H_K$), and the weak nearest-neighbor $f$-electron repulsion $H_{U_2}$ at the Hartree-Fock level, grouping them into a one-body symmetric term denoted by $h^{\text{MF} \prime} \left( \vec{k} \right)$. In contrast, the onsite $f$-electron repulsion $H_{U_1}$ is treated dynamically within DMFT. Under the DMFT approximation, the dynamical part of the self-energy is nonzero only for the $f$-electrons, originates solely from $H_{U_1}$, and is site-diagonal (momentum independent). In this combined DMFT and Hartree-Fock framework~\cite{RAI23a}, we solve the following Hamiltonian
\begin{widetext}
\begin{equation}
	\label{eqn:dmft_hf_ham}
	H^{\text{DMFT+HF}} =  \sum_{\substack{\vec{k}, i, \eta, s \\ i', \eta', s'}} h^{\text{MF} \prime}_{i \eta s; i' \eta' s'} \left( \vec{k} \right) \hat{\gamma}^\dagger_{\vec{k},i,\eta,s} \hat{\gamma}_{\vec{k},i',\eta',s'} + \frac{U_1}{2}  \sum_{\substack{\vec{R}, \alpha, \eta, s \\ \alpha', \eta', s'}} :\mathrel{ \hat{f}^\dagger_{\vec{R},\alpha, \eta, s} \hat{f}_{\vec{R},\alpha, \eta, s} }: :\mathrel{ \hat{f}^\dagger_{\vec{R},\alpha', \eta', s'} \hat{f}_{\vec{R},\alpha', \eta', s'}}:.
\end{equation}
\end{widetext}
The symmetries of the THF model -- moir\'e translation, $C_{6z}$ rotations, time reversal, and $\mathrm{SU} \left(2\right) \times \mathrm{SU} \left(2\right)$ spin-valley rotation~\cite{BER21a,CAL21,SON22} -- impose constraints on the Green's functions, spectral function, and dynamical self-energy in the symmetric phase, as detailed in \cref{app:sec:se_symmetric:definition:sym_constr}. In particular, these symmetries require the $f$-electron dynamical self-energy matrix to be proportional to the identity, $\Sigma^{f} \left( i \omega_n \right) = \tilde{\Sigma}^{f} \left( i \omega_n \right) \mathbb{1}$, where, throughout, diagonal entries of a matrix $M$ proportional to the identity are denoted by $\tilde{M}$. Since the dominant contribution to the $f$-electron Green's function is also site-diagonal~\cite{HU23i}, we neglect the off-site $f$-electron density matrix when computing $h^{\text{MF} \prime} \left( \vec{k} \right)$, which corresponds to dropping the Fock contribution from $H_{U_2}$. Consequently, the $f$-electron block of $h^{\text{MF} \prime} \left( \vec{k} \right)$ is $\vec{k}$-independent and proportional to the identity, making the full $f$-electron self-energy -- both the static and dynamic parts -- $\vec{k}$-independent and proportional to the identity.

The same symmetries also require that the site-diagonal part of the $f$-electron Green's function be proportional to the identity. For convenience, we denote the interacting $f$-electron Green's function by
{\small \begin{equation}
	\label{eqn:gf_f_lattice}
	-\left\langle \mathcal{T}_{\tau} \hat{f}_{\vec{R},\alpha, \eta, s} \left( \tau \right) \hat{f}^\dagger_{\vec{0},\alpha', \eta', s'} \left( 0 \right)  \right\rangle  = \mathcal{G}^{f}_{\alpha \eta s; \alpha' \eta' s'} \left(\tau, \vec{R} \right),
\end{equation}}such that, in the symmetric phase,
\begin{equation}
	\eval{ \mathcal{G}^{f} \left(i \omega_n, \vec{R} \right)}_{\vec{R} = \vec{0}} = \tilde{\mathcal{G}}^{f} \left(i \omega_n \right) \mathbb{1},
\end{equation}
where $\mathcal{G}^{f} \left(i \omega_n \right)$ is the site-diagonal $f$-electron lattice Green's function. We further define the relative $f$-electron filling as
{\small \begin{equation}
	\label{eqn:def_n}
	n \equiv \frac{\nu_f + 4}{N_f} =  -\frac{1}{\pi} \int_{-\infty}^{\infty} \dd{\omega} n_{\mathrm{F}} \left( \omega \right) \Im \left( \tilde{\mathcal{G}}^{f} \left( \omega + i 0^{+} \right) \right).
\end{equation}}

\subsubsection{Brief review of DMFT as applied to the THF model}\label{sec:symmetric:dmft_hf_theory:dmft_review}

\Cref{app:sec:se_symmetric:DMFT_overview} reviews the application of DMFT to our multi-orbital model. Following the standard approach~\cite{GEO96}, we single out the $f$-electrons at a given site, $\hat{f}^\dagger_{\alpha, \eta, s} \equiv \hat{f}^\dagger_{\vec{R}_0,\alpha, \eta, s}$, and decompose the system's action into three parts: the contribution from $\hat{f}^\dagger_{\alpha, \eta, s}$ ($S^{\text{bare}}_{\text{ss}}$), the contribution from all other fermions -- termed the ``cavity'' ($S_{\text{cavity}}$) -- and a coupling term between $\hat{f}^\dagger_{\alpha, \eta, s}$ and the remaining fermionic degrees of freedom ($\Delta S$). In our Hartree-Fock treatment of all interactions except $H_{U_1}$, $\Delta S$ reduces to a one-body hybridization term, which can be interpreted as a source term coupling to $\hat{f}^\dagger_{\alpha, \eta, s}$ in $S^{\text{bare}}_{\text{ss}}$. Integrating out the cavity degrees of freedom~\cite{GEO96} while retaining only the quadratic term in $\hat{f}^\dagger_{\alpha, \eta, s}$ yields the single-site action
\begin{widetext}
{\small\begin{equation}
	\label{eqn:single_site_action}
	S_{\text{ss}} = \int_{0}^{\beta} \dd{\tau} \left( -\int_{0}^{\beta} \dd{\tau'} \sum_{\alpha, \eta, s} \hat{f}^\dagger_{\alpha, \eta, s} \left( \tau \right) G^{-1}_0 \left( \tau - \tau' \right) \hat{f}_{\alpha, \eta, s} \left( \tau' \right) +  \frac{U_1}{2} \hspace{-0.5em} \sum_{\substack{\alpha, \eta, s \\ \alpha', \eta', s'}} :\mathrel{\hat{f}^\dagger_{\alpha, \eta, s} \left( \tau \right) \hat{f}_{\alpha, \eta, s} \left( \tau \right)}: :\mathrel{\hat{f}^\dagger_{\alpha', \eta', s'}\left( \tau \right)  \hat{f}_{\alpha', \eta', s'} \left( \tau \right)}: \right).
\end{equation}}\end{widetext}
The first term in \cref{eqn:single_site_action} represents the ``kinetic'' part of the single-site action, with $G_0(\tau)$ denoting the non-interacting Green's function of the single-site problem, while the second term is simply the onsite Hubbard repulsion at site $\vec{R}_0$. The diagonal form of the first term follows from the symmetries of the THF model. In general, $G_0(\tau)$ is a complicated function whose exact expression requires knowledge of the full solution of the entire lattice problem ({\it i.e.}{}, single site plus cavity). For a given $G_0(\tau)$, the single-site action can be solved -- for example, using QMC -- to obtain the interacting single-site Green's function, 
\begin{equation}
	\label{eqn:definition_ss_interacting_gf}
	G \left(\tau \right) = - \left\langle \mathcal{T}_{\tau} \hat{f}_{\alpha, \eta, s} \left( \tau \right) \hat{f}^\dagger_{\alpha, \eta, s} \left( 0 \right) \right\rangle^{\text{ss}},
\end{equation}
and, from it, the dynamical self-energy of the single-site problem, $ \Sigma_{\text{ss}} \left( i \omega_n \right) = G_0^{-1} \left( i \omega_n \right) - G^{-1} \left( i \omega_n \right)- U_1 \left( n_{\text{ss}} - \frac{1}{2} \right) \left( N_f -1 \right)$, where the last term denotes the static Hartree-Fock contribution. Here, $\left\langle \dots \right\rangle^{\text{ss}}$ indicates that expectations are computed with respect to the single-site action in \cref{eqn:single_site_action}, and $n_{\text{ss}} \equiv \left\langle \hat{f}^\dagger_{\alpha, \eta, s} \hat{f}_{\alpha, \eta, s} \right\rangle^{\text{ss}}$.

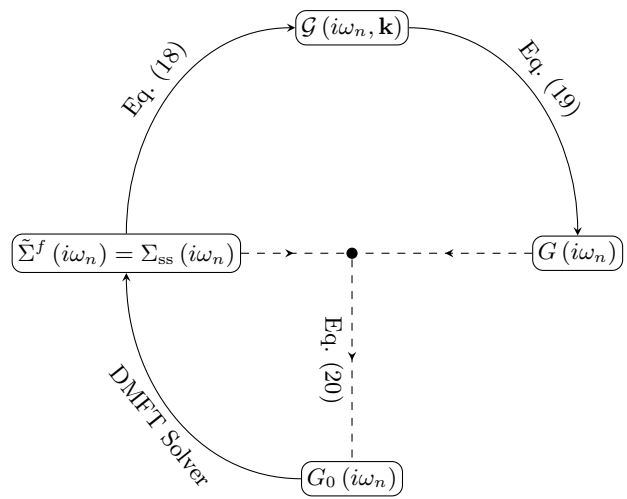
\begin{figure}[!t]
	\begin{tikzpicture}[
		>=stealth,
		boxed/.style={draw, rounded corners, inner sep=2pt},
		scale=0.75,
		decoration={markings, mark=at position 0.5 with {\arrow{>}}}
		]
		
		\node (A) at (0,0) [boxed] {$\tilde{\Sigma}^{f} \left( i \omega_n \right) = \Sigma_{\text{ss}} \left( i \omega_n \right)$};
		\node (B) at (4,4) [boxed] {$\mathcal{G} \left( i \omega_n, \vec{k} \right)$};
		\node (C) at (8,0) [boxed] {$G \left( i \omega_n\right)$};
		\node (D) at (4,-4) [boxed] {$G_0 \left( i \omega_n\right)$};
		\node (ct) at (4,0) [circle,fill,inner sep=1.5pt] {};
		
		\draw[->] (A) to[out=90,in=180] node[midway, above, sloped] {\cref*{eqn:DMFT_GF_symmetric}} (B);
		\draw[->] (B) to[out=0,in=90] node[midway, above, sloped] {\cref*{eqn:DMFT_self_consistent_eq}} (C);
		\draw[->] (D) to[out=180,in=270] node[midway, below, sloped] {DMFT Solver} (A);
		\draw[dashed,postaction={decorate}] (A) -- (ct); 
		\draw[dashed,postaction={decorate}] (C) -- (ct); 
		\draw[dashed,postaction={decorate}] (ct) -- node[midway, below, sloped] {\cref*{eqn:DMFT_G0_expression}} (D); 
	\end{tikzpicture}
	\caption{The DMFT self-consistent problem. The $f$-electron self-energy is computed in the single-site model from the non-interacting Green's function of the latter, $G_{0} \left( i \omega_n \right)$. The lattice problem is ascribed the same $f$-electron self-energy as the single-site one. At the same time, the non-interacting Green's function of the single-site model $G_{0} \left( i \omega_n \right)$ is obtained from the \emph{interacting} Green's function of the lattice model $\mathcal{G} \left( i \omega_n, \vec{k} \right)$ and the $f$-electron self-energy. As a result, $G_{0} \left( i \omega_n \right)$ and $\tilde{\Sigma}^{f} \left( i \omega_n \right)$ are two \emph{interdependent} quantities which need to be determined self-consistently.}
	\label{fig:self_consistent_DMFT}
\end{figure}

The first key DMFT assumption is that the (assumed site-diagonal) self-energy of the \emph{entire} lattice problem is equal to that of the single-site one
\begin{equation}
	\label{eqn:DMFT_assumption_1}
	\tilde{\Sigma}^{f} \left( i \omega_n \right) = \Sigma_{\text{ss}} \left( i \omega_n \right) \qq{and} n = n_{\text{ss}}.
\end{equation}
Under this assumption, the interacting Green's function of the full lattice is given by
\begin{align}
	\mathcal{G}^{-1} \left( i \omega_n , \vec{k} \right) = &\left( i \omega_n + \mu \right) \mathbb{1} - h^{\text{MF} \prime} \left( \vec{k} \right) - \tilde{\Sigma}^f \left( i \omega_n \right) \mathbb{1}_f \nonumber \\
	&- U_1 \left( n_{\text{ss}} - \frac{1}{2} \right) \left( N_f -1 \right) \mathbb{1}_f, \label{eqn:DMFT_GF_symmetric}
\end{align}
where $\mathbb{1}_f$ is the identity projector onto the $f$-electron block and $N_f = 8$ is the number of $f$-electron flavors. Nevertheless, one still needs to determine the complicated function $G_0(\tau)$. The second DMFT assumption, which enables this, requires that the site-diagonal interacting Green's function of the system defined in \cref{eqn:gf_f_lattice} equals the interacting single-site Green's function~\cite{GEO96}
\begin{equation}
	\label{eqn:DMFT_self_consistent_eq}
	G \left( i \omega_n \right) = \tilde{\mathcal{G}}^{f} \left( i \omega_n \right).
\end{equation}
This relation allows us to obtain $G_0 \left( i \omega_n \right)$ via
{\small \begin{equation}
	\label{eqn:DMFT_G0_expression}
	\frac{1}{G_{0} \left(i \omega_n \right)} = \frac{1}{\tilde{\mathcal{G}}^{f} \left( i \omega_n \right)} + U_1 \left( n - \frac{1}{2} \right) \left( N_f - 1 \right) + \tilde{\Sigma}^{f} \left( i \omega_n \right)
\end{equation}}Consequently, solving the THF model in the symmetric phase with DMFT becomes a self-consistent problem: $G_{0} \left( i \omega_n \right)$ and $\Sigma^f \left( i \omega_n \right)$ are mutually dependent, as illustrated schematically in \cref{fig:self_consistent_DMFT}. 

Moreover, unlike in the original Anderson model, our model includes Hartree-Fock contributions from interactions other than $H_{U_1}$. Specifically, the $h^{\text{MF} \prime} \left( \vec{k} \right)$ Hamiltonian from \cref{eqn:dmft_hf_ham} depends on the lattice density matrix $\varrho \left( \vec{k} \right)$, which in turn is determined by the interacting lattice Green's function from \cref{eqn:DMFT_GF_symmetric}. Consequently, achieving self-consistency requires simultaneously varying $\varrho \left( \vec{k} \right)$, $G_0 \left( \tau \right)$, and $\Sigma^f \left( i \omega_n \right)$. Further details are provided in \cref{app:sec:se_symmetric}.

\subsubsection{The IPT impurity solver}\label{sec:symmetric:dmft_hf_theory:mipt}

In principle, the single-site action from \cref{eqn:single_site_action} can be solved using either QMC~\cite{HAU19,HU23,DAT23,RAI23a,CAL25a} or NRG~\cite{ZHO24,WAN24,YOU24} methods, each with its own advantages and drawbacks. On the one hand, QMC impurity solvers are numerically exact but computationally expensive, and operate in imaginary time, which often necessitates analytical continuation from Matsubara to real frequencies to extract physical observables such as the spectral function. On the other hand, NRG solvers work directly in the real-frequency domain, but are practically restricted to $N_f \leq 4$ fermionic flavors per site. As a result, solving the THF model with NRG typically requires artificially breaking system symmetries~\cite{ZHO24,WAN24,YOU24}, for instance by freezing certain $f$-electron flavors.

In this work, we take a different approach by adapting the IPT impurity solver~\cite{MAR86,GEO92,YEY93,KAJ96,POT97,ANI97,LIC98,YEY99,MEY99,YEY00,SAS01,SAV01,FUJ03,LAA03,KUS06,ARS12,DAS16,WAG21,MIZ21,VAN22,CAN24,CAN25} to solve the THF model. Although approximative in nature, IPT is significantly less computationally demanding than both QMC and NRG, is not limited by the number of fermionic flavors, and operates directly in the real-frequency domain, thus avoiding the numerically ill-posed problem of analytical continuation.

The core idea of the IPT impurity solver is to solve the single-site problem from \cref{eqn:single_site_action} to second order in perturbation theory, using the non-dressed fermionic propagator $G_{0} \left( i\omega_n \right)$. At first sight, such an approach may appear unsuitable for the strongly correlated THF model, where a perturbative expansion in $U_1$ would generally be unreliable. Remarkably, however, while the second-order approximation is strictly valid in the \emph{weakly interacting} limit ({\it i.e.}{}, $U_1 \to 0$), it also becomes \emph{exact} in the atomic limit ({\it i.e.}{}, $U_1 \to \infty$) for a single-orbital spinful Hubbard model ({\it i.e.}{}, $N_f = 2$) at particle-hole symmetric filling ({\it i.e.}{}, half-filling)~\cite{GEO92,ROZ94,ROZ95}. Away from particle-hole symmetry in a single-orbital Hubbard model, the second-order treatment is only strictly valid in the \emph{weakly interacting} limit, but \emph{not} in the \emph{strongly interacting} or \emph{atomic} regime. Nevertheless, as shown in \cref{app:sec:se_symmetric_details:IPT}, the second-order self-energy of \cref{eqn:single_site_action} in the atomic limit has a functional form closely resembling that of the \emph{exact} atomic-limit dynamical self-energy derived in \cref{app:sec:se_symmetric_details:atomic_se}. The IPT method exploits this observation by constructing a self-energy ansatz that \emph{interpolates} between the numerically computed second-order self-energy of the single-site model, obtained from $G_{0} \left(i \omega_n \right)$, and the analytically derived atomic-limit self-energy. The validity of this interpolation is assessed in \cref{sec:symmetric:benchmark} by benchmarking against results from a QMC impurity solver.

The implementation details of the IPT method~\cite{MAR86,GEO92,YEY93,KAJ96,POT97,ANI97,LIC98,YEY99,MEY99,YEY00,SAS01,SAV01,FUJ03,LAA03,KUS06,ARS12,DAS16,WAG21,MIZ21,VAN22,CAN24,CAN25}, as well as rigorous proofs of the identities presented here, are given in \cref{app:sec:se_symmetric,app:sec:se_symmetric_details}. Here, we only outline the key steps. We begin by introducing a variational parameter $\tilde{\mu}$, referred to as the \emph{fictitious} chemical potential, together with a modified non-interacting single-site Green's function
\begin{equation}
	\frac{1}{G^{\tilde{\mu}}_0 \left( \omega^+ \right)}=\frac{1}{G_0 \left( \omega^+ \right)} - \mu + \tilde{\mu},
\end{equation}
where, as in what follows, the shorthand notation $\omega^+ = \omega + i 0^{+}$ emphasizes that the IPT solution is obtained directly in real frequency. The parameter $\tilde{\mu}$, whose determination will be discussed later, is introduced to improve the agreement between the IPT dynamical self-energy and the exact result. Using $G^{\tilde{\mu}}_0 \left( \omega^+ \right)$, the single-site problem from \cref{eqn:single_site_action} is solved to second order in perturbation theory, yielding the following dynamical self-energy
\begin{widetext}
{\small \begin{align}
	\frac{\tilde{\Sigma}^{f,(2)} \left(\omega^+ \right)}{U_1^2 \left(N_f - 1 \right)} =&  \int_{-\infty}^{\infty} \int_{-\infty}^{\infty} \int_{-\infty}^{\infty} \prod_{i=1}^3 \left( \rho^{\tilde{\mu}}_0 \left(\omega_i \right) \dd{\omega_i} \right) \frac{n_{\mathrm{F}} \left( \omega_{1} \right) \left( 1 - n_{\mathrm{F}} \left( \omega_{2} \right) \right) n_{\mathrm{F}} \left( \omega_{3} \right) + \left( 1 - n_{\mathrm{F}} \left( \omega_{1} \right) \right) n_{\mathrm{F}} \left( \omega_{2} \right) \left( 1 - n_{\mathrm{F}} \left( \omega_{3} \right) \right)}{\omega^+ - \omega_1 + \omega_2 - \omega_3},
	\label{eqn:second_order_f_symmetric}
\end{align}}
\end{widetext}
with $\rho^{\tilde{\mu}}_0 \left(\omega \right) = -\frac{1}{\pi}\Im G^{\tilde{\mu}}_0 \left( \omega^+ \right)$ being the spectral function of $G^{\tilde{\mu}}_0 \left( \omega^+ \right)$. In IPT, $\tilde{\Sigma}^{f,(2)} \left(\omega^+ \right)$ is not used directly. Instead, the dynamical self-energy is approximated by the following \emph{interpolated} ansatz~\cite{MAR86,GEO92,YEY93,KAJ96,POT97,ANI97,LIC98,YEY99,MEY99,YEY00,SAS01,SAV01,FUJ03,LAA03,KUS06,ARS12,DAS16,WAG21,MIZ21,VAN22,CAN24,CAN25}
\begin{equation}
	\label{eqn:interpolated_sigma}
	\tilde{\Sigma}_{\text{ss}} \left( \omega^+  \right) \approx \tilde{\Sigma}^{f,\text{Int}} \left( \omega^+  \right) = \frac{a \tilde{\Sigma}^{f,(2)}\left( \omega^+ \right)}{1 - b \left( \omega^+ \right) \tilde{\Sigma}^{f,(2)}\left( \omega^+ \right)}.
\end{equation}
In \cref{eqn:interpolated_sigma}, the constant $a$ is fixed to ensure that $\tilde{\Sigma}^{f,\text{Int}} \left( \omega^+  \right)$ satisfies the correct $\omega \to \infty$ behavior, shown in \cref{app:sec:se_symmetric_details:se_exact} to be equivalent to reproducing the first two nontrivial moments of the single-site interacting spectral function. The constant $a$ is given by
\begin{equation}
	\label{eqn:IPT_a_constant}
	a = \frac{ n + \left(N_f - 2\right) \left\langle nn \right\rangle^{\text{ss}}  - \left(N_f -1 \right) n^2}{n^{\tilde{\mu}}_0 \left(1 - n^{\tilde{\mu}}_0 \right)},
\end{equation}
where $n$ was defined in \cref{eqn:def_n}, $n_0^{\tilde{\mu}} = \int_{-\infty}^{\infty} \dd{\omega} \rho^{\tilde{\mu}}_0 \left(\omega \right)$ is the filling of the non-interacting single-site problem at $\tilde{\mu}$, and $\left\langle nn \right\rangle^{\text{ss}}$ denotes the double occupation of the single site
\begin{equation}
	\left\langle nn \right\rangle^{\text{ss}} = \left\langle \hat{f}^\dagger_{\alpha, \eta, s} \hat{f}_{\alpha, \eta, s} \hat{f}^\dagger_{\alpha', \eta', s'} \hat{f}_{\alpha', \eta', s'} \right\rangle^{\text{ss}},  
\end{equation}
for $\left(\alpha, \eta, s\right) \neq \left(\alpha', \eta', s'\right)$. We note that $\left\langle nn \right\rangle^{\text{ss}}$ can also be expressed in terms of the interacting single-site Green's function and dynamical self-energy, as derived in \cref{app:sec:se_symmetric_details:se_exact}.

The function $b\left( \omega^+ \right)$, which admits an analytical (though lengthy and not particularly illuminating) expression, is chosen so that $\tilde{\Sigma}^{f,\text{Int}} \left( \omega^+  \right)$ reproduces the exact analytical dynamical self-energy in the atomic limit. Specifically,
\begin{equation}
	\label{eqn:b_function_ipt}
	b \left( \omega^+  \right) = \frac{1}{\tilde{\Sigma}^{f,(2),\text{At}} \left( \omega^+ \right)} - \frac{a^{\text{At}}}{\tilde{\Sigma}^{\text{At}} \left( \omega^+ \right)}.
\end{equation}
In \cref{eqn:b_function_ipt}, $\tilde{\Sigma}^{f,(2),\text{At}} \left( \omega^+ \right)$ is the second-order self-energy from \cref{eqn:second_order_f_symmetric} evaluated in the atomic limit, $\tilde{\Sigma}^{\text{At}} \left( \omega^+ \right)$ is the \emph{exact} atomic-limit dynamical self-energy at chemical potential $\mu$ (derived in \cref{app:sec:se_symmetric_details:atomic_se}), and $a^{\text{At}}$ is the constant $a$ from \cref{eqn:IPT_a_constant} evaluated in the atomic limit. The resulting $b \left( \omega^+ \right)$ can be expressed in closed analytical form in terms of $n$, $\mu$, $n_0$, and $\tilde{\mu}$. 

We note that \cref{eqn:interpolated_sigma} automatically reduces to the correct second-order perturbative result $\tilde{\Sigma}^{f,(2)}$ in the weak-coupling limit $U_1 \to 0$. Since it interpolates between the weakly and strongly correlated limits while preserving the correct $\omega \to \infty$ behavior, the ansatz is expected to perform well across a range of interaction strengths, including the intermediate-coupling regime.

Finally, we address the determination of the fictitious chemical potential $\tilde{\mu}$. This parameter is introduced to ensure that the IPT solution satisfies Luttinger's theorem~\cite{KAJ96, DAS16, FUJ03, SAS01, LIC98, ANI97, POT97, LAA03, YEY99, YEY00}. While this prescription is strictly valid only at zero temperature, in this work we fix $\tilde{\mu}$ by requiring $n^{\tilde{\mu}}_0 = n$~\cite{MAR86, YEY93, POT97, MEY99, YEY00, ARS12}. This approach can be applied at finite temperature, has been validated against exact diagonalization, and \emph{approximately} satisfies Luttinger's theorem in the zero-temperature limit.

\subsubsection{Numerical implementation }\label{sec:symmetric:dmft_hf_theory:num_impl}

\begin{figure*}[!t]\includegraphics[width=\textwidth]{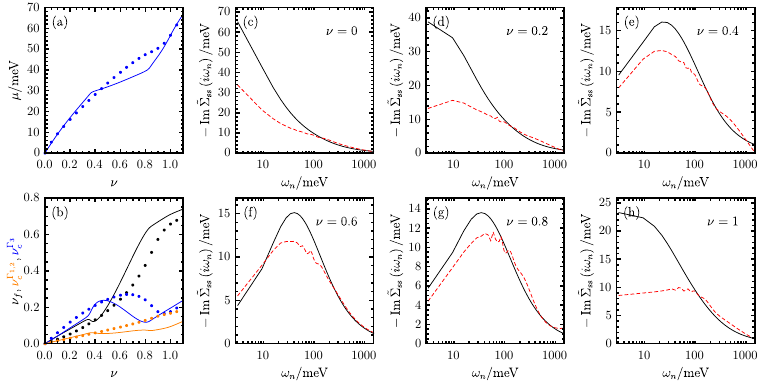}\subfloat{\label{fig:IPT_vs_QMC:a}}\subfloat{\label{fig:IPT_vs_QMC:b}}\subfloat{\label{fig:IPT_vs_QMC:c}}\subfloat{\label{fig:IPT_vs_QMC:d}}\subfloat{\label{fig:IPT_vs_QMC:e}}\subfloat{\label{fig:IPT_vs_QMC:f}}\subfloat{\label{fig:IPT_vs_QMC:g}}\subfloat{\label{fig:IPT_vs_QMC:h}}\caption{Benchmarking the IPT impurity solver against QMC results for fillings $0 \leq \nu \leq 1$. (a) shows the chemical potential as a function of filling, with IPT results given by the blue line and QMC results by the blue dots. (b) compares the filling of the $f$-electrons, the $\Gamma_1 \oplus \Gamma_2$ $c$-electrons, and the $\Gamma_3$ $c$-electrons as a function of the total filling, where lines correspond to IPT and dots to QMC. (c)-(h) present the Matsubara self-energy from QMC (red dashed lines) and from IPT (black line), with the filling indicated in each panel. All calculations are performed at $\beta^{-1} = \SI{1}{\milli\electronvolt}$.}\label{fig:IPT_vs_QMC}\end{figure*}

As outlined in \cref{sec:symmetric:dmft_hf_theory:mipt}, solving the single-site action with the IPT method requires satisfying a set of coupled nonlinear integral equations involving the single-site Green's functions $G\left( \omega^+\right)$ and $G_0\left( \omega^+\right)$, the corresponding second-order and interpolated self-energies $\tilde{\Sigma}^{f,\text{Int}} \left( \omega^+  \right)$ and $\tilde{\Sigma}^{f,(2)}\left( \omega^+ \right)$, the real and fictitious chemical potentials $\mu$ and $\tilde{\mu}$, the fillings $n$ and $n_0^{\tilde{\mu}}$, the density matrix $\varrho \left( \vec{k} \right)$, and the double occupation $\left\langle nn \right\rangle^{\text{ss}}$. The self-consistent solution at a given filling $\nu$ is obtained iteratively, with each step updating all quantities based on the current estimates.  

The setup of the self-consistent problem, along with the numerical algorithm used to compute the symmetric-state spectral functions and self-energy, is described in \cref{app:sec:se_symmetric:IPT:sc_and_numerical}. At a given filling, the symmetric solution is fully specified by the density matrix and the $\vec{k}$-independent dynamical self-energy. These determine all other quantities required in the IPT procedure, which in turn yield improved estimates for $\varrho \left( \vec{k} \right)$ and the $f$-electron dynamical self-energy. Since the IPT calculation is performed in real frequency, we include, as in \cref{eqn:total_self_sym_br}, a small broadening factor $\Gamma_c$ for the $c$-electrons.  

The iteration continues until the convergence criteria in \cref{app:sec:se_correction_beyond_HF:sc_problem_and_numerics} are met. The initial conditions depend on the target filling $\nu$. At charge neutrality ($\nu=0$), we start from $\varrho \left( \vec{k} \right) = \mathbb{0}$ and $\tilde{\Sigma}_{\text{ss}} \left( \omega^+ \right) = -i \Gamma_c$. The $\nu=0$ state is then incrementally doped in steps of $\delta \nu = \frac{1}{50}$ to obtain solutions across a broad range of dopings.

\DeclareSIUnit{\core}{CPU\,core}

\subsection{Benchmarking the IPT impurity solver}\label{sec:symmetric:benchmark}

\begin{figure*}[!t]\includegraphics[width=\textwidth]{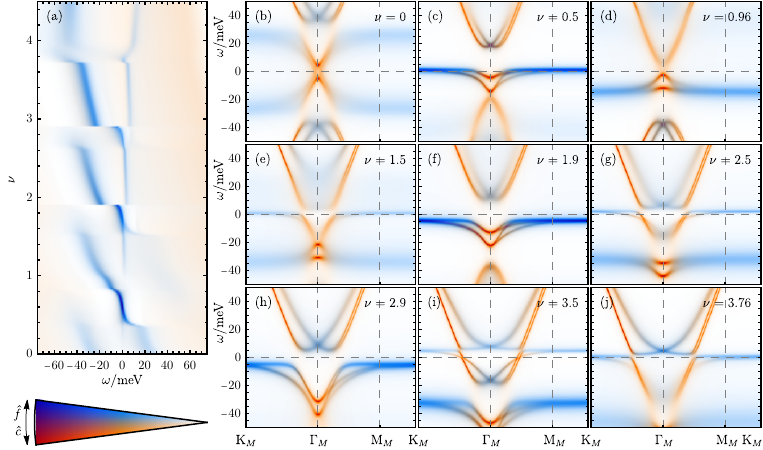}\subfloat{\label{fig:sym_bs:a}}\subfloat{\label{fig:sym_bs:b}}\subfloat{\label{fig:sym_bs:c}}\subfloat{\label{fig:sym_bs:d}}\subfloat{\label{fig:sym_bs:e}}\subfloat{\label{fig:sym_bs:f}}\subfloat{\label{fig:sym_bs:g}}\subfloat{\label{fig:sym_bs:h}}\subfloat{\label{fig:sym_bs:i}}\subfloat{\label{fig:sym_bs:j}}\caption{Spectral functions of TBG within the THF model in the symmetric phase, computed using the IPT impurity solver at $T=\SI{10}{\kelvin}$. (a) shows the DOS $\mathcal{A}(\omega)$ as a function of filling $\nu$ and energy $\omega$, while panels (b)-(j) present the momentum-resolved spectral function $\mathcal{A}(\omega, \vec{k})$ at the fillings indicated in each panel. The same color scale as in \cref{fig:sym_br_results} is used throughout.}    \label{fig:sym_bs}\end{figure*}

The IPT method for solving the single-site problem is inherently approximate, so benchmarking against numerically exact methods is essential. To this end, we performed simulations using a QMC impurity solver, as described in \cref{app:sec:se_symmetric:QMC}. For simplicity, and in contrast to \cref{eqn:dmft_hf_ham}, the QMC simulations omit the $H_{\tilde{J}}$ and $H_{K}$ interactions, as well as the Fock term from $H_{V}$ (all of which are otherwise included in our IPT simulations). To enable a direct comparison, we also drop these terms in our IPT simulations for this section only.

\Cref{fig:IPT_vs_QMC} compares the DMFT solutions obtained with our IPT solver to those from the numerically exact QMC solver for $0 \leq \nu \leq 1$. The chemical potential in \cref{fig:IPT_vs_QMC:a} shows good quantitative agreement between the two methods, except perhaps for $0.6 \lesssim \nu \lesssim 0.9$, where deviations of up to $\SI{10}{\milli\electronvolt}$ occur. We also compare the individual fillings of the $f$-electrons, the $\Gamma_1 \oplus \Gamma_2$ $c$-electrons, and the $\Gamma_3$ $c$-electrons, defined respectively as
\begin{align}
	\nu_f &= \frac{1}{N_0} \sum_{\vec{k}} \sum_{i=5}^6 \varrho_{i \eta s;i \eta s} \left( \vec{k} \right), \nonumber \\
	\nu^{\Gamma_{1,2}}_c &= \frac{1}{N_0} \sum_{\vec{k}} \sum_{i=3}^4 \varrho_{i \eta s;i \eta s} \left( \vec{k} \right), \\
	\nu^{\Gamma_3}_c &= \frac{1}{N_0} \sum_{\vec{k}} \sum_{i=1}^2 \varrho_{i \eta s;i \eta s} \left( \vec{k} \right). \nonumber 
\end{align}
\Cref{fig:IPT_vs_QMC:b} indicates overall good qualitative agreement, with quantitative agreement away from $0.6 \lesssim \nu \lesssim 0.9$. Deviations in the individual fillings are no larger than $0.1$. The deviations in both the fillings and the chemical potential may stem from the fact that IPT only approximately satisfies Luttinger’s theorem, as explained in \cref{sec:symmetric:dmft_hf_theory:mipt}. As will be shown in \cref{sec:symmetric:results}, these fillings lie in the intermediate-valence regime, where a pronounced $f$-electron peak forms near the Fermi level. The associated strong charge fluctuations could make the system particularly sensitive to this approximate treatment of Luttinger’s theorem, thereby amplifying the discrepancies relative to QMC.

Finally, we compare the dynamical self-energies at different fillings in \crefrange{fig:IPT_vs_QMC:c}{fig:IPT_vs_QMC:h}. To avoid uncertainties from analytical continuation of QMC data from Matsubara to real frequencies, we instead compare directly in Matsubara frequency. The IPT results are analytically continued to Matsubara frequencies via the spectral representation of the self-energy,
\begin{equation}
	\tilde{\Sigma}_{\text{ss}} \left( i \omega_n \right) = -\frac{1}{\pi} \int_{-\infty}^{\infty} \dd{\omega} \frac{\Im \tilde{\Sigma}_{\text{ss}} \left( \omega + i 0^{+} \right)}{i \omega_n - \omega}.
\end{equation}
Following Ref.~\cite{DAS16}, we compare $\Im \tilde{\Sigma}_{\text{ss}} \left( i \omega_n \right)$ between the two solvers. The large-frequency tails are accurately reproduced by IPT across all fillings considered. Near integer fillings, as in \cref{fig:IPT_vs_QMC:c,fig:IPT_vs_QMC:d,fig:IPT_vs_QMC:h}, IPT shows a stronger tendency toward Mott-like behavior: at low $\omega_n$, $\Im \tilde{\Sigma} \left( i \omega_n \right)$ has a larger magnitude than in QMC. This arises because, at integer fillings and large $U_1$, the IPT self-energy reduces (by design) to the atomic one, which diverges at low frequencies. Away from integer fillings, the IPT and QMC results are quantitatively consistent. Combined with the fact that each IPT simulation requires only \SI{8}{\core\minute} per filling -- compared to roughly \SI{1000}{\core\hour} for QMC -- this establishes IPT as a practical tool for broad explorations of the THF parameter space. Once such a scan is performed, various regions of the phase diagram can then be selected for more detailed DMFT studies with numerically exact impurity solvers.

\subsection{Results}\label{sec:symmetric:results}

\Cref{fig:sym_bs} presents the spectral function of TBG within the THF model, obtained using the IPT solver at a fixed temperature $T = \SI{10}{\kelvin}$. Results for other temperatures, as well as for TSTG, are provided in \cref{app:sec:results_symmetry}.

\begin{figure*}[t]\includegraphics[width=\textwidth]{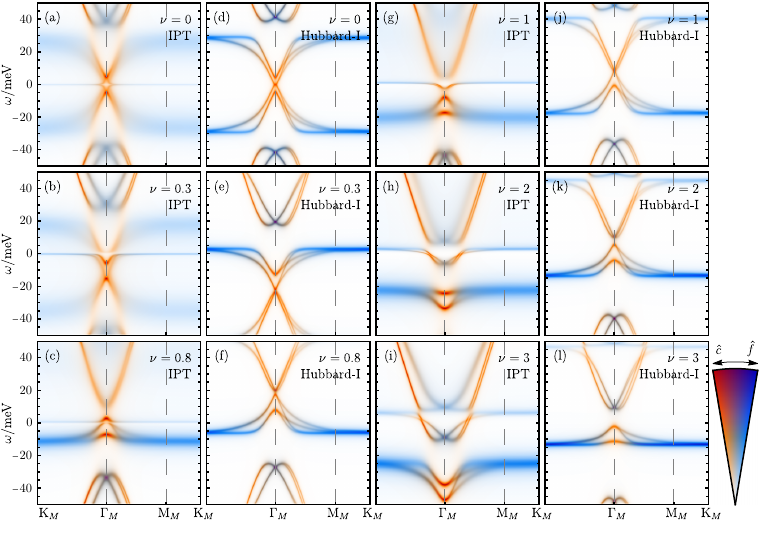}\subfloat{\label{fig:hubbard_i:a}}\subfloat{\label{fig:hubbard_i:b}}\subfloat{\label{fig:hubbard_i:c}}\subfloat{\label{fig:hubbard_i:d}}\subfloat{\label{fig:hubbard_i:e}}\subfloat{\label{fig:hubbard_i:f}}\subfloat{\label{fig:hubbard_i:g}}\subfloat{\label{fig:hubbard_i:h}}\subfloat{\label{fig:hubbard_i:i}}\subfloat{\label{fig:hubbard_i:j}}\subfloat{\label{fig:hubbard_i:k}}\subfloat{\label{fig:hubbard_i:l}}\caption{Spectral function of TBG obtained using DMFT with the IPT impurity solver and the Hubbard-I approximation. Results are shown for six integer and non-integer fillings, as indicated in each panel, at $T = \SI{3}{\kelvin}$. The approximation used to compute each spectral function is also indicated within the panels: the first and third columns correspond to IPT, while the second and fourth columns correspond to the Hubbard-I approximation.}\label{fig:hubbard_i}\end{figure*}

The dominant feature of the DOS shown in \cref{fig:sym_bs:a} is the emergence of Hubbard bands formed by the $f$-electrons. As the filling is varied, the energy of these bands exhibits a cascade of transitions. These transitions originate from changes between distinct $f$-electron occupation states and can be understood within the zero-hybridization limit of the model~\cite{HU23i}. As explained in \cref{app:sec:results_symmetry}, the Hubbard bands persist up to higher temperatures. A Kondo resonance (or zero bias-peak) can be clearly seen away from integer fillings (where they appear even at larger temperatures $T=\SI{20}{\kelvin}$). Combining the Hubbard bands and the Kondo peaks leads to a tail-feather feature of the spectral functions, which is also broadly consistent with various spectroscopy experiments~\cite{WON20,CHO21,KIM22,NUC23,ZHA25a,XIA25}. 

Near integer fillings, \cref{fig:sym_bs:b,fig:sym_bs:d,fig:sym_bs:f,fig:sym_bs:h,fig:sym_bs:j} reveal gap openings, or at least a suppression of spectral weight near the Fermi level. Due to the approximate nature of the IPT impurity solver (in particular its only approximate enforcement of Luttinger’s theorem), these gaps do not appear exactly \emph{at} integer fillings, unlike in the numerically exact QMC results~\cite{DAT23,RAI23a,CAL25a}. We also note that at charge neutrality ($\nu = 0$), the $f$-electron Hubbard bands are more widely separated in the IPT results compared to QMC~\cite{RAI23a}, reflecting the stronger tendency toward Mott-like behavior at integer fillings in the IPT method, as discussed in \cref{sec:symmetric:benchmark}. Finally, the apparent discontinuities in the spectral function around $\nu \approx 1.9$ and $\nu \approx 2.9$ arise from the solution changing rapidly with filling; these features vanish at higher temperatures, as shown in \cref{app:sec:results_symmetry}.

\subsection{Hubbard-I Approximation}\label{sec:symmetric:hubbard_i}

To gain analytical insight into the gapped phases occurring at nonzero integer fillings, as well as the semi-metallic phase at $\nu = 0$, we employ the Hubbard-I approximation for the $f$-electron self-energy~\cite{HUB63,HUB64,HU25,LED25,LED25a}. As detailed in \cref{app:sec:results_symmetry:hubbard_I}, this approach approximates the dynamical self-energy of the system by its atomic-limit form, derived analytically in \cref{app:sec:se_symmetric_details:atomic_se}. This allows one to compute the interacting Green's function of the THF model and the corresponding single-particle spectrum in closed form.

While the Hubbard-I approximation fails to capture Kondo physics, it is expected to reproduce the Mott behavior of the $f$-electrons in the regime where the hybridization strength is small compared to the on-site Hubbard interaction $U_1$~\cite{HU25}. Near integer fillings -- where the Kondo effect is less relevant~\cite{HU23} -- it provides a reasonable and analytically tractable description of the interacting spectrum.

In the atomic limit, when $r \in \mathbb{Z}$ out of the $N_f$ $f$-electron flavors are filled ($0 \leq r \leq N_f$), the atomic $f$-electron self-energy can be approximated by a single pole 
\begin{equation}
	\tilde{\Sigma}^{\text{At}} \left( i \omega_n \right) 
	\approx \frac{1}{N_f} \frac{ U_1^2\, r \left( N_f - r \right)}{ i \omega_n N_f - U_1 \left( r - \frac{N_f}{2} \right)},	\label{eqn:self_energy_at_integer_filling}
\end{equation}
which opens a gap for the $f$-electrons. The resulting Mott-Hubbard bands then hybridize with the $c$-electron bands, yielding the spectra shown in \cref{fig:hubbard_i}. Thanks to the analytical tractability of the Hubbard-I approximation, an explicit formula for the single-particle excitation spectrum can be obtained; we refer the reader to Ref.~\cite{HU25} for a detailed derivation and for a discussion of the evolution of the Dirac node as the filling is tuned. A brief review is also provided in \cref{app:sec:results_symmetry:hubbard_I}.

In \cref{fig:hubbard_i}, we compare the Hubbard-I approximation with IPT results for both integer and non-integer fillings. Away from integer fillings, we generalize the Hubbard-I approximation by replacing the interpolated dynamical self-energy from \cref{eqn:interpolated_sigma} with the exact atomic-limit self-energy at the corresponding chemical potential $\mu$. At low temperatures $T = \SI{3}{\kelvin}$, a Kondo peak develops not only at fractional fillings but also at integer fillings; at $\nu = 0$, the Kondo peak arises due to a nonzero $M$ (the single particle parameter characterizing the splitting of the $\Gamma_1$ and $\Gamma_2$ $c$-electrons). The Hubbard-I approximation does not capture the Kondo physics, as expected. At fractional fillings, the Mott-Hubbard bands obtained within the Hubbard-I approximation can appear near the Fermi energy, but they do not constitute a Kondo peak. The IPT results also exhibit lifetime broadening, which is absent in the Hubbard-I approximation. When a Kondo peak appears at integer fillings, the lower Hubbard band in the IPT results shifts to slightly lower energies compared to the Hubbard-I prediction, with more pronounced spectral changes for $\nu = 3$. We also note that within DMFT the upper Hubbard band at $\nu = 1,2,3$ is significantly broadened -- almost invisible at $\nu = 3$ -- making the Hubbard-I and DMFT spectra differ substantially even at integer fillings.

\section{Conclusions}\label{sec:conclusions}

In this work, we have introduced and applied two complementary approaches based on second-order perturbation theory to study the THF models of TBG and TSTG beyond the Hartree-Fock approximation. In the symmetry-broken phases, we computed the spectral functions of the system at finite temperature and doping using self-consistent second-order perturbation theory~\cite{SCH89,SCH90,SCH91}. In the symmetric phase, we employed a combined DMFT and Hartree-Fock framework~\cite{HAU19,HU23,DAT23,RAI23a,CAL25a} and adapted and benchmarked an approximate, yet computationally inexpensive, IPT impurity solver~\cite{MAR86,GEO92,YEY93,KAJ96,POT97,ANI97,LIC98,YEY99,MEY99,YEY00,SAS01,SAV01,FUJ03,LAA03,KUS06,ARS12,DAS16,WAG21,MIZ21,VAN22,CAN24,CAN25,VAN22}. Benchmarking the IPT solutions against numerically exact QMC results revealed good agreement, highlighting the utility of IPT for rapidly exploring extensive parameter regimes prior to applying more computationally demanding methods.

Still within the symmetric phase, we compared the Hubbard-I approximation~\cite{HU25,LED25} with the IPT method. At integer fillings -- in particular at charge neutrality -- the spectral function obtained within the Hubbard-I approximation only \emph{broadly} matches the IPT results. Despite its analytical tractability, the Hubbard-I approximation fails to capture either the Kondo physics or the lifetime broadening of the $f$-electron bands.

With the IPT impurity solver now benchmarked for the ideal parameters of the THF model~\cite{SON22,YU23a}, an important direction for future work is its generalization to incorporate additional perturbations, such as strain effects~\cite{CRI25}. Strain generically breaks the $C_{3z}$ symmetry, rendering the $f$-electron site-diagonal self-energy no longer proportional to the identity, but instead block diagonal with $2\times 2$ blocks. Developing an IPT-based treatment in this more general setting would enable the exploration of phase diagrams across a wider range of parameter regimes, including those with relaxation effects not considered in the present study.

\begin{acknowledgments}
We are very grateful to Lin Lin, Jonah Herzog-Arbeitman, Jiewen Xiao, Shahal Ilani, Andrew J. Millis, Antoine Geroges, Piers Coleman, Siddharth A. Parameswaran, Francisco Guinea-Lopez, Leonid I. Glazman, Elena Bascones, and Patrick J. Ledwidth for useful discussions. Additionally, we thank Rafael Luque-Merino, Dmitri K. Efetov, Sergi Batlle-Porro, Petr Stepanov, Frank L. Koppens, Eva Y. Andrei, Hyunjin Kim, and Stevan Nadj-Perge for collaborations on related projects~\cite{MER24,BAT24,ZHA25a,KIM25a}. D.C. was supported by a European Research Council (ERC) under the European Union's Horizon 2020 research and innovation program (Grant Agreement No.~101020833). D.C. acknowledges the hospitality of the Donostia International Physics Center and of the Aspen Center for Physics, at which part of this work was carried out. D.C. also gratefully acknowledges the support provided by the Leverhulme Trust, as well as additional support from the UKRI Horizon Europe Guarantee Grant No. EP/Z002419/1 (for an ERC Consolidator Grant to Siddharth A. Parameswaran). H.H. was supported by the Gordon and Betty Moore Foundation through Grant No.~GBMF8685 towards the Princeton theory program, the Gordon and Betty Moore Foundation's EPiQS Initiative (Grant No.~GBMF11070), the Office of Naval Research (ONR Grant No.~N00014-20-1-2303), the Global Collaborative Network Grant at Princeton University, the Simons Investigator Grant No.~404513, the Princeton Catalysis Initiative, the NSF-MERSEC (Grant No.~MERSEC DMR 2011750), Simons Collaboration on New Frontiers in Superconductivity (SFI-MPS-NFS-00006741-01), and the Schmidt Foundation at the Princeton University. B.A.B. was supported by Office of Basic Energy Sciences, Material Sciences and Engineering Division, U.S. Department of Energy (DOE) under Contracts No. DE-SC0016239. G.R. and T.W. gratefully acknowledge funding from the cluster of excellence ``CUI: Advanced Imaging of Matter" of the Deutsche Forschungsgemeinschaft  (DFG EXC 2056, Project ID 390715994). G.R, L.C, T.W., G.S. and R.V. thank the DFG for funding through the research unit QUAST FOR 5249 (project ID: 449872909; projects P4 and P5).
\end{acknowledgments}
 
\let\oldaddcontentsline\addcontentsline
\renewcommand{\addcontentsline}[3]{}

\let\addcontentsline\oldaddcontentsline

\renewcommand{\thetable}{S\arabic{table}}
\renewcommand{\thefigure}{S\arabic{figure}}
\renewcommand{\theequation}{S\arabic{section}.\arabic{equation}}
\onecolumngrid
\pagebreak
\thispagestyle{empty}
\newpage
\begin{center}
	\textbf{\large Supplemental Material: Obtaining the Spectral Function of Moir\'e Graphene Heavy-Fermions Using Iterative Perturbation Theory}\\[.2cm]
\end{center}
\appendix
\renewcommand{\thesection}{\Roman{section}}
\tableofcontents
\let\oldaddcontentsline\addcontentsline
\newpage

\section{Review of the topological heavy fermion models for twisted graphene heterostructures}\label{app:sec:HF_review}

This \siSection{} succinctly summarizes the topological heavy fermion (THF) models for twisted bilayer graphene (TBG)~\cite{SON22} and twisted symmetric trilayer graphene (TSTG)~\cite{YU23a}. We first discuss the single-particle THF Hamiltonians for the two twisted graphene system. We then give a brief overview of the interaction THF Hamiltonian. Additionally, we specify the single-particle and interaction THF parameters utilized in this work, which are identical with those adopted in Ref.~\cite{CAL24}.

\subsection{The single-particle THF model}
\label{app:sec:HF_review:single_particle}

\subsubsection{The single-particle THF model of TBG}
\label{app:sec:HF_review:single_particle:TBG}

\begin{figure}[!t]
	\centering
	\begin{tikzpicture}[scale=1, transform shape]
		\tikzset{myarrow/.style={-{Triangle[length=16pt,width=2mm]}, black, fill=black}}
\coordinate (Gamma) at (0,0);
		\coordinate (A) at ($(Gamma)+(30:2)$);
		\coordinate (B) at ($(Gamma)+(90:2)$);
		\coordinate (C) at ($(Gamma)+(150:2)$);
		\coordinate (D) at ($(Gamma)+(210:2)$);
		\coordinate (E) at ($(Gamma)+(270:2)$);
		\coordinate (F) at ($(Gamma)+(330:2)$);
	
		\coordinate (M) at ($(C)!.5!(D)$);
		\coordinate (Kp) at ($(A)$);
		\coordinate (K) at ($(F)$);
\filldraw[fill=blue!20, draw=black, thick] (A) -- (B) -- (C) -- (D) -- (E) -- (F) -- cycle;
	
\foreach \point/\position/\name in {Gamma/below left/\Gamma, M/left/\mathrm{M}, K/below right/\mathrm{K}, Kp/above right/\mathrm{K}'}
		\draw[fill=black] (\point) circle (2pt) node[\position=3pt] {$\name_M$};
		
		\draw[myarrow, thick] (Gamma) -- node[left] {$\vec{q}_1$} (E);
		\draw[myarrow, thick] (Gamma) -- node[below] {$\vec{q}_2$} (A);
		\draw[myarrow, thick] (Gamma) -- node[below] {$\vec{q}_3$} (C);
	\end{tikzpicture}
	\caption{Moir\'e Brillouin zone (BZ) of the THF model. The first BZ denoted by the blue hexagon is centered at the $\Gamma_M$ point. We also explicitly label the $\mathrm{K}_M$, $\mathrm{K}'_M$ and $\mathrm{M}_M$ high-symmetry points. Additionally, the three vectors $\vec{q}_i$ (for $1 \leq i \leq 3$) are indicated by the black arrows.}
	\label{app:fig:moire_BZ}
\end{figure}
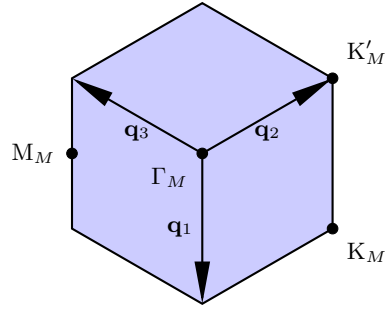

TBG comprises two stacked graphene layers rotated at an angle $\theta$ relative to one another~\cite{BIS11}. The single-particle THF model Hamiltonian for TBG is given by~\cite{SON22}
\begin{align}
	H^{\text{TBG}}_{0} =& \sum_{\substack{\abs{\vec{k}} \leq \Lambda_c\\ \eta, s}} \left[ \sum_{a,a'} h^{cc,\eta}_{a a'} \left( \vec{k} \right) \hat{c}^\dagger_{\vec{k},a,\eta,s} \hat{c}_{\vec{k},a,\eta,s} + \left( \sum_{a,\alpha} h^{cf,\eta}_{a \alpha} \left( \vec{k} \right) \hat{c}^\dagger_{\vec{k},a,\eta,s} \hat{f}_{\vec{k},\alpha,\eta,s}  + \text{h.c.} \right) \right] \nonumber \\
	+&  \sum_{\substack{\vec{k},\alpha,\alpha' \\ \eta,s}} h^{ff,\eta}_{\alpha \alpha'} \left( \vec{k} \right) \hat{f}^\dagger_{\vec{k},\alpha,\eta,s} \hat{f}_{\vec{k},\alpha',\eta,s} ,
	\label{app:eqn:single_part_THF_TBG}
\end{align}
where ``$+\text{h.c.}$'' denotes the addition of the Hermitian conjugate. The THF model for TBG features two types of fermions denoted by the $f$- and $c$-electron operators in \cref{app:eqn:single_part_THF_TBG}. The $f$-fermions are localized at the moir\'e AA-sites and transform as a pair of $p_x \pm i p_y$ orbitals. The operator $\hat{f}^\dagger_{\vec{k},\alpha, \eta, s}$ creates an $f$-fermion of spin $s=\uparrow,\downarrow$ at moir\'e momentum $\vec{k}$ in valley $\eta = \pm$, having an orbital index $\alpha=1,2$, with $\alpha = 1$ ($\alpha = 2$) corresponding to the $f$-electron state transforming as a $p_x+ip_y$ ($p_x-ip_y$) orbital. The real-space $f$-electron operators are defined according to 
\begin{equation}
	\label{app:eqn:f_fermions_real_def}
	\hat{f}^\dagger_{\vec{R},\alpha,\eta,s} = \frac{1}{\sqrt{N_0}}\sum_{\vec{k}} \hat{f}^\dagger_{\vec{k},\alpha,\eta,s} e^{-i \vec{k} \cdot \vec{R}},
\end{equation} 
where $N_0$ is the total number of moir\'e unit cells, which are labeled by their position lattice vector $\vec{R}$. The $c$-fermions form semimetallic, highly dispersive, and anomalous electronic bands. At a given $\vec{k}$ point, the $c$-electron states are labeled by their spin $s=\uparrow,\downarrow$, valley $\eta=\pm$, and orbital numbers $1 \leq a \leq 4$. At the $\Gamma_M$ point ($\vec{k} = 0$) of the moir\'e Brillouin zone (BZ) shown in \cref{app:fig:moire_BZ}, the $\hat{c}^\dagger_{\vec{k},a,\eta,s}$ form a $\Gamma_3$ ($\Gamma_1 \oplus \Gamma_2$) irreducible (reducible) representation for $a=1,2$ ($a=3,4$). 

The matrix blocks appearing in \cref{app:eqn:single_part_THF_TBG} are given by~\cite{SON22}
\begin{align}
	h^{cc,\eta} \left( \vec{k} \right) =& \begin{pmatrix}
		\mathbb{0} & v_{\star} \left(\eta k_x \sigma_0 + i k_y \sigma_z \right) \\ 
		v_{*} \left(\eta k_x \sigma_0 - i k_y \sigma_z \right) & M \sigma_z
	\end{pmatrix} \label{app:eqn:cc_thf_block},\\
	h^{cf,\eta} \left( \vec{k} \right) =& \begin{pmatrix}
		\gamma \sigma_0 + v_{*}' \left( \eta k_x \sigma_x + k_y \sigma_y \right) \\ 
		v_{*}'' \left( \eta k_x \sigma_x - k_y \sigma_y \right)
	\end{pmatrix} e^{-\frac{\abs{\vec{k}}^2 \lambda^2}{2}} \label{app:eqn:cf_thf_block}, \\
	h^{ff,\eta} \left( \vec{k} \right) =& \mathcal{O} \left( t \right) \approx \mathbb{0} \label{app:eqn:f_thf_block},
\end{align}
with $\mathbb{0}$ denoting the zero matrix. Note that a cutoff $\Lambda_c$ exists for the momenta of the $c$-electrons, and $\sigma_{i}$ ($i=0,x,y,z$) denoting the Pauli matrices acting in orbital space. The choice of $\Lambda_c$ will be discussed in \cref{app:sec:hartree_fock:generic_not}. Throughout this work and Ref.~\cite{CAL24}, we consider the following numerical values for the single-particle parameters appearing in \cref{app:eqn:single_part_THF_TBG}: $\lambda = 0.3375 \abs{\vec{a}_{M1}}$, $\gamma = \SI{-24.75}{\milli\electronvolt}$, $M = \SI{3.697}{\milli\electronvolt}$, $v_{\star} = \SI{-4.303}{\electronvolt\angstrom}$, $v'_{\star} = \SI{1.623}{\electronvolt\angstrom}$, and $v''_{\star} = \SI{-0.0332}{\electronvolt\angstrom}$. These THF parameters correspond to $\theta = 1.05 \degree$, $w_1 = \SI{110}{\milli\electronvolt}$, $v_F = \SI{5.944}{\electronvolt \angstrom}$, $\abs{\vec{K}_+} = \SI{1.703}{\angstrom^{-1}}$, and $w_0/w_1 = 0.8$~\cite{SON22}, where $\theta$ denotes the TBG twist angle, $v_F$ and $\vec{K}_+$ are the single layer graphene Fermi velocity and wave vector of the $\mathrm{K}$ point, while $w_0$ and $w_1$ are, respectively, the interlayer tunneling amplitudes at the AA and AB sites. Near the magic angle, which will be the focus of this work, the $f$-electrons are dispersionless, which implies that their nearest neighbor hopping amplitude $t$ vanishes. As a result, we will approximate the matrix block from \cref{app:eqn:f_thf_block} to zero. 

\subsubsection{The single-particle THF model of TSTG}
\label{app:sec:HF_review:single_particle:TSTG}
TSTG consists of three graphene layers: the top and bottom layers are located directly above one another and are rotated by an angle $\theta$ relative to the middle layer~\cite{LI19,KHA19,CAR20,CAL21}. The THF model of TSTG can be obtained form the TBG one in \cref{app:eqn:single_part_THF_TBG} by the addition of a high-velocity Dirac ($d$) fermion to the latter~\cite{YU23a}
\begin{equation}
	\label{app:eqn:single_part_THF_TSTG}
	H_0^{\text{TSTG}} = H_0^{\text{TBG}} + H_0^{\text{D}} + H_0^{\mathcal{E}}.
\end{equation}
The second and third terms of \cref{app:eqn:single_part_THF_TSTG} are given by~\cite{YU23a}
\begin{equation}
	H_0^{\text{D}} = \sum_{\substack{\vec{p} \leq \Lambda_d \\ \alpha,\alpha',\eta,s}} h^{dd,\eta}_{\alpha \alpha'} \left( \vec{p} \right) \hat{d}^\dagger_{\vec{p},\alpha,\eta,s} \hat{d}_{\vec{p},\alpha',\eta,s} \qq{and}
	H_0^{\mathcal{E}} = \sum_{\substack{\vec{p} \leq \Lambda_d \\ \alpha,\eta,s}} h^{df,\eta}_{\alpha \alpha'} \left( \vec{p} \right)  \hat{d}^\dagger_{\vec{p},\alpha',\eta,s} \hat{f}_{\vec{p}+\eta \vec{q}_1,\alpha,\eta,s} + \text{h.c.},
\end{equation}
and correspond, respectively, to the Dirac Hamiltonian of the $d$-fermions and the $f$-$d$ single-particle hybridization. The operator $\hat{d}^\dagger_{\vec{p},\alpha, \eta, s}$ creates a high-velocity Dirac fermion at of momentum $\vec{p}$ and spin $s=\uparrow,\downarrow$, in valley $\eta = \pm$, belonging to the single-layer graphene sublattice $\alpha=1,2$. Note that the momenta of the $d$-fermions are measured from $\eta \vec{q}_1$, which, as depicted in \cref{app:fig:moire_BZ}, corresponds to the $\mathrm{K}'_{M}$ ($\mathrm{K}_M$) point in valley $\eta = +$ ($\eta = -$). At the single-particle level, the $d$- and $c$-electrons are uncoupled, whereas the $d$- and $f$-fermions only couple in the presence of a non-zero perpendicular displacement field $\mathcal{E}$. The matrix blocks of the last two terms in \cref{app:eqn:single_part_THF_TSTG} are
\begin{align}
	h^{dd,\eta} \left( \vec{p} \right) = & v_F \left( \eta p_x \sigma_x + p_y \sigma_y \right) \label{app:eqn:dd_thf_block},\\
	h^{df,\eta} \left( \vec{p} \right) =& M_1 \mathcal{E} \left( \sigma_0 - i \eta \sigma_z \right) e^{-\frac{\abs{\vec{p}}^2 \lambda^2}{2}} \label{app:eqn:df_thf_block},
\end{align}
where $v_F$ is the graphene Fermi velocity, while $M_1$ is the $d$-$f$ coupling constant. Similarly, to the $c$-electrons in \cref{app:eqn:single_part_THF_TBG}, a momentum cutoff $\Lambda_d$ was introduced for the $d$-electrons' momenta, which will be fixed in \cref{app:sec:hartree_fock:generic_not}. 

The first term of \cref{app:eqn:single_part_THF_TSTG}, $H_{0}^{\text{TBG}}$ assumes the same form as in \cref{app:eqn:single_part_THF_TSTG}, but the values of the $\gamma$, $v''_{*}$, and $M$ parameters now depend on the displacement field $\mathcal{E}$ according to~\cite{YU23a}
\begin{equation}
	\label{app:eqn:change_gamma_v_m}
	\gamma \to \gamma + B_{\gamma} \mathcal{E}^2, \quad
	v''_{*} \to v''_{*} + B_{v''_{*}} \mathcal{E}^2,
	\quad
	M \to M + B_{M} \mathcal{E}^2.
\end{equation}
The numerical values of the parameters that we employ for TSTG are consistent with those of Refs.~\cite{CAL24,YU23a} and correspond to a twist angle of $\theta = 1.4703 \degree$ (with the same values of the interlayer tunneling amplitudes, graphene Fermi velocity and $K$-point wave vector as in the case of TBG from \cref{app:sec:HF_review:single_particle:TBG}). Specifically, we employ $\lambda = 0.3359 \abs{\vec{a}_{M1}}$, $\gamma = \SI{-33.16}{\milli\electronvolt}$, $M = \SI{7.01}{\milli\electronvolt}$, $v_{\star} = \SI{-4.301}{\electronvolt\angstrom}$, $v'_{\star} = \SI{1.625}{\electronvolt\angstrom}$, $v''_{\star} = \SI{-0.0346}{\electronvolt\angstrom}$, $M_1 = -0.1394$, $B_{\gamma} = \SI{-3.75e-4}{\milli\electronvolt^{-1}}$, $B_{v''_{*}} = \SI{7.65e3}{\electronvolt^{-1}\angstrom}$, and $B_{M} = \SI{3.28e-4}{\milli\electronvolt^{-1}}$.

\subsection{The interaction THF Hamiltonian}
\label{app:sec:HF_review:interaction}

\subsubsection{The interaction THF Hamiltonian for TBG}
\label{app:sec:HF_review:interaction:TBG}

The THF interaction Hamiltonian for TBG is given by~\cite{SON22}
\begin{equation}
	\label{app:eqn:THF_interaction_TBG}
	H^{\text{TBG}}_I = H_{U_1} + H_{U_2} + H_V + H_W + H_J + H_{\tilde{J}} + H_K,
\end{equation}
where
\begin{align}
	H_{U_1} &= \frac{U_1}{2} \sum_{\vec{R}} \sum_{\substack{\alpha,\eta,s \\ \alpha',\eta',s'}} :\mathrel{\hat{f}^\dagger_{\vec{R},\alpha,\eta,s} \hat{f}_{\vec{R},\alpha,\eta,s}}: :\mathrel{\hat{f}^\dagger_{\vec{R},\alpha',\eta',s'} \hat{f}_{\vec{R},\alpha',\eta',s'}}:, \label{app:eqn:THF_int:U1} \\
H_{U_2} &= \frac{U_2}{2} \sum_{\left\langle \vec{R}, \vec{R}' \right\rangle} \sum_{\substack{\alpha,\eta,s \\ \alpha',\eta',s'}} :\mathrel{\hat{f}^\dagger_{\vec{R}',\alpha,\eta,s} \hat{f}_{\vec{R}',\alpha,\eta,s}}: :\mathrel{\hat{f}^\dagger_{\vec{R},\alpha',\eta',s'} \hat{f}_{\vec{R},\alpha',\eta',s'}}:, \label{app:eqn:THF_int:U2} \\
H_V &= \frac{1}{2 \Omega_0 N_0} \sum_{\abs{\vec{k}_1},\abs{\vec{k}_2} \leq \Lambda_c} \sum_{\substack{\vec{q} \\ \abs{\vec{k}_1+\vec{q}},\abs{\vec{k}_2+\vec{q}} \leq \Lambda_c}} \sum_{\substack{a,\eta,s \\ a',\eta',s'}} V \left( \vec{q} \right):\mathrel{\hat{c}^\dagger_{\vec{k}_1 + \vec{q} ,a,\eta,s} \hat{c}_{\vec{k}_1,a,\eta,s}}: :\mathrel{\hat{c}^\dagger_{\vec{k}_2 - \vec{q},a',\eta',s'} \hat{c}_{\vec{k}_2,a',\eta',s'}}:, \label{app:eqn:THF_int:V} \\
H_W &= \frac{1}{N_0} \sum_{\substack{\vec{k}_1 \\ \abs{\vec{k}_2} \leq \Lambda_c}} \sum_{\substack{\vec{q} \\ \abs{\vec{k}_2-\vec{q}} \leq \Lambda_c}} \sum_{\substack{\alpha,\eta,s \\ a',\eta',s'}} W_{a'} :\mathrel{\hat{f}^\dagger_{\vec{k}_1 + \vec{q},\alpha,\eta,s} \hat{f}_{\vec{k}_1,\alpha,\eta,s}}: :\mathrel{\hat{c}^\dagger_{\vec{k}_2 - \vec{q},a',\eta',s'} \hat{c}_{\vec{k}_2,a',\eta',s'}}:, \label{app:eqn:THF_int:W} \\
H_J &= -\frac{J}{2N_0} \sum_{\substack{\vec{k}_1 \\ \abs{\vec{k}_2} \leq \Lambda_c}} \sum_{\substack{\vec{q} \\ \abs{\vec{k}_2+\vec{q}} \leq \Lambda_c}} \sum_{\substack{\alpha,\eta,s \\ a',\eta',s'}} \left[ \eta \eta' + (-1)^{\alpha + \alpha'} \right] :\mathrel{\hat{f}^\dagger_{\vec{k}_1 + \vec{q},\alpha,\eta,s} \hat{f}_{\vec{k}_1,\alpha',\eta',s'}}: :\mathrel{\hat{c}^\dagger_{\vec{k}_2 - \vec{q},\alpha'+2,\eta',s'} \hat{c}_{\vec{k}_2,\alpha+2,\eta,s}}:, \label{app:eqn:THF_int:J} \\
H_{\tilde{J}} &= -\frac{J}{4N_0} \sum_{\abs{\vec{k}_1}, \abs{\vec{k}_2} \leq \Lambda_c} \sum_{\vec{q}} \sum_{\substack{\alpha,\eta,s \\ a',\eta',s'}} \left[ \eta \eta' - (-1)^{\alpha + \alpha'} \right] \hat{f}^\dagger_{\vec{k}_1 + \vec{q},\alpha,\eta,s} \hat{f}^\dagger_{\vec{k}_2 - \vec{q},\alpha',\eta',s'} \hat{c}_{\vec{k}_2,\alpha'+2,\eta',s'} \hat{c}_{\vec{k}_1,\alpha+2,\eta,s} + \text{h.c.} \label{app:eqn:THF_int:Jtilde} \\	
H_{K} &= \frac{K}{N_0 \Omega_0} \sum_{\abs{\vec{k}_1},\abs{\vec{k}_2},\abs{\vec{k}_3} \leq \Lambda_c} \sum_{\substack{\alpha,\eta,s \\ \eta',s'}} \eta \eta' \left(
	\hat{c}^\dagger_{\vec{k}_1,\bar{\alpha},\eta,s} \hat{c}_{\vec{k}_3,\alpha+2,\eta,s} \hat{f}^\dagger_{\vec{k}_1 - \vec{k}_2 - \vec{k}_3,\alpha,\eta',s'} \hat{c}_{\vec{k}_2,\alpha+2,\eta',s'} 
	\right. \nonumber \\
	&\left. -\hat{f}^\dagger_{\vec{k}_2 + \vec{k}_3 - \vec{k}_1, \alpha, \eta', s'} \hat{c}_{\vec{k}_2,\alpha+2,\eta',s'} \hat{c}^\dagger_{\vec{k}_1,\bar{\alpha}+2,\eta,s} \hat{c}_{\vec{k}_3,\alpha,\eta,s} 
	\right)
	+ \text{h.c.}, \label{app:eqn:THF_int:K}
 \end{align}
which correspond, respectively, to the onsite ($H_{U_1}$) and nearest-neighbor ($H_{U_2}$) $f$-electron repulsions, the $c$-electron Coulomb interaction ($H_V$), the $f$-$c$ density-density interaction ($H_W$), the $f$-$c$ exchange interaction ($H_J$), the double-hybridzation interaction ($H_{\tilde{J}}$), and the density-hybridization interaction ($H_{K}$). Additionally, the number of moir\'e unit cells and their area are given, respectively, by $N_0$ and $\Omega_0$. The THF interaction Hamiltonian was obtained assuming a double-gated experimental setup, for which the Fourier transformation of the inter-electron Coulomb interaction potential is given by
\begin{equation}
	\label{app:eqn:double_gate_interaction}
	V \left( \vec{q} \right) = \left( \pi U_{\xi} \xi^2 \right) \frac{\tanh \left( \abs{\vec{q}} \xi/2 \right)}{\abs{\vec{q}}\xi/2},
\end{equation}
In \cref{app:eqn:double_gate_interaction} $U_{\xi} = \frac{e^2}{4 \pi \epsilon_0 \epsilon \xi}$ is the interaction energy scale (in SI units), with $\epsilon$ being the dielectric constant and $\xi$, the distance between the two screening gates. Throughout this work and Ref.~\cite{CAL24}, we consider $\epsilon = 6$, $\xi = \SI{10}{\nano\meter}$, and $U_{\xi} = \SI{24}{\milli\electronvolt}$. 

In \crefrange{app:eqn:THF_int:U1}{app:eqn:THF_int:J}, we have also defined the normal-ordered form of an operator $\mathcal{O}$ to be $:\mathrel{\mathcal{O}}: = \mathcal{O} - \ev{\mathcal{O}}{\mathrm{G}_0}$, where $\ket{\mathrm{G}_0}$ denotes a state at the charge neutrality point. We take this state to be such that 
\begin{align}
	\ev{\hat{c}^\dagger_{\vec{k},a,\eta,s} \hat{c}_{\vec{k}',a',\eta',s'}}{\mathrm{G}_0} &= \frac{1}{2} \delta_{\vec{k}, \vec{k}'} \delta_{a a'} \delta_{\eta \eta'} \delta_{s s'}, \nonumber \\
	\ev{\hat{f}^\dagger_{\vec{k},\alpha,\eta,s} \hat{f}_{\vec{k}',\alpha',\eta',s'}}{\mathrm{G}_0} &= \frac{1}{2} \delta_{\vec{k}, \vec{k}'} \delta_{\alpha \alpha'} \delta_{\eta \eta'} \delta_{s s'}, \nonumber \\
	\ev{\hat{f}^\dagger_{\vec{k},\alpha,\eta,s} \hat{c}_{\vec{k}',a',\eta',s'}}{\mathrm{G}_0} &= \ev{\hat{c}^\dagger_{\vec{k},a,\eta,s} \hat{f}_{\vec{k}',\alpha',\eta',s'}}{\mathrm{G}_0} = 0.
\end{align}
Additionally, in \cref{app:eqn:THF_int:U2}, $\left\langle \vec{R}, \vec{R}' \right\rangle$ denotes nearest-neighbor lattice sites. For the twist angle and tunneling ratio chosen in \cref{app:sec:HF_review:single_particle:TBG}, as well as the relative permittivity and screening length chosen below \cref{app:eqn:double_gate_interaction}, the six interaction parameters ($W_1 = W_2$ and $W_3 = W_4$ follows by symmetry~\cite{SON22}) are given by~\cite{SON22}
\begin{align}
	U_1 &= \SI{57.95}{\milli\electronvolt}, &
	U_2 &= \SI{2.239}{\milli\electronvolt}, &
	W_1 &= W_2 = \SI{44.03}{\milli\electronvolt}, \nonumber \\ 
	J &= \SI{16.38}{\milli\electronvolt}, &
	K &= \SI{4.887}{\milli\electronvolt}, &
	W_3 &= W_4 = \SI{50.2}{\milli\electronvolt}, \label{app:eqn:THF_interaction_params_TBG}
\end{align}
while the $c$-$c$ interaction potential is the same as in \cref{app:eqn:double_gate_interaction}.

\subsubsection{The interaction THF Hamiltonian for TSTG}
\label{app:sec:HF_review:interaction:TSTG}

The mirror-even and mirror-odd fermions of TSTG are always coupled through density-density interactions, even in the absence of a perpendicularly applied displacement field~\cite{CAL21}. As a result the TSTG interaction Hamiltonian of TSTG ($H^{\text{TSTG}}_{I}$) can be constructed from the one of TBG by simply adding three terms to the latter,
\begin{equation}
	\label{app:eqn:THF_interaction_TSTG}
	H^{\text{TSTG}}_{I} = H^{\text{TBG}}_{I} + H_V^{d} + H_V^{cd} + H_W^{fd},
\end{equation}
which correspond to the Coulomb repulsion terms between the $d$-electrons ($H_V^{d}$), the $c$- and $d$-electrons ($H_V^{cd}$), as well as the $f$- and $d$-electrons ($H_W^{fd}$)~\cite{YU23a}. These three additional terms are given respectively by~\cite{YU23a}
\begin{align}
	H_V^{d} &= \frac{1}{2 \Omega_0 N_0} \sum_{\abs{\vec{p}_1},\abs{\vec{p}_1} \leq \Lambda_d} \sum_{\substack{\vec{q} \\ \abs{\vec{p}_1+\vec{q}},\abs{\vec{p}_2+\vec{q}} \leq \Lambda_d}} \sum_{\substack{\alpha,\eta,s \\ \alpha',\eta',s'}} V \left( \vec{q} \right):\mathrel{\hat{d}^\dagger_{\vec{p}_1 + \vec{q} ,\alpha ,\eta,s} \hat{d}_{\vec{p}_1,\alpha,\eta,s}}: :\mathrel{\hat{d}^\dagger_{\vec{p}_2 - \vec{q},\alpha',\eta',s'} \hat{d}_{\vec{p}_2,\alpha',\eta',s'}}:, \label{app:eqn:TSTG_int:Vd} \\
H_V^{cd} &= \frac{1}{\Omega_0 N_0} \sum_{\substack{\abs{\vec{k}_1} \leq \Lambda_c \\ \abs{\vec{p}_2}< \Lambda_d}} \sum_{\substack{\vec{q} \\ \abs{\vec{k}_1+\vec{q}} \leq \Lambda_c \\ \abs{\vec{p}_2+\vec{q}} \leq \Lambda_d}} \sum_{\substack{a,\eta,s \\ \alpha',\eta',s'}} V \left( \vec{q} \right):\mathrel{\hat{c}^\dagger_{\vec{k}_1 + \vec{q} ,a,\eta,s} \hat{c}_{\vec{k}_1,a,\eta,s}}: :\mathrel{\hat{d}^\dagger_{\vec{p}_2 - \vec{q},\alpha',\eta',s'} \hat{d}_{\vec{p}_2,\alpha',\eta',s'}}:, \label{app:eqn:TSTG_int:Vcd} \\
H_W^{fd} &= \frac{1}{N_0} \sum_{\substack{\vec{k}_1 \\ \abs{\vec{p}_2} \leq \Lambda_d}} \sum_{\substack{\vec{q} \\ \abs{\vec{p}_2-\vec{q}} \leq \Lambda_d}} \sum_{\substack{\alpha,\eta,s \\ \alpha',\eta',s'}} W_{fd} :\mathrel{\hat{f}^\dagger_{\vec{k}_1 + \vec{q},\alpha,\eta,s} \hat{f}_{\vec{k}_1,\alpha,\eta,s}}: :\mathrel{\hat{d}^\dagger_{\vec{p}_2 - \vec{q},\alpha',\eta',s'} \hat{d}_{\vec{p}_2,\alpha',\eta',s'}}:, \label{app:eqn:TSTG_int:Vfd}
 \end{align} 
where we note that the state $\ket{\mathrm{G}_0}$ at charge neutrality relative to which the normal ordering is defined obeys
\begin{align}
	\ev{\hat{d}^\dagger_{\vec{p},\alpha,\eta,s} \hat{d}_{\vec{p}',\alpha',\eta',s'}}{\mathrm{G}_0} &= \frac{1}{2} \delta_{\vec{p}, \vec{p}'} \delta_{\alpha \alpha'} \delta_{\eta \eta'} \delta_{s s'}, \nonumber \\
	\ev{\hat{d}^\dagger_{\vec{p},\alpha,\eta,s} \hat{c}_{\vec{k}',a',\eta',s'}}{\mathrm{G}_0} &= \ev{\hat{d}^\dagger_{\vec{p},\alpha,\eta,s} \hat{f}_{\vec{k}',\alpha',\eta',s'}}{\mathrm{G}_0} = 0, \nonumber \\
	\ev{\hat{c}^\dagger_{\vec{k},a,\eta,s} \hat{d}_{\vec{p}',\alpha',\eta',s'}}{\mathrm{G}_0} &= \ev{\hat{f}^\dagger_{\vec{k},\alpha,\eta,s} \hat{d}_{\vec{p}',\alpha',\eta',s'}}{\mathrm{G}_0} = 0.
\end{align}
Finally, for the single-particle parameters chosen in \cref{app:sec:HF_review:single_particle:TSTG} and the Coulomb repulsion assumed in \cref{app:eqn:double_gate_interaction}, the THF interaction parameters for TSTG are given by~\cite{YU23a}
\begin{align}
	U_1 &= \SI{91.5}{\milli\electronvolt}, &
	U_2 &= \SI{6.203}{\milli\electronvolt}, &
	W_1 &= W_2 = \SI{88.54}{\milli\electronvolt}, \nonumber \\ 
	J &= \SI{24.25}{\milli\electronvolt}, &
	K &= \SI{7.061}{\milli\electronvolt}, &
	W_3 &= W_4 = \SI{97.67}{\milli\electronvolt}, &
	W_{fd}& = \SI{94.71}{\milli\electronvolt}. \label{app:eqn:THF_interaction_params_TSTG}
\end{align}
As explained in \cref*{Seebeck:app:sec:HF_review:interaction:TSTG} of Ref.~\cite{CAL24}, the approximate increase by a factor of $\sqrt{2}$ in the interaction parameters of TSTG relative to the ones of TBG from \cref{app:eqn:THF_interaction_params_TBG} can be traced back to the magic twist angle of TSTG, which is itself larger by a factor of $\sqrt{2}$ than the one of TBG~\cite{YU23a}.

\section{Hartree-Fock theory for the THF model}\label{app:sec:hartree_fock}

In this \siSection{}, we give a brief overview of the Hartree-Fock theory for the THF model, with the main goal of formalizing the notation that will be used in subsequent sections. The reader is referred to Refs.~\cite{SON22,YU23a} for a more in-depth treatment of the subject. We start by introducing a useful notation that treats the $f$-, $c$-, and (in the case of TSTG) $d$-electrons on equal footing and define the Green's function for the system. Next, we derive the Hartree-Fock Hamiltonian of the system assuming a generic ground state. In addition to the zero-temperature regime already considered in Refs.~\cite{SON22,YU23a}, we also consider the finite temperature case. Finally, we list the symmetry-breaking correlated ground state candidates that we consider in this work and in Ref.~\cite{CAL24} and derive the implications of the many-body charge-conjugation symmetry on the THF model spectral function.

\subsection{Generic notation}\label{app:sec:hartree_fock:generic_not}

The $c$-fermions (for both TBG and TSTG) and $d$-fermions (for TSTG) are only defined within a limited region around the $\Gamma_M$ and $\mathrm{K}_M$ (or $\mathrm{K}'_M$, depending on the valley) points, respectively. In order to restore the system's symmetries, Refs.~\cite{SON22,YU23a} set the momentum cutoffs, $\Lambda_c$ and $\Lambda_d$, to infinity (or to a sufficiently large value in the numerical implementation). In our numerical approach, we will extend these two cutoffs so that the $\hat{c}^\dagger_{\vec{k},a,\eta,s}$ ($\hat{d}^\dagger_{\vec{p},\alpha,\eta,s}$) fermion covers exactly one BZ around the $\Gamma_M$ ($\eta \vec{q}_1$) point. 

We introduce a notation that treats all the fermionic species on equal footing. As such, for TBG, we define the $\hat{\gamma}^\dagger_{\vec{k},\eta,i,s}$ fermions (for $1 \leq i \leq 6$)
\begin{equation}
	\label{app:eqn:shorthand_gamma_not}
	\hat{\gamma}^\dagger_{\vec{k},\eta,i,s} \equiv \begin{cases}
		\hat{c}^\dagger_{\vec{k},\eta,i,s}, & \qq{for} 1 \leq i \leq 4 \\
		\hat{f}^\dagger_{\vec{k},\eta,i-4,s}, & \qq{for} 5 \leq i \leq 6
	\end{cases}.
\end{equation}
For TSTG, we introduce two additional fermions
\begin{equation}
	\label{app:eqn:shorthand_gamma_not_TSTG}
	\hat{\gamma}^\dagger_{\vec{k},\eta,i,s} \equiv \hat{d}^\dagger_{\vec{P}_{\vec{k}},\eta,i-6,s} \qq{for} 7 \leq  i \leq 8,
\end{equation}
where here and in what follows we will let $\vec{P}_{\vec{k}}$ denote the image of $\vec{k} - \eta \vec{q}_1$ in the first BZ. For simplicity, we also define the following single-particle Hamiltonian matrices corresponding respectively to the THF models of TBG and TSTG, whose blocks are given by \cref{app:eqn:cc_thf_block,app:eqn:cf_thf_block,app:eqn:f_thf_block,app:eqn:dd_thf_block,app:eqn:df_thf_block},
\begin{equation}
	\label{app:eqn:full_THF_Hamiltonian_gamma_bas}
	h^{\text{TBG},\eta} \left( \vec{k} \right) = \begin{pNiceMatrix}[last-col=3,first-row]
		4 & 2 \\
		h^{cc,\eta} \left( \vec{k} \right) & h^{cf,\eta} \left( \vec{k} \right) & 4 \\
		h^{\dagger cf,\eta} \left( \vec{k} \right) & h^{ff,\eta} \left( \vec{k} \right)  & 2 \\
	\end{pNiceMatrix},
	\quad
	h^{\text{TSTG},\eta} \left( \vec{k} \right) = \begin{pNiceMatrix}[last-col=4,first-row]
		4 & 2 & 2\\
		h^{cc,\eta} \left( \vec{k} \right) & h^{cf,\eta} \left( \vec{k} \right) & \mathbb{0} & 4 \\
		h^{\dagger cf,\eta} \left( \vec{k} \right) & h^{ff,\eta} \left( \vec{k} \right) & h^{df,\eta} \left( \vec{P}_{\vec{k}} \right) & 2 \\
		\mathbb{0} & h^{\dagger df,\eta} \left( \vec{P}_{\vec{k}} \right) & h^{dd,\eta} \left( \vec{P}_{\vec{k}} \right) & 2 \\
	\end{pNiceMatrix}.
\end{equation}
In \cref{app:eqn:full_THF_Hamiltonian_gamma_bas}, the dimensions of the blocks are explicitly indicated outside each matrix. To make the expressions valid even away from the magic angle, we have included the $f$-electron single-particle Hamiltonian $h^{ff,\eta} \left( \vec{k} \right)$. At the magic angle, $h^{ff,\eta} \left( \vec{k} \right) \approx \mathbb{0}$. 

With the new notation at hand, we now consider the Green's function of the system. First, we let $K$ denote the grand canonical Hamiltonian for the THF model with $K^{\text{TBG}} = H^{\text{TBG}} - \mu \hat{N}$ ($ K^{\text{TSTG}} = H^{\text{TSTG}} - \mu \hat{N}$) for TBG (TSTG). The many-body THF Hamiltonians for TBG and TSTG are given, respectively by 
\begin{equation}
	H^{\text{TBG}} = H^{\text{TBG}}_0 + H^{\text{TBG}}_I, \quad
	H^{\text{TSTG}} = H^{\text{TSTG}}_0 + H^{\text{TSTG}}_I,
\end{equation} 
while the total number operator is $\hat{N} = \sum_{\vec{k},\eta,s} \sum_{i=1}^{6} \hat{\gamma}^\dagger_{\vec{k},i,\eta,s} \hat{\gamma}_{\vec{k},i,\eta,s}$ for TBG and $\hat{N} = \sum_{\vec{k},\eta,s} \sum_{i=1}^{8} \hat{\gamma}^\dagger_{\vec{k},i,\eta,s} \hat{\gamma}_{\vec{k},i,\eta,s}$ for TSTG. Throughout this work, we restrict to states that preserve moir\'e translation symmetry. The Matsubara Green's function of the system reads as
\begin{equation}
	\label{app:eqn:matsubara_gf_THF_tau}
	-\left\langle \mathcal{T}_{\tau} \hat{\gamma}_{\vec{k},i,\eta,s} \left( \tau \right) \hat{\gamma}^\dagger_{\vec{k}',i',\eta',s'} \left( 0 \right)  \right\rangle  = \delta_{\vec{k},\vec{k}'} \mathcal{G}_{i \eta s; i' \eta' s'} \left(\tau, \vec{k} \right),
\end{equation}
where $\tau$ is the imaginary time, $\mathcal{T}_{\tau}$ enforces the ordering of the operators that follow with respect to the imaginary time, and $\left\langle \hat{\mathcal{O}} \right\rangle$ denotes the expectation value of the operator $\hat{\mathcal{O}}$ in the grand canonical ensemble
\begin{equation}
	\left\langle \hat{\mathcal{O}} \right\rangle  = \frac{1}{Z} \Tr\left[e^{-\beta K} \hat{\mathcal{O}} \right],
\end{equation}
where $Z \equiv \Tr\left[e^{-\beta K}\right]$ is the partition function of the system. The fermion operators are also evolved using the grand canonical ensemble Hamiltonian $K$\footnote{In this work, we use the notation $K$ to denote the grand canonical Hamiltonian, following the convention in Ref.~\cite{MAH00}. This should not be confused with the density hybridization interaction constant defined in \cref{app:eqn:THF_int:K}, which will not be referenced explicitly hereafter.}
\begin{equation}
	\label{app:eqn:gr_can_evolved_gamma_ops}
	\hat{\gamma}^\dagger_{\vec{k},i,\eta,s} \left( \tau \right) = e^{K \tau} \hat{\gamma}^\dagger_{\vec{k},i,\eta,s} \left( 0 \right) e^{-K\tau}.
\end{equation}

The Fourier transformation of the Green's function is given by 
\begin{equation}
	\label{app:eqn:matsubara_gf_THF_ft}
	\mathcal{G}_{i \eta s; i' \eta' s'} \left(i \omega_n, \vec{k} \right) = \int_{0}^{\beta} \dd{\tau} e^{i \omega_n \tau} 	\mathcal{G}_{i \eta s; i' \eta' s'} \left(\tau, \vec{k} \right),
\end{equation}
with $\beta = 1/T$ being the inverse temperature and $\omega_n = \frac{(2 n + 1)\pi}{\beta}$, the fermionic Matsubara frequencies. It is also useful to define the \emph{non-interacting} Matsubara Green's function
\begin{equation}
	\label{app:eqn:matsubara_gf_THF_tau_noninteracting}
	-\left\langle \mathcal{T}_{\tau} \hat{\gamma}_{\vec{k},i,\eta,s} \left( \tau \right) \hat{\gamma}^\dagger_{\vec{k}',i',\eta',s'} \left( 0 \right)  \right\rangle_0  = \delta_{\vec{k},\vec{k}'} \mathcal{G}^{0}_{i \eta s; i' \eta' s'} \left(\tau, \vec{k} \right),
\end{equation}
whose definition is similar to \cref{app:eqn:matsubara_gf_THF_tau}, with the only exception being that the imaginary time-evolution and the averaging $\left\langle \dots \right\rangle_0$ in \cref{app:eqn:matsubara_gf_THF_tau_noninteracting} is performed within the \emph{non-interacting} grand canonical ensemble Hamiltonian $K_0 = H_0 - \mu \hat{N}$. The non-interacting Green's function can readily be expressed in terms of the single-particle Hamiltonian
from \cref{app:eqn:full_THF_Hamiltonian_gamma_bas}, 
\begin{equation}
	\label{app:eqn:gf_within_non_int}
	\mathcal{G}^{0} \left(i\omega_n,  \vec{k} \right) = \left[\left( i\omega_n + \mu \right) \mathbb{1} - h \left( \vec{k} \right)  \right]^{-1}, \qq{where} h \left( \vec{k} \right) = h^{\text{TBG}} \left( \vec{k} \right), h^{\text{TSTG}} \left( \vec{k} \right).
\end{equation}
The fully-interacting Green's function and the non-interacting one are connected via the Dyson equation~\cite{MAH00}
\begin{equation}
	\label{app:eqn:dyson_equation}
	\mathcal{G} \left( i \omega_n, \vec{k} \right) = \left[ \left( \mathcal{G}^{0} \left( i \omega_n, \vec{k} \right) \right)^{-1} - \Sigma \left(i \omega_n, \vec{k} \right) \right]^{-1}, 
\end{equation}
where $\Sigma \left(i \omega_n, \vec{k} \right)$ is by definition the self-energy matrix of the system. 

The Matsubara Green's function is defined at discrete values along the imaginary axes ({\it i.e.}{}, the Matsubara frequencies). Nevertheless, it can also be analytically continued across the entire complex plane~\cite{MAH00}. Of special interest to us will be the retarded Green's function $	\mathcal{G}_{i \eta s; i' \eta' s'} \left(\omega + i 0^{+}, \vec{k} \right)$, which is defined along the real $\omega$-axis with a small positive imaginary offset (hereby denoted by ``$+ i 0^{+}$''). To relate the Matsubara and retarded Green's functions, we introduce the Lehman representation of the Green's function~\cite{MAH00}. We let $\ket{m}$ and $E_m$ denote, respectively, the exact many-body eigenstates and eigenenergies of the Grand canonical Hamiltonian $K$. Inserting resolutions of the identity operator in terms of the eigenbasis of $K$ in \cref{app:eqn:matsubara_gf_THF_tau}, we can express the imaginary-time Matsubara Green's function as 
\begin{equation}
	\mathcal{G}_{i \eta s;i' \eta' s'} \left( \tau, \vec{k} \right)= - \frac{1}{Z} \sum_{n,m} e^{ \left( \tau - \beta \right)E_n} \mel**{n}{\hat{\gamma}_{\vec{k},i, \eta, s}}{m} e^{-\tau E_m } \mel**{m}{\hat{\gamma}^\dagger_{\vec{k},i', \eta', s'}}{n}, \qq{for} \tau > 0.
\end{equation}
We then perform a Fourier transformation according to \cref{app:eqn:matsubara_gf_THF_ft} and obtain
\begin{equation}
	\mathcal{G}_{i \eta s;i' \eta' s'} \left( i\omega_n, \vec{k} \right)= \frac{1}{Z}\sum_{n,m}  \mel**{n}{\hat{\gamma}_{\vec{k},i, \eta, s}}{m} \mel**{m}{\hat{\gamma}^\dagger_{\vec{k},i', \eta', s'}}{n} \frac{e^{ - \beta E_n} - e^{- \beta E_m}}{i \omega_n + E_n - E_m},
\end{equation}
such that the retarded Green's function is given by
\begin{equation}
	\mathcal{G}_{i \eta s;i' \eta' s'} \left( \omega + i 0^{+}, \vec{k} \right)= \frac{1}{Z}\sum_{n,m}  \mel**{n}{\hat{\gamma}_{\vec{k},i, \eta, s}}{m} \mel**{m}{\hat{\gamma}^\dagger_{\vec{k},i', \eta', s'}}{n} \frac{e^{ - \beta E_n} - e^{- \beta E_m}}{\omega + i 0^{+} + E_n - E_m},
\end{equation}
We also introduce the Fermion spectral function of the system, $A_{i \eta s; i' \eta' s'} \left( \omega, \vec{k} \right)$, which is defined in terms of the retarded Green's function as
\begin{align}
	A_{i \eta s; i' \eta' s'} \left( \omega, \vec{k} \right) &= \frac{-1}{2 \pi i} \left( \mathcal{G}_{i \eta s; i' \eta' s'} \left(\omega + i 0^{+}, \vec{k} \right) - \mathcal{G}^{*}_{i' \eta' s',i \eta s} \left(\omega + i 0^{+}, \vec{k} \right) \right) \nonumber \\
	&= \frac{1}{Z}\sum_{n,m}  \mel**{n}{\hat{\gamma}_{\vec{k},i, \eta, s}}{m} \mel**{m}{\hat{\gamma}^\dagger_{\vec{k},i', \eta', s'}}{n} \left( e^{ - \beta E_n} - e^{- \beta E_m} \right) \delta \left( \omega + E_n - E_m \right),  \label{app:eqn:spectral_function}
\end{align}
and which is Hermitian and positive semi-definite~\cite{LUT61,PAV19}. The Green's function for generic complex frequencies $z$ can then be recovered from the spectral function
\begin{equation}
	\label{app:eqn:spectral_rep_of_GF}
	\mathcal{G}_{i \eta s; i' \eta' s'} \left(z, \vec{k} \right) = \int_{-\infty}^{\infty} \frac{\dd{\omega}}{z-\omega} A_{i \eta s; i' \eta' s'} \left( \omega, \vec{k} \right).
\end{equation} 

Finally, the density matrix of the system is given by
\begin{equation}
	\label{app:eqn:def_rho_HF}
	\varrho_{i \eta s; i' \eta' s'} \left(\vec{k} \right) = \left\langle  :\mathrel{\hat{\gamma}^\dagger_{\vec{k}, i, \eta, s} \hat{\gamma}_{\vec{k}, i' \eta' s'}}: \right\rangle = \int_{-\infty}^{\infty} \dd{\omega} n_{\mathrm{F}} \left( \omega \right) A_{i' \eta' s'; i \eta s} \left( \omega, \vec{k} \right) - \frac{1}{2} \delta_{ii'}\delta_{\eta \eta'}\delta_{s s'},
\end{equation}
where $n_{\mathrm{F}} \left( \omega \right)$ denotes the Fermi-Dirac distribution function 
\begin{equation}
	\label{app:eqn:fd_distribution}
	n_{\mathrm{F}} \left( \omega \right) = \frac{1}{e^{\beta \omega} + 1}.
\end{equation}
Additionally, we define the total fillings of the $c$-, $f$-, and $d$-electrons to be 
\begin{equation}
	\label{app:eqn:flavor_filing}
	\nu_c = \frac{1}{N_0} \sum_{i=1}^{4} \sum_{\vec{k},\eta,s} \varrho_{i \eta s; i \eta s} \left(\vec{k} \right), \quad
	\nu_f = \frac{1}{N_0} \sum_{i=5}^{6} \sum_{\vec{k},\eta,s} \varrho_{i \eta s; i \eta s} \left(\vec{k} \right), \quad
	\nu_d = \frac{1}{N_0} \sum_{i=7}^{8} \sum_{\vec{k},\eta,s} \varrho_{i \eta s; i \eta s} \left(\vec{k} \right),
\end{equation} 
as well as the total electron filling $\nu = \nu_c + \nu_f + \nu_d$.

\subsection{The Hartree-Fock Hamiltonian}\label{app:sec:hartree_fock:hamiltonian}

In this section we briefly review the Hartree-Fock theory for the THF model. We start from the THF interaction Hamiltonian, which, for TBG, can be rewritten as 
{
	\small
	\begin{equation}
		\label{app:eqn:general_TBG_int_forHF}
		H^{\text{TBG}}_{I} = \sum_{\substack{i,\eta_1,s_1 \\ j,\eta_2,s_2}}  \sum_{\substack{k,\eta_3,s_3 \\ l,\eta_4,s_4}}  \sum_{\substack{\vec{k}_1,\vec{k}_2,\vec{q} \\ \vec{k}'_1, \vec{k}'_2}} V_{i \eta_1 s_1; j \eta_2 s_2; k \eta_3 s_3; l \eta_4 s_4} \left( \vec{q} \right) \delta_{\vec{k}'_2,\vec{k}_2 - \vec{q}} \delta_{\vec{k}'_1,\vec{k}_1 + \vec{q}} 
		:\mathrel{\hat{\gamma}^\dagger_{\vec{k}'_1 ,i,\eta_1,s_1} \hat{\gamma}_{\vec{k}_1,j,\eta_2,s_2}}:
		:\mathrel{\hat{\gamma}^\dagger_{\vec{k}'_2,k,\eta_3,s_3} \hat{\gamma}_{\vec{k}_2,l,\eta_4,s_4}}:.
\end{equation}}The interaction tensor from \cref{app:eqn:general_TBG_int_forHF} can be directly obtained by casting \crefrange{app:eqn:THF_int:U1}{app:eqn:THF_int:K} into the notation introduced in \cref{app:eqn:shorthand_gamma_not}, and equating \cref{app:eqn:general_TBG_int_forHF} with \cref{app:eqn:THF_interaction_TBG}. For example, the $H_{U_1}$ and $H_{U_2}$ contributions from \cref{app:eqn:THF_int:U1,app:eqn:THF_int:U2} give rise to the following terms in the interaction tensor
\begin{align}
	V_{i \eta_1 s_1; j \eta_2 s_2; k \eta_3 s_3; l \eta_4 s_4} \left( \vec{q} \right) =& \sum_{\alpha,\alpha'=1}^{2} \left( \frac{U_1}{2} + \frac{U_2}{2} \sum_{n=0}^{5} \cos \left( \vec{q} \cdot C^n_{6z} \vec{a}_{M_1} \right) \right) \nonumber\\
	\times& \delta_{\eta_1 \eta_2} \delta_{\eta_3 \eta_4} \delta_{s_1 s_2} \delta_{s_3 s_4} \delta_{i(\alpha+4)} \delta_{j(\alpha+4)} \delta_{k(\alpha'+4)} \delta_{l(\alpha'+4)} + \dots,
\end{align}
with the ``$\dots$'' denoting contributions from the other interaction terms. We note that in \cref{app:eqn:general_TBG_int_forHF}, the summation over $\vec{k}_1$, $\vec{k}_2$, $\vec{k}'_1$, and $\vec{k}'_2$ is performed within the first BZ, whereas $1 \leq i,j,k,l \leq 6$. Using the standard Hartree-Fock decoupling procedure, the Hartree-Fock (or mean-field) interaction Hamiltonian can be written as
\begin{equation}
	\label{app:eqn:full_TBG_HF_Hamiltonian}
	H^{\text{TBG}}_{I,\text{MF}} = \sum_{\substack{i,\eta,s \\ i',\eta',s'}} h^{\text{TBG},I,\text{MF}}_{i \eta s; i' \eta' s'} \left( \vec{k} \right) \hat{\gamma}^\dagger_{\vec{k},i,\eta,s} \hat{\gamma}_{\vec{k},i',\eta',s'},
\end{equation}
where, for a given state of the system characterized by the density matrix $\varrho_{m \eta_1 s_1; n \eta_2 s_2} \left(\vec{k} \right)$, the Hartree-Fock matrix is given by
\begin{align}
	\label{app:eqn:genera_TBG_int_HF}
	h^{\text{TBG},I,\text{MF}}_{i \eta s; i' \eta' s'} \left( \vec{k} \right) 
	=& 2 \sum_{\vec{k}'} \sum_{\substack{m,\eta_1,s_1 \\ n,\eta_2,s_2}} \left( V_{n \eta_1 s_1; m \eta_2 s_2; i \eta s; i' \eta' s'} \left( \vec{0} \right) 	\varrho_{n \eta_1 s_1; m \eta_2 s_2} \left(\vec{k}' \right) \right. \nonumber\\
	-& \left. V_{i \eta s; m \eta_2 s_2; n \eta_1 s_1; i' \eta' s'} \left( \vec{k} - \vec{k}' \right) \varrho_{n \eta_1 s_1; m \eta_2 s_2} \left(\vec{k}' \right) \right).
\end{align}
The \emph{total} mean-field Hamiltonian is obtained by summing the single-particle Hamiltonian and the Hartree-Fock interaction one, $H^{\text{TBG}}_{\text{MF}} = H^{\text{TBG}}_{0} + H^{\text{TBG}}_{I,\text{MF}}$, with the corresponding matrix reading as
\begin{equation}
	\label{app:eqn:TBG_HF_Hamiltonian}
	h^{\text{TBG},\text{MF}}_{i \eta s; i' \eta' s'} \left( \vec{k} \right) = h^{\text{TBG},\eta}_{i i'} \left(\vec{k} \right) \delta_{\eta \eta'} \delta_{s s'}+  h^{\text{TBG},I,\text{MF}}_{i \eta s; i' \eta' s'} \left( \vec{k} \right).
\end{equation}

The THF interaction Hamiltonian of TSTG contains three additional contributions given by \cref{app:eqn:TSTG_int:Vd,app:eqn:TSTG_int:Vcd,app:eqn:TSTG_int:Vfd} and which involve the $d$-electrons. For simplicity, we will follow Ref.~\cite{YU23a} and decouple $H_V^{d}$ and $H_V^{cd}$ only in the Hartree channel, but decouple $H_W^{fd}$ in both the Hartree and the Fock channels\footnote{In general, symmetry breaking is promoted by the Fock term~\cite{SON22}. For the TBG and TSTG correlated states, the symmetry breaking mainly occurs within the $f$-electron modes. As such, it is reasonable to ingore the Fock channels of the $H_V^{d}$ and $H_V^{cd}$ terms, which do \emph{not} involve any $f$-electrons~\cite{YU23a}.}. The corresponding Hartree-Fock matrix will be given by
\begin{align}
	h^{\text{TSTG},I,\text{MF}}_{i \eta s; i' \eta' s'} \left( \vec{k} \right) =& h^{\text{TBG},I,\text{MF}}_{i \eta s; i' \eta' s'} \left( \vec{k} \right) + \left[\frac{V \left( \vec{0} \right)}{\Omega_0} \left( \nu_d + \nu_c \right) + W_{fd} \nu_f \right] \sum_{\alpha=1}^{2} \delta_{i (\alpha + 6)} \delta_{i' (\alpha + 6)} \delta_{\eta \eta'} \delta_{s s'} \nonumber \\
	+& \frac{V \left( \vec{0} \right)}{\Omega_0} \nu_d \sum_{\alpha=1}^{2} \delta_{i (\alpha + 4)} \delta_{i' (\alpha + 4)} \delta_{\eta \eta'} \delta_{s s'} + W_{fd} \nu_d \sum_{\alpha=1}^{2} \delta_{i \alpha} \delta_{i' \alpha } \delta_{\eta \eta'} \delta_{s s'} \nonumber \\
	-& \frac{W_{fd}}{N_0} \sum_{\vec{k}'} \sum_{\alpha,\alpha'=1}^{2} \left( \varrho_{(\alpha'+6) \eta' s'; \alpha \eta s} \left( \vec{k}' \right) \delta_{\alpha i} \delta_{(\alpha'+6)i'} + \varrho_{\alpha' \eta' s'; (\alpha + 6) \eta s} \left( \vec{k}' \right) \delta_{(\alpha+6) i} \delta_{\alpha'i'} \right),	\label{app:eqn:TSTG_HF_Hamiltonian}
\end{align}
while the total mean-field Hamiltonian for TSTG is given by
\begin{equation}
	\label{app:eqn:full_TSTG_HF_Hamiltonian}
	h^{\text{TSTG},\text{MF}}_{i \eta s; i' \eta' s'} \left( \vec{k} \right) = h^{\text{TSTG},\eta}_{i i'} \left(\vec{k} \right) \delta_{\eta \eta'} \delta_{s s'} +  h^{\text{TSTG},I,\text{MF}}_{i \eta s; i' \eta' s'} \left( \vec{k} \right).
\end{equation}

We note that at self-consistency, the Hartree-Fock Hamiltonian matrix from \cref{app:eqn:TSTG_HF_Hamiltonian} or \cref{app:eqn:full_TSTG_HF_Hamiltonian} and the density matrix defined in \cref{app:eqn:def_rho_HF} are related by 
\begin{equation}
	\label{app:eqn:hf_scf_fin_temp}
	\varrho^{T} \left( \vec{k} \right) = \left\lbrace \exp \left[  \beta \left( h^{\text{MF}} \left( \vec{k} \right) - \mu \mathbb{1} \right) \right] + \mathbb{1} \right\rbrace ^{-1} - \frac{1}{2} \mathbb{1},
\end{equation}
where $\mathbb{1}$ denotes the identity matrix. Because $h^{\text{MF}} \left( \vec{k} \right)$ itself depends on the density matrix $\varrho \left( \vec{k} \right)$, \cref{app:eqn:hf_scf_fin_temp} must be solved self-consistently, as we will explain in \cref{app:sec:hartree_fock:ground_states}. \Cref{app:eqn:hf_scf_fin_temp} follows straightforwardly by noting that within the Hartree-Fock approximation ({\it i.e.}{} with only the Hartree-Fock interaction contribution being included), the Green's function of the system is given simply by
\begin{equation}
	\label{app:eqn:gf_within_hf}
	\mathcal{G} \left(i\omega_n,  \vec{k} \right) = \left[\left( i\omega_n + \mu \right) \mathbb{1} - h^{\text{MF}} \left( \vec{k} \right)  \right]^{-1},
\end{equation}
which implies that the spectral function of the system is a Dirac $\delta$-function,
\begin{equation}
	\label{app:eqn:hf_spectral_func}
	A (\omega,\vec{k}) = \delta \left( \left( \omega + \mu \right) \mathbb{1} - h^{\text{MF}} \left( \vec{k} \right) \right).
\end{equation}
By substituting \cref{app:eqn:hf_spectral_func} into \cref{app:eqn:def_rho_HF}, one directly recovers \cref{app:eqn:hf_scf_fin_temp}. The chemical potential is fixed at each step by requiring that the filling of the system equals some target value. 

Finally, it is also worth noting that from \cref{app:eqn:gf_within_non_int,app:eqn:gf_within_hf,app:eqn:dyson_equation}, one can obtain the Hartree-Fock self-energy of the system (which is first order in the interaction), 
\begin{equation}
	\Sigma \left( i\omega_n, \vec{k} \right) = h^{I,\text{MF}} \left( \vec{k} \right).
\end{equation}
The Hartree-Fock self-energy is none other than the Hartree-Fock interaction Hamiltonian. Being frequency-independent, the former is often called the \emph{static} contribution to the self-energy. 

\subsection{Correlated ground state candidates}\label{app:sec:hartree_fock:ground_states}

\subsubsection{Model states}\label{app:sec:hartree_fock:ground_states:model}

\begin{table}[t]
	\centering
	\begin{tabular}{|c|l|r|}
		\hline
		$\nu$ & State & Parent state \\
		\hline\hline
		$-4$ & $\IfStrEqCase{1}{{1}{\ket{\nu={-}4} }
		{2}{\ket{\nu={-}3, \mathrm{IVC}}}
		{3}{\ket{\nu={-}3, \mathrm{VP}}}
		{4}{\ket{\nu={-}2, \mathrm{K-IVC}}}
		{5}{\ket{\nu={-}2, \mathrm{VP}}}
		{6}{\ket{\nu={-}1, (\mathrm{K-IVC}+\mathrm{VP})}}
		{7}{\ket{\nu={-}1, \mathrm{VP}}}
		{8}{\ket{\nu=0, \mathrm{K-IVC}}}
		{9}{\ket{\nu=0, \mathrm{VP}}}
	}
	[nada]
$ & $\ket{\mathrm{FS}}$ \\
		\hline
		\multirow[b]{2}{*}{$-3$} & $\IfStrEqCase{2}{{1}{\ket{\nu={-}4} }
		{2}{\ket{\nu={-}3, \mathrm{IVC}}}
		{3}{\ket{\nu={-}3, \mathrm{VP}}}
		{4}{\ket{\nu={-}2, \mathrm{K-IVC}}}
		{5}{\ket{\nu={-}2, \mathrm{VP}}}
		{6}{\ket{\nu={-}1, (\mathrm{K-IVC}+\mathrm{VP})}}
		{7}{\ket{\nu={-}1, \mathrm{VP}}}
		{8}{\ket{\nu=0, \mathrm{K-IVC}}}
		{9}{\ket{\nu=0, \mathrm{VP}}}
	}
	[nada]
$ & $\displaystyle \prod_{\vec{k}} \frac{1}{\sqrt{2}} \left( \hat{f}^\dagger_{\vec{k},1,+,\uparrow} - i\hat{f}^\dagger_{\vec{k},2,-,\uparrow} \right)\ket{\mathrm{FS}}$ \\
		& $\IfStrEqCase{3}{{1}{\ket{\nu={-}4} }
		{2}{\ket{\nu={-}3, \mathrm{IVC}}}
		{3}{\ket{\nu={-}3, \mathrm{VP}}}
		{4}{\ket{\nu={-}2, \mathrm{K-IVC}}}
		{5}{\ket{\nu={-}2, \mathrm{VP}}}
		{6}{\ket{\nu={-}1, (\mathrm{K-IVC}+\mathrm{VP})}}
		{7}{\ket{\nu={-}1, \mathrm{VP}}}
		{8}{\ket{\nu=0, \mathrm{K-IVC}}}
		{9}{\ket{\nu=0, \mathrm{VP}}}
	}
	[nada]
$ & $\displaystyle \prod_{\vec{k}} \hat{f}^\dagger_{\vec{k},1,+,\uparrow} \ket{\mathrm{FS}}$ \\
		\hline
		\multirow[b]{2}{*}{$-2$} & $\IfStrEqCase{4}{{1}{\ket{\nu={-}4} }
		{2}{\ket{\nu={-}3, \mathrm{IVC}}}
		{3}{\ket{\nu={-}3, \mathrm{VP}}}
		{4}{\ket{\nu={-}2, \mathrm{K-IVC}}}
		{5}{\ket{\nu={-}2, \mathrm{VP}}}
		{6}{\ket{\nu={-}1, (\mathrm{K-IVC}+\mathrm{VP})}}
		{7}{\ket{\nu={-}1, \mathrm{VP}}}
		{8}{\ket{\nu=0, \mathrm{K-IVC}}}
		{9}{\ket{\nu=0, \mathrm{VP}}}
	}
	[nada]
$ & $\displaystyle \prod_{\vec{k}} \frac{1}{2} \left( \hat{f}^\dagger_{\vec{k},1,+,\uparrow} - i\hat{f}^\dagger_{\vec{k},2,-,\uparrow} \right) \left( \hat{f}^\dagger_{\vec{k},2,+,\uparrow} + i\hat{f}^\dagger_{\vec{k},1,-,\uparrow} \right) \ket{\mathrm{FS}}$ \\
		& $\IfStrEqCase{5}{{1}{\ket{\nu={-}4} }
		{2}{\ket{\nu={-}3, \mathrm{IVC}}}
		{3}{\ket{\nu={-}3, \mathrm{VP}}}
		{4}{\ket{\nu={-}2, \mathrm{K-IVC}}}
		{5}{\ket{\nu={-}2, \mathrm{VP}}}
		{6}{\ket{\nu={-}1, (\mathrm{K-IVC}+\mathrm{VP})}}
		{7}{\ket{\nu={-}1, \mathrm{VP}}}
		{8}{\ket{\nu=0, \mathrm{K-IVC}}}
		{9}{\ket{\nu=0, \mathrm{VP}}}
	}
	[nada]
$ & $\displaystyle \prod_{\vec{k}} \hat{f}^\dagger_{\vec{k},1,+,\uparrow} \hat{f}^\dagger_{\vec{k},2,+,\uparrow} \ket{\mathrm{FS}}$ \\
		\hline
		\multirow[b]{2}{*}{$-1$} & $\IfStrEqCase{6}{{1}{\ket{\nu={-}4} }
		{2}{\ket{\nu={-}3, \mathrm{IVC}}}
		{3}{\ket{\nu={-}3, \mathrm{VP}}}
		{4}{\ket{\nu={-}2, \mathrm{K-IVC}}}
		{5}{\ket{\nu={-}2, \mathrm{VP}}}
		{6}{\ket{\nu={-}1, (\mathrm{K-IVC}+\mathrm{VP})}}
		{7}{\ket{\nu={-}1, \mathrm{VP}}}
		{8}{\ket{\nu=0, \mathrm{K-IVC}}}
		{9}{\ket{\nu=0, \mathrm{VP}}}
	}
	[nada]
$ & $\displaystyle \prod_{\vec{k}} \frac{1}{2} \hat{f}^\dagger_{\vec{k},1,+,\downarrow} \left( \hat{f}^\dagger_{\vec{k},1,+,\uparrow} - i\hat{f}^\dagger_{\vec{k},2,-,\uparrow} \right) \left( \hat{f}^\dagger_{\vec{k},2,+,\uparrow} + i\hat{f}^\dagger_{\vec{k},1,-,\uparrow} \right) \ket{\mathrm{FS}}$ \\
		& $\IfStrEqCase{7}{{1}{\ket{\nu={-}4} }
		{2}{\ket{\nu={-}3, \mathrm{IVC}}}
		{3}{\ket{\nu={-}3, \mathrm{VP}}}
		{4}{\ket{\nu={-}2, \mathrm{K-IVC}}}
		{5}{\ket{\nu={-}2, \mathrm{VP}}}
		{6}{\ket{\nu={-}1, (\mathrm{K-IVC}+\mathrm{VP})}}
		{7}{\ket{\nu={-}1, \mathrm{VP}}}
		{8}{\ket{\nu=0, \mathrm{K-IVC}}}
		{9}{\ket{\nu=0, \mathrm{VP}}}
	}
	[nada]
$ & $\displaystyle \prod_{\vec{k}} \hat{f}^\dagger_{\vec{k},1,+,\downarrow} \hat{f}^\dagger_{\vec{k},1,+,\uparrow} \hat{f}^\dagger_{\vec{k},2,+,\uparrow} \ket{\mathrm{FS}}$ \\
		\hline
		\multirow[b]{2}{*}{$0$} & $\IfStrEqCase{8}{{1}{\ket{\nu={-}4} }
		{2}{\ket{\nu={-}3, \mathrm{IVC}}}
		{3}{\ket{\nu={-}3, \mathrm{VP}}}
		{4}{\ket{\nu={-}2, \mathrm{K-IVC}}}
		{5}{\ket{\nu={-}2, \mathrm{VP}}}
		{6}{\ket{\nu={-}1, (\mathrm{K-IVC}+\mathrm{VP})}}
		{7}{\ket{\nu={-}1, \mathrm{VP}}}
		{8}{\ket{\nu=0, \mathrm{K-IVC}}}
		{9}{\ket{\nu=0, \mathrm{VP}}}
	}
	[nada]
$ & $\displaystyle \prod_{\vec{k}} \prod_{s} \frac{1}{4} \left( \hat{f}^\dagger_{\vec{k},1,+,s} - i\hat{f}^\dagger_{\vec{k},2,-,s} \right) \left( \hat{f}^\dagger_{\vec{k},2,+,s} + i\hat{f}^\dagger_{\vec{k},1,-,s} \right) \ket{\mathrm{FS}}$ \\
		& $\IfStrEqCase{9}{{1}{\ket{\nu={-}4} }
		{2}{\ket{\nu={-}3, \mathrm{IVC}}}
		{3}{\ket{\nu={-}3, \mathrm{VP}}}
		{4}{\ket{\nu={-}2, \mathrm{K-IVC}}}
		{5}{\ket{\nu={-}2, \mathrm{VP}}}
		{6}{\ket{\nu={-}1, (\mathrm{K-IVC}+\mathrm{VP})}}
		{7}{\ket{\nu={-}1, \mathrm{VP}}}
		{8}{\ket{\nu=0, \mathrm{K-IVC}}}
		{9}{\ket{\nu=0, \mathrm{VP}}}
	}
	[nada]
$ & $\displaystyle \prod_{\vec{k}} \prod_{\alpha,s}\hat{f}^\dagger_{\vec{k},\alpha,+,s} \ket{\mathrm{FS}}$ \\
		\hline
		$\nu' > 0$ & $\ket{\nu = \nu',\text{Type}}$ & $\mathcal{P}^{(\prime)} \ket{\nu = -\nu',\text{Type}}$ \\
		\hline
	\end{tabular}
 	\caption{Correlated ground state candidates considered in this work. For each filling $\nu$ we list the correlated ground state candidate, as well as the wave functions of the corresponding parent states~\cite{SON22,YU23a}. Each correlated state is obtained using zero-temperature Hartree-Fock starting from the corresponding parent state. In turn, the parent states are obtained by occupying $(\nu+4)$ $f$-electron bands on top of a half-filled Fermi sea of $c$- (for TBG) or $c$- and $d$- (for TSTG) electrons (denoted as $\ket{\mathrm{FS}}$). For all integer fillings $-4 < \nu \leq 0$, we consider valley-polarized (VP) states. At $\nu = -2,0$, we also consider the Kramers intervalley-coherent (K-IVC) states~\cite{BUL20a}. At $\nu=-3$ there is only one maximal intervalley-coherent (IVC) state, while for $\nu=-1$, the ground state has been shown to have two filled K-IVC $f$-electron bands and one filled VP one~\cite{LIA21,SON22}. Finally, we note that for TBG, these correlated states are insulating~\cite{SON22}, while for TSTG they are metallic~\cite{YU23a}. The ground state candidates at positive integer fillings are obtained from the ones at negative integer fillings using the charge conjugation operator $\mathcal{P}$ (for TBG~\cite{SON19,SON21,BER21a,CAL21,SON22}) or $\mathcal{P}'$ (for TSTG~\cite{CAL21,YU23a}), which are defined, respectively, in \cref{app:eqn:ph_sym_act_TBG,app:eqn:ph_sym_act_TSTG}.}
	\label{app:tab:model_states}
\end{table}

For both TBG and TSTG, in this work, we consider nine different correlated ground state candidates at different integer fillings (together with their charge conjugated counterparts). These are summarized in \cref{app:tab:model_states}. Within Hartree-Fock, the wave function of each state (or equivalently its self-consistent density matrix) is obtained at zero temperature by starting from the indicated parent states~\cite{SON22}, as will be explained below. The parent states themselves are obtained by fully-populating $f$-fermion bands on top of a half-filled Fermi sea ($\ket{\mathrm{FS}}$) of just $c$-electrons in the case of TBG or both $c$- and $d$-electrons in the case of TSTG. The states we consider here have been shown to be ground states or low-energy states in TBG~\cite{SON22} or TSTG at low values of the displacement field $\mathcal{E}$ and in the absence of strain~\cite{YU23a}. 

To obtain the self-consistent density matrix for a given type of ground state candidate at filling $\nu_0$, we start from an initial guess of the density matrix ({\it i.e.}{}, obtained directly from the parent states in \cref{app:tab:model_states}) and form the Hartree-Fock Hamiltonian according to \cref{app:eqn:full_TBG_HF_Hamiltonian} or \cref{app:eqn:full_TSTG_HF_Hamiltonian}. The latter is then diagonalized as
\begin{equation}
	\label{app:eqn:diag_hf_ham}
	\sum_{i',\eta',s'} h^{\text{MF}}_{i \eta s; i' \eta' s'} 
	\left( \vec{k} \right) \varphi_{n;i' \eta' s'} \left( \vec{k} \right) = \epsilon_n \left( \vec{k} \right) \varphi_{n;i \eta s} \left( \vec{k} \right),
\end{equation}
where $\epsilon_n \left( \vec{k} \right)$ and $\varphi_{n;i' \eta' s'} \left( \vec{k} \right)$ denote the $n$-th Hartree-Fock energy and eigenvector. A new mean-field density matrix is then constructed according to 
\begin{equation}
	\label{app:eqn:assemble_rho_hf}
	\varrho_{i \eta s; i' \eta' s'} \left(\vec{k} \right) = \sum_{n} \frac{1}{e^{\beta \left(\epsilon_n \left( \vec{k} \right) - \mu \right)}+1}\varphi^{*}_{n;i \eta s} \left( \vec{k} \right) \varphi_{n;i' \eta' s'} \left( \vec{k} \right),
\end{equation}
where the chemical potential is fixed by requiring that the total filling of the system equals the initial filling, {\it i.e.}{} $\nu = \nu_0$. With the new density matrix obtained, the Hartree-Fock Hamiltonian is then recomputed. The two steps in \cref{app:eqn:diag_hf_ham,app:eqn:assemble_rho_hf} are repeated until self-consistency is achieved. 

Numerically, we accelerate convergence using the direct inversion in the iterative subspace (DIIS) algorithm~\cite{PUL80,PUL82,GAR12} at both zero and finite temperature. Additionally, at zero temperature, we employ the relaxed constrained algorithm (RCA)~\cite{CAN00,GAR12}, as well as the so-called ``energy-DIIS'' (EDIIS) algorithm~\cite{KUD02,GAR12}. Convergence is ascertained by noting that at self-consistency, \cref{app:eqn:hf_scf_fin_temp} implies that the Hartree-Fock Hamiltonian computed from \cref{app:eqn:TBG_HF_Hamiltonian,app:eqn:full_TSTG_HF_Hamiltonian} and the transpose of the density matrix computed from \cref{app:eqn:diag_hf_ham,app:eqn:assemble_rho_hf} commute.

We note that at a given integer filling $\nu_0$, the self-consistent Hartree-Fock density matrices obtained starting from different parent states will generically be different, as they will respect the symmetries of the corresponding parent states\footnote{This is because the self-consistent density matrix is obtained iteratively. The Hartree-Fock Hamiltonian obtained from a density matrix obeying some of the symmetries of the system will also obey those symmetries. Conversely, diagonalizing a Hartree-Fock Hamiltonian obeying some of the symmetries of the system will lead to a density matrix obeying those symmetries. A more in-depth discussion is provided in Ref.~\cite{RAI23a}.}. For example, at $\nu=-2$, the self-consistent state $\IfStrEqCase{5}{{1}{\ket{\nu={-}4} }
		{2}{\ket{\nu={-}3, \mathrm{IVC}}}
		{3}{\ket{\nu={-}3, \mathrm{VP}}}
		{4}{\ket{\nu={-}2, \mathrm{K-IVC}}}
		{5}{\ket{\nu={-}2, \mathrm{VP}}}
		{6}{\ket{\nu={-}1, (\mathrm{K-IVC}+\mathrm{VP})}}
		{7}{\ket{\nu={-}1, \mathrm{VP}}}
		{8}{\ket{\nu=0, \mathrm{K-IVC}}}
		{9}{\ket{\nu=0, \mathrm{VP}}}
	}
	[nada]
$ will be valley polarized, while the self-consistent one $\IfStrEqCase{4}{{1}{\ket{\nu={-}4} }
		{2}{\ket{\nu={-}3, \mathrm{IVC}}}
		{3}{\ket{\nu={-}3, \mathrm{VP}}}
		{4}{\ket{\nu={-}2, \mathrm{K-IVC}}}
		{5}{\ket{\nu={-}2, \mathrm{VP}}}
		{6}{\ket{\nu={-}1, (\mathrm{K-IVC}+\mathrm{VP})}}
		{7}{\ket{\nu={-}1, \mathrm{VP}}}
		{8}{\ket{\nu=0, \mathrm{K-IVC}}}
		{9}{\ket{\nu=0, \mathrm{VP}}}
	}
	[nada]
$ will be Kramers intervalley-coherent. For both TBG~\cite{BRI22, BUL20a, BUL20b, CEA20, CHA21, CHE21, CHI20b, CHR20, CLA19, DA19, DA21, DOD18, EUG20, HOF22, HUA19, KAN19, KAN20a, KAN21, KEN18, KOS18, KWA21, LIA21, LIU19, LIU21a, OCH18, PO18a, REP20, SEO19, SON22, THO18, VAF20, VEN18, WAG22, WU19, WU20, XIE20b, XIE21, XIE23a, XU18b, YUA18, ZHA20, ZHA23a} and TSTG~\cite{CHR22, XIE21b, YU23a}, a myriad of correlated ground states have been proposed. Because the different states at a given integer filling are close in energy (with the precise hierarchy being determined by effects such as strain, relaxation, {\it etc.}{}), we consider multiple correlated ground state candidates. Nevertheless, the differences in their thermoelectric transport~\cite{CAL24} are primarily quantitative rather than qualitative.

\subsubsection{Many-body charge conjugation}\label{app:sec:hartree_fock:ground_states:ph_symmetry}

Given that the THF models for both TBG and TSTG without lattice relaxation possess many-body (unitary) charge conjugation symmetry~\cite{SON19,SON21,BER21a,CAL21,SON22,YU23a}, we can restrict, without loss of generality, to positive fillings with $\nu \geq 0$. For concreteness, the many-body charge conjugation operator $\mathcal{P}$ is defined for TBG as the single-particle transformation $C_{2z}TP$ (where $P$, $T$, and $C_{2z}$ are the unitary particle-hole, the antiunitary time-reversal, and two-fold $z$-directed unitary rotation symmetry operations) followed by an exchange of the creation and annihilation operators~\cite{BER21a,SON22},
\begin{equation}
	\label{app:eqn:ph_sym_act_TBG}
	\mathcal{P} \hat{\gamma}^\dagger_{\vec{k},i,\eta,s} \mathcal{P}^{-1} = \sum_{i',\eta'} \left[ D \left( \mathcal{P} \right) \right]_{i' \eta'; i \eta} \hat{\gamma}_{-\vec{k},i',\eta',s},
\end{equation}
with the representation matrix of $\mathcal{P}$ being
\begin{equation}
	\label{app:eqn:ph_sym_rep_TBG}
	D \left( \mathcal{P} \right) = \left[ \sigma_y \oplus \sigma_y \oplus \left( - \sigma_y \right) \right] \tau_z.
\end{equation}
In \cref{app:eqn:ph_sym_rep_TBG}, $\sigma_{\mu}$ ($\mu=0,x,y,z$) represent the Pauli and identity matrices acting on the $i=1,2$, $i=3,4$, and $i=5,6$, orbital subspaces (and additionally in the $i=7,8$ orbital subspace in the case of TSTG), while $\tau_{\mu}$ ($\mu=0,x,y,z$) are the Pauli and identity matrices operating on the valley subspace. For TBG, the many-body THF Hamiltonian is symmetric under the action of $\mathcal{P}$ ({\it i.e.}{}, $\commutator{H^{\text{TBG}}}{\mathcal{P}} = 0$). In contrast, TSTG at a finite displacement field does not maintain this symmetry. Instead, it exhibits a \emph{spatial} many-body charge conjugation symmetry, represented by $\mathcal{P}'$~\cite{CAL21}. The latter is defined as the combined antiunitary single-particle transformation $m_z C_{2x} C_{2z} T P$ (where $C_{2x}$ denotes a two-folder rotation around the $x$ axis and $m_z$ is the mirror symmetry operator perpendicular to the $z$ axis) followed by an interchange between the creation and annihilation operators~\cite{CAL21}. Its action on the TSTG THF fermions is described by
\begin{equation}
	\label{app:eqn:ph_sym_act_TSTG}
	\mathcal{P}' \hat{\gamma}^\dagger_{\vec{k},i,\eta,s} \mathcal{P}^{\prime-1} = \sum_{i',\eta'} \left[ D \left( \mathcal{P}' \right) \right]_{i' \eta'; i \eta} \hat{\gamma}_{-C_{2x} \vec{k},i',\eta',s},
\end{equation}
where the representation matrix of $\mathcal{P}'$ reads as
\begin{equation}
	\label{app:eqn:ph_sym_rep_TSTG}
	D \left( \mathcal{P}' \right) = \left[ i\sigma_z \oplus i\sigma_z \oplus \left( - i\sigma_z \right) \oplus \left( - \sigma_0 \right) \right] \tau_z.
\end{equation}
The spatial many-body charge conjugation operator commutes with the TSTG many-body Hamiltonian both in the absence or in the presence of a displacement field, $\commutator{H^{\text{TSTG}}}{\mathcal{P}'} = 0$. 

The two charge-conjugation operators $\mathcal{P}$ and $\mathcal{P}'$ map states at filling $\nu$ to states at filling $-\nu$. Alternatively, in the grand canonical ensemble, the charge conjugation operators map phases at chemical potential $\mu = \mu_0$ to ones at chemical potential $\mu = -\mu_0$. Let the Green's function of the system at chemical potential $\mu = \pm \mu_0$ be given by $\mathcal{G}^{\pm \mu_0}_{i \eta s; i' \eta' s'} \left( i \omega_n, \vec{k} \right)$, such that  
\begin{equation}
	\label{app:eqn:matsubara_gf_THF_mu0}
	-\left\langle \mathcal{T}_{\tau} \hat{\gamma}_{\vec{k},i,\eta,s} \left( \tau \right) \hat{\gamma}^\dagger_{\vec{k}',i',\eta',s'} \left( 0 \right)  \right\rangle_{\pm \mu_0}  = \delta_{\vec{k},\vec{k}'} \mathcal{G}^{\pm \mu_0}_{i \eta s; i' \eta' s'} \left(\tau, \vec{k} \right),
\end{equation}
where the subscript $\left\langle \dots \right\rangle_{\pm \mu_0}$ implies that the averaging is performed within the grand canonical ensemble at chemical potential $\mu = \pm \mu_0$. Using the fact that $\mathcal{P}^{(\prime)} \hat{N} \mathcal{P}^{(\prime)-1} = N_{\gamma} - \hat{N}$ (with $N_{\gamma}$ denoting the total number of fermionic species: $N_{\gamma} = 24 N_0$ for TBG and $N_{\gamma} = 32 N_0$ for TSTG), the partition function computed at chemical potential $\mu$ obeys 
\begin{equation}
	\label{app:eqn:PH_of_Z}
	Z \left( \mu \right) = \Tr \left[ e^{-\beta K } \right]  = \Tr \left[ e^{-\beta K } \mathcal{P}^{(\prime)} \mathcal{P}^{(\prime)-1} \right] = \Tr \left[ \mathcal{P}^{(\prime)} e^{-\beta \left( H + \mu \hat{N} - \mu N_{\gamma} \right) }  \mathcal{P}^{(\prime)-1} \right] = e^{\beta \mu N_{\gamma}} Z \left( - \mu \right).
\end{equation}
Using \cref{app:eqn:PH_of_Z}, we can show that for $\beta > \tau > 0$
\begin{align}
	\mathcal{G}^{\mu_0}_{i \eta s; i' \eta' s'} \left( \tau, \vec{k} \right) =& \frac{-1}{Z \left( \mu_0 \right)} \Tr \left[ e^{-\beta \left( H - \mu_0 \hat{N} \right)} e^{\left( H - \mu_0 \hat{N} \right) \tau} \hat{\gamma}_{\vec{k},i,\eta,s} e^{-\left( H - \mu_0 \hat{N} \right) \tau} \hat{\gamma}^\dagger_{\vec{k},i',\eta',s'} \right] \nonumber \\
	=& -\frac{e^{-\beta \mu_{0} N_{\gamma}}}{Z \left( -\mu_0 \right)} \Tr \left[ e^{-\beta \left( H - \mu_0 \hat{N} \right)} e^{\left( H - \mu_0 \hat{N} \right) \tau} \hat{\gamma}_{\vec{k},i,\eta,s} e^{-\left( H - \mu_0 \hat{N} \right) \tau} \hat{\gamma}^\dagger_{\vec{k},i',\eta',s'} \mathcal{P}^{(\prime)-1} \mathcal{P}^{(\prime)} \right] \nonumber \\
	=& -\sum_{\substack{i_1, \eta_1 \\ i_2, \eta_2}} \Tr \left[ \mathcal{P}^{(\prime)-1} e^{-\beta \left( H + \mu_0 \hat{N} - \mu_0 N_{\gamma} \right)} e^{\left( H + \mu_0 \hat{N} \right) \tau} \hat{\gamma}^\dagger_{- \left( C_{2x} \right) \vec{k},i_1,\eta_1,s} e^{-\left( H + \mu_0 \hat{N} \right) \tau} \hat{\gamma}_{- \left( C_{2x} \right) \vec{k},i_2,\eta_2,s'} \mathcal{P}^{(\prime)} \right] \nonumber \\
	&\times \frac{e^{-\beta \mu_{0} N_{\gamma}}}{Z \left( -\mu_0 \right) } \left[ D \left( \mathcal{P}^{(\prime)} \right) \right]^{*}_{i_1 \eta_1; i \eta} \left[ D \left( \mathcal{P}^{(\prime)} \right) \right]_{i_2 \eta_2; i' \eta'}\nonumber \\
	=& -\sum_{\substack{i_1, \eta_1 \\ i_2, \eta_2}} \Tr \left[ e^{- \beta \left(H +\mu_0 \hat{N} \right)} e^{\left( \beta - \tau \right)\left( H + \mu_0 \hat{N} \right) } \hat{\gamma}_{- \left( C_{2x} \right) \vec{k},i_2,\eta_2,s'}  e^{- \left(\beta - \tau \right) \left( H + \mu_0 \hat{N} \right)} \hat{\gamma}^\dagger_{- \left( C_{2x} \right) \vec{k},i_1,\eta_1,s} \right] \nonumber \\
	&\times \frac{1}{Z \left( -\mu_0 \right) } \left[ D \left( \mathcal{P}^{(\prime)} \right) \right]^{*}_{i_1 \eta_1; i \eta} \left[ D \left( \mathcal{P}^{(\prime)} \right) \right]_{i_2 \eta_2; i' \eta'}\nonumber \\
	=& \sum_{\substack{i_1, \eta_1 \\ i_2, \eta_2}} \left[ D \left( \mathcal{P}^{(\prime)} \right) \right]^{*}_{i_1 \eta_1; i \eta} \left[ D \left( \mathcal{P}^{(\prime)} \right) \right]_{i_2 \eta_2; i' \eta'} \mathcal{G}^{-\mu_0}_{i_2 \eta_2 s'; i_1 \eta_1 s} \left( \beta - \tau, - \left( C_{2x} \right) \vec{k} \right), \label{app:eqn:ph_on_imag_time_gf}
\end{align}
where $\left( C_{2x} \right)$ denotes the action of $C_{2x}$ on the momentum $\vec{k}$ in the case of TSTG. Using \cref{app:eqn:matsubara_gf_THF_ft}, we find that in frequency space \cref{app:eqn:ph_on_imag_time_gf} implies that 
\begin{equation}
	\mathcal{G}^{\mu_0}_{i \eta s; i' \eta' s'} \left( i \omega_n, \vec{k} \right) = - \sum_{\substack{i_1, \eta_1 \\ i_2, \eta_2}} \left[ D \left( \mathcal{P}^{(\prime)} \right) \right]^{*}_{i_1 \eta_1; i \eta} \left[ D \left( \mathcal{P}^{(\prime)} \right) \right]_{i_2 \eta_2; i' \eta'} \mathcal{G}^{-\mu_0}_{i_2 \eta_2 s'; i_1 \eta_1 s} \left( - i \omega_n, - \left( C_{2x} \right) \vec{k} \right),
\end{equation}
which, after analytical continuation $i \omega_n \to \omega + i 0^{+}$, becomes
\begin{equation}
	\label{app:eqn:ph_on_ret_gf}
	\mathcal{G}^{\mu_0}_{i \eta s; i' \eta' s'} \left( \omega + i 0^{+} , \vec{k} \right) = - \sum_{\substack{i_1, \eta_1 \\ i_2, \eta_2}} \left[ D \left( \mathcal{P}^{(\prime)} \right) \right]^{*}_{i_1 \eta_1; i \eta} \left[ D \left( \mathcal{P}^{(\prime)} \right) \right]_{i_2 \eta_2; i' \eta'} \mathcal{G}^{-\mu_0}_{i_2 \eta_2 s'; i_1 \eta_1 s} \left( - \omega - i 0^{+}, - \left( C_{2x} \right) \vec{k} \right).
\end{equation}
Using the spectral representation of the Green's function from \cref{app:eqn:spectral_rep_of_GF}, as well as the Hermiticity of the spectral function, we can obtain the analogue of \cref{app:eqn:spectral_function} in terms of the \emph{advanced} Green's function,
\begin{equation}
	\label{app:eqn:spectral_function_from_advanced}
	A_{i \eta s; i' \eta' s'} \left( -\omega, \vec{k} \right) = \frac{1}{2 \pi i} \left( \mathcal{G}_{i \eta s; i' \eta' s'} \left(-\omega - i 0^{+}, \vec{k} \right) - \mathcal{G}^{*}_{i' \eta' s',i \eta s} \left(- \omega - i 0^{+}, \vec{k} \right) \right).
\end{equation}
If we combine \cref{app:eqn:spectral_function,app:eqn:spectral_function_from_advanced,app:eqn:ph_on_ret_gf} and denote by $A^{\pm \mu_0} \left( \omega, \vec{k} \right)$ the spectral function of the system at chemical potential $\pm \mu_0$, then we obtain
\begin{align}
	A^{\mu_0}_{i \eta s; i' \eta' s'} \left( \omega, \vec{k} \right) &= \sum_{\substack{i_1, \eta_1 \\ i_2, \eta_2}} \left[ D \left( \mathcal{P} \right) \right]^{*}_{i_1 \eta_1; i \eta} \left[ D \left( \mathcal{P} \right) \right]_{i_2 \eta_2; i' \eta'} A^{-\mu_0}_{i_2 \eta_2 s'; i_1 \eta_1 s} \left( - \omega, - \vec{k} \right), \label{app:eqn:TBG_sp_func_ph_sym} \\
	A^{\mu_0}_{i \eta s; i' \eta' s'} \left( \omega, \vec{k} \right) &= \sum_{\substack{i_1, \eta_1 \\ i_2, \eta_2}} \left[ D \left( \mathcal{P}' \right) \right]^{*}_{i_1 \eta_1; i \eta} \left[ D \left( \mathcal{P}' \right) \right]_{i_2 \eta_2; i' \eta'} A^{-\mu_0}_{i_2 \eta_2 s'; i_1 \eta_1 s} \left( - \omega, - C_{2x} \vec{k} \right), \label{app:eqn:TSTG_sp_func_ph_sym}
\end{align} 
for TBG and TSTG, respectively. This proves that it is sufficient to consider positive chemical potentials, since the physics of the system at negative chemical potentials follows from the one at positive ones via the many-body chanrge conjugation symmetry. As a result and without loss of generality, we will only consider the positive integer-filled insulators from \cref{app:tab:model_states}.

\section{Self-energy correction in the THF model beyond Hartree-Fock}\label{app:sec:se_correction_beyond_HF}

The transport properties of electronic systems are governed not only by the dispersion of the charge-one excitations, but also by their \emph{lifetime}~\cite{MAH00}. While the former can be readily determined using the Hartree-Fock method ({\it i.e.}{}, to first-order in the interaction), as succinctly discussed in \cref{app:sec:hartree_fock}, the latter manifests as \emph{second-order} effects in the interaction. This \siSection{} focuses on obtaining the corrections to the electrons' self-energy within the THF model, thereby incorporating finite-lifetime effects. Throughout this work, we will only consider elastic electron-electron scattering and ignore the effects of disorder.

For this \siSection{}, we will assume that the system is in an ordered phase, specifically one of the integer-filled correlated ground state candidates described in \cref{app:tab:model_states}, or an ordered phase achieved by doping these states away from integer fillings. The discussion of the fully-symmetric (unordered) phase will be addressed separately in \cref{app:sec:se_symmetric}.

We start by outlining the second-order interaction contributions to the self-energy of the system. A basic familiarity with Feynmann diagrams will be assumed in this \siSection{}~\cite{MAH00}. A self-contained alternative derivation that does not rely on Feynman diagrams, but instead uses the path integral formulation is provided in \cref{app:sec:additional_mb_results:self-energy}. Within a \emph{self-consistent} treatment, we will demonstrate that only two diagrams contribute at the second order. Because they are dispersionless and their onsite repulsion is the largest energy scale of the system in TBG~\cite{SON22} (and among the largest energy-scales in TSTG~\cite{YU23a}), we focus solely on the second-order corrections to the $f$-electrons' self-energy, which arise from the interaction of the $f$-electrons between themselves ({\it i.e.}{}, the $H_{U_1}$ and $H_{U_2}$ terms of the Hamiltonian). We expect that the strongly-dispersing $c$-electrons (and $d$-electrons in the case of TSTG) will have a much longer lifetime. As a result, we only consider a small broadening of their spectral function which is added by hand. As we explain in \cref{app:sec:se_correction_beyond_HF:sc_problem_and_numerics}, provided that this broadening factor is smaller than any other energy scale of the problem, the exact value is not important. 

Following Refs.~\cite{SCH89,SCH90,SCH91}, we will then assume that the second-order self-energy correction for the $f$-electrons is site-diagonal (to be defined rigorously in \cref{app:sec:se_correction_beyond_HF:all_so}), or, alternatively, momentum-independent. This will enable us to self-consistently compute the second-order correction to the self-energy of the $f$-electrons. After deriving the formula of the second-order self-energy correction, we will also give a brief overview of our numerical implementation. 

\subsection{General remarks}\label{app:sec:se_correction_beyond_HF:all_so_corrections}

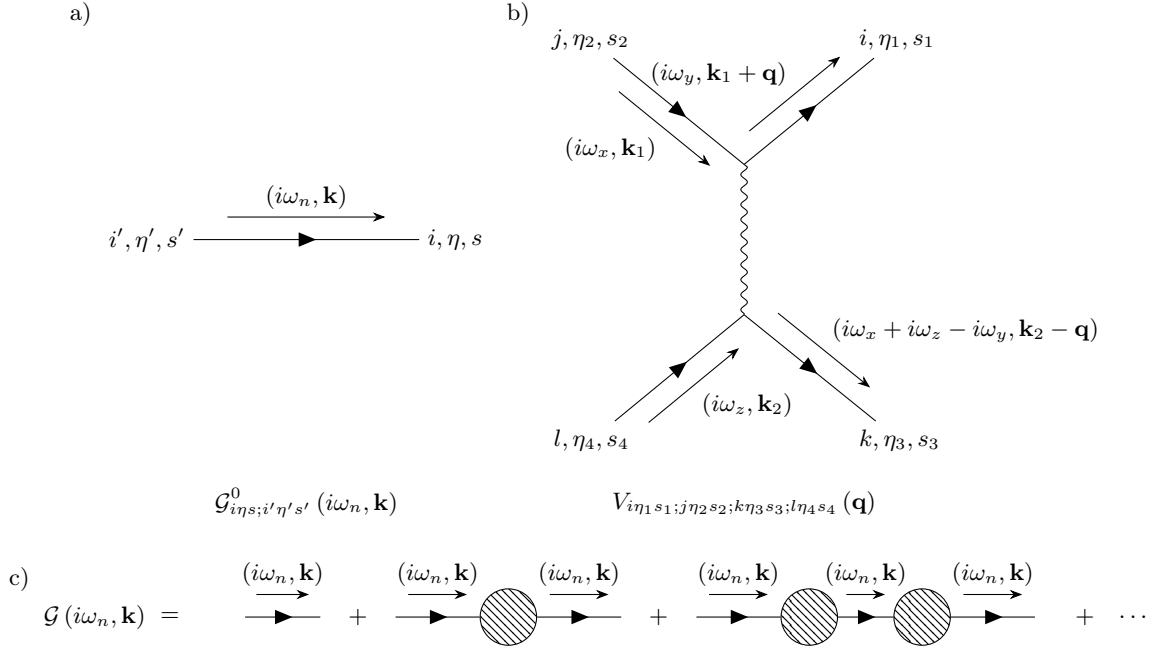
\begin{figure}[!t]
	\centering
	\begin{tikzpicture}[baseline=(current bounding box.north)]
		\begin{feynman}
			\vertex (mid);
			\vertex [left=1.5 cm of mid] (a) {$i', \eta', s'{}$};
			\vertex [right=1.5 cm of mid] (c) {$i, \eta, s{}$};
			\diagram*{
				(a) -- [fermion, momentum={$\left( i \omega_n, \vec{k} \right)$}]  (c),
			};
		\end{feynman}
		\path (mid) ++(-3cm,3cm) node{a)};
		\path (mid) ++(0,-3.5cm) node[anchor=center]{$\displaystyle \mathcal{G}^{0}_{i \eta s; i' \eta' s'} \left(i \omega_n, \vec{k} \right) $};
	\end{tikzpicture}
\begin{tikzpicture}[baseline=(current bounding box.north)]
		\begin{feynman}
			\vertex (b);
			\vertex [above=2 cm of b] (d);
			\vertex [above=1 cm of b] (mid);
			\vertex [below left=2 cm of b] (a) {$l,\eta_4,s_4$};
			\vertex [below right=2 cm of b] (c) {$k,\eta_3,s_3$};
			\vertex [above left=2 cm of d] (e) {$j,\eta_2,s_2$};
			\vertex [above right=2 cm of d] (f) {$i,\eta_1,s_1$};
			\diagram*{
				(a) -- [fermion, momentum'={$\left( i \omega_z, \vec{k}_2 \right)$}]  (b) -- [fermion, momentum={$\left( i \omega_x +  i \omega_z - i \omega_y , \vec{k}_2 - \vec{q} \right)$}] (c),
				(b) -- [photon] (d),
				(e) -- [fermion, momentum'={$\left( i \omega_x, \vec{k}_1 \right)$}]  (d) -- [fermion, momentum={$\left( i \omega_y, \vec{k}_1 + \vec{q} \right)$}] (f),
			};
			\path (mid) ++(0,-3.5cm) node[anchor=center]{$\displaystyle V_{i \eta_1 s_1; j \eta_2 s_2; k \eta_3 s_3; l \eta_4 s_4} \left( \vec{q} \right)$};
		\end{feynman}
		\path (mid) ++(-3cm,3cm) node{b)};
	\end{tikzpicture}\\
	\begin{tikzpicture}[baseline=(current bounding box.north)]
		\begin{feynman}
			\vertex (a) {$\mathcal{G} \left(i \omega_n, \vec{k} \right)$};
\vertex [right=2 cm of a] (s1);
			\vertex [right=1 cm of s1] (f1);
\vertex [right=1 cm of f1] (s2);
			\vertex [right=3 cm of s2] (f2);
			\vertex[blob] (sigma1) at ($(s2)!0.5!(f2)$) {};
\vertex [right=1 cm of f2] (s3);
			\vertex [right=4.5 cm of s3] (f3);
			\vertex[blob] (sigma2) at ($(s3)!0.333333!(f3)$) {};
			\vertex[blob] (sigma3) at ($(s3)!0.666667!(f3)$) {};
\vertex [right=1 cm of f3] (s4) {$\dots$};
\vertex at ($(a)!0.5!(s1)$) {$=$};
			\vertex at ($(f1)!0.5!(s2)$) {$+$};
			\vertex at ($(f2)!0.5!(s3)$) {$+$};
			\vertex at ($(f3)!0.5!(s4)$) {$+$};
			\diagram*{
				(s1) -- [fermion,momentum={$\left( i \omega_n, \vec{k} \right)$}] (f1),
				(s2) -- [fermion,momentum={$\left( i \omega_n, \vec{k} \right)$}] (sigma1) -- [fermion,momentum={$\left( i \omega_n, \vec{k} \right)$}] (f2),
				(s3) -- [fermion,momentum={$\left( i \omega_n, \vec{k} \right)$}] (sigma2) -- [fermion,momentum={$\left( i \omega_n, \vec{k} \right)$}] (sigma3) -- [fermion,momentum={$\left( i \omega_n, \vec{k} \right)$}] (f3)
			};
		\end{feynman}
		\path (a) ++(-1,0.5cm) node{c)};
	\end{tikzpicture}
\subfloat{\label{app:fig:general_diagramatology:a}}\subfloat{\label{app:fig:general_diagramatology:b}}\subfloat{\label{app:fig:general_diagramatology:c}}\caption{Feynman diagrams for the THF model of TBG. The non-interacting Matsubara Green's function ({\it i.e.}{}, the bare propagator) of the THF model for TBG is represented schematically in (a) by a line. The process corresponds to the propagation of a $\hat{\gamma}^\dagger_{\vec{k},i', \eta', s'}$ fermion into a $\hat{\gamma}^\dagger_{\vec{k},i, \eta, s}$ one as indicated by the arrow on the line (which therefore indicates the ``flow'' of $\mathrm{U} \left(1\right)$ charge). Additionally, the momentum and imaginary frequency are indicated above the propagator. The THF interaction Hamiltonian for TBG can be written formally as in \cref{app:eqn:general_TBG_int_forHF}. The corresponding four-fermion interaction vertex is shown in (b). It corresponds to the scattering of two fermions, $\hat{\gamma}^\dagger_{\vec{k}_2,l,\eta_4,s_4}$ and $\hat{\gamma}^\dagger_{\vec{k}_1,j,\eta_2,s_2}$, into the $\hat{\gamma}^\dagger_{\vec{k}_2 - \vec{q},k,\eta_3,s_3}$ and $\hat{\gamma}^\dagger_{\vec{k}_1 + \vec{q},i,\eta_1,s_1}$ fermions. The amplitude of the scattering process is indicated below the diagram. Note that even though the interaction happens instantaneously, conventionally it is represented by a ``photon'' line~\cite{MAH00}, as explained in text. Finally, (c) shows the infinite summation of diagrams equivalent to Dyson's equation. The hatched blob represents the sum of all the 1-particle irreducible diagrams ({\it i.e.}{}, the self-energy diagrams).}
	\label{app:fig:general_diagramatology}
\end{figure}

We begin by considering the THF interaction Hamiltonian for TBG on general grounds (the discussion for TSTG follows analogously). The (matrix) fermion propagator of the THF model is shown schematically by a line, as depicted in \cref{app:fig:general_diagramatology:a}, where the flow of $\mathrm{U} \left(1\right)$ charge is indicated by an arrow. At the same time, the four-fermion TBG interaction Hamiltonian in the THF model can be represented by the four-fermion vertex from \cref{app:fig:general_diagramatology:a}, whose amplitude can be related to the rank-four tensor from \cref{app:eqn:general_TBG_int_forHF}. Even though the interaction is assumed to happen instantaneously, the four-fermion vertex is still represented by a ``photon'' line connecting the end of four fermion propagators. This is a matter of convention~\cite{MAH00}, as a point-like representation of this scattering process would be ambiguous and would need to be supplemented by indices (causing unnecessary clutter).

\subsubsection{Diagramatic contributions to the second-order self-energy correction}\label{app:sec:se_correction_beyond_HF:all_so_corrections:diagrams}

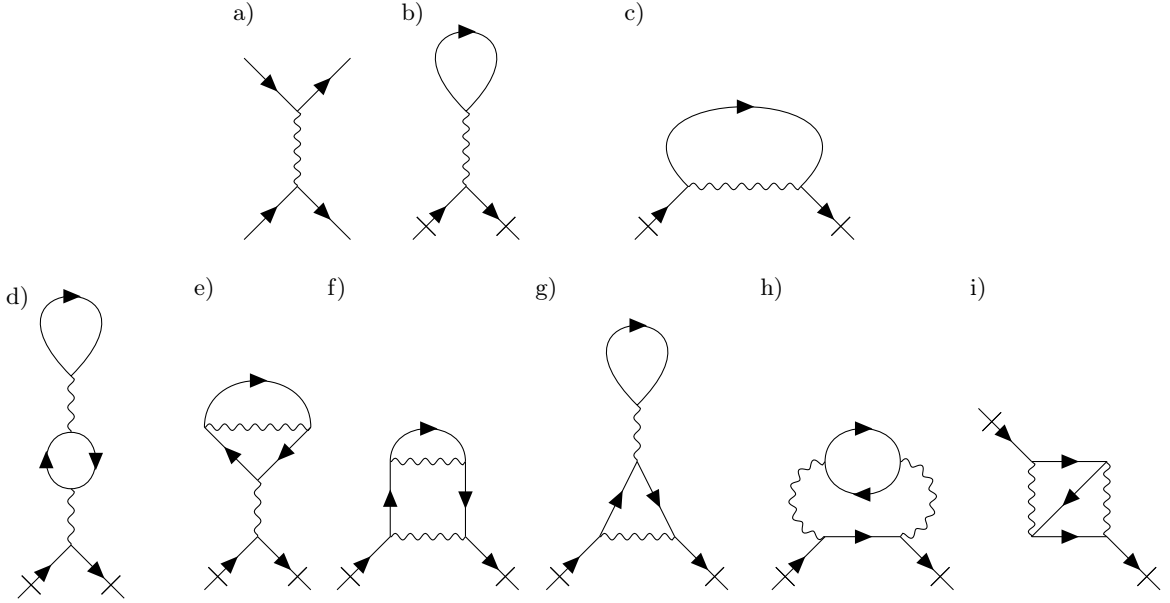
\begin{figure}[!t]
	\centering
	\begin{tikzpicture}[baseline=(current bounding box.north)]
		\begin{feynman}
			\vertex (b);
			\vertex [above=1 cm of b] (d);
			\vertex [below left=1 cm of b] (a);
			\vertex [below right=1 cm of b] (c);
			\vertex [above left=1 cm of d] (e);
			\vertex [above right=1 cm of d] (f);
			\diagram*{
				(a) -- [fermion]  (b) -- [fermion] (c),
				(b) -- [photon] (d),
				(e) -- [fermion]  (d) -- [fermion] (f),
			};
		\end{feynman}
		\path (a) ++(0,3cm) node{a)};
	\end{tikzpicture}
\begin{tikzpicture}[baseline=(current bounding box.north)]
		\begin{feynman}
			\vertex (b);
			\vertex [above=1 cm of b] (d);
			\vertex [below left=1 cm of b] (a);
			\vertex [below right=1 cm of b] (c);
			\diagram*{
				(a) -- [fermion, strike through pre]  (b) -- [fermion, strike through post] (c),
				(b) -- [photon] (d),
				d --[fermion, out=135, in=45, loop, min distance=2cm] d,
			};
		\end{feynman}
		\path (a) ++(0,3cm) node{b)};
	\end{tikzpicture}
\begin{tikzpicture}[baseline=(current bounding box.north)]
		\begin{feynman}
			\vertex (b);
			\vertex [right=of b] (c);
			\vertex [below left=1 cm of b] (a);
			\vertex [below right=1 cm of c] (d);
			\diagram* {
				(a) -- [fermion, strike through pre] (b) -- [photon] (c) -- [fermion, strike through post] (d),
				(b) -- [fermion, out=135, in=45,min distance=2cm] (c)
			};
		\end{feynman}
		\path (a) ++(0,3cm) node{c)};
	\end{tikzpicture} \\
\begin{tikzpicture}[baseline=(current bounding box.north)]
		\begin{feynman}
			\vertex (b);
			\vertex [above=0.75 cm of b] (d);
			\vertex [above=0.75 cm of d] (e);
			\vertex [above=0.75 cm of e] (f);
			\vertex [below left=1 cm of b] (a);
			\vertex [below right=1 cm of b] (c);
			\diagram*{
				(a) -- [fermion, strike through pre]  (b) -- [fermion, strike through post] (c),
				(b) -- [photon] (d),
				(e) -- [photon] (f),
				(d) -- [fermion, half left] (e),
				(e) -- [fermion, half left] (d),
				f --[fermion, out=135, in=45, loop, min distance=2cm] f,
			};
		\end{feynman}
		\path (a) ++(0,4cm) node{d)};
	\end{tikzpicture}
\begin{tikzpicture}[baseline=(current bounding box.north)]
		\begin{feynman}
			\vertex (b);
			\vertex [above=0.75 cm of b] (d);
			\vertex [below left=1 cm of b] (a);
			\vertex [below right=1 cm of b] (c);
			\vertex [above left=1cm of d] (e);
			\vertex [above right=1cm of d] (f);
			\diagram*{
				(a) -- [fermion, strike through pre]  (b) -- [fermion, strike through post] (c),
				(b) -- [photon] (d),
				(d) -- [fermion] (e),
				(f) -- [fermion] (d),
				(e) -- [photon] (f),
				(e) -- [fermion, half left] (f),
			};
		\end{feynman}
		\path (a) ++(0,4cm) node{e)};
	\end{tikzpicture}
\begin{tikzpicture}[baseline=(current bounding box.north)]
		\begin{feynman}
			\vertex (b);
			\vertex [right=1 cm of b] (c);
			\vertex [below left=1 cm of b] (a);
			\vertex [below right=1 cm of c] (d);
			\vertex [above=1 cm of b] (bb);
			\vertex [above=1 cm of c] (cc);
			\diagram* {
				(a) -- [fermion, strike through pre] (b) -- [photon] (c) -- [fermion, strike through post] (d),
				(b) -- [fermion] (bb),
				(cc) -- [fermion] (c),
				(bb) -- [photon] (cc),
				(bb) -- [fermion, half left] (cc),
			};
		\end{feynman}
		\path (a) ++(0,4cm) node{f)};
	\end{tikzpicture}
\begin{tikzpicture}[baseline=(current bounding box.north)]
		\begin{feynman}
			\vertex (b);
			\vertex [right=1 cm of b] (c);
			\vertex [below left=1 cm of b] (a);
			\vertex [below right=1 cm of c] (d);
			\vertex [right=0.5 cm of b] (mid);
			\vertex [above=1 cm of mid] (e);
			\vertex [above=0.75 cm of e] (f);
			\vertex [above=0.75 cm of f] (g);
			\diagram* {
				(a) -- [fermion, strike through pre] (b) -- [photon] (c) -- [fermion, strike through post] (d),
				(b) -- [fermion] (e),
				(e) -- [fermion] (c),
				(e) -- [photon] (f),
				f --[fermion, out=135, in=45, loop, min distance=2cm] f,
			};
		\end{feynman}
		\path (a) ++(0,4cm) node{g)};
	\end{tikzpicture}
\begin{tikzpicture}[baseline=(current bounding box.north)]
		\begin{feynman}
			\vertex (b);
			\vertex [right=1 cm of b] (c);
			\vertex [below left=1 cm of b] (a);
			\vertex [below right=1 cm of c] (d);
			\vertex [above=1 cm of b] (bb);
			\vertex [above=1 cm of c] (cc);
			\diagram* {
				(a) -- [fermion, strike through pre] (b) -- [fermion] (c) -- [fermion, strike through post] (d),
				(b) -- [photon, half left] (bb),
				(cc) -- [photon, half left] (c),
				(bb) -- [fermion, half left] (cc),
				(cc) -- [fermion, half left] (bb),
			};
		\end{feynman}
		\path (a) ++(0,4cm) node{h)};
	\end{tikzpicture}
\begin{tikzpicture}[baseline=(current bounding box.north)]
		\begin{feynman}
			\vertex (b);
			\vertex [left=1 cm of b] (a);
			\vertex [below=1 cm of b] (c);
			\vertex [left=1 cm of c] (d);
			\vertex [above left=1 cm of a] (aa);
			\vertex [below right=1 cm of c] (cc);
			\vertex [below left=1 cm of d] (dd);
			\diagram* {
				(aa) -- [fermion, strike through pre] (a),
				(a) -- [fermion] (b)  -- [fermion] (d) -- [fermion] (c),
				(b) -- [photon] (c),
				(d) -- [photon] (a),
				(c) -- [fermion, strike through post] (cc),
			};
		\end{feynman}
		\path (dd) ++(0,4cm) node{i)};
	\end{tikzpicture}
	\subfloat{\label{app:fig:general_self_en_diags:a}}\subfloat{\label{app:fig:general_self_en_diags:b}}\subfloat{\label{app:fig:general_self_en_diags:c}}\subfloat{\label{app:fig:general_self_en_diags:d}}\subfloat{\label{app:fig:general_self_en_diags:e}}\subfloat{\label{app:fig:general_self_en_diags:f}}\subfloat{\label{app:fig:general_self_en_diags:g}}\subfloat{\label{app:fig:general_self_en_diags:h}}\subfloat{\label{app:fig:general_self_en_diags:i}}\caption{Diagrams that contribute to the self-energy of the THF model up to second order in the interactions. We consider a general four-fermion interaction (that potentially involves all the fermionic species) and whose four-fermion vertex is shown in (a). As explained in the text, it is conventional to represent the interaction with a ``photon'' line~\cite{MAH00}, even if the interaction is assumed to happen instantaneously. The remaining diagrams (b)-(i) denote the diagramatic contributions to the electron self-energy. The insertion points are shown as slashed (amputated) fermion propagators. The last two diagrams on the first row are the first-order corrections to the electron self-energy, with (b) denoting the Hartree contribution and (c) denoting the Fock one. The second row shows the second-order contributions to the electron self-energy. It is worth noting that within a self-consistent perturbation scheme, wherein all the internal edges of the diagrams are replaced by \emph{dressed}, as opposed to \emph{bare} electron propagators, the second-order diagrams from (d)-(g) are already included in the Hartree-Fock self-consistent self-energy. 
For instance, the diagram in (d) can be interpreted as replacing the propagators forming a loop in (b) with a propagator dressed by Hartree contributions.
    As a result, in a self-consistent second-order self-energy scheme, one only needs to consider the diagrams in (h) and (i) at the second-order level.}
	\label{app:fig:general_self_en_diags}
\end{figure}

Starting from Dyson's equation in \cref{app:eqn:dyson_equation}, the fully-interacting Green's function of the THF model can be written as an infinite series
{\small\begin{equation}
	\label{app:eqn:dyson_series}
	\mathcal{G} \left( i \omega_n, \vec{k} \right) = \mathcal{G}^{0} \left( i \omega_n, \vec{k} \right) + \mathcal{G}^{0} \left( i \omega_n, \vec{k} \right) \Sigma \left( i \omega_n, \vec{k} \right) \mathcal{G}^{0} \left( i \omega_n, \vec{k} \right) + \mathcal{G}^{0} \left( i \omega_n, \vec{k} \right) \Sigma \left( i \omega_n, \vec{k} \right) \mathcal{G}^{0} \left( i \omega_n, \vec{k} \right)  \Sigma \left( i \omega_n, \vec{k} \right) \mathcal{G}^{0} \left( i \omega_n, \vec{k} \right) + \dots
\end{equation}}The infinite summation is represented diagramatically in \cref{app:fig:general_diagramatology:c}. The self-energy $\Sigma \left( i \omega_n, \vec{k} \right)$ (which is represented by the hatched blob) can be understood as the sum of all \emph{1-particle-irreducible} Feynman diagrams. The 1-particle-irreducible Feynman diagrams cannot be separated into two diagrams by cutting an electron propagator~\cite{ABR75,PES95,MAH00}.

Apart from being 1-particle-irreducible, the diagrams making up the self-energy correction have no external lines (or otherwise stated, their external lines have been \emph{amputated}), meaning that they can be inserted in a propagator. To understand what the ``amputation'' of the diagram entails, consider the second diagram in the sum from \cref{app:fig:general_diagramatology:c}. This diagram corresponds to the second term in \cref{app:eqn:dyson_series}, namely $\mathcal{G}^{0} \left( i \omega_n, \vec{k} \right) \Sigma \left( i \omega_n, \vec{k} \right) \mathcal{G}^{0} \left( i \omega_n, \vec{k} \right)$. The two bare propagators at the ends of the diagram are \emph{not} part of the self-energy contribution. Formally, we say that the self-energy diagram (shown schematically as the hatched blob) is amputated, meaning that it has no external lines. However, by attaching two bare propagators at the end of a self-energy diagram, one obtains a \textit{bona fide} Feynman diagram describing a propagating fermion. As a result, we say that the self-energy diagrams can be inserted in a propagator, effectively replacing $\mathcal{G}^{0} \left( i \omega_n, \vec{k} \right)$ by $\mathcal{G}^{0} \left( i \omega_n, \vec{k} \right) \Sigma \left( i \omega_n, \vec{k} \right) \mathcal{G}^{0} \left( i \omega_n, \vec{k} \right)$. Finally, we mention that when representing the self-energy Feynman diagrams, it is conventional to depict the ends of the diagram to which bare electron propagators can be added to obtain an electron propagation diagram (known also as the insertion points of the self-energy diagram) by slashed fermion propagators. 

We want to determine the Feynman diagrams that contribute to the electron self-energy up to and including second order in the interaction. For a generic four-fermion interaction shown in \cref{app:fig:general_self_en_diags:a}, the only such diagrams are shown in \crefrange{app:fig:general_self_en_diags:b}{app:fig:general_self_en_diags:i}~\cite{ABR75}. 

The two diagrams in \cref{app:fig:general_self_en_diags:b,app:fig:general_self_en_diags:c} are, respectively, the Hartree and Fock contributions to the self-energy (which were derived by elementary means in \cref{app:sec:hartree_fock:hamiltonian}) and together form the first-order self-energy correction. The second row of \cref{app:fig:general_self_en_diags} shows the six diagrams that contribute to the second order in the interaction to the fermion self-energy. We note, however, that, within a self-consistent calculation, only the last two diagrams from \cref{app:fig:general_self_en_diags:h,app:fig:general_self_en_diags:i} need to be considered, as the other four are already included in the Hartree and Fock contributions, as we explain below.

To see why this is so, we note that the self-energy diagrams ``dress'' the fermion propagators by inserting themselves recursively to infinite order. In a self-consistent calculation, all the propagators used in computing the diagrams are dressed. As such, one can readily see that the diagram in \cref{app:fig:general_self_en_diags:d} is really only the Hartree self-energy correction from \cref{app:fig:general_self_en_diags:b} where the fermionic propagator in the loop was dressed by one Hartree self-energy correction. Similarly, the diagram in \cref{app:fig:general_self_en_diags:e} is the Hartree contribution with the propagator dressed by a Fock correction, the one in \cref{app:fig:general_self_en_diags:f} is the Fock contribution with the propagator dressed by another Fock correction, while the one in \cref{app:fig:general_self_en_diags:g} is the Fock contribution where the propagator was dressed by another Hartree correction. Since our calculation is self-consistent (and thus employs dressed propagators), only the diagrams in \cref{app:fig:general_self_en_diags:h,app:fig:general_self_en_diags:i} need to be considered at second order.

\subsubsection{Leading contributions to the second-order self-energy}\label{app:sec:se_correction_beyond_HF:all_so_corrections:particularization}

The main goal of this \siSection{} is to compute the lifetime of the fermions of the THF model from perturbation theory. The finite lifetime of quasiparticles is linked to the imaginary part of the self-energy correction, which is only non-vanishing beyond first order in perturbation theory~\cite{MAH00}. However, whereas the Hartree and Fock self-energies can be determined by means of a simple tensor contraction, as shown in \cref{app:eqn:TBG_HF_Hamiltonian,app:eqn:TSTG_HF_Hamiltonian}, and are functions of momentum \emph{only}, the second-order self-energy correction is generically a function of both momentum \emph{and} Matsubara frequency, and its computation is much more involved.  As we will show explicitly in \cref{app:sec:se_correction_beyond_HF:all_so}, the second-order self-energy also depends on the Green's function of the systems as whole, rather than just on the density matrix~\cite{SCH90}.

Just like the Green's function, in general, the dynamical self-energy depends on both momentum and Matsubara frequency. Moreover, as we will show explicitly in \cref{app:eqn:se_2a_complicated,app:eqn:se_2b_complicated}, the second-order self-energy involves double summations over momenta (or, equivalently, lattice sites) on top of double summations over Matsubara frequencies. As such, in order to make our self-consistent second-order perturbation scheme numerically tractable, a number of approximations are in order:
\begin{itemize}
	\item We first note that the $f$-electrons are dispersionless and have a large onsite repulsion term (which is the largest -- in TBG -- or among the largest -- in TSTG -- energy scales of the system). Consequently, their scattering rate is expected to be significantly higher than that of the strongly dispersing $c$- and $d$-electrons, leading to a much shorter lifetime. A good approximation is to account for second-order self-energy effects \emph{only} for the $f$-electrons. For the other fermionic species, we introduce a small broadening factor in their spectral function, effectively assigning them a finite but large lifetime. As long as this artificial broadening remains much smaller than that of the $f$-electrons, its precise value is unimportant.
	
	\item When computing the second-order self-energy correction to the $f$-electrons using the Feynman diagrams in \cref{app:fig:general_self_en_diags:h,app:fig:general_self_en_diags:i}, all interaction terms in \cref{app:eqn:THF_interaction_TBG,app:eqn:THF_interaction_TSTG}, except for $H_{V}$, $H^{cd}_{V}$, and $H^{d}_{V}$ (which do not involve $f$-electrons), could, in principle, contribute. However, interaction terms that couple both $f$- and $c$- (or $f$- and $d$-) electrons, such as $H_{W}$, would also generate second-order self-energy corrections for the $c$- (or $d$-) fermions through the same type of diagrams as in \crefrange{app:fig:general_self_en_diags:d}{app:fig:general_self_en_diags:i}. To maintain consistency in neglecting second-order self-energy corrections for the $c$- and $d$-electrons, we must also discard contributions to the $f$-electron self-energy arising from $H_{W}$, $H_{J}$, $H_{\tilde{J}}$, $H_{K}$, and $H^{fd}_{W}$. In other words, performing a second-order perturbation expansion in, for instance, $W_{1,3}$ would require computing and incorporating self-energy corrections to the $c$-electron propagator, which we omit in our calculations. 
	
	As a result, we only consider the second-order self-energy corrections from $H_{U_1}$ and $H_{U_2}$\footnote{Since $U_2 \ll U_1$, corrections from $H_{U_2}$ are much smaller than those from $H_{U_1}$. However, as shown in \cref{app:sec:se_correction_beyond_HF:all_so}, the computation of the self-energy contribution from $H_{U_2}$ follows the same procedure as for $H_{U_1}$. Thus, we can determine the second-order self-energy contributions from both $H_{U_1}$ and $H_{U_2}$ simultaneously without additional computational effort.}, with the dominant contribution arising from $H_{U_1}$.

	\item Focusing on $f$-electrons, we then make another mean-field-type of approximation and take the corresponding second-order self-energy correction to be \emph{site-diagonal} in real space, or otherwise stated, $\vec{k}$-independent in momentum space~\cite{SCH90}. Such an approximation is at the heart of dynamical mean-field theory methods~\cite{MET89,GEO96,DAT23,RAI23a}, and is exact in an infinite number of spatial dimensions~\cite{SCH90}. In the same spirit we will also neglect correlations between $f$-electrons from different lattice sites, which are expected to be small due their localized nature, as explained around \cref{app:eqn:correlation_between_f_different_sites_ignored}. We note that in the atomic limit of the problem, which will be defined and solved analytically in \cref{app:sec:se_symmetric_details:atomic_se}, both the $f$-electron Green's function and self-energy are site-diagonal. 
\end{itemize}
In summary, we will take the second-order contribution to the self-energy to be $\vec{k}$-independent and compute it from $H_{U_1}$ and $H_{U_2}$ using the Feynman diagrams from \cref{app:fig:general_self_en_diags:h,app:fig:general_self_en_diags:i} and ignoring correlation between $f$-electrons not residing on the same lattice site.

\subsection{Feynman rules}\label{app:sec:se_correction_beyond_HF:feynman_rules}

\begin{figure}[!t]
	\centering
	\begin{tikzpicture}[baseline=(current bounding box.north)]
		\begin{feynman}
			\vertex (b);
			\vertex [above=1 cm of b] (d);
			\vertex [above=0.5 cm of b] (mid);
			\vertex [below left=1 cm of b] (a);
			\vertex [below right=1 cm of b] (c);
			\vertex [above left=1 cm of d] (e);
			\vertex [above right=1 cm of d] (f);
			\diagram*{
				(a) -- [fermion, momentum={$i \omega_z$}]  (b) -- [fermion, momentum={$i \omega_x +  i \omega_z - i \omega_y$}] (c),
				(b) -- [photon] (d),
				(e) -- [fermion, momentum'={$i \omega_x$}]  (d) -- [fermion, momentum'={$i \omega_y$}] (f),
			};
		\end{feynman}
		\path (mid) ++(-2cm,2cm) node{a)};
		\path (b) ++(0cm,-1cm) node{$\vec{R},\alpha,\eta,s$};
		\path (d) ++(0cm,1cm) node{$\vec{R}',\alpha',\eta',s'$};
		\path (mid) ++(0,-2.5cm) node[anchor=center]{$\displaystyle U_1 \delta_{\vec{R},\vec{R}'} \left(1 - \delta_{\alpha \alpha'} \delta_{\eta \eta'} \delta_{s s'} \right) + U_2 \sum_{i=0}^{5} \delta_{\vec{R},\vec{R}'+C^i_{6z} \vec{a}_{M1}} $};
	\end{tikzpicture}
\begin{tikzpicture}[baseline=(current bounding box.north)]
		\begin{feynman}
			\vertex (mid);
			\vertex [left=1.5 cm of mid] (a) {$\vec{R}',\alpha',\eta',s'$};
			\vertex [right=1.5 cm of mid] (c) {$\vec{R},\alpha,\eta,s$};
			\diagram*{
				(a) -- [fermion, momentum=$i \omega_n$]  (c),
			};
		\end{feynman}
		\path (mid) ++(-2cm,2cm) node{b)};
		\path (mid) ++(0,-2.5cm) node[anchor=center]{$\displaystyle \mathcal{G}^{f,0}_{\alpha \eta s; \alpha' \eta' s'} \left(i \omega_n, \vec{R}-\vec{R}' \right) $};
	\end{tikzpicture}
\subfloat{\label{app:fig:feynmann_rules:a}}\subfloat{\label{app:fig:feynmann_rules:b}}\caption{Feynman rules for the $f$-electron interaction and bare $f$-electron propagator. The vertex contribution is shown in (a) and includes both the onsite and the nearest neighbor $f$-electron interaction (stemming from the $H_{U_1}$ and $H_{U_2}$ interaction terms, respectively). The interaction vertex is shown as a ``photon'' line linking two $f$-electron states. Each end of the ``photon'' line serves as the starting and ending point of a propagator line. The $f$-electron bare propagator in real space is shown in (b). The flow of charge is indicated by the arrow inside the propagator, while the ``flow'' of imaginary frequency is indicated by the arrow above the propagator.}
	\label{app:fig:feynmann_rules}
\end{figure}
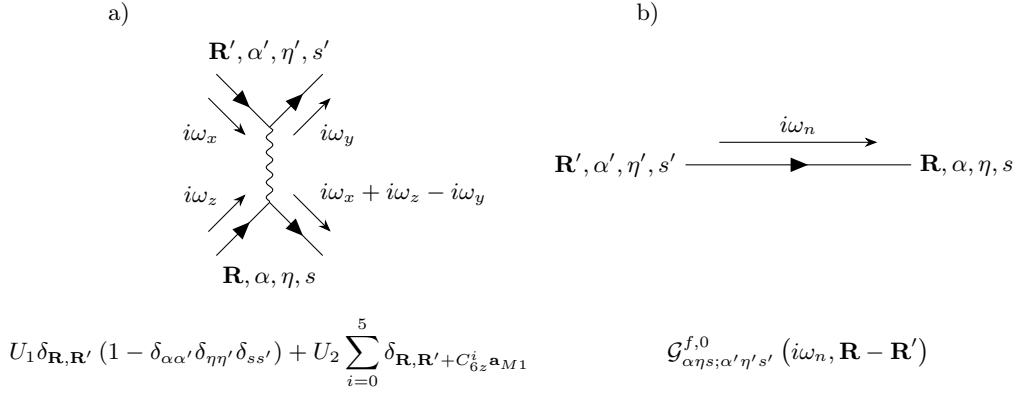

We are now ready to derive the Feynman rules for the interactions and propagators relevant for computing the $f$-electron second-order self-energy correction. As the $f$-electron interaction is local in real-space we will find it useful to define the non-interacting ($\mathcal{G}^{f,0}$)\footnote{The non-interacting $f$-electron Green's function still includes the effects of $f$-$c$ hybridization, and therefore $\mathcal{G}^{f,0} \left( i \omega_n ,\vec{R} \right) \neq \frac{\delta_{\vec{R},\vec{0}}}{i \omega_n}$ } and the fully-interacting ($\mathcal{G}^{f}$) $f$-electron Matsubara Green's functions in real space, as well,
\begin{equation}
	\label{app:eqn:matsubara_gf_f}
	-\left\langle \mathcal{T}_{\tau} \hat{f}_{\vec{R},\alpha,\eta,s} \left( \tau \right) \hat{f}^\dagger_{\vec{R}',\alpha',\eta',s'} \left( 0 \right)  \right\rangle_{(0)}  = \mathcal{G}^{f,(0)}_{\alpha \eta s; \alpha' \eta' s'} \left(\tau, \vec{R}-\vec{R}' \right),
\end{equation}
where, in keeping the discussion general for now, we have not yet ignored the correlations between $f$-electrons residing at different lattice sites ({\it i.e.}{}, site-off-diagonal elements of the Green's function). The real-space $f$-electron Green's functions are related to the ones introduced in \cref{app:eqn:matsubara_gf_THF_tau,app:eqn:matsubara_gf_THF_tau_noninteracting} by a Fourier transformation over the spatial variable
\begin{equation}
	\label{app:eqn:spatial_fourier_trafo_gf_f}
	\mathcal{G}^{f,(0)}_{\alpha \eta s; \alpha' \eta' s'} \left(\tau, \vec{R} \right) = \frac{1}{N_0} \sum_{\vec{k}} \mathcal{G}^{(0)}_{(\alpha+4) \eta s; (\alpha'+4) \eta' s'} \left(\tau, \vec{k} \right) e^{i \vec{k} \cdot \vec{R}}. 
\end{equation}
Similarly to \cref{app:eqn:spatial_fourier_trafo_gf_f}, we can also define the real-space self-energy of the $f$-electrons to be
\begin{equation}
	\label{app:eqn:spatial_fourier_trafo_sigma_f}
	\Sigma^{f}_{\alpha \eta s; \alpha' \eta' s'} \left( i \omega_n, \vec{R} \right) = \frac{1}{N_0} \sum_{\vec{k}} \Sigma_{(\alpha+4) \eta s; (\alpha'+4) \eta' s'} \left(i \omega_n, \vec{k} \right) e^{i \vec{k} \cdot \vec{R}},
\end{equation}
as well as the real-space spectral function of the $f$-electrons
\begin{equation}
	\label{app:eqn:spatial_fourier_trafo_spectral_f}
	A^{f}_{\alpha \eta s; \alpha' \eta' s'} \left(\omega, \vec{R} \right) = \frac{1}{N_0} \sum_{\vec{k}} A_{(\alpha+4) \eta s; (\alpha'+4) \eta' s'} \left(\omega, \vec{k} \right) e^{i \vec{k} \cdot \vec{R}}. 
\end{equation}

The $f$-electron onsite and nearest-neighbor repulsion terms from \cref{app:eqn:THF_int:U1,app:eqn:THF_int:U2} can be rewritten as
\begin{align}
	H_{U_1} &= \frac{U_1}{2}  \sum_{\substack{\vec{R} \\ \left(\alpha,\eta,s\right) \neq \left(\alpha',\eta',s' \right)}} \hat{f}^\dagger_{\vec{R},\alpha,\eta,s} \hat{f}_{\vec{R},\alpha,\eta,s} \hat{f}^\dagger_{\vec{R},\alpha',\eta',s'} \hat{f}_{\vec{R},\alpha',\eta',s'} - \frac{7 U_1}{2}  \sum_{\substack{\vec{R} \\ \alpha,\eta,s}} \hat{f}^\dagger_{\vec{R},\alpha,\eta,s} \hat{f}_{\vec{R},\alpha,\eta,s} + 8 U_1, \label{app:eqn:THF_int:U1:subtracted} \\
	H_{U_2} &= \frac{U_2}{2}  \sum_{\substack{\left\langle \vec{R}, \vec{R}' \right\rangle \\ \alpha,\eta,s, \\ \alpha',\eta',s'}} \hat{f}^\dagger_{\vec{R},\alpha,\eta,s} \hat{f}_{\vec{R},\alpha,\eta,s} \hat{f}^\dagger_{\vec{R}',\alpha',\eta',s'} \hat{f}_{\vec{R}',\alpha',\eta',s'} - 24 U_2  \sum_{\substack{\vec{R} \\ \alpha,\eta,s}} \hat{f}^\dagger_{\vec{R},\alpha,\eta,s} \hat{f}_{\vec{R},\alpha,\eta,s} + 96 U_2. \label{app:eqn:THF_int:U2:subtracted}
\end{align}
In this form, it is easy to see that each of $H_{U_1}$ and $H_{U_2}$ is a sum between a purely interacting term, a single-particle contribution and a constant\footnote{The difference between the first term in each of \cref{app:eqn:THF_int:U1:subtracted,app:eqn:THF_int:U2:subtracted} and $H_{U_1}$ and $H_{U_2}$, respectively, is akin to the difference between the ``standard Hubard model'' and the ``full Hubbard model''~\cite{MAH00}.}. It is worth noting that in our self-consistent Hartree-Fock treatment, the single-particle and constant contributions in \cref{app:eqn:THF_int:U1:subtracted,app:eqn:THF_int:U2:subtracted} are already treated exactly. Therefore, for the second-order correction to the self-energy one should only consider the purely interacting terms, {\it i.e.}{} the first terms of each of \cref{app:eqn:THF_int:U1:subtracted,app:eqn:THF_int:U2:subtracted}. 

The bare $f$-electron propagator line and the $f$-electron interaction vertex are shown in \cref{app:fig:feynmann_rules}. Although the Coulomb interaction between the $f$-electrons is assumed to happen instantaneously, we still find it useful to draw the interaction vertex with a ``photon'' line. The Feynman rules for constructing and computing the $f$-electron self-energy contribution $\Sigma_{\alpha \eta s; \alpha' \eta' s'} \left(i \omega_n, \vec{R}-\vec{R}' \right)$ can be summarized below (they were adapted from Ref.~\cite{MAH00} to which the reader is pointed for their complete derivation):
\begin{enumerate}
	\item Draw the Feynman diagram for the $f$-electron self-energy with all the corresponding vertices and propagators. The Feynman diagrams are obtained by connecting together interaction vertices with propagator edges. The self-energy diagram should be 1-particle-irreducible and should contain exactly two amputated propagator edges labeled as ``$\vec{R},\alpha,\eta,s$'' (on the amputated edge for which the charge ``leaves'' the diagram) and ``$\vec{R}',\alpha',\eta',s'$'' (on the amputated edge for which the charge ``enters'' the diagram). The amputated edges should also carry imaginary frequency $i \omega_n$ is the same direction as the charge.
	\item Label every end point of the interaction ``photon'' line by a quadruplet of lattice site, orbital type, valley, and spin. The ``photon'' end points connected to the amputated edges should be labeled with the same quadruplet as the corresponding edge, as shown in \cref{app:fig:feynmann_rules:a}.
	\item Assign an imaginary frequency to each internal electron propagator together with an arrow indicating the ``flow'' thereof. The Matsubara frequency should be conserved at each interaction vertex, as shown in \cref{app:fig:feynmann_rules:a}.
	\item Write down the amplitude corresponding to the diagram in the following way:
	\begin{enumerate}
		\item For each \emph{internal} propagator line of imaginary frequency $i \omega_x$ where the flow of charge points from ``$\vec{R}_2, \alpha_2, \eta_2, s_2$'' to ``$\vec{R}_1, \alpha_1, \eta_1, s_1$'', associate a factor $\mathcal{G}^{f,0}_{\alpha_1 \eta_1 s_1; \alpha_2 \eta_2 s_2} \left(i \omega_x, \vec{R}_1-\vec{R}_2 \right) $.
		\item For each interaction vertex where the ``photon'' line connects two $f$-states labeled, respectively, by ``$\vec{R}_2, \alpha_2, \eta_2, s_2$'' to ``$\vec{R}_1, \alpha_1, \eta_1, s_1$'', associate a factor $U_1 \delta_{\vec{R}_1,\vec{R}_2} \left(1 - \delta_{\alpha_1 \alpha_2} \delta_{\eta_1 \eta_2} \delta_{s_1 s_2} \right) + U_2 \sum_{i=0}^{5} \delta_{\vec{R}_1,\vec{R}_2+C^i_{6z} \vec{a}_{M1}} $.
		\item Sum over internal degrees of freedom: lattice sites, orbital numbers, valleys, spins, and imaginary frequencies.
		\item Finally, as explained in Ref.~\cite{MAH00}, multiply the expression by 
		\begin{equation}
			\label{app:eqn:feynman_rule_sign}
			\frac{\left( -1 \right)^{m+F}}{\beta^m},
		\end{equation}
		where $F$ is the number of closed $f$-fermion loops, and $m$ is the number of interaction vertices in the diagram ({\it i.e.}{}, the order of the diagram). As a result of this rule, the two diagrams in \cref{app:fig:general_self_en_diags:h,app:fig:general_self_en_diags:i} differ by an relative minus sign, as a result of the first one having a closed fermion loop.
	\end{enumerate}
\end{enumerate}

\subsection{The site-diagonal second-order correction of the \texorpdfstring{$f$}{f}-electrons' self-energy}\label{app:sec:se_correction_beyond_HF:all_so}

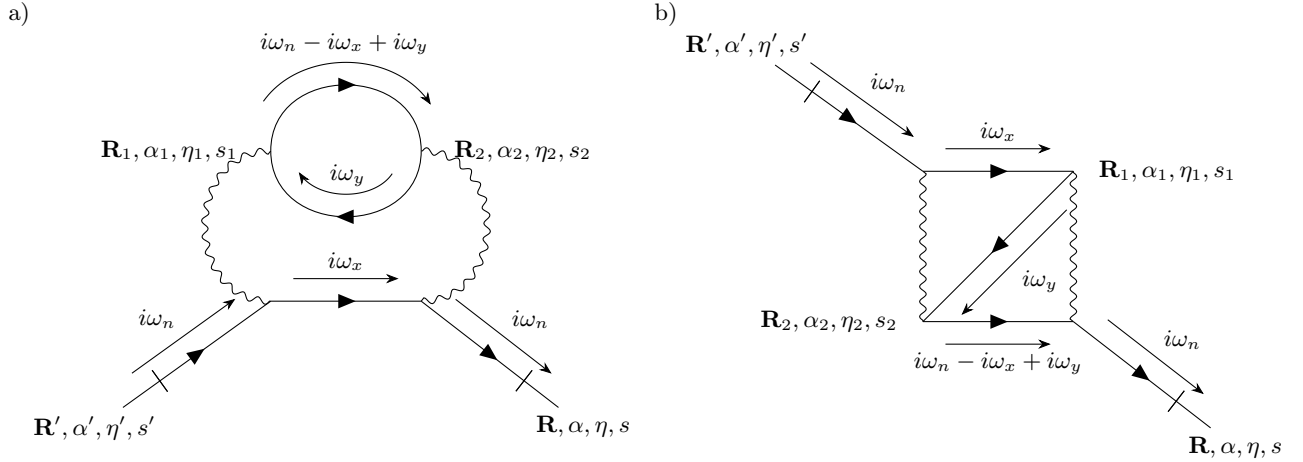
\begin{figure}[!t]
	\centering
	\begin{tikzpicture}[baseline=(current bounding box.north)]
		\begin{feynman}
			\vertex (b);
			\vertex [right=2 cm of b] (c);
			\vertex [below left=2 cm of b] (a) {$\vec{R}',\alpha',\eta',s'$};
			\vertex [below right=2 cm of c] (d) {$\vec{R},\alpha,\eta,s$};
			\vertex [above=2 cm of b] (bb); 
			\vertex [above=2 cm of c] (cc);
			\diagram* {
				(a) -- [fermion, strike through pre, momentum=$i \omega_n$] (b) -- [fermion, momentum=$i \omega_x$] (c) -- [fermion, strike through post, momentum=$i \omega_n$] (d),
				(b) -- [photon, half left] (bb),
				(cc) -- [photon, half left] (c),
				(bb) -- [fermion, half left, momentum=$i \omega_n - i \omega_x + i \omega_y$] (cc),
				(cc) -- [fermion, half left, momentum'=$i \omega_y$] (bb),
			};
		\end{feynman}
		\path (a) ++(-1,5.5cm) node{a)};
		\path (bb) ++(-1.35cm,0cm) node{$\vec{R}_1,\alpha_1,\eta_1,s_1$};
		\path (cc) ++(+1.35cm,0cm) node{$\vec{R}_2,\alpha_2,\eta_2,s_2$};
	\end{tikzpicture}
\begin{tikzpicture}[baseline=(current bounding box.north)]
		\begin{feynman}
			\vertex (b) ;
			\vertex [left=2 cm of b] (a);
			\vertex [below=2 cm of b] (c);
			\vertex [left=2 cm of c] (d);
			\vertex [above left=2 cm of a] (aa) {$\vec{R}',\alpha',\eta',s'$};
			\vertex [below right=2 cm of c] (cc) {$\vec{R},\alpha,\eta,s$};
			\vertex [below left=2 cm of d] (dd);
			\diagram* {
				(aa) -- [fermion, strike through pre, momentum=$i \omega_n$] (a),
				(a) -- [fermion, momentum=$i \omega_x$] (b)  -- [fermion, momentum=$i \omega_y$] (d) -- [fermion, momentum'=$i \omega_n - i \omega_x + i \omega_y$] (c),
				(b) -- [photon] (c),
				(d) -- [photon] (a),
				(c) -- [fermion, strike through post, momentum=$i \omega_n$] (cc),
			};
		\end{feynman}
		\path (dd) ++(-2cm,5.5cm) node{b)};
		\path (b) ++(+1.25cm,+0cm) node{$\vec{R}_1,\alpha_1,\eta_1,s_1$};
		\path (d) ++(-1.25cm,-0cm) node{$\vec{R}_2,\alpha_2,\eta_2,s_2$};
	\end{tikzpicture}
	\subfloat{\label{app:fig:f_self_en_diags:a}}\subfloat{\label{app:fig:f_self_en_diags:b}}\caption{Feynman diagrams contributing to the $f$-electron self-energy at the second order.}
	\label{app:fig:f_self_en_diags}
\end{figure}

In \cref{app:fig:f_self_en_diags}, we show the two diagrams which contribute to the $f$-electron self-energy at the second order within a self-consistent perturbation scheme, as discussed in \cref{app:sec:se_correction_beyond_HF:all_so_corrections:diagrams}. Using the Feynman rules derived in \cref{app:sec:se_correction_beyond_HF:feynman_rules}, we compute the self-energy contributions from each of the two diagrams to be, respectively, given by
\begin{align}
	& \Sigma^{f,(2a)}_{\alpha \eta s; \alpha' \eta' s'} \left(i \omega_n, \vec{R}-\vec{R}' \right) = -\frac{1}{\beta^{2}}\sum_{i \omega_x, i \omega_y} \sum_{\substack{\vec{R}_1,\alpha_1,\eta_1,s_1 \\ \vec{R}_2,\alpha_2,\eta_2,s_2}} \left[ U_1 \delta_{\vec{R}',\vec{R}_1} \left(1 - \delta_{\alpha' \alpha_1} \delta_{\eta' \eta_1} \delta_{s' s_1} \right) + U_2 \sum_{i=0}^{5} \delta_{\vec{R}',\vec{R}_1+C^i_{6z} \vec{a}_{M1}} \right] \nonumber \\
	\times & \left[ U_1 \delta_{\vec{R},\vec{R}_2} \left(1 - \delta_{\alpha \alpha_2} \delta_{\eta \eta_2} \delta_{s s_2} \right) + U_2 \sum_{i=0}^{5} \delta_{\vec{R},\vec{R}_2+C^i_{6z} \vec{a}_{M1}} \right] \nonumber \\ 
	\times & \mathcal{G}^{f,0}_{\alpha \eta s; \alpha' \eta' s'} \left(i \omega_x, \vec{R}-\vec{R}' \right) \mathcal{G}^{f,0}_{\alpha_{2} \eta_{2} s_{2};\alpha_{1} \eta_{1} s_{1}} \left(i \omega_n - i \omega_x + i \omega_y, \vec{R}_2-\vec{R}_1 \right) \mathcal{G}^{f,0}_{\alpha_{1} \eta_{1} s_{1};\alpha_{2} \eta_{2} s_{2}} \left(i \omega_y, \vec{R}_1-\vec{R}_2 \right), \label{app:eqn:se_2a_complicated}\\
& \Sigma^{f,(2b)}_{\alpha \eta s; \alpha' \eta' s'} \left(i \omega_n, \vec{R}-\vec{R}' \right) = \frac{1}{\beta^{2}}\sum_{i \omega_x, i \omega_y} \sum_{\substack{\vec{R}_1,\alpha_1,\eta_1,s_1 \\ \vec{R}_2,\alpha_2,\eta_2,s_2}} \left[ U_1 \delta_{\vec{R}',\vec{R}_2} \left(1 - \delta_{\alpha' \alpha_2} \delta_{\eta' \eta_2} \delta_{s' s_2} \right) + U_2 \sum_{i=0}^{5} \delta_{\vec{R}',\vec{R}_2+C^i_{6z} \vec{a}_{M1}} \right] \nonumber \\
	\times & \left[ U_1 \delta_{\vec{R},\vec{R}_1} \left(1 - \delta_{\alpha \alpha_1} \delta_{\eta \eta_1} \delta_{s s_1} \right) + U_2 \sum_{i=0}^{5} \delta_{\vec{R},\vec{R}_1+C^i_{6z} \vec{a}_{M1}} \right] \nonumber \\ 
	\times & \mathcal{G}^{f,0}_{\alpha_{1} \eta_{1} s_{1}; \alpha' \eta' s'} \left(i \omega_x, \vec{R}_1-\vec{R}' \right) \mathcal{G}^{f,0}_{\alpha_{2} \eta_{2} s_{2};\alpha_{1} \eta_{1} s_{1}} \left(i \omega_y, \vec{R}_2-\vec{R}_1 \right) \mathcal{G}^{f,0}_{\alpha \eta s;\alpha_{2} \eta_{2} s_{2}} \left(i \omega_n - i \omega_x + i \omega_y, \vec{R}-\vec{R}_2 \right). \label{app:eqn:se_2b_complicated}
\end{align}
The letters in the superscripts $(2a)$ and $(2b)$ point to the panel of \cref{app:fig:f_self_en_diags} in which the corresponding Feynman diagram is shown. Both diagrams have the same number of interaction vertices ({\it i.e.}{} two), but the diagram from \cref{app:fig:f_self_en_diags:a} has one closed fermion loop (while the one in \cref{app:fig:f_self_en_diags:b} has none). As a result of the rule above \cref{app:eqn:feynman_rule_sign}, both diagrams will have a $\frac{1}{\beta^2}$ prefactor, but the diagram in \cref{app:fig:f_self_en_diags:a} will have an additional minus sign. 

To simplify the expressions in \cref{app:eqn:se_2a_complicated,app:eqn:se_2b_complicated}, we now take the last approximation discussed in \cref{app:sec:se_correction_beyond_HF:all_so_corrections:particularization} and neglect any off-site second-order self-energy contributions and also ignore the lattice off-diagonal $f$-electron Green's function when computing the second-order self-energy corrections from \cref{app:eqn:se_2a_complicated,app:eqn:se_2b_complicated}\footnote{We note that because the $f$-electrons are coupled with the $c$-electrons, the off-site $f$-electron Green's function is not necessarily zero even if the $f$-electron second-order self-energy is approximated to be.}. This amounts to making the approximation in \cref{app:eqn:se_2a_complicated,app:eqn:se_2b_complicated}
\begin{equation}
	\label{app:eqn:correlation_between_f_different_sites_ignored}
	\mathcal{G}^{f}_{\alpha \eta s;\alpha' \eta' s'} \left( i \omega_n , \vec{R} \right) \approx \mathcal{G}^{f}_{\alpha \eta s;\alpha' \eta' s'} \left( i \omega_n , \vec{0} \right) \delta_{\vec{R}, \vec{0}}.
\end{equation}

As we will be interested in a self-consistent perturbation scheme, we also replace all the non-interacting Green's functions with their fully-interacting counterparts. Furthermore, for brevity of notation, we also suppress the lattice-site dependence of the $f$-electron Green's function, self-energy, and spectral function defining
\begin{equation}
	\begin{split}
		\mathcal{G}^{f}_{\alpha \eta s; \alpha' \eta' s'} \left(i \omega_n \right) &\equiv \eval{\mathcal{G}^{f}_{\alpha \eta s; \alpha' \eta' s'} \left(i \omega_n, \vec{R} \right)}_{\vec{R}=\vec{0}}, \\
		\Sigma^{f}_{\alpha \eta s; \alpha' \eta' s'} \left(i \omega_n \right) &\equiv \eval{\Sigma^{f}_{\alpha \eta s; \alpha' \eta' s'} \left(i \omega_n, \vec{R} \right)}_{\vec{R}=\vec{0}}, \\
		A^{f}_{\alpha \eta s; \alpha' \eta' s'} \left(\omega \right) &\equiv \eval{A^{f}_{\alpha \eta s; \alpha' \eta' s'} \left(\omega, \vec{R} \right)}_{\vec{R}=\vec{0}}.
	\end{split}
\end{equation} 
With these simplifications, the second-order self-energy contributions become
\begin{align}
	& \Sigma^{f,(2a)}_{\alpha \eta s; \alpha' \eta' s'} \left(i \omega_n \right) = -\frac{1}{\beta^{2}}\sum_{i \omega_x, i \omega_y} \sum_{\substack{\alpha_1,\eta_1,s_1 \\ \alpha_2,\eta_2,s_2}} \left( U_1^2 \left(1 - \delta_{\alpha' \alpha_1} \delta_{\eta' \eta_1} \delta_{s' s_1} \right) \left(1 - \delta_{\alpha \alpha_2} \delta_{\eta \eta_2} \delta_{s s_2} \right)  + 6 U_2^2 \right) \nonumber \\
	\times & \mathcal{G}^{f}_{\alpha \eta s; \alpha' \eta' s'} \left(i \omega_x \right) \mathcal{G}^{f}_{\alpha_{1} \eta_{1} s_{1};\alpha_{2} \eta_{2} s_{2}} \left(i \omega_y \right) \mathcal{G}^{f}_{\alpha_{2} \eta_{2} s_{2};\alpha_{1} \eta_{1} s_{1}} \left(i \omega_n - i \omega_x + i \omega_y \right), \label{app:eqn:se_2a_simple}\\
& \Sigma^{f,(2b)}_{\alpha \eta s; \alpha' \eta' s'} \left(i \omega_n \right) = \frac{1}{\beta^{2}}\sum_{i \omega_x, i \omega_y} \sum_{\substack{\alpha_1,\eta_1,s_1 \\ \alpha_2,\eta_2,s_2}} U_1^2 \left(1 - \delta_{\alpha' \alpha_2} \delta_{\eta' \eta_2} \delta_{s' s_2} \right) \left(1 - \delta_{\alpha \alpha_1} \delta_{\eta \eta_1} \delta_{s s_1} \right)  \nonumber \\
	\times & \mathcal{G}^{f}_{\alpha_{1} \eta_{1} s_{1}; \alpha' \eta' s'} \left(i \omega_x \right) \mathcal{G}^{f}_{\alpha_{2} \eta_{2} s_{2};\alpha_{1} \eta_{1} s_{1}} \left(i \omega_y \right) \mathcal{G}^{f}_{\alpha \eta s;\alpha_{2} \eta_{2} s_{2}} \left(i \omega_n - i \omega_x + i \omega_y \right). \label{app:eqn:se_2b_simple}
\end{align}
Note that \cref{app:eqn:se_2a_simple} is the generalization of the result of Ref.~\cite{SCH90} for an eight-band Hubbard model \emph{with} symmetry-breaking. The contribution in \cref{app:eqn:se_2b_simple} vanishes whenever one assumes a paramagnetic solution, and, as a result, was not considered in Ref.~\cite{SCH90}\footnote{The vanishing of $\Sigma^{f,(2b)}_{\alpha \eta s; \alpha' \eta' s'} \left(i \omega_n \right)$ in the paramagnetic solution can be seen immediately by substituting $\mathcal{G}^{f}_{\alpha \eta s;\alpha' \eta' s'} \left( i \omega_n \right) = \mathcal{G}^{f}_{\alpha \eta s;\alpha \eta s} \left( i \omega_n \right) \delta_{\alpha \alpha'} \delta_{\eta \eta'} \delta_{s s'}$ in \cref{app:eqn:se_2b_simple}.}. We will discuss the paramagnetic ({\it i.e.}{} fully--symmetric) case in \cref{app:sec:se_symmetric}.

In order to compute the Matsubara frequency summations in \cref{app:eqn:se_2a_simple,app:eqn:se_2b_simple}, we must employ the spectral representation of the Green's function from \cref{app:eqn:spectral_rep_of_GF}. This is because $\mathcal{G}^{f} \left( i \omega_n \right)$ is the \emph{interacting} Green's function which does not have a simple expression in terms of the non-interacting Hamiltonian. Using the spectral representation, we find that
\begin{align}
	& \Sigma^{f,(2a)}_{\alpha \eta s; \alpha' \eta' s'} \left(i \omega_n \right) = - \int \dd{\omega_1} \dd{\omega_2} \dd{\omega_3} \sum_{\substack{\alpha_1,\eta_1,s_1 \\ \alpha_2,\eta_2,s_2}} \left( U_1^2 \left(1 - \delta_{\alpha' \alpha_1} \delta_{\eta' \eta_1} \delta_{s' s_1} \right) \left(1 - \delta_{\alpha \alpha_2} \delta_{\eta \eta_2} \delta_{s s_2} \right)  + 6 U_2^2 \right) \nonumber \\
	\times & A^{f}_{\alpha \eta s; \alpha' \eta' s'} \left(\omega_1 \right) A^{f}_{\alpha_{1} \eta_{1} s_{1};\alpha_{2} \eta_{2} s_{2}} \left(\omega_2 \right) A^{f}_{\alpha_{2} \eta_{2} s_{2};\alpha_{1} \eta_{1} s_{1}} \left(\omega_3 \right) \mathcal{I} \left( i\omega_n; \omega_1, \omega_2, \omega_3 \right), \label{app:eqn:se_2a_simple_sum_1}\\
& \Sigma^{f,(2b)}_{\alpha \eta s; \alpha' \eta' s'} \left(i \omega_n \right) = \int \dd{\omega_1} \dd{\omega_2} \dd{\omega_3} \sum_{\substack{\alpha_1,\eta_1,s_1 \\ \alpha_2,\eta_2,s_2}} U_1^2 \left(1 - \delta_{\alpha' \alpha_2} \delta_{\eta' \eta_2} \delta_{s' s_2} \right) \left(1 - \delta_{\alpha \alpha_1} \delta_{\eta \eta_1} \delta_{s s_1} \right)  \nonumber \\
	\times & A^{f}_{\alpha_{1} \eta_{1} s_{1}; \alpha' \eta' s'} \left(\omega_1 \right) A^{f}_{\alpha_{2} \eta_{2} s_{2};\alpha_{1} \eta_{1} s_{1}} \left(\omega_2 \right) A^{f}_{\alpha \eta s;\alpha_{2} \eta_{2} s_{2}} \left(\omega_3 \right) \mathcal{I} \left( i\omega_n; \omega_1, \omega_2, \omega_3 \right). \label{app:eqn:se_2b_simple_sum_1}
\end{align}
In \cref{app:eqn:se_2a_simple_sum_1,app:eqn:se_2b_simple_sum_1}, we have introduced the following Matsubara sum
\begin{equation}
	\label{app:eqn:three_matsubara_sum}
	\mathcal{I} \left( i\omega_n; \omega_1, \omega_2, \omega_3 \right) = \frac{1}{\beta^2} \sum_{i\omega_x, i\omega_y} \frac{1}{i \omega_x - \omega_1} \frac{1}{i \omega_y - \omega_2} \frac{1}{i \omega_n - i\omega_x + i \omega_y - \omega_3}.
\end{equation}
The summations in \cref{app:eqn:three_matsubara_sum} can be evaluated using the identity~\cite{WIK23}
\begin{equation}
	\frac{1}{\beta}\sum_{i \omega_n} \frac{1}{\left( i\omega_n - \epsilon_1 \right) \left( i\omega_n - \epsilon_2 \right)} = \frac{n_{\mathrm{F}} \left(\epsilon_1 \right) - n_{\mathrm{F}} \left(\epsilon_2 \right)}{\epsilon_1 - \epsilon_2},
\end{equation}
where $\epsilon_1$, $\epsilon_2$ can be complex and the sumation is performed over fermionic Matsubara frequencies. The Matsubara sum in \cref{app:eqn:three_matsubara_sum} thus becomes
\begin{align}
	\mathcal{I} \left( i\omega_n; \omega_1, \omega_2, \omega_3 \right) =& \frac{1}{\beta} \sum_{i\omega_x} \frac{1}{i \omega_x - \omega_1} \frac{1}{\omega_2 + i\omega_n - i \omega_x - \omega_3} \left( n_{\mathrm{F}} \left( \omega_{2} \right) - n_{\mathrm{F}} \left( \omega_3 - i\omega_n - i\omega_x \right) \right) \nonumber \\
	=& \frac{1}{\beta} \sum_{i\omega_x} \frac{1}{i \omega_x - \omega_1} \frac{-1}{i \omega_x - \omega_2 - i\omega_n + \omega_3} \left( n_{\mathrm{F}} \left( \omega_{2} \right) - n_{\mathrm{F}} \left( \omega_{3} \right) \right) \nonumber \\
=& -\frac{ n_{\mathrm{F}} \left( \omega_{1} \right) - n_{\mathrm{F}} \left( \omega_2 + i \omega_n - \omega_3 \right)}{\omega_1 - \omega_2 - i \omega_n + \omega_3} \left( n_{\mathrm{F}} \left( \omega_{2} \right) - n_{\mathrm{F}} \left( \omega_{3} \right) \right) \nonumber \\
	=& \frac{1}{i \omega_n - \omega_1 + \omega_2 - \omega_3} \left( n_{\mathrm{F}} \left( \omega_{1} \right) - \frac{1}{1-e^{\beta \omega_2} e^{-\beta \omega_3} } \right) \left( n_{\mathrm{F}} \left( \omega_{2} \right) - n_{\mathrm{F}} \left( \omega_{3} \right) \right) \nonumber \\
	=& \frac{1}{i \omega_n - \omega_1 + \omega_2 - \omega_3} \left[ n_{\mathrm{F}} \left( \omega_{1} \right) - \frac{1}{1- \frac{n_{\mathrm{F}} \left( \omega_{3} \right) \left( 1 - n_{\mathrm{F}} \left( \omega_{2} \right) \right)}{n_{\mathrm{F}} \left( \omega_{2} \right) \left( 1 - n_{\mathrm{F}} \left( \omega_{3} \right) \right)} } \right] \left( n_{\mathrm{F}} \left( \omega_{2} \right) - n_{\mathrm{F}} \left( \omega_{3} \right) \right) \nonumber \\
	=& \frac{1}{i \omega_n - \omega_1 + \omega_2 - \omega_3} \left[ n_{\mathrm{F}} \left( \omega_{1} \right) - \frac{n_{\mathrm{F}} \left( \omega_{2} \right)-n_{\mathrm{F}} \left( \omega_{2} \right) n_{\mathrm{F}} \left( \omega_{3} \right)}{n_{\mathrm{F}} \left( \omega_{2} \right)-n_{\mathrm{F}} \left( \omega_{3} \right)} \right] \left( n_{\mathrm{F}} \left( \omega_{2} \right) - n_{\mathrm{F}} \left( \omega_{3} \right) \right) \nonumber \\
	=& \frac{n_{\mathrm{F}} \left( \omega_{1} \right)n_{\mathrm{F}} \left( \omega_{2} \right) - n_{\mathrm{F}} \left( \omega_{1} \right) n_{\mathrm{F}} \left( \omega_{3} \right) - n_{\mathrm{F}} \left( \omega_{2} \right) + n_{\mathrm{F}} \left( \omega_{2} \right) n_{\mathrm{F}} \left( \omega_{3} \right)}{i \omega_n - \omega_1 + \omega_2 - \omega_3}  \nonumber \\
	=& - \frac{n_{\mathrm{F}} \left( \omega_{1} \right) \left( 1 - n_{\mathrm{F}} \left( \omega_{2} \right) \right) n_{\mathrm{F}} \left( \omega_{3} \right) + \left( 1 - n_{\mathrm{F}} \left( \omega_{1} \right) \right) n_{\mathrm{F}} \left( \omega_{2} \right) \left( 1 - n_{\mathrm{F}} \left( \omega_{3} \right) \right)}{i \omega_n - \omega_1 + \omega_2 - \omega_3}. \label{app:eqn:matsubara_sum_I}
\end{align}

Even with the analytical expression of the Matsubara sum $\mathcal{I} \left( i\omega_n; \omega_1, \omega_2, \omega_3 \right)$, \cref{app:eqn:se_2a_simple_sum_1,app:eqn:se_2b_simple_sum_1} are still cumbersome to evaluate, since they involve a nested triple integral over frequency. To move forward, we follow Ref.~\cite{SCH90} and perform an analytical continuation to the real axis by substituting $i \omega_n \to \omega + i 0^{+}$. The denominator of $\mathcal{I} \left( i\omega_n; \omega_1, \omega_2, \omega_3 \right)$ can then be separated by means of the integral
\begin{equation}
	\frac{1}{z - \omega_1 + \omega_2 - \omega_3} = -i \int_{0}^{\infty} \dd{\lambda} \exp \left[i \lambda \left( z - \omega_1 + \omega_2 - \omega_3 \right)  \right], \qq{for any} z \in \mathbb{C}, \quad \Im{z} > 0.
\end{equation}
In turn, this allows us to express the second-order self-energy corrections in \cref{app:eqn:se_2a_simple_sum_1} as a single integral
{\small \begin{align}
		& \Sigma^{f,(2a)}_{\alpha \eta s; \alpha' \eta' s'} \left(\omega + i 0^{+} \right) = - i \sum_{\substack{\alpha_1,\eta_1,s_1 \\ \alpha_2,\eta_2,s_2}} \int_{0}^{\infty} \dd{\lambda} e^{i \lambda \omega} \left( U_1^2 \left(1 - \delta_{\alpha' \alpha_1} \delta_{\eta' \eta_1} \delta_{s' s_1} \right) \left(1 - \delta_{\alpha \alpha_2} \delta_{\eta \eta_2} \delta_{s s_2} \right)  + 6 U_2^2 \right) \nonumber \\
		\times & 
		\left( B_{\alpha \eta s; \alpha' \eta' s'} \left(\lambda \right) C_{\alpha_{1} \eta_{1} s_{1};\alpha_{2} \eta_{2} s_{2}} \left(-\lambda \right) B_{\alpha_{2} \eta_{2} s_{2};\alpha_{1} \eta_{1} s_{1}} \left(\lambda \right) 
		+ C_{\alpha \eta s; \alpha' \eta' s'} \left(\lambda \right) B_{\alpha_{1} \eta_{1} s_{1};\alpha_{2} \eta_{2} s_{2}} \left(-\lambda \right) C_{\alpha_{2} \eta_{2} s_{2};\alpha_{1} \eta_{1} s_{1}} \left(\lambda \right) \right), \label{app:eqn:se_2a_simple_sum_2}\\
& \Sigma^{f,(2b)}_{\alpha \eta s; \alpha' \eta' s'} \left(\omega + i 0^{+} \right) = i \sum_{\substack{\alpha_1,\eta_1,s_1 \\ \alpha_2,\eta_2,s_2}} \int_{0}^{\infty} \dd{\lambda} e^{i \lambda \omega} U_1^2 \left(1 - \delta_{\alpha' \alpha_2} \delta_{\eta' \eta_2} \delta_{s' s_2} \right) \left(1 - \delta_{\alpha \alpha_1} \delta_{\eta \eta_1} \delta_{s s_1} \right)  \nonumber \\
		\times & 
		\left( B_{\alpha_{1} \eta_{1} s_{1}; \alpha' \eta' s'} \left(\lambda \right) C_{\alpha_{2} \eta_{2} s_{2};\alpha_{1} \eta_{1} s_{1}} \left(-\lambda \right) B_{\alpha \eta s;\alpha_{2} \eta_{2} s_{2}} \left(\lambda \right) 
		+ C_{\alpha_{1} \eta_{1} s_{1}; \alpha' \eta' s'} \left(\lambda \right) B_{\alpha_{2} \eta_{2} s_{2};\alpha_{1} \eta_{1} s_{1}} \left(-\lambda \right) C_{\alpha \eta s;\alpha_{2} \eta_{2} s_{2}} \left(\lambda \right)
		\right), \label{app:eqn:se_2b_simple_sum_2}
\end{align}}where we have introduced the following auxiliary functions
\begin{align}
	B_{\alpha \eta s; \alpha' \eta' s'} \left( \lambda \right) &= \int_{-\infty}^{\infty} \dd{\omega} n_{\mathrm{F}} \left( \omega \right) A^{f}_{\alpha \eta s; \alpha' \eta' s'} \left(\omega \right) e^{-i \lambda \omega}, \label{app:eqn:aux_func_B_self_en} \\
	C_{\alpha \eta s; \alpha' \eta' s'} \left( \lambda \right) &= \int_{-\infty}^{\infty} \dd{\omega} \left( 1 - n_{\mathrm{F}} \left( \omega \right) \right) A^{f}_{\alpha \eta s; \alpha' \eta' s'} \left(\omega \right) e^{-i \lambda \omega}
	\label{app:eqn:aux_func_C_self_en}.
\end{align}
Note that \cref{app:eqn:se_2a_simple_sum_2,app:eqn:se_2b_simple_sum_2} only involve a sequence of one-dimensional integrals (one to compute the auxiliary functions in \cref{app:eqn:aux_func_B_self_en,app:eqn:aux_func_C_self_en} and one to compute the self-energy contributions from \cref{app:eqn:se_2a_simple_sum_2,app:eqn:se_2b_simple_sum_2}). In contrast, \cref{app:eqn:se_2a_simple_sum_1,app:eqn:se_2a_simple_sum_2} involve a nested triple integral. As a result, \cref{app:eqn:se_2a_simple_sum_2,app:eqn:se_2b_simple_sum_2} are much more amenable to numerical evaluation and, as we will show in \cref{app:sec:se_correction_beyond_HF:sc_problem_and_numerics}, can be used to self-consistently solve the THF model within second-order perturbation theory. The total $f$-electron self-energy at the second-order level can be found by summing the two contributions from \cref{app:eqn:se_2a_simple_sum_2,app:eqn:se_2b_simple_sum_2}
\begin{equation}
	\label{app:eqn:full_f_second_order_sigma}
	\Sigma^{f,(2)} \left(\omega + i 0^{+} \right) \equiv \Sigma^{f,(2a)} \left(\omega + i 0^{+} \right) + \Sigma^{f,(2b)} \left(\omega + i 0^{+} \right),
\end{equation}
which correspond, respectively, to the two diagrams from \cref{app:fig:f_self_en_diags}. 

\subsection{The self-consistent problem and numerical implementation}\label{app:sec:se_correction_beyond_HF:sc_problem_and_numerics}

Having derived the second-order self-energy correction stemming from the $H_{U_1}$ and $H_{U_2}$ terms of the interaction Hamiltonian, we now show how the THF model can be solved numerically in the symmetry-broken phase using self-consistent second-order perturbation theory. 

For a given density matrix $\varrho \left( \vec{k} \right)$ and a local ({\it i.e.}{}, $\vec{k}$-independent) second-order self-energy correction $\Sigma^{(2)} \left( \omega + i 0^{+} \right)$, the fully-interacting (up to second order in the $H_{U_1}$ and $H_{U_2}$ interactions and up to first order in all the other interaction terms) retarded Green's function of the system can be found from the Dyson equation
\begin{equation}
	\label{app:eqn:second_order_GF}
	\mathcal{G} \left( \omega + i 0^{+}, \vec{k} \right) = \left[\left( \omega + i 0^{+} + \mu \right) \mathbb{1} - h^{\text{MF}} \left( \vec{k} \right) - \Sigma^{(2)} \left( \omega + i 0^{+} \right) \right]^{-1},
\end{equation}
where the mean-field Hamiltonian $h^{\text{MF}} \left( \vec{k} \right)$ is determined from the density matrix via \cref{app:eqn:TBG_HF_Hamiltonian} or \cref{app:eqn:full_TSTG_HF_Hamiltonian}. At the same time, $\mathcal{G} \left( \omega + i 0^{+}, \vec{k} \right)$ determines the spectral function of the system via \cref{app:eqn:spectral_function}, and, through the latter, the density matrix of the system $\varrho \left( \vec{k} \right)$ from \cref{app:eqn:def_rho_HF} as well as the second-order $f$-electron self-energy $\Sigma^{f,(2)} \left(\omega + i 0^{+} \right)$ from \cref{app:eqn:se_2a_simple_sum_2,app:eqn:se_2b_simple_sum_2,app:eqn:aux_func_B_self_en,app:eqn:aux_func_C_self_en,app:eqn:full_f_second_order_sigma}. Assuming that the $c$- and $d$-fermions have a constant broadening $\Gamma_{c}$, the \emph{entire} second-order self-energy correction $\Sigma^{(2)} \left( \omega + i 0^{+} \right)$ can also be determined from 
\begin{equation}
	\label{app:eqn:full_second_order_sigma}
	\Sigma^{(2)}_{i \eta s;i' \eta' s'} \left( \omega + i 0^{+} \right) = \begin{cases}
		\Sigma^{f,(2)}_{(i-4) \eta s; (i'-4) \eta' s'} \left(\omega + i 0^{+} \right) & \qq{if} 5 \leq i,i' \leq 6 \\
		- i \Gamma_{c} \delta_{i i'} \delta_{\eta \eta'} \delta_{s s'} & \qq{otherwise}
	\end{cases}.
\end{equation}

We employ $\Gamma_{c} = \SI{1}{\milli\electronvolt}$ ($\Gamma_{c} = \SI{1.5}{\milli\electronvolt}$) for TBG (TSTG), which is kept constant during the iterative algorithm described below\footnote{The fact that $\Gamma^{\text{TSTG}}_{c}/ \Gamma^{\text{TBG}}_{c} =  3/2  \approx{\sqrt{2}}$ reflects the scaling of the TSTG band structure by a factor of approximately $\sqrt{2}$ relative to the TBG one~\cite{YU23a}.}. The broadening factor assigned to the $c$- and $d$-electrons is added to avoid numerical issues associated with the matrix inversion used to compute the retarded Green's function of the system from \cref{app:eqn:second_order_GF}. If the dynamical self-energy of the $c$- and $f$-electrons is set to zero, then $\mathcal{G} \left( \omega + i 0^{+}, \vec{k} \right)$ will diverge at frequencies equal to the charge-one excitations at the edge of the moir\'e BZ, where the remote bands are mostly $c$-electron-like in character (or in the case of TSTG at the Dirac cone energy and momenta). $\Gamma_{c}$ can alternatively be understood as the small ``$+ i 0^{+}$'' term added to the frequency (for the $c$- and $d$-electron block) when computing the retarded Green's function. No such factor is needed in the $f$-electron block, since the diagonal part of the $f$-electron self-energy already contains a positive imaginary part. It is worth mentioning that since $\Gamma_{c}$ is much smaller than any other energy-scale of the problem (including $\Im{\Sigma^{f,(2)}_{\alpha \eta s;\alpha \eta s} \left( \omega + i 0^{+} \right)}$ around the $f$-electron excitation energies), the exact choice of $\Gamma_{c}$ is inconsequential, a fact which we checked numerically. 

Taken together \cref{app:eqn:second_order_GF,app:eqn:TBG_HF_Hamiltonian,app:eqn:full_TSTG_HF_Hamiltonian,app:eqn:spectral_function,app:eqn:def_rho_HF,app:eqn:se_2a_simple_sum_2,app:eqn:se_2b_simple_sum_2,app:eqn:aux_func_B_self_en,app:eqn:aux_func_C_self_en,app:eqn:full_f_second_order_sigma,app:eqn:full_second_order_sigma} form a self-consistent system of integral equations for the density matrix $\varrho \left(\vec{k} \right)$ and the second-order local self-energy $\Sigma^{(2)} \left( \omega + i 0^{+} \right)$. We solve the system using the following iterative algorithm: first, we denote by $\varrho_{n} \left( \vec{k} \right)$ and $\Sigma^{(2)}_{n} \left( \omega + i 0^{+} \right)$ the density matrix and second-order local self-energy of the system at the $n$-th step of the algorithm. To obtain $\varrho_{n+1} \left( \vec{k} \right)$ and $\Sigma^{(2)}_{n+1} \left( \omega + i 0^{+} \right)$, we employ the following sequence:
\begin{enumerate}
	\item From $\varrho_{n} \left( \vec{k} \right)$, we determine the Hartree-Fock Hamiltonian of the system ({\it i.e.}{} the first-order contribution to the self-energy) through \cref{app:eqn:TBG_HF_Hamiltonian,app:eqn:full_TSTG_HF_Hamiltonian}. From the Hartree-Fock Hamiltonian and $\Sigma^{(2)}_{n} \left( \omega + i 0^{+} \right)$, we determine the retarded Green's function of the system $\mathcal{G} \left( \omega + i 0^{+}, \vec{k} \right)$ using \cref{app:eqn:second_order_GF}.
	\item We then determine the spectral function of the system $A \left( \omega, \vec{k} \right)$ through \cref{app:eqn:spectral_function}. At this step, we also fix the chemical potential of the system by requiring that the desired filling $\nu_0$ obeys
	\begin{equation}
		\label{app:eqn:fixing_filling_self_energy_algo}
		\nu_0 = \frac{1}{N_0} \sum_{\vec{k}} \sum_{i,\eta,s} \left (\int_{-\infty}^{\infty} \dd{\omega} n_{\mathrm{F}} \left( \omega \right) A_{i \eta s; i \eta s} \left( \omega, \vec{k} \right) - \frac{1}{2} \right).
	\end{equation}
	We find that working at fixed filling is more stable numerically than working at a fixed chemical potential.
	\item Once the chemical potential $\mu$ has been fixed, we determine $\varrho_{n+1} \left( \vec{k} \right)$ from the spectral function via \cref{app:eqn:def_rho_HF}.
	\item Also from the spectral function, we compute $\Sigma^{(2)}_{n+1} \left( \omega + i 0^{+} \right)$ via \cref{app:eqn:se_2a_simple_sum_2,app:eqn:se_2b_simple_sum_2,app:eqn:aux_func_B_self_en,app:eqn:aux_func_C_self_en,app:eqn:full_f_second_order_sigma,app:eqn:full_second_order_sigma}. 
\end{enumerate}

The iterations are repeated until convergence is achieved. Numerically, we employ a mesh of $48 \times 48 = 2304$ $\vec{k}$-points. Along the discretized real frequency axis we use $2^{12}$ $\omega$-points. As shown in \cref{app:eqn:second_order_GF}, the algorithm entails inverting $48 \times 48 \times 2^{12}$ matrices at each step in order to obtain the retarded Green's function. To accelerate the code, we use the Intel MKL library. Furthermore, the continuous Fourier integrals over $\omega$ and $\lambda$ from \cref{app:eqn:se_2a_simple_sum_2,app:eqn:se_2b_simple_sum_2,app:eqn:aux_func_B_self_en,app:eqn:aux_func_C_self_en} are approximated by a Discrete Fourier Transformation (which is then computed using the Fast Fourier transform algorithm) according to the method devised by Ref.~\cite{BAI06}. Finally, we consider the solution to be converged when the relative change in the density matrix and second-order self-energy are no larger than $5 \times 10^{-4}$ and $5 \times 10^{-3}$, respectively. In practice, we also dampen any potential oscillations around the converged solution by linearly mixing the $n$-th and $(n+1)$-th solutions at each step.

For the various correlated states from \cref{app:tab:model_states}, the second-order self-consistent results can be obtained by starting the iterative algorithm with the zero-temperature Hartree-Fock density matrix of the given integer-filled correlated state. The initial guess for the second-order self-energy correction is simply given by a small broadening
\begin{equation}
	\Sigma^{(2)}_{i \eta s;i' \eta' s'} \left( \omega + i 0^{+} \right)  = - i \Gamma_{c} \delta_{i i'} \delta_{\eta \eta'} \delta_{s s'}.
\end{equation}

Once the second-order self-consistent solution of a given correlated state at integer filling $\nu_0$ is obtained, the symmetry-broken phases at small doping \emph{around} $\nu_0$ can be approached by doping the integer-filled solution in small increments. To be precise, we consider the solutions at fillings $\nu_0 + n \delta\nu$, where $n\in \mathbb{Z}$ and $\abs{n} \leq \frac{1}{2 \delta\nu}$ ({\it i.e.}{} the integer-filled correlated phase is doped in increments of $\delta\nu$ by at most $1/2$ away from the integer filling). The solution at $\nu_0 + n \delta \nu$ is found using the self-consistent $\varrho \left( \vec{k} \right)$ and $\Sigma^{(2)}\left( \omega + i 0^{+} \right)$ obtained for $\nu_0 + (n-1) \delta \nu$ (if $n>0$) or $\nu_0 + (n+1) \delta \nu$ (if $n<0$) as initial conditions. In practice we take $\delta \nu = 1/50$.

\section{The symmetric phase of the THF model beyond Hartree-Fock}\label{app:sec:se_symmetric}

This \siSection{} offers a detailed overview of the symmetric phase of the THF model and its characterization. To effectively capture the correlation effects within this phase, we employ a combination of dynamical mean-field theory (DMFT)~\cite{GEO96,KOT06a}, Hartree-Fock theory, and iterated perturbation theory (IPT)~\cite{MAR86,GEO92,YEY93,KAJ96,POT97,ANI97,LIC98,YEY99,MEY99,YEY00,SAS01,SAV01,FUJ03,LAA03,KUS06,ARS12,DAS16,WAG21,MIZ21,VAN22,CAN24,CAN25}. After briefly introducing the notation, our approach begins with a characterization of the THF model's symmetric phase. Subsequently, we review the DMFT method, focusing on its application to the THF model in conjunction with Hartree-Fock theory. We then detail the IPT method, emphasizing its numerical implementation. Finally, we mention that solving the THF model in the symmetric phase using the aforementioned techniques relies on several rather technical intermediate results from many-body perturbation theory. Therefore, this \siSection{} will present the methodology without proof, with the more complex technical derivations being deferred to \cref{app:sec:se_symmetric_details}.

\subsection{Summary of notation}\label{app:sec:se_symmetric:notation_summary}

Considering the technical depth of the discussions in this \siSection{} and in \cref{app:sec:se_symmetric_details}, it will be practical to begin with a summary of the functions and symbols that will be frequently used. Their definitions and the specific equations where they are introduced are listed in \cref{app:DMFT_gfs}. This concise notation summary is intended as a quick reference to facilitate easier navigation and understanding of the subsequent sections. First time readers might choose to skip this table and then consult it as needed.

{\renewcommand{\arraystretch}{1.5}
	\begin{longtable}{|l|p{0.4\linewidth}|p{0.35\linewidth}|l|}
		\hline
		\makecell{\makecell[l]{Function or \\Symbol}} & Expression & Interpretation & Introduced \\ 
		\hline \endhead
		$\tilde{\mathcal{G}}^{f} \left(i \omega_n \right)$ & $\tilde{\mathcal{G}}^{f} \left(\tau \right) \equiv -\left\langle \mathcal{T}_{\tau} \hat{f}_{\vec{R},\alpha, \eta, s} \left( \tau \right) \hat{f}^\dagger_{\vec{R},\alpha, \eta, s} \left( 0 \right)  \right\rangle$ & Site- and flavor-diagonal interacting Green's function of the $f$-electrons in the lattice problem. & \makecell[l]{\cref{app:eqn:f_elec_gf_site_flav_diag}} \\
		\hline
		$n$ & \makecell[l]{$ n \equiv \left\langle \hat{f}^\dagger_{\vec{R},\alpha, \eta, s} \hat{f}_{\vec{R},\alpha, \eta, s} \right\rangle$, \\ $n = -\frac{1}{\pi} \int_{-\infty}^{\infty} \dd{\omega} n_{\mathrm{F}} \left( \omega \right) \Im \left( \tilde{\mathcal{G}}^{f} \left( \omega + i 0^{+} \right) \right)$} & Relative $f$-electron occupation of the lattice problem. & \makecell[l]{\cref{app:eqn:int_rel_f_filling}} \\
		\hline
		$\tilde{\Sigma}^{f} \left( i \omega_n \right)$ & N/A & Site- and flavor-diagonal DMFT ansatz for the dynamical $f$-electron self-energy of the lattice problem. & \makecell[l]{\cref{app:eqn:DMFT_diagonal_self_energy}} \\
		\hline
		$G_0 \left(i \omega_n \right)$ & $G^{-1}_{0} \left( i \omega_n \right) = \left( \tilde{\mathcal{G}}^{f} \left( i \omega_n \right) \right)^{-1} + U_1 \left( n - \frac{1}{2} \right) \left( N_f - 1 \right) + \tilde{\Sigma}^{f} \left( i \omega_n \right)$ & Non-interacting Green's function of the single-site model. Formally defined in \cref{app:eqn:definition_of_g0_cavity_DMFT}, but practically computed through \cref{app:eqn:DMFT_G0_expression}. & \makecell[l]{\cref{app:eqn:definition_of_g0_cavity_DMFT}} \\
		\hline
		$\rho_{0} (\omega)$ & $\rho_{0} (\omega) \equiv -\frac{1}{\pi} \Im G_{0} \left( \omega + i 0^{+} \right)$ & Spectral function of the non-interacting single-site problem. & \makecell[l]{\cref{app:eqn:causal_g0}} \\
		\hline
		$G \left(i \omega_n \right)$ & $G \left(\tau \right) \equiv - \left\langle \mathcal{T}_{\tau} \hat{f}_{\alpha, \eta, s} \left( \tau \right) \hat{f}^\dagger_{\alpha, \eta, s} \left( 0 \right) \right\rangle^{\text{ss}}$ & Interacting Green's function of the single-site model. Per the DMFT assumption in \cref{app:eqn:DMFT_self_consistent_eq}, $G \left(i \omega_n \right) = \tilde{\mathcal{G}}^{f} \left(i \omega_n \right)$. & \makecell[l]{\cref{app:eqn:definition_ss_interacting_gf}} \\
		\hline
		$\rho (\omega)$ & $\rho (\omega) \equiv -\frac{1}{\pi} \Im G \left( \omega + i 0^{+} \right)$ & Spectral function of the interacting single-site problem. & \makecell[l]{\cref{app:eqn:causal_g}} \\
		\hline
		$n_{\text{ss}}$ & \makecell[l]{$ n_{\text{ss}} \equiv \left\langle \hat{f}^\dagger_{\alpha, \eta, s} \hat{f}_{\alpha, \eta, s} \right\rangle^{\text{ss}}$, \\ $n_{\text{ss}} = \int_{-\infty}^{\infty} \dd{\omega} n_{\mathrm{F}} \left( \omega \right) \rho (\omega)$} & Relative $f$-electron occupation of the interacting single-site problem. Per the DMFT assumption in \cref{app:eqn:DMFT_self_consistent_eq}, $n_{\text{ss}} = n$. & \makecell[l]{\cref{app:eqn:ss_relative_f_interacting}} \\
		\hline
		$\Sigma_{\text{ss}} \left( i \omega_n \right)$ & \makecell[l]{$\Sigma_{\text{ss}} \left( i \omega_n \right) = G_0^{-1} \left( i \omega_n \right) - G^{-1} \left( i \omega_n \right) $\\$ - U_1 \left( n_{\text{ss}} - \frac{1}{2} \right) \left( N_f -1 \right)$} & Dynamic $f$-electron self-energy in the single-site model. Per the DMFT assumption in \cref{app:eqn:dmft_ass_1}, $\tilde{\Sigma}^{f} \left( i \omega_n \right) = \Sigma_{\text{ss}} \left( i \omega_n \right)$. & \makecell[l]{\cref{app:eqn:all_self_energy_of_ss_model}} \\
		\hline
		$\left\langle nn \right\rangle^{\text{ss}}$ & \makecell[l]{$\left\langle nn \right\rangle^{\text{ss}} \equiv \left\langle \hat{f}^\dagger_{\alpha, \eta, s} \hat{f}_{\alpha, \eta, s} \hat{f}^\dagger_{\alpha', \eta', s'} \hat{f}_{\alpha', \eta', s'} \right\rangle^{\text{ss}}$, \\ for any $\left(\alpha, \eta, s \right) \neq \left(\alpha', \eta', s' \right)$, \\ $\left\langle nn \right\rangle^{\text{ss}} = n_{\text{ss}}^2 - \frac{\int_{-\infty}^{\infty} \dd{\omega} \Im{\Sigma_{\text{ss}} \left( \omega + i 0^{+} \right) G \left( \omega + i 0^{+} \right) } n_{\mathrm{F}} \left( \omega \right)}{\pi U_1 \left(N_f - 1 \right)}$, \\ $\left\langle nn \right\rangle^{\text{ss}} = n^2 - \frac{\int_{-\infty}^{\infty} \dd{\omega} \Im{\tilde{\Sigma}^{f} \left( \omega + i 0^{+} \right) \tilde{\mathcal{G}}^{f} \left( \omega + i 0^{+} \right) } n_{\mathrm{F}} \left( \omega \right)}{\pi U_1 \left(N_f - 1 \right)}$} & Double occupation of the $f$-electrons in the single-site problem. Per the DMFT assumptions in \cref{app:eqn:dmft_ass_1,app:eqn:DMFT_self_consistent_eq}, it can be expressed in terms of both the lattice and single-site Green's function, self-energy, and $f$-electron occupation. & \makecell[l]{\cref{app:eqn:nn_correlator_from_self_energy}} \\
		\hline
		$\tilde{\mu}$ & \textit{Computed self-consistently} & Fictitious chemical potential chosen for the single-site problem in the IPT method. & \makecell[l]{\cref{app:eqn:def_of_g_tilde_mu}} \\
		\hline
		$G^{\tilde{\mu}}_0 \left(i \omega_n \right)$ & $G^{\tilde{\mu}}_0 \left( i \omega_n \right) = \frac{1}{G_0^{-1} \left( i \omega_n \right) - \mu + \tilde{\mu}}$ & Non-interacting Green's function of the single-site model at chemical potential $\tilde{\mu}$. & \makecell[l]{\cref{app:eqn:def_of_g_tilde_mu}} \\
		\hline
		$\rho^{\tilde{\mu}}_{0} \left( \omega \right)$ & $\rho^{\tilde{\mu}}_{0} \left( \omega \right) \equiv -\frac{1}{\pi} \Im G^{\tilde{\mu}}_{0} \left( \omega + i 0^{+} \right)$ & Spectral function of the non-interacting single-site problem at chemical potential $\tilde{\mu}$. & \makecell[l]{\cref{app:eqn:spectral_gf_relation_tilde}} \\
		\hline
		$n^{\tilde{\mu}}_0$ & $n^{\tilde{\mu}}_0 \equiv \int_{-\infty}^{\infty} \dd{\omega} n_{\mathrm{F}} \left( \omega \right) \rho^{\tilde{\mu}}_0 \left( \omega \right)$ & Relative $f$-electron filling in the non-interacting single-site problem at chemical potential $\tilde{\mu}$. & \makecell[l]{\cref{app:eqn:int_rel_f0_filling}} \\
		\hline
		$G^{\text{HF}}_0 \left(i \omega_n \right)$ & $G^{\text{HF}}_0 \left( i \omega_n \right) = \frac{1}{G_0^{-1} \left( i \omega_n \right) - U_1 \left(n_{\text{ss}} - \frac{1}{2} \right) \left( N_f - 1 \right)}$ & Hartree-Fock $f$-electron Green's function in the single-site model. & \makecell[l]{\cref{app:eqn:hf_gf_single_site}} \\
		\hline	
		$\tilde{\mathcal{G}}^{\text{At}} \left( i \omega_n \right)$ & $\tilde{\mathcal{G}}^{\text{At}} \left( \tau \right) \equiv -\left\langle \mathcal{T}_{\tau} \hat{f}_{\vec{R},\alpha, \eta, s} \left( \tau \right) \hat{f}^\dagger_{\vec{R},\alpha, \eta, s} \left( 0 \right)  \right\rangle^{\text{At}}$ & Interacting Green's function of the lattice problem in the atomic limit and at chemical potential $\mu$. & \makecell[l]{\cref{app:eqn:atomic_gf_diagonal_part_with_prob}} \\
		\hline
		$n_{\text{At}}$ & $ n_{\text{At}} \equiv \left\langle \hat{f}^\dagger_{\alpha, \eta, s} \hat{f}_{\alpha, \eta, s} \right\rangle^{\text{At}}$ & Relative $f$-electron occupation of the lattice problem in the atomic limit at chemical potential $\mu$. & \makecell[l]{\cref{app:eqn:def_n_at}} \\
		\hline
		$\left\langle nn \right\rangle^{\text{At}}$ & \makecell[l]{$\left\langle nn \right\rangle^{\text{At}} \equiv \left\langle \hat{f}^\dagger_{\vec{R},\alpha, \eta, s} \hat{f}_{\vec{R},\alpha, \eta, s} \hat{f}^\dagger_{\vec{R},\alpha', \eta', s'} \hat{f}_{\vec{R},\alpha', \eta', s'}  \right\rangle^{\text{At}}$, \\ for any $\left(\alpha, \eta, s \right) \neq \left(\alpha', \eta', s' \right)$} & Double occupation of the $f$-electrons in the lattice problem in the atomic limit. & \makecell[l]{\cref{app:eqn:def_nn_at}} \\
		\hline
		$\tilde{\Sigma}^{f,\text{At}} \left( i \omega_n \right)$ & \makecell[l]{$\tilde{\Sigma}^{f,\text{At}} \left( i \omega_n  \right) \equiv i \omega_n + \mu - \left(\tilde{\mathcal{G}}^{\text{At}} \left( i \omega_n \right) \right)^{-1} $ \\ $ - U_1 \left(n_{\text{At}} -\frac{1}{2} \right) \left( N_f - 1 \right)$} & Dynamical $f$-electron self-energy of the lattice problem in the atomic limit. & \makecell[l]{\cref{app:eqn:def_sigma_at}} \\
		\hline
		\caption{Functions and symbols used in this \siSection{} and in \cref{app:sec:se_symmetric_details}. For each entry, we provide the definition, its interpretation, as well as the equation where it will be introduced. We let $\left\langle \dots \right\rangle$, $\left\langle \dots \right\rangle^{\text{ss}}$, and $\left\langle \dots \right\rangle^{\text{At}}$ denote the expectation values of a quantity in the lattice problem, single-site problem, and lattice problem within the atomic limit ($U_1 \to \infty$), respectively. The explanations for these symbols, alongside the concepts of lattice and single-site problems, as well as the atomic limit, will be elaborated upon in the subsequent parts of this \siSection{}. We use the convention in which Green's functions of the single-site problem are denote by a normal ({\it i.e.}{}, non-caligraphic) ``$G$''.}
		\label{app:DMFT_gfs}
	\end{longtable}
}

\subsection{The symmetric phase of the THF model}\label{app:sec:se_symmetric:definition}

The main focus of this \siSection{} is the symmetric phase of the THF model. At low temperatures (usually below $\SI{10}{\kelvin}$ in experiments or DMFT simulations~\cite{RAI23a}) and around integer fillings, TBG and TSTG order, {\it i.e.}{} they form phases that break some of the symmetries of the system. For example, the $\ket{\nu=-2, \text{K-IVC}}$ correlated phase from \cref{app:tab:model_states} breaks (among other symmetries) the time-reversal, valley $\mathrm{U} \left(1\right)$, and spin $\mathrm{SU} \left(2\right)$ symmetries of the system. On the other hand, at higher temperatures ({\it i.e.}{} above approximately $\SI{10}{\kelvin}$), the system is likely to be in an unordered or \emph{symmetric} phase. 

While the system \emph{is not} ordered in the symmetric phase, interaction effects are still expected to play an important role. This is because the relevant temperatures, which are high with respect to the order-disorder critical temperatures (approximately $\SI{10}{\kelvin}$)~\cite{RAI23a}, are still much lower than the characteristic temperature of the leading interaction scale (which is of the order of hundreds of Kelvin). Moreover, as opposed to the ordered phases discussed in \cref{app:sec:hartree_fock,app:sec:se_correction_beyond_HF}, for which the excitations' dispersion was primarily determined by the static (Hartree-Fock) self-energy contribution, while the dynamic one mainly affected their lifetimes, for the symmetric phases, the dynamical contribution to the self-energy will play a significant role for \emph{both} their lifetime \emph{and} their dispersion ({\it e.g.}{}, the Hubbard bands appearing in the symmetric phase arise from the \emph{dynamic} part of the self-energy)~\cite{RAI23a}.

In what follows, we will first discuss the constraints that various symmetries impose on the THF Green's function and associated quantities and then define and characterize the symmetric phase of the THF model. 

\subsubsection{Symmetry constraints on the THF interacting Green's function}\label{app:sec:se_symmetric:definition:sym_constr}

We begin by considering a symmetry operation $g$, whose action on the THF fermions is given by 
\begin{equation}
	\label{app:eqn:sym_act_THF}
	g \hat{\gamma}^\dagger_{\vec{k},i, \eta, s} g^{-1} = \sum_{i', \eta', s'} \left[ D (g) \right]_{i' \eta' s';i \eta s} \hat{\gamma}^\dagger_{(-) g \vec{k}, i' \eta' s'},
\end{equation}
where $D (g)$ is the unitary representation matrix of $g$ and $(-) g \vec{k}$ denotes the action of $g$ on the Cartesian momentum vector $\vec{k}$. An additional minus sign, which was denoted by $(-)$, is introduced if $g$ is an antiunitary symmetry. In the notation introduced in \cref{app:sec:hartree_fock:generic_not}, it is easy to show, using the cyclic property of the trace, as well as the fact that $g$ is a symmetry of the THF Hamiltonian (and hence commutes with it), that for any operator $\mathcal{O}$
\begin{equation}
	\label{app:eqn:symmetry_expec_val}
	\left\langle \mathcal{O} \right\rangle = \Tr \left( e^{-\beta K} \mathcal{O} \right) = \Tr \left(g e^{-\beta K} \mathcal{O} g^{-1} \right)^{(*)} =  \Tr \left( e^{-\beta K} g \mathcal{O} g^{-1}\right)^{(*)} = 	\left\langle g \mathcal{O} g^{-1}  \right\rangle^{(*)},
\end{equation}
with ${}^{(*)}$ denoting complex conjugation for the cases when $g$ is antiunitary. Through \cref{app:eqn:symmetry_expec_val}, \cref{app:eqn:sym_act_THF} imposes the following constraints on the non-interacting and interacting Green's functions of the THF model
\begin{equation}
	\label{app:eqn:constr_gf_interacting}
	D^{-1} (g) \mathcal{G}^{0,(*)} \left( \tau ,(-) g \vec{k} \right) D(g) = \mathcal{G}^{0} \left( \tau , \vec{k} \right), \qq{and}
	D^{-1} (g) \mathcal{G}^{(*)} \left( \tau ,(-) g \vec{k} \right) D(g) = \mathcal{G} \left( \tau , \vec{k} \right).
\end{equation}
In \cref{app:eqn:constr_gf_interacting}, the constraint on the interacting Green's function $\mathcal{G} \left(\tau , \vec{k} \right)$ only holds for the cases when the symmetry $g$ is \emph{not} spontaneously broken in the corresponding phase. Fourier-transforming in the Matsubara frequency according to \cref{app:eqn:matsubara_gf_THF_ft}, we find that 
\begin{equation}
	\label{app:eqn:constr_gf_interacting_freq_1}
	D^{-1} (g) \mathcal{G}^{0,(*)} \left( (-) i \omega_n ,(-) g \vec{k} \right) D(g) = \mathcal{G}^{0} \left( i \omega_n , \vec{k} \right), \qq{and}
	D^{-1} (g) \mathcal{G}^{(*)} \left( (-) i\omega_n ,(-) g \vec{k} \right) D(g) = \mathcal{G} \left( i \omega_n , \vec{k} \right).
\end{equation}
Using the spectral representation of the interacting Green's function from \cref{app:eqn:spectral_rep_of_GF}, as well as the Hermiticity of the spectral function, one can easily show that  
\begin{equation}
	\label{app:eqn:greens_func_conj_arg}
	\mathcal{G}^{\dagger} \left(z, \vec{k} \right) = \int_{-\infty}^{\infty} \frac{\dd{\omega}}{z^*-\omega} A^{\dagger} \left( \omega, \vec{k} \right)  = \mathcal{G} \left(z^*, \vec{k} \right),
\end{equation} 
and similarly for the non-interacting Green's function. In turn \cref{app:eqn:greens_func_conj_arg} allows us to recast \cref{app:eqn:constr_gf_interacting_freq_1} to 
\begin{equation}
	\label{app:eqn:constr_gf_interacting_freq_2}
	D^{-1} (g) \mathcal{G}^{0,(T)} \left( i \omega_n ,(-) g \vec{k} \right) D(g) = \mathcal{G}^{0} \left( i \omega_n , \vec{k} \right), \qq{and}
	D^{-1} (g) \mathcal{G}^{(T)} \left( i\omega_n ,(-) g \vec{k} \right) D(g) = \mathcal{G} \left( i \omega_n , \vec{k} \right),
\end{equation}
which, upon analytical continuation becomes
\begin{equation}
	\label{app:eqn:constr_gf_interacting_freq_3}
	D^{-1} (g) \mathcal{G}^{0,(T)} \left( z, (-) g \vec{k} \right) D(g) = \mathcal{G}^{0} \left( z , \vec{k} \right), \qq{and}
	D^{-1} (g) \mathcal{G}^{(T)} \left( z ,(-) g \vec{k} \right) D(g) = \mathcal{G} \left( z , \vec{k} \right),
\end{equation}
where the reader is reminded that the transposition and the minus sign in front of the momentum argument are only to be added for the cases where $g$ is antiunitary. 

In this \siSection{}, we will find it useful to explicitly separate the static ({\it i.e.}{} Hartree-Fock) and dynamic contributions to the self-energy and recast Dyson's equation as
\begin{equation}
	\label{app:eqn:dyson_equation_for_DMFT}
	\mathcal{G}^{-1} \left( z, \vec{k} \right) =  \left( \mathcal{G}^{0} \left( z, \vec{k} \right) \right)^{-1} - h^{\text{MF}} \left( \vec{k} \right) - \Sigma \left(z, \vec{k} \right).
\end{equation} 
In \cref{app:eqn:dyson_equation_for_DMFT}, $h^{\text{MF}} \left( \vec{k} \right)$ denote the static contribution to the self-energy and are first order in the interaction, while $\Sigma \left(z, \vec{k} \right)$ is the dynamical self-energy contribution and incorporates the contributions which are beyond first order in the interaction. Using \cref{app:eqn:constr_gf_interacting_freq_3}, we can directly show that:
\begin{align}
	D^{-1} (g) A^{(*)} \left( \omega, (-) g \vec{k} \right) D(g) &= A \left( \omega , \vec{k} \right), \label{app:eqn:sym_constr_A} \\
	D^{T} (g) \varrho^{(*)} \left( (-) g \vec{k} \right) D^{*}(g) &= \varrho \left( \vec{k} \right), \label{app:eqn:sym_constr_dens_mat} \\
	D^{-1} (g) h^{\text{MF} (*)} \left( (-) g \vec{k} \right) D(g) &= h^{\text{MF}} \left( \vec{k} \right), \label{app:eqn:sym_constr_mf_ham} \\
	D^{-1} (g) \Sigma^{(T)} \left( z ,(-) g \vec{k} \right) D(g) &= \Sigma \left( z , \vec{k} \right). \label{app:eqn:sym_constr_sigma}
\end{align}
In proving \cref{app:eqn:sym_constr_A}, we have additionally employed \cref{app:eqn:spectral_function}. \Cref{app:eqn:sym_constr_dens_mat} then follows by integrating the spectral function according to \cref{app:eqn:def_rho_HF}. Using Dyson's equation from \cref{app:eqn:dyson_equation_for_DMFT}, we can impose the symmetry constraints on the THF model's self-energy and show that \cref{app:eqn:sym_constr_mf_ham,app:eqn:sym_constr_sigma} have to hold independently, since $h^{\text{MF}} \left( \vec{k} \right)$ and $\Sigma \left( z , \vec{k} \right)$ are first order and beyond first order in the interaction. 

In this appendix, as in \cref{app:sec:se_correction_beyond_HF:feynman_rules}, we will incorporate only $f$-electron correlation effects. To this end, we particularize \crefrange{app:eqn:sym_constr_A}{app:eqn:sym_constr_sigma} for the $f$-electron block. By Fourier-transforming the corresponding block of the Green’s function and dynamical self-energy according to \cref{app:eqn:spatial_fourier_trafo_sigma_f,app:eqn:spatial_fourier_trafo_gf_f}, we directly obtain
\begin{align}
	D_f^{-1} (g) \mathcal{G}^{f,(T)} \left( z, g \vec{R}  \right) D_f(g) &= \mathcal{G}^{f} \left( z , \vec{R} \right), \label{app:eqn:sym_constr_gf_re} \\
	D_f^{-1} (g) \Sigma^{f,(T)} \left( z , g \vec{R} \right) D_f(g) &= \Sigma^{f} \left( z , \vec{R} \right), \label{app:eqn:sym_constr_sigma_re}
\end{align}
where the unitary matrix $D_f (g)$ represents the $f$-electron block of the $D (g)$ representation matrix
\begin{equation}
	\left[ D_{f} (g) \right]_{\alpha \eta s;\alpha' \eta' s'} = \left[ D(g) \right]_{(\alpha + 4) \eta s; (\alpha' + 4) \eta' s'}, \qq{for} 1 \leq \alpha, \alpha' \leq 2.
\end{equation}

\subsubsection{The symmetric phase of the THF model within DMFT and Hartree-Fock theory}\label{app:sec:se_symmetric:definition:approximations}

The interaction Hamiltonian of the THF model is given in \cref{app:eqn:THF_interaction_TBG,app:eqn:THF_interaction_TSTG} and is also shown below
\begin{equation}
	H_I = H_{U_1} + H_{U_2} + H_V + H_W + H_J + H_{\tilde{J}} + H_K \left( + H_V^{d} + H_V^{cd} + H_W^{fd} \right),
\end{equation}
where the terms in the parenthesis only appear in the case of TSTG. Due to their zero-kinetic energy and large onsite Hubbard repulsion, the $f$-electrons are expected to be strongly correlated. Conversely, the $c$-electrons (and $d$-electrons in TSTG) are highly dispersive, suggesting that they are only weakly correlated. Therefore, it is sufficient to treat the interactions between the $c$-electrons ($H_V$), as well as between the $c$- and $f$-electrons ($H_W + H_J + H_{\tilde{J}} + H_K$) at the Hartree-Fock level. For TSTG, this also applies to $H_V^{d}$, $H_V^{cd}$, and $H_W^{fd}$. Finally, since the nearest-neighbor Hubbard interaction between $f$-electrons ($H_{U_2}$) is much weaker than the on-site one, we expect the correlation physics to be mostly driven by the $H_{U_1}$ term. As a result, we will also treat the $H_{U_2}$ term via Hartree-Fock theory. 

Therefore, in the symmetric phase, we will assume that the dynamical part of the $f$-electron interaction self-energy arises exclusively from the $H_{U_1}$ term (with all the other terms being treated at the Hartree-Fock level). Compared to the symmetry-broken calculations from \cref{app:sec:se_correction_beyond_HF}, the $H_{U_2}$ term of the Hamiltonian is treated at the mean-field level. In what follows, we will determine the dynamical $f$-electron self-energy using DMFT, as will be outlined \cref{app:sec:se_symmetric:DMFT_overview}. To do so, we will first approximate the THF Hamiltonian by decoupling all interactions \emph{except} $H_{U_1}$ at the Hartree-Fock level and obtain
\begin{equation}
	\label{app:eqn:DMFT_partial_mf_all}
	H^{\text{DMFT+HF}} =  \sum_{\substack{\vec{k}, i, \eta, s \\ i', \eta', s'}} h^{\text{MF} \prime}_{i \eta s; i' \eta' s'} \left( \vec{k} \right) \hat{\gamma}^\dagger_{\vec{k},i,\eta,s} \hat{\gamma}_{\vec{k},i',\eta',s'} + \frac{U_1}{2}  \sum_{\substack{\vec{R}, \alpha, \eta, s \\ \alpha', \eta', s'}} :\mathrel{ \hat{f}^\dagger_{\vec{R},\alpha, \eta, s} \hat{f}_{\vec{R},\alpha, \eta, s} }: :\mathrel{ \hat{f}^\dagger_{\vec{R},\alpha', \eta', s'} \hat{f}_{\vec{R},\alpha', \eta', s'}}:.
\end{equation}
In \cref{app:eqn:DMFT_partial_mf_all}, $h^{\text{MF} \prime} \left( \vec{k} \right)$ is the quadratic mean-field Hamiltonian, which incorporates both the single-particle contribution and all the THF interactions except $H_{U_1}$ decoupled at the Hartree-Fock level and whose expression will be given in \cref{app:eqn:hartree_fock_hamiltonian_for_DMFT}.  

Because the $f$-electrons are dispersionless and have a large onsite repulsion, we expect their onsite inter-orbital, valley, or spin correlations to dominate over the offsite ones. As a result, and in the same spirit as the last approximation of \cref{app:sec:se_correction_beyond_HF:all_so_corrections:particularization}, we will take the $f$-electron dynamical self-energy to be site-diagonal, {\it i.e.}{}
\begin{equation}
	\label{app:eqn:DMFT_diagonal_self_energy_interm}
	\Sigma^{f}_{\alpha \eta s; \alpha' \eta' s'} \left( i \omega_n, \vec{R} \right) = 0 \qq{if} \vec{R}\neq \vec{0},
\end{equation} 
in the notation of \cref{app:eqn:spatial_fourier_trafo_sigma_f}. Moreover, within the symmetric phase, which is the focus of this \siSection{}, the ground state candidate (or thermal ensemble of low-energy states) preserves all the symmetries of the system, namely moir\'e translation symmetry, $C_{6z}$ rotation symmetry, time-reversal symmetry, and $\mathrm{SU} \left(2\right) \times \mathrm{SU} \left(2\right)$ spin-valley rotation symmetry~\cite{BER21a,CAL21}. By imposing all these symmetries on the $f$-electron self-energy according to \cref{app:eqn:sym_constr_sigma_re}, we can show that site-diagonal self-energy ansatz from \cref{app:eqn:DMFT_diagonal_self_energy_interm} becomes
\begin{equation}
	\label{app:eqn:DMFT_diagonal_self_energy}
	\Sigma^{f}_{\alpha \eta s; \alpha' \eta' s'} \left( i \omega_n, \vec{R} \right) = \tilde{\Sigma}^{f} \left( i \omega_n \right) \delta_{\alpha \alpha'} \delta_{\eta \eta'} \delta_{s s'} \delta_{\vec{R},\vec{0}},
\end{equation} 
In \cref{app:eqn:DMFT_diagonal_self_energy}, as well as throughout this \siSection{} and \cref{app:sec:se_symmetric_details}, the diagonal entries of a matrix $M$ are denoted with a tilde as the scalar $\tilde{M}$. 

In addition to the approximation from \cref{app:eqn:DMFT_diagonal_self_energy}, we also expect the largest contribution to the $f$-electron Green's function to be site-diagonal. Such an approximation would be exact if the $f$-electrons were not hybridized with the $c$- and/or $d$-electrons, as it happens in the zero-hybridization limit considered by Ref.~\cite{HU23i}. In computing the Hartree-Fock Hamiltonian $h^{\text{MF} \prime} \left( \vec{k} \right)$ from \cref{app:eqn:DMFT_partial_mf_all}, we will, therefore, ignore any terms that depend on the correlation between $f$-electrons belonging to different sites. The only such contribution is the Fock term arising from $H_{U_2}$, which will therefore be dropped in the symmetric phase calculations. As a result, the Hartree-Fock Hamiltonian matrix from \cref{app:eqn:DMFT_partial_mf_all} is related to the fully-decoupled one defined in \cref{app:eqn:TBG_HF_Hamiltonian,app:eqn:full_TSTG_HF_Hamiltonian} for TBG and TSTG, respectively, by subtracting from the latter the Hartree-Fock contribution of $H_{U_1}$, as well as the Fock contribution of $H_{U_2}$, {\it i.e.}{} 
\begin{align}
	\label{app:eqn:hartree_fock_hamiltonian_for_DMFT}
	h^{\text{MF} \prime}_{i \eta s;i' \eta' s'} \left( \vec{k} \right) =\begin{cases}
		h^{\text{MF}}_{i \eta s;i' \eta' s'} \left( \vec{k} \right) - U_1 \nu_f \delta_{ii'} \delta_{\eta \eta'} \delta_{s s'} \\ \quad + \sum_{\vec{k}'} \left[ U_1 + U_2 \sum_{n=0}^{5} \cos \left( \left( \vec{k} - \vec{k}' \right) \cdot C^n_{6z} \vec{a}_{M_1} \right) \right] \varrho_{i' \eta' s';i \eta s} \left( \vec{k}' \right), & \qq{if} 5\leq i,i' \leq 6 \\
		h^{\text{MF}}_{i \eta s;i' \eta' s'} \left( \vec{k} \right), & \qq{otherwise}
	\end{cases}.
\end{align}
As a result, it is important to note that the $f$-electron block of $h^{\text{MF} \prime} \left( \vec{k} \right)$ is $\vec{k}$-independent and proportional to the identity matrix, {\it i.e.}{}
\begin{equation}
	h^{\text{MF} \prime}_{(\alpha + 4) \eta s;(\alpha' + 4) \eta' s'} \left( \vec{k} \right) = h^{\text{MF} \prime}_{(\alpha + 4) \eta s;(\alpha' + 4) \eta' s'} \left( \vec{0} \right) = h^{\text{MF} \prime}_{5 + \uparrow; 5 + \uparrow} \left( \vec{0} \right) \delta_{\alpha \alpha'} \delta_{\eta \eta'} \delta_{s s'},
\end{equation}
where the last equality follows from \cref{app:eqn:sym_constr_mf_ham}, in the symmetric phase.
 
From \cref{app:eqn:sym_constr_gf_re}, we can also show that in the symmetric phase, the site-diagonal part of the $f$-electron Green's function is also diagonal in the $f$-electron indices. Using the notation from \cref{app:eqn:spatial_fourier_trafo_gf_f}, this is equivalent to 
\begin{equation}
	\label{app:eqn:f_elec_gf_site_flav_diag}
	\eval{ \mathcal{G}^{f}_{\alpha \eta s; \alpha' \eta' s'} \left(i \omega_n, \vec{R} \right)}_{\vec{R} = \vec{0}} = \tilde{\mathcal{G}}^{f} \left(i \omega_n \right) \delta_{\alpha \alpha'} \delta_{\eta \eta'} \delta_{s s'},
\end{equation}
where $\tilde{\mathcal{G}}^{f} \left(i \omega_n \right)$ is the site- and flavor-diagonal part of the $f$-electron Green's function. For convenience, we also introduce the relative filling of the $f$-electrons in the system $n$, which can be related to the site- and flavor-diagonal $f$-electron Green's function by employing \cref{app:eqn:spectral_function,app:eqn:def_rho_HF,app:eqn:flavor_filing,app:eqn:spatial_fourier_trafo_gf_f,app:eqn:f_elec_gf_site_flav_diag}
\begin{align}
	n &\equiv \frac{\nu_f + 4}{N_f} = \frac{1}{N_0 N_f} \sum_{\vec{k}, \alpha, \eta, s} \int_{-\infty}^{\infty} \dd{\omega} n_{\mathrm{F}} \left( \omega \right) A_{(\alpha + 4) \eta s; (\alpha + 4) \eta s} \left( \omega, \vec{k} \right) \nonumber \\
	&= -\frac{1}{\pi N_0 N_f} \sum_{\vec{k}, \alpha, \eta, s} \int_{-\infty}^{\infty} \dd{\omega} n_{\mathrm{F}} \left( \omega \right) \Im \left( \mathcal{G}_{(\alpha + 4) \eta s; (\alpha + 4) \eta s} \left( \omega + i 0^{+}, \vec{k} \right) \right) \nonumber \\
	&= -\frac{1}{\pi} \int_{-\infty}^{\infty} \dd{\omega} n_{\mathrm{F}} \left( \omega \right) \Im \left( \tilde{\mathcal{G}}^{f} \left( \omega + i 0^{+} \right) \right), \label{app:eqn:int_rel_f_filling}
\end{align}
where $N_f = 8$ is the number of $f$-electron fermion flavors.

\subsection{Brief overview of the DMFT and Hartree-Fock theory of the symmetric phase}\label{app:sec:se_symmetric:DMFT_overview}

Having discussed the properties of the symmetric phase of the THF model in \cref{app:sec:se_symmetric:definition:approximations}, we now turn to solving the Hamiltonian from \cref{app:eqn:DMFT_partial_mf_all} using DMFT~\cite{GEO96}. The corresponding partition function reads as 
\begin{equation}
	\label{app:eqn:partition_lattice_problem_DMFT}
	Z^{\text{DMFT+HF}}=\int \mathcal{D} \left[ \hat{\gamma}_{}, \hat{\gamma}^\dagger_{} \right] e^{-S_{\text{DMFT+HF}}},
\end{equation}
with the action being given by
\begin{align}
	S_{\text{DMFT+HF}} = \int_{0}^{\beta} \dd{\tau} &
 \left[ \sum_{\vec{k},i, \eta, s} \hat{\gamma}^\dagger_{\vec{k},i, \eta, s} (\tau) \left( \partial_\tau - \mu \right) \hat{\gamma}_{\vec{k},i, \eta, s} (\tau) +
 H^{\text{DMFT+HF}}
 \right] 
 \nonumber \\
	= \int_{0}^{\beta} \dd{\tau} &\left[ \sum_{\substack{\eta, s, \eta', s' \\ i,i' \neq 5, 6}}  h^{\text{MF} \prime}_{i \eta s; i' \eta' s'} \left( \vec{k} \right) \hat{\gamma}^\dagger_{\vec{k},i,\eta,s} ( \tau ) \hat{\gamma}_{\vec{k},i',\eta',s'} ( \tau ) + \sum_{\vec{k},i, \eta, s} \hat{\gamma}^\dagger_{\vec{k},i, \eta, s} (\tau) \left( \partial_\tau - \mu \right) \hat{\gamma}_{\vec{k},i, \eta, s} (\tau) \right. \nonumber \\
	+ & \sum_{\vec{R},\alpha, \eta, s} h^{\text{MF} \prime}_{(\alpha + 4) \eta s; (\alpha + 4) \eta s} \left( \vec{0} \right) \hat{f}^\dagger_{\vec{R},\alpha, \eta, s} ( \tau ) \hat{f}_{\vec{R}, \alpha, \eta, s} ( \tau ) \nonumber \\
	+ &  \frac{U_1}{2} \sum_{\substack{\vec{R}, \alpha, \eta, s \\ \alpha', \eta', s'}} :\mathrel{ \hat{f}^\dagger_{\vec{R},\alpha, \eta, s} ( \tau ) \hat{f}_{\vec{R},\alpha, \eta, s}  ( \tau ) }: :\mathrel{ \hat{f}^\dagger_{\vec{R},\alpha', \eta', s'}  ( \tau ) \hat{f}_{\vec{R},\alpha', \eta', s'}  ( \tau ) }: \nonumber \\
	+ & \left. \frac{1}{\sqrt{N_0}} \sum_{\vec{k}} \sum_{\substack{i'\neq 5,6 \\ \eta', s'}} \sum_{\vec{R},\alpha, \eta, s}  \left( h^{\text{MF} \prime}_{(\alpha + 4) \eta s; i' \eta' s'} \left( \vec{k} \right) e^{i \vec{k} \cdot \vec{R} } \hat{f}^\dagger_{\vec{R},\alpha,\eta,s} (\tau) \hat{\gamma}_{\vec{k},i',\eta',s'} (\tau) + \text{h.c.} \right) \right]. \label{app:eqn:action_lattice_problem_DMFT}
\end{align}
In deriving \cref{app:eqn:action_lattice_problem_DMFT}, we have used the Fourier transformation of the $f$-electron operators from \cref{app:eqn:f_fermions_real_def}, as well as the $\vec{k}$-independence of the $f$-electron block of $h^{\text{MF} \prime} \left( \vec{k} \right)$. 

\subsubsection{The single-site problem}\label{app:sec:se_symmetric:DMFT_overview:ss_problem}
 
The main idea of DMFT is to map the lattice problem from \cref{app:eqn:partition_lattice_problem_DMFT} to a single-site one. We start by focusing on the $f$-electrons from the site $\vec{R}_0$ and letting 
\begin{equation}
	\label{app:eqn:def_focus_f}
	\hat{f}^\dagger_{\alpha, \eta, s} \equiv \hat{f}^\dagger_{\vec{R}_0,\alpha, \eta, s}.
\end{equation}
As we are interested in translationally invariant solutions, the choice of $\vec{R}_0$ will not affect the final result. We then split the action from \cref{app:eqn:action_lattice_problem_DMFT} into three separate contributions
\begin{equation}
	\label{app:eqn:action_DMFT_separation}
	S_{\text{DMFT+HF}} = S^{\text{bare}}_{\text{ss}} + S_{\text{cavity}} + \Delta S,
\end{equation}
where
\begin{align}
	S^{\text{bare}}_{\text{ss}} = \int_{0}^{\beta} \dd{\tau} &\left[ \sum_{\alpha, \eta, s} \hat{f}^\dagger_{\alpha, \eta, s} ( \tau ) \left[ h^{\text{MF} \prime}_{(\alpha + 4) \eta s; (\alpha + 4) \eta s} \left( \vec{0} \right) + \partial_{\tau} - \mu \right]  \hat{f}_{\alpha, \eta, s} ( \tau ) \right. \nonumber \\
	+ & \left. \frac{U_1}{2} \sum_{\substack{\alpha, \eta, s \\ \alpha', \eta', s'}} :\mathrel{ \hat{f}^\dagger_{\alpha, \eta, s} ( \tau ) \hat{f}_{\alpha, \eta, s}  ( \tau ) }: :\mathrel{ \hat{f}^\dagger_{\alpha', \eta', s'}  ( \tau ) \hat{f}_{\alpha', \eta', s'}  ( \tau ) }: \right], \label{app:eqn:action_focus_site} \\
	S_{\text{cavity}} = \int_{0}^{\beta} \dd{\tau} &\left[ \sum_{\substack{\eta, s, \eta', s' \\ i,i' \neq 5, 6}} \hat{\gamma}^\dagger_{\vec{k},i,\eta,s} ( \tau ) \left[ h^{\text{MF} \prime}_{i \eta s; i' \eta' s'} \left( \vec{k} \right) + \left( \partial_{\tau} - \mu \right) \delta_{i i'} \delta_{\eta \eta'} \delta_{s s'} \right]  \hat{\gamma}_{\vec{k},i',\eta',s'} ( \tau ) \right. \nonumber \\
	+ & \sum_{\substack{\vec{R} \neq \vec{R}_0 \\ \alpha, \eta, s}} \hat{f}^\dagger_{\vec{R},\alpha, \eta, s} ( \tau ) \left[ h^{\text{MF} \prime}_{(\alpha + 4) \eta s; (\alpha + 4) \eta s} \left( \vec{0} \right) + \partial_{\tau} - \mu \right]  \hat{f}_{\vec{R}, \alpha, \eta, s} ( \tau ) \nonumber \\
	+ &  \frac{U_1}{2} \sum_{\substack{\vec{R} \neq \vec{R}_0 \\ \alpha, \eta, s \\ \alpha', \eta', s'}} :\mathrel{ \hat{f}^\dagger_{\vec{R},\alpha, \eta, s} ( \tau ) \hat{f}_{\vec{R},\alpha, \eta, s}  ( \tau ) }: :\mathrel{ \hat{f}^\dagger_{\vec{R},\alpha', \eta', s'}  ( \tau ) \hat{f}_{\vec{R},\alpha', \eta', s'}  ( \tau ) }: \nonumber \\
	+ & \left. \frac{1}{\sqrt{N_0}} \sum_{\vec{k}} \sum_{\substack{i'\neq 5,6 \\ \eta', s'}} \sum_{\substack{ \vec{R} \neq \vec{R}_0 \\ \alpha, \eta, s }} \left( h^{\text{MF} \prime}_{(\alpha + 4) \eta s; i' \eta' s'} \left( \vec{k} \right) e^{i \vec{k} \cdot \vec{R} } \hat{f}^\dagger_{\vec{R},\alpha,\eta,s} (\tau) \hat{\gamma}_{\vec{k},i',\eta',s'} (\tau) + \text{h.c.} \right) \right], \label{app:eqn:action_rest_site} \\
	\Delta S = \int_{0}^{\beta} \dd{\tau} & \frac{1}{\sqrt{N_0}} \sum_{\vec{k}} \sum_{\substack{i'\neq 5,6 \\ \eta', s'}} \sum_{ \alpha, \eta, s } \left(  h^{\text{MF} \prime}_{(\alpha + 4) \eta s; i' \eta' s'} \left( \vec{k} \right) e^{i \vec{k} \cdot \vec{R}_0} \hat{f}^\dagger_{\alpha, \eta, s} (\tau) \hat{\gamma}_{\vec{k},i',\eta',s'} (\tau) + \text{h.c.} \right) . \label{app:eqn:action_coupling_with_rest_site}
\end{align}
The three terms in \cref{app:eqn:action_DMFT_separation} correspond to the action of the $f$-electrons at the lattice site $\vec{R}_0$ ($S^{\text{bare}}_{\text{ss}}$), the action of all  the other degrees of freedom, ($S_{\text{cavity}}$), and a term that couples the $f$-electrons at lattice site $\vec{R}_0$ with all the other fermionic degrees of freedom ($\Delta S$). The system formed by all the fermions of the model \emph{except} the $f$-electrons at lattice site $\vec{R}_0$ is termed the ``cavity''. The partition function of the lattice problem from \cref{app:eqn:partition_lattice_problem_DMFT} can then be rewritten as 
\begin{equation}
	Z^{\text{DMFT+HF}} = \int \mathcal{D} \left[ \hat{f}_{}, \hat{f}^\dagger_{} \right]_{\vec{R}_0} e^{-S^{0}_{\text{ss}}} \int \mathcal{D} \left[ \hat{\gamma}_{}, \hat{\gamma}^\dagger_{} \right]_{\text{cavity}} e^{-S_{\text{cavity}} - \Delta S},
\end{equation}
where
\begin{align}
	\mathcal{D} \left[ \hat{f}_{}, \hat{f}^\dagger_{} \right]_{\vec{R}_0} &\equiv \prod_{\alpha, \eta, s} \mathcal{D} \left[ \hat{f}_{\alpha, \eta, s}, \hat{f}^\dagger_{\alpha, \eta, s} \right], \\
	\mathcal{D} \left[ \hat{\gamma}_{}, \hat{\gamma}^\dagger_{} \right]_{\text{cavity}} &\equiv \prod_{\substack{\vec{k},\eta,s \\ i \neq 5,6}} \mathcal{D} \left[ \hat{\gamma}_{\vec{k},i, \eta, s}, \hat{\gamma}^\dagger_{\vec{k},i, \eta, s} \right] \prod_{\substack{\vec{R} \neq \vec{R}_0  \\ \alpha, \eta, s}} \mathcal{D} \left[ \hat{f}_{\vec{R},\alpha, \eta, s}, \hat{f}^\dagger_{\vec{R},\alpha, \eta, s} \right].
\end{align}

We now formally integrate out all the $f$-, $c$-, or $d$-electrons of the system except $\hat{f}^\dagger_{\alpha,\eta,s}$. To do so, we can interpret $\Delta S$ as a source term for the fermionic terms that are to be integrated out (where $\hat{f}^\dagger_{\alpha, \eta, s}$ constitutes the source field). This allows us to formally obtain
\begin{equation}
	\int \mathcal{D} \left[ \hat{\gamma}_{}, \hat{\gamma}^\dagger_{} \right]_{\text{cavity}} e^{-S_{\text{cavity}} - \Delta S} = Z_{\text{cavity}} e^{-S_{\text{eff}}},
    \quad Z_{\text{cavity}} = \int \mathcal{D} \left[ \hat{\gamma}_{}, \hat{\gamma}^\dagger_{} \right]_{\text{cavity}} e^{-S_{\text{cavity}}}
\end{equation} 
where $Z_{\text{cavity}}$ is the partition function of the $S_{\text{cavity}}$ action, while the effective action $S_{\text{eff}}$ is given by
\begin{align}
	S_{\text{eff}} = \sum_{n=1}^{\infty} \sum_{\substack{\alpha_{i}, \eta_{i}, s_{i} \\ \alpha'_{i}, \eta'_{i}, s'_{i} \\ 1 \leq i \leq n}}  \prod_{i=1}^{n} \left( \int_{0}^{\beta}\dd{\tau_i} \int_{0}^{\beta}\dd{\tau'_i} \right)& G^{\text{cavity}}_{\alpha_{1} \eta_{1} s_{1}, \dots, \alpha_{n} \eta_{n} s_{n};\alpha'_{1} \eta'_{1} s'_{1}, \dots, \alpha'_{n} \eta'_{n} s'_{n}} \left( \tau_1, \dots, \tau_n; \tau'_1, \dots, \tau'_n \right) \nonumber \\
	\times& \hat{f}^\dagger_{\alpha_{1}, \eta_{1}, s_{1}} \left(\tau_1 \right) \dots \hat{f}^\dagger_{\alpha_{n}, \eta_{n}, s_{n}} \left( \tau_n \right)
	\hat{f}_{\alpha'_{1}, \eta'_{1}, s'_{1}} \left( \tau'_1 \right) \dots \hat{f}_{\alpha'_{n}, \eta'_{n}, s'_{n}} \left( \tau'_n \right). \label{app:eqn:DMFT_effective_action_source}
\end{align}
where we have expanded the action in powers of $f$ operators, with $n$ corresponding to $2n$ $f$-electron interactions.  
In \cref{app:eqn:DMFT_effective_action_source}, $G^{\text{cavity}}_{\alpha_{1} \eta_{1} s_{1}, \dots, \alpha_{n} \eta_{n} s_{n};\alpha'_{1} \eta'_{1} s'_{1}, \dots, \alpha'_{n} \eta'_{n} s'_{n}} \left( \tau_1, \dots, \tau_n; \tau'_1, \dots, \tau'_n \right)$ are related to the connected Green's functions of the cavity action $S_{\text{cavity}}$~\cite{GEO96}. For terms with $n>1$ in \cref{app:eqn:DMFT_effective_action_source}, the exact relation is given in Ref.~\cite{GEO96}, but irrelevant for the present discussion, since the former will be dropped in what follows. For $n=1$, we have that~\cite{GEO96}
\begin{align}\label{app:eqn:DMFT_cavity_G_function}
	G^{\text{cavity}}_{\alpha \eta s;\alpha' \eta' s'} \left( \tau; \tau' \right) = -\frac{1}{N_0} \sum_{\substack{\vec{k},\vec{k}' \\ i_{1}, \eta_{1}, s_{1} \\ i_{2}, \eta_{2}, s_{2} \\ i_1, i_2 \neq 5,6}} &\left\langle \mathcal{T}_{\tau} \hat{\gamma}_{\vec{k}',i_{1}, \eta_{1}, s_{1}} \left( \tau \right) \hat{\gamma}^\dagger_{\vec{k},i_{2}, \eta_{2}, s_{2}} \left( \tau' \right) \right\rangle_{\text{cavity}} \nonumber \\ 
	\times & h^{\text{MF} \prime}_{(\alpha + 4) \eta s; i_{1} \eta_{1} s_{1} } \left( \vec{k}' \right) h^{\text{MF} \prime}_{i_{2} \eta_{2} s_{2}; (\alpha' + 4) \eta' s' } \left( \vec{k} \right) e^{i \left(\vec{k} - \vec{k}' \right) \cdot \vec{R}_0},
\end{align}
where the Green's function of the $\hat{\gamma}^\dagger_{\vec{k},i, \eta, s}$ fermions is evaluated within the cavity (whose action is given by $S_{\text{cavity}}$). Because the latter only depends on the time difference $\tau - \tau'$, we must have that 
\begin{equation}
	G^{\text{cavity}}_{\alpha \eta s;\alpha' \eta' s'} \left( \tau; \tau' \right) = G^{\text{cavity}}_{\alpha \eta s;\alpha' \eta' s'} \left( \tau - \tau'; 0 \right) \equiv  G^{\text{cavity}}_{\alpha \eta s;\alpha' \eta' s'} \left( \tau - \tau' \right).
\end{equation} 
Moreover, we note that the cavity system with the $\hat{f}^\dagger_{\alpha, \eta, s}$ fermion source field (for which the action is given by $S_{\text{cavity}}+\Delta S$) will still have $C_{6z}$ rotation symmetry around the lattice site $\vec{R}_0$, time-reversal symmetry, and $\mathrm{SU} \left(2\right) \times \mathrm{SU} \left(2\right)$ spin-valley rotation symmetry (these symmetries are preserved even when the $S^{\text{bare}}_{\text{ss}}$ part of the action is removed). As a result, the effective action from \cref{app:eqn:DMFT_effective_action_source} will also obey these symmetries, which implies that $G^{\text{cavity}}_{\alpha \eta s;\alpha' \eta' s'} \left( \tau - \tau' \right)$ is diagonal
\begin{equation}
	G^{\text{cavity}}_{\alpha \eta s;\alpha' \eta' s'} \left( \tau \right) = \tilde{G}^{\text{cavity}} \left( \tau \right) \delta_{\alpha \alpha'} \delta_{\eta \eta'} \delta_{s s'}.
\end{equation} 
Finally, we note that the $\tilde{G}^{\text{cavity}} \left( \tau \right)$ function is a complicated object that depends on the \emph{fully-interacting} Green's function of the cavity system, which cannot be (in general) computed explicitly. 

To move forward, we use one of the key simplifying assumptions of the DMFT method and ignore all the terms in \cref{app:eqn:DMFT_effective_action_source} that are not quadratic in the $\hat{f}^\dagger_{\alpha, \eta, s}$ fermions (where quadratic term corresponds to the $n=1$ term), which allows us to approximate
\begin{equation}
	\label{app:eqn:approximation_s_eff}
	S_{\text{eff}} \approx \sum_{\alpha, \eta, s} \int_{0}^{\beta}\dd{\tau} \int_{0}^{\beta}\dd{\tau'} \tilde{G}^{\text{cavity}} \left( \tau- \tau' \right) \hat{f}^\dagger_{\alpha, \eta, s} \left(\tau \right) \hat{f}_{\alpha, \eta, s} \left( \tau' \right).
\end{equation}
This approximation is valid for the Hubbard model in infinite dimensions~\cite{GEO96}. As a result, the partition function of the lattice problem from \cref{app:eqn:partition_lattice_problem_DMFT} can be approximated as 
\begin{equation}
	Z^{\text{DMFT+HF}} = Z_{\text{cavity}} \int \mathcal{D} \left[ \hat{f}_{}, \hat{f}^\dagger_{} \right]_{\vec{R}_0} e^{-S^{\text{bare}}_{\text{ss}} - S_{\text{eff}}} \approx Z_{\text{cavity}} \int \mathcal{D} \left[ \hat{f}_{}, \hat{f}^\dagger_{} \right]_{\vec{R}_0} e^{-S_{\text{ss}}},
\end{equation}
where the single-site action is defined as 
\begin{align}
	S_{\text{ss}} &= - \int_{0}^{\beta} \dd{\tau} \int_{0}^{\beta} \dd{\tau'} \sum_{\alpha, \eta, s} \hat{f}^\dagger_{\alpha, \eta, s} \left( \tau \right) \left[ \tilde{G}^{\text{cavity}} \left( \tau - \tau' \right) - \delta \left(\tau - \tau' \right) \left( \partial_{\tau'} - \mu \right) \right] \hat{f}_{\alpha, \eta, s} \left( \tau' \right) \nonumber \\
	&+\int_{0}^{\beta} \dd{\tau} \frac{U_1}{2}  \sum_{\substack{\alpha, \eta, s \\ \alpha', \eta', s'}} :\mathrel{\hat{f}^\dagger_{\alpha, \eta, s} \left( \tau \right) \hat{f}_{\alpha, \eta, s} \left( \tau \right)}: :\mathrel{\hat{f}^\dagger_{\alpha', \eta', s'}\left( \tau \right)  \hat{f}_{\alpha', \eta', s'} \left( \tau \right)}:. \label{app:eqn:single_site_action_interm}
\end{align}
We can interpret the first term of \cref{app:eqn:single_site_action_interm} as the inverse of the ``non-interacting'' Green's function $G_{0} \left( \tau \right)$
\footnote{For the single-site action from \cref{app:eqn:single_site_action_interm}, the non-interacting limit refers to limit when $U_1$ is set to zero while keeping $\tilde{G}^{\text{cavity}} \left( \tau - \tau' \right)$ constant ({\it i.e.}{}, such that $U_1$ is set to zero only at the lattice site $\vec{R}_0$).} of the single-site action $S_{\text{ss}}$ and define 
\begin{equation}
	\label{app:eqn:definition_of_g0_cavity_DMFT}
	G^{-1}_{0} \left( \tau - \tau' \right) \equiv \tilde{G}^{\text{cavity}} \left( \tau - \tau' \right) - \left( - \partial_{\tau'} - \mu \right) \delta \left(\tau - \tau' \right),
\end{equation}
in terms of which the single-site action becomes
\begin{align}
	S_{\text{ss}} &= - \int_{0}^{\beta} \dd{\tau} \int_{0}^{\beta} \dd{\tau'} \sum_{\alpha, \eta, s} \hat{f}^\dagger_{\alpha, \eta, s} \left( \tau \right) G^{-1}_0 \left( \tau - \tau' \right) \hat{f}_{\alpha, \eta, s} \left( \tau' \right) \nonumber \\
	&+\int_{0}^{\beta} \dd{\tau} \frac{U_1}{2}  \sum_{\substack{\alpha, \eta, s \\ \alpha', \eta', s'}} :\mathrel{\hat{f}^\dagger_{\alpha, \eta, s} \left( \tau \right) \hat{f}_{\alpha, \eta, s} \left( \tau \right)}: :\mathrel{\hat{f}^\dagger_{\alpha', \eta', s'}\left( \tau \right)  \hat{f}_{\alpha', \eta', s'} \left( \tau \right)}:.
	\label{app:eqn:single_site_action}
\end{align}
The relation between the $G_{0} \left(\tau \right)$ and $G^{-1}_{0} \left( \tau \right)$ is \emph{defined} to be 
\begin{equation}
	\label{app:eqn:def_inverse_of_g0_dmft}
	\int_{0}^{\beta} \dd{\tau} G_0 \left(\tau_1 - \tau \right) G^{-1}_0 \left(\tau - \tau_2 \right) = \delta \left( \tau_1 - \tau_2 \right).
\end{equation}

To see that $G_{0} \left( \tau \right)$ is the non-interacting Green's function of the single-site model (in the sense of taking $U_1 \to 0$ only at the lattice site $\vec{R}_0$), we can define the partition function of the latter in the presence of the source terms $\hat{\eta}^\dagger_{\alpha, \eta, s}$ and $\hat{\eta}_{\alpha, \eta, s}$
\begin{equation}
	\label{app:eqn:non_int_part_func_DMFT}
	Z^{0}_{\text{ss}}\left[ \hat{\eta}^\dagger_{}, \hat{\eta}_{} \right] = \int \mathcal{D} \left[ \hat{f}_{}, \hat{f}^\dagger_{} \right] e^{-S^{0}_{\text{ss}}},
\end{equation}
where 
\begin{align}
	S^{0}_{\text{ss}} &= - \int_{0}^{\beta} \dd{\tau} \int_{0}^{\beta} \dd{\tau'} \sum_{\alpha, \eta, s} \hat{f}^\dagger_{\alpha, \eta, s} \left( \tau \right) G^{-1}_0 \left( \tau - \tau' \right) \hat{f}_{\alpha, \eta, s} \left( \tau' \right) \nonumber \\
	&+\int_{0}^{\beta} \dd{\tau} \sum_{\alpha, \eta, s} \left( \hat{f}^\dagger_{\alpha, \eta, s} (\tau) \hat{\eta}_{\alpha, \eta, s} (\tau) +  \hat{\eta}^\dagger_{\alpha, \eta, s} (\tau) \hat{f}_{\alpha, \eta, s} (\tau) \right) 
	\nonumber \\
&= - \int_{0}^{\beta} \dd{\tau} \int_{0}^{\beta} \dd{\tau'} \sum_{\alpha, \eta, s} \hat{f}^\dagger_{\alpha, \eta, s} \left( \tau \right) G^{-1}_0 \left( \tau - \tau' \right) \hat{f}_{\alpha, \eta, s} \left( \tau' \right) \nonumber \\
	&+\int_{0}^{\beta} \dd{\tau} \int_{0}^{\beta} \dd{\tau'} \int_{0}^{\beta} \dd{\tau''} G^{-1}_{0} \left( \tau - \tau' \right) G_{0} \left( \tau'- \tau'' \right) \sum_{\alpha, \eta, s} \hat{f}^\dagger_{\alpha, \eta, s} (\tau) \hat{\eta}_{\alpha, \eta, s} \left(\tau'' \right) \nonumber \\
	&+\int_{0}^{\beta} \dd{\tau} \int_{0}^{\beta} \dd{\tau'} \int_{0}^{\beta} \dd{\tau''} G^{-1}_{0} \left( \tau - \tau' \right) G_{0} \left( \tau'' - \tau \right) \sum_{\alpha, \eta, s} \hat{\eta}^\dagger_{\alpha, \eta, s} \left(\tau'' \right) \hat{f}_{\alpha, \eta, s} \left( \tau' \right) .
\end{align}
The non-interacting Green's function of the single-site model is then given by
\begin{equation}
	\label{app:eqn:def_non_int_DMFT_gf}
	-\left\langle \mathcal{T}_{\tau} \hat{f}_{\alpha, \eta, s} \left(\tau_1 \right) \hat{f}^\dagger_{\alpha, \eta, s} \left( \tau_2 \right) \right\rangle_{0} = \eval{ \left( \frac{1}{Z^{0}_{\text{ss}} \left[ \eta^{\dagger}_{}, \eta_{} \right]} \fdv{\eta^{\dagger}_{i}(\tau_2)} \fdv{\eta_{j}(\tau_1)} Z^{0}_{\text{ss}} \left[ \eta^{\dagger}_{}, \eta_{} \right] \right)}_{\eta^{\dagger}_{},\eta_{}=0}.
\end{equation}
Performing a change of variables in \cref{app:eqn:non_int_part_func_DMFT}
\begin{equation}
	\hat{g}^\dagger_{\alpha, \eta, s}\left( \tau \right) \equiv \hat{f}^\dagger_{\alpha, \eta, s} \left( \tau \right) - \int_{0}^{\beta} \dd{\tau''} G_0 \left( \tau'' - \tau \right) \hat{\eta}^\dagger_{\alpha, \eta, s} \left( \tau'' \right),
\end{equation}
we find that the partition function becomes
\begin{equation}
	\label{app:eqn:non_int_part_func_DMFT_2}
	Z^{0}_{\text{ss}}\left[ \hat{\eta}^\dagger_{}, \hat{\eta}_{} \right] = \int \mathcal{D} \left[ \hat{g}_{}, \hat{g}^\dagger_{} \right] e^{-S^{0}_{\text{ss}}},
\end{equation}
with the corresponding action being given by
\begin{align}
	S^{0}_{\text{ss}} &= - \int_{0}^{\beta} \dd{\tau} \int_{0}^{\beta} \dd{\tau'} \sum_{\alpha, \eta, s} \hat{g}^\dagger_{\alpha, \eta, s} \left( \tau \right) G^{-1}_0 \left( \tau - \tau' \right) \hat{g}_{\alpha, \eta, s} \left( \tau' \right) \nonumber \\
	&-\int_{0}^{\beta} \dd{\tau_1} \int_{0}^{\beta} \dd{\tau_2} \int_{0}^{\beta} \dd{\tau_3}  \int_{0}^{\beta} \dd{\tau_4} \sum_{\alpha, \eta, s} G_{0} \left( \tau_3-\tau_1 \right) G^{-1}_{0} \left( \tau_1 - \tau_2 \right) 
	G_{0} \left( \tau_2-\tau_4 \right) \hat{\eta}^\dagger_{\alpha \eta s} \left( \tau_3 \right) \hat{\eta}_{\alpha \eta s} \left( \tau_4 \right) \nonumber \\ 
	&= - \int_{0}^{\beta} \dd{\tau} \int_{0}^{\beta} \dd{\tau'} \sum_{\alpha, \eta, s} \hat{g}^\dagger_{\alpha, \eta, s} \left( \tau \right) G^{-1}_0 \left( \tau - \tau' \right) \hat{g}_{\alpha, \eta, s} \left( \tau' \right) \nonumber \\
	&- \int_{0}^{\beta} \dd{\tau} \int_{0}^{\beta} \dd{\tau'} \sum_{\alpha, \eta, s} \hat{\eta}^\dagger_{\alpha, \eta, s} \left( \tau \right) G_0 \left( \tau - \tau' \right) \hat{\eta}_{\alpha, \eta, s} \left( \tau' \right). \label{app:eqn:single_site_action_non_int}
\end{align}
Using \cref{app:eqn:def_non_int_DMFT_gf,app:eqn:non_int_part_func_DMFT_2,app:eqn:single_site_action_non_int}, we can immediately conclude that 
\begin{equation}
	-\left\langle \mathcal{T}_{\tau} \hat{f}_{\alpha, \eta, s} \left(\tau_1 \right) \hat{f}^\dagger_{\alpha, \eta, s} \left( \tau_2 \right) \right\rangle^{\text{ss}}_{0} = G_{0} \left( \tau_1 - \tau_2 \right),
\end{equation}
which shows that, indeed, $G_{0} \left( \tau_1 - \tau_2 \right)$ is the Green's function of the single-site action when $U_1 \to 0$ at the lattice site $\vec{R}_0$.

\Cref{app:eqn:single_site_action} forms the central focus of the DMFT method. The reader is reminded that \cref{app:eqn:single_site_action} was obtained by integrating out all the fermions except the $f$-electrons from the lattice site $\vec{R}_0$. Any terms other than quadratic in the $\hat{f}^\dagger_{\alpha, \eta, s}$ fermions arising from the integration procedure were then ignored. We also note that the expression of $G^{-1}_{0} \left( \tau \right)$ from \cref{app:eqn:definition_of_g0_cavity_DMFT} is only of formal importance: in practice $G^{-1}_{0} \left( \tau \right)$ \emph{cannot} be computed this way, as it requires knowledge on the exact solution of the cavity system. Instead, $G^{-1}_{0} \left( \tau \right)$ needs to be obtained self-consistently, as we will show in \cref{app:sec:se_symmetric:DMFT_overview:dmft_sol}.

\subsubsection{The DMFT solution}\label{app:sec:se_symmetric:DMFT_overview:dmft_sol}

Within DMFT, for a given $G_0 \left( \tau \right)$ (which will be determined self-consistently, as shown later), one solves the action $S_{\text{ss}}$ via some exact or approximate method and obtains the interacting Green's function for the $f$-electrons of the single-site model 
\begin{equation}
	\label{app:eqn:definition_ss_interacting_gf}
	G \left(\tau \right) = - \left\langle \mathcal{T}_{\tau} \hat{f}_{\alpha, \eta, s} \left( \tau \right) \hat{f}^\dagger_{\alpha, \eta, s} \left( 0 \right) \right\rangle^{\text{ss}},
\end{equation}
where both the imaginary time evolution and the thermodynamic averaging are done within the single-site action from \cref{app:eqn:single_site_action}. Once $G \left( \tau \right)$ is computed, the self-energy of the single-site $f$-electrons can be obtained (by definition) from $G \left(\tau \right)$ and $G_0 \left( \tau \right)$ using Dyson's equation,
\begin{equation}
	\label{app:eqn:all_self_energy_of_ss_model}
	U_1 \left( n_{\text{ss}} - \frac{1}{2} \right) \left( N_f -1 \right) + \Sigma_{\text{ss}} \left( i \omega_n \right) = G_0^{-1} \left( i \omega_n \right) - G^{-1} \left( i \omega_n \right),
\end{equation} 
where the Fourier transformations of $G (\tau)$ and $G_0 (\tau)$ Green's function in the Matsubara frequency are defined analogously to \cref{app:eqn:matsubara_gf_THF_ft}. The first and second terms on the left hand side of \cref{app:eqn:all_self_energy_of_ss_model} are, respectively, the static (Hartree-Fock) and dynamic contributions to the self-energy of the single-site model, with 
\begin{equation}
	\label{app:eqn:ss_relative_f_interacting}
	n_{\text{ss}} \equiv \left\langle \hat{f}^\dagger_{\alpha, \eta, s} \hat{f}_{\alpha, \eta, s} \right\rangle^{\text{ss}}. 
\end{equation}

In DMFT, there are two further \emph{assumptions} (both of which are exactly satisfied for the infinite-dimensional Hubbard model)~\cite{GEO96}, which simultaneously (1) provide a solution to the original lattice problem from \cref{app:eqn:DMFT_partial_mf_all}, and (2) enable one to compute $G_{0} \left( \tau \right)$:
\begin{itemize}
	\item First, we take the $f$-electron self-energy of the lattice problem to be $\vec{k}$-independent \emph{and} equal to the $f$-electron self-energy of the single-site model. Specifically, letting 
	\begin{equation}
		\Sigma_{i \eta s; i' \eta' s'} \left( i \omega_n \right) = \begin{cases}
			\tilde{\Sigma}^{f} \left( i \omega_n \right) \delta_{ii'} \delta_{\eta \eta'} \delta_{s s'} & \qq{if} i=5,6 \\
			0 & \qq{otherwise}
		\end{cases},
	\end{equation}
	with 
	\begin{equation}
		\label{app:eqn:dmft_ass_1}
		\tilde{\Sigma}^{f} \left( i \omega_n \right) = \Sigma_{\text{ss}} \left( i \omega_n \right),
	\end{equation} 
	the $\vec{k}$-dependent Green's function of the lattice problem can be calculated via Dyson's equation
	\begin{equation}
		\label{app:eqn:DMFT_GF_symmetric}
		\mathcal{G} \left( i \omega_n , \vec{k} \right) = \left[\left( i \omega_n + \mu \right) \mathbb{1} - h^{\text{MF} \prime} \left( \vec{k} \right) - U_1 \left( n_{\text{ss}} - \frac{1}{2} \right) \left( N_f -1 \right) \mathbb{1}_{f} - \Sigma \left( i \omega_n \right) \right]^{-1},
	\end{equation}
	where $\mathbb{1}_f$ is the identity matrix in the $f$-electron block
	\begin{equation}
		\left[\mathbb{1}_f \right]_{i \eta s;i' \eta' s'}=\begin{cases}
			\delta_{ii'} \delta_{\eta \eta'} \delta_{s s'} & \qq{if} i=5,6 \\
			0 & \qq{otherwise}
		\end{cases}.
	\end{equation} 
	Note that there is no dynamical contribution to the $c$- and/or $d$-electrons' self-energy, as the corresponding interaction terms are treated at the Hartree-Fock level in \cref{app:eqn:DMFT_partial_mf_all}.
	
	\item Second, we take the \emph{full} Green's function of the single-site model to be equal to the site-diagonal interacting Green's function of the lattice model~\cite{GEO96}, which was defined in \cref{app:eqn:f_elec_gf_site_flav_diag},
	\begin{equation}
		\label{app:eqn:DMFT_self_consistent_eq}
		G \left( i \omega_n \right) = \tilde{\mathcal{G}}^{f} \left( i \omega_n \right). 
	\end{equation} 
	As a result of \cref{app:eqn:DMFT_self_consistent_eq}, we must have that $n=n_{\text{ss}}$. This implies that for a given lattice Green's function $\mathcal{G} \left(i\omega_n, \vec{k} \right)$ and $f$-electron self-energy $\tilde{\Sigma}^{f} \left( i \omega_n \right)$ (assumed to be $\vec{k}$-independent), the non-interacting single-site Green's function $G_{0} \left( i \omega_n \right)$ can be recovered from \cref{app:eqn:all_self_energy_of_ss_model,app:eqn:dmft_ass_1,app:eqn:DMFT_self_consistent_eq},
	\begin{equation}
		\label{app:eqn:DMFT_G0_expression}
		G^{-1}_{0} \left(i \omega_n \right) = \left( \tilde{\mathcal{G}}^{f} \left( i \omega_n \right) \right)^{-1} + U_1 \left( n - \frac{1}{2} \right) \left( N_f - 1 \right) + \tilde{\Sigma}^{f} \left( i \omega_n \right).
	\end{equation} 
	It is \cref{app:eqn:DMFT_G0_expression}, as opposed to \cref{app:eqn:definition_of_g0_cavity_DMFT}, which is used in practice to obtain $G_{0} \left( i \omega_n \right)$.
\end{itemize}

Taken together, \cref{app:eqn:DMFT_GF_symmetric,app:eqn:DMFT_G0_expression} form a self-consistent problem, as illustrated in \cref{fig:self_consistent_DMFT}. The two quantities $G_{0} \left( i \omega_n \right)$ and $\Sigma_{0} \left( i \omega_n \right)$ are interdependent and need to be determined at the same time so as to satisfy both \cref{app:eqn:all_self_energy_of_ss_model} -- which implies that $\tilde{\Sigma}^{f} \left( i \omega_n \right) = \Sigma_{\text{ss}} \left( i \omega_n \right)$ is the \emph{exact} self-energy of the single-site problem -- and \cref{app:eqn:DMFT_G0_expression} -- which relates back the dynamical $f$-electron self-energy to the non-interacting Green's function of the single-site problem.

In our problem, as we also have a Hartree-Fock contribution stemming from the interactions other than $H_{U_1}$, we will vary the density matrix $\varrho \left(\vec{k} \right)$, the Green's function $G_0 \left( \tau \right)$, and the dynamical $f$-electron self energy  until all the self-consistent equations, including \cref{app:eqn:def_rho_HF} which relates the density matrix back to the lattice Green's function, are satisfied. The detailed description and implementation of the self-consistent procedure will be given in \cref{app:sec:se_symmetric:IPT:sc_and_numerical}. 

\subsubsection{Analytical properties of the $G_{0} (z)$, $G (z)$, and $\Sigma_{\text{ss}} (z)$}\label{app:sec:se_symmetric:DMFT_overview:anal_props}

\textit{Bona fide} Green's functions admit a spectral representation. For example, the full ({\it i.e.}{} unapproximated) Green's function of the THF model admits the spectral representation from \cref{app:eqn:spectral_rep_of_GF}. The proof and implication of a spectral representation are discussed at lengths in Ref.~\cite{LUT61,PAV19} and will not be repeated here. 

On the other hand, the ``Green's functions'' of the single-site model and the corresponding dynamical self-energies are more complicated objects and it is \emph{not} immediately obvious whether they admit a spectral representation. Setting aside issues related to the uniqueness of the DMFT solution~\cite{KOZ15,GUN17}, Ref.~\cite{LIN19} has shown that if the DMFT self-consistent cycle in \cref{fig:self_consistent_DMFT} is solved iteratively starting from a guess of the lattice Green's function $\mathcal{G} \left(i \omega_n, \vec{k} \right)$ which \emph{admits} a spectral representation, then at each iteration, $G_{0} (z)$, $G (z)$, and $\Sigma_{\text{ss}} (z)$ will all admit a spectral representation
\begin{alignat}{9}
	G_{0} (z) &&= &\int_{-\infty}^{\infty}  \frac{\rho_{0} (\omega) }{z-\omega} \dd{\omega}, &\qq{with}&& \rho_{0} (\omega) &&\equiv& -\frac{1}{\pi} \Im G_{0} \left( \omega + i 0^{+} \right),& \quad \rho_{0} (\omega) &&\geq& 0,& \quad \int_{-\infty}^{\infty} \rho_{0} (\omega) \dd{\omega}&& =& 1 \label{app:eqn:causal_g0}\\
	G (z) &&= &\int_{-\infty}^{\infty}  \frac{\rho (\omega) }{z-\omega} \dd{\omega}, &\qq{with}&& \rho (\omega) &&\equiv& -\frac{1}{\pi} \Im G \left( \omega + i 0^{+} \right),& \quad \rho (\omega) &&\geq& 0,& \quad \int_{-\infty}^{\infty} \rho (\omega) \dd{\omega}&& =& 1,\label{app:eqn:causal_g}
\end{alignat}
and similarly for the self-energy
\begin{equation}
	\label{app:eqn:causal_se}
	\Sigma_{\text{ss}} (z) = - \frac{1}{\pi} \int_{-\infty}^{\infty}  \frac{\dd{\omega}}{z-\omega} \Im \Sigma_{\text{ss}} \left( \omega + i 0^{+} \right),  \qq{with} \Im \Sigma_{\text{ss}} \left( \omega + i 0^{+} \right) \leq 0. 
\end{equation}
Rigorous mathematical results have also been proven in Ref.~\cite{CAN24,CAN25}. In what follows, we will take \cref{app:eqn:causal_g0,app:eqn:causal_g,app:eqn:causal_se} to hold for our DMFT self-consistent solution. As a corollary of \cref{app:eqn:causal_g0,app:eqn:causal_g}, we have that 
\begin{equation}
	\label{app:eqn:DMFT_gf_to_z_infty}
	G_{0} (z) = \frac{1}{z}, \quad G (z) = \frac{1}{z}, \qq{as} z \to \infty,
\end{equation}
while 
\begin{equation}
	\label{app:eqn:asymptote_of_sigma_general}
	\Sigma_{\text{ss}} (z) \propto \frac{1}{z}, \qq{as} z \to \infty,
\end{equation}
where the proportionality constant will be derived in \cref{app:sec:se_symmetric_details:se_exact}. Additionally, it is easy to see that from \cref{app:eqn:causal_g0,app:eqn:causal_g,app:eqn:causal_se}, that $G_{0} (z)$, $G (z)$, and $\Sigma_{\text{ss}} (z)$ are analytical above and below the real axis, with 
\begin{equation}
	\label{app:eqn:DMFT_gf_cc_of_argument}
	G_{0} \left( z^{*} \right) = G^{*}_{0} \left( z \right), \quad
	G \left( z^{*} \right) = G^{*} \left( z \right), \quad
	\Sigma_{\text{ss}} \left( z^{*} \right) = \Sigma_{\text{ss}}^{*} \left( z \right).
\end{equation}

The final question concerns the existence of a spectral representation for the lattice Green's function $\mathcal{G} \left(i \omega_n , \vec{k} \right)$ (given that it is \emph{not} the exact interacting Green's function of the THF model, but rather \emph{some} Green's function constructed with the DMFT $f$-electron self-energy). Ref.~\cite{LIN19} also proves that if $\Sigma_{\text{ss}} \left( z \right)$ admits a spectral representation akin to the one in \cref{app:eqn:causal_se}, then 
\begin{equation}
	\label{app:eqn:DMFT_GF_symmetric_in_z}
	\mathcal{G} \left( z , \vec{k} \right) = \left[\left( z + \mu \right) \mathbb{1} - h^{\text{MF} \prime} \left( \vec{k} \right) - U_1 \left( n - \frac{1}{2} \right) \left( N_f -1 \right) \mathbb{1}_{f} - \Sigma \left( z \right) \right]^{-1},
\end{equation}
will also admit the usual spectral representation of a Green's function from \cref{app:eqn:spectral_rep_of_GF,app:eqn:spectral_function}. As a result, we will assume that all the Green's function computed within the DMFT and Hartree-Fock theory will obey the usual analytic properties and have \emph{bona fide} spectral representations. 

\subsection{Solving the DMFT and Hartree-Fock problem with IPT}\label{app:sec:se_symmetric:introduction:second_order_perturbation}

As we described in the \cref{app:sec:se_symmetric:DMFT_overview}, within the DMFT approximation, the lattice model is mapped to a single-site problem described by the action $S_{\text{ss}}$ introduced in \cref{app:eqn:single_site_action}. The dynamical self-energy of the lattice problem is then approximated by the dynamical self-energy of the single-site problem, which still needs to be solved exactly. Throughout this work, we will employ the approximate IPT method~\cite{ANI97,ARS12,DAS16,FUJ03,KAJ96,LAA03,YEY00,LIC98,MAR86,MEY99,MIZ21,POT97,SAS01,SAV01,YEY93,YEY99,VAN22} to obtain the dynamical self-energy of the single-site problem. Compared to exact quantum Monte Carlo (QMC) methods~\cite{DAT23,RAI23a}, this method offers two significant advantages: time-efficiency and, as we will see, the avoidance of numerical analytical continuation from Matsubara to real frequency (which is a numerically ill-posed problem). In this section, we will provide an overview of IPT method, stating all the results without proof, with all the detailed derivations being relegated to \cref{app:sec:se_symmetric_details}.

\subsubsection{Overview of the IPT method}\label{app:sec:se_symmetric:introduction:second_order_perturbation:overview} 

To start with, we introduce a modified non-interacting Green's function for the single-site problem at a different chemical potential $\tilde{\mu}$ 
\begin{equation}
	\label{app:eqn:def_of_g_tilde_mu}
	G^{\tilde{\mu}}_0 \left( i \omega_n \right) = \frac{1}{G_0^{-1} \left( i \omega_n \right) - \mu + \tilde{\mu}},
\end{equation}
together with its spectral representation\footnote{Since $G_0 \left( i \omega_n \right)$ has a spectral representation, as mentioned in \cref{app:eqn:causal_g0}, $G^{\tilde{\mu}}_0 \left( i \omega_n \right)$ will also have one.}
\begin{equation}
	\label{app:eqn:spectral_gf_relation_tilde}
	\rho^{\tilde{\mu}}_{0} \left( \omega \right) \equiv -\frac{1}{\pi} \Im G^{\tilde{\mu}}_{0} \left( \omega + i 0^{+} \right), \qq{such that}
	G^{\tilde{\mu}}_{0} \left( z \right) = \int_{-\infty}^{\infty} \frac{\dd{\omega}}{z-\omega} \rho^{\tilde{\mu}}_0 \left( \omega \right).
\end{equation}
As we explain in \cref{app:sec:se_symmetric_details:IPT:fix_mu_tilde}, the so-called \emph{fictitious} chemical potential $\tilde{\mu}$ can be thought of as variational parameter which will be fixed in such a way so as to render the IPT approximation for the dynamical self-energy as close as possible to the exact result.  

The first idea of the IPT method is to solve the single-site problem to second order in perturbation theory using the $G^{\tilde{\mu}}_0 \left( i \omega_n \right)$ propagator. The corresponding diagonal part of the dynamical self-energy is given by
\begin{align}
	\tilde{\Sigma}^{f,(2)} \left(\omega + i 0^{+} \right) =&  U_1^2 \left(N_f - 1 \right) \int_{-\infty}^{\infty} \dd{\omega_1} \int_{-\infty}^{\infty} \dd{\omega_2} \int_{-\infty}^{\infty} \dd{\omega_3} \rho^{\tilde{\mu}}_0 \left(\omega_1 \right) \rho^{\tilde{\mu}}_0 \left(\omega_2 \right) \rho^{\tilde{\mu}}_0 \left(\omega_3 \right) \nonumber \\
	&\times \frac{n_{\mathrm{F}} \left( \omega_{1} \right) \left( 1 - n_{\mathrm{F}} \left( \omega_{2} \right) \right) n_{\mathrm{F}} \left( \omega_{3} \right) + \left( 1 - n_{\mathrm{F}} \left( \omega_{1} \right) \right) n_{\mathrm{F}} \left( \omega_{2} \right) \left( 1 - n_{\mathrm{F}} \left( \omega_{3} \right) \right)}{\omega - \omega_1 + \omega_2 - \omega_3 + i 0^{+}}.
	\label{app:eqn:second_order_f_symmetric}
\end{align}
However, for reasons that will be explained in detail in \cref{app:sec:se_symmetric_details}, within IPT, $\tilde{\Sigma}^{f,(2)} \left(\omega + i 0^{+} \right)$ is not used directly as the dynamical self-energy of the single-site problem. Instead, the latter is approximated by the following \emph{interpolated} ansatz
\begin{equation}
	\label{app:eqn:interpolated_sigma}
	\tilde{\Sigma}^{f,\text{Int}} \left( \omega + i 0^{+}  \right) = \frac{a \tilde{\Sigma}^{f,(2)}\left( \omega + i 0^{+}  \right)}{1-b \left( \omega + i 0^{+} \right) \tilde{\Sigma}^{f,(2)}\left( \omega + i 0^{+}  \right)}.
\end{equation}
The constant $a$ and the function $b \left( \omega + i 0^{+} \right)$ are respectively chosen such that the interpolated self-energy $\tilde{\Sigma}^{f,\text{Int}} \left( \omega + i 0^{+}  \right)$ obeys the correct (exact) asymptotic form in the infinite frequency ($\omega \to \infty$) limit and reduces to the exact analytical \emph{atomic limit}~\cite{HUB64} expression $\tilde{\Sigma}^{f,\text{At}}  \left( \omega + i 0^{+}  \right)$ in the $U_1 \to \infty$ limit (the atomic limit will be briefly discussed above \cref{app:eqn:def_n_at} and then solved in \cref{app:sec:se_symmetric_details:atomic_se}). Additionally, it can also be shown that $\tilde{\Sigma}^{f,\text{Int}} \left( \omega + i 0^{+}  \right)$ reduces to the second-order perturbation result in the weakly interacting limit ($U_1 \to 0$). Because $\tilde{\Sigma}^{f,\text{Int}} \left( \omega + i 0^{+}  \right)$ reproduces the exact results in both the large ($U_1 \to \infty$) and small ($U_1 \to 0$) interaction limits, while preserving the exact $\omega \to \infty$ asymptotic form, the interpolated self-energy from \cref{app:eqn:interpolated_sigma} is expected to correctly capture the correlation physics of the $f$-electrons even in the intermediate coupling regime.

For a given chemical potential $\mu$ and fictitious chemical potential $\tilde{\mu}$, the constant $a$ of \cref{app:eqn:interpolated_sigma} is given by
\begin{equation}
	\label{app:eqn:a_constant_ipt}
	a = \frac{ n + \left(N_f - 2\right) \left\langle nn \right\rangle^{\text{ss}}  - \left(N_f -1 \right) n^2}{n^{\tilde{\mu}}_0 \left(1 - n^{\tilde{\mu}}_0 \right)},
\end{equation}
where $n$ is the relative $f$-electron filling defined in \cref{app:eqn:int_rel_f_filling}, $n^{\tilde{\mu}}_0$ is the filling of the non-interacting single-site problem at the fictitious chemical potential $\tilde{\mu}$
\begin{equation}
	\label{app:eqn:int_rel_f0_filling}
	n^{\tilde{\mu}}_0 \equiv \int_{-\infty}^{\infty} \dd{\omega} n_{\mathrm{F}} \left( \omega \right) \rho^{\tilde{\mu}}_0 \left( \omega \right),
\end{equation}
and $\left\langle nn \right\rangle^{\text{ss}}$ is the $f$-electron double-occupation of the single-site problem and which can be related back to the $f$-electron dynamical self-energy and interacting site-diagonal Green's function according to  
{\small
	\begin{equation}
	\label{app:eqn:nn_correlator_from_self_energy}
	\left\langle nn \right\rangle^{\text{ss}} \equiv \left\langle \hat{f}^\dagger_{\alpha, \eta, s} \hat{f}_{\alpha, \eta, s} \hat{f}^\dagger_{\alpha', \eta', s'} \hat{f}_{\alpha', \eta', s'}  \right\rangle^{\text{ss}} = n^2 - \frac{1}{\pi U_1 \left(N_f - 1 \right)} \int_{-\infty}^{\infty} \dd{\omega} \Im{\tilde{\Sigma}^{f} \left( \omega + i 0^{+} \right) \tilde{\mathcal{G}}^{f} \left( \omega + i 0^{+} \right) } n_{\mathrm{F}} \left( \omega \right),
\end{equation}}where the definition holds for any $ \left(\alpha, \eta, s\right) \neq \left(\alpha', \eta', s' \right)$. 

The function $b \left( \omega + i 0^{+} \right)$ from \cref{app:eqn:interpolated_sigma} is determined from the atomic limit~\cite{HUB64} of the THF model, which is defined and solved analytically in \cref{app:sec:se_symmetric_details:atomic_se}. In the atomic limit, both the many-body and single-particle coupling terms between the $f$-electrons at a given site and all the other fermions of the system are set to zero. At a given chemical potential $\mu$ of the lattice problem, we denote the $f$-electron occupation, double occupation, and dynamical self-energy in the atomic limit by 
\begin{align}
	n_{\text{At}} & \equiv \left\langle \hat{f}^\dagger_{\alpha, \eta, s} \hat{f}_{\alpha, \eta, s} \right\rangle^{\text{At}}, \qq{for any}  \left(\alpha, \eta, s\right), \label{app:eqn:def_n_at}\\
	\left\langle nn \right\rangle^{\text{At}} & \equiv \left\langle \hat{f}^\dagger_{\alpha, \eta, s} \hat{f}_{\alpha, \eta, s} \hat{f}^\dagger_{\alpha', \eta', s'} \hat{f}_{\alpha', \eta', s'} \right\rangle^{\text{At}} \neq  n_{\text{At}}^2, \qq{for any} \left(\alpha, \eta, s\right) \neq \left(\alpha', \eta', s' \right), \label{app:eqn:def_nn_at} \\
	\tilde{\Sigma}^{f,\text{At}} \left( \omega + i 0^{+}  \right) & \equiv \omega + i 0^{+} + \mu - U_1 \left(n_{\text{At}} -\frac{1}{2} \right) \left( N_f - 1 \right) - \left(\tilde{\mathcal{G}}^{\text{At}} \left( \omega + i 0^{+} \right) \right)^{-1}, \label{app:eqn:def_sigma_at}
\end{align}
where $\left\langle \dots \right\rangle^{\text{At}}$ denotes the expectation value in the atomic limit and $\tilde{\mathcal{G}}^{\text{At}} \left( \omega + i 0^{+} \right)$ denotes the diagonal part of the full $f$-electron Green's function in the atomic limit. The analytical expressions for \cref{app:eqn:def_n_at,app:eqn:def_nn_at,app:eqn:def_sigma_at} as a function of the chemical potential $\mu$ will be given in \cref{app:sec:se_symmetric_details:atomic_se}. For a chemical potential $\mu$ and fictitious chemical potential $\tilde{\mu}$, the function $b \left( \omega + i 0^{+} \right)$ \cref{app:eqn:interpolated_sigma} is given by
\begin{equation}
	\label{app:eqn:b_function_ipt}
	b \left(\omega + i 0^{+}\right) = \frac{1}{\tilde{\Sigma}^{f,(2),\text{At}} \left(\omega + i 0^{+} \right)} - \frac{a^{\text{At}}}{\tilde{\Sigma}^{f,\text{At}} \left(\omega + i 0^{+} \right)},
\end{equation}
with the constant $a^{\text{At}}$ being obtained by evaluating \cref{app:eqn:a_constant_ipt} in the atomic limit 
\begin{equation}
	\label{app:eqn:aAt_constant_ipt}
	a^{\text{At}} = \frac{ n_{\text{At}} + \left(N_f - 2\right) \left\langle nn \right\rangle^{\text{At}}  - \left(N_f -1 \right) n_{\text{At}}^2}{n^{\tilde{\mu}}_0 \left(1 - n^{\tilde{\mu}}_0 \right)},
\end{equation}
$\tilde{\Sigma}^{f,\text{At}} \left(\omega + i 0^{+} \right)$ being the exact $f$-electron dynamical self-energy in the atomic limit, and $\tilde{\Sigma}^{f,(2),\text{At}} \left(\omega + i 0^{+} \right)$ being the dynamical self-energy of the single-site model obtained to second-order in perturbation theory in the atomic limit
\begin{equation}
	\label{app:eqn:second_order_se_imp_atomic}
	\tilde{\Sigma}^{f,(2),\text{At}} \left(\omega + i 0^{+} \right) \approx \frac{U_1^2 \left(N_f - 1 \right) n^{\tilde{\mu}}_0 \left( 1 - n^{\tilde{\mu}}_0 \right)}{\omega + \tilde{\mu} + i 0^{+}}.
\end{equation}
Finally, we note that the fictitious chemical potential $\tilde{\mu}$ is fixed by ensuring that the $f$-electron filling of the non-interacting single-site problem at chemical potential $\tilde{\mu}$ and the $f$-electron filling of the interacting problem (at chemical potential $\mu$) are equal, {\it i.e.}{} $n^{\tilde{\mu}}_0 = n$~\cite{MAR86, YEY93, POT97, MEY99, YEY00, ARS12}. The reasons for this choice will be explained in more detail in \cref{app:sec:se_symmetric_details:IPT:fix_mu_tilde}.

\subsubsection{The self-consistent problem and numerical implementation}\label{app:sec:se_symmetric:IPT:sc_and_numerical}

\newcommand{\interText}[5]{
	\draw[postaction={decorate}] (#3) -- node[midway, #5, sloped] {\scalebox{0.5}{\rotatebox{#1}{\cref*{#2}}}} (#4);
}

In \cref{app:sec:se_symmetric:introduction:second_order_perturbation:overview}, we have outlined the IPT method by effectively listing a series of equations that the dynamical self-energy solution needs to satisfy at a given chemical potential $\mu$. All these equations need to be concurrently satisfied with the DMFT self-consistent conditions from \cref{app:eqn:DMFT_GF_symmetric,app:eqn:DMFT_G0_expression}, along with the Hartree-Fock ones from \cref{app:eqn:def_rho_HF,app:eqn:DMFT_GF_symmetric,app:eqn:hartree_fock_hamiltonian_for_DMFT}. 

Because the resulting system of equations is nonlinear, it cannot be solved directly, and its solution must be constructed numerically in an iterative fashion. To do so, we first note that at a given filling, the symmetric solution of the THF model is fully specified by the density matrix and the $\vec{k}$-independent dynamical self-energy of the system. Indeed, the retarded THF Green's function corresponding to a certain density matrix $\rho \left( \vec{k} \right)$ and to a dynamical self-energy correction $\Sigma \left( \omega + i 0^{+} \right)$ can be found from Dyson's equation to be 
\begin{equation}
	\label{app:eqn:second_order_GF_symmetric}
	\mathcal{G} \left( \omega + i 0^{+}, \vec{k} \right) =  \left[\left( \omega + i 0^{+} + \mu \right) \mathbb{1} - h^{\text{MF} \prime} \left( \vec{k} \right) - U_1 \left( n - \frac{1}{2} \right) \left( N_f -1 \right) \mathbb{1}_{f} - \Sigma \left( \omega + i 0^{+} \right) \right]^{-1},
\end{equation}
with the density-matrix-dependent mean-field Hamiltonian $h^{\text{MF} \prime} \left( \vec{k} \right)$ from \cref{app:eqn:hartree_fock_hamiltonian_for_DMFT} containing all the interactions terms besides $H_{U_1}$ decoupled at the mean-field level. As mentioned above \cref{app:eqn:hartree_fock_hamiltonian_for_DMFT}, in the symmetric phase, we ignore the Fock contribution stemming from $H_{U_2}$, which implies that $h^{\text{MF} \prime} \left( \vec{k} \right)$ only depends on the \emph{total} $f$-electron filling and \emph{not} on the detailed momentum dependence of the $f$-electron block of the density matrix\footnote{This is consistent with our approximation in \cref{app:eqn:hartree_fock_hamiltonian_for_DMFT}, under which we ignore any contributions to the Hartree-Fock Hamiltonian arising on the correlation between $f$-electron belonging to different sites.}. For the purpose of this calculation, the latter can therefore be taken without loss of generality to be $\vec{k}$-independent [as well as flavor and site diagonal, as implied by \cref{app:eqn:f_elec_gf_site_flav_diag}], {\it i.e.}{}
\begin{equation}
	\label{app:eqn:symmetric_rho_f_space}
	\rho_{i \eta s; i' \eta' s'} \left( \vec{k} \right) = \left(n - \frac{1}{2} \right) \delta_{ii'} \delta_{\eta \eta'} \delta_{ss'}, \qq{for} 5 \leq i,i' \leq 6,
\end{equation}
in the notation introduced in \cref{app:sec:hartree_fock:generic_not}. At the same time, the dynamical part of the self-energy is given by 
\begin{equation}
	\label{app:eqn:full_ipt_sigma}
	\Sigma_{i \eta s;i' \eta' s'} \left( \omega + i 0^{+} \right) = \begin{cases}
		\tilde{\Sigma}^{f,\text{Int}} \left( \omega + i 0^{+} \right) \delta_{ii'} \delta_{\eta \eta'} \delta_{ss'} & \qq{if} 5 \leq i,i' \leq 6 \\
		- i \Gamma_{c} \delta_{i i'} \delta_{\eta \eta'} \delta_{s s'} & \qq{otherwise}
	\end{cases},
\end{equation}
where, identically to \cref{app:sec:se_correction_beyond_HF}, we have ascribed a constant broadening $\Gamma_c$ to the $c$- and $d$-fermions (with $\Gamma_c = \SI{1}{\milli\electronvolt}$ for TBG and $\Gamma_c = \SI{1.5}{\milli\electronvolt}$ for TSTG). The dynamical self-energy of the $f$-electrons is given by the interpolated self-energy from \cref{app:eqn:interpolated_sigma}. The choice of broadening factors for the $c$- and/or $d$-electrons was justified around \cref{app:eqn:full_second_order_sigma}.

\begin{figure}[!t]
	\centering
	\begin{tikzpicture}[
		scale=0.5,
		decoration={markings, mark=at position 0.5 with {\arrow{>}}},
		boxed/.style={draw, rounded corners, inner sep=2pt},
		>=stealth
		]
		\coordinate (A) at (3,0);
		\coordinate (STA) at (18,0);
		\coordinate (B) at (6,-2);
		\node at (-1,-2) [boxed, fill=green!20] (initSE) {$\Sigma_{m} \left(\omega^+ \right)$};
		\node at (3,-2) [boxed] (C) {$\tilde{\Sigma}_{m}^{f,\text{Int}} \left(\omega^+ \right)$};
		\node at (-1,-6) [boxed, fill=green!20] (D) {$\varrho_m \left( \vec{k} \right)$};
		\node at (3,-6) [boxed] (E) {$h^{\text{MF}\prime} \left( \vec{k} \right)$};
		\coordinate (F) at (6,-6);
		\coordinate (midBF) at ($(B)!0.5!(F)$);
		\coordinate (F) at (6,-6);
		\node at (8,-4) [boxed] (fullGF) {$\mathcal{G} \left(\omega^+ , \vec{k} \right)$};
		\node at (14,-14) [boxed, fill=green!60] (G) {$\varrho_{m+1} \left( \vec{k} \right)$}; \node at (2,-14) [boxed] (H) {$\left\langle nn \right\rangle^{\text{At}}$};\node at (6,-14) [boxed] (I) {$\tilde{\Sigma}^{f,\text{At}} \left( \omega^+ \right)$};\coordinate (K) at (2,-16);
		\coordinate (L) at (6,-16);
		\node at (8,-8) [boxed] (spec) {$A \left( \omega, \vec{k} \right)$};
		\node at (6,-10) [boxed] (mu) {$\mu$};\coordinate (J) at (12,-16);
		\node at (12,-18) [boxed] (M) {$b \left( \omega^+ \right)$};
		\coordinate (N) at (14,-8);
		\coordinate (O) at (18,-8);
		\node at (14,-10) [boxed] (P) {$n$};
		\node at (14,-4) [boxed] (Q) {$G \left( \omega^+ \right)$};
		\node at (18,-6) [boxed] (R) {$\left\langle nn \right\rangle^{\text{ss}}$};
		\coordinate (S) at (18,-4);
		\coordinate (ST) at (18,-2);
		\node at (22,-2) [boxed] (T) {$G_{0} \left( \omega^+ \right)$};
		\node at (26,-2) [boxed] (U) {$\tilde{\mu}$};
		\coordinate (V) at (30,-2);
		\node at (22,-6) [boxed] (W) {$G^{\tilde{\mu}}_{0} \left( \omega^+ \right)$};
		\node at (26,-6) [boxed] (X) {$n^{\tilde{\mu}}_0$};
		\coordinate (Z) at (30,-6);
		\coordinate (AB) at (26,-8);
		\node at (10,-14) [boxed] (AE) {$n_{\text{At}}$};\coordinate (AEL) at (10,-16) {};
		\node at (22,-18) [boxed] (AF) {$\tilde{\Sigma}^{f,(2)} \left( \omega^+ \right)$};
		\node at (26,-18) [boxed] (AG) {$a$};
		\coordinate (AH) at (26,-20);
		\node at (30,-14) [boxed] (AI) {$\tilde{\Sigma}^{f,(2),\text{At}} \left( \omega^+ \right)$};
		\coordinate (AJ) at (30,-16);
		\node at (8,-20) [boxed] (AK) {$\tilde{\Sigma}_{m+1}^{f} \left( \omega^+ \right)$};
		\node at (2,-20) [boxed, fill=green!60] (finSE) {$\Sigma_{m+1} \left( \omega^+ \right)$};
		\coordinate (AL) at (12,-20);
		\coordinate (AM) at (22,-20);
		\coordinate (jump1) at (intersection of W--AF and K--AJ);
		\coordinate (jump2) at (intersection of AB--AG and K--AJ);
		\coordinate (jump3) at (intersection of W--AF and N--AB);
		
		\interText{90}{app:eqn:hartree_fock_hamiltonian_for_DMFT}{D}{E}{above} \node[anchor=east] at (midBF) {\scalebox{0.5}{\rotatebox{90}{\cref*{app:eqn:second_order_GF_symmetric}}}};
		\draw[-] (C) -- (A);
		\draw[postaction={decorate}] (B) -- (midBF);\draw[postaction={decorate}] (F) -- (midBF);\draw[-] (E) -- (F);
\interText{90}{app:eqn:full_ipt_sigma}{initSE}{C}{above}
		\draw[-] (C) -- (B);
		\draw[-] (midBF) -- (fullGF) ;
		\interText{0}{app:eqn:DMFT_self_consistent_eq}{fullGF}{Q}{above}
		\draw[-] (Q) -- (S) -- (ST);
		\interText{0}{app:eqn:all_self_energy_of_ss_model}{ST}{T}{above}
		\interText{0}{app:eqn:def_of_g_tilde_mu}{T}{W}{above}
		\draw[postaction={decorate}] (U) -- (V);
		\interText{0}{app:eqn:second_order_se_imp_atomic}{V}{AI}{above}	
		\interText{0}{app:eqn:nf_from_density_matrix}{G}{P}{above}
		\draw[dashed,<->] (P) -- node[pos=0.25, below, sloped] {\scalebox{0.5}{\rotatebox{0}{Fixes $\tilde{\mu}$ by requiring to be equal}}} (X);
		\interText{90}{app:eqn:nn_correlator_from_self_energy}{S}{R}{above}
		\draw[postaction={decorate}] (P) -- (N); \draw[postaction={decorate}] (R) -- (O); \interText{0}{app:eqn:fixing_filling_self_energy_algo_ipt}{spec}{mu}{below}
		\interText{0}{app:eqn:def_nn_at}{mu}{H}{above}
		\interText{0}{app:eqn:def_sigma_at}{mu}{I}{above}
		\interText{0}{app:eqn:def_n_at}{mu}{AE}{above}
		\draw[postaction={decorate}] (H) -- (K); \draw[postaction={decorate}] (I) -- (L); \draw[postaction={decorate}] (AE) -- (AEL); \interText{0}{app:eqn:def_rho_HF}{spec}{G}{above}
		\interText{0}{app:eqn:spectral_function}{fullGF}{spec}{above}
		\draw[postaction={decorate}] (AI) -- (AJ); \draw (K) -- ($(jump1)!0.5cm!(K)$) to[bend right=90] ($(jump1)!0.5cm!(jump2)$) -- ($(jump2)!0.5cm!(jump1)$) to[bend right=90] ($(jump2)!0.5cm!(AJ)$) -- (AJ);
		\draw[postaction={decorate}] (U) -- (W);
		\interText{0}{app:eqn:int_rel_f0_filling}{W}{X}{above}
		\draw[postaction={decorate}] (X) -- (AB); \interText{90}{app:eqn:b_function_ipt}{J}{M}{above}
		\interText{0}{app:eqn:second_order_f_symmetric}{W}{AF}{above}
		\interText{0}{app:eqn:a_constant_ipt}{AB}{AG}{above}
		\draw (N) -- ($(jump3)!0.5cm!(N)$) to[bend right=90] ($(jump3)!0.5cm!(AB)$) -- (AB);
		\draw[-] (AH) -- (AL); 
		\interText{0}{app:eqn:interpolated_sigma}{AL}{AK}{above}
		\draw[postaction={decorate}] (M) -- (AL); \draw[postaction={decorate}] (AF) -- (AM); \draw[postaction={decorate}] (AG) -- (AH); \draw[postaction={decorate}] (A) -- (STA); \draw[-] (STA) -- (ST); 
		\draw[postaction={decorate}] (X) -- (Z); \interText{0}{app:eqn:full_ipt_sigma}{AK}{finSE}{above}
	\end{tikzpicture}
	\caption{Iterative cycle employed for obtaining the symmetric phase solution of the THF model within DMFT and Hartree-Fock theory using the IPT approach. For brevity, we have denoted $\omega^{+} \equiv \omega + i 0^{+}$. In the DMFT and Hartree-Fock theory approach the symmetric solution is fully specified by the $\vec{k}$-independent dynamical self-energy of the system and its density matrix. Starting from an initial guess for the symmetric solution's density matrix and dynamical self-energy (light green cells) an updated solution is obtained (dark green cells) during a single iteration. The cycle then repeats until convergence. The arrows in the flowchart illustrate the calculation sequence of the intermediate quantities integral to the IPT method (represented by the white rectangular cells), as detailed in \cref{app:sec:se_symmetric:introduction:second_order_perturbation:overview}. Above each arrow, the corresponding equations that relate the connected quantities are provided. The dashed arrow indicates that the fictitious chemical potential is fixed by ensuring that $n = n^{\tilde{\mu}}_0$. }
	\label{app:fig:self_consistent_mipt}
\end{figure}

\Cref{app:fig:self_consistent_mipt} illustrates the iteration process used to construct the symmetric state solution. Just as in \cref{app:sec:se_correction_beyond_HF:sc_problem_and_numerics}, we let $\varrho_{m} \left( \vec{k} \right)$ and $\Sigma_{m} \left( \omega + i 0^{+} \right)$ be the density matrix and dynamical self-energy of the system at the $m$-th step of the algorithm. We also let $\tilde{\Sigma}^{f,\text{Int}}_m \left(\omega + i 0^{+} \right) = \left[ \Sigma_{m} \left( \omega + i 0^{+} \right) \right]_{(\alpha + 4 ) \eta s; (\alpha + 4 ) \eta s}$ denote the interpolated dynamical $f$-electron self energy at step $m$. We postpone the discussion on how $\varrho \left( \vec{k} \right)$ and $\Sigma \left( \omega + i 0^{+} \right)$ are initialized to the end of this section. To obtain $\varrho_{m+1} \left( \vec{k} \right)$ and $\Sigma_{m+1} \left( \omega + i 0^{+} \right)$, we employ the following sequence (note that the first two steps are identical to the algorithm used to obtain the second-order self-consistent self-energy from \cref{app:sec:se_correction_beyond_HF:sc_problem_and_numerics}):
\begin{enumerate}
	\item From $\varrho_{m} \left( \vec{k} \right)$, we determine the Hartree-Fock Hamiltonian $h^{\text{MF} \prime} \left( \vec{k} \right)$ ({\it i.e.}{} the first-order contribution to the self-energy) through \cref{app:eqn:TBG_HF_Hamiltonian,app:eqn:full_TSTG_HF_Hamiltonian,app:eqn:hartree_fock_hamiltonian_for_DMFT}. From the Hartree-Fock Hamiltonian and $\Sigma_{m} \left( \omega + i 0^{+} \right)$, we determine the retarded Green's function of the system $\mathcal{G} \left( \omega + i 0^{+}, \vec{k} \right)$ using \cref{app:eqn:second_order_GF_symmetric}.
	
	\item We then determine the spectral function of the system $A \left( \omega, \vec{k} \right)$ through \cref{app:eqn:spectral_function}. At this step, we also fix the chemical potential $\mu$ of the system by requiring that the desired filling $\nu_0$ obeys
	\begin{equation}
		\label{app:eqn:fixing_filling_self_energy_algo_ipt}
		\nu_0 = \frac{1}{N_0} \sum_{\vec{k}} \sum_{i,\eta,s} \left (\int_{-\infty}^{\infty} \dd{\omega} n_{\mathrm{F}} \left( \omega \right) A_{i \eta s; i \eta s} \left( \omega, \vec{k} \right) - \frac{1}{2} \right).
	\end{equation}
	We find that working at fixed filling is more stable numerically ({\it i.e.}{} the solution converges within a smaller number of iterations) than working at a fixed chemical potential.
	
	\item Once the chemical potential $\mu$ has been fixed, we determine the new density matrix $\varrho'_{m+1} \left( \vec{k} \right)$ from the spectral function via \cref{app:eqn:def_rho_HF}. Generally, the new density matrix $\varrho'_{m+1} \left( \vec{k} \right)$ is only \emph{approximately} symmetric and $\vec{k}$-independent in the $f$-electron block~\footnote{With infinite precision arithmetic, $\varrho'_{m+1} \left( \vec{k} \right)$ will inherit the symmetries of the Green's function $\mathcal{G} \left( \omega + i 0^{+} \right)$. The latter is computed from the Hartree-Fock matrix of the symmetric phase and a symmetric self-energy, and hence will obey all the symmetries of TBG or TSTG. However, finite precision arithmetic causes small symmetry-breaking errors which, over the course of many iterations, might lead to the self-consistent solution \emph{not} obeying the all the symmetries of the system. To avoid these numerical instabilities, we re-symmetrize the density matrix at every step. }. Therefore, we symmetrize it by computing the relative $f$-electron filling $n$,
	\begin{equation}
		\label{app:eqn:nf_from_density_matrix}
		n = \frac{1}{2} + \frac{1}{N_0 N_f} \sum_{\vec{k}} \sum_{\alpha, \eta, s} \left[ \varrho'_{m+1} \left( \vec{k} \right) \right]_{(\alpha+4) \eta s; (\alpha+4) \eta s} ,
	\end{equation}  
	and reconstructing the explicitly \emph{symmetric} density matrix $\varrho_{m+1} \left( \vec{k} \right)$ via
	\begin{equation}
		\left[ \varrho_{m+1} \left( \vec{k} \right) \right]_{i \eta s; i' \eta' s'} = \begin{cases}
			\left(n - \frac{1}{2} \right) \delta_{ii'} \delta_{\eta \eta'} \delta_{ss'} & \qq{if} 5 \leq i,i' \leq 6 \\ 
			\left[ \varrho'_{m+1} \left( \vec{k} \right) \right]_{i \eta s; i' \eta' s'} & \qq{otherwise} 
		\end{cases},
	\end{equation}
	thus ensuring that \cref{app:eqn:symmetric_rho_f_space} is satisfied (meaning that $\varrho_{m+1} \left( \vec{k} \right)$ is diagonal in the $f$-electrons block). We find that re-symmetrizing only in the $f$-electron block is enough to prevent the accumulation of symmetry-breaking errors arising from finite precision arithmetic.
	
	\item From the Green's function of the system $\mathcal{G} \left( \omega + i 0^{+}, \vec{k} \right)$, we determine the site-diagonal $f$-electron Green's function through
	\begin{equation}
		\tilde{\mathcal{G}}^{f} \left( \omega + i 0^{+}\right) = \frac{1}{N_0 N_f} \sum_{\vec{k}} \sum_{\alpha, \eta, s} \mathcal{G}_{(\alpha+4) \eta s; (\alpha+4) \eta s} \left( \omega + i 0^{+}, \vec{k} \right).
	\end{equation}
	Using $\tilde{\mathcal{G}}^{f} \left( \omega + i 0^{+}\right)$, we then compute the following quantities:
	\begin{enumerate}
		\item We determine $G^{\tilde{\mu}}_{0} \left( \omega + i 0^{+} \right)$ through \cref{app:eqn:DMFT_G0_expression,app:eqn:def_of_g_tilde_mu}, using the previously computed $\tilde{\Sigma}_m^{f,\text{Int}} \left( \omega + i 0^{+} \right)$ as an estimate of the $f$-electron dynamical self-energy
		\begin{equation}
			G^{\tilde{\mu}}_{0} \left( \omega + i 0^{+} \right) = \frac{1}{ \left( \tilde{\mathcal{G}}^{f} \left( \omega + i 0^{+} \right) \right)^{-1} - \mu + \tilde{\mu} + U_1 \left(n -\frac{1}{2} \right) \left(N_f - 1 \right) + \tilde{\Sigma}_m^{f,\text{Int}} \left( \omega + i 0^{+} \right)}.
		\end{equation}
		The chemical potential of the non-interacting impurity problem $\tilde{\mu}$ is fixed from \cref{app:eqn:int_rel_f0_filling} by requiring that $n = n_0$~\cite{MAR86, YEY93, POT97, MEY99, YEY00, ARS12}.
		\item We also compute the correlator $\left\langle nn \right\rangle^{\text{ss}}$ from \cref{app:eqn:nn_correlator_from_self_energy}, using $\tilde{\Sigma}_m^{f,\text{Int}} \left( \omega + i 0^{+} \right)$ as an approximation of the $f$-electron dynamical self-energy.
		\item We then obtain the $a$ constant and the $b \left( \omega + i 0^{+} \right)$ function from \cref{app:eqn:a_constant_ipt,app:eqn:b_function_ipt}, respectively. 
	\end{enumerate}
	
	\item Using the Green's function $G^{\tilde{\mu}}_0 \left( \omega + i 0^{+} \right)$, we compute the (non-interpolated) second-order self-energy correction $\tilde{\Sigma}^{f,(2)} \left( \omega + i 0^{+} \right)$ using \cref{app:eqn:second_order_f_symmetric}.
	
	\item Finally, we use the second-order self-energy correction $\tilde{\Sigma}^{f,(2)} \left( \omega + i 0^{+} \right)$ to construct the new diagonal part of the interpolated self-energy $\tilde{\Sigma}_{m+1}^{f,\text{Int}} \left( \omega + i 0^{+} \right)$ from \cref{app:eqn:interpolated_sigma} and, with it, the updated dynamical part of the self-energy from \cref{app:eqn:full_ipt_sigma}. 
\end{enumerate}

The algorithm is iterated until convergence is achieved, where convergence is assessed by the same criteria as in \cref{app:sec:se_correction_beyond_HF:sc_problem_and_numerics}. We use the same $\vec{k}$- and $\omega$-point discretizations as in \cref{app:sec:se_correction_beyond_HF:sc_problem_and_numerics}. The slowest part of the algorithm is constructing the retarded Green's function of the system $\mathcal{G} \left( \omega + i 0^{+}, \vec{k} \right)$ using \cref{app:eqn:second_order_GF_symmetric}, as that entails numerous matrix inversions (one for each $\vec{k}$ and $\omega$). We improve the efficiency of our algorithm by only computing $\mathcal{G} \left( \omega + i 0^{+}, \vec{k} \right)$ for one spin sector and for one sixth of the BZ and then recovering the other matrix components and $\vec{k}$-points using the $\mathrm{SU} \left(2\right)$ spin and $C_{6z}$ rotation symmetries. As in \cref{app:sec:se_correction_beyond_HF:sc_problem_and_numerics}, we also efficiently compute the Fourier integral of \cref{app:eqn:second_order_f_symmetric} by means of a Fast Fourier transformation~\cite{BAI06}. Any potential oscillations in the self-consistent iterations are dampened by mixing the old and updated solution at every iteration: at the end of the self-consistent cycle, we admix the old solution to the new one $\tilde{\Sigma}_{m+1}^{f,\text{Int}} \left( \omega + i 0^{+} \right) \to \alpha \tilde{\Sigma}_{m+1}^{f,\text{Int}} \left( \omega + i 0^{+} \right) + \left( 1 - \alpha \right) \tilde{\Sigma}_{m}^{f,\text{Int}} \left( \omega + i 0^{+} \right)$ and $\varrho_{m+1} \left( \vec{k} \right) \to \alpha \varrho_{m+1} \left( \vec{k} \right) + \left( 1 - \alpha \right) \varrho_{m} \left( \vec{k} \right)$, where $\alpha$ is a random number $\alpha \in \left[0.1, 0.5 \right]$ independently drawn at each iteration.

To obtain results in the symmetric phase, we always start from charge neutrality and move towards doping the remote bands in increments of $\delta \nu = \frac{1}{50}$. To obtain the charge-neutral solution (corresponding to the target filling $\nu_0 = 0$), we take the following initial conditions for the density matrix and the dynamical self-energy:
\begin{equation}
	\Sigma_{i \eta s;i' \eta' s'} \left( \omega + i 0^{+} \right)  = - i \Gamma_{c} \delta_{i i'} \delta_{\eta \eta'} \delta_{s s'} \qq{and}
	\varrho_{i \eta s;i' \eta' s'} \left( \vec{k} \right) = 0, \qq{for all} i,i', 
\end{equation}
the first of which entails a constant broadening factor for all electron species.

Once the self-consistent symmetric solution at charge neutrality is constructed, we obtain the solutions at $\nu_0 = m \delta_{\nu}$ (where $m \in \mathbb Z$ and $1 \leq m \leq \frac{4.5}{\delta \nu}$) incrementally using the self-consistent density matrix and dynamical self-energy obtained at filling $\nu_0 = \left(m-1 \right) \delta_{\nu}$ as initial condition. 

\subsection{Benchmarking the IPT method}\label{app:sec:se_symmetric:QMC} 

In \cref{sec:symmetric:benchmark}, we benchmark the IPT method against a continuous-time quantum Monte Carlo (CT-QMC) impurity solver. Here we outline the details of the latter simulations. In all CT-QMC runs, following Ref.~\cite{SON21}, we omit the $H_{\tilde{J}}$ and $H_K$ interaction terms from \cref{app:eqn:THF_int:Jtilde,app:eqn:THF_int:K}, respectively. As reviewed in \cref{app:sec:se_symmetric:DMFT_overview}, DMFT captures the local correlations generated by the onsite interaction $H_{U_1}$, but it cannot directly incorporate offsite interactions such as the nearest-neighbor $f$-$f$ repulsion $H_{U_2}$ or two-body operators involving momentum-dependent $c$-electrons, including the density-density interactions $H_W$, $H_V$, and the magnetic exchange $H_J$. 

To handle the interaction Hamiltonian from \cref{app:eqn:THF_interaction_TBG} within DMFT, we adopt the combined DMFT and Hartree-Fock approach described in \cref{app:sec:se_symmetric:DMFT_overview}, where $H_{U_1}$ is treated dynamically and all other terms are mean-field decoupled. At each DMFT iteration, the non-interacting Hamiltonian from \cref{app:eqn:single_part_THF_TBG} is supplemented by $H_{U_1}$ and a Hartree-Fock term of the type in \cref{app:eqn:full_TBG_HF_Hamiltonian}, but which only arises from $H_{U_2}$ and $H_{V}$ (both treated only at the Hartree level), as well as $H_{W}$ and $H_{J}$ (treated at the Hartree-Fock level). Explicitly, the interacting Hamiltonian reads
\begin{align}
	H_{\text{CT-QMC}}^{\text{DMFT+HF}} =& H^{\text{TBG}}_0 + H_{U_1} + 6 U_{2} \nu_{f} \sum_{\vec{R},\alpha, \eta, s} \hat{f}^\dagger_{\vec{R},\alpha, \eta, s} \hat{f}_{\vec{R},\alpha, \eta, s} +  \frac{V \left( \vec{0} \right)}{2 \Omega_0 N_0} \nu_c \sum_{\vec{k},a,\eta,s} \hat{c}^\dagger_{\vec{k},a,\eta,s} \hat{c}_{\vec{k},a,\eta,s} \nonumber\\ 
	&+ \sum_{\vec{k},a,\eta,s} W_a \nu_f \hat{c}^\dagger_{\vec{k},a,\eta,s} \hat{c}_{\vec{k},a,\eta,s} +     
	\sum_{\substack{\vec{R},\alpha, \eta, s \\ a',\eta',s'}} W_a \mathcal{O}_{a \eta s; a \eta s} \hat{f}^\dagger_{\vec{R},\alpha, \eta, s} \hat{f}_{\vec{R},\alpha, \eta, s} \nonumber\\
	&- \sum_{\vec{k}} \sum_{\substack{\alpha, \eta, s \\ a' \eta' s'}} \left( W_{a'} O_{a' \eta' s'; (\alpha+4) \eta s} \hat{f}^\dagger_{\vec{k},\alpha, \eta, s} \hat{c}_{\vec{k},a',\eta',s'} + \text{h.c.} \right) \nonumber\\
	&- \frac{J}{2} \sum_{\vec{R}} \sum_{\substack{\alpha, \eta, s \\ \alpha', \eta', s'}} \left[ \eta \eta' + (-1)^{\alpha+\alpha'} \right] \hat{f}^\dagger_{\vec{R},\alpha, \eta, s} \hat{f}_{\mathbf{R},\alpha', \eta', s'} \mathcal{O}_{(\alpha'+2)\eta' s'; (\alpha+2)\eta s} \nonumber\\
	&- \frac{J}{2} \sum_{\vec{k}} \sum_{\substack{\alpha, \eta, s \\ \alpha', \eta', s'}} \left[ \eta \eta' + (-1)^{\alpha+\alpha'} \right] \hat{c}^\dagger_{\vec{k},(\alpha+2) e s} \hat{c}_{\vec{k},(\alpha'+2) \eta' s'} \mathcal{O}_{ (\alpha' + 4) \eta' s'; (\alpha + 4) \eta \sigma} \nonumber\\
	&+  \frac{J}{2} \sum_{\vec{k}} \sum_{\substack{\alpha, \eta, s \\ \alpha', \eta', s'}} \left[ \eta \eta' + (-1)^{\alpha+\alpha'} \right] \left[\mathcal{O}_{(\alpha'+2) \eta'\sigma'; (\alpha'+4) \eta' s'} \hat{f}^\dagger_{\vec{k},\alpha, \eta, s} \hat{c}_{\vec{k}, (\alpha+2),\eta,s} + \text{h.c.} \right],
\end{align}
where 
\begin{equation}
	\mathcal{O} = \frac{1}{N_0} \sum_{\vec{k}} \varrho \left( \vec{k} \right),
\end{equation}
is the onsite density matrix, updated at every DMFT iteration. Consistent with Ref.~\cite{SON21}, we neglect the Fock term of $H_V$ for these simulations (and in the corresponding IPT calculations used for comparison). We also find that including $H_J$ has negligible influence above the ordering temperature (approximately $\SI{10}{\kelvin}$).

The CT-QMC simulations are performed with the DMFT-CTHYB solver \textit{w2dynamics}~\cite{PAR12,WAL19}. We set $\beta^{-1} = \SI{1}{\milli\electronvolt}$, above the ordering temperature~\cite{RAI23a}, and carry out fixed-density simulations for $-1 \leq \nu \leq +1$ around charge neutrality, adjusting the chemical potential dynamically within the DMFT loop to match the target filling. Numerical parameters follow closely those of Ref.~\cite{RAI23a}: runs are executed on 144 CPU cores, with ${\texttt{Nmeas}} = 1.5 \times 10^{5}$ measurements per core per DMFT step, and ${\texttt{Ncorr}} = 1500$ updates per core between measurements to suppress autocorrelation effects. A total of ${\texttt{Nwarmup}} = 2 \cdot 10^{5}$ thermalization steps per core are performed prior to measurements. Analytical continuation is carried out starting from the final DMFT iteration using the MaxEnt method in the \textit{ana\_cont} package~\cite{KAU23}.

\section{Details on the IPT method as applied to the THF model}\label{app:sec:se_symmetric_details}

This \siSection{} provides the full derivation of all intermediate results used in \cref{app:sec:se_symmetric}, where we discussed the symmetric phase of the THF model and outlined its solution via a combination of DMFT and Hartree-Fock theory. In particular, we previously introduced the IPT method as a tool to solve the DMFT equations, presenting only the algorithmic steps without proof. Here, we supply the complete analytical derivations underpinning the identities employed there. We begin by defining and solving the atomic limit of the THF model. Next, we derive two exact properties of the dynamical self-energy in the symmetric phase, which are essential to the IPT framework. Finally, we present a detailed derivation of the IPT ansatz for the dynamical self-energy.

\subsection{The atomic limit of the THF model}\label{app:sec:se_symmetric_details:atomic_se}

In this section, we consider the atomic problem for the THF $f$-electrons and derive the dynamical part of the corresponding atomic self-energy. In the atomic limit, the $f$-electrons at different lattice sites are decoupled from one another and also decoupled from the $c$- or $d$-electrons. Because in this limit, the $f$-electrons at a given lattice site are not hybridizing or interacting with any other fermions, we can study the atomic problem via the following Hamiltonian
\begin{align}
	H^{\text{At}} &= \frac{U_1}{2} \sum_{\substack{\alpha, \eta, s \\ \alpha', \eta', s'}} :\mathrel{\hat{f}^\dagger_{\alpha, \eta, s} \hat{f}_{\alpha, \eta, s}}: :\mathrel{\hat{f}^\dagger_{\alpha', \eta', s'} \hat{f}_{\alpha', \eta', s'}}: \nonumber \\
	&= \frac{U_1}{2}  \sum_{\left(\alpha, \eta, s\right) \neq \left( \alpha', \eta', s' \right)} \hat{f}^\dagger_{\alpha, \eta, s} \hat{f}_{\alpha, \eta, s} \hat{f}^\dagger_{\alpha', \eta', s'} \hat{f}_{\alpha', \eta', s'} - \frac{\left( N_f - 1 \right) U_1}{2}  \sum_{\alpha, \eta, s} \hat{f}^\dagger_{\alpha, \eta, s} \hat{f}_{\alpha, \eta, s} + \frac{N_f^2 U_1}{8}, \label{app:eqn:atomic_Hamiltonian}
\end{align}
which is just the $H_{U_1}$ interaction Hamiltonian defined for a single-site. In \cref{app:eqn:atomic_Hamiltonian}, $N_f=8$ denotes the number of $f$-fermion flavors and we have suppressed the position indices of the $f$-electron operators, as $H^{\text{At}}$ is defined only for a single-site.

\subsubsection{Computing the exact atomic self-energy}\label{app:sec:se_symmetric_details:atomic_se:exact}

To obtain the self-energy associated with $H^{\text{At}}$, we will first find the exact interacting and non-interacting Green's functions of the system using elementary means, and then derive the self-energy employing Dyson's equation. We define the interacting and non-interacting Green's functions of the atomic problem to be, respectively,
\begin{align}
	-\left\langle \mathcal{T}_{\tau} \hat{f}_{\alpha, \eta, s} \left( \tau \right) \hat{f}^\dagger_{\alpha', \eta', s'} \left( 0 \right)  \right\rangle^{\text{At}}  &= \mathcal{G}^{\text{At}}_{\alpha \eta s;\alpha' \eta' s'} \left(\tau \right), \label{app:eqn_gf_THF_at_int} \\
	-\left\langle \mathcal{T}_{\tau} \hat{f}_{\alpha, \eta, s} \left( \tau \right) \hat{f}^\dagger_{\alpha', \eta', s'} \left( 0 \right)  \right\rangle^{\text{At}}_{0}  &= \mathcal{G}^{\text{At},0}_{\alpha \eta s;\alpha' \eta' s'} \left(\tau \right),  \label{app:eqngf_THF_at_nonint}
\end{align}
where the time evolution and thermodynamic averaging is done with respect to the interacting grand canonical atomic Hamiltonian $K^{\text{At}} = H^{\text{At}} - \mu N^{\text{At}}$ in \cref{app:eqn_gf_THF_at_int}, and with respect to the non-interacting one $K_0^{\text{At}} = - \mu N^{\text{At}}$ in \cref{app:eqngf_THF_at_nonint}. The number operator in the atomic limit is simply $N^{\text{At}} = \sum_{\alpha, \eta, s} \hat{f}^\dagger_{\alpha, \eta, s}\hat{f}_{\alpha, \eta, s}$. 

The non-interacting Green's function in the atomic limit is trivially given by
\begin{equation}
	\mathcal{G}^{\text{At},0}_{\alpha \eta s;\alpha' \eta' s'} \left(i \omega_n \right) = \tilde{\mathcal{G}}^{\text{At},0} \left( i \omega_n \right) \delta_{\alpha \alpha'} \delta_{\eta \eta'} \delta_{s s'} , \qq{with} \tilde{\mathcal{G}}^{\text{At},0} \left( i \omega_n \right) = \frac{1}{i \omega_n - \mu}.
\end{equation}
To determine the interacting one, we will employ its Lehmann representation~\cite{MAH00}. First, we note that the atomic Hamiltonian can be diagonalized exactly. The eigenstates of $H^{\text{At}}$ are simply Slater determinants of $\hat{f}^\dagger_{\alpha, \eta, s}$ electrons occupying different orbital, valley, and spin flavors. There are $d_{m} \equiv \binom{N_f}{m}$ different eigenstates with $m$ $f$-electrons, where $0 \leq m \leq N_f$ and $N_f$ is the total number of $f$-fermion flavors, $N_f = 8$. We label the $i$-th eigenstate containing $m$ $f$-electrons as $\ket{m;i}$, where $1 \leq i \leq d_{m}$. The energy of the $\ket{m;i}$ state depends only on the number of occupied $f$-fermions
\begin{equation}
	\label{app:eqn:atomic_eigenbasis}
	K^{\text{At}} \ket{m;i} = E_{m} \ket{m;i}, \qq{where} E_{m} \equiv \frac{U_1 \left(m - N_f/2 \right)^2}{2} - \mu m \qq{and} 1 \leq i \leq d_{m}.
\end{equation} 

Letting 
\begin{equation}
	\label{app:eqn:partition_function}
	Z = \Tr \left( e^{-\beta K^{\text{At}}} \right) = \sum_{m=0}^{N_f} d_m e^{-\beta E_m}
\end{equation} 
be the partition function of the atomic system, the interacting Green's function $\mathcal{G}^{\text{At}} \left(\tau \right)$ for $\tau>0$ can be expressed as 
\begin{align}
	\mathcal{G}^{\text{At}}_{\alpha \eta s;\alpha' \eta' s'} \left(\tau \right) &= -\frac{1}{Z} \Tr \left( e^{ \left(\tau - \beta\right) K^{\text{At}}} \hat{f}_{\alpha, \eta, s} e^{-\tau K^{\text{At}}} \hat{f}^\dagger_{\alpha', \eta', s'} \right) \nonumber \\
	&= -\frac{1}{Z} \sum_{m_1,m_2=0}^{N_f} \sum_{i=1}^{d_{m_1}} \sum_{j=1}^{d_{m_2}} \bra{m_1;i} \hat{f}_{\alpha, \eta, s} \ket{m_2;j} \bra{m_2;j} \hat{f}^\dagger_{\alpha', \eta', s'} \ket{m_1;i} e^{ \left(\tau - \beta\right)E_{m_1}} e^{ -\tau E_{m_2}}, \label{app:eqn:at_gf_time}
\end{align}
which, using the Fourier transformation convention from \cref{app:eqn:matsubara_gf_THF_ft}, leads to the following expression of the interacting Green's function in Matsubara frequency
\begin{align}
	\mathcal{G}^{\text{At}}_{\alpha \eta s;\alpha' \eta' s'} \left(i \omega_n \right) &= \int_{0}^{\beta} \dd{\tau} e^{i \omega_n \tau} 	\mathcal{G}^{\text{At}}_{\alpha \eta s;\alpha' \eta' s'} \left(\tau \right) \nonumber \\
	&= \frac{1}{Z} \sum_{m_1,m_2=0}^{N_f} \sum_{i=1}^{d_{m_1}} \sum_{j=1}^{d_{m_2}} \bra{m_1;i} \hat{f}_{\alpha, \eta, s} \ket{m_2;j} \bra{m_2;j} \hat{f}^\dagger_{\alpha', \eta', s'} \ket{m_1;i}  
	\frac{e^{-\beta E_{m_1}} + e^{-\beta E_{m_2}} }{i \omega_n - E_{m_2} + E_{m_1}}. \label{app:eqn:at_gf_mats_1}
\end{align}
Since the eigenbasis of $H^{\text{At}}$ is formed by Slater determinant states of $f$-electrons, we note that $\bra{m_2;j} \hat{f}^\dagger_{\alpha', \eta', s'} \ket{m_1;i}$ in \cref{app:eqn:at_gf_mats_1} can only be non-zero whenever $\ket{m_2;j} = \hat{f}^\dagger_{\alpha', \eta', s'} \ket{m_1;i}$. Similarly, $\bra{m_1;i} \hat{f}_{\alpha, \eta, s} \ket{m_2;j}$ is only non-zero if $\ket{m_2;j} = \hat{f}^\dagger_{\alpha' \eta' s'} \ket{m_1;i}$. As such, we can simplify the expression in \cref{app:eqn:at_gf_mats_1} and obtain 
\begin{equation}
	\label{app:eqn:at_gf_mats_2}
	\mathcal{G}^{\text{At}}_{\alpha \eta s;\alpha' \eta' s'} \left(i \omega_n \right) = \frac{1}{Z} \sum_{m=0}^{N_f-1} \sum_{i=1}^{d_{m_1}} \abs{\bra{m;i} \hat{f}_{\alpha, \eta, s} \hat{f}^\dagger_{\alpha, \eta, s}  \ket{m;i}}^2 \delta_{\alpha \alpha'} \delta_{\eta \eta'} \delta_{s s'}  
	\frac{e^{-\beta E_{m}} + e^{-\beta E_{m+1}} }{i \omega_n - E_{m+1} + E_{m}}. 
\end{equation}
We now note that the matrix element $\bra{m;i} \hat{f}_{\alpha, \eta, s} \hat{f}^\dagger_{\alpha, \eta, s}  \ket{m;i}$ is exactly one for those Slater-determinant states containing $m$ $f$-electrons of which none are in the $\left( \alpha, \eta, s \right)$ orbital, valley, and spin flavor (being zero otherwise). There are exactly $\binom{N_f-1}{m}$ such states, which allows us to show that the Green's function in the atomic limit is diagonal the orbital, valley, and spin basis (as expected from the $\mathrm{U} \left(N_f\right)$ symmetry of the atomic limit Hamiltonian), and reads as
\begin{equation}
	\label{app:eqn:at_gf_matrix_form}
	\mathcal{G}^{\text{At}}_{\alpha \eta s;\alpha' \eta' s'} \left(i \omega_n \right) = \tilde{\mathcal{G}}^{\text{At}} \left( i \omega_n \right) \delta_{\alpha \alpha'} \delta_{\eta \eta'} \delta_{s s'},
\end{equation} 
where [using also $E_{m+1} = E_{m} + U_1 \left( m - \frac{N_f}{2} \right) + \frac{U_1}{2} - \mu$, which follows from \cref{app:eqn:atomic_eigenbasis}]
\begin{equation}
	\label{app:eqn:atomic_gf_diagonal_part}
	\tilde{\mathcal{G}}^{\text{At}} \left( i \omega_n \right) = \frac{1}{Z} \sum_{m=0}^{N_f-1} \binom{N_f-1}{m} \frac{e^{-\beta E_{m}} + e^{-\beta E_{m+1}} }{i \omega_n + \mu - U_1 \left(m - \frac{N_f}{2} \right) - \frac{U_1}{2}}.
\end{equation}

One can obtain a more intuitive form of the atomic Green's function in \cref{app:eqn:atomic_gf_diagonal_part} by defining the classical probabilities of occupation for the $(N_f+1)$ atomic energy levels ({\it i.e.}{} corresponding to the states with $0$ up to $N_f$ electrons). Specifically, we let $p_m$ ($0 \leq m \leq N_f$) be the probability of the system being in a state with $m$ $f$-electrons. One obtains that 
\begin{equation}
	\label{app:eqn:atomic_occ_probabilities}
	p_m = \frac{d_m e^{-\beta E_m}}{Z},
\end{equation}
which then implies that $\frac{e^{-\beta E_m}}{Z}= p_m/ \binom{N_f}{m}$. Plugging this expression of the Boltzmann factor into \cref{app:eqn:atomic_gf_diagonal_part} allows one to rewrite it as~\cite{FUJ03}
\begin{equation}
	\label{app:eqn:atomic_gf_diagonal_part_with_prob}
	\tilde{\mathcal{G}}^{\text{At}} \left( i \omega_n \right) = \frac{1}{N_f} \sum_{m=0}^{N_f-1} \frac{ \left(N_f - m \right) p_m + \left(m+1 \right) p_{m+1} }{i \omega_n + \mu - U_1 \left(m - \frac{N_f}{2} \right) - \frac{U_1}{2}}.
\end{equation}

The atomic self-energy can be extracted from the interacting atomic Green's function. It is customary to separate the self-energy into the static (Hartree-Fock) and the dynamical one (all the corrections beyond first-order in perturbation theory). To be specific, we assume that the atomic system is at a certain chemical potential $\mu$. As a result of the $\mathrm{U} \left(N_f\right)$ symmetry of the atomic problem, the density matrix of the system is proportional to identity (since the Green's function itself is diagonal) and is given by 
\begin{equation}
	\label{app:eqn:rel_fil_of_f}
	\left\langle \hat{f}^\dagger_{\alpha, \eta, s} \hat{f}_{\alpha', \eta', s'} \right\rangle^{\text{At}} = n_{\text{At}} \delta_{\alpha \alpha'} \delta_{\eta \eta'} \delta_{s s'}, 
\end{equation} 
where $0 \leq n_{\text{At}} \leq 1$ denotes the relative filling of the system, such that $n_{\text{At}}=0$ ($n_{\text{At}}=1$) corresponds to the fully-empty (fully-filled) case\footnote{Note that because we are considering an interacting system at finite temperature, the total filling of the system can assume fractional values, implying that $n_{\text{At}}$ can take any value in the interval $\left[0, 1\right]$.}. We also define the correlator between the fillings of different orbital, spin, and valley flavors to be 
\begin{equation}
	\label{app:eqn:nn_of_f}
	\left\langle nn \right\rangle^{\text{At}} = \left\langle \hat{f}^\dagger_{\alpha, \eta, s} \hat{f}_{\alpha, \eta, s} \hat{f}^\dagger_{\alpha', \eta', s'} \hat{f}_{\alpha', \eta', s'} \right\rangle^{\text{At}}\neq n_{\text{At}}^2, \qq{for any} \left(\alpha, \eta, s\right) \neq \left(\alpha', \eta', s' \right),
\end{equation}
where the reader is reminded that a repeated index does not imply summation. Note that as a result of the system's $\mathrm{U} \left(N_f\right)$ symmetry $\left\langle \hat{f}^\dagger_{\alpha, \eta, s} \hat{f}_{\alpha, \eta, s} \hat{f}^\dagger_{\alpha', \eta', s'} \hat{f}_{\alpha', \eta', s'} \right\rangle^{\text{At}}$ is independent on $\left( \alpha, \eta, s \right) \neq \left( \alpha', \eta', s' \right)$. Also note that $\left\langle \hat{f}^\dagger_{\alpha, \eta, s} \hat{f}_{\alpha, \eta, s} \hat{f}^\dagger_{\alpha, \eta, s} \hat{f}_{\alpha, \eta, s} \right\rangle^{\text{At}} = n_{\text{At}} $.
It is now easy to show that 
\begin{align}
	\left\langle N^{\text{At}} \right\rangle^{\text{At}} =& \sum_{\alpha, \eta, s} \left\langle \hat{f}^\dagger_{\alpha, \eta, s} \hat{f}_{\alpha, \eta, s} \right\rangle^{\text{At}} = N_f n_{\text{At}}, \\
	\left\langle \left( N^{\text{At}} \right)^2 \right\rangle^{\text{At}} =& \sum_{\substack{\alpha, \eta, s \\ \alpha', \eta', s'}} \left\langle \hat{f}^\dagger_{\alpha, \eta, s} \hat{f}_{\alpha, \eta, s} \hat{f}^\dagger_{\alpha', \eta', s'} \hat{f}_{\alpha', \eta', s'}  \right\rangle^{\text{At}}  \nonumber\\
	=&\sum_{\left(\alpha, \eta, s \right) \neq \left(\alpha', \eta', s' \right)} \left\langle \hat{f}^\dagger_{\alpha, \eta, s} \hat{f}_{\alpha, \eta, s} \hat{f}^\dagger_{\alpha', \eta', s'} \hat{f}_{\alpha', \eta', s'}  \right\rangle^{\text{At}} + \sum_{\alpha, \eta, s} \left\langle \hat{f}^\dagger_{\alpha, \eta, s} \hat{f}_{\alpha, \eta, s} \hat{f}^\dagger_{\alpha, \eta, s} \hat{f}_{\alpha, \eta, s} \right\rangle^{\text{At}} \nonumber \\
	=& N_f \left(N_f - 1 \right) \left\langle nn \right\rangle^{\text{At}} + N_f n_{\text{At}},
\end{align}
which enables us to express in terms of $n_{\text{At}}$ and $\left\langle nn_{\text{At}} \right\rangle$ the first two nontrivial moments (in addition to the normalization condition) of the discrete probability distribution $p_m$ (for $0 \leq m \leq N_f$), which characterizes the $f$-electron occupations
\begin{equation}
	\sum_{m=0}^{N_f} p_m = 1, \qquad
	\sum_{m=0}^{N_f} m p_m = N_f n_{\text{At}} ,\qquad
	\sum_{m=0}^{N_f} m^2 p_m =  N_f \left(N_f - 1 \right) \left\langle nn \right\rangle^{\text{At}} + N_f n_{\text{At}} . \label{app:eqn:f_electron_prob_moments}
\end{equation}

The \emph{exact} dynamical self-energy can be extracted directly from the atomic Green's function. To do so, we first decouple the atomic Hamiltonian $H^{\text{At}}$ at the Hartree-Fock level to obtain
\begin{equation}
	\label{app:eqn:mf_of_at_H}
	H_{\text{MF}}^{\text{At}} =  U_1 \left( n_{\text{At}} - \frac{1}{2} \right) \left( N_f - 1 \right) \sum_{\alpha, \eta, s} \hat{f}^\dagger_{\alpha, \eta, s} \hat{f}_{\alpha, \eta, s} + \frac{N_f^2 U_1}{8}, 
\end{equation}
which follows straight-forwardly from the second row of \cref{app:eqn:atomic_Hamiltonian} using the density matrix in \cref{app:eqn:rel_fil_of_f}. From Dyson's equation, we can re-express the atomic Green's function as
\begin{equation}
	\label{app:eqn:at_self_energy_pre_def}
	\left[ {\mathcal{G}^{\text{At}}}^{-1} \right]_{\alpha \eta s;\alpha' \eta' s'}\left( i\omega_n \right) = \left[i \omega_n + \mu - U_1 \left(n_{\text{At}} -\frac{1}{2} \right) \left(N_f - 1 \right) \right] \delta_{\alpha \alpha'} \delta_{\eta \eta'} \delta_{s s'} - \Sigma^{\text{At}}_{\alpha \eta s;\alpha' \eta' s'} \left( i \omega_n \right),
\end{equation}
where the third term corresponds to the static (Hartree-Fock) self-energy -- {\it i.e.}{} the prefactor of the quadratic term in \cref{app:eqn:mf_of_at_H} -- and $\Sigma^{\text{At}}_{\alpha \eta s;\alpha' \eta' s'} \left( i \omega_n \right)$ is the dynamical ({\it i.e.}{} $\omega$-dependent) contribution to the atomic self-energy\footnote{The total self-energy is obtained by summing the static (Hartree-Fock) and dynamic contributions.}. The latter is defined through \cref{app:eqn:at_self_energy_pre_def}, is diagonal in the $f$-electron flavors (since both $\mathcal{G}^{\text{At}} \left( i \omega_n \right)$ and $\mathcal{G}^{\text{At},0} \left( i \omega_n \right)$ are), and is given explicitly by
\begin{equation}
	\label{app:eqn:at_self_energy}
	\Sigma^{\text{At}}_{\alpha \eta s;\alpha' \eta' s'} \left( i \omega_n \right) = \tilde{\Sigma}^{\text{At}} \left( i \omega_n \right) \delta_{\alpha \alpha'} \delta_{\eta \eta'} \delta_{s s'},
\end{equation}
where the diagonal part reads as
\begin{equation}
	\label{app:eqn:at_self_energy_diagonal}
	\tilde{\Sigma}^{\text{At}} \left( i \omega_n \right) \equiv i \omega_n + \mu - U_1 \left(n_{\text{At}} -\frac{1}{2} \right) \left( N_f - 1 \right) - \left(\tilde{\mathcal{G}}^{\text{At}} \left( i \omega_n \right) \right)^{-1}.
\end{equation}

\subsubsection{Analytical approximation in the $\beta U_1 \to \infty$ limit}\label{app:sec:se_symmetric_details:atomic_se:approximation_low_t}

The expression in \cref{app:eqn:at_self_energy_diagonal} is not particularly illuminating. To obtain an analytical formula for the dynamical part of the atomic self-energy $\tilde{\Sigma}^{\text{At}} \left( i \omega_n \right)$, we can consider the $\beta U_1 \to \infty$ limit ({\it i.e.}{}, the low-temperature limit)~\cite{HUB64,FUJ03}. Starting from \cref{app:eqn:atomic_occ_probabilities}, we determine the ratio between the probabilities of the system having $m_1$ and $m_2$ particles (with $m_1 \neq m_2$) to be 
\begin{equation}
	\label{app:eqn:at_ratio_of_prob}
	\frac{p_{m_1}}{p_{m_2}} = \frac{m_2! \left(N_f - m_2 \right)!}{m_1 ! \left( N_f - m_1 \right)!} \exp \left\lbrace \beta \left( m_2 - m_1 \right) \left[ \frac{U_1}{2} \left( m_1 + m_2 - N_f \right) - \mu \right] \right\rbrace.
\end{equation}
For generic values of the chemical potential, $\frac{U_1}{2} \left( m_1 + m_2 - N_f \right) - \mu$ will never be zero, meaning that one of the probabilities $p_m$ will be exponentially (in $\beta U_1$) larger than all the other ones, which implies through the first equality of \cref{app:eqn:f_electron_prob_moments} that $p_m \approx 1$, and through the second equality of \cref{app:eqn:f_electron_prob_moments} that the system will be at some integer filling ({\it i.e.}{}, $n_{\text{At}} N_f = m \in \mathbb{N}$). For the system to be at a generic \emph{fractional} filling, at least two of the probabilities $p_m$ should be non-vanishing and, in this sense, have comparable values (as we show later, \emph{exactly} two probabilities have comparable values). Assume without loss of generality that $0 \leq m_1 < m_2 \leq N_f$ are the occupation numbers corresponding to the largest two probabilities $p_{m_1}$ and $p_{m_2}$ and that $p_{m_1}$ and $p_{m_2}$ are both non-vanishing ($p_{m_1} \sim p_{m_2}$). From \cref{app:eqn:at_ratio_of_prob}, in the limit $\beta U_1 \to \infty$, the only way for $p_{m_1} \sim p_{m_2}$ to be true is if the exponent in \cref{app:eqn:at_ratio_of_prob} is close to zero, which only happens for
\begin{equation}
	\label{app:eqn:chem_pot_beta_infty}
	\mu = \frac{U_1}{2} \left(m_1 + m_2 - N_f \right) + \mathcal{O} \left( \frac{1}{\beta} \right).
\end{equation}
Consider now $m' \neq m_1, m_2$. By plugging \cref{app:eqn:chem_pot_beta_infty} into \cref{app:eqn:at_ratio_of_prob}, we find that 
\begin{equation}
	\label{app:eqn:at_ratio_of_prob_third}
	\frac{p_{m'}}{p_{m_i}} = \frac{m_i! \left(N_f - m_i \right)!}{m' ! \left( N_f - m' \right)!} \exp  \left[ \frac{\beta U_1}{2} \left( m' - m_{\bar{i}} \right) \left( m_i - m' \right)  \right],
\end{equation}
where $\bar{i}=2,1$ for $i=1,2$. The only way $p_{m'} \sim p_{m_1}, p_{m_2}$ in the limit $\beta U_1 \to \infty$ is if the exponent in \cref{app:eqn:at_ratio_of_prob_third} is zero, which it cannot be since $m' \neq m_1, m_2$ by assumption. It follows that at fractional filling in the $\beta U_1 \to \infty$ limit only two occupation numbers have non-vanishing probabilities of comparable size ({\it i.e.}{}, $m_1$ and $m_2$). Moreover, if we further assume that $m_1 < m' < m_2$, then \cref{app:eqn:at_ratio_of_prob_third} implies that 
\begin{equation}
	\frac{p_{m'}}{p_{m_2}} = \frac{m_2! \left(N_f - m_2 \right)!}{m' ! \left( N_f - m' \right)!} \exp  \left[ \frac{\beta U_1}{2} \left( m' - m_1 \right) \left( m_2 - m' \right)  \right] \gg 1,
\end{equation}
which contradicts our initial assumption that $p_{m_1}$ and $p_{m_2}$ are the largest two probabilities in the distribution. As such, we further conclude that in the fractional filling case and for the $\beta U_1 \to \infty$ limit, only two occupation probabilities are non-vanishing: $p_{m}$ and $p_{m+1}$ for some integer $0 \leq m < N_f$.  

In the limit $\beta U_1 \to \infty$, we will take the relative filling of the system to be $\frac{r}{N_f} < n_{\text{At}}  < \frac{r+1}{N_f}$ (where $r \in \mathbb{Z}$ and $0 \leq r < N_f$). As discussed in the above paragraph, only two occupation numbers ($r$ and $r+1$) will have non-vanishing probabilities. Using the second equality of \cref{app:eqn:f_electron_prob_moments}, as well as \cref{app:eqn:chem_pot_beta_infty}, we can obtain
\begin{equation}
	\label{app:eqn:solution_probabilities_atomic_generic}
	p_{m} \approx \left. \begin{cases}
		1-\left(n_{\text{At}} N_f - r \right), & \qq{if} m=r \\
		n_{\text{At}} N_f -r, & \qq{if} m=r+1 \\
		0, & \qq{otherwise}
	\end{cases} \right| \qq{and}
	\mu \approx U_1 \left( r - \frac{N_f}{2} \right) + \frac{U_1}{2}.
\end{equation}
Within this approximation, the diagonal part of the atomic Green's function will be given by only three terms
\begin{align}
	\tilde{\mathcal{G}}^{\text{At}} \left( i \omega_n \right) \approx& \frac{1}{N_f} \left[ 
	\frac{ \left(N_f - r \right) p_r + \left(r+1 \right) p_{r+1} }{i \omega_n + \mu - U_1 \left(r - \frac{N_f}{2} \right) - \frac{U_1}{2}} 
	+\frac{ r p_{r} }{i \omega_n + \mu - U_1 \left(r -1 - \frac{N_f}{2} \right) - \frac{U_1}{2}} \right. \nonumber \\
	&\left. +\frac{ \left(N_f - r - 1 \right) p_{r+1} }{i \omega_n + \mu - U_1 \left(r + 1 - \frac{N_f}{2} \right) - \frac{U_1}{2}} 
	\right] \nonumber \\
	\approx& \frac{1}{N_f} \left[ 
	\frac{ \left(N_f - r \right) p_r + \left(r+1 \right) p_{r+1} }{i \omega_n} 
	+\frac{ r p_{r} }{i \omega_n + U_1} + \frac{ \left(N_f - r - 1 \right) p_{r+1} }{i \omega_n - U_1} 
	\right].
\end{align}
The diagonal part of the atomic self-energy can then be computed from \cref{app:eqn:at_self_energy_diagonal}. In the general $N_f$ case, the formula is more algebraically tedious, but for $N_f = 8$, the atomic self-energy is given by 
{\small
	\begin{equation}
		\label{app:eqn:atomic_self_energy_low_temp_integer_dop}
		\tilde{\Sigma}^{\text{At}} \left( i \omega_n \right) \approx U_1^2 \frac{i\omega_n \left[ 196 n_{\text{At}}^2-4 n_{\text{At}} (12 r+7)+3 r (r+1) \right]-U_1 (7 n_{\text{At}}-r) \left[4 n_{\text{At}} (2r-7)-(r-4) (r+1) \right]}{U_1^2 \left[4 n_{\text{At}} (2 r-7)-(r-4) (r+1) \right]+4 U_1 i\omega_n  (r-7 n_{\text{At}})-4 \left( i\omega_n \right)^2}.
	\end{equation}}In the limit when $n_{\text{At}} N_f$ is close to integer filling, but otherwise $\frac{r}{N_f} < n_{\text{At}}  < \frac{r+1}{N_f}$, the atomic self-energy can be simplified further by expanding in the small deviations from the integer filling
\begin{equation}
	\label{app:eqn:atomic_self_energy_around_integer_filling}
	\tilde{\Sigma}^{\text{At}} \left( i \omega_n \right) \approx \begin{cases}
		\frac{1}{N_f} \frac{U_1^2 r \left( N_f - r \right)}{i \omega_n N_f- U_1 \left(r - N_f \right) } + \mathcal{O} \left[ U_1 \left( n_{\text{At}} N_f - r \right) \right], & \qq{if} n_{\text{At}} \approx \frac{r}{N_f}, n_{\text{At}} > \frac{r}{N_f} \\
		\frac{1}{N_f} \frac{U_1^2 \left( r +1  \right) \left( N_f - r - 1 \right)}{i \omega_n N_f- U_1 \left(r + 1 \right) } + \mathcal{O} \left[ U_1 \left( n_{\text{At}} N_f - r - 1 \right) \right], & \qq{if} n_{\text{At}} \approx \frac{r + 1}{N_f}, n_{\text{At}} < \frac{r + 1}{N_f} \\
	\end{cases}.
\end{equation}
Finally, we note that the correlator between the $f$-electron filling of different orbital, spin, and valley flavors can also be computed analytically by inserting the solution for $p_m$ from \cref{app:eqn:solution_probabilities_atomic_generic} into \cref{app:eqn:f_electron_prob_moments} and is given by
\begin{equation}
	\label{app:eqn:nn_in_at_limit}
	\left\langle nn \right\rangle^{\text{At}} \approx \frac{r \left( 2 n_{\text{At}} N_f - r -1 \right)}{N_f \left(N_f - 1 \right)}.
\end{equation}

Another interesting case in the $\beta U_1 \to \infty$ limit is the \emph{exact} integer filling $n_{\text{At}} = \frac{r}{N_f}$, where $r \in \mathbb{Z}$ and $0 < r < N_f$\footnote{The limiting cases $n_{\text{At}}=0$ and $n_{\text{At}}=1$ correspond to $\mu = -\infty$ or $\mu = + \infty$ and will not be discussed further.}. Based on our previous analysis, when $0 \neq \abs{n_{\text{At}}-\frac{r}{N_f}} \ll 1$ and $n_{\text{At}} \gtrsim \frac{r}{N_f}$ ($n_{\text{At}} \lesssim \frac{r}{N_f}$), $p_r$ and $p_{r+1}$ ($p_{r-1}$) will be nonvanishing with 
\begin{equation}
	\mu \approx \begin{cases}
		U_1 \left( r - \frac{N_f}{2} \right) + \frac{U_1}{2} & \qq{for} n_{\text{At}} \gtrsim \frac{r}{N_f} \\
		U_1 \left( r - 1 - \frac{N_f}{2} \right) + \frac{U_1}{2} & \qq{for} n_{\text{At}} \lesssim \frac{r}{N_f}
	\end{cases}.
\end{equation}
As such, when $n_{\text{At}} = \frac{r}{N_f}$ exactly, we expect that $	U_1 \left( r - 1 - \frac{N_f}{2} \right) + \frac{U_1}{2} \lesssim \mu \lesssim  U_1 \left( r - \frac{N_f}{2} \right) + \frac{U_1}{2} $, which implies through \cref{app:eqn:at_ratio_of_prob} that the largest probabilities will be $p_{r-1}$, $p_{r}$ and $p_{r+1}$ (with the other ones being at least $\mathcal{O} \left( e^{-\beta U_1} \right)$ smaller). Neglecting all the other occupation probabilities, we can obtain from the first two equalities of \cref{app:eqn:f_electron_prob_moments} 
\begin{align}
	p_{r-1} + p_r + p_{r+1} &= 1, \\
	(r-1) p_{r-1} + r p_r + (r+1) p_{r+1} &= r,
\end{align}
which imply that 
\begin{equation}
	\label{app:eqn:equality_of_prob_integer_filling}
	p_{r-1} = p_{r+1}.
\end{equation}
From \cref{app:eqn:at_ratio_of_prob}, \cref{app:eqn:equality_of_prob_integer_filling} can only be realized in the $\beta U_1 \to \infty$ limit whenever 
\begin{equation}
	\mu \approx U_1 \left( r - \frac{N_f}{2} \right), \qq{which immediately implies that}
	p_{m} \approx \begin{cases}
		1, & \qq{if} m=r \\
		0, & \qq{otherwise}
	\end{cases}.
\end{equation}
Using \cref{app:eqn:at_self_energy_pre_def,app:eqn:atomic_gf_diagonal_part_with_prob}, we find that exactly at the integer filling, the dynamical self-energy of the atomic problem is given by
\begin{equation}
	\label{app:eqn:atomic_self_energy_low_temp_integer}
	\tilde{\Sigma}^{\text{At}} \left( i \omega_n \right) \approx \frac{1}{N_f}\frac{ U_1^2 r\left(N_f-r \right)}{ i \omega_n N_f - U_1 (r - \frac{N_f}{2})}, \qq{for} n_{\text{At}} = \frac{r}{N_f},
\end{equation}
and has a simple pole\footnote{Strictly speaking, the exact interacting Green's function from \cref{app:eqn:atomic_gf_diagonal_part_with_prob} has exactly $N_f$ poles with $\left( N_f - 1 \right)$ zeroes in between. As a result, the corresponding exact dynamical self-energy from \cref{app:eqn:at_self_energy_diagonal} will have exactly $\left(N_f - 1 \right)$ poles. In \cref{app:eqn:atomic_self_energy_low_temp_integer} the strengths of all the other poles are vanishingly small, allowing one to approximate the dynamical self-energy by a simple-pole expression.} for $i \omega_n = \frac{U_1 (r - \frac{N_f}{2})}{N_f}$.

Finally, we note that the only approximation in \cref{app:eqn:atomic_self_energy_low_temp_integer,app:eqn:atomic_self_energy_low_temp_integer_dop} is $\beta U_1 \to \infty$, meaning that the corresponding expressions become \emph{exact} in the limit of zero temperature.

\subsubsection{Analytical approximation in the $U_1 \to 0$ limit}\label{app:sec:se_symmetric_details:atomic_se:approximation_low_u}

Another limit in which the expression of the atomic self-energy simplifies is the $U_1 \to 0$ limit. In this limit we will work at fixed chemical potential $\mu$ and determine $n_{\text{At}}$ and $\left\langle nn \right\rangle^{\text{At}}$ through \cref{app:eqn:f_electron_prob_moments}. To obtain $\tilde{\Sigma}^{\text{At}} \left( i \omega_n \right)$ in this limit, we first expand \cref{app:eqn:atomic_gf_diagonal_part_with_prob} to second order in $U_1$. Letting $z=i \omega_n + \mu$, we find that 
{\small
\begin{align}
	\tilde{\mathcal{G}}^{\text{At}} \left( i \omega_n \right) &= \frac{1}{N_f} \sum_{m=0}^{N_f-1} \left[ \left(N_f  - m\right) p_m + \left( m + 1 \right) p_{m+1} \right] \left[ \frac{1}{z} + \frac{1+2m-N_f}{2z^2} U_1 + \frac{\left(1+2m-N_f \right)^2}{4 z^3} U_1^2 + \mathcal{O} \left( U_1^3 \right)\right] \nonumber \\
	&= \frac{1}{N_f} \sum_{m=0}^{N_f} \left(N_f  - m\right) p_m \left[ \frac{1}{z} + \frac{1+2m-N_f}{2z^2} U_1 + \frac{\left(1+2m-N_f \right)^2}{4 z^3} U_1^2 + \mathcal{O} \left( U_1^3 \right)\right] \nonumber \\
	&+ \frac{1}{N_f} \sum_{m=0}^{N_f}   m  p_m  \left[ \frac{1}{z} + \frac{2m-1-N_f}{2z^2} U_1 + \frac{\left(2m-1-N_f \right)^2}{4 z^3} U_1^2 + \mathcal{O} \left( U_1^3 \right)\right] \nonumber \\
	&=\frac{ \left(N_f - 2 \right) U_1^2}{z^3} \frac{1}{N_f} \sum_{m=0}^{N_f} p_m m^2 - \frac{\left[ \left( N_f - 2 \right) N_f U_1 + z - N_f z \right] U_1}{z^3} \frac{1}{N_f} \sum_{m=0}^{N_f} p_m m \nonumber \\
	&+\frac{ N_f \left[ \left(N_f -1 \right)^2 U_1^2 - 2 \left( N_f -1 \right) U_1 z + 4 z^2 \right]}{4 z^3} \frac{1}{N_f} \sum_{m=0}^{N_f} p_m + \mathcal{O} \left( U_1^3 \right)\nonumber \\
	&=\frac{4z^2 - \left(N_f - 1 \right) U_1^2 \left[ 4 n_{\text{At}} \left( N_f-2 \right)- 4 N_f \left\langle nn \right\rangle^{\text{At}} - N_f + 8 \left\langle nn \right\rangle^{\text{At}} + 1 \right] + \left( 4 n_{\text{At}} - 2 \right) \left(N_f - 1\right) U_1 z}{4 z^3} + \mathcal{O} \left( U_1^3\right).
\end{align}}Using \cref{app:eqn:at_self_energy_diagonal}, we then obtain
\begin{equation}
	\label{app:eqn:low_u_asymptote_of_sigma_at}
	\tilde{\Sigma}^{\text{At}} \left( i \omega_n \right) = \frac{\left(N_f - 1 \right) U_1^2 \left[n_{\text{At}} + \left( N_f-2 \right) \left\langle nn \right\rangle^{\text{At}} -n_{\text{At}}^2 \left(N_f - 1 \right) \right]}{i\omega_n + \mu} + \mathcal{O} \left(U_1^3 \right).
\end{equation}
We will employ \cref{app:eqn:at_self_energy_diagonal} in \cref{app:sec:se_symmetric_details:IPT:low_u_limit} to verify that the interpolated dynamical self-energy introduced in \cref{app:eqn:interpolated_sigma} correctly reproduces the dynamical self-energy in the low interaction strength limit, up to second order in $U_1$.

\subsubsection{The single-site model in the atomic limit}\label{app:sec:se_symmetric_details:atomic_se:atomic_limit}
By definition, in the atomic limit, the $f$-electrons at a given site are decoupled from all the other fermionic species of the problem. As such, the latter cavity degrees of freedom can be trivially integrated out resulting in the following single-site action 
\begin{align}
	\label{app:eqn:single_site_atomic_limit}
	S^{\text{At}}_{\text{ss}} &= \int_{0}^{\beta} \dd{\tau} \left[ \sum_{\alpha, \eta, s} \hat{f}^\dagger_{\alpha, \eta, s} \left( \tau \right) \left( \partial_\tau - \mu \right) \hat{f}_{\alpha, \eta, s} \left( \tau \right) + \frac{U_1}{2}  \sum_{\substack{\alpha, \eta, s \\ \alpha', \eta', s'}} :\mathrel{\hat{f}^\dagger_{\alpha, \eta, s} \left( \tau \right) \hat{f}_{\alpha, \eta, s} \left( \tau \right)}: :\mathrel{\hat{f}^\dagger_{\alpha', \eta', s'}\left( \tau \right)  \hat{f}_{\alpha', \eta', s'} \left( \tau \right)}: \right] \nonumber \\
	&= \int_{0}^{\beta} \dd{\tau} \left[ \sum_{\alpha, \eta, s} \hat{f}^\dagger_{\alpha, \eta, s} \left( \tau \right) \left( \partial_\tau - \mu \right) \hat{f}_{\alpha, \eta, s} \left( \tau \right) + H^{\text{At}} \left( \tau \right) \right].
\end{align}
In the atomic limit, one can therefore identify 
\begin{equation}
	\label{app:eqn:single_site_action_identification_at_limit}
	G_0 \left( i \omega_n \right) \to \tilde{\mathcal{G}}^{\text{At},0} \left( i \omega_n \right), \quad
	G \left( i \omega_n \right) \to \tilde{\mathcal{G}}^{\text{At}}  \left( i \omega_n \right), \quad
	\Sigma_{\text{ss}} \left( i \omega_n \right) \to \tilde{\Sigma}^{\text{At}} \left( i \omega_n \right),
\end{equation}
which also exactly satisfy the DMFT self-consistency conditions from \cref{app:eqn:DMFT_self_consistent_eq,app:eqn:dmft_ass_1}.

\subsection{Exact properties of the dynamical self-energy of the single-site THF model}\label{app:sec:se_symmetric_details:se_exact}

In this section, we derive two \emph{exact} properties of the dynamical part of the self-energy of the single-site THF model introduced in \cref{app:sec:se_symmetric:DMFT_overview}. We will employ these two properties to show that the \emph{interpolated} dynamical self-energy $\tilde{\Sigma}^{f,\text{Int}} \left( i \omega_n \right)$ postulated in \cref{app:eqn:interpolated_sigma} and further elaborated upon in \cref{app:sec:se_symmetric_details:IPT} reproduces the correct asymptotic form of the exact dynamical self-energy $\tilde{\Sigma}^{f} \left( i \omega_n \right)$ as $i \omega_n \to \infty$. After briefly summarizing the results of this section, we will cast the single-site problem from the Lagrangian formulation of \cref{app:eqn:single_site_action} to an equivalent Hamiltonian description. The latter will be employed to prove the exact properties of the dynamical self-energy of the single-site model.

\subsubsection{Results}\label{app:sec:se_symmetric_details:se_exact:results}

Given the technical nature of this section, we begin by presenting the main results which will be derived therein. The $\omega \to \infty$ asymptotic form of the \emph{exact} dynamical self-energy of the single-site model defined in \cref{app:eqn:all_self_energy_of_ss_model} will be shown to be given by 
\begin{equation}
	\label{app:eqn:exact_se_asymptote}
	\Sigma_{\text{ss}} \left( \omega + i 0^{+} \right) = \frac{U_1^2 \left(N_f - 1 \right) \left[ n_{\text{ss}} + \left(N_f - 2\right) \left\langle nn \right\rangle^{\text{ss}}  - \left(N_f -1 \right) n_{\text{ss}}^2 \right]}{\omega} + \mathcal{O} \left( \frac{1}{\omega^2} \right),
\end{equation}
where $n_{\text{ss}}$ and $\left\langle nn \right\rangle^{\text{ss}}$ are the relative $f$-electron occupation and the correlator between the occupation of different $f$-electron flavors in the single-site model. As $\Sigma_{\text{ss}} \left( \omega + i 0^{+} \right)$ is the \emph{exact}, rather than the second-order dynamical self-energy, it generically contains terms which are proportional to higher powers of $U_1$, but its leading $\omega \to \infty$ term is proportional to $U^2_1$. The two quantities $n_{\text{ss}}$ and $\left\langle nn \right\rangle^{\text{ss}}$ were introduced in \cref{app:eqn:ss_relative_f_interacting,app:eqn:nn_correlator_from_self_energy}, respectively, and also defined in \cref{app:eqn:rel_fil_of_f,app:eqn:nn_of_f} for the atomic limit. In this section, we also show that for the single-site model in the general case, the $\left\langle nn \right\rangle^{\text{ss}}$ correlator can be related back to the single-site interacting Green's function via the second of the following equations 
{\small
	\begin{align}
		n_{\text{ss}} &\equiv \left\langle \hat{f}^\dagger_{\alpha, \eta, s} \hat{f}_{\alpha, \eta, s} \right\rangle^{\text{ss}} = \int_{-\infty}^{\infty} \dd{\omega} n_{\mathrm{F}} \left( \omega \right) \rho \left( \omega \right), \label{app:eqn:n_occ_from_dmft_int_gf} \\
		\left\langle nn \right\rangle^{\text{ss}} &\equiv \left\langle \hat{f}^\dagger_{\alpha, \eta, s} \hat{f}_{\alpha, \eta, s} \hat{f}^\dagger_{\alpha', \eta', s'} \hat{f}_{\alpha', \eta', s'}  \right\rangle^{\text{ss}} = n_{\text{ss}}^2 - \frac{1}{\pi U_1 \left(N_f - 1 \right)} \int_{-\infty}^{\infty} \dd{\omega} \Im{\Sigma_{\text{ss}} \left( \omega + i 0^{+} \right) G \left( \omega + i 0^{+} \right) } n_{\mathrm{F}} \left( \omega \right), \label{app:eqn:nn_correlator_from_self_energy_ss}	
	\end{align}}for any $\left(\alpha, \eta, s \right) \neq \left( \alpha', \eta', s' \right)$. In \cref{app:eqn:nn_correlator_from_self_energy_ss}, we have used the interacting spectral function for the single-site model defined in \cref{app:eqn:causal_g} and repeated here for convenience,
\begin{equation*}
	\rho \left( \omega \right) \equiv -\frac{1}{\pi} \Im G \left( \omega + i 0^{+} \right), \qq{such that}
	G \left( z \right) = \int_{-\infty}^{\infty} \frac{\dd{\omega}}{z-\omega} \rho \left( \omega \right),
\end{equation*}
where the interacting Green's function of the single-site model $G \left( i \omega_n \right)$ was defined in \cref{app:eqn:definition_ss_interacting_gf}. As we show explicitly in \cref{app:sec:se_symmetric_details:se_exact:infinite_omega}, requiring that $\Sigma_{\text{ss}} \left( \omega + i 0^{+} \right)$ has the exact $\omega \to \infty$ asymptotic form, as imposed by \cref{app:eqn:exact_se_asymptote,app:eqn:nn_correlator_from_self_energy_ss} is equivalent to requiring that the first three moments of the interacting spectral function of the single-site model [which will be defined in \cref{app:eqn:def_of_spectral_function_moments}] are exactly reproduced~\cite{POT97}. 

\subsubsection{The single-site problem in the Hamiltonian formulation}\label{app:sec:se_symmetric_details:se_exact:single_site_ham}

As explained in \cref{app:sec:se_symmetric:DMFT_overview:ss_problem}, within DMFT~\cite{MET89,GEO96}, the dynamical part of the lattice self-energy $\tilde{\Sigma}^{f} \left( i \omega_n \right)$ is site-diagonal and identical to $\Sigma_{\text{ss}} \left(i \omega_n \right)$, the dynamical self-energy computed from the single-site action of \cref{app:eqn:single_site_action} and repeated here for convenience,
\begin{align}
	S_{\text{ss}} &= - \int_{0}^{\beta} \dd{\tau} \int_{0}^{\beta} \dd{\tau'} \sum_{\alpha, \eta, s} \hat{f}^\dagger_{\alpha, \eta, s} \left( \tau \right) G^{-1}_0 \left( \tau - \tau' \right) \hat{f}_{\alpha, \eta, s} \left( \tau' \right) \nonumber \\
	&+\int_{0}^{\beta} \dd{\tau} \frac{U_1}{2}  \sum_{\substack{\alpha, \eta, s \\ \alpha', \eta', s'}} :\mathrel{\hat{f}^\dagger_{\alpha, \eta, s} \left( \tau \right) \hat{f}_{\alpha, \eta, s} \left( \tau \right)}: :\mathrel{\hat{f}^\dagger_{\alpha', \eta', s'}\left( \tau \right)  \hat{f}_{\alpha', \eta', s'} \left( \tau \right)}:. \nonumber
\end{align}
$S_{\text{ss}}$ was obtained from the Hamiltonian of \cref{app:eqn:DMFT_partial_mf_all} by integrating out all the hybridization and mean-field interaction terms between the eight $f$-fermions at a given site and the rest of the electrons (including all the $c$- and/or $d$-electrons), and dropping the higher-order terms (which would give a contribution to the single-site action beyond the fermion bilinear term). The Green's function $G_{0} \left( \tau \right)$ appearing in \cref{app:eqn:single_site_action} is thus \emph{not} the non-interacting $f$-electron Green's function of the \emph{lattice} problem, but rather an \emph{effective} Green's function that contains information about all the other electronic degrees of freedom that were integrated out~\cite{GEO92}.

For the purpose of deriving \cref{app:eqn:exact_se_asymptote,app:eqn:nn_correlator_from_self_energy_ss}, we will find it easier to work within the Hamiltonian, rather than the Lagrangian formulation of the single-site problem. Note that the former is not trivial to obtain, since the single-particle term of the action takes the retarded form 
\begin{equation} 
	- \int_{0}^{\beta} \dd{\tau} \int_{0}^{\beta} \dd{\tau'} \sum_{\alpha, \eta, s} \hat{f}^\dagger_{\alpha, \eta, s} \left( \tau \right) G^{-1}_0 \left( \tau - \tau' \right) \hat{f}_{\alpha, \eta, s} \left( \tau' \right),
\end{equation} 
which differs from the canonical form $- \int_{0}^{\beta} \dd{\tau} \sum_{\alpha, \eta, s} \hat{f}^\dagger_{\alpha, \eta, s} \left( \tau \right) (\partial_\tau + \epsilon_f)\hat{f}_{\alpha, \eta, s} \left( \tau \right)$ of a non-interacting fermion.  

To obtain the Hamiltonian of the single-site problem, we first define the $\Delta\left( i\omega_n \right)$ hybridization field 
\begin{equation} 
	\label{app:eqn:def_of_hyb_field}
	\Delta\left( i\omega_n \right) \equiv i\omega_n + \mu - G_0^{-1} \left( i\omega_n \right).
\end{equation} 
In terms of $\Delta\left( i\omega_n \right)$, the action of the system can then be rewritten as
\begin{align} 
	S_{\text{ss}} &= \sum_{i\omega_n}\sum_{\alpha, \eta, s} \hat{f}^\dagger_{\alpha, \eta, s} \left( i\omega_n\right) \left(-i\omega_n  - \mu  + \Delta \left(i\omega_n \right) \right) \hat{f}_{\alpha, \eta, s} \left( i\omega_n \right) \nonumber \\
	&+\int_{0}^{\beta} \dd{\tau} \frac{U_1}{2}  \sum_{\substack{\alpha, \eta, s \\ \alpha', \eta', s'}} :\mathrel{\hat{f}^\dagger_{\alpha, \eta, s} \left( \tau \right) \hat{f}_{\alpha, \eta, s} \left( \tau \right)}: :\mathrel{\hat{f}^\dagger_{\alpha', \eta', s'}\left( \tau \right)  \hat{f}_{\alpha', \eta', s'} \left( \tau \right)}:, \label{app:eqn:ss_action_with_hyb_func}
\end{align} 
where the single-particle term has been expressed in Matsubara frequency~\cite{ALT10},
\begin{equation}
	\hat{f}_{\alpha, \eta, s} (\tau) = \frac{1}{\sqrt{\beta}} \sum_{i \omega_n} \hat{f}_{\alpha, \eta, s} \left( i \omega_n \right) e^{-i \omega_n \tau}, \quad
	\hat{f}_{\alpha, \eta, s} \left( i \omega_n \right) \equiv \frac{1}{\sqrt{\beta}} \int_{0}^{\beta} \dd{\tau} \hat{f}_{\alpha, \eta, s} (\tau) e^{i \omega_n \tau}. \\
\end{equation}
Up to the presence of the hybridization term $\Delta \left( i \omega_n \right)$, the kinetic term of \cref{app:eqn:ss_action_with_hyb_func} resembles the canonical form of the action of a non-interacting fermion. The $i \omega_n$-dependent hybridization function can be removed by introducing additional fermionic degrees of freedom. This can be shown by employing the spectral representation of $\Delta\left(i\omega_n \right)$,
\begin{equation} 
	\label{app:eqn:def_of_effective_GF_spec_decomp}
	\Delta \left( i\omega_n \right) = \Delta_0 + \int_{-\infty}^{\infty} \dd{\omega} \frac{\rho_{\Delta} \left( \omega \right)}{i\omega_n - \omega},
\end{equation}
where $\Delta_0 = \lim_{z \to \infty} \Delta \left( z \right)$ is a real constant and the spectral function $\rho_{\Delta} \left( \omega \right)$ is non-negative and defined by the following analytical continuation
\begin{align} 
	\label{app:eqn:def_of_effective_GF_spec_approx}
	\rho_\Delta(\omega) =- \frac{1}{\pi}\Delta \left(\omega + i 0^{+} \right).
\end{align} 
The mathematically rigorous proof for the existence of a spectral representation for $\Delta \left( i\omega_n \right)$ can be found in Ref.~\cite{LIN19}, where the analytical properties of $G (z)$ and $G_{0} (z)$ from \cref{app:sec:se_symmetric:DMFT_overview:anal_props} are also proved. In fact, from \cref{app:eqn:DMFT_gf_to_z_infty}, we can show that $\Delta (z) = \Delta_0 + \mathcal{O} \left( \frac{1}{z} \right)$, with $\Delta_0 \in \mathbb{R}$ following from \cref{app:eqn:DMFT_gf_cc_of_argument}. Additionally, from \cref{app:eqn:causal_g0}, it follows that $\Im G_{0} (z) < 0$ ($\Im G_{0} (z) >0$) for $\Im z > 0 $ ($\Im z < 0 $), meaning that $G_0 (z)$ has no zeros above or below the real axis. As a result, from its definition in \cref{app:eqn:def_of_hyb_field}, it follows that $\Delta(z)$ has the same analytical properties as $G_{0} \left( z \right)$ -- it is analytical below and above the real axis and also obeys $\Delta \left( z^{*} \right) = \Delta^{*} \left( z \right)$, from \cref{app:eqn:DMFT_gf_cc_of_argument}. As a result of these properties and according to the lemma stated and proved in \cref{app:sec:se_symmetric_details:se_exact:lemma_spectral_rep}, $\Delta (z)$ is endowed with a spectral representation akin to the one introduced in \cref{app:eqn:def_of_effective_GF_spec_decomp,app:eqn:def_of_effective_GF_spec_approx}.

To move forward, we approximate the integral in \cref{app:eqn:def_of_effective_GF_spec_decomp} by a Riemann sum. Formally, as we show in \cref{app:eqn:approximation_of_delta}, this is equivalent to approximating the spectral function $\rho_{\Delta} \left( \omega \right)$ by a series of Dirac $\delta$-functions
\begin{equation} 
	\label{app:eqn:approximation_of_spectral}
	\rho_{\Delta} \left( \omega \right) \approx \sum_{i=1}^{N_a} V_i^2 \delta \left( \omega - \epsilon_i +\mu \right),
\end{equation} 
where $V_i$ and $\epsilon_i$ are real numbers determined by fitting to $\rho_{\Delta} \left( \omega \right)$. $V_i^2 \delta(\omega - \epsilon_i +\mu )$ corresponds to a peak of density in the spectral function $\rho_{\Delta} \left( \omega \right)$ at $\omega= \epsilon_i -\mu $ with weight $V_i^2$. We reiterate that the purpose of introducing the $V_i$ and $\epsilon_i$ constants is to obtain the Hamiltonian reformulation of the single-site problem, within which certain exact results can be more easily derived. Because the final results already summarized in \cref{app:sec:se_symmetric_details:se_exact:results} do not depend on $V_i$ and $\epsilon_i$ explicitly, we will leave them unspecified. Moreover, the number $N_a$ of $\delta$-functions we use to approximate the spectral function $\rho_{\Delta} \left( \omega \right)$ is also arbitrary -- the larger $N_a$ is, the more accurate the approximation in \cref{app:eqn:approximation_of_spectral} is. 

Combining \cref{app:eqn:def_of_effective_GF_spec_decomp,app:eqn:approximation_of_spectral,app:eqn:def_of_effective_GF_spec_approx}, we can now write $\Delta\left( i\omega_n \right)$ as 
\begin{equation} 
	\label{app:eqn:approximation_of_delta}
	\Delta\left( i\omega_n \right) = \Delta_0 + \int_{-\infty}^{\infty} \frac{ \dd{\omega} }{i\omega_n- \omega} \sum_{i=1}^{N_a} V_i^2 \delta\left( \omega - \epsilon_i +\mu \right) = \Delta_0 + \sum_{i=1}^{N_a} \frac{V_i^2}{i\omega_n- \epsilon_i+\mu},
\end{equation} 
which shows that, indeed \cref{app:eqn:approximation_of_spectral} is equivalent to approximating the integral in \cref{app:eqn:def_of_effective_GF_spec_decomp} by a Riemann sum. Using \cref{app:eqn:approximation_of_delta}, we can rewrite the partition function and the action of the single-site problem as 
\begin{align} 
	Z& = \int \mathcal{D} \left[ \hat{f}_{},\hat{f}^\dagger_{} \right] e^{-S_{\text{ss}}}, \\ 
	S_{\text{ss}} &= \sum_{i\omega_n} \sum_{\alpha, \eta, s} \hat{f}^\dagger_{\alpha, \eta, s} \left( i\omega_n\right) \left(-i\omega_n - \mu 
	+ \Delta_0 + \sum_{i=1}^{N_a}\frac{V_i^2}{i\omega_n-\epsilon_i +\mu} \right)\hat{f}_{\alpha, \eta, s} \left( i\omega_n \right) \nonumber \\
	&+\int_{0}^{\beta} \dd{\tau} \frac{U_1}{2}  \sum_{\substack{\alpha, \eta, s \\ \alpha', \eta', s'}} :\mathrel{\hat{f}^\dagger_{\alpha, \eta, s} \left( \tau \right) \hat{f}_{\alpha, \eta, s} \left( \tau \right)}: :\mathrel{\hat{f}^\dagger_{\alpha', \eta', s'}\left( \tau \right)  \hat{f}_{\alpha', \eta', s'} \left( \tau \right)}:. 
\end{align} 
Next, we employ the Gaussian integral of the fermionic field to obtain
{\small
	\begin{align} 
		&\exp\left( -{ \sum_{i\omega_n}\sum_{\alpha, \eta, s}\hat{f}^\dagger_{\alpha, \eta, s} \left( i\omega_n\right) \sum_{ i=1 }^{N_a}\frac{V_i^2}{i\omega_n-\epsilon_i + \mu }\hat{f}_{\alpha, \eta, s} \left( i\omega_n \right) } \right) \nonumber\\ 
		=&\int \mathcal{D} \left[ \hat{a}_{},\hat{a}^\dagger_{} \right] \exp\left\{ \sum_{\substack{\alpha, \eta, s \\ i\omega_n}} \sum_{i=1}^{N_a}\left[ \hat{a}^\dagger_{i,\alpha, \eta, s}  \left( i\omega_n \right) \left( i\omega_n - \epsilon_i +\mu \right) \hat{a}_{i,\alpha, \eta, s} \left( i\omega_n \right) - \left( V_i \hat{a}^\dagger_{i,\alpha, \eta, s} \left( i\omega_n \right) \hat{f}_{\alpha, \eta, s} \left( i\omega_n \right) + \text{h.c.} \right) \right]
		\right\},
\end{align}}where we have introduced the auxiliary fermionic fields $\hat{a}^\dagger_{i,\alpha, \eta, s}$ ($1 \leq i \leq N_a$)\footnote{We ignore multiplicative factors appearing in the expression of the partition function. Note also the additional minus sign appearing in the Gaussian integral over Grassman variables.}. Now we can rewrite the partition function as follows
\begin{align} 
	Z& = \int \mathcal{D} \left[ \hat{f},\hat{f}^\dag,\hat{a},\hat{a}^\dag \right] e^{-S_{\text{ss}}^{\hat{f},\hat{a}}} \label{app:eqn:partition_with_c_electrons} \\ 
S^{\hat{f},\hat{a}}_{\text{ss}} &= \sum_{i\omega_n} \sum_{\alpha, \eta, s} \hat{f}^\dagger_{\alpha, \eta, s} \left( i\omega_n\right) \left(-i\omega_n - \mu + \Delta_{0} \right) \hat{f}_{\alpha, \eta, s} \left( i\omega_n \right) \nonumber \\
&+ \sum_{i\omega_n}\sum_{\alpha, \eta, s} \sum_{i=1}^{N_a} \hat{a}^\dagger_{i,\alpha,\eta,s}\left( i\omega_n \right)  \left(-i\omega_n - \mu + \epsilon_i \right)\hat{a}_{i,\alpha,\eta,s}\left( i\omega_n \right) \nonumber \\
&+ \sum_{i\omega_n} \sum_{\alpha, \eta, s} \sum_{i=1}^{N_a}\left(V_i \hat{a}_{i,\alpha,\eta, s}^\dag(i\omega)\hat{f}_{\alpha, \eta, s}\left( i\omega_n \right)+ \text{h.c.} \right)\nonumber\\ 
& +\int_{0}^{\beta} \dd{\tau} \frac{U_1}{2}  \sum_{\substack{\alpha, \eta, s \\ \alpha', \eta', s'}} :\mathrel{\hat{f}^\dagger_{\alpha, \eta, s} \left( \tau \right) \hat{f}_{\alpha, \eta, s} \left( \tau \right)}: :\mathrel{\hat{f}^\dagger_{\alpha', \eta', s'}\left( \tau \right)  \hat{f}_{\alpha', \eta', s'} \left( \tau \right)}: \label{app:action_single_site_in_all_freq}\\ 
&= \sum_{i\omega_n}\sum_{\alpha, \eta, s} \left[ \hat{f}^\dagger_{\alpha, \eta, s} \left( i\omega_n\right) \left(-i\omega_n \right)\hat{f}_{\alpha, \eta, s} \left( i\omega_n \right) + \sum_{i=1}^{N_a} \hat{a}^\dagger_{i,\alpha, \eta, s} \left( i\omega_n \right) \left(-i\omega_n\right) \hat{a}_{i,\alpha,\eta,s} \left( i\omega_n \right) \right] \nonumber \\
	&+\int_0^\beta \dd{\tau} H_{\text{ss}} (\tau),  \label{app:action_single_site_with_c}
\end{align} 
where, in the last equality, we have introduced the Hamiltonian corresponding to the single-site problem\footnote{Although the $H_{\text{ss}}$ features additional $a$-fermions, we will still refer to the corresponding system as the single-site problem.} evaluated at imaginary time $\tau$, where
\begin{align} 
	H_{\text{ss}} &\equiv \sum_{\alpha, \eta, s}  \left( \Delta_{0} - \mu \right) \hat{f}^\dagger_{\alpha, \eta, s} \hat{f}_{\alpha, \eta, s} + \sum_{i=1}^{N_a} \sum_{\alpha, \eta, s}\left( \epsilon_i - \mu \right) \hat{a}^\dagger_{i,\alpha, \eta, s}\hat{a}_{i,\alpha, \eta, s} 
	+ \sum_{\alpha, \eta, s} \sum_{i=1}^{N_a} V_i \left( \hat{a}^\dagger_{i,\alpha, \eta, s} \hat{f}_{\alpha, \eta, s} + \hat{f}^\dagger_{\alpha, \eta, s} \hat{a}_{i,\alpha, \eta, s} \right) \nonumber \\
	&+ \frac{U_1}{2} \sum_{\substack{\alpha, \eta, s \\ \alpha', \eta', s'}} :\mathrel{\hat{f}^\dagger_{\alpha, \eta, s} \hat{f}_{\alpha, \eta, s}}: :\mathrel{\hat{f}^\dagger_{\alpha', \eta', s'} \hat{f}_{\alpha', \eta', s'}}:. \label{app:eqn:Hamiltonian_single_site_equivalent}
\end{align} 
Using the Hamiltonian formulation from \cref{app:eqn:Hamiltonian_single_site_equivalent}, the exact properties of $\Sigma_{\text{ss}} \left(i \omega_n \right)$ summarized in \cref{app:sec:se_symmetric_details:se_exact:results} can be more easily derived, as we will show in \cref{app:sec:se_symmetric_details:se_exact}.

The dynamics of the $f$-electrons within the single-site action $S^{\hat{f},\hat{a}}_{\text{ss}}$ (or, equivalently, the single-site Hamiltonian $H_{\text{ss}}$) is identical to the $f$-electron's dynamics within the original single-site action \emph{without} $\hat{a}$ electrons from \cref{app:eqn:single_site_action}. This is because by reversing the process in \cref{app:eqn:partition_with_c_electrons,app:action_single_site_in_all_freq,app:action_single_site_with_c} and integrating out the auxiliary electrons, one can directly obtain $S_{\text{ss}}$ from $S^{\hat{f},\hat{a}}_{\text{ss}}$. In other words, in the Hamiltonian formulation, the $N_{a} \times N_{f}$ auxiliary fermions have been introduced to ``emulate'' the effective Green's function $G_{0} \left( \tau \right)$. Note also that the $N_a$ auxiliary fermions introduced in the Hamiltonian formulation of the single-site problem render it impossible to solve exactly, due to the prohibitively large Hilbert space dimension. 

\subsubsection{Additional results for the Hamiltonian formulation of the single-site problem}\label{app:sec:se_symmetric_details:se_exact:definitions}

In terms of the effective hybridization function $\Delta\left( i\omega_n \right)$, the interacting $f$-electron Green's function of the single-site model can be written as 
\begin{equation}
	\label{app:eqn:impurity_interacting_gf}
	G \left( i \omega_n \right) = \frac{1}{i \omega_n + \mu - \Delta \left( i \omega_n \right) - U_1 \left(n_{\text{ss}} -\frac{1}{2} \right) \left(N_f - 1 \right) - \Sigma_{\text{ss}} \left( i \omega_n \right)},
\end{equation}
which follows directly from \cref{app:eqn:all_self_energy_of_ss_model} and the definition in \cref{app:eqn:def_of_effective_GF_spec_decomp}. In \cref{app:sec:se_symmetric_details:se_exact:nn}, we will also employ the Green's function corresponding to an $f$-electron propagating into an $a$-electron
\begin{equation}
	\label{app:eqn:af_correlation_matsubara}
	G^{af}_{i} \left( i \omega_n \right) \equiv - \left\langle \hat{a}_{i,\alpha, \eta, s} \left( i \omega_n \right) \hat{f}^\dagger_{\alpha, \eta, s} \left( i \omega_n \right) \right\rangle^{\text{ss}},
\end{equation}
where the expectation value is taken within the action $S^{\hat{f},\hat{a}}_{\text{ss}}$. The easiest way to compute this correlation function is by introducing a source term to the action. The partition function of the problem, taking the source term into account, is given by 
\begin{equation}
	Z\left[ \eta, \eta^{\dagger} \right] \equiv \int \mathcal{D} \left[ \hat{f},\hat{f}^\dag,\hat{a},\hat{a}^\dag \right] e^{-S_{\text{ss}}^{\hat{f},\hat{a}}-S_{\eta}} \label{app:eqn:partition_with_source},
\end{equation}
where the source term reads as
\begin{equation}
	S_{\eta} \equiv \sum_{i \omega_n} \sum_{\alpha, \eta, s} \sum_{i=1}^{N_a} \left( \eta^{\dagger}_{i,\alpha, \eta, s} \left( i \omega_n \right) \hat{a}_{i,\alpha, \eta, s} \left(i \omega_n \right) +  \hat{a}^\dagger_{i,\alpha, \eta, s} \left(i \omega_n \right) \eta_{i,\alpha, \eta, s} \left( i \omega_n \right) \right).
\end{equation}

Using the new action, the correlation function from \cref{app:eqn:af_correlation_matsubara} can be expressed as 
\begin{align}
	G^{af}_{i} \left( i \omega_n \right) &=  \frac{1}{	Z\left[ \eta, \eta^{\dagger} \right]} \int \mathcal{D} \left[ \hat{f},\hat{f}^\dag,\hat{a},\hat{a}^\dag \right] \pdv{\eta^{\dagger}_{i,\alpha, \eta, s} \left( i \omega_n \right)} \hat{f}^\dagger_{\alpha, \eta, s} \left( i\omega_n \right) e^{-S_{\text{ss}}^{\hat{f},\hat{a}}-S_{\eta}} \eval_{\eta,\eta^{\dagger} = 0} \nonumber \\
	&=   \frac{1}{Z} \int \mathcal{D} \left[ \hat{f},\hat{f}^\dag \right] \pdv{\eta^{\dagger}_{i,\alpha, \eta, s} \left( i \omega_n \right)} \hat{f}^\dagger_{\alpha, \eta, s} \left( i\omega_n \right) e^{-S_{\text{ss}}^{\eta}} \eval_{\eta,\eta^{\dagger} = 0}. \label{app:eqn:correlation_with_source}
\end{align}
In the last line of \cref{app:eqn:correlation_with_source}, we have integrated out the $a$-fermions (but \emph{kept} the $\eta$ source fields) to obtain the following action 
\begin{align}
	S_{\text{ss}}^{\eta} &= \sum_{i\omega_n} \sum_{\alpha, \eta, s} \hat{f}^\dagger_{\alpha, \eta, s} \left( i\omega_n\right) \left(-i\omega_n - \mu + \Delta_{0} \right) \hat{f}_{\alpha, \eta, s} \left( i\omega_n \right) \nonumber \\
&+ \sum_{i\omega_n}\sum_{\alpha, \eta, s} \sum_{i=1}^{N_a} \left( \eta^{\dagger}_{i,\alpha, \eta, s} \left( i\omega_n \right) + V_i \hat{f}^\dagger_{\alpha, \eta, s} \left( i\omega_n \right)  \right) \frac{1}{i\omega_n + \mu - \epsilon_i} \left( \eta_{i,\alpha, \eta, s} \left( i\omega_n \right) + V_i \hat{f}_{\alpha, \eta, s} \left( i\omega_n \right)  \right) \nonumber \\
& +\int_{0}^{\beta} \dd{\tau} \frac{U_1}{2}  \sum_{\substack{\alpha, \eta, s \\ \alpha', \eta', s'}} :\mathrel{\hat{f}^\dagger_{\alpha, \eta, s} \left( \tau \right) \hat{f}_{\alpha, \eta, s} \left( \tau \right)}: :\mathrel{\hat{f}^\dagger_{\alpha', \eta', s'}\left( \tau \right)  \hat{f}_{\alpha', \eta', s'} \left( \tau \right)}:. \label{app:eqn:action_with_source} 
\end{align}
Substituting \cref{app:eqn:action_with_source} into \cref{app:eqn:correlation_with_source} and computing the derivative, we can derive the desired correlator in terms of the interacting $f$-electron Green's function of the single-site model
\begin{align}
	G^{af}_{i} \left( i \omega_n \right) &=  \frac{1}{Z} \int \mathcal{D} \left[ \hat{f},\hat{f}^\dag \right] \frac{1}{i\omega_n + \mu - \epsilon_i} \left( \eta_{i,\alpha, \eta, s} \left( i\omega_n \right) + V_i \hat{f}_{\alpha, \eta, s} \left( i\omega_n \right)  \right) \hat{f}^\dagger_{\alpha, \eta, s} \left( i\omega_n \right) e^{-S_{\text{ss}}^{\eta}} \eval_{\eta,\eta^{\dagger} = 0} \nonumber \\ 
	&=  \frac{1}{Z} \int \mathcal{D} \left[ \hat{f},\hat{f}^\dag \right] \frac{V_i}{i\omega_n + \mu - \epsilon_i}  \hat{f}_{\alpha, \eta, s} \left( i\omega_n \right) \hat{f}^\dagger_{\alpha, \eta, s} \left( i\omega_n \right) e^{-S_{\text{ss}}} \nonumber \\
	&=  \frac{V_i G \left( i \omega_n \right)}{i\omega_n + \mu - \epsilon_i}. \label{app:eqn:ss_gf_f_to_a_propagator}
\end{align}

\subsubsection{The $\omega \to \infty$ behavior of $\Sigma_{\text{ss}} \left(\omega + i 0^{+} \right)$}\label{app:sec:se_symmetric_details:se_exact:infinite_omega}

We are now in the position of proving the exact properties of the single-site dynamical self-energy summarized in \cref{app:sec:se_symmetric_details:se_exact:results}. First, we consider the \emph{exact} asymptotic form of the dynamical self-energy of the single-site model at infinite frequency. To do so, we introduce the moments $M_m$ of the interacting spectral function~\cite{POT97}, which are defined according to
\begin{equation}
	\label{app:eqn:def_of_spectral_function_moments}
	M_m = \int_{-\infty}^{\infty} \dd{\omega} \omega^{m} \rho \left( \omega \right), \qq{for} m \in \mathbb{Z}, m \geq 0.
\end{equation}
The spectral function moments can be employed to construct an expansion around infinity of the interacting single-site Green's function using \cref{app:eqn:causal_g}~\cite{POT97}. More precisely,
\begin{equation}
	\label{app:eqn:1_over_z_impurity_gf}
	G \left( z \right) = \int_{-\infty}^{\infty} \frac{\dd{\omega}}{z-\omega} \rho \left( \omega \right) = \int_{-\infty}^{\infty} \frac{\dd{\omega}}{z} \sum_{m=0}^{\infty} \left( \frac{\omega}{z} \right)^{m} \rho \left( \omega \right) = \sum_{m=0}^{\infty} \frac{M_m}{z^{m+1}}.
\end{equation}
Since the dynamical self-energy vanishes at complex infinity ({\it i.e.}{}, $\lim_{z \to \infty} \Sigma_{\text{ss}} \left( z \right) = 0$), as discussed in \cref{app:eqn:asymptote_of_sigma_general}, one can also construct a similar expansion of $ \Sigma_{\text{ss}} \left( z \right)$,
\begin{equation}
	\label{app:eqn:expansion_sigma}
	\Sigma_{\text{ss}} \left( z \right) = \sum_{m=1}^{\infty} \frac{\mathcal{C}_m}{z^{m}},
\end{equation}
where $\mathcal{C}_m$ are complex coefficients. To obtain the leading $\omega \to \infty$ behavior of $\Sigma_{\text{ss}} \left(\omega + i 0^{+} \right)$, we will solely be interested in the $\mathcal{C}_1$ coefficient. Plugging \cref{app:eqn:expansion_sigma,app:eqn:approximation_of_delta} into \cref{app:eqn:impurity_interacting_gf}, performing an expansion in inverse frequency and equating with the series from \cref{app:eqn:1_over_z_impurity_gf}, we can obtain the first three moments of the spectral function
\begin{align}
	M_0 &= 1, \label{app:eqn:mom_0_from_spectral}\\
	M_1 &= U_1 \left(n_{\text{ss}} -\frac{1}{2} \right) \left(N_f - 1 \right) - \mu + \Delta_{0}, \label{app:eqn:mom_1_from_spectral} \\
	M_2 &= M_1^2 + \mathcal{C}_1 + \sum_{i=1}^{N_a} V_i^2, \label{app:eqn:mom_2_from_spectral}
\end{align}
where the zeroth moment $M_0$ also follows from the normalization of the spectral function. The coefficient $\mathcal{C}_{1}$ can thus be obtained by 
\begin{equation}
	\label{app:eqn:obtaining_c_1_self_en}
	\mathcal{C}_1 = M_2 - M_1^2 - \sum_{i=1}^{N_a} V_i^2.
\end{equation}

The spectral function moments also have an \emph{elementary} expression involving expectation values of commutators and anticommutators of the $f$-electrons with the $H_{\text{ss}}$ Hamiltonian~\cite{POT97}. To derive this form, we first consider the Lehmann representation of the fully-interacting $f$-electron spectral function. Letting $\ket{m}$ denote the exact eigenvectors of the grand canonical Hamiltonian $H_{\text{ss}}$ with energy $E_m$, 
\begin{equation}
	H_{\text{ss}} \ket{m} = E_m \ket{m},
\end{equation}
and using the fact that the $f$-electron Green's function is flavor diagonal ({\it i.e.}{}, there is no symmetry-breaking in the system), the interacting spectral function can be expressed as~\cite{MAH00}
\begin{equation}
	\label{app:eqn:lehmann_interacting_sp_single_site}
	\rho \left( \omega \right) = \frac{1}{Z} \sum_{m_1,m_2} \abs{\bra{m_1} \hat{f}_{\alpha, \eta, s} \ket{m_2}}^2 \left[ e^{-\beta E_{m_1}} + e^{-\beta E_{m_2}} \right] \delta \left(\omega + E_{m_1} - E_{m_2}\right),
\end{equation}
where $Z = \sum_{m} e^{-\beta E_m}$ is the partition function of the single-site system (note that we are \emph{not} summing over the $\alpha$, $\eta$, and $s$ indices). \Cref{app:eqn:lehmann_interacting_sp_single_site} is the analogue of \cref{app:eqn:spectral_function} for the single-site model and can be derived in an identical fashion. Defining the operator $\mathcal{L}$ such that
\begin{equation}
	\mathcal{L} \mathcal{\hat{O}} = \left[H_{\text{ss}},\hat{O}\right],
\end{equation}
we now consider the expression 
\begin{align}
	\left\langle \anticommutator{\mathcal{L}^m \hat{f}_{\alpha, \eta, s}}{\hat{f}^\dagger_{\alpha, \eta, s}} \right\rangle^{\text{ss}} =& \frac{1}{Z} \sum_{m_1,m_2} \left( \bra{m_1} \mathcal{L}^m \hat{f}_{\alpha, \eta, s} \ketbra{m_2} \hat{f}^\dagger_{\alpha, \eta, s} \ket{m_1}  +
	\bra{m_1} \hat{f}^\dagger_{\alpha, \eta, s} \ketbra{m_2} \mathcal{L}^m \hat{f}_{\alpha, \eta, s} \ket{m_1} \right) e^{-\beta E_{m_1}} \nonumber \\
	=& \frac{1}{Z} \sum_{m_1,m_2} \left[ \bra{m_1} \hat{f}_{\alpha, \eta, s} \ketbra{m_2} \hat{f}^\dagger_{\alpha, \eta, s} \ket{m_1} \left( E_{m_1} - E_{m_2} \right)^m \right. \nonumber \\
	&\left.+\bra{m_1} \hat{f}^\dagger_{\alpha, \eta, s} \ketbra{m_2} \hat{f}_{\alpha, \eta, s} \ket{m_1} 
	\left( E_{m_2} - E_{m_1} \right)^m \right] e^{-\beta E_{m_1}} \nonumber \\
	=& \frac{1}{Z} \sum_{m_1,m_2} \abs{\bra{m_1} \hat{f}_{\alpha, \eta, s} \ket{m_2}}^2  \left( e^{-\beta E_{m_1}} + e^{-\beta E_{m_2}} \right) \left( E_{m_1} - E_{m_2} \right)^m \nonumber \\
	=& \int_{-\infty}^{\infty} \dd{\omega} \left(-\omega \right)^{m} \rho \left( \omega \right) = (-1)^m M_m, \label{app:spectral_moment_commutator}
\end{align}
where $\anticommutator{\cdot}{\cdot}$ denotes the anticommutator and we have employed the Lehmann representation of the interacting spectral function from \cref{app:eqn:lehmann_interacting_sp_single_site}. This allows us to express the first two nontrivial moments of the spectral function as 
\begin{align}
	M_1 =& - \left\langle \anticommutator{\commutator{H_{\text{ss}}}{\hat{f}_{\alpha, \eta, s}}}{\hat{f}^\dagger_{\alpha, \eta, s}} \right\rangle^{\text{ss}}, \label{app:eqn:mom_com_generic_1}\\
	M_2 =& \left\langle \anticommutator{\commutator{H_{\text{ss}}}{\commutator{H_{\text{ss}}}{\hat{f}_{\alpha, \eta, s}}}}{\hat{f}^\dagger_{\alpha, \eta, s}} \right\rangle^{\text{ss}}. \label{app:eqn:mom_com_generic_2}
\end{align}

Note that the Hamiltonian formulation of the single-site problem from \cref{app:eqn:Hamiltonian_single_site_equivalent} greatly simplifies the evaluation of these commutators. We find that 
\begin{align}
	\commutator{H_{\text{ss}}}{\hat{f}_{\alpha, \eta, s}} =& -\hat{A}_{\alpha, \eta, s} \hat{f}_{\alpha, \eta, s} -\sum_{i=1}^{N_a} V_i \hat{a}_{i,\alpha, \eta, s}, \label{app:eqn:mom_com_1}\\
	\commutator{H_{\text{ss}}}{\commutator{H_{\text{ss}}}{\hat{f}_{\alpha, \eta, s}}} =& \hat{A}_{\alpha, \eta, s}^2 \hat{f}_{\alpha, \eta, s} + \sum_{i=1}^{N_a} V_i^2 \hat{f}_{\alpha, \eta, s} + \sum_{i=1}^{N_a} V_i \left(\epsilon_i - \mu \right) \hat{a}_{i,\alpha, \eta, s} \nonumber \\
	&- U_1 \sum_{i=1}^{N_a} \sum_{\left(\alpha', \eta', s' \right) \neq \left( \alpha, \eta, s \right)} V_i \left( - \hat{f}^\dagger_{\alpha', \eta', s'}\hat{a}_{i,\alpha', \eta', s'} + \hat{a}^\dagger_{i,\alpha', \eta', s'}\hat{f}_{\alpha', \eta', s'} \right) \hat{f}_{\alpha, \eta, s}, \label{app:eqn:mom_com_2}
\end{align}
where we have defined $\hat{n}_{\alpha' \eta' s'}=\hat{f}^\dagger_{\alpha' \eta' s'}\hat{f}_{\alpha' \eta' s'}$ and 
\begin{equation}
	\hat{A}_{\alpha, \eta, s} \equiv U_1 \sum_{\left(\alpha', \eta', s'\right) \neq \left(\alpha, \eta, s \right)} \hat{n}_{\alpha', \eta', s'} - \left[\frac{\left(N_f - 1 \right) U_1}{2} + \mu - \Delta_{0} \right].
\end{equation}
By virtue of $H_{\text{ss}}$ having time-reversal symmetry, $\left\langle \hat{f}^\dagger_{\alpha, \eta, s} \hat{a}_{i,\alpha, \eta, s} \right\rangle^{\text{ss}}$ must be real, and, as a result,
\begin{equation}
	\left\langle \hat{f}^\dagger_{\alpha, \eta, s} \hat{a}_{i,\alpha, \eta, s} \right\rangle^{\text{ss}}= \left( \left\langle \hat{f}^\dagger_{\alpha, \eta, s} \hat{a}_{i,\alpha, \eta, s} \right\rangle^{\text{ss}} \right)^{*} = \left\langle \hat{a}^\dagger_{i,\alpha, \eta, s} \hat{f}_{\alpha, \eta, s} \right\rangle^{\text{ss}}.
\end{equation}
Using this together with \cref{app:eqn:mom_com_generic_1,app:eqn:mom_com_generic_2,app:eqn:mom_com_1,app:eqn:mom_com_2}, we find that the first two nontrivial spectral moments are given by 
\begin{align}
	M_1 =& \left\langle \anticommutator{\hat{A}_{\alpha, \eta, s} \hat{f}_{\alpha, \eta, s} + \sum_{i=1}^{N_a} V_i \hat{a}_{i,\alpha, \eta, s}}{\hat{f}^\dagger_{\alpha, \eta, s}} \right\rangle^{\text{ss}} \nonumber \\
	=& \left\langle \hat{A}_{\alpha, \eta, s} \right\rangle^{\text{ss}} \nonumber \\
	=& U_1 \left(n_{\text{ss}} -\frac{1}{2} \right) \left(N_f - 1 \right) - \mu + \Delta_{0}, \label{app:eqn:mom_1_result}\\
	M_2 =& \left\langle  \anticommutator{\left( \hat{A}^{2}_{\alpha, \eta, s} + \sum_{i=1}^{N_a} V_i \right) \hat{f}_{\alpha, \eta, s}}{\hat{f}^\dagger_{\alpha, \eta, s}}  \right\rangle^{\text{ss}} + \left\langle  \anticommutator{\sum_{i=1}^{N_a} V_i \left(\epsilon_i - \mu \right) \hat{a}_{i,\alpha, \eta, s} }{\hat{f}^\dagger_{\alpha, \eta, s}}  \right\rangle^{\text{ss}} \nonumber \\
	&- \left\langle  \anticommutator{ U_1 \sum_{i=1}^{N_a} \sum_{\left(\alpha', \eta', s' \right) \neq \left( \alpha, \eta, s \right)} V_i \left( - \hat{f}^\dagger_{\alpha', \eta', s'}\hat{a}_{i,\alpha', \eta', s'} + \hat{a}^\dagger_{i,\alpha', \eta', s'}\hat{f}_{\alpha', \eta', s'} \right) \hat{f}_{\alpha, \eta, s}}{\hat{f}^\dagger_{\alpha, \eta, s}} \right\rangle^{\text{ss}} \nonumber \\
	=& \left\langle  \hat{A}^{2}_{\alpha, \eta, s}  \right\rangle^{\text{ss}} + \sum_{i}^{N_a} V_i^{2}- U_1 \sum_{i=1}^{N_a} \sum_{\left(\alpha', \eta', s' \right) \neq \left( \alpha, \eta, s \right)} V_i \left( - \left\langle \hat{f}^\dagger_{\alpha', \eta', s'}\hat{a}_{i,\alpha', \eta', s'} \right\rangle^{\text{ss}} + \left\langle \hat{a}^\dagger_{i,\alpha', \eta', s'}\hat{f}_{\alpha', \eta', s'} \right\rangle^{\text{ss}} \right)  \nonumber \\
	=& \left[\frac{\left(N_f - 1 \right) U_1}{2} + \mu - \Delta_{0} \right]^2 - 2 U_1 \left[\frac{\left(N_f - 1 \right) U_1}{2} + \mu - \Delta_{0} \right] \sum_{\left(\alpha', \eta', s'\right) \neq \left(\alpha, \eta, s \right)} \left\langle  \hat{n}_{\alpha', \eta', s'} \right\rangle^{\text{ss}} \nonumber \\
	& + U_1^2 \sum_{\substack{\left(\alpha_{1}, \eta_{1}, s_{1}\right) \neq \left(\alpha, \eta, s \right) \\ \left(\alpha_{2}, \eta_{2}, s_{2}\right) \neq \left(\alpha, \eta, s \right)}} \left\langle  \hat{n}_{\alpha_{1}, \eta_{1}, s_{1}} \hat{n}_{\alpha_{2}, \eta_{2}, s_{2}}  \right\rangle^{\text{ss}} + \sum_{i}^{N_a} V_i^{2} \nonumber \\
	=& U_1^2 \left[\left(N_f - 1 \right) n_{\text{ss}} + \left(N_f - 1\right)\left(N_f - 2\right) \left\langle nn \right\rangle^{\text{ss}} \right] - 2 U_1 \left(N_f -1 \right) n_{\text{ss}} \left[\frac{\left(N_f - 1 \right) U_1}{2} + \mu - \Delta_{0}\right] \nonumber \\ 
	& + \left[\frac{\left(N_f - 1 \right) U_1}{2} + \mu - \Delta_{0} \right]^2 + \sum_{i=1}^{N_a} V_i^2, \label{app:eqn:mom_2_result_partial}
\end{align}
where the relative filling of the $f$-electrons $n_{\text{ss}}$, as well as the correlator between the fillings of $f$-electrons belonging to different orbital, valley, and spin flavors, $\left\langle nn \right\rangle^{\text{ss}}$, were introduced in \cref{app:eqn:rel_fil_of_f,app:eqn:nn_of_f}, respectively.
The expression for the first nontrivial spectral moment agrees between \cref{app:eqn:mom_1_from_spectral,app:eqn:mom_1_result}. The second moment from \cref{app:eqn:mom_2_result_partial} can be simplified to 
\begin{equation}
	\label{app:eqn:mom_2_result} 
	M_2 = M_1^2 + U_1^2 \left[\left(N_f - 1 \right) n_{\text{ss}} + \left(N_f - 1\right)\left(N_f - 2\right) \left\langle nn \right\rangle^{\text{ss}}  - \left(N_f -1 \right)^2 n_{\text{ss}}^2 \right] + \sum_{i=1}^{N_a} V_i^2.
\end{equation}
In turn, through \cref{app:eqn:obtaining_c_1_self_en}, \cref{app:eqn:mom_2_result} allows us to obtain the first coefficient of the dynamical self-energy expansion from \cref{app:eqn:expansion_sigma},
\begin{equation}
	\mathcal{C}_1 = U_1^2 \left[\left(N_f - 1 \right) n_{\text{ss}} + \left(N_f - 1\right)\left(N_f - 2\right) \left\langle nn \right\rangle^{\text{ss}}  - \left(N_f -1 \right)^2 n_{\text{ss}}^2 \right],
\end{equation}
thereby determining the \emph{exact} behavior of the dynamical self-energy at infinite frequency 
\begin{equation}
	\Sigma_{\text{ss}} \left( \omega + i 0^{+} \right) = \frac{U_1^2 \left(N_f - 1 \right) \left[ n_{\text{ss}} + \left(N_f - 2\right) \left\langle nn \right\rangle^{\text{ss}}  - \left(N_f -1 \right) n_{\text{ss}}^2 \right]}{\omega} + \mathcal{O} \left( \frac{1}{\omega^2} \right).
\end{equation}
The large-frequency asymptotic form of $\Sigma_{\text{ss}} \left( \omega + i 0^{+} \right)$ depends on the filling of the $f$-electrons $n_{\text{ss}}$, as well as on the correlator between the filling of $f$-electrons belonging to different orbital, valley, and spin flavors, $\left\langle nn \right\rangle^{\text{ss}}$. In the next \cref{app:sec:se_symmetric_details:se_exact:nn}, we will derive another way through which $\left\langle nn \right\rangle^{\text{ss}}$ can be extracted from the interacting Green's function of the single-site model. 

Finally, we note that requiring that the IPT dynamical self-energy of the single-site model $\Sigma_{\text{ss}} \left( \omega + i 0^{+} \right)$ obeys the exact $\omega \to \infty$ asymptotic behavior is equivalent to requiring that the first three moments of the interacting spectral function of the single-site model are accurately captured within our IPT method. 

\subsubsection{Determining $\left\langle nn \right\rangle^{\text{ss}}$}\label{app:sec:se_symmetric_details:se_exact:nn}

In order to relate $\left\langle nn \right\rangle^{\text{ss}}$ to the Matsubara Green's function, we start from the definition of the latter in imaginary time from \cref{app:eqn:definition_ss_interacting_gf}, and perform a time derivative approaching $\tau \to 0$ from below. For any $\left( \alpha, \eta, s \right)$, we have
\begin{align}
	- \lim_{\tau \to 0^{-}} \pdv{G \left( \tau \right)}{\tau} &= - \lim_{\tau \to 0^{-}} \left\langle  \hat{f}^\dagger_{\alpha, \eta, s} \left( 0 \right) \pdv{\tau} \hat{f}_{\alpha, \eta, s} \left( \tau \right) \right\rangle^{\text{ss}} \nonumber \\
	&= - \left\langle  \hat{f}^\dagger_{\alpha, \eta, s} \left( 0 \right) \commutator{H_{\text{ss}}}{\hat{f}_{\alpha, \eta, s} \left( \tau \right)} \right\rangle^{\text{ss}} \nonumber \\
&=U_1 \lim_{\tau \to 0^{-}} \left\langle \sum_{\left(\alpha', \eta', s'\right) \neq \left(\alpha, \eta, s \right) } \hat{f}^\dagger_{\alpha', \eta', s'} \left(\tau \right) \hat{f}_{\alpha', \eta', s'} \left(\tau \right) \hat{f}^\dagger_{\alpha, \eta, s} \left( 0 \right) \hat{f}_{\alpha, \eta, s} \left( \tau \right) \right\rangle^{\text{ss}} \nonumber \\
	&- \left[\frac{\left(N_f - 1 \right) U_1}{2} + \mu - \Delta_{0} \right] \lim_{\tau \to 0^{-}} \left\langle \hat{f}^\dagger_{\alpha, \eta, s} \left( 0 \right) \hat{f}_{\alpha, \eta, s} \left( \tau \right) \right\rangle^{\text{ss}} \nonumber \\
	&+ \lim_{\tau \to 0^{-}} \sum_{i=1}^{N_a} V_i \left\langle \hat{f}^\dagger_{\alpha, \eta, s} \left( 0 \right) \hat{a}_{i} \left(\tau \right) \right\rangle^{\text{ss}} \nonumber \\
&= U_1 \left(N_f -1 \right) \left\langle nn \right\rangle^{\text{ss}} + \lim_{\tau \to 0^{-}} \sum_{i=1}^{N_a} V_i G_i^{af} \left( \tau \right) - \lim_{\tau \to 0^{-}} \left[\frac{\left(N_f - 1 \right) U_1}{2} + \mu - \Delta_{0} \right] G \left( \tau \right),
\end{align}
where we have used the fact that 
\begin{equation}
	\lim_{\tau \to 0^{-}} \left\langle  \hat{f}^\dagger_{\alpha', \eta', s'} \left(\tau \right) \hat{f}_{\alpha', \eta', s'} \left(\tau \right) \hat{f}^\dagger_{\alpha, \eta, s} \left( 0 \right) \hat{f}_{\alpha, \eta, s} \left( \tau \right) \right\rangle^{\text{ss}} = \left\langle  \hat{f}^\dagger_{\alpha', \eta', s'} \hat{f}_{\alpha', \eta', s'} \hat{f}^\dagger_{\alpha, \eta, s} \hat{f}_{\alpha, \eta, s} \right\rangle^{\text{ss}} = \left\langle nn \right\rangle^{\text{ss}},
\end{equation}
for any $\left( \alpha, \eta, s \right) \neq \left( \alpha', \eta', s' \right)$ and the reader is reminded that repeated indices are not summed over. We now Fourier transform to the frequency domain according to \cref{app:eqn:matsubara_gf_THF_ft} in order to obtain 
\begin{align}
	U_1 \left(N_f -1 \right) \left\langle nn \right\rangle^{\text{ss}} &= - \lim_{\tau \to 0^{-}} \left( \pdv{G \left( \tau \right)}{\tau} + \sum_{i=1}^{N_a} V_i G^{af}_i \left( \tau \right) - \left[\frac{\left(N_f - 1 \right) U_1}{2} + \mu - \Delta_{0} \right] G \left( \tau \right) \right) \nonumber \\ 
	&= \frac{1}{\beta} \sum_{i \omega_n} \left\lbrace \left[ i \omega_n + \mu - \Delta_{0} + \frac{\left(N_f - 1 \right) U_1}{2} \right] G \left( i\omega_n \right) -  \sum_{i=1}^{N_a} V_i G^{af}_i \left( i \omega_n \right) \right\rbrace e^{-i \omega_n 0^-} \nonumber \\ 
	&= \frac{1}{\beta} \sum_{i \omega_n} \left\lbrace \left[ i \omega_n + \mu - \Delta_{0} + \frac{\left(N_f - 1 \right) U_1}{2} \right]  -  \sum_{i=1}^{N_a} \frac{V_i^2}{i \omega_n + \mu - \epsilon_i} \right\rbrace G \left( i\omega_n \right) e^{i \omega_n 0^+} \nonumber \\ 
	&= \frac{1}{\beta} \sum_{i \omega_n} \left[ i \omega_n + \mu - \Delta \left( i \omega_n \right)+ \frac{\left(N_f - 1 \right) U_1}{2} \right] G \left( i\omega_n \right) e^{i \omega_n 0^+} \nonumber \\ 
	&= \frac{1}{\beta} \sum_{i \omega_n} \left[ G^{-1} \left( i \omega_n \right) + U_1 n_{\text{ss}} \left(N_f - 1 \right) + \Sigma_{\text{ss}} \left( i \omega_n \right) \right] G \left( i\omega_n \right) e^{i \omega_n 0^+}
	\nonumber \\ 
	&= \frac{1}{\beta} \sum_{i \omega_n} e^{i \omega_n 0^{+}} + \frac{1}{\beta} \sum_{i \omega_n} U_1 n_{\text{ss}} \left(N_f - 1 \right) G \left( i\omega_n \right) e^{i \omega_n 0^+} + \frac{1}{\beta} \sum_{i \omega_n} \Sigma_{\text{ss}} \left( i \omega_n \right) G \left( i\omega_n \right) e^{i \omega_n 0^+}. \label{app:matsubara_sum_with_phase_all}  
\end{align}

To evaluate the three Matsubara summations from \cref{app:matsubara_sum_with_phase_all}, we consider the following sum for a general complex function $g(z)$,
\begin{equation}
	\label{app:eqn:matsubara_sum_s_g}
	S_g = \frac{1}{\beta} \sum_{i \omega_n} g \left( i \omega_n \right) e^{i \omega_n 0^{+}} = \frac{1}{2 \pi i \beta} \oint \dd{z} \frac{- \beta}{1 + e^{\beta z}} g(z) e^{z 0^{+}},
\end{equation}
where the contour integral encloses the imaginary axis. We now note that on the right half plane ({\it i.e.}{} for $\Re(z) \geq 0$), as $\abs{z} \to \infty$, the integrand in \cref{app:eqn:matsubara_sum_s_g} tends to zero at least as fast as $e^{- \left( \beta - 0^{+} \right) z}$, whereas on the left half plane, the integrand tends to zero at least as quickly as $e^{z 0^{+}}$. By Jordan's lemma, the integration contour can be deformed so that it only encircles the poles of $g(z)$ (clockwise), which allows us to evaluate
\begin{equation}
	\label{app:eqn:matsubara_sum_s_g_evaluated}
	S_g = \sum_{z_0 \in \text{poles of $z_0$}} \Res g\left(z_0 \right) n_{\mathrm{F}} \left( z_0 \right),
\end{equation}
where $\Res g\left(z_0 \right)$ denotes the residue of $g(z)$ at $z_0$. In particular, \cref{app:eqn:matsubara_sum_s_g_evaluated} implies that 
\begin{align}
	\frac{1}{\beta} \sum_{i \omega_n} e^{i \omega_n 0^{+}} &= 0, \label{app:eqn:matsubara_sum_phase_1} \\
	\frac{1}{\beta} \sum_{i \omega_n} \frac{e^{i \omega_n 0^{+}}}{i \omega_n - z} &= n_{\mathrm{F}} \left( z \right), \label{app:eqn:matsubara_sum_phase_2} 
\end{align}
thus evaluating the first Matsubara sum of \cref{app:matsubara_sum_with_phase_all}.

Using \cref{app:eqn:matsubara_sum_phase_2}, we can simplify the second term of \cref{app:matsubara_sum_with_phase_all}
\begin{equation}
	\label{app:eqn:matsubara_sum_phase_2ndterm}
	\frac{1}{\beta} \sum_{i \omega_n} G \left( i \omega_n \right) e^{i \omega_n 0^{+}} 
	= \int_{-\infty}^{\infty} \dd{\omega}  \rho \left( \omega \right) \frac{1}{\beta} \sum_{i \omega_n} \frac{e^{i \omega_n 0^{+}}}{i \omega_n -\omega}
	= \int_{-\infty}^{\infty} \dd{\omega}  \rho \left( \omega \right) n_{\mathrm{F}} \left( \omega \right) = n_{\text{ss}},
\end{equation}
where we have employed \cref{app:eqn:n_occ_from_dmft_int_gf}. Finally, to evaluate the last term of \cref{app:matsubara_sum_with_phase_all}, we first note that from the analytical properties of $G(z)$ and $\Sigma_{\text{ss}} \left( z \right)$ reviewed in \cref{app:sec:se_symmetric:DMFT_overview:anal_props}, we have that $\Sigma_{\text{ss}} \left( z \right) G\left(z \right)$ is an analytical complex function above and below the real axis with $\Sigma_{\text{ss}} \left( z^* \right) G\left(z^* \right) = \left[ \Sigma_{\text{ss}} \left( z \right) G\left(z \right) \right]^*$ and $\Sigma_{\text{ss}} \left( z \right) G\left(z \right) = \mathcal{O} \left( z^{-2}\right)$ as $z \to \infty$. Per the lemma stated and proved in \cref{app:sec:se_symmetric_details:se_exact:lemma_spectral_rep}, it follows that $\Sigma_{\text{ss}} \left( z \right) G\left(z \right)$ is endowed with a spectral representation~\cite{LUT61,PAV19}
\begin{equation}
	\label{app:eqn:spectral_rep_prod_self_en_gf}
	\Sigma_{\text{ss}} \left( z \right) G \left( z \right) = -\frac{1}{\pi} \int_{-\infty}^{\infty} \frac{\dd{\omega}}{z - \omega} \Im{\Sigma_{\text{ss}} \left( \omega + i 0^{+} \right) G \left( \omega + i 0^{+} \right) }.
\end{equation}
In turn, \cref{app:eqn:spectral_rep_prod_self_en_gf} enables us to evaluate the Matsubara sum in the third term of \cref{app:matsubara_sum_with_phase_all} similarly to \cref{app:eqn:matsubara_sum_phase_2ndterm} 
\begin{equation}
	\label{app:eqn:matsubara_sum_phase_3rdterm}
	\frac{1}{\beta} \sum_{i \omega_n} \Sigma_{\text{ss}} \left( i \omega_n \right) G \left( i\omega_n \right) e^{i \omega_n 0^+} = -\frac{1}{\pi} \int_{-\infty}^{\infty} \dd{\omega} \Im{\Sigma_{\text{ss}} \left( \omega + i 0^{+} \right) G \left( \omega + i 0^{+} \right) } n_{\mathrm{F}} \left( \omega \right).
\end{equation}

Using \cref{app:matsubara_sum_with_phase_all,app:eqn:matsubara_sum_phase_1,app:eqn:matsubara_sum_phase_2ndterm,app:eqn:matsubara_sum_phase_3rdterm}, we conclude that 
\begin{equation}
	\left\langle nn \right\rangle^{\text{ss}} = n_{\text{ss}}^2 - \frac{1}{\pi U_1 \left(N_f - 1 \right)} \int_{-\infty}^{\infty} \dd{\omega} \Im{\Sigma_{\text{ss}} \left( \omega + i 0^{+} \right) G \left( \omega + i 0^{+} \right) } n_{\mathrm{F}} \left( \omega \right).
\end{equation}

\subsubsection{A useful lemma on spectral representation}\label{app:sec:se_symmetric_details:se_exact:lemma_spectral_rep}

\begin{figure}
	\begin{tikzpicture}[scale=0.75]
\draw[thick, -Latex] (-3,0) -- (3,0) node[anchor=north west] {Re};
		\draw[thick, -Latex] (0,-4) -- (0,4) node[anchor=south east] {Im};
		
\draw[-<-=.25,-<-=.75] (-3,0.2) arc (180:0:3);
		\draw[-<-=.75] (3,0.2) -- (-3,0.2);
		\node at (1.255,2.5) {$\mathcal{C}_1$};
		
\draw[ -<-=.25, -<-=.75] (-3,-0.2) arc (-180:0:3);
		\draw[ -<-=.25] (3,-0.2) -- (-3,-0.2);
		\node at (1.25,-2.5) {$\mathcal{C}_2$};
		
\filldraw (0.5,1.5) circle (2pt) node[anchor=south west] {$z'$};
	\end{tikzpicture}
	\caption{Integration contours used in \cref{app:eqn:c1_cauchy_lemma,app:eqn:c2_cauchy_lemma}.}
	\label{app:fig:contour_for_anal_lemma}
\end{figure}

We now state and prove a useful lemma used repeatedly throughout this section on the existence of a spectral representation. Let $f(z)$ be a complex function obeying the following properties:
\begin{enumerate}
	\item $f (z)$ is analytical above ($\Im (z) > 0$) and below ($\Im (z) < 0$) the real axis.
	\item $f\left( z^{*} \right) = f^{*} \left( z \right)$.
	\item $f (z)$ tends to zero at least as fast as $\frac{1}{z}$ for $z \to \infty$. 
\end{enumerate}
It follows that $f(z)$ is endowed with a real spectral representation $\rho_{f} \left( \omega \right)$
\begin{equation}
	f(z) = \int_{-\infty}^{\infty} \dd{\omega} \frac{\rho_{f} \left( \omega \right)}{z-\omega}, \qq{with}
	\rho_{f} \left( \omega \right) = -\frac{1}{\pi} \Im f \left( \omega + i 0^{+} \right).
\end{equation}

To prove the existence of a spectral representation for $f(z)$, we start by defining $\mathcal{C}_1$ ($\mathcal{C}_2$) to be the counterclockwise (clockwise) contour encircling the upper (lower) complex half plane, as shown in \cref{app:fig:contour_for_anal_lemma}. For $z'$ in the upper complex half plane ({\it i.e.}{} $\Im \left( z' \right) > 0$), we must have from Cauchy's theorem that
\begin{align}
	\frac{1}{2 \pi i} \oint_{\mathcal{C}_1} \dd{z} \frac{f(z)}{z-z'} &= f \left( z' \right), \label{app:eqn:c1_cauchy_lemma} \\
	\frac{1}{2 \pi i} \oint_{\mathcal{C}_2} \dd{z} \frac{f(z)}{z-z'} &= 0. \label{app:eqn:c2_cauchy_lemma}
\end{align}
Because $f(z)$ tends to zero at least as fast as $\frac{1}{z}$ for $z \to \infty$, the contribution stemming from the semicircular arcs in \cref{app:eqn:c1_cauchy_lemma,app:eqn:c2_cauchy_lemma} vanishes. As such, \cref{app:eqn:c1_cauchy_lemma,app:eqn:c2_cauchy_lemma} become equivalent with
\begin{align}
	-\frac{1}{2 \pi i} \int_{-\infty}^{\infty} \dd{\omega} \frac{\Re f\left( \omega + i 0^{+} \right) + i \Im f\left( \omega + i 0^{+} \right) }{z'-\omega- i 0^{+}} &= f \left( z' \right), \label{app:eqn:c1_cauchy_lemma_equiv} \\
	-\frac{1}{2 \pi i} \int_{-\infty}^{\infty} \dd{\omega} \frac{\Re f\left( \omega - i 0^{+} \right) + i \Im f\left( \omega - i 0^{+} \right)}{z'-\omega + i 0^{+}} &= 0. \label{app:eqn:c2_cauchy_lemma_equiv}
\end{align}
Since $\Im \left( z' \right)>0$ ({\it i.e.}{} $z'$ is away from the real axis), we can safely ignore the small imaginary offset in the denominators of \cref{app:eqn:c1_cauchy_lemma_equiv,app:eqn:c2_cauchy_lemma_equiv}. Next, we employ $f \left( z^{*} \right) = f^{*} \left( z \right)$ to cast \cref{app:eqn:c2_cauchy_lemma_equiv} to 
\begin{equation}
	\frac{1}{2 \pi i} \int_{-\infty}^{\infty} \dd{\omega} \frac{\Re f\left( \omega + i 0^{+} \right) - i \Im f\left( \omega + i 0^{+} \right)}{z'-\omega + i 0^{+}} = 0, \label{app:eqn:c2_cauchy_lemma_final}
\end{equation}
By substituting \cref{app:eqn:c2_cauchy_lemma_final,app:eqn:c1_cauchy_lemma_equiv} we directly obtain~\cite{LUT61,PAV19}
\begin{equation}
	\label{app:eqn:lemma_final_result}
	f \left( z' \right) = -\frac{1}{\pi} \int_{-\infty}^{\infty} \dd{\omega} \frac{\Im f\left( \omega + i 0^{+} \right) }{z' - \omega}, \qq{for} \Im\left( z' \right) > 0.
\end{equation}
By repeating the proof for $z'$ in the lower half plane, one can prove \cref{app:eqn:lemma_final_result} for any $z'$ away from the real axis. This shows that $f(z)$ is indeed endowed with a real spectral representation.

\subsection{Details on the IPT method as applied to the THF model}\label{app:sec:se_symmetric_details:IPT}

In \cref{app:sec:se_symmetric:introduction:second_order_perturbation}, we outlined how the DMFT single-site problem corresponding to the THF model is solved using the approximate IPT method~\cite{MAR86,GEO92,YEY93,KAJ96,POT97,ANI97,LIC98,YEY99,MEY99,YEY00,SAS01,SAV01,FUJ03,LAA03,KUS06,ARS12,DAS16,WAG21,MIZ21,VAN22,CAN24,CAN25}. By building on the exact properties of the single-site self-energy obtained in \cref{app:sec:se_symmetric_details:se_exact,app:sec:se_symmetric_details:atomic_se}, this section briefly reviews the IPT method in general and details its derivation in the context of the THF model in the symmetric phase. 

The crux of the IPT approximation is to solve the single-site problem to second order in perturbation theory. By definition, such an approximation is valid in the \emph{weakly interacting} limit ({\it i.e.}{} the $U_1 \to 0$ limit), but is \emph{not} generally expected to hold in the \emph{strongly interacting} or \emph{atomic} limit ({\it i.e.}{} in the $U_1 \to \infty$ limit), which was discussed in \cref{app:sec:se_symmetric_details:atomic_se}. However, as we will show in this section, both the second-order self-energy correction in the atomic limit (which will be derived below) and the \emph{exact} atomic-limit dynamical self-energy from \cref{app:eqn:at_self_energy_diagonal} have similar functional forms. Based on this insight, the IPT method constructs an \emph{interpolated} self-energy ansatz starting from the second-order self-energy of the single-site model -- which can be computed numerically from the effective single-site Green's function $G_{0} \left(i \omega_n \right)$ according to \cref{app:eqn:second_order_f_symmetric}, as we will show explicitly below -- and the atomic-limit self-energy -- which was derived analytically in \cref{app:eqn:at_self_energy_diagonal}.

The interpolated self-energy ansatz used in IPT depends on one constant and one function of $\omega$, which are determined, respectively, by requiring that the interpolated self-energy has the exact $\omega \to \infty$ asymptotic form dictated by \cref{app:eqn:exact_se_asymptote,app:eqn:nn_correlator_from_self_energy_ss}\footnote{As mentioned in \cref{app:sec:se_symmetric_details:se_exact:results}, this is equivalent to correctly reproducing the first three moments of the interacting spectral function of the single-site model.} and that it recovers the exact atomic-limit self-energy in the $U_1 \to \infty$ limit. We will then show that in the weakly interacting $U_1 \to 0$ limit this ansatz reduces to the second-order perturbative result from \cref{app:eqn:second_order_f_symmetric}. By interpolating between two exact limits and having the correct $\omega \to \infty$ behavior, we expect the interpolated self-energy to correctly recover the intermediate interaction strength regime, as well. 

As in \cref{app:sec:se_symmetric:introduction:second_order_perturbation}, in this section, we will derive a series of relations between the dynamical $f$-electron self-energy and the effective Green's function of the single-site model, $G_{0} \left( i \omega_n \right)$. Neither one of these two functions is known \textit{a priori} and thus need to be determined self-consistently. The algorithm that we use in our numerical implementation was already detailed in \cref{app:sec:se_symmetric:IPT:sc_and_numerical}. 

\subsubsection{Second-order self-energy of the single-site model}\label{app:sec:se_symmetric_details:IPT:so_se_ss_model}

We start by considering the solution of the single-site model to second-order in perturbation theory. The calculation of the corresponding second-order self-energy correction proceeds analogously to its computation in the symmetry-broken THF phases from \cref{app:sec:se_correction_beyond_HF:all_so}. There are only three differences:
\begin{enumerate}
	\item There is no $H_{U_2}$ second-order self-energy contribution in the single-site model, since we focus on the correlation driven by the on-site interaction and this term is not included in \cref{app:eqn:single_site_action}.

	\item Instead of using the fully dressed $f$-electron propagator $G \left( i \omega_n \right)$ to compute the second-order contribution, as was done in \cref{app:sec:se_correction_beyond_HF:all_so}, one employs the Hartree-Fock propagator of the single-site model~\cite{GEO92}. Anticipating our derivation of the IPT dynamical self-energy in \cref{app:sec:se_symmetric_details:IPT:se_interpolated}, we will find it useful to work in terms of the family of propagators first defined in \cref{app:eqn:spectral_gf_relation_tilde} and repeated here for convenience
	\begin{equation}
		G^{\tilde{\mu}}_{0} \left( i \omega_n \right) = \frac{1}{G_{0}^{-1} \left(i \omega_n \right) + \tilde{\mu} - \mu} = \frac{1}{i \omega_n + \tilde{\mu} - \Delta \left( i \omega_n \right)}, \nonumber
	\end{equation}
	which is parameterized by the \emph{effective} chemical potential $\tilde{\mu}$. $G^{\tilde{\mu}}_{0} \left( i \omega_n \right)$ can be thought of as the non-interacting Green's function of the single-site model if the chemical potential of the latter was set to $\tilde{\mu}$, as opposed to $\mu$ [while keeping the same effective hybridization function $\Delta \left( i\omega_n \right)$]. The family of propagators $G^{\tilde{\mu}}_{0} \left( i \omega_n \right)$ includes both the non-interacting and the Hartree-Fock Green's functions of the single-site model
	\begin{alignat}{3}
		G^{\tilde{\mu}}_{0} \left( i \omega_n \right) &&= &G_{0} \left( i \omega_n \right), &&\qq{for} \tilde{\mu} = \mu, \label{app:eqn:case_tilde_mu_nonint}\\ 
		G^{\tilde{\mu}}_{0} \left( i \omega_n \right) &&= &G^{\text{HF}}_{0} \left( i \omega_n \right), &&\qq{for} \tilde{\mu}= \mu - U_1 \left(n_{\text{ss}} -\frac{1}{2} \right) \left(N_f - 1 \right). \label{app:eqn:case_tilde_mu_HF}
	\end{alignat} 
	where the latter is the Hartree-Fock Green's function of the single-site model
	\begin{equation}
		\label{app:eqn:hf_gf_single_site}
		G^{\text{HF}}_0 \left( i \omega_n \right) \equiv \frac{1}{G_0^{-1} \left( i \omega_n \right) - U_1 \left(n_{\text{ss}} - \frac{1}{2} \right) \left( N_f - 1 \right)}.
	\end{equation}
	When solving the single site model to second-order in perturbation theory, one should fix $\tilde{\mu}= \mu - U_1 \left(n_{\text{ss}} -\frac{1}{2} \right) \left(N_f - 1 \right)$, thus ensuring that the Hartree-Fock single-site Green's function is employed in computing the subsequent diagrammatic corrections.

	\item Of the two diagrams in \cref{app:fig:f_self_en_diags}, only the one in \cref{app:fig:f_self_en_diags:a} contributes. The \cref{app:fig:f_self_en_diags:b} contribution vanishes in the symmetric phase. To see this, we simply replace $\mathcal{G}^{f}_{\alpha \eta s;\alpha' \eta' s'} \left(i \omega_n \right)$ by $G^{\tilde{\mu}}_{0} \left( i \omega_n \right) \delta_{\alpha \alpha'} \delta_{\eta \eta'} \delta_{s s'}$ (coupled with setting $U_2 \to 0$) in \cref{app:eqn:se_2a_simple,app:eqn:se_2b_simple} to obtain
	\begin{align}
		& \Sigma^{f,(2a)}_{\alpha \eta s; \alpha' \eta' s'} \left(i \omega_n \right) = -\frac{1}{\beta^{2}}\sum_{i \omega_x, i \omega_y} \sum_{\substack{\alpha_1,\eta_1,s_1 \\ \alpha_2,\eta_2,s_2}} U_1^2 \left(1 - \delta_{\alpha' \alpha_1} \delta_{\eta' \eta_1} \delta_{s' s_1} \right) \left(1 - \delta_{\alpha \alpha_2} \delta_{\eta \eta_2} \delta_{s s_2} \right) \nonumber \\
		\times & G^{\tilde{\mu}}_{0} \left(i \omega_x \right) G^{\tilde{\mu}}_{0} \left(i \omega_y \right) G^{\tilde{\mu}}_{0} \left(i \omega_n - i \omega_x + i \omega_y \right) \delta_{\alpha \alpha'} \delta_{\eta \eta'} \delta_{s s'} \delta_{\alpha_1 \alpha_2} \delta_{\eta_1 \eta_2} \delta_{s_1 s_2} \delta_{\alpha_2 \alpha_1} \delta_{\eta_2 \eta_1} \delta_{s_2 s_1},  \label{app:eqn:se_2a_dmft}\\
& \Sigma^{f,(2b)}_{\alpha \eta s; \alpha' \eta' s'} \left(i \omega_n \right) = \frac{1}{\beta^{2}}\sum_{i \omega_x, i \omega_y} \sum_{\substack{\alpha_1,\eta_1,s_1 \\ \alpha_2,\eta_2,s_2}} U_1^2 \left(1 - \delta_{\alpha' \alpha_2} \delta_{\eta' \eta_2} \delta_{s' s_2} \right) \left(1 - \delta_{\alpha \alpha_1} \delta_{\eta \eta_1} \delta_{s s_1} \right)  \nonumber \\
		\times & G^{\tilde{\mu}}_{0} \left(i \omega_x \right) G^{\tilde{\mu}}_{0} \left(i \omega_y \right) G^{\tilde{\mu}}_{0} \left(i \omega_n - i \omega_x + i \omega_y \right) \delta_{\alpha_1 \alpha'} \delta_{\eta_1 \eta'} \delta_{s_1 s'} \delta_{\alpha_2 \alpha_1} \delta_{\eta_2 \eta_1} \delta_{s_2 s_1} \delta_{\alpha \alpha_2} \delta_{\eta \eta_2} \delta_{s s_2}, \label{app:eqn:se_2b_dmft}
	\end{align}
	from which it is easy to check that $\Sigma^{f,(2b)}_{\alpha \eta s; \alpha' \eta' s'} \left(i \omega_n \right) = 0$. 
\end{enumerate}

Using \cref{app:eqn:se_2a_dmft,app:eqn:se_2b_dmft}, the dynamical self-energy correction of the single-site model to second-order in $U_1$ is given by 
\begin{align}
	\tilde{\Sigma}^{f,(2)} \left(i \omega_n \right) &= \Sigma^{f,(2a)} \left(i \omega_n \right) + \Sigma^{f,(2b)} \left(i \omega_n \right)  = \Sigma^{f,(2a)} \left(i \omega_n \right) \nonumber \\
	&= -\frac{U_1^2 \left( N_f - 1 \right)}{\beta^{2}}\sum_{i \omega_x, i \omega_y} G^{\tilde{\mu}}_{0} \left(i \omega_x \right) G^{\tilde{\mu}}_{0} \left(i \omega_y \right) G^{\tilde{\mu}}_{0} \left(i \omega_n - i \omega_x + i \omega_y \right). 	\label{app:eqn:sec_ord_impurity_unsimplified}
\end{align}
By employing the spectral function of the $G^{\tilde{\mu}}_{0} \left( i \omega_n \right)$ propagator defined in \cref{app:eqn:spectral_gf_relation_tilde}, one can manipulate \cref{app:eqn:sec_ord_impurity_unsimplified} similarly to \crefrange{app:eqn:se_2a_simple}{app:eqn:full_f_second_order_sigma} and obtain the dynamical self-energy correction of the single-site model at the second-order level
\begin{equation}
	\label{app:sec:sec_ord_impurity_simplified}
	\tilde{\Sigma}^{f,(2)} \left(\omega + i 0^{+} \right) = - i U_1^2 \left( N_f - 1 \right) \int_{0}^{\infty} \dd{\lambda} e^{i \lambda \omega} \left( B^{\tilde{\mu}}_{0} \left(\lambda \right) C^{\tilde{\mu}}_{0} \left(-\lambda \right) B^{\tilde{\mu}}_{0} \left(\lambda \right) + C^{\tilde{\mu}}_{0} \left(\lambda \right) B^{\tilde{\mu}}_{0} \left(-\lambda \right) C^{\tilde{\mu}}_{0} \left(\lambda \right) \right).
\end{equation}
In \cref{app:sec:sec_ord_impurity_simplified}, we have introduced the following auxiliary functions
\begin{equation}
	\label{app:eqn:aux_func_impurity}
	B^{\tilde{\mu}}_{0} \left( \lambda \right) = \int_{-\infty}^{\infty} \dd{\omega} n_{\mathrm{F}} \left( \omega \right) \rho^{\tilde{\mu}}_{0} \left( \omega \right) e^{-i \lambda \omega},  \quad
	C^{\tilde{\mu}}_{0} \left( \lambda \right) = \int_{-\infty}^{\infty} \dd{\omega} \left( 1 - n_{\mathrm{F}} \left( \omega \right) \right) \rho^{\tilde{\mu}}_{0} \left( \omega \right) e^{-i \lambda \omega}
\end{equation}
which are analogous to the ones defined in \cref{app:eqn:aux_func_B_self_en,app:eqn:aux_func_C_self_en} and in which $n_{\mathrm{F}} \left(\omega \right)$ is the Fermi-Dirac distribution defined in \cref{app:eqn:fd_distribution}. Finally, we mention that in this method, the second-order self-energy from \cref{app:sec:sec_ord_impurity_simplified} is already computed above the real axis, meaning that no analytical continuation is necessary. 

In principle, one can directly solve the single-site model to second-order in perturbation theory [by taking $G^{\tilde{\mu}}_{0} \left(i \omega_n \right) $ to be the Hartree-Fock propagator of the single-site model $G^{\text{HF}}_{0} \left(i \omega_n \right)$], thereby approximating its exact dynamical self-energy by the second-order contribution $\Sigma_{\text{ss}} \left( i \omega_n \right) \approx \tilde{\Sigma}^{f,(2)} \left( i \omega_n \right)$. Within such an approach, the DMFT self-consistency conditions are identical to ones depicted in \cref{fig:self_consistent_DMFT}. Due to its perturbative character, such a method is, by definition, valid in the small-interaction regime.  

The second-order perturbative method for solving the single-site model has been successfully applied to the standard ({\it i.e.}{}, spinful, single-band) Hubbard model in infinite dimensions at \emph{exactly} half-filling~\cite{GEO92,GEO92a}. In this spinful, single-band, particle-hole symmetric case, $\tilde{\Sigma}^{f,(2)} \left( i \omega_n \right)$ captures the correct behavior of the self-energy in both the weakly and strongly interacting regimes~\cite{ROZ94,KAJ96}, accurately reproducing the associated Mott transition. However, a naive extension of this approach away from particle-hole symmetry or beyond the single-band case yields unphysical results in the strongly interacting limit~\cite{KAJ96}. We elaborate on these issues in \cref{app:sec:se_symmetric_details:IPT}. To address this shortcoming, Refs.~\cite{MAR86,GEO92,YEY93,KAJ96,POT97,ANI97,LIC98,YEY99,MEY99,YEY00,SAS01,SAV01,FUJ03,LAA03,KUS06,ARS12,DAS16,WAG21,MIZ21,VAN22,CAN24,CAN25} have refined the original method of Ref.~\cite{GEO92} by introducing an \emph{interpolated} self-energy constructed from the second-order perturbative result and the atomic self-energy solution in \cref{app:eqn:at_self_energy_diagonal}. In the following \cref{app:sec:se_symmetric_details:IPT:se_interpolated}, we discuss both the exact and second-order self-energies of the single-site model in the atomic limit and present the construction and justification of the interpolated self-energy ansatz.

\subsubsection{The interpolated self-energy}\label{app:sec:se_symmetric_details:IPT:se_interpolated}

In \cref{app:sec:se_symmetric_details:IPT:so_se_ss_model}, we outlined how the single-site model can be solved using perturbation theory in $U_1$. While this approach works well for the spinful single-band Hubbard model at the particle-hole symmetric filling $n = \frac{1}{2}$, it fails to capture the correct physics in the strongly interacting regime for either $n \neq \frac{1}{2}$~\cite{KAJ96} or generically in the multi-band case. In this section, we briefly justify this empirical observation and explain how IPT~\cite{MAR86,GEO92,YEY93,KAJ96,POT97,ANI97,LIC98,YEY99,MEY99,YEY00,SAS01,SAV01,FUJ03,LAA03,KUS06,ARS12,DAS16,WAG21,MIZ21,VAN22,CAN24,CAN25} addresses this limitation.

We begin with the expression for the second-order self-energy correction of the single-site model from \cref{app:eqn:second_order_f_symmetric}, written in terms of the spectral function of the propagator $G^{\tilde{\mu}}_{0} \left( i \omega_n \right)$, and reproduced here for convenience
\begin{align*}
	\tilde{\Sigma}^{f,(2)} \left(\omega + i 0^{+} \right) =&  U_1^2 \left(N_f - 1 \right) \int_{-\infty}^{\infty} \dd{\omega_1} \int_{-\infty}^{\infty} \dd{\omega_2} \int_{-\infty}^{\infty} \dd{\omega_3} \rho^{\tilde{\mu}}_0 \left(\omega_1 \right) \rho^{\tilde{\mu}}_0 \left(\omega_2 \right) \rho^{\tilde{\mu}}_0 \left(\omega_3 \right) \\
	&\times \frac{n_{\mathrm{F}} \left( \omega_{1} \right) \left( 1 - n_{\mathrm{F}} \left( \omega_{2} \right) \right) n_{\mathrm{F}} \left( \omega_{3} \right) + \left( 1 - n_{\mathrm{F}} \left( \omega_{1} \right) \right) n_{\mathrm{F}} \left( \omega_{2} \right) \left( 1 - n_{\mathrm{F}} \left( \omega_{3} \right) \right)}{\omega - \omega_1 + \omega_2 - \omega_3 + i 0^{+}}.
\end{align*}
\Cref{app:eqn:second_order_f_symmetric} was derived from \cref{app:eqn:sec_ord_impurity_unsimplified} by performing the Matsubara summation as in \cref{app:eqn:matsubara_sum_I}, followed by analytic continuation. Our goal is to assess the applicability of the second-order perturbative solution in the $U_1 \to \infty$ limit (noting that in the $U_1 \to 0$ limit, the solution is valid by definition). 

In the strongly interacting regime, the detailed energy dependence of the effective Green’s function $G_{0} \left(i \omega_n \right)$ becomes unimportant. At energy scales on the order of $U_1$, the non-interacting spectral function of the single-site model can be approximated as a broadened Dirac $\delta$-function ({\it i.e.}{}, a Lorentzian). This behavior can be modeled by setting the effective hybridization of the $f$-electrons, as defined in \cref{app:eqn:def_of_hyb_field}, to $\Delta \left( i \omega_n \right) = i \Gamma_f$, where $\Gamma_f$ is a real broadening parameter. With this choice, the effective spectral function from \cref{app:eqn:spectral_gf_relation_tilde} takes the Lorentzian form
\begin{equation}
	\label{app:eqn:lorentz_ss_rho}
	\rho^{\tilde{\mu}}_0 \left(\omega \right) = \frac{\Gamma_f}{\pi} \frac{1}{\left( \omega + \tilde{\mu} \right)^2 + \Gamma_f^2}.
\end{equation} 

In what follows, we compare the second-order self-energy correction in the atomic limit with the \emph{exact} atomic-limit self-energy from \cref{app:eqn:at_self_energy_diagonal}. To do so, we first observe that $\rho^{\tilde{\mu}}_0 \left(\omega \right)$ decays rapidly outside a small region of width $\Gamma_f$ centered around $-\tilde{\mu}$. As a result, in the region where the product $\rho^{\tilde{\mu}}_0 \left(\omega_1 \right) \rho^{\tilde{\mu}}_0 \left(\omega_2 \right) \rho^{\tilde{\mu}}_0 \left(\omega_3 \right)$ contributes significantly, one must have
\begin{equation}
	\omega - \omega_1 + \omega_2 - \omega_3 = \omega + \tilde{\mu} + \mathcal{O} \left( \Gamma_f \right).
\end{equation}
It follows that for $\abs{\frac{\omega + \tilde{\mu}}{\Gamma_f}} \to \infty$, the denominator in \cref{app:eqn:second_order_f_symmetric} can be approximated as $\omega + \tilde{\mu}$. Consequently, in this same limit we obtain
\begin{equation}
	\label{app:eqn:self_energy_2nd_in_small_Gamma}
	\tilde{\Sigma}^{f,(2)} \left(\omega + i 0^{+} \right) \approx \frac{U_1^2 \left(N_f - 1 \right) n^{\tilde{\mu}}_0 \left( 1 - n^{\tilde{\mu}}_0 \right)}{\omega + \tilde{\mu} + i 0^{+}}, \qq{for} \abs{\frac{\omega + \tilde{\mu}}{\Gamma_f}} \to \infty.
\end{equation}
In \cref{app:eqn:self_energy_2nd_in_small_Gamma}, $n^{\tilde{\mu}}_0$ can be interpreted as the $f$-electron filling of the single-site model with chemical potential set to $\tilde{\mu}$. Its expression, originally given in \cref{app:eqn:int_rel_f0_filling} and repeated here for convenience, reads as
\begin{equation*}
	n^{\tilde{\mu}}_0 = \int_{-\infty}^{\infty} \dd{\omega} n_{\mathrm{F}} \left( \omega \right) \rho^{\tilde{\mu}}_0 \left( \omega \right),
\end{equation*}
We now note that taking the atomic limit for the $f$-electrons is equivalent to sending $\Gamma_f \to 0$. It follows that the second-order self-energy correction computed from the single-site model in this limit -- denoted henceforth by $\tilde{\Sigma}^{f,(2),\text{At}} \left(\omega + i 0^{+} \right)$ -- must be given by 
\begin{equation}
	\label{app:eqn:self_energy_2nd_in_atomic}
	\tilde{\Sigma}^{f,(2),\text{At}} \left(\omega + i 0^{+} \right) = \frac{U_1^2 \left(N_f - 1 \right) n^{\tilde{\mu}}_0 \left( 1 - n^{\tilde{\mu}}_0 \right)}{\omega + \tilde{\mu} + i 0^{+}},
\end{equation} 
which holds for all $\omega$. The functional form of \cref{app:eqn:self_energy_2nd_in_atomic} can also be justified on general grounds: the expression must be proportional to $U_1^2$ since it arises from a second-order correction; the factor $\frac{1}{\omega + \tilde{\mu}}$ follows from dimensional analysis and the requirement that the self-energy vanishes at infinite frequency; and the prefactor $n^{\tilde{\mu}}_0 \left( 1 - n^{\tilde{\mu}}_0 \right)$ ensures that dynamical fluctuations vanish in the fully filled or completely empty limit. 

We now observe that solving the single-site model using second-order perturbation theory, as described in \cref{app:sec:se_symmetric_details:IPT:so_se_ss_model}, is equivalent to choosing $\tilde{\mu} = \mu - U \left(n_{\text{ss}}  - \frac{1}{2} \right) \left(N_f - 1 \right)$, as implied by \cref{app:eqn:case_tilde_mu_HF}. In this case, \cref{app:eqn:self_energy_2nd_in_atomic} yields the following expression for the second-order self-energy in the atomic limit
\begin{equation}
	\label{app:eqn:self_energy_atomic_HF_prop}
	\tilde{\Sigma}^{f,(2),\text{At}} \left(\omega + i 0^{+} \right) = \frac{U_1^2 \left(N_f - 1 \right) n^{\tilde{\mu}}_0 \left( 1 - ^{\tilde{\mu}}n_0 \right)}{\omega + \mu - U_1 \left(n_{\text{ss}}  - \frac{1}{2} \right) \left(N_f - 1 \right) + i 0^{+}}.
\end{equation}
\Cref{app:eqn:self_energy_atomic_HF_prop} can be compared with the exact atomic-limit dynamical self-energy derived in \cref{app:sec:se_symmetric_details:atomic_se:approximation_low_t} for $\beta U_1 \to \infty$. At \emph{general} $f$-electron filling, the exact atomic self-energy from \cref{app:eqn:atomic_self_energy_low_temp_integer_dop} features two poles in $\omega$ with non-vanishing residues, while \cref{app:eqn:self_energy_atomic_HF_prop} contains only a single pole. Near integer fillings, where the exact atomic self-energy reduces to a single pole, we find from \cref{app:eqn:atomic_self_energy_around_integer_filling,app:eqn:atomic_self_energy_low_temp_integer} that
\begin{equation}
	\label{app:eqn:compare_self_energies_at}
	\tilde{\Sigma}^{\text{At}} \left( \omega + i 0^{+} \right) \approx \begin{cases}
		\frac{U_1^2 n^{\text{At}} \left( 1 - n^{\text{At}} \right)}{ \omega - U_1 \left( n^{\text{At}} - 1 \right) + i 0^{+} } + \mathcal{O} \left[ U_1 \left( n^{\text{At}} N_f - r \right) \right], 
		& \qq{if} n^{\text{At}} \approx \frac{r}{N_f}, n^{\text{At}} > \frac{r}{N_f}, 0 \leq r \leq N_f \\
		\frac{U_1^2 n^{\text{At}} \left( 1 - n^{\text{At}} \right)}{ \omega - U_1 n^{\text{At}}  + i 0^{+} } + \mathcal{O} \left[ U_1 \left( n^{\text{At}} N_f - r - 1 \right) \right], 
		& \qq{if} n^{\text{At}} \approx \frac{r + 1}{N_f}, n^{\text{At}} < \frac{r + 1}{N_f}, 0 \leq r \leq N_f \\
		\frac{ U_1^2 n^{\text{At}} \left( 1 - n^{\text{At}} \right)}{ \omega - U_1 (n^{\text{At}} - \frac{1}{2}) + i 0^{+}}, 
		& \qq{if} n^{\text{At}} = \frac{r}{N_f}, 0 < r < N_f
	\end{cases},
\end{equation}
where the expressions become exact in the zero-temperature limit. On the other hand, using the exact expressions for the chemical potential as a function of filling in the atomic limit at low temperature, as derived in \cref{app:sec:se_symmetric_details:atomic_se:approximation_low_t}, we obtain for the second-order self-energy from \cref{app:eqn:self_energy_atomic_HF_prop}
\begin{equation}
	\label{app:eqn:compare_self_energies_perturbation}
	\tilde{\Sigma}^{f,(2),\text{At}} \left(\omega + i 0^{+} \right) = 
	\begin{cases}
		\frac{U_1^2 \left(N_f - 1 \right) n^{\tilde{\mu}}_0 \left( 1 - n^{\tilde{\mu}}_0 \right)}{\omega + U_1 n_{\text{ss}} + i 0^{+}}, 
		& \qq{if} n_{\text{ss}} \approx \frac{r}{N_f}, n_{\text{ss}} > \frac{r}{N_f}, 0 \leq r \leq N_f \\
		\frac{U_1^2 \left(N_f - 1 \right) n^{\tilde{\mu}}_0 \left( 1 - n^{\tilde{\mu}}_0 \right)}{\omega + U_1 \left( n_{\text{ss}} - 1 \right) + i 0^{+}}, 
		& \qq{if} n_{\text{ss}} \approx \frac{r + 1}{N_f}, n_{\text{ss}} < \frac{r + 1}{N_f}, 0 \leq r \leq N_f \\
		\frac{U_1^2 \left(N_f - 1 \right) n^{\tilde{\mu}}_0 \left( 1 - n^{\tilde{\mu}}_0 \right)}{\omega + U_1 \left(n_{\text{ss}}  - \frac{1}{2} \right) + i 0^{+}}, 
		& \qq{if} n_{\text{ss}} = \frac{r}{N_f}, 0 < r < N_f
	\end{cases}.
\end{equation}

Comparing \cref{app:eqn:compare_self_energies_at,app:eqn:compare_self_energies_perturbation}, we find that near integer filling, the second-order perturbative result fails to reproduce the correct pole location in the self-energy -- except at the particle-hole symmetric point $n = n_{\text{ss}}= \frac{1}{2}$. At this special filling, we have $\mu = \tilde{\mu} = 0$, and due to the particle-hole symmetry of $G_{0} \left( i \omega_n \right)$\footnote{The effective Green's function of the single-site model is obtained by integrating out all lattice electrons except for the $f$-electrons at a single site. At $\nu = 0$, the full system exhibits particle-hole symmetry, implying that $G_{0} \left( i\omega_n \right)$ also possesses this symmetry.}, it follows from \cref{app:eqn:int_rel_f0_filling} that $n^{\tilde{\mu}}_0 = \frac{1}{2}$. As a result, even the at particle-hole symmetric filling, the second-order self-energy $\tilde{\Sigma}^{f,(2),\text{At}} \left(i \omega_n \right) = \frac{U_1^2 \left(N_f - 1 \right)}{4 i \omega_n}$ differs from the exact result $\tilde{\Sigma}^{\text{At}} \left( i \omega_n \right) = \frac{ U_1^2}{4 i \omega_n}$ by a prefactor when $N_f \neq 2$. These observations explain why solving the impurity problem perturbatively at $n = \frac{1}{2}$ in the single-band case ($N_f = 2$) is remarkably successful for all $U_1$: in the $U_1 \to \infty$ limit, the exact and perturbative self-energies coincide~\cite{GEO92,KAJ96}. Furthermore, \cref{app:eqn:compare_self_energies_at,app:eqn:compare_self_energies_perturbation} show that away from particle-hole symmetry and in the strongly interacting regime, second-order perturbation theory fails to reproduce the correct self-energy pole location, resulting in unphysical outcomes~\cite{KAJ96}.

Although the second-order perturbative expression does not correctly reproduce the pole locations and residues near integer fillings, the exact and perturbative self-energies \emph{do} share the same functional form. In particular, rewriting \cref{app:eqn:self_energy_2nd_in_atomic} in imaginary frequency
\begin{equation}
	i \omega_n = \frac{U_1^2 \left(N_f -1 \right)n^{\tilde{\mu}}_0 \left( 1 - n^{\tilde{\mu}}_0 \right)}{\tilde{\Sigma}^{f,(2),\text{At}} \left( i \omega_n \right)} - \tilde{\mu},
\end{equation}
one can express the exact atomic self-energy at integer $f$-electron filling $\nu_f = r - 4$ (for $0 < r < 8$, $r \in \mathbb{Z}$) as 
\begin{equation}
	\label{app:eqn:intution_interpolation}
	\tilde{\Sigma}^{\text{At}} \left( i \omega_n \right) \approx \frac{1}{N_f}\frac{ U_1^2 r\left(N_f-r \right)}{ N_f \frac{U_1^2 \left(N_f -1 \right)n^{\tilde{\mu}}_0 \left( 1 - n^{\tilde{\mu}}_0 \right)}{\tilde{\Sigma}^{f,(2),\text{At}} \left( i \omega_n \right)} - \tilde{\mu} - U_1 (r - \frac{N_f}{2})}.
\end{equation}
\Cref{app:eqn:intution_interpolation} has motivated Refs.~\cite{ANI97, ARS12, DAS16, FUJ03, KAJ96, LAA03, LIC98, MAR86, MEY99, MIZ21, POT97, SAS01, SAV01, YEY00, YEY93, YEY99, VAN22} to generalize this ansatz to arbitrary filling and interaction strength. The result is an \textit{interpolated} self-energy, valid in both the weakly and strongly interacting regimes, whose expression is given in \cref{app:eqn:interpolated_sigma} and repeated here for convenience
\begin{equation}
	\label{app:eqn:interpolated_sigma_def}
	\tilde{\Sigma}^{f,\text{Int}} \left( \omega + i 0^{+} \right) = \frac{a }{ \frac{1}{\tilde{\Sigma}^{f,(2)}\left( \omega + i 0^{+} \right)} - b \left( \omega + i 0^{+} \right)} = \frac{a \tilde{\Sigma}^{f,(2)}\left( \omega + i 0^{+} \right)}{1-b \left( \omega + i 0^{+} \right) \tilde{\Sigma}^{f,(2)}\left( \omega + i 0^{+} \right)}.
\end{equation}

In \cref{app:eqn:interpolated_sigma_def}, the constant $a$ and the function $b \left( \omega + i 0^{+} \right)$ are determined by requiring that $\tilde{\Sigma}^{f,\text{Int}} \left( \omega + i 0^{+} \right)$ exhibits the correct asymptotic behavior as $\omega \to \infty$, as dictated by \cref{app:eqn:exact_se_asymptote}, and that $\tilde{\Sigma}^{f,\text{Int}} \left( \omega + i 0^{+} \right)$ reproduces the exact atomic-limit self-energy as $U_1 \to \infty$. Once $a$ and $b \left( \omega + i 0^{+} \right)$ are fixed, we will also demonstrate in \cref{app:sec:se_symmetric_details:IPT:low_u_limit} that $\tilde{\Sigma}^{f,\text{Int}} \left( \omega + i 0^{+} \right)$ correctly reproduces the $U_1 \to 0$ limit~\cite{ANI97, DAS16, FUJ03, LAA03, LIC98, MIZ21, SAS01, SAV01}. The self-energy \emph{ansatz} in \cref{app:eqn:interpolated_sigma_def} is referred to as interpolated because $\tilde{\Sigma}^{f,\text{Int}} \left( \omega + i 0^{+} \right)$ reduces to the exact atomic limit for $U_1 \to \infty$ \emph{and} to the second-order perturbative result for $U_1 \to 0$. As an interpolation between two exact limits, it is expected to provide a reliable approximation to the dynamical self-energy in the \emph{intermediate} interaction strength regime. Therefore, rather than using the second-order self-energy $\tilde{\Sigma}^{f,(2)} \left( \omega + i 0^{+} \right)$, we construct our self-consistent solution based on the interpolated form $\tilde{\Sigma}^{f,\text{Int}} \left( \omega + i 0^{+} \right)$.

Finally, following Refs.~\cite{ANI97, ARS12, DAS16, FUJ03, KAJ96, LAA03, LIC98, MAR86, MEY99, MIZ21, POT97, SAS01, SAV01, YEY00, YEY93, YEY99, VAN22}, we compute the second-order self-energy correction $\tilde{\Sigma}^{f,(2)}\left( i \omega_n \right)$ not using the propagator $G_0 \left(i \omega_n \right)$ or $G^{\text{HF}}_0 \left( i \omega_n \right)$, but rather the newly introduced Green's function $G_0^{\tilde{\mu}} \left( i \omega_n \right)$. The fictitious chemical potential $\tilde{\mu}$ thus serves as an additional free ({\it i.e.}{} variational) parameter, which is determined in simulations as described in \cref{app:sec:se_symmetric_details:IPT:fix_mu_tilde}.

\subsubsection{Determining the $a$ constant}\label{app:sec:se_symmetric_details:IPT:a}

To determine the constant $a$ in \cref{app:eqn:interpolated_sigma_def}, we examine the behavior of the dynamical $f$-electron self-energy in the $\omega \to \infty$ limit. We first \emph{assume} that $b \left( \omega + i 0^{+} \right)$ remains bounded as $\omega \to \infty$, a result that will be demonstrated explicitly in \cref{app:eqn:asymptote_b_infty}. In this limit, $\omega$ far exceeds the characteristic width of the spectral function $\rho^{\tilde{\mu}}_0 \left( \omega \right)$. Therefore, we can use the asymptotic form of the second-order self-energy from \cref{app:eqn:self_energy_2nd_in_small_Gamma} to show that 
\begin{equation}
\label{app:eqn:int_se_asymptote}
\tilde{\Sigma}^{f,\text{Int}} \left( \omega + i 0^{+} \right) = \frac{\frac{U_1^2 a \left(N_f - 1 \right) n^{\tilde{\mu}}_0 \left( 1 - n^{\tilde{\mu}}_0 \right)}{\omega} + \mathcal{O} \left( \frac{1}{\omega^2} \right) }{1-b \left( \omega + i 0^{+} \right) \mathcal{O} \left( \frac{1}{\omega} \right)} =  \frac{U_1^2 a \left(N_f - 1 \right) n^{\tilde{\mu}}_0 \left( 1 - n^{\tilde{\mu}}_0 \right)}{\omega} + \mathcal{O} \left( \frac{1}{\omega^2} \right),
\end{equation}
where we have used the assumption that $b \left( \omega + i 0^{+} \right)$ is bounded as $\omega \to \infty$. Requiring that the $\omega \to \infty$ asymptotic behavior of $\tilde{\Sigma}^{f,\text{Int}} \left( \omega + i 0^{+} \right)$ in \cref{app:eqn:int_se_asymptote} matches the exact result from \cref{app:eqn:exact_se_asymptote}, we find 
\begin{equation}
\label{app:eqn:IPT_a_constant}
	a = \frac{ n_{\text{ss}} + \left(N_f - 2\right) \left\langle nn \right\rangle^{\text{ss}}  - \left(N_f -1 \right) n_{\text{ss}}^2}{n^{\tilde{\mu}}_0 \left(1 - n^{\tilde{\mu}}_0 \right)},
\end{equation}
In \cref{app:eqn:IPT_a_constant}, the relative $f$-electron fillings of the interacting ($n_{\text{ss}}$) and non-interacting ($n^{\tilde{\mu}}_0$) single-site problems are defined in \cref{app:eqn:int_rel_f_filling,app:eqn:int_rel_f0_filling}, respectively. For the $\left\langle nn \right\rangle^{\text{ss}}$ correlator, we make one further approximation by evaluating it using the \emph{interpolated} self-energy rather than the exact one, as given by
\begin{equation}
\label{app:eqn:nn_correlator_from_ipt_self_energy}
\left\langle nn \right\rangle^{\text{ss}} = n_{\text{ss}}^2 - \frac{1}{\pi U_1 \left(N_f - 1 \right)} \int_{-\infty}^{\infty} \dd{\omega} \Im{\tilde{\Sigma}^{f,\text{Int}} \left( \omega + i 0^{+} \right) G \left( \omega + i 0^{+} \right) } n_{\mathrm{F}} \left( \omega \right).
\end{equation}

\subsubsection{Determining the $b \left( \omega + i 0^{+} \right)$ function}\label{app:sec:se_symmetric_details:IPT:b}

The $b \left( \omega + i 0^{+} \right)$ function is found by requiring that in the atomic limit, the interpolated dynamical self-energy from \cref{app:eqn:interpolated_sigma_def} reproduces the exact result from \cref{app:eqn:at_self_energy_diagonal}. This is equivalent with
\begin{equation}
\tilde{\Sigma}^{\text{At}} \left(\omega + i 0^{+} \right) = \frac{a^{\text{At}}}{\frac{1}{\tilde{\Sigma}^{f,(2),\text{At}} \left(\omega + i 0^{+} \right)} - b \left(\omega + i 0^{+}\right)},
\end{equation}
and allows us to determine the function 
\begin{equation}
b \left(\omega + i 0^{+}\right) = \frac{1}{\tilde{\Sigma}^{f,(2),\text{At}} \left(\omega + i 0^{+} \right)} - \frac{a^{\text{At}}}{\tilde{\Sigma}^{\text{At}} \left(\omega + i 0^{+} \right)},
\end{equation}
where the ${}^{\text{At}}$ superscript indicates that $a^{\text{At}}$ is obtained by evaluating \cref{app:eqn:IPT_a_constant} within the atomic limit, whereas $\tilde{\Sigma}^{f,(2),\text{At}} \left(\omega + i 0^{+} \right)$ and $\tilde{\Sigma}^{\text{At}} \left(\omega + i 0^{+} \right)$ are given, respectively, by \cref{app:eqn:at_self_energy_diagonal,app:eqn:self_energy_2nd_in_atomic}. The coefficient $a^{\text{At}}$ is explicitly given by
\begin{equation}
	\label{app:eqn:IPT_a_constant_at_limit}
	a^{\text{At}} = \frac{ n_{\text{At}} + \left(N_f - 2\right) \left\langle nn \right\rangle^{\text{At}}  - \left(N_f -1 \right) n_{\text{At}}^2}{n^{\tilde{\mu}}_0 \left(1 - n^{\tilde{\mu}}_0 \right)},
\end{equation}
where the $\left\langle nn \right\rangle^{\text{At}}$ correlator is computed within the atomic-limit Hamiltonian from \cref{app:eqn:atomic_Hamiltonian} at chemical potential $\mu$. \Cref{app:eqn:b_function_ipt} also inherently assumes that $b \left(\omega + i 0^{+}\right)$ is fully determined by $n_{\text{ss}}$, $\mu$, $n^{\tilde{\mu}}_0$, $\tilde{\mu}$, and $\left\langle nn \right\rangle^{\text{ss}}$ and has the same functional form both \emph{at} and \emph{away from} the atomic limit. Finally, we mention that both $\tilde{\Sigma}^{f,(2),\text{At}} \left(\omega + i 0^{+} \right)$ and $\tilde{\Sigma}^{\text{At}} \left(\omega + i 0^{+} \right)$ have analytical expressions in terms of $n_{\text{ss}}$, $\mu$, $n^{\tilde{\mu}}_0$, and $\tilde{\mu}$. As a result, $b \left(\omega + i 0^{+}\right) $ itself has an analytical (although tedious) expression.

We also note that as far as the $\omega$-dependence is concerned,
\begin{equation}
a^{\text{At}} = \frac{\tilde{\Sigma}^{\text{At}} \left(\omega + i 0^{+} \right)}{\tilde{\Sigma}^{f,(2),\text{At}} \left(\omega + i 0^{+} \right)} + \mathcal{O} \left( \frac{1}{\omega} \right),
\end{equation}
which follows from \cref{app:eqn:IPT_a_constant_at_limit} by considering the infinite frequency limit of $\tilde{\Sigma}^{\text{At}} \left(\omega + i 0^{+} \right)$ and $\tilde{\Sigma}^{f,(2),\text{At}} \left(\omega + i 0^{+} \right)$. As a result, we find that in the large $\omega$ limit, we must have that
\begin{equation}
\label{app:eqn:asymptote_b_infty}
b \left(\omega + i 0^{+}\right) = \frac{1}{\tilde{\Sigma}^{\text{At}} \left(\omega + i 0^{+} \right)} \left( \frac{\tilde{\Sigma}^{\text{At}} \left(\omega + i 0^{+} \right)}{\tilde{\Sigma}^{f,(2),\text{At}} \left(\omega + i 0^{+} \right)} - a^{\text{At}} \right) = \mathcal{O} \left( 1 \right),
\end{equation}
thus confirming our assumption from \cref{app:sec:se_symmetric_details:IPT:a} that $b \left(\omega + i 0^{+}\right)$ is bounded as $\omega \to \infty$. 

\subsubsection{Fixing the $\tilde{\mu}$ fictitious chemical potential}\label{app:sec:se_symmetric_details:IPT:fix_mu_tilde}

As discussed at the end of \cref{app:sec:se_symmetric_details:IPT:se_interpolated},  within the IPT method, the second-order self-energy correction $\tilde{\Sigma}^{f,(2)}\left( i \omega_n \right)$, which forms the basis for the self-energy ansatz in \cref{app:eqn:interpolated_sigma_def}, is derived from $G^{\tilde{\mu}}_{0}\left( i \omega_n \right)$ -- the effective Green's function of the single-site model at chemical potential $\tilde{\mu}$. The interpolated self-energy thus depends on the fictitious chemical potential $\tilde{\mu}$ and the $f$-electron filling $n^{\tilde{\mu}}_0$ of the non-interacting single-site model at $\tilde{\mu}$, in adition to the \emph{actual} $f$-electron filling $n$ and chemical potential $\mu$. The $\mu$ chemical potential is fixed at a given total filling of the THF model, which, in turn, determines $n$. Similarly, $n^{\tilde{\mu}}_0$ is fully specified from $G^{\tilde{\mu}}_{0} \left( i \omega_n \right)$ through \cref{app:eqn:int_rel_f0_filling}, once a certain $\tilde{\mu}$ chemical potential is assumed. The latter is thus a variational parameter of the problem: for fixed $G_{0} \left( i \omega_n \right)$ and $\mu$ there exist an \emph{entire} family of solutions parameterized by the fictitious chemical potential $\tilde{\mu}$. A method for fixing $\tilde{\mu}$ in a way that renders the interpolated self-energy as close as possible to the exact result is thus necessary.

Multiple methods have been proposed for fixing $\tilde{\mu}$, with the most important ones being:
\begin{enumerate}
\item Refs.~\cite{KAJ96, DAS16, FUJ03, SAS01, LIC98, ANI97, POT97, LAA03, YEY99, YEY00} fix $\tilde{\mu}$ by requiring that the interpolated self-energy $\tilde{\Sigma}^{f,\text{Int}} \left( \omega + i 0^{+} \right)$ obeys Luttinger's theorem~\cite{DAS16}
\begin{equation}
	-\frac{1}{\pi} \Im{\int_{-\infty}^{\infty} \dd{\omega} n_{\mathrm{F}} \left( \omega \right) \dv{\tilde{\Sigma}^{f,\text{Int}} \left( \omega + i 0^{+} \right)}{\omega} G \left( \omega + i 0^{+} \right)} = 0.
\end{equation}
Strictly speaking, Luttinger's theorem is only valid at zero temperatures. As a result, previous studies either restrict to the zero temperature regime~\cite{KAJ96, SAS01, LAA03}, impose Luttinger's theorem at finite temperature~\cite{FUJ03, LIC98, ANI97, POT97, YEY99, YEY00}, or use the $\tilde{\mu}$ chemical potential determined at zero temperature for finite temperature calculations~\cite{DAS16}.

\item Another method is to fix $\tilde{\mu}$ by requiring that $n^{\tilde{\mu}}_0 = n$~\cite{MAR86, YEY93, POT97, MEY99, YEY00, ARS12}. It has been shown that imposing $n^{\tilde{\mu}}_0 = n$ approximately satisfies Luttinger's theorem at zero temperature~\cite{YEY93, YEY00}, but has the added benefit of being usable at finite temperatures, as well. Moreover, this method has also been validated against exact diagonalization results~\cite{POT97}.

\item Finally, the simplest method is to take $\tilde{\mu} = \mu$. While this prescription does yield satisfactory results in the low interaction strength limit, it does not agree with exact diagonalization results for more general values of the interaction~\cite{POT97}.  
\end{enumerate}

In this work, we are interested exclusively in finite temperature results. Owing to its proven reliability in this context~\cite{POT97}, we will employ the second method of fixing $\tilde{\mu}$, which relies on imposing $n = n^{\tilde{\mu}}_0$. 

\subsubsection{Recovering the $U_1 \to 0$ limit}\label{app:sec:se_symmetric_details:IPT:low_u_limit}

So far we have verified that $\tilde{\Sigma}^{f,\text{Int}} \left( \omega + i 0^{+} \right)$ correctly recovers the large frequency behavior of the \emph{exact} self-energy, as well as the $U_1 \to \infty$ limit, wherein $\tilde{\Sigma}^{f,\text{Int}} \left( \omega + i 0^{+} \right)$ reduces to the dynamical self-energy of the atomic limit. One still needs to show that $\tilde{\Sigma}^{f,\text{Int}} \left( \omega + i 0^{+} \right)$ correctly reproduces the $U_1 \to 0$ limit.

As discussed in \cref{app:sec:se_symmetric_details:IPT:fix_mu_tilde}, in our implementation of IPT, $\tilde{\mu}$ is fixed such that $n = n^{\tilde{\mu}}_0$. At the same time, in the limit of vanishing interaction ($U_1 \to 0$), $G \left(i \omega_n \right) \to G_{0} \left(i \omega_n \right)$, which results in $\tilde{\mu} \to \mu$. In this limit, the $b \left(\omega + i 0^{+}\right)$ function can be computed by noting that 
\begin{align}
\tilde{\Sigma}^{\text{At}} \left( \omega + i 0^{+} \right) &= \frac{\left(N_f - 1 \right) U_1^2 \left[n_{\text{At}} + \left( N_f-2 \right) \left\langle nn \right\rangle^{\text{At}} -n_{\text{At}}^2 \left(N_f - 1 \right) \right]}{\omega + \mu + i 0^{+}} + \mathcal{O} \left(U_1^3 \right), \label{app:lim_low_u_1}\\ 
\tilde{\Sigma}^{f,(2),\text{At}} \left(\omega + i 0^{+} \right) &= \frac{U_1^2 \left(N_f - 1 \right) n \left( 1 - n \right)}{\omega + \mu + i 0^{+}}, \label{app:lim_low_u_2}\\ 
a^{\text{At}} &= \frac{ n_{\text{At}} + \left(N_f - 2\right) \left\langle nn \right\rangle^{\text{At}}  - \left(N_f -1 \right) n_{\text{At}}^2}{n \left(1 - n \right)},\label{app:lim_low_u_3}
\end{align}
which follow, respectively, from \cref{app:eqn:self_energy_2nd_in_atomic,app:eqn:low_u_asymptote_of_sigma_at,app:eqn:IPT_a_constant_at_limit}, by imposing $n^{\tilde{\mu}}_0 \to n$ and $\tilde{\mu} \to \mu$. Substituting \cref{app:lim_low_u_1,app:lim_low_u_2,app:lim_low_u_3} in \cref{app:eqn:b_function_ipt}, we obtain (for the $U_1$ dependence)
\begin{equation}
b \left( \omega + i 0^{+} \right) = \mathcal{O} \left( \frac{1}{U_1} \right).
\end{equation}
Since $\tilde{\Sigma}^{f,(2)}\left( \omega + i 0^{+} \right) = \mathcal{O} \left(U_1^2 \right)$, we find from \cref{app:eqn:interpolated_sigma_def} that 
\begin{equation}
\tilde{\Sigma}^{f,\text{Int}} \left( \omega + i 0^{+} \right) = a \tilde{\Sigma}^{f,(2)}\left( \omega + i 0^{+} \right) + \mathcal{O}\left( U_1^3 \right).
\end{equation}
We now note that in the vanishing interaction limit, $\left\langle nn \right\rangle_{\text{ss}} = n^2$, and as a result, from \cref{app:eqn:IPT_a_constant}, we find that
\begin{equation}
a = \frac{ n + \left(N_f - 2\right) n^2  - \left(N_f -1 \right) n^2}{n^{\tilde{\mu}}_0 \left(1 - n^{\tilde{\mu}}_0 \right)} = \frac{ n - n^2}{n^{\tilde{\mu}}_0 \left(1 - n^{\tilde{\mu}}_0 \right)} = 1,
\end{equation}
which then implies that indeed the exact low-$U_1$ dynamical self-energy is recovered up to third-order corrections ({\it i.e.}{}, $\tilde{\Sigma}^{f,\text{Int}} \left( \omega + i 0^{+} \right)$ is as accurate as $\tilde{\Sigma}^{f,(2)}\left( \omega + i 0^{+} \right)$ in the low interaction strength limit)
\begin{equation}
\tilde{\Sigma}^{f,\text{Int}} \left( \omega + i 0^{+} \right) = \tilde{\Sigma}^{f,(2)}\left( \omega + i 0^{+} \right) + \mathcal{O}\left( U_1^3 \right).
\end{equation}

\section{Additional result in many-body perturbation theory}\label{app:sec:additional_mb_results}

In this \siSection{}, we derive two useful results for many-body perturbation theory. We start by introducing a general fermionic Hamiltonian with quartic interaction. The interaction term is normal-ordered with respect to half-filling, akin to the THF model interaction Hamiltonian from \cref{app:sec:HF_review:interaction}. We then express the internal energy of the model in terms of its density matrix and self-energy correction. Finally, we compute the self-energy correction of this model to second order in perturbation theory using the path integral formalism. Taking the interacting Hamiltonian to be given by the $H_{U_1}$ and $H_{U_2}$ terms of the THF model from \cref{app:eqn:THF_int:U1,app:eqn:THF_int:U2}, respectively, we re-derive the second-order self-energy correction of the $f$-electrons, obtained already using a Feynman diagrammatic approach in \cref{app:sec:se_correction_beyond_HF:all_so}. 

\subsection{Model}\label{app:sec:additional_mb_results:model}

We will consider a general Hamiltonian of interacting fermions
\begin{equation}
	\label{app:eqn:gr_can_Ham}
	\mathcal{H} = H_0 + H_I,
\end{equation}
where 
\begin{align}
	H_0 &= h_{ij} \hat{\gamma}^\dagger_{i} \hat{\gamma}_{i} \label{app:eqn:non_int_h}\\
	H_I &= \frac{1}{2} V_{ijkl} \left(\hat{\gamma}^\dagger_{i} \hat{\gamma}_{j} - \frac{1}{2} \delta_{ij} \right) \left(\hat{\gamma}^\dagger_{k} \hat{\gamma}_{l} - \frac{1}{2} \delta_{kl} \right) = \frac{1}{8} V_{ijkl} \left(\hat{\gamma}^\dagger_{i} \hat{\gamma}_{j} - \hat{\gamma}_{j} \hat{\gamma}^\dagger_{i} \right) \left(\hat{\gamma}^\dagger_{k} \hat{\gamma}_{l} - \hat{\gamma}_{l} \hat{\gamma}^\dagger_{k} \right). \label{app:eqn:int_h}
\end{align}
In \cref{app:eqn:non_int_h,app:eqn:int_h} and throughout this \siSection{}, we will employ Einstein's summation convention, whereby repeated indices are summed over. We consider $N$ fermionic degrees of freedom denoted by $\hat{\gamma}^\dagger_{i}$ (where $1 \leq i \leq N$). The hermitian Hamiltonian matrix $h$ includes both the single-particle ({\it i.e.}{} kinetic) and chemical potential contributions, given respectively by the two terms in 
\begin{equation}
	h_{ij} = h'_{ij} - \mu \delta_{ij}.
\end{equation}
Additionally, $V$ is a four-fermion interaction tensor. Expanding the interacting Hamiltonian as 
\begin{equation}
	H_I = \frac{1}{2} V_{ijkl} \hat{\gamma}^\dagger_{i} \hat{\gamma}^\dagger_{k} \hat{\gamma}_{l} \hat{\gamma}_{j} - \frac{1}{4} V_{kkij} \hat{\gamma}^\dagger_{i} \hat{\gamma}_{j}  - \frac{1}{4} V_{ijkk} \hat{\gamma}^\dagger_{i} \hat{\gamma}_{j} + \frac{1}{8} V_{iijj},
\end{equation}
we see that the four-fermion interaction tensor $V_{ijkl}$ can be chosen without loss of generality to obey $V_{ijkl} = V_{klij}$, such that an equivalent expression for $H_I$ is given by
\begin{equation}
	\label{app:eqn:int_h_norm_ord}
	H_I = \frac{1}{2} V_{ijkl} \hat{\gamma}^\dagger_{i} \hat{\gamma}^\dagger_{k} \hat{\gamma}_{l} \hat{\gamma}_{j} - \frac{1}{2} V_{ijkk} \hat{\gamma}^\dagger_{i} \hat{\gamma}_{j} + \frac{1}{8} V_{iijj}. 
\end{equation} 
Finally, we note that because the chemical potential contribution is included implicitly in $H_0$, $\mathcal{H}$ is the grand canonical Hamiltonian of the system.

The Green's function of the system is defined as 
\begin{equation}
	\label{app:eqn:interacting_gf}
	-\left\langle \mathcal{T}_{\tau} \hat{\gamma}_{i} ( \tau ) \hat{\gamma}^\dagger_{j} ( 0 )  \right\rangle  = \mathcal{G}_{i j} ( \tau ),
\end{equation}
where $\tau$ is the imaginary time, $\mathcal{T}_{\tau}$ enforces the ordering of the operators that follow with respect to the imaginary time, and $\left\langle \hat{\mathcal{O}} \right\rangle$ denotes the expectation value of the operator $\hat{\mathcal{O}}$ in the grand canonical ensemble
\begin{equation}
	\left\langle \hat{\mathcal{O}} \right\rangle = \frac{\Tr\left[e^{-\beta \mathcal{H}} \hat{\mathcal{O}} \right]}{\Tr\left[e^{-\beta \mathcal{H}}\right]}.
\end{equation}
The fermion operators are also evolved using the grand canonical ensemble Hamiltonian $\mathcal{H}$
\begin{equation}
	\label{app:eqn:gr_can_evolved_gamma_ops_simple}
	\hat{\gamma}^\dagger_{i} ( \tau ) = e^{\mathcal{H} \tau} \hat{\gamma}^\dagger_{i} ( 0 ) e^{- \mathcal{H} \tau}.
\end{equation} 

The Fourier transformation of the Green's function is given by 
\begin{equation}
	\label{app:eqn:matsubara_gf_THF_ft_simple}
	\mathcal{G}_{i j} \left(i \omega_n \right) = \int_{0}^{\beta} \dd{\tau} e^{i \omega_n \tau} 	\mathcal{G}_{ij} ( \tau ) \qq{and}
	\mathcal{G}_{i j} ( \tau ) = \frac{1}{\beta} \sum_{i \omega_n}  \mathcal{G}_{ij} \left(i \omega_n \right) e^{ - i \omega_n \tau},
\end{equation}
with $\beta = 1/T$ being the inverse temperature and $\omega_n = \frac{(2 n + 1)\pi}{\beta}$, the fermionic Matsubara frequencies. It is also useful to define the \emph{non-interacting} Matsubara Green's function
\begin{equation}
	\label{app:eqn:non_interacting_gf_simple}
	-\left\langle \mathcal{T}_{\tau} \hat{\gamma}_{i}  ( \tau ) \hat{\gamma}^\dagger_{j} ( 0 )  \right\rangle_0  = \mathcal{G}^{0}_{ij} ( \tau ),
\end{equation}
whose definition is similar to \cref{app:eqn:interacting_gf}, with the only exception being that the imaginary time-evolution and the averaging $\left\langle \dots \right\rangle_0$ in \cref{app:eqn:non_interacting_gf_simple} are performed within the \emph{non-interacting} grand canonical ensemble Hamiltonian $H_0$. The non-interacting Green's function can readily be expressed in terms of the single-particle Hamiltonian
from \cref{app:eqn:non_int_h}, 
\begin{equation}
	\label{app:eqn:gf_within_non_int_simple}
	\mathcal{G}^{0} \left(i\omega_n \right) = \left[ i \omega_n \mathbb{1} - h \right]^{-1} .
\end{equation}

We also define the density matrix of the system
\begin{equation}
	\label{app:eqn:def_rho_HF_simple}
	\varrho_{ij} = \left\langle  \hat{\gamma}^\dagger_{i} \hat{\gamma}_{j} \right\rangle - \frac{1}{2} \delta_{ij} = \lim_{\tau \to 0^{-}} \mathcal{G}_{ji} (\tau) - \frac{1}{2} \delta_{ij}  = \lim_{\tau \to 0^{+}} \mathcal{G}_{ji} (\tau) + \frac{1}{2} \delta_{ij} = \frac{1}{2} \left( \mathcal{G}_{ji} (0^+) + \mathcal{G}_{ji} (0^-) \right).
\end{equation}
In terms of the density matrix, the Hartree-Fock Hamiltonian matrix reads as
\begin{equation}
	h^{\text{MF}}_{ij} = V_{ijkl} \varrho_{kl} - V_{ilkj} \varrho_{kl}.
\end{equation}
This allows to connect the interaction and the non-interacting Green's function of the system via Dyson's equation~\cite{MAH00}
\begin{equation}
	\label{app:eqn:dyson_equation_simple}
	\mathcal{G} \left( i \omega_n \right) = \left[ \left( \mathcal{G}^{0} \left( i \omega_n \right) \right)^{-1} - h^{\text{MF}} - \Sigma \left(i \omega_n \right) \right]^{-1}, 
\end{equation}
where $\Sigma \left(i \omega_n \right)$ is the dynamical self-energy matrix of the system.

\subsection{Internal energy}\label{app:sec:additional_mb_results:internal_u}

In this section, we will compute the internal energy of the system, which is defined as 
\begin{equation}
	\mathcal{U} = \left\langle \mathcal{H} + \mu \hat{N} \right\rangle = \left\langle H_0 + \mu \hat{N} \right\rangle + \left\langle H_I \right\rangle,
\end{equation}
where $\hat{N} = \hat{\gamma}^\dagger_{i}\hat{\gamma}_{i}$ is the total number operator and we remind the reader that the internal energy is the expectation value of the canonical, rather than the \emph{grand} canonical Hamiltonian ({\it i.e.}{} without the chemical potential contribution). The expectation value of the single-particle Hamiltonian can be readily expressed in terms of the density matrix of the system
\begin{equation}
	\left\langle H_0 + \mu \hat{N} \right\rangle = h'_{ij} \left\langle \hat{\gamma}^\dagger_{i} \hat{\gamma}_{j} \right\rangle = h'_{ij} \left( \varrho_{ij} + \frac{1}{2} \delta_{ij} \right). \label{app:eqn:expec_sp_en_final}
\end{equation}
The expectation value of the interacting Hamiltonian requires a more careful treatment. We first note that at the Hartree-Fock level, $\left\langle H_I \right\rangle$ can be readily computed using Wick's theorem
\begin{align}
	\left\langle H_I \right\rangle_{\text{MF}}&=\frac{1}{2} V_{ijkl} \left\langle \left( \hat{\gamma}^\dagger_{i} \hat{\gamma}_{j} -\frac{1}{2} \delta_{ij} \right) \left( \hat{\gamma}^\dagger_{k} \hat{\gamma}_{l} -\frac{1}{2} \delta_{kl} \right) \right\rangle_{\text{MF}} \nonumber \\
	&=\frac{1}{2} V_{ijkl} \left\langle \hat{\gamma}^\dagger_{i} \hat{\gamma}_{j} \hat{\gamma}^\dagger_{k} \hat{\gamma}_{l}  \right\rangle_{\text{MF}} - \frac{1}{2}V_{ijkk} \left\langle \hat{\gamma}^\dagger_{i} \hat{\gamma}_{j} - \frac{1}{2} \delta_{ij}  \right\rangle_{\text{MF}} - \frac{1}{8}V_{iijj} \nonumber \\
	&=- \frac{1}{2} V_{ijkl} \left\langle \hat{\gamma}^\dagger_{i} \hat{\gamma}^\dagger_{k} \hat{\gamma}_{j} \hat{\gamma}_{l}  \right\rangle_{\text{MF}} + \frac{1}{2} V_{ikkj} \left\langle \hat{\gamma}^\dagger_{i} \hat{\gamma}_{j} \right\rangle_{\text{MF}} - \frac{1}{2}V_{ijkk} \left\langle \hat{\gamma}^\dagger_{i} \hat{\gamma}_{j} - \frac{1}{2} \delta_{ij}  \right\rangle_{\text{MF}} - \frac{1}{8}V_{iijj} \nonumber \\
	&=\frac{1}{2} V_{ijkl} \left( \varrho_{ij} + \frac{1}{2} \delta_{ij} \right) \left( \varrho_{kl} + \frac{1}{2} \delta_{kl} \right)
	-\frac{1}{2} V_{ijkl} \left( \varrho_{il} + \frac{1}{2} \delta_{il} \right) \left( \varrho_{kj} + \frac{1}{2} \delta_{kj} \right) \nonumber \\
	&+ \frac{1}{2} V_{ikkj} \left( \varrho_{ij} + \frac{1}{2} \delta_{ij} \right) - \frac{1}{2}V_{ijkk} \varrho_{ij} - \frac{1}{8}V_{iijj} \nonumber \\
	&=\frac{1}{2} V_{ijkl} \varrho_{ij} \varrho_{kl}
	-\frac{1}{2} V_{ijkl} \left( \varrho_{il} + \frac{1}{2} \delta_{il} \right) \left( \varrho_{kj} + \frac{1}{2} \delta_{kj} \right) + \frac{1}{2} V_{ikkj} \varrho_{ij} + \frac{1}{4} V_{ijji} \nonumber \\
	&=\frac{1}{2} V_{ijkl} \varrho_{ij} \varrho_{kl}
	-\frac{1}{2} V_{ijkl} \varrho_{il} \varrho_{kj} + \frac{1}{8} V_{ijji} = \frac{1}{2} h^{\text{MF}}_{ij} \varrho_{ij} + \frac{1}{8} V_{ijji}. \label{app:eqn:interaction_mean_field}
\end{align}
In \cref{app:eqn:interaction_mean_field} and in what follows, $\left\langle \dots \right\rangle_{\text{MF}}$ indicates that the expectation value is computed at the mean-field ({\it i.e.}{}, Hartree-Fock) level. To evaluate $\left\langle H_I \right\rangle$ \emph{beyond} the Hartree-Fock level, we need to employ the Green's function equation of motion. Starting from the defintion in \cref{app:eqn:interacting_gf}, we find that 
\begin{equation}
	\label{app:eqn:eom_gf_interacting}
	- \lim_{\tau \to 0^{-}} \pdv{\mathcal{G}_{aa} (\tau)}{\tau} = - \lim_{\tau \to 0^{-}} \left\langle \hat{\gamma}^\dagger_{a} (0) \pdv{\tau} \hat{\gamma}_{a} (\tau) \right\rangle = - \lim_{\tau \to 0^{-}} \left\langle \hat{\gamma}^\dagger_{a} (0) \commutator{\mathcal{H}}{ \hat{\gamma}_{a} (\tau)} \right\rangle.
\end{equation} 
The commutator can be readily evaluated and gives
\begin{equation}
	\label{app:eqn:commutator_for_gf}
	\commutator{\mathcal{H}}{ \hat{\gamma}_{a}} = - h_{aj} \hat{\gamma}_{j} - \frac{1}{2} V_{ajkl} \hat{\gamma}_{j} \left( \hat{\gamma}^\dagger_{k} \hat{\gamma}_{l} - \frac{1}{2} \delta_{kl} \right) - \frac{1}{2} V_{ijal} \left( \hat{\gamma}^\dagger_{i} \hat{\gamma}_{j} - \frac{1}{2}\delta_{ij} \right) \hat{\gamma}_{l}.
\end{equation}
Substituting \cref{app:eqn:commutator_for_gf} into \cref{app:eqn:eom_gf_interacting}, we obtain 
\begin{align}
	- \lim_{\tau \to 0^{-}} \pdv{\mathcal{G}_{aa} (\tau)}{\tau} &= h_{aj} \lim_{\tau \to 0^{-}} \left\langle \hat{\gamma}^\dagger_{a} (0) \hat{\gamma}_{j} (\tau) \right\rangle + \frac{1}{2} V_{ajkl} \lim_{\tau \to 0^{-}} \left\langle \hat{\gamma}^\dagger_{a} (0) \hat{\gamma}_{j} ( \tau ) \left( \hat{\gamma}^\dagger_{k}  ( \tau ) \hat{\gamma}_{l}  ( \tau ) - \frac{1}{2} \delta_{kl} \right) \right\rangle \nonumber \\
	&+ \frac{1}{2} V_{ijal} \lim_{\tau \to 0^{-}} \left\langle \hat{\gamma}^\dagger_{a} (0) \left( \hat{\gamma}^\dagger_{i} (\tau) \hat{\gamma}_{j} (\tau) - \frac{1}{2}\delta_{ij} \right) \hat{\gamma}_{l} (\tau)  \right\rangle \nonumber \\
	&= h_{aj} \lim_{\tau \to 0^{-}} \mathcal{G}_{ja} (\tau) + \frac{1}{2} V_{ijkl} \left[ \left\langle \hat{\gamma}^\dagger_{i} \hat{\gamma}_{j} \left( \hat{\gamma}^\dagger_{k} \hat{\gamma}_{l} - \frac{1}{2} \delta_{kl} \right) \right\rangle + \left\langle \hat{\gamma}^\dagger_{k} \left( \hat{\gamma}^\dagger_{i} \hat{\gamma}_{j} - \frac{1}{2}\delta_{ij} \right) \hat{\gamma}_{l}  \right\rangle \right] \nonumber  \\
	&= h_{aj} \lim_{\tau \to 0^{-}} \mathcal{G}_{ja} (\tau) + \frac{1}{2} V_{ijkl} \left[ \left\langle \hat{\gamma}^\dagger_{i} \hat{\gamma}_{j} \left( \hat{\gamma}^\dagger_{k} \hat{\gamma}_{l} - \frac{1}{2} \delta_{kl} \right) \right\rangle + \left\langle  \left( \hat{\gamma}^\dagger_{i} \hat{\gamma}_{j} - \frac{1}{2}\delta_{ij} \right) \hat{\gamma}^\dagger_{k} \hat{\gamma}_{l}  \right\rangle + \left\langle \commutator{ \hat{\gamma}^\dagger_{k}}{ \hat{\gamma}^\dagger_{i} \hat{\gamma}_{j}} \hat{\gamma}_{l}  \right\rangle  \right] \nonumber  \\
	&= h_{aj} \lim_{\tau \to 0^{-}} \mathcal{G}_{ja} (\tau) + 2 \left\langle H_I \right\rangle + \frac{1}{2} V_{ijkl} \left[ \frac{1}{2} \delta_{ij} \varrho_{kl} + \frac{1}{2} \delta_{kl} \varrho_{ij} - \delta_{jk} \left( \varrho_{il} + \frac{1}{2} \delta_{il} \right)  \right]  \nonumber  \\
	&= h_{aj} \lim_{\tau \to 0^{-}} \mathcal{G}_{ja} (\tau) + 2 \left\langle H_I \right\rangle + \frac{1}{2} V_{iijk} \varrho_{jk} - \frac{1}{2} V_{ijjl} \varrho_{il} - \frac{1}{4}V_{ijji}   \nonumber  \\
	&= h_{ij} \lim_{\tau \to 0^{-}} \mathcal{G}_{ji} (\tau) + 2 \left\langle H_I \right\rangle + \frac{1}{2} \left( V_{kkij} - V_{ikkj} \right) \lim_{\tau \to 0^{-}} \left( \mathcal{G}_{ji}(\tau) - \frac{1}{2} \delta_{ij} \right)  - \frac{1}{4}V_{ijji}   \nonumber  \\
	&= \left[ h_{ij} + \frac{1}{2} \left( V_{kkij} - V_{ikkj} \right) \right] \lim_{\tau \to 0^{-}} \mathcal{G}_{ji} (\tau) + 2 \left\langle H_I \right\rangle - \frac{1}{4}V_{iijj}. \label{app:eqn:eom_HI}
\end{align}
By Fourier-transforming \cref{app:eqn:eom_HI} in Matsubara frequency according to \cref{app:eqn:matsubara_gf_THF_ft_simple}, we obtain
\begin{align}
	2 \left\langle H_I \right\rangle - \frac{1}{4} V_{iijj} &= \frac{1}{\beta} \sum_{i \omega_n} \left[ i \omega_n \delta_{ij} - h_{ij} - \frac{1}{2} \left( V_{kkij} - V_{ikkj} \right) \right] \mathcal{G}_{ji} \left( i\omega_n \right) e^{i \omega_n 0^{+}} \nonumber \\
	&= \frac{1}{\beta} \sum_{i \omega_n} \left[ \left[ \mathcal{G}^{-1} \left( i \omega_n \right) \right]_{ij} - \frac{1}{2} \left( V_{kkij} - V_{ikkj} \right) + \Sigma_{ij} \left( i \omega_n \right) + h^{\text{MF}}_{ij} \right] \mathcal{G}_{ji} \left( i\omega_n \right) e^{i \omega_n 0^{+}} \nonumber \\
	&= \frac{1}{\beta} \sum_{i \omega_n} \delta_{ii} e^{i \omega_n 0^{+}} + \frac{1}{\beta} \sum_{i \omega_n} \left[  \Sigma_{ij} \left( i \omega_n \right) + h^{\text{MF}}_{ij} - \frac{1}{2} \left( V_{kkij} - V_{ikkj} \right) \right] \mathcal{G}_{ji} \left( i\omega_n \right) e^{i \omega_n 0^{+}}
	\nonumber \\
	&= \frac{1}{\beta} \sum_{i \omega_n} \delta_{ii} e^{i \omega_n 0^{+}} + \frac{1}{\beta} \sum_{i \omega_n} \Sigma_{ij} \left( i \omega_n \right) \mathcal{G}_{ji} \left( i\omega_n \right) e^{i \omega_n 0^{+}} \nonumber \\
	&+ \frac{1}{\beta} \sum_{i \omega_n} \left[  h^{\text{MF}}_{ij} - \frac{1}{2} \left( V_{kkij} - V_{ikkj} \right) \right] \mathcal{G}_{ji} \left( i\omega_n \right) e^{i \omega_n 0^{+}}. \label{app:eqn:expec_interact_en_1}
\end{align}
To move forward, we can compute the last term of \cref{app:eqn:expec_interact_en_1} by deriving 
\begin{equation}
	\sum_{i\omega_n} \mathcal{G}_{ji} \left( i\omega_n \right) e^{i \omega_n 0^{+}} = \lim_{\tau \to 0^{-}} \mathcal{G}_{ji} \left( \tau \right) = \varrho_{ij} + \frac{1}{2} \delta_{ij},
\end{equation}
which follows straightforwardly from   \cref{app:eqn:matsubara_gf_THF_ft_simple,app:eqn:def_rho_HF_simple}. For the other two terms, we employ \cref{app:eqn:matsubara_sum_s_g,app:eqn:matsubara_sum_s_g_evaluated}, which, for a general function $g(z)$, lead to
\begin{equation}
	\label{app:eqn:general_exp_zero_plus}
	\frac{1}{\beta} \sum_{i \omega_n} g \left( i \omega_n \right) e^{i \omega_n 0^{+}} = \sum_{z_0 \in \text{poles of $z_0$}} \Res g\left(z_0 \right) n_{\mathrm{F}} \left( z_0 \right),
\end{equation}
as was shown in \cref{app:eqn:matsubara_sum_s_g,app:eqn:matsubara_sum_s_g_evaluated}. This immediately implies that the first term in \cref{app:eqn:expec_interact_en_1} vanishes. For the third term, we note that, since the both the self-energy and Green's function are analytical above and below the real axis~\cite{LUT61,PAV19}, the trace of their matrix product is also analytical above and below the real axis, and is therefore endowed with a spectral representation~\cite{LUT61,PAV19}
\begin{equation}
	\label{app:eqn:spect_rep_of_tr_sigma_gf}
	\Tr \left[ \Sigma (z) \mathcal{G} (z) \right] = -\frac{1}{\pi} \int_{-\infty}^{\infty} \frac{\dd{\omega}}{z - \omega} \Im \left\lbrace \Tr \left[ \Sigma \left( \omega + i 0^{+} \right) \mathcal{G} \left( \omega + i 0^{+} \right) \right] \right\rbrace .
\end{equation}
\Cref{app:eqn:general_exp_zero_plus,app:eqn:spect_rep_of_tr_sigma_gf} then leads to 
\begin{equation}
	\frac{1}{\beta} \sum_{i \omega_n} \Sigma_{ij} \left( i \omega_n \right) \mathcal{G}_{ji} \left( i\omega_n \right) e^{i \omega_n 0^{+}} = -\frac{1}{\pi} \int_{-\infty}^{\infty} \dd{\omega} \Im \left\lbrace \Tr \left[ \Sigma \left( \omega + i 0^{+} \right) \mathcal{G} \left( \omega + i 0^{+} \right) \right] \right\rbrace n_{\mathrm{F}} \left( \omega \right).
\end{equation}
As such, \cref{app:eqn:expec_interact_en_1} can be simplified to 
\begin{align}
	2 \left\langle H_I \right\rangle - \frac{1}{4} V_{iijj} &= -\frac{1}{\pi} \int_{-\infty}^{\infty} \dd{\omega} \Im \left\lbrace \Tr \left[ \Sigma \left( \omega + i 0^{+} \right) \mathcal{G} \left( \omega + i 0^{+} \right) \right] \right\rbrace n_{\mathrm{F}} \left( \omega \right) + \left[ h^{\text{MF}}_{ij} - \frac{1}{2} \left( V_{kkij} - V_{ikkj} \right) \right] \left( \varrho_{ij} + \frac{1}{2} \delta_{ij} \right)
	\nonumber \\
	&= \left[ h^{\text{MF}}_{ij} - \frac{1}{2} \left( V_{kkij} - V_{ikkj} \right) \right] \left( \varrho_{ij} + \frac{1}{2} \delta_{ij} \right) - \frac{1}{\pi} \int_{-\infty}^{\infty} \dd{\omega} \Im \left\lbrace \Tr \left[ \Sigma \left( \omega + i 0^{+} \right) \mathcal{G} \left( \omega + i 0^{+} \right) \right] \right\rbrace n_{\mathrm{F}} \left( \omega \right) \nonumber \\
	&= h^{\text{MF}}_{ij} \rho_{ij} + \frac{1}{2} h^{\text{MF}}_{ii} - \frac{1}{2} h^{\text{MF}}_{ii} - \frac{1}{4} \left( V_{iijj} - V_{ijji} \right) - \frac{1}{\pi} \int_{-\infty}^{\infty} \dd{\omega} \Im \left\lbrace \Tr \left[ \Sigma \left( \omega + i 0^{+} \right) \mathcal{G} \left( \omega + i 0^{+} \right) \right] \right\rbrace n_{\mathrm{F}} \left( \omega \right)  \nonumber \\
	&= h^{\text{MF}}_{ij} \rho_{ij}  - \frac{1}{4} \left( V_{iijj} - V_{ijji} \right) - \frac{1}{\pi} \int_{-\infty}^{\infty} \dd{\omega} \Im \left\lbrace \Tr \left[ \Sigma \left( \omega + i 0^{+} \right) \mathcal{G} \left( \omega + i 0^{+} \right) \right] \right\rbrace n_{\mathrm{F}} \left( \omega \right). \label{app:eqn:expec_interact_en_2}
\end{align}
In turn, \cref{app:eqn:expec_interact_en_2} allows us to obtain the \emph{exact} expectation value of the interaction Hamiltonian
\begin{align}
	\left\langle H_I \right\rangle &= \frac{1}{2} h^{\text{MF}}_{ij} \varrho_{ij} + \frac{1}{8} V_{ijji} - \frac{1}{2\pi} \int_{-\infty}^{\infty} \dd{\omega} \Im \left\lbrace \Tr \left[ \Sigma \left( \omega + i 0^{+} \right) \mathcal{G} \left( \omega + i 0^{+} \right) \right] \right\rbrace n_{\mathrm{F}} \left( \omega \right) \nonumber \\
	&= \left\langle H_I \right\rangle_{\text{MF}} - \frac{1}{2\pi} \int_{-\infty}^{\infty} \dd{\omega} \Im \left\lbrace \Tr \left[ \Sigma \left( \omega + i 0^{+} \right) \mathcal{G} \left( \omega + i 0^{+} \right) \right] \right\rbrace n_{\mathrm{F}} \left( \omega \right). \label{app:eqn:expec_interact_en_final}
\end{align}

The total internal energy of the system can be obtained by combining \cref{app:eqn:expec_sp_en_final,app:eqn:expec_interact_en_final}.

\subsection{Second-order self-energy correction}\label{app:sec:additional_mb_results:self-energy}

In this section, we obtain the self-energy of the model from \cref{app:eqn:gr_can_Ham} to second order in the interaction. To do so, we first define the action of the system
\begin{align}
	S \left[ \hat{\gamma}^\dagger_{}, \hat{\gamma}_{}, \eta^{\dagger}_{}, \eta_{} \right] &= \int_{0}^{\beta} \dd{\tau} \left( \hat{\gamma}^\dagger_{i} (\tau) \partial_{\tau} \hat{\gamma}_{i} (\tau) + \mathcal{H} + \hat{\gamma}^\dagger_{i} ( \tau ) \eta_{i} ( \tau ) + \eta^{\dagger}_{i} (\tau)  \hat{\gamma}_{i} (\tau) \right) \nonumber \\
	&= S_0 \left[ \hat{\gamma}^\dagger_{}, \hat{\gamma}_{}, \eta^{\dagger}_{}, \eta_{} \right] + S_I \left[ \hat{\gamma}^\dagger_{}, \hat{\gamma}_{} \right], \label{app:eqn:general_total_action}
\end{align}
where the single-particle and interaction contributions are respectively given by
\begin{align}
	S_0 \left[ \hat{\gamma}^\dagger_{}, \hat{\gamma}_{}, \eta^{\dagger}_{}, \eta_{} \right] &= \int_{0}^{\beta} \dd{\tau} \left[ \hat{\gamma}^\dagger_{i} (\tau) \left( \partial_{\tau} \delta_{ij} + h_{ij} \right) \hat{\gamma}_{j} (\tau) + \hat{\gamma}^\dagger_{i} ( \tau ) \eta_{i} ( \tau ) + \eta^{\dagger}_{i} (\tau)  \hat{\gamma}_{i} (\tau) \right], \label{app:eqn:general_nonint_action} \\
	S_I \left[ \hat{\gamma}^\dagger_{}, \hat{\gamma}_{} \right] &= \int_{0}^{\beta} \dd{\tau} \frac{1}{8} V_{ijkl} \left( \hat{\gamma}^\dagger_{i} (\tau + 0^{+}) \hat{\gamma}_{j} (\tau) + \hat{\gamma}^\dagger_{i} (\tau) \hat{\gamma}_{j} (\tau + 0^+) \right) \left( \hat{\gamma}^\dagger_{k} (\tau + 0^{+}) \hat{\gamma}_{l} (\tau) + \hat{\gamma}^\dagger_{k} (\tau) \hat{\gamma}_{l} (\tau + 0^+) \right). \label{app:eqn:general_int_action}
\end{align}
In \cref{app:eqn:general_total_action}, we have introduced a source term $\eta_{i} (\tau)$ for the $\hat{\gamma}^\dagger_{i}$ fermions. Note also that the interaction contribution to the action was written using the second form of the interaction of $H_I$ from \cref{app:eqn:int_h}. To control the relative ordering of the fermionic operators appearing in the interaction, a small positive offset was added to the imaginary time for the operator that is to be acted last.

The partition function of the (interacting) system in the presence of a source reads as
\begin{equation}
	\label{app:eqn:path_integral_full_int}
	Z \left[ \eta^{\dagger}_{}, \eta_{} \right] \equiv \int \mathcal{D}\left[ \hat{\gamma}_{}, \hat{\gamma}^\dagger_{} \right] e^{-S  \left[ \hat{\gamma}^\dagger_{}, \hat{\gamma}_{}, \eta^{\dagger}_{}, \eta_{} \right]}.
\end{equation}
In turn, this allows us to express the full Green's function of the system as a functional derivative
\begin{equation}
	\label{app:eqn:path_integral_full_gf}
	\mathcal{G}_{ij} \left( \tau_2 - \tau_1 \right) =  \eval{ \left( \frac{1}{Z \left[ \eta^{\dagger}_{}, \eta_{} \right]} \fdv{\eta^{\dagger}_{i}(\tau_2)} \fdv{\eta_{j}(\tau_1)} Z \left[ \eta^{\dagger}_{}, \eta_{} \right] \right)}_{\eta^{\dagger}_{},\eta_{}=0} = \eval{ \left( \fdv{\eta^{\dagger}_{i}(\tau_2)} \fdv{\eta_{j}(\tau_1)} \log Z \left[ \eta^{\dagger}_{}, \eta_{} \right] \right)}_{\eta^{\dagger}_{},\eta_{}=0}.
\end{equation}
Additionally, we also define the partition of the \emph{non-interacting} system as 
\begin{equation}
	Z_0 \left[ \eta^{\dagger}_{}, \eta_{} \right] \equiv \int \mathcal{D}\left[ \hat{\gamma}_{}, \hat{\gamma}^\dagger_{} \right] e^{-S_0  \left[ \hat{\gamma}^\dagger_{}, \hat{\gamma}_{}, \eta^{\dagger}_{}, \eta_{} \right]}.
\end{equation}
The partition function of the non-interacting system can be computed directly by Fourier-transforming the source and the Fermion operators in frequency space~\cite{ALT10}
\begin{alignat}{4}
	\hat{\gamma}_{i} (\tau) &&=& \frac{1}{\sqrt{\beta}} \sum_{i \omega_n} \hat{\gamma}_{i} \left( i \omega_n \right) e^{-i \omega_n \tau},& \quad
	\hat{\gamma}_{i} \left( i \omega_n \right) &&= &\frac{1}{\sqrt{\beta}} \int_{0}^{\beta} \dd{\tau} \hat{\gamma}_{i} (\tau) e^{i \omega_n \tau}, \\
	\eta_{i} (\tau) &&=& \frac{1}{\sqrt{\beta}} \sum_{i \omega_n} \eta_{i} \left( i \omega_n \right) e^{-i \omega_n \tau},& \quad
	\eta_{i} \left( i \omega_n \right) &&= &\frac{1}{\sqrt{\beta}} \int_{0}^{\beta} \dd{\tau} \eta_{i} (\tau) e^{i \omega_n \tau}, \\
\end{alignat}
in terms of which, the non-interacting action becomes
\begin{align}
	S_0 \left[ \hat{\gamma}^\dagger_{}, \hat{\gamma}_{}, \eta^{\dagger}_{}, \eta_{} \right] = -\sum_{i \omega_n} \left[ \hat{\gamma}^\dagger_{i} \left( i\omega_n \right) \left( i\omega_n \delta_{ij} - h_{ij} \right) \hat{\gamma}_{j} \left( i\omega_n \right) + \hat{\gamma}^\dagger_{i} \left( i\omega_n \right) \eta_{i} \left( i\omega_n \right) + \eta^{\dagger}_{i} \left( i\omega_n \right)  \hat{\gamma}_{i} \left( i\omega_n \right)\right].
\end{align}
Using the Gaussian integral over Grassman variables~\cite{ALT10}, we obtain
\begin{align}
	Z_{0} \left[ \eta^{\dagger}_{}, \eta_{} \right] &= Z_{0} \left[0, 0 \right] \exp \left( - \sum_{i \omega_n} \eta^{\dagger}_{i} \left( i\omega_n \right) \mathcal{G}^{0}_{ij} \left( i\omega_n \right) \eta_{j} \left( i \omega_n \right) \right), \nonumber \\
	&= Z_{0} \left[0, 0 \right] \exp \left( - \int_{0}^{\beta} \dd{\tau_1} \int_{0} ^{\beta} \dd{\tau_2} \eta^{\dagger}_{i} (\tau_2) \mathcal{G}^{0}_{ij} \left( \tau_2 - \tau_1 \right)  \eta_{j} (\tau_1) \right),
\end{align}
where we have employed the non-interacting Green's function of the system defined in \cref{app:eqn:non_interacting_gf_simple}.

Our strategy for finding the second-order self-energy correction will be the following:
\begin{enumerate}
	\item Starting from Dyson's equation from \cref{app:eqn:dyson_equation_simple}, we express the dynamical self-energy of the problem in powers of the interaction tensor and truncate it to second order
	\begin{equation}
		\label{app:eqn:expansion_sigma_path_int}
		\Sigma \left( i \omega_n \right) = \Sigma^{(2)} \left( i \omega_n \right) + \mathcal{O} \left( V^3 \right),
	\end{equation}
	where $\Sigma^{(2)} \left( i \omega_n \right)$ denotes the second-order contribution to the self-energy. Note that the static contribution to the self-energy $h^{\text{MF}}$ is first order in the interaction. Substituting \cref{app:eqn:expansion_sigma_path_int} into \cref{app:eqn:dyson_equation_simple}, we find that 
	\begin{align}
		\mathcal{G} \left( i \omega_n \right) &= \mathcal{G}^{0} \left( i \omega_n \right) + \mathcal{G}^{0} \left( i \omega_n \right) h^{\text{MF}} \mathcal{G}^{0} \left( i \omega_n \right)  + \mathcal{G}^{0} \left( i \omega_n \right) h^{\text{MF}} \mathcal{G}^{0} \left( i \omega_n \right) h^{\text{MF}} \mathcal{G}^{0} \left( i \omega_n \right)  \nonumber \\ 
		&+ \mathcal{G}^{0} \left( i \omega_n \right) \Sigma^{(2)} \left( i \omega_n \right) \mathcal{G}^{0} \left( i \omega_n \right)  + \mathcal{O} \left( V^3 \right), \label{app:eqn:expansion_full_gf_self_energy}
	\end{align}
	which is just the expression of the \emph{full} Green's function of the system up to second order in the interaction.
	
	\item Using \cref{app:eqn:path_integral_full_gf}, we compute the full Green's function of the system perturbatively to second order in the interaction. 
	
	\item By comparing the expression of the full Green's function with the one in \cref{app:eqn:expansion_full_gf_self_energy}, we can directly read off the first and second-order self-energy contributions.
\end{enumerate}
A fully rigorous derivation also proves Dyson's equation, which we assume to hold without proof. The reader is pointed to Ref.~\cite{KOP10} for such a complete derivation. We now derive the interacting Green's function of the system to second order in the interaction.

\subsubsection{Second-order interacting Green's function}\label{app:sec:additional_mb_results:self-energy:interacting_gf}

To find the interaction Green's function of the system, we use \cref{app:eqn:path_integral_full_gf} and compute the partition function of the interacting system in the presence of the source perturbatively in powers of the interaction tensor $V_{ijkl}$. This can be done by noting that from \cref{app:eqn:general_total_action,app:eqn:path_integral_full_int}
\begin{align}
	Z \left[ \eta^{\dagger}_{}, \eta_{} \right] =& \int \mathcal{D}\left[ \hat{\gamma}_{}, \hat{\gamma}^\dagger_{} \right] e^{-S_I \left[ \hat{\gamma}^\dagger_{}, \hat{\gamma}_{} \right]} e^{-S_0 \left[ \hat{\gamma}^\dagger_{}, \hat{\gamma}_{}, \eta^{\dagger}_{}, \eta_{} \right]} \nonumber \\
	=& \int \mathcal{D}\left[ \hat{\gamma}_{}, \hat{\gamma}^\dagger_{} \right] \exp \left[ - \int_{0}^{\beta} \dd{\tau} \frac{1}{8} V_{ijkl} \left( \fdv{\eta_{i} (\tau + 0^+)} \fdv{\eta^{\dagger}_{j} (\tau)} + \fdv{\eta_{i} (\tau)} \fdv{\eta^{\dagger}_{j} (\tau + 0^+)} \right) 
	\right. \nonumber \\ 
	&\left. \left( \fdv{\eta_{k} (\tau + 0^+)} \fdv{\eta^{\dagger}_{l} (\tau)} + \fdv{\eta_{k} (\tau)} \fdv{\eta^{\dagger}_{l} (\tau + 0^+)} \right) \right] e^{-S_0 \left[ \hat{\gamma}^\dagger_{}, \hat{\gamma}_{}, \eta^{\dagger}_{}, \eta_{} \right]} \nonumber \\
	=& Z_{0} \left[0, 0 \right] \exp \left[ - \int_{0}^{\beta} \dd{\tau} \frac{1}{8} V_{ijkl} \left( \fdv{\eta_{i} (\tau + 0^+)} \fdv{\eta^{\dagger}_{j} (\tau)} + \fdv{\eta_{i} (\tau)} \fdv{\eta^{\dagger}_{j} (\tau + 0^+)} \right) 
	\right. \nonumber \\ 
	&\left. \left( \fdv{\eta_{k} (\tau + 0^+)} \fdv{\eta^{\dagger}_{l} (\tau)} + \fdv{\eta_{k} (\tau)} \fdv{\eta^{\dagger}_{l} (\tau + 0^+)} \right) \right] \exp \left( - \int_{0}^{\beta} \dd{\tau_1} \dd{\tau_2} \eta^{\dagger}_{i} (\tau_2) \mathcal{G}^{0}_{ij} \left( \tau_2 - \tau_1 \right)  \eta_{j} (\tau_1) \right). \label{app:eqn:exact_partition_function_expansion}
\end{align} 
We now expand the first exponential in powers of the interaction tensor. The functional derivatives with respect to the source field will ``pull down'' Green's functions from the exponential of the non-interacting partition function. The functional derivatives with respect to the source fields contain small positive imaginary time offsets, which will result in offsets in the argument of the non-interacting Green's function $\mathcal{G}^{0}_{ij} (\tau) $. The latter is continuous for any value of the imaginary time, \emph{except} for $\tau=0$, where it is discontinuous. As such, any small offset appearing in the Green's function can be safely ignored whenever the argument of the Green's function is non-zero, but needs to be taken into account when the Green's function's argument \emph{is} vanishing. 

Expanding the exponential to second order in the interactions, acting the functional derivatives, and using the symmetry of the interaction tensor $V_{ijkl} = V_{klij}$ assumed above \cref{app:eqn:int_h_norm_ord}, we find that 
\begin{align}
	Z \left[ \eta^{\dagger}_{}, \eta_{} \right] =&  Z_{0} \left[0, 0 \right] \exp \left( - \int_{0}^{\beta} \dd{\tau_1} \dd{\tau_2} \eta^{\dagger}_{a} (\tau_2) \mathcal{G}^{0}_{ab} \left( \tau_2 - \tau_1 \right)  \eta_{b} (\tau_1) \right) \nonumber \\
	\times& \left[
	1 - \frac{1}{2} V_{ijkl} \beta \left( \mathcal{G}^{0}_{lk} (0) \mathcal{G}^{0}_{ji} (0) - \mathcal{G}^{0}_{li} (0) \mathcal{G}^{0}_{jk} (0) \right) 
	\right. \nonumber \\
	& - V_{ijkl} \int_{0}^{\beta} \dd{\tau} \dd{\tau_1}  \dd{\tau_2} \eta^{\dagger}_{m} (\tau_2) \eta_{n} (\tau_1) \left( \mathcal{G}^{0}_{mi} (\tau_2 - \tau) \mathcal{G}^{0}_{jn} (\tau - \tau_1) \mathcal{G}^{0}_{lk} (0) - \mathcal{G}^{0}_{mi} (\tau_2 - \tau) \mathcal{G}^{0}_{ln} (\tau - \tau_1) \mathcal{G}^{0}_{jk} (0) \right) \nonumber \\
& - V_{ijkl} V_{mnop} \int_{0}^{\beta} \dd{\tau} \dd{\tau'} \dd{\tau_1} \dd{\tau_2} \eta^{\dagger}_{q} (\tau_2) \eta_{r} (\tau_1) \mathcal{G}^{0}_{qm} (\tau_2 - \tau) \mathcal{G}^{0}_{nr} (\tau - \tau_1) \mathcal{G}^{0}_{jo} (\tau' - \tau) \mathcal{G}^{0}_{pi} (\tau - \tau') \mathcal{G}^{0}_{lk} (0) \nonumber \\
	& + V_{ijkl} V_{mnop} \int_{0}^{\beta} \dd{\tau} \dd{\tau'} \dd{\tau_1} \dd{\tau_2} \eta^{\dagger}_{q} (\tau_2) \eta_{r} (\tau_1) \mathcal{G}^{0}_{qm} (\tau_2 - \tau) \mathcal{G}^{0}_{nr} (\tau - \tau_1) \mathcal{G}^{0}_{lo} (\tau' - \tau) \mathcal{G}^{0}_{pi} (\tau - \tau') \mathcal{G}^{0}_{jk} (0) \nonumber \\
	& - V_{ijkl} V_{mpon} \int_{0}^{\beta} \dd{\tau} \dd{\tau'} \dd{\tau_1} \dd{\tau_2} \eta^{\dagger}_{q} (\tau_2) \eta_{r} (\tau_1) \mathcal{G}^{0}_{qm} (\tau_2 - \tau) \mathcal{G}^{0}_{nr} (\tau - \tau_1) \mathcal{G}^{0}_{pi} (\tau - \tau') \mathcal{G}^{0}_{lo} (\tau' - \tau) \mathcal{G}^{0}_{jk} (0) \nonumber \\
	& + V_{ijkl} V_{mpon} \int_{0}^{\beta} \dd{\tau} \dd{\tau'} \dd{\tau_1} \dd{\tau_2} \eta^{\dagger}_{q} (\tau_2) \eta_{r} (\tau_1) \mathcal{G}^{0}_{qm} (\tau_2 - \tau) \mathcal{G}^{0}_{nr} (\tau - \tau_1) \mathcal{G}^{0}_{jo} (\tau' - \tau) \mathcal{G}^{0}_{pi} (\tau - \tau') \mathcal{G}^{0}_{lk} (0) \nonumber \\
	& + V_{ijkl} V_{mnop} \int_{0}^{\beta} \dd{\tau} \dd{\tau'} \dd{\tau_1} \dd{\tau_2} \eta^{\dagger}_{q} (\tau_2) \eta_{r} (\tau_1) \mathcal{G}^{0}_{qm} (\tau_2 - \tau') \mathcal{G}^{0}_{jr} (\tau - \tau_1) \mathcal{G}^{0}_{ni} (\tau' - \tau) \mathcal{G}^{0}_{lo} (\tau - \tau') \mathcal{G}^{0}_{pk} (\tau' - \tau) \nonumber \\
	& - V_{ijkl} V_{mnop} \int_{0}^{\beta} \dd{\tau} \dd{\tau'} \dd{\tau_1} \dd{\tau_2} \eta^{\dagger}_{q} (\tau_2) \eta_{r} (\tau_1) \mathcal{G}^{0}_{qm} (\tau_2 - \tau') \mathcal{G}^{0}_{lr} (\tau - \tau_1) \mathcal{G}^{0}_{ni} (\tau' - \tau) \mathcal{G}^{0}_{jo} (\tau - \tau') \mathcal{G}^{0}_{pk} (\tau' - \tau) \nonumber \\
& - V_{ijkl} V_{mnop} \int_{0}^{\beta} \dd{\tau} \dd{\tau'} \dd{\tau_1} \dd{\tau_2} \eta^{\dagger}_{q} (\tau_2) \eta_{r} (\tau_1) \mathcal{G}^{0}_{qm} (\tau_2 - \tau') \mathcal{G}^{0}_{ni} (\tau' - \tau) \mathcal{G}^{0}_{jr} (\tau - \tau_1) \mathcal{G}^{0}_{po} (0) \mathcal{G}^{0}_{lk} (0) \nonumber \\
	& + V_{ilkj} V_{mpon} \int_{0}^{\beta} \dd{\tau} \dd{\tau'} \dd{\tau_1} \dd{\tau_2} \eta^{\dagger}_{q} (\tau_2) \eta_{r} (\tau_1) \mathcal{G}^{0}_{qm} (\tau_2 - \tau') \mathcal{G}^{0}_{ni} (\tau' - \tau) \mathcal{G}^{0}_{jr} (\tau - \tau_1) \mathcal{G}^{0}_{po} (0) \mathcal{G}^{0}_{lk} (0) \nonumber \\
	& - V_{ijkl} V_{mnop} \int_{0}^{\beta} \dd{\tau} \dd{\tau'} \dd{\tau_1} \dd{\tau_2} \eta^{\dagger}_{q} (\tau_2) \eta_{r} (\tau_1) \mathcal{G}^{0}_{qm} (\tau_2 - \tau') \mathcal{G}^{0}_{ni} (\tau' - \tau) \mathcal{G}^{0}_{jr} (\tau - \tau_1) \mathcal{G}^{0}_{po} (0) \mathcal{G}^{0}_{lk} (0) \nonumber \\
	& + V_{ilkj} V_{mpon} \int_{0}^{\beta} \dd{\tau} \dd{\tau'} \dd{\tau_1} \dd{\tau_2} \eta^{\dagger}_{q} (\tau_2) \eta_{r} (\tau_1) \mathcal{G}^{0}_{qm} (\tau_2 - \tau') \mathcal{G}^{0}_{ni} (\tau' - \tau) \mathcal{G}^{0}_{jr} (\tau - \tau_1) \mathcal{G}^{0}_{po} (0) \mathcal{G}^{0}_{lk} (0) \nonumber \\
& + \frac{1}{2} V_{mnop} \beta \mathcal{G}^{0}_{po} (0) \mathcal{G}^{0}_{nm} (0) V_{ijkl} \int_{0}^{\beta} \dd{\tau} \dd{\tau_1}  \dd{\tau_2} \eta^{\dagger}_{q} (\tau_2) \eta_{r} (\tau_1) \mathcal{G}^{0}_{qi} (\tau_2 - \tau) \mathcal{G}^{0}_{jr} (\tau - \tau_1) \mathcal{G}^{0}_{lk} (0)  \nonumber \\
	& - \frac{1}{2} V_{mnop} \beta \mathcal{G}^{0}_{pm} (0) \mathcal{G}^{0}_{no} (0) V_{ijkl} \int_{0}^{\beta} \dd{\tau} \dd{\tau_1}  \dd{\tau_2} \eta^{\dagger}_{q} (\tau_2) \eta_{r} (\tau_1) \mathcal{G}^{0}_{qi} (\tau_2 - \tau) \mathcal{G}^{0}_{jr} (\tau - \tau_1) \mathcal{G}^{0}_{lk} (0)  \nonumber \\
	& - \frac{1}{2} V_{mnop} \beta \mathcal{G}^{0}_{po} (0) \mathcal{G}^{0}_{nm} (0) V_{ijkl} \int_{0}^{\beta} \dd{\tau} \dd{\tau_1}  \dd{\tau_2} \eta^{\dagger}_{q} (\tau_2) \eta_{r} (\tau_1) \mathcal{G}^{0}_{qi} (\tau_2 - \tau) \mathcal{G}^{0}_{lr} (\tau - \tau_1) \mathcal{G}^{0}_{jk} (0)  \nonumber \\
	&\left. + \frac{1}{2} V_{mnop} \beta \mathcal{G}^{0}_{pm} (0) \mathcal{G}^{0}_{no} (0) V_{ijkl} \int_{0}^{\beta} \dd{\tau} \dd{\tau_1}  \dd{\tau_2} \eta^{\dagger}_{q} (\tau_2) \eta_{r} (\tau_1) \mathcal{G}^{0}_{qi} (\tau_2 - \tau) \mathcal{G}^{0}_{jr} (\tau - \tau_1) \mathcal{G}^{0}_{lk} (0) + \dots \right].
	\label{app:eqn:second_order_partition_function_expansion} 
\end{align}
where, in an abuse of notation, we have \emph{defined}
\begin{equation}
	\label{app:eqn:simple_shorthand_gf_0}
	\mathcal{G}^{0}_{ij} (0) \equiv \frac{1}{2} \left( \mathcal{G}^{0}_{ij} (0^+) + \mathcal{G}^{0}_{ij} (0^-) \right).
\end{equation}
In \cref{app:eqn:second_order_partition_function_expansion}, the dots ``$\dots$'' denote terms which are higher order in the interaction tensor, as well as terms that, upon taking the logarithm, do \emph{not} give rise to contributions that contain \emph{exactly} two source fields $\eta^{\dagger}_{} \eta_{}$ (which would otherwise vanish upon taking the functional derivative in \cref{app:eqn:path_integral_full_gf} and setting the source fields to zero). Schematically, the partition function from \cref{app:eqn:second_order_partition_function_expansion} can be expressed as 
\begin{equation}
	Z \left[ \eta^{\dagger}_{}, \eta_{} \right] = \left( 1 + a V + b V^2 + \mathcal{O} \left( V^3 \right) \right) Z_{0} \left[0, 0 \right] \exp \left( - \int_{0}^{\beta} \dd{\tau_1} \dd{\tau_2} \eta^{\dagger}_{i} (\tau_2) \mathcal{G}^{0}_{ij} \left( \tau_2 - \tau_1 \right)  \eta_{j} (\tau_1) \right),
\end{equation}
where $a V$ and $b V^2$ denote the terms from \cref{app:eqn:second_order_partition_function_expansion} that are first and second order in the interaction. Taking the logarithm, we obtain 
\begin{equation}
	\log Z \left[ \eta^{\dagger}_{}, \eta_{} \right] = \log Z_{0} \left[0, 0 \right] - \int_{0}^{\beta} \dd{\tau_1} \dd{\tau_2} \eta^{\dagger}_{i} (\tau_2) \mathcal{G}^{0}_{ij} \left( \tau_2 - \tau_1 \right)  \eta_{j} (\tau_1) + a V + \left( b - \frac{a^2}{2} \right) V^2 + \mathcal{O} \left( V^3 \right). 
\end{equation}

Substituting \cref{app:eqn:second_order_partition_function_expansion} into \cref{app:eqn:path_integral_full_gf} and setting $\tau_1 = 0$ (without loss of generality), we find 
\begin{align}
	\mathcal{G}_{qr} (\tau_2) =&  \mathcal{G}^{0}_{qr} (\tau_2) 
	+ V_{ijkl} \int_{0}^{\beta} \dd{\tau} \left( \mathcal{G}^{0}_{qi} (\tau_2 - \tau) \mathcal{G}^{0}_{jr} (\tau) \mathcal{G}^{0}_{lk} (0) - \mathcal{G}^{0}_{qi} (\tau_2 - \tau) \mathcal{G}^{0}_{lr} (\tau) \mathcal{G}^{0}_{jk} (0) \right) \nonumber \\
	& + V_{ijkl} V_{mnop} \int_{0}^{\beta} \dd{\tau} \dd{\tau'} \mathcal{G}^{0}_{qm} (\tau_2 - \tau) \mathcal{G}^{0}_{nr} (\tau) \mathcal{G}^{0}_{jo} (\tau' - \tau) \mathcal{G}^{0}_{pi} (\tau - \tau') \mathcal{G}^{0}_{lk} (0) \nonumber \\
	& - V_{ijkl} V_{mnop} \int_{0}^{\beta} \dd{\tau} \dd{\tau'} \mathcal{G}^{0}_{qm} (\tau_2 - \tau) \mathcal{G}^{0}_{nr} (\tau) \mathcal{G}^{0}_{lo} (\tau' - \tau) \mathcal{G}^{0}_{pi} (\tau - \tau') \mathcal{G}^{0}_{jk} (0) \nonumber \\
	& + V_{ijkl} V_{mpon} \int_{0}^{\beta} \dd{\tau} \dd{\tau'} \mathcal{G}^{0}_{qm} (\tau_2 - \tau) \mathcal{G}^{0}_{nr} (\tau) \mathcal{G}^{0}_{pi} (\tau - \tau') \mathcal{G}^{0}_{lo} (\tau' - \tau) \mathcal{G}^{0}_{jk} (0) \nonumber \\
	& - V_{ijkl} V_{mpon} \int_{0}^{\beta} \dd{\tau} \dd{\tau'} \mathcal{G}^{0}_{qm} (\tau_2 - \tau) \mathcal{G}^{0}_{nr} (\tau) \mathcal{G}^{0}_{jo} (\tau' - \tau) \mathcal{G}^{0}_{pi} (\tau - \tau') \mathcal{G}^{0}_{lk} (0) \nonumber \\
	& - V_{ijkl} V_{mnop} \int_{0}^{\beta} \dd{\tau} \dd{\tau'} \mathcal{G}^{0}_{qm} (\tau_2 - \tau') \mathcal{G}^{0}_{jr} (\tau) \mathcal{G}^{0}_{ni} (\tau' - \tau) \mathcal{G}^{0}_{lo} (\tau - \tau') \mathcal{G}^{0}_{pk} (\tau' - \tau) \nonumber \\
	& + V_{ijkl} V_{mnop} \int_{0}^{\beta} \dd{\tau} \dd{\tau'} \mathcal{G}^{0}_{qm} (\tau_2 - \tau') \mathcal{G}^{0}_{lr} (\tau) \mathcal{G}^{0}_{ni} (\tau' - \tau) \mathcal{G}^{0}_{jo} (\tau - \tau') \mathcal{G}^{0}_{pk} (\tau' - \tau) \nonumber \\
& + V_{ijkl} V_{mnop} \int_{0}^{\beta} \dd{\tau} \dd{\tau'} \mathcal{G}^{0}_{qm} (\tau_2 - \tau') \mathcal{G}^{0}_{ni} (\tau' - \tau) \mathcal{G}^{0}_{jr} (\tau) \mathcal{G}^{0}_{po} (0) \mathcal{G}^{0}_{lk} (0) \nonumber \\
	& - V_{ilkj} V_{mpon} \int_{0}^{\beta} \dd{\tau} \dd{\tau'} \mathcal{G}^{0}_{qm} (\tau_2 - \tau') \mathcal{G}^{0}_{ni} (\tau' - \tau) \mathcal{G}^{0}_{jr} (\tau) \mathcal{G}^{0}_{po} (0) \mathcal{G}^{0}_{lk} (0) \nonumber \\
	& + V_{ijkl} V_{mnop} \int_{0}^{\beta} \dd{\tau} \dd{\tau'} \mathcal{G}^{0}_{qm} (\tau_2 - \tau') \mathcal{G}^{0}_{ni} (\tau' - \tau) \mathcal{G}^{0}_{jr} (\tau) \mathcal{G}^{0}_{po} (0) \mathcal{G}^{0}_{lk} (0) \nonumber \\
	& - V_{ilkj} V_{mpon} \int_{0}^{\beta} \dd{\tau} \dd{\tau'} \mathcal{G}^{0}_{qm} (\tau_2 - \tau') \mathcal{G}^{0}_{ni} (\tau' - \tau) \mathcal{G}^{0}_{jr} (\tau) \mathcal{G}^{0}_{po} (0) \mathcal{G}^{0}_{lk} (0) + \mathcal{O} \left( V^3 \right). \label{app:eqn:second_gf_expansion_time} 
\end{align}
Fourier transforming with the aid of \cref{app:eqn:matsubara_gf_THF_ft_simple}, we obtain
\begin{align}
	\mathcal{G}_{qr} \left( i \omega_n \right) =&  \mathcal{G}^{0}_{qr}  \left( i \omega_n \right)
	+ V_{ijkl} \left( \mathcal{G}^{0}_{qi}  \left( i \omega_n \right) \mathcal{G}^{0}_{jr}  \left( i \omega_n \right) \mathcal{G}^{0}_{lk} (0) - \mathcal{G}^{0}_{qi}  \left( i \omega_n \right) \mathcal{G}^{0}_{lr}  \left( i \omega_n \right) \mathcal{G}^{0}_{jk} (0) \right) \nonumber \\
	& + V_{ijkl} V_{mnop} \frac{1}{\beta} \sum_{i \omega_x} \mathcal{G}^{0}_{qm} \left( i \omega_n \right) \mathcal{G}^{0}_{nr} \left( i \omega_n \right) \mathcal{G}^{0}_{jo} \left( i \omega_x \right) \mathcal{G}^{0}_{pi} \left( i \omega_x \right) \mathcal{G}^{0}_{lk} (0) \nonumber \\
	& - V_{ijkl} V_{mnop} \frac{1}{\beta} \sum_{i \omega_x} \mathcal{G}^{0}_{qm} \left( i \omega_n \right) \mathcal{G}^{0}_{nr} \left( i \omega_n \right) \mathcal{G}^{0}_{lo} \left( i \omega_x \right) \mathcal{G}^{0}_{pi} \left( i \omega_x \right) \mathcal{G}^{0}_{jk} (0) \nonumber \\
	& + V_{ijkl} V_{mpon} \frac{1}{\beta} \sum_{i \omega_x} \mathcal{G}^{0}_{qm} \left( i \omega_n \right) \mathcal{G}^{0}_{nr} \left( i \omega_n \right) \mathcal{G}^{0}_{pi} \left( i \omega_x \right) \mathcal{G}^{0}_{lo} \left( i \omega_x \right) \mathcal{G}^{0}_{jk} (0) \nonumber \\
	& - V_{ijkl} V_{mpon} \frac{1}{\beta} \sum_{i \omega_x} \mathcal{G}^{0}_{qm} \left( i \omega_n \right) \mathcal{G}^{0}_{nr} \left( i \omega_n \right) \mathcal{G}^{0}_{jo} \left( i \omega_x \right) \mathcal{G}^{0}_{pi} \left( i \omega_x \right) \mathcal{G}^{0}_{lk} (0) \nonumber \\
	& - V_{ijkl} V_{mnop} \frac{1}{\beta^2} \sum_{i \omega_x, i\omega_y} \mathcal{G}^{0}_{qm} \left( i \omega_n \right) \mathcal{G}^{0}_{jr}  \left( i \omega_n \right) \mathcal{G}^{0}_{ni} \left( i \omega_x \right) \mathcal{G}^{0}_{lo} \left( i \omega_x + i \omega_y - i \omega_n \right) \mathcal{G}^{0}_{pk} \left( i \omega_y \right) \nonumber \\
	& + V_{ijkl} V_{mnop}  \frac{1}{\beta^2}\sum_{i \omega_x, i\omega_y} \mathcal{G}^{0}_{qm} \left( i \omega_n \right) \mathcal{G}^{0}_{lr} \left( i \omega_n \right) \mathcal{G}^{0}_{ni} \left( i \omega_x \right) \mathcal{G}^{0}_{jo} \left( i \omega_x + i \omega_y - i \omega_n \right) \mathcal{G}^{0}_{pk} \left( i \omega_y \right) \nonumber \\
& + V_{ijkl} V_{mnop} \mathcal{G}^{0}_{qm} \left( i \omega_n \right) \mathcal{G}^{0}_{ni} \left( i \omega_n \right) \mathcal{G}^{0}_{jr} \left( i \omega_n \right) \mathcal{G}^{0}_{po} (0) \mathcal{G}^{0}_{lk} (0) \nonumber \\
	& - V_{ilkj} V_{mpon} \mathcal{G}^{0}_{qm} \left( i \omega_n \right) \mathcal{G}^{0}_{ni} \left( i \omega_n \right) \mathcal{G}^{0}_{jr} \left( i \omega_n \right) \mathcal{G}^{0}_{po} (0) \mathcal{G}^{0}_{lk} (0) \nonumber \\
	& + V_{ijkl} V_{mnop} \mathcal{G}^{0}_{qm} \left( i \omega_n \right) \mathcal{G}^{0}_{ni} \left( i \omega_n \right) \mathcal{G}^{0}_{jr} \left( i \omega_n \right) \mathcal{G}^{0}_{po} (0) \mathcal{G}^{0}_{lk} (0) \nonumber \\
	& - V_{ilkj} V_{mpon} \mathcal{G}^{0}_{qm} \left( i \omega_n \right)) \mathcal{G}^{0}_{ni} \left( i \omega_n \right) \mathcal{G}^{0}_{jr} \left( i \omega_n \right) \mathcal{G}^{0}_{po} (0) \mathcal{G}^{0}_{lk} (0)  + \mathcal{O} \left( V^3 \right) .\label{app:eqn:second_gf_expansion_freq} 
\end{align}

\begin{figure}[!t]
	\centering
	\begin{tikzpicture}[baseline=(current bounding box.north)]
		\begin{feynman}
			\vertex (b);
			\vertex [above=1 cm of b] (d);
			\vertex [below left=1 cm of b] (a);
			\vertex [below right=1 cm of b] (c);
			\vertex [above left=1 cm of d] (e);
			\vertex [above right=1 cm of d] (f);
			\diagram*{
				(a) -- [fermion]  (b) -- [fermion] (c),
				(b) -- [photon] (d),
				(e) -- [fermion]  (d) -- [fermion] (f),
			};
		\end{feynman}
		\path (a) ++(-0.25cm,0cm) node{$j$};
		\path (c) ++(+0.25cm,0cm) node{$i$};
		\path (e) ++(-0.25cm,0cm) node{$l$};
		\path (f) ++(0.25cm,0cm) node{$k$};
	\end{tikzpicture}
	\caption{Representation of the interaction vertex of $H_I$ from \cref{app:eqn:int_h} for Feynman diagrams. The amplitude of this interaction vertex is given by $V_{ijkl}$.}
	\label{app:fig:simple_feyn_vertex}
\end{figure}
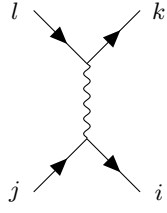

\subsubsection{Second-order self-energy}\label{app:sec:additional_mb_results:self-energy:interacting_sf}

Comparing \cref{app:eqn:expansion_full_gf_self_energy} with \cref{app:eqn:second_gf_expansion_freq}, we can immediately identify
\begin{align}
	h^{\text{MF}}_{ij} &= \left( V_{ijkl} - V_{ilkj} \right) \mathcal{G}^{0}_{lk} (0), \label{app:eqn:simple_perturbation_result_1} \\
	\Sigma^{(2)}_{mn} \left( i \omega_n \right) &= V_{ijkl} V_{mnop} \frac{1}{\beta} \sum_{i \omega_x} \mathcal{G}^{0}_{jo} \left( i \omega_x \right) \mathcal{G}^{0}_{pi} \left( i \omega_x \right) \mathcal{G}^{0}_{lk} (0) \nonumber \\
	& - V_{ijkl} V_{mnop} \frac{1}{\beta} \sum_{i \omega_x} \mathcal{G}^{0}_{lo} \left( i \omega_x \right) \mathcal{G}^{0}_{pi} \left( i \omega_x \right) \mathcal{G}^{0}_{jk} (0) \nonumber \\
	& + V_{ijkl} V_{mpon} \frac{1}{\beta} \sum_{i \omega_x} \mathcal{G}^{0}_{pi} \left( i \omega_x \right) \mathcal{G}^{0}_{lo} \left( i \omega_x \right) \mathcal{G}^{0}_{jk} (0) \nonumber \\
	& - V_{ijkl} V_{mpon} \frac{1}{\beta} \sum_{i \omega_x} \mathcal{G}^{0}_{jo} \left( i \omega_x \right) \mathcal{G}^{0}_{pi} \left( i \omega_x \right) \mathcal{G}^{0}_{lk} (0) \nonumber \\
	& - V_{inkl} V_{mjop} \frac{1}{\beta^2} \sum_{i \omega_x, i\omega_y} \mathcal{G}^{0}_{ji} \left( i \omega_x \right) \mathcal{G}^{0}_{lo} \left( i \omega_x + i \omega_y - i \omega_n \right) \mathcal{G}^{0}_{pk} \left( i \omega_y \right) \nonumber \\
	& + V_{inkl} V_{mjop}  \frac{1}{\beta^2}\sum_{i \omega_x, i\omega_y} \mathcal{G}^{0}_{jk} \left( i \omega_x \right) \mathcal{G}^{0}_{lo} \left( i \omega_x + i \omega_y - i \omega_n \right) \mathcal{G}^{0}_{pi} \left( i \omega_y \right) \label{app:eqn:simple_perturbation_result_2}.
\end{align}
As expected, the static self-energy contribution is precisely the Hartree-Fock Hamiltonian, although computed with the \emph{non-interacting} density matrix since, from \cref{app:eqn:simple_shorthand_gf_0}, we find that 
\begin{equation}
	\mathcal{G}^{0}_{ji} \left( 0 \right) = \frac{1}{2} \left( \mathcal{G}^{0}_{ji} \left( 0^+ \right) + \mathcal{G}^{0}_{ji} \left( 0^- \right) \right) = \left\langle \hat{\gamma}^\dagger_{i} \hat{\gamma}_{j} - \frac{1}{2} \delta_{ij} \right\rangle_0.
\end{equation}
The non-interacting rather than the interacting density matrix of the system appears in \cref{app:eqn:simple_perturbation_result_1,app:eqn:simple_perturbation_result_2} because we are perturbing the \emph{non-interacting} solution. If we represent the interaction vertex of the Hamiltonian $H_I$ as shown in \cref{app:fig:simple_feyn_vertex}, then the two contribution to the static self-energy from \cref{app:eqn:simple_perturbation_result_1} correspond to the Feynman diagrams from \cref{app:fig:general_self_en_diags:b,app:fig:general_self_en_diags:c}, respectively. Similarly, the six terms of the dynamical self-energy from \cref{app:eqn:simple_perturbation_result_2} correspond, respectively, to the six Feynman diagrams from \crefrange{app:fig:general_self_en_diags:d}{app:fig:general_self_en_diags:i}.

\subsubsection{The second-order self-energy correction using dressed propagators}\label{app:sec:additional_mb_results:self-energy:dressed_gf}

In \cref{app:sec:se_correction_beyond_HF:all_so_corrections:diagrams}, we argued that if the dressed (as opposed to the bare) propagators are to be used, then only the Feynman diagrams from \cref{app:fig:general_self_en_diags:h,app:fig:general_self_en_diags:i} should be considered, as the other ones are already included at the Hartree-Fock level. To see why this is so, we consider Green's function of the system at the Hartree-Fock level ({\it i.e.}{}, dressed with the Hartree-Fock contribution)
\begin{equation}
	\mathcal{G}^{\text{MF}} \left( i \omega_n \right) = \left[ i\omega_n \mathbb{1} - h - h^{\text{MF}} \right]^{-1},
\end{equation}
where $h^{\text{MF}}$ is given by \cref{app:eqn:simple_perturbation_result_1}. By expanding $\mathcal{G}$ to first order in the interaction, it is easy to see that 
\begin{equation}
	\label{app:eqn:simple_hf_dressed_expansion_matrix}
	\mathcal{G}^{\text{MF}} \left( i \omega_n \right) = \mathcal{G}^{0} \left( i \omega_n \right) + \mathcal{G}^{0} \left( i \omega_n \right) h^{\text{MF}} \mathcal{G}^{0} \left( i \omega_n \right) + \mathcal{O} \left( V^2 \right).
\end{equation}
Multiplying \cref{app:eqn:simple_hf_dressed_expansion_matrix} by $e^{-i \omega_n 0^+}$ and Fourier transforming to imaginary time, we obtain
\begin{equation}
	\label{app:eqn:simple_hf_dressed_expansion_1}
	\mathcal{G}^{\text{MF}} (0^+) = \mathcal{G}^{0} (0^+) + \frac{1}{\beta} \sum_{i \omega_n} \mathcal{G}^{0} \left( i \omega_n \right) h^{\text{MF}} \mathcal{G}^{0} \left( i \omega_n \right) e^{-i \omega_n 0^+} + \mathcal{O} \left( V^2 \right).
\end{equation}
For large $\abs{\omega_n}$, $\mathcal{G}^{0} \left( i \omega_n \right) \sim \frac{1}{i \omega_n}$, meaning that the Matsubara sum in \cref{app:eqn:simple_hf_dressed_expansion_1} converges even \emph{without} the exponential factor, so it can be dropped. Similarly, for small negative imaginary times, we have
\begin{equation}
	\label{app:eqn:simple_hf_dressed_expansion_2}
	\mathcal{G}^{\text{MF}} (0^-) = \mathcal{G}^{0} (0^-) + \frac{1}{\beta} \sum_{i \omega_n} \mathcal{G}^{0} \left( i \omega_n \right) h^{\text{MF}} \mathcal{G}^{0} \left( i \omega_n \right) + \mathcal{O} \left( V^2 \right).
\end{equation}
\emph{Defining}
\begin{equation}
	\label{app:eqn:simple_shorthand_gf_hf_0}
	\mathcal{G}^{\text{MF}}_{ij} (0) \equiv \frac{1}{2} \left( \mathcal{G}^{\text{MF}}_{ij} (0^+) + \mathcal{G}^{\text{MF}}_{ij} (0^-) \right)
\end{equation}
in the same spirit as \cref{app:eqn:simple_shorthand_gf_0}, \cref{app:eqn:simple_hf_dressed_expansion_1,app:eqn:simple_hf_dressed_expansion_2} imply that
\begin{equation}
	\label{app:eqn:simple_hf_dressed_expansion_index}
	\mathcal{G}_{ij}^{\text{MF}} (0) = \mathcal{G}_{ij}^{0} (0) + \frac{1}{\beta} \sum_{i \omega_n }V_{mnop} \mathcal{G}_{im}^{0} \left( i \omega_n \right) \mathcal{G}_{nj}^{0} \left( i \omega_n \right) \mathcal{G}^{0}_{po} (0) -  \frac{1}{\beta} \sum_{i \omega_n} V_{mpon} \mathcal{G}_{im}^{0} \left( i \omega_n \right) \mathcal{G}_{nj}^{0} \left( i \omega_n \right) \mathcal{G}^{0}_{po} (0) + \mathcal{O} \left( V^2 \right).
\end{equation}
Using \cref{app:eqn:simple_hf_dressed_expansion_index}, we can rewrite the first and second-order self-energy contributions from \cref{app:eqn:simple_perturbation_result_1,app:eqn:simple_perturbation_result_2} in a simpler form
\begin{align}
	h^{\text{MF}}_{mn} + \Sigma^{(2)}_{mn} \left( i \omega_n \right) &= \left( V_{mnop} - V_{mpon} \right) \mathcal{G}^{\text{MF}}_{po} (0) \nonumber \\
	& - V_{inkl} V_{mjop} \frac{1}{\beta^2} \sum_{i \omega_x, i\omega_y} \mathcal{G}^{\text{MF}}_{ji} \left( i \omega_x \right) \mathcal{G}^{\text{MF}}_{lo} \left( i \omega_x + i \omega_y - i \omega_n \right) \mathcal{G}^{\text{MF}}_{pk} \left( i \omega_y \right) \nonumber \\
	& + V_{inkl} V_{mjop}  \frac{1}{\beta^2}\sum_{i \omega_x, i\omega_y} \mathcal{G}^{\text{MF}}_{jk} \left( i \omega_x \right) \mathcal{G}^{\text{MF}}_{lo} \left( i \omega_x + i \omega_y - i \omega_n \right) \mathcal{G}^{\text{MF}}_{pi} \left( i \omega_y \right) + \mathcal{O} \left( V^3 \right)
	 \label{app:eqn:simple_perturbation_result_dressed}.
\end{align}
Thus, by ``dressing'' Green's function with the Hartree-Fock self-energy, the second-order contributions to the self-energy from the first four terms in \cref{app:eqn:simple_perturbation_result_2} have been effectively absorbed into the first term of \cref{app:eqn:simple_perturbation_result_dressed}. This confirms that when using ``dressed'' propagators, the diagrams from \crefrange{app:fig:general_self_en_diags:d}{app:fig:general_self_en_diags:g} are already included in the Hartree-Fock contributions from \cref{app:fig:general_self_en_diags:b,app:fig:general_self_en_diags:c}.

\subsubsection{The $f$-electron self-energy}\label{app:sec:additional_mb_results:self-energy:f_electron_self_energy}

The second-order $f$-electron self-energy corrections corresponding to the diagrams in \cref{app:fig:general_self_en_diags:h,app:fig:general_self_en_diags:i} were computed in \cref{app:eqn:se_2a_complicated,app:eqn:se_2b_complicated}, and their sum can be \emph{simplified} to 
\begin{align}
	& \Sigma^{f,(2a)}_{\alpha \eta s; \alpha' \eta' s'} \left(i \omega_n, \vec{R}-\vec{R}' \right) + \Sigma^{f,(2b)}_{\alpha \eta s; \alpha' \eta' s'} \left(i \omega_n, \vec{R}-\vec{R}' \right) = \nonumber \\
	= & \frac{1}{\beta^{2}}\sum_{i \omega_x, i \omega_y} \sum_{\substack{\vec{R}_1,\alpha_1,\eta_1,s_1 \\ \vec{R}_2,\alpha_2,\eta_2,s_2}} \left[ U_1 \delta_{\vec{R}',\vec{R}_1} + U_2 \sum_{i=0}^{5} \delta_{\vec{R}',\vec{R}_1+C^i_{6z} \vec{a}_{M1}} \right] \left[ U_1 \delta_{\vec{R},\vec{R}_2} + U_2 \sum_{i=0}^{5} \delta_{\vec{R},\vec{R}_2+C^i_{6z} \vec{a}_{M1}} \right] \nonumber \\ 
	\times & \left[ 
	\mathcal{G}^{f}_{\alpha_{1} \eta_{1} s_{1}; \alpha' \eta' s'} \left(i \omega_x, \vec{R}_1-\vec{R}' \right) \mathcal{G}^{f}_{\alpha_{2} \eta_{2} s_{2};\alpha_{1} \eta_{1} s_{1}} \left(i \omega_y, \vec{R}_2-\vec{R}_1 \right) \mathcal{G}^{f}_{\alpha \eta s;\alpha_{2} \eta_{2} s_{2}} \left(i \omega_n - i \omega_x + i \omega_y, \vec{R}-\vec{R}_2 \right) \right. \nonumber \\
	-&\left. \mathcal{G}^{f}_{\alpha \eta s; \alpha' \eta' s'} \left(i \omega_x, \vec{R}-\vec{R}' \right) \mathcal{G}^{f}_{\alpha_{2} \eta_{2} s_{2};\alpha_{1} \eta_{1} s_{1}} \left(i \omega_n - i \omega_x + i \omega_y, \vec{R}_2-\vec{R}_1 \right) \mathcal{G}^{f}_{\alpha_{1} \eta_{1} s_{1};\alpha_{2} \eta_{2} s_{2}} \left(i \omega_y, \vec{R}_1-\vec{R}_2 \right) \right],
	\label{app:eqn:second_order_from_feynman_simple} 
\end{align}
where we have also replaced the bare $f$-electron Green's function with the dressed one.
In \cref{app:eqn:second_order_from_feynman_simple} and for the rest of this \siSection{}, we will no longer employ Einstein's summation convention. To use the result in \cref{app:eqn:simple_perturbation_result_dressed} for the $f$-electron second-order self-energy correction, we must convert between the notation introduced in \cref{app:sec:additional_mb_results:model} and the notation employed in \cref{app:sec:se_correction_beyond_HF}
\begin{equation}
	\label{app:eqn:dictionary_of_notation}	
	\begin{split}
		\hat{\gamma}^\dagger_{i} &\to \hat{f}^\dagger_{\vec{R}_i,\alpha_i, \eta_i, s_i}, \\
		V_{ijkl} & \to \left( U_1 \delta_{\vec{R}_i,\vec{R}_k} + U_2 \sum_{n=0}^{5} \delta_{\vec{R}_i,\vec{R}_k+C^n_{6z} \vec{a}_{M1} } \right) \delta_{\vec{R}_i, \vec{R}_j} \delta_{\vec{R}_k, \vec{R}_l} \delta_{\alpha_i,\alpha_j} \delta_{\eta_i,\eta_j} \delta_{s_i,s_j} \delta_{\alpha_k,\alpha_l} \delta_{\eta_k,\eta_l} \delta_{s_k,s_l}, \\
		\Sigma_{ij} \left( i \omega_n \right) &\to \Sigma^{f}_{\alpha_{i} \eta_{i} s_{i};\alpha_{j} \eta_{j} s_{j}} \left( i \omega_n, \vec{R}_i - \vec{R}_j \right), \\
		\mathcal{G}_{ij} \left( i \omega_n \right) &\to \mathcal{G}^{f}_{\alpha_{i} \eta_{i} s_{i};\alpha_{j} \eta_{j} s_{j}} \left( i \omega_n, \vec{R}_i - \vec{R}_j \right).
	\end{split}
\end{equation}
Using the ``dictionary'' in \cref{app:eqn:dictionary_of_notation}, as well as the result obtain using perturbation theory from \cref{app:eqn:simple_perturbation_result_dressed}, we can obtain the second-order self-energy of the $f$-electrons corresponding to the diagrams from \cref{app:fig:general_self_en_diags:h,app:fig:general_self_en_diags:i} as 
\begin{align}
	& \Sigma^{f,(2a)}_{\alpha_{m} \eta_{m} s_{m}; \alpha_{n} \eta_{n} s_{n}} \left(i \omega_n, \vec{R}_m-\vec{R}_n \right) + \Sigma^{f,(2b)}_{\alpha_{m} \eta_{m} s_{m}; \alpha_{n} \eta_{n} s_{n}} \left(i \omega_n, \vec{R}_m-\vec{R}_n \right) \nonumber \\
	= & \frac{1}{\beta^{2}}\sum_{i \omega_x, i \omega_y} \sum_{\substack{\vec{R}_k,\alpha_{k} \eta_{k} s_{k} \\ \vec{R}_o,\alpha_{o} \eta_{o} s_{o}}} \left[ U_1 \delta_{\vec{R}_m,\vec{R}_o} + U_2 \sum_{i=0}^{5} \delta_{\vec{R}_m,\vec{R}_o+C^i_{6z} \vec{a}_{M1}} \right] \left[ U_1 \delta_{\vec{R}_n,\vec{R}_k} + U_2 \sum_{i=0}^{5} \delta_{\vec{R}_n,\vec{R}_k+C^i_{6z} \vec{a}_{M1}} \right] \nonumber \\ 
	\times & \left[ 
	\mathcal{G}^{f}_{\alpha_{m} \eta_{m} s_{m};\alpha_{k} \eta_{k} s_{k}} \left( i \omega_x, \vec{R}_m - \vec{R}_k \right) \mathcal{G}^{f}_{\alpha_{k} \eta_{k} s_{k};\alpha_{o} \eta_{o} s_{o}} \left( i \omega_x + i \omega_y - i \omega_n, \vec{R}_k - \vec{R}_o \right) \mathcal{G}^{f}_{\alpha_{o} \eta_{o} s_{o};\alpha_{n} \eta_{n} s_{n}} \left( i \omega_y, \vec{R}_o - \vec{R}_n \right) \right. \nonumber \\
	-&\left. \mathcal{G}^{f}_{\alpha_{m} \eta_{m} s_{m};\alpha_{n} \eta_{n} s_{n}} \left( i \omega_x, \vec{R}_m - \vec{R}_n \right) \mathcal{G}^{f}_{\alpha_{k} \eta_{k} s_{k};\alpha_{o} \eta_{o} s_{o}} \left( i \omega_x + i \omega_y - i \omega_n, \vec{R}_k - \vec{R}_o \right) \mathcal{G}^{f}_{\alpha_{o} \eta_{o} s_{o};\alpha_{k} \eta_{k} s_{k}} \left( i \omega_y, \vec{R}_o - \vec{R}_k \right) \right].
	\label{app:eqn:second_order_from_path_integral_simple}
\end{align}
Relabeling the dummy indices, we find that indeed \cref{app:eqn:second_order_from_path_integral_simple,app:eqn:second_order_from_feynman_simple} are identical.

\section{Numerical band structure results in the symmetry-broken states}\label{app:sec:results_corr_ins}

{\renewcommand{\arraystretch}{1.2}
\begin{table}[t]
	\centering
	\begin{tabular}{|c|l|l|c|c|c|c|}
		\hline
		\multirow[b]{2}{*}{$\nu$} & \multirow[b]{2}{*}{Correlated state} & \multirow[b]{2}{*}{\SiSection{}} & \multicolumn{2}{c|}{Low temperature} & \multicolumn{2}{c|}{High temperature} \\\cline{4-7}
		& & & TBG & TSTG & TBG & TSTG \\
		\hline
		$4$ & $\IfStrEqCase{1}{{1}{\ket{\nu={}4} }
		{2}{\ket{\nu={}3, \mathrm{IVC}}}
		{3}{\ket{\nu={}3, \mathrm{VP}}}
		{4}{\ket{\nu={}2, \mathrm{K-IVC}}}
		{5}{\ket{\nu={}2, \mathrm{VP}}}
		{6}{\ket{\nu={}1, (\mathrm{K-IVC}+\mathrm{VP})}}
		{7}{\ket{\nu={}1, \mathrm{VP}}}
		{8}{\ket{\nu=0, \mathrm{K-IVC}}}
		{9}{\ket{\nu=0, \mathrm{VP}}}
	}
	[nada]
$ & \cref{app:sec:results_corr_ins:1} & $T = \SI{5}{\kelvin}$ & $T = \SI{7}{\kelvin}$ & $T = \SI{40}{\kelvin}$ & $T = \SI{56}{\kelvin}$ \\ 
		\hline
		\multirow{2}{*}{$3$} & $\IfStrEqCase{2}{{1}{\ket{\nu={}4} }
		{2}{\ket{\nu={}3, \mathrm{IVC}}}
		{3}{\ket{\nu={}3, \mathrm{VP}}}
		{4}{\ket{\nu={}2, \mathrm{K-IVC}}}
		{5}{\ket{\nu={}2, \mathrm{VP}}}
		{6}{\ket{\nu={}1, (\mathrm{K-IVC}+\mathrm{VP})}}
		{7}{\ket{\nu={}1, \mathrm{VP}}}
		{8}{\ket{\nu=0, \mathrm{K-IVC}}}
		{9}{\ket{\nu=0, \mathrm{VP}}}
	}
	[nada]
$ & \cref{app:sec:results_corr_ins:2} & $T = \SI{4}{\kelvin}$ & $T = \SI{5.6}{\kelvin}$ & $T = \SI{6}{\kelvin}$ & $T = \SI{8.4}{\kelvin}$ \\ 
		\cline{2-7}
		& $\IfStrEqCase{3}{{1}{\ket{\nu={}4} }
		{2}{\ket{\nu={}3, \mathrm{IVC}}}
		{3}{\ket{\nu={}3, \mathrm{VP}}}
		{4}{\ket{\nu={}2, \mathrm{K-IVC}}}
		{5}{\ket{\nu={}2, \mathrm{VP}}}
		{6}{\ket{\nu={}1, (\mathrm{K-IVC}+\mathrm{VP})}}
		{7}{\ket{\nu={}1, \mathrm{VP}}}
		{8}{\ket{\nu=0, \mathrm{K-IVC}}}
		{9}{\ket{\nu=0, \mathrm{VP}}}
	}
	[nada]
$ & \cref{app:sec:results_corr_ins:3} & $T = \SI{4}{\kelvin}$ & $T = \SI{5.6}{\kelvin}$ & $T = \SI{6}{\kelvin}$ & $T = \SI{8.4}{\kelvin}$ \\ 
		\hline
		\multirow{2}{*}{$2$} & $\IfStrEqCase{4}{{1}{\ket{\nu={}4} }
		{2}{\ket{\nu={}3, \mathrm{IVC}}}
		{3}{\ket{\nu={}3, \mathrm{VP}}}
		{4}{\ket{\nu={}2, \mathrm{K-IVC}}}
		{5}{\ket{\nu={}2, \mathrm{VP}}}
		{6}{\ket{\nu={}1, (\mathrm{K-IVC}+\mathrm{VP})}}
		{7}{\ket{\nu={}1, \mathrm{VP}}}
		{8}{\ket{\nu=0, \mathrm{K-IVC}}}
		{9}{\ket{\nu=0, \mathrm{VP}}}
	}
	[nada]
$ & \cref{app:sec:results_corr_ins:4} & $T = \SI{5}{\kelvin}$ & $T = \SI{7}{\kelvin}$ & $T = \SI{17}{\kelvin}$ & $T = \SI{23.8}{\kelvin}$ \\ 
		\cline{2-7}
		& $\IfStrEqCase{5}{{1}{\ket{\nu={}4} }
		{2}{\ket{\nu={}3, \mathrm{IVC}}}
		{3}{\ket{\nu={}3, \mathrm{VP}}}
		{4}{\ket{\nu={}2, \mathrm{K-IVC}}}
		{5}{\ket{\nu={}2, \mathrm{VP}}}
		{6}{\ket{\nu={}1, (\mathrm{K-IVC}+\mathrm{VP})}}
		{7}{\ket{\nu={}1, \mathrm{VP}}}
		{8}{\ket{\nu=0, \mathrm{K-IVC}}}
		{9}{\ket{\nu=0, \mathrm{VP}}}
	}
	[nada]
$ & \cref{app:sec:results_corr_ins:5} & $T = \SI{5}{\kelvin}$ & $T = \SI{7}{\kelvin}$ & $T = \SI{17}{\kelvin}$ & $T = \SI{23.8}{\kelvin}$ \\ 
		\hline
		\multirow{2}{*}{$1$} & $\IfStrEqCase{6}{{1}{\ket{\nu={}4} }
		{2}{\ket{\nu={}3, \mathrm{IVC}}}
		{3}{\ket{\nu={}3, \mathrm{VP}}}
		{4}{\ket{\nu={}2, \mathrm{K-IVC}}}
		{5}{\ket{\nu={}2, \mathrm{VP}}}
		{6}{\ket{\nu={}1, (\mathrm{K-IVC}+\mathrm{VP})}}
		{7}{\ket{\nu={}1, \mathrm{VP}}}
		{8}{\ket{\nu=0, \mathrm{K-IVC}}}
		{9}{\ket{\nu=0, \mathrm{VP}}}
	}
	[nada]
$ & \cref{app:sec:results_corr_ins:6} & $T = \SI{7}{\kelvin}$ & $T = \SI{9.8}{\kelvin}$ & $T = \SI{27}{\kelvin}$ & $T = \SI{37.8}{\kelvin}$ \\ 
		\cline{2-7}
		& $\IfStrEqCase{7}{{1}{\ket{\nu={}4} }
		{2}{\ket{\nu={}3, \mathrm{IVC}}}
		{3}{\ket{\nu={}3, \mathrm{VP}}}
		{4}{\ket{\nu={}2, \mathrm{K-IVC}}}
		{5}{\ket{\nu={}2, \mathrm{VP}}}
		{6}{\ket{\nu={}1, (\mathrm{K-IVC}+\mathrm{VP})}}
		{7}{\ket{\nu={}1, \mathrm{VP}}}
		{8}{\ket{\nu=0, \mathrm{K-IVC}}}
		{9}{\ket{\nu=0, \mathrm{VP}}}
	}
	[nada]
$ & \cref{app:sec:results_corr_ins:7} & $T = \SI{7}{\kelvin}$ & $T = \SI{9.8}{\kelvin}$ & $T = \SI{27}{\kelvin}$ & $T = \SI{37.8}{\kelvin}$ \\ 
		\hline
		\multirow{2}{*}{$0$} & $\IfStrEqCase{8}{{1}{\ket{\nu={}4} }
		{2}{\ket{\nu={}3, \mathrm{IVC}}}
		{3}{\ket{\nu={}3, \mathrm{VP}}}
		{4}{\ket{\nu={}2, \mathrm{K-IVC}}}
		{5}{\ket{\nu={}2, \mathrm{VP}}}
		{6}{\ket{\nu={}1, (\mathrm{K-IVC}+\mathrm{VP})}}
		{7}{\ket{\nu={}1, \mathrm{VP}}}
		{8}{\ket{\nu=0, \mathrm{K-IVC}}}
		{9}{\ket{\nu=0, \mathrm{VP}}}
	}
	[nada]
$ & \cref{app:sec:results_corr_ins:8} & $T = \SI{8}{\kelvin}$ & $T = \SI{11.2}{\kelvin}$ & $T = \SI{32}{\kelvin}$ & $T = \SI{44.8}{\kelvin}$ \\ 
		\cline{2-7}
		& $\IfStrEqCase{9}{{1}{\ket{\nu={}4} }
		{2}{\ket{\nu={}3, \mathrm{IVC}}}
		{3}{\ket{\nu={}3, \mathrm{VP}}}
		{4}{\ket{\nu={}2, \mathrm{K-IVC}}}
		{5}{\ket{\nu={}2, \mathrm{VP}}}
		{6}{\ket{\nu={}1, (\mathrm{K-IVC}+\mathrm{VP})}}
		{7}{\ket{\nu={}1, \mathrm{VP}}}
		{8}{\ket{\nu=0, \mathrm{K-IVC}}}
		{9}{\ket{\nu=0, \mathrm{VP}}}
	}
	[nada]
$ & \cref{app:sec:results_corr_ins:9} & $T = \SI{8}{\kelvin}$ & $T = \SI{11.2}{\kelvin}$ & $T = \SI{32}{\kelvin}$ & $T = \SI{44.8}{\kelvin}$ \\ 
		\hline
	\end{tabular}
	\caption{Overview of numerical results detailing the band structures of the TBG and TSTG correlated ground state candidates. We restrict to the positive integer-filled states from \cref{app:tab:model_states}. The band structures of the negative integer-filled states can be deduced from their positive counterparts, via the many-body particle-hole symmetry shared by TBG and TSTG~\cite{CAL21,SON22}. For each ground state candidate, we list the \siSection{} where the results are presented. The band structures have been computed at both low and high temperatures, as detailed in the table. For all states in the table with $\nu \leq 2$, the high temperature is close to the peak temperature where both the correlated integer-filled state and the states obtained by doping it up to $\Delta \nu = \pm 0.5$ remain stable. The $\nu=3$ states are not stable for electron or hole doping larger than $\Delta \nu \approx \pm 0.15$ at any temperature. The high temperature of the $\nu = 3$ states is approximately the highest temperature at which the integer filled state is stable. For the $\nu = 4$ state, the high temperature we choose has no physical significance, and is simply the highest temperature we have simulated.}
	\label{app:tab:summary_band_structures}
	\end{table}}
	
	In this \siSection{}, we present comprehensive numerical results on the band structures of the correlated ground state candidates of TBG and TSTG from \cref{app:tab:model_states}. The band structure calculations are performed using the second-order self-consistent perturbation theory outlined in \cref{app:sec:se_correction_beyond_HF}. The results are summarized in \cref{app:tab:summary_band_structures} and then presented in detail in the subsequent \crefrange{app:sec:results_corr_ins:1}{app:sec:results_corr_ins:9}.

	For every positively filled correlated ground state candidate listed in \cref{app:tab:summary_band_structures}, we determine the second-order self-consistent solution for the integer-filled state. We also obtain the self-consistent solutions of the correlated phases that result by doping the integer filled states, up to a maximum of $\Delta \nu = \pm 0.5$. This methodology was explained in \cref{app:sec:se_correction_beyond_HF:sc_problem_and_numerics}. The band structures of each of the correlated ground state candidates are obtained at both low and high temperatures and within TBG, TSTG with no displacement field ($\mathcal{E} = \SI{0}{\milli\electronvolt}$), and TSTG with a small value of the displacement field ($\mathcal{E} = \SI{25}{\milli\electronvolt}$). 
	
	For each graphene heterostructure we consider (TBG and TSTG), temperature, value of the displacement field (in the case of TSTG), and correlated state, the results are shown in a single figure ({\it e.g.}{}, \cref{app:fig:sym_br_bs_1_TBG_low}). To enable easy comparisons between different band structures, we maintain consistent figure layouts. First, the \emph{total} spectral function, or the density of states (DOS) 
	\begin{equation}
		\label{app:eqn:definition_total_spectral_function}
		\mathcal{A} \left( \omega \right) = \frac{1}{N_0} \sum_{\vec{k}} \sum_{i, \eta, s} A_{i \eta s; i \eta s} \left(\omega, \vec{k} \right),
	\end{equation} 
is plotted in panel (a) of each figure as a function of $\omega$ and of the filling $\nu_0 - \Delta\nu \leq  \nu \leq \nu_0 + \Delta \nu$ (where $\nu_0$ is the filling of the correlated state depicted in the figure and $\Delta \nu= 0.5$ is the maximal doping we consider). The $\vec{k}$-resolved spectral function 
\begin{equation}
\mathcal{A} \left( \omega, \vec{k} \right) = \sum_{i, \eta, s} A_{i \eta s; i \eta s} \left(\omega, \vec{k} \right),
\end{equation}
is then plotted in panels (b)-(d) of each figure along the high-symmetry lines of the moir\'e BZ for the integer-filled correlated state at $\nu = \nu_0$ and for the correlated phases obtained by doping it to $\nu = \nu_0 \pm 0.2$ (for $\nu_0 \neq 3$), or $\nu = \nu_0 \pm 0.1$ (for $\nu_0 = 3$). We consider a small energy range around the Fermi level such that $ \omega / \si{\milli\electronvolt} \in \left[-50, 50 \right]$ for TBG and $ \omega / \si{\milli\electronvolt} \in \left[ -75, 75 \right]$ for TSTG. Both the $\vec{k}$-resolved spectral function and the total density of states are plotted using a color map where the saturation of the color is proportional to either $\mathcal{A} \left( \omega \right)$, for panel (a), or $\mathcal{A} \left( \omega, \vec{k} \right)$, for panel (b). The hue of the color map is related to the $f$-character $\%_{\hat{f}}$ of the spectral function in that region, with bluer regions corresponding to a larger $f$-character. The $f$-character is defined as 
\begin{equation}
\%_{\hat{f}} = \frac{ \frac{1}{N_0} \sum_{\vec{k}} \sum_{\alpha, \eta, s} A_{(\alpha + 4) \eta s; (\alpha + 4) \eta s} \left(\omega, \vec{k} \right)}{	\mathcal{A} \left( \omega \right)}, \qq{or} \%_{\hat{f}} = \frac{ \sum_{\alpha, \eta, s} A_{(\alpha + 4) \eta s; (\alpha + 4) \eta s} \left(\omega, \vec{k} \right)}{	\mathcal{A} \left( \omega, \vec{k} \right)},
\end{equation} 
for the DOS and $\vec{k}$-resolved spectral function plots, respectively. 

Finally, we note that the results are shown only for the $\nu \geq 0$ correlated ground state candidates from \cref{app:tab:model_states}. The results for the $\nu < 0$ correlated phases can be inferred directly from the $\nu \geq 0$ ones using the many-body charge conjugation symmetries of TBG and TSTG discussed in \cref{app:sec:hartree_fock:ground_states:ph_symmetry}.

\subsection{General discussion}\label{app:sec:results_corr_ins:general_remarks}
Our comprehensive numerical results of the correlated ground state candidates of the THF model reveal several general features, which we summarize below, starting with TBG:
\begin{itemize}

\item Near integer fillings $0 \leq \nu \leq 3$, we observe that the gap between the flat bands above and below the Fermi energy near the $\mathrm{K}_M$ point, when computed using second-order self-consistent perturbation theory, is reduced in comparison to the value obtained from Hartree-Fock calculations~\cite{SON22}. The gap at $\mathrm{K}_M$ acts as a proxy for the magnitude of the order parameter of the $f$-electrons~\cite{SON22}. Due to the more accurate representation of interaction-driven fluctuations in the second-order approach, there is a consequent decrease in the $f$-electron order parameter and in the associated gap at the $\mathrm{K}_M$ point. 

\item The shrinking of the gap in the second-order perturbative approach is more pronounced the further a state is from charge neutrality. For example, within Hartree-Fock theory, the $\IfStrEqCase{5}{{1}{\ket{\nu={}4} }
		{2}{\ket{\nu={}3, \mathrm{IVC}}}
		{3}{\ket{\nu={}3, \mathrm{VP}}}
		{4}{\ket{\nu={}2, \mathrm{K-IVC}}}
		{5}{\ket{\nu={}2, \mathrm{VP}}}
		{6}{\ket{\nu={}1, (\mathrm{K-IVC}+\mathrm{VP})}}
		{7}{\ket{\nu={}1, \mathrm{VP}}}
		{8}{\ket{\nu=0, \mathrm{K-IVC}}}
		{9}{\ket{\nu=0, \mathrm{VP}}}
	}
	[nada]
$ state shows an indirect gap~\cite{SON22}. However, such a gap is not present in second-order perturbative calculations. Considering the $\nu=3$ states, self-consistent Hartree-Fock calculations indicate a gap of approximately $\Delta \omega \approx \SI{30}{\milli\electronvolt}$ at the $\mathrm{K}_M$ point~\cite{SON22}. This is in contrast to the second-order perturbation theory, which predicts a notably smaller gap of $\Delta\omega \approx \SI{15}{\milli\electronvolt}$. The reason for this difference lies in the behavior of the $f$-electrons for the $1 \leq \nu \leq 3$ integer-filled insulators. These electrons, which largely constitute the hole excitation bands, are nearer to the Fermi energy (or chemical potential) the more $\nu$ deviates from charge neutrality. This observation can be directly seen from the momentum-resolved spectral functions at integer filling $1 \leq \nu \leq 3$. Consequently, the $f$-electrons are more prone to charge fluctuations, which can potentially reduce the formation of $f$-electron local moments and thus reduce the order parameter formed by the $f$-electrons.

\item For $0 \leq \nu <4$ and close to integer fillings, the flat $f$-electron excitations at the moir\'e BZ boundary lead to sharp peaks in the DOS.

\item At a given filling $0 \leq \nu <4$, the spectra of the valley- polarized and intervalley-coherent ground state candidates are broadly similar, as they are related by the \emph{approximate} nonchiral flat $\mathrm{U} \left(4\right)$ symmetry~\cite{BER21a,SON22}. The latter is an exact symmetry of the system in the \emph{flat} $M=0$ limit. One notable exception already revealed both numerically and analytically within  Hartree-Fock theory~\cite{SON22}, for $0 \leq \nu \leq 2$, the intervalley-coherent states have a larger gap at the $\Gamma_M$ point. This larger gap is responsible for their energy being lower than that of valley-polarized states in the absence of strain.	

\item The spectrum computed around $\nu=0$ is particle hole-symmetric, due to the $\mathcal{P}$ symmetry of TBG. For small doping around the charge neutrality point, both the electron and the hole excitations consist of light fermions with predominantly $c$-character. Being far from the Fermi energy (chemical potential), the $f$-electrons do not experience interaction-driven fluctuations at small levels of doping. In effect, the doped band structure is almost identical to the undoped one, up to a shift in energy. This is confirmed in the DOS plots, which reveal a two-peak structure stemming from the $f$-electron flat bands near the moir\'e BZ boundary: for doping up to $\abs{\Delta \nu} \lesssim 0.1$ at low temperatures or $\abs{\Delta \nu} \lesssim 0.05$ at high temperatures, the two $f$-electron peaks move rigidly with doping. As doping is increased, the $f$-electron excitations are brought closer to the Fermi energy and thus experience larger interaction-driven fluctuations. This results in a shrinking of the gap at $\mathrm{K}_M$ between the $f$-electron flat bands and a more incoherent peak of the $f$-electron bands in the $\vec{k}$-resolved spectral function and the DOS. This effect is more pronounced at higher temperatures. The resulting DOS plots are similar in shape to the local DOS measured by scanning tunneling microscopy experiments in the ultra-low strain device of Ref.~\cite{NUC23}.

\item Near $\nu=1$ or $\nu=2$, when doping is small, the electron excitations consist of light fermions with predominantly $c$-character near the $\Gamma_M$ point. In contrast, the hole excitations are mostly given by heavy electrons with predominantly $f$-character at the edge of the moir\'e BZ. We also point out that the light-fermion bands at $\nu=2$ cross the Fermi energy and create a small Fermi surface near $\Gamma_M$ point, which is different from the Hartree-Fock result~\cite{SON22}.

\item Due to the markedly different nature of the electron and hole charge-one excitations around the $1 \leq \nu \leq 2$ insulators, the electron and hole doped regions are very different:
\begin{itemize}
	\item The electron doping case is qualitatively similar to the charge neutrality case: for small electron doping, the electrons near the Fermi energy have $c$-character and experience little interaction-driven fluctuations, leaving the $f$-electron order parameter relatively unaffected. In this case, the doped bands are simply an energy-shifted version of the undoped ones. The gap at $\mathrm{K}_M$ between the $f$-electron flat bands remains relatively constant and the DOS plots show a two-peak structure with relatively sharp $f$-electron peaks which move parallel to one another as a function of doping.  
	
	\item As the electron doping is further increased, $f$-electrons are brought close to the Fermi level. As a result, they experience larger interaction-driven fluctuations which diminish their order parameter leading to increased incoherence of their spectral signal and a shrinking of the gap at $\mathrm{K}_M$.
	
	\item In the hole doping case, the $f$-electrons from the hole band are brought close to the Fermi energy, which leads to large charge fluctuations of the $f$-electrons. Their spectral signal becomes incoherent and the gap at the $\mathrm{K}_M$ point shrinks with doping. In the DOS plots, the hole $f$-electron peak remains pinned close to zero energy, while the electron one moves towards zero energy as hole doping is increased.
	
	\item The effects of interaction-driven fluctuations increases with increasing temperatures.
\end{itemize}

\item For the doped $\nu=1$ correlated insulators, at low temperature, a three-peaked structure can be seen in the DOS plots for $\nu \approx 0.6$. This is due to a phase transition where the $f$-electron order parameter changes between different orders (one more $f$-electron per unit cell becomes filled). 

\item For the $\nu = 3$ correlated state, we find that:
\begin{itemize}
	\item The system does not have an indirect gap, with $c$-electrons already forming gapless charge-one excitations around the Fermi energy, which is also similar to $\nu=2$. As a result, for very small electron or hole doping $\abs{\Delta \nu} \lesssim 0.1$, the main fermionic species around the Fermi energy are still $c$-electrons and the DOS dependence with $\nu$ is consistent with a rigid doping behavior.
	\item For larger doping $\abs{\Delta \nu} \gtrsim 0.1$, the $f$-electron bands are brought closer to the Fermi energy. As a result, large interaction-driven fluctuations diminish the order parameter of the system, the gap at $\mathrm{K}_M$ decreases, and the two $f$-electron peaks in the DOS merge together. At around $\abs{\Delta \nu} \sim 0.3$, the system undergoes an order-disorder transition and the order is lost. 
\end{itemize}

\item As already pointed out in \cref{app:tab:summary_band_structures}, the high-temperature chosen for the $0 \leq \nu \leq 2$ correlated insulator is close to the critical temperature at which the system start transitioning to the disordered (symmetric) phase within the doping range we consider. The critical temperatures obtained through second-order perturbation are close to the DMFT results~\cite{RAI23a}. This is in stark contrast with the finite-temperature Hartree-Fock calculations, which overestimate the critical temperature by an order of magnitude~\cite{RAI23a}.  

\item The band structure calculations at low and high temperatures are qualitatively similar, besides the fact that the $f$-electron excitations become more incoherent at higher temperatures.  

\item The band structure of the $\nu = 4$ band insulator is largely unaffected by the addition of the second-order perturbative correction. The latter only leads to more incoherent $f$-electron excitations.
\end{itemize}

The spectral functions of TSTG in the symmetry-broken case, whether a displacement field is present or not, closely resemble those of TBG. The primary difference is the inclusion of the highly dispersive $d$-electron bands:
\begin{itemize}
\item The Dirac cone formed by the $d$-electrons is not pinned at charge neutrality and, therefore, slightly changes the relative filling of the $f$- and $c$-electrons away from charge neutrality.
\item For $\nu>0$, the filling of the $d$-electrons $\nu_d$ is slightly larger than zero (with $\nu_d$ increasing with $\nu$). Relative to the TBG case at the same overall filling, the TBG bands within TSTG are slightly hole doped.
\item In the presence of displacement field, the $d$- and $f$-electrons hybridize slightly near the $\mathrm{K}_M$ point. 
\end{itemize}

\clearpage
\FloatBarrier

\subsection{The \texorpdfstring{$\protect\IfStrEqCase{1}{{1}{\ket{\nu={}4} }
		{2}{\ket{\nu={}3, \mathrm{IVC}}}
		{3}{\ket{\nu={}3, \mathrm{VP}}}
		{4}{\ket{\nu={}2, \mathrm{K-IVC}}}
		{5}{\ket{\nu={}2, \mathrm{VP}}}
		{6}{\ket{\nu={}1, (\mathrm{K-IVC}+\mathrm{VP})}}
		{7}{\ket{\nu={}1, \mathrm{VP}}}
		{8}{\ket{\nu=0, \mathrm{K-IVC}}}
		{9}{\ket{\nu=0, \mathrm{VP}}}
	}
	[nada]
$}{nu=4} correlated ground state candidate}
\label{app:sec:results_corr_ins:1}
\subsubsection{Low temperature}\label{app:sec:results_corr_ins_1_low}
\begin{figure}[!h]\includegraphics[width=\textwidth]{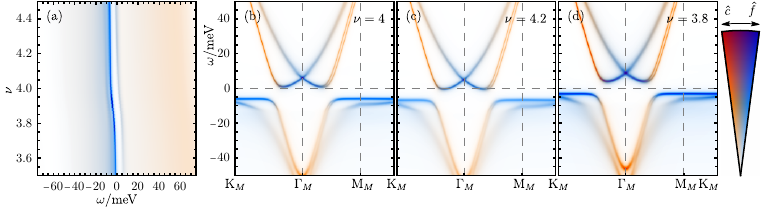}\subfloat{\label{app:fig:sym_br_bs_1_TBG_low:a}}\subfloat{\label{app:fig:sym_br_bs_1_TBG_low:b}}\subfloat{\label{app:fig:sym_br_bs_1_TBG_low:c}}\subfloat{\label{app:fig:sym_br_bs_1_TBG_low:d}}\caption{Band structure of the $\protect\IfStrEqCase{1}{{1}{\ket{\nu={}4} }
		{2}{\ket{\nu={}3, \mathrm{IVC}}}
		{3}{\ket{\nu={}3, \mathrm{VP}}}
		{4}{\ket{\nu={}2, \mathrm{K-IVC}}}
		{5}{\ket{\nu={}2, \mathrm{VP}}}
		{6}{\ket{\nu={}1, (\mathrm{K-IVC}+\mathrm{VP})}}
		{7}{\ket{\nu={}1, \mathrm{VP}}}
		{8}{\ket{\nu=0, \mathrm{K-IVC}}}
		{9}{\ket{\nu=0, \mathrm{VP}}}
	}
	[nada]
$ ground state candidate of TBG at $T=\SI{5}{\kelvin}$. The total spectral function $\mathcal{A} \left( \omega \right)$ of the system is shown in (a) as a function of filling. (b)-(d) show the $\vec{k}$-resolved spectral function $\mathcal{A} \left( \omega, \vec{k}\right)$ of the correlated state at integer filling and for small hole and electron doping around it, respectively.}\label{app:fig:sym_br_bs_1_TBG_low}\end{figure}\begin{figure}[!h]\includegraphics[width=\textwidth]{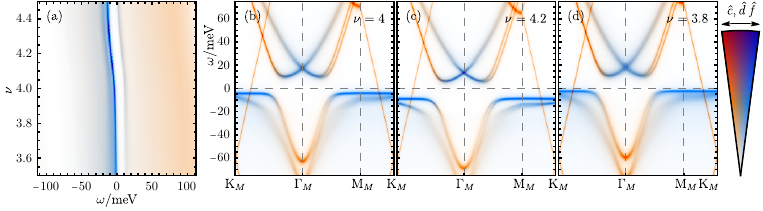}\subfloat{\label{app:fig:sym_br_bs_1_TSTG_low:a}}\subfloat{\label{app:fig:sym_br_bs_1_TSTG_low:b}}\subfloat{\label{app:fig:sym_br_bs_1_TSTG_low:c}}\subfloat{\label{app:fig:sym_br_bs_1_TSTG_low:d}}\caption{Band structure of the $\protect\IfStrEqCase{1}{{1}{\ket{\nu={}4} }
		{2}{\ket{\nu={}3, \mathrm{IVC}}}
		{3}{\ket{\nu={}3, \mathrm{VP}}}
		{4}{\ket{\nu={}2, \mathrm{K-IVC}}}
		{5}{\ket{\nu={}2, \mathrm{VP}}}
		{6}{\ket{\nu={}1, (\mathrm{K-IVC}+\mathrm{VP})}}
		{7}{\ket{\nu={}1, \mathrm{VP}}}
		{8}{\ket{\nu=0, \mathrm{K-IVC}}}
		{9}{\ket{\nu=0, \mathrm{VP}}}
	}
	[nada]
$ ground state candidate of TSTG for $\mathcal{E}=\SI{0}{\milli\electronvolt}$ at $T=\SI{7}{\kelvin}$. The total spectral function $\mathcal{A} \left( \omega \right)$ of the system is shown in (a) as a function of filling. (b)-(d) show the $\vec{k}$-resolved spectral function $\mathcal{A} \left( \omega, \vec{k}\right)$ of the correlated state at integer filling and for small hole and electron doping around it, respectively.}\label{app:fig:sym_br_bs_1_TSTG_low}\end{figure}\begin{figure}[!h]\includegraphics[width=\textwidth]{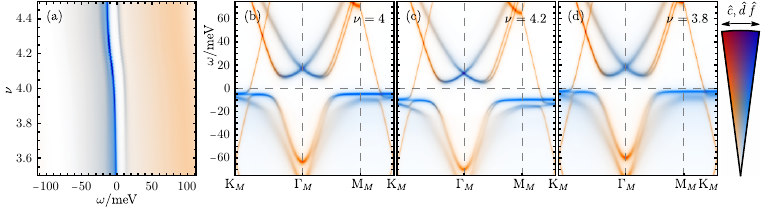}\subfloat{\label{app:fig:sym_br_bs_1_TSTGu_low:a}}\subfloat{\label{app:fig:sym_br_bs_1_TSTGu_low:b}}\subfloat{\label{app:fig:sym_br_bs_1_TSTGu_low:c}}\subfloat{\label{app:fig:sym_br_bs_1_TSTGu_low:d}}\caption{Band structure of the $\protect\IfStrEqCase{1}{{1}{\ket{\nu={}4} }
		{2}{\ket{\nu={}3, \mathrm{IVC}}}
		{3}{\ket{\nu={}3, \mathrm{VP}}}
		{4}{\ket{\nu={}2, \mathrm{K-IVC}}}
		{5}{\ket{\nu={}2, \mathrm{VP}}}
		{6}{\ket{\nu={}1, (\mathrm{K-IVC}+\mathrm{VP})}}
		{7}{\ket{\nu={}1, \mathrm{VP}}}
		{8}{\ket{\nu=0, \mathrm{K-IVC}}}
		{9}{\ket{\nu=0, \mathrm{VP}}}
	}
	[nada]
$ ground state candidate of TSTG for $\mathcal{E}=\SI{25}{\milli\electronvolt}$ at $T=\SI{7}{\kelvin}$. The total spectral function $\mathcal{A} \left( \omega \right)$ of the system is shown in (a) as a function of filling. (b)-(d) show the $\vec{k}$-resolved spectral function $\mathcal{A} \left( \omega, \vec{k}\right)$ of the correlated state at integer filling and for small hole and electron doping around it, respectively.}\label{app:fig:sym_br_bs_1_TSTGu_low}\end{figure}

\subsubsection{High temperature}\label{app:sec:results_corr_ins_1_high}
\begin{figure}[!h]\includegraphics[width=\textwidth]{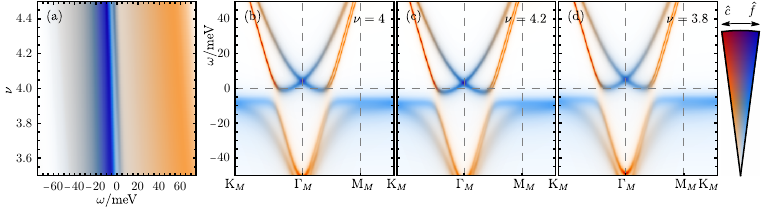}\subfloat{\label{app:fig:sym_br_bs_1_TBG_high:a}}\subfloat{\label{app:fig:sym_br_bs_1_TBG_high:b}}\subfloat{\label{app:fig:sym_br_bs_1_TBG_high:c}}\subfloat{\label{app:fig:sym_br_bs_1_TBG_high:d}}\caption{Band structure of the $\protect\IfStrEqCase{1}{{1}{\ket{\nu={}4} }
		{2}{\ket{\nu={}3, \mathrm{IVC}}}
		{3}{\ket{\nu={}3, \mathrm{VP}}}
		{4}{\ket{\nu={}2, \mathrm{K-IVC}}}
		{5}{\ket{\nu={}2, \mathrm{VP}}}
		{6}{\ket{\nu={}1, (\mathrm{K-IVC}+\mathrm{VP})}}
		{7}{\ket{\nu={}1, \mathrm{VP}}}
		{8}{\ket{\nu=0, \mathrm{K-IVC}}}
		{9}{\ket{\nu=0, \mathrm{VP}}}
	}
	[nada]
$ ground state candidate of TBG at $T=\SI{40}{\kelvin}$. The total spectral function $\mathcal{A} \left( \omega \right)$ of the system is shown in (a) as a function of filling. (b)-(d) show the $\vec{k}$-resolved spectral function $\mathcal{A} \left( \omega, \vec{k}\right)$ of the correlated state at integer filling and for small hole and electron doping around it, respectively.}\label{app:fig:sym_br_bs_1_TBG_high}\end{figure}\begin{figure}[!h]\includegraphics[width=\textwidth]{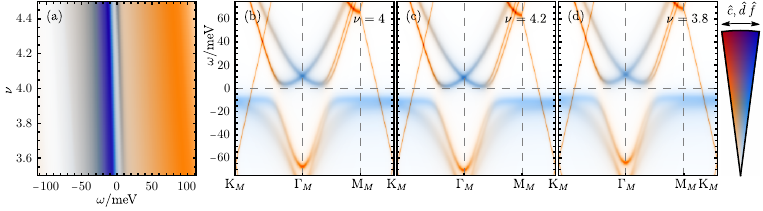}\subfloat{\label{app:fig:sym_br_bs_1_TSTG_high:a}}\subfloat{\label{app:fig:sym_br_bs_1_TSTG_high:b}}\subfloat{\label{app:fig:sym_br_bs_1_TSTG_high:c}}\subfloat{\label{app:fig:sym_br_bs_1_TSTG_high:d}}\caption{Band structure of the $\protect\IfStrEqCase{1}{{1}{\ket{\nu={}4} }
		{2}{\ket{\nu={}3, \mathrm{IVC}}}
		{3}{\ket{\nu={}3, \mathrm{VP}}}
		{4}{\ket{\nu={}2, \mathrm{K-IVC}}}
		{5}{\ket{\nu={}2, \mathrm{VP}}}
		{6}{\ket{\nu={}1, (\mathrm{K-IVC}+\mathrm{VP})}}
		{7}{\ket{\nu={}1, \mathrm{VP}}}
		{8}{\ket{\nu=0, \mathrm{K-IVC}}}
		{9}{\ket{\nu=0, \mathrm{VP}}}
	}
	[nada]
$ ground state candidate of TSTG for $\mathcal{E}=\SI{0}{\milli\electronvolt}$ at $T=\SI{56}{\kelvin}$. The total spectral function $\mathcal{A} \left( \omega \right)$ of the system is shown in (a) as a function of filling. (b)-(d) show the $\vec{k}$-resolved spectral function $\mathcal{A} \left( \omega, \vec{k}\right)$ of the correlated state at integer filling and for small hole and electron doping around it, respectively.}\label{app:fig:sym_br_bs_1_TSTG_high}\end{figure}\begin{figure}[!h]\includegraphics[width=\textwidth]{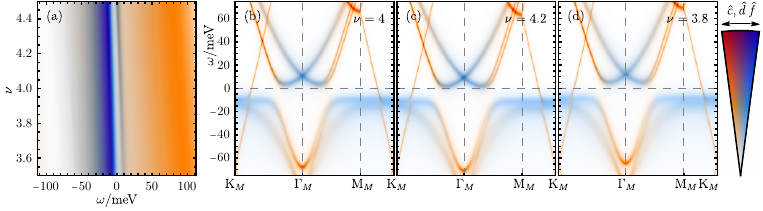}\subfloat{\label{app:fig:sym_br_bs_1_TSTGu_high:a}}\subfloat{\label{app:fig:sym_br_bs_1_TSTGu_high:b}}\subfloat{\label{app:fig:sym_br_bs_1_TSTGu_high:c}}\subfloat{\label{app:fig:sym_br_bs_1_TSTGu_high:d}}\caption{Band structure of the $\protect\IfStrEqCase{1}{{1}{\ket{\nu={}4} }
		{2}{\ket{\nu={}3, \mathrm{IVC}}}
		{3}{\ket{\nu={}3, \mathrm{VP}}}
		{4}{\ket{\nu={}2, \mathrm{K-IVC}}}
		{5}{\ket{\nu={}2, \mathrm{VP}}}
		{6}{\ket{\nu={}1, (\mathrm{K-IVC}+\mathrm{VP})}}
		{7}{\ket{\nu={}1, \mathrm{VP}}}
		{8}{\ket{\nu=0, \mathrm{K-IVC}}}
		{9}{\ket{\nu=0, \mathrm{VP}}}
	}
	[nada]
$ ground state candidate of TSTG for $\mathcal{E}=\SI{25}{\milli\electronvolt}$ at $T=\SI{56}{\kelvin}$. The total spectral function $\mathcal{A} \left( \omega \right)$ of the system is shown in (a) as a function of filling. (b)-(d) show the $\vec{k}$-resolved spectral function $\mathcal{A} \left( \omega, \vec{k}\right)$ of the correlated state at integer filling and for small hole and electron doping around it, respectively.}\label{app:fig:sym_br_bs_1_TSTGu_high}\end{figure}

\subsection{The \texorpdfstring{$\protect\IfStrEqCase{2}{{1}{\ket{\nu={}4} }
		{2}{\ket{\nu={}3, \mathrm{IVC}}}
		{3}{\ket{\nu={}3, \mathrm{VP}}}
		{4}{\ket{\nu={}2, \mathrm{K-IVC}}}
		{5}{\ket{\nu={}2, \mathrm{VP}}}
		{6}{\ket{\nu={}1, (\mathrm{K-IVC}+\mathrm{VP})}}
		{7}{\ket{\nu={}1, \mathrm{VP}}}
		{8}{\ket{\nu=0, \mathrm{K-IVC}}}
		{9}{\ket{\nu=0, \mathrm{VP}}}
	}
	[nada]
$}{nu=3 IVC} correlated ground state candidate}
\label{app:sec:results_corr_ins:2}
\subsubsection{Low temperature}\label{app:sec:results_corr_ins_2_low}
\begin{figure}[!h]\includegraphics[width=\textwidth]{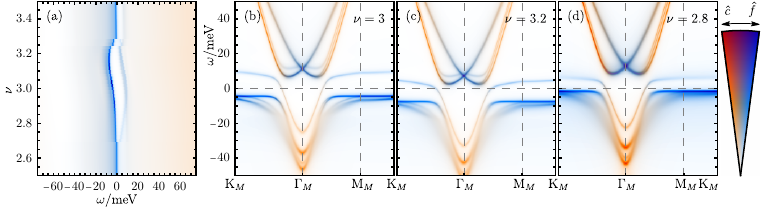}\subfloat{\label{app:fig:sym_br_bs_2_TBG_low:a}}\subfloat{\label{app:fig:sym_br_bs_2_TBG_low:b}}\subfloat{\label{app:fig:sym_br_bs_2_TBG_low:c}}\subfloat{\label{app:fig:sym_br_bs_2_TBG_low:d}}\caption{Band structure of the $\protect\IfStrEqCase{2}{{1}{\ket{\nu={}4} }
		{2}{\ket{\nu={}3, \mathrm{IVC}}}
		{3}{\ket{\nu={}3, \mathrm{VP}}}
		{4}{\ket{\nu={}2, \mathrm{K-IVC}}}
		{5}{\ket{\nu={}2, \mathrm{VP}}}
		{6}{\ket{\nu={}1, (\mathrm{K-IVC}+\mathrm{VP})}}
		{7}{\ket{\nu={}1, \mathrm{VP}}}
		{8}{\ket{\nu=0, \mathrm{K-IVC}}}
		{9}{\ket{\nu=0, \mathrm{VP}}}
	}
	[nada]
$ ground state candidate of TBG at $T=\SI{4}{\kelvin}$. The total spectral function $\mathcal{A} \left( \omega \right)$ of the system is shown in (a) as a function of filling. (b)-(d) show the $\vec{k}$-resolved spectral function $\mathcal{A} \left( \omega, \vec{k}\right)$ of the correlated state at integer filling and for small hole and electron doping around it, respectively.}\label{app:fig:sym_br_bs_2_TBG_low}\end{figure}\begin{figure}[!h]\includegraphics[width=\textwidth]{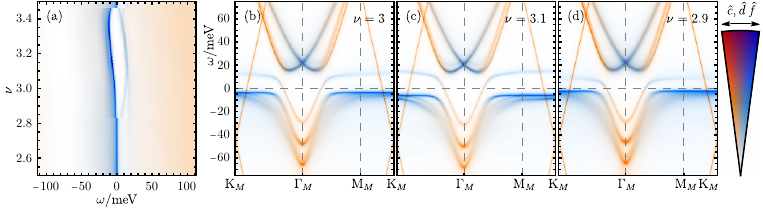}\subfloat{\label{app:fig:sym_br_bs_2_TSTG_low:a}}\subfloat{\label{app:fig:sym_br_bs_2_TSTG_low:b}}\subfloat{\label{app:fig:sym_br_bs_2_TSTG_low:c}}\subfloat{\label{app:fig:sym_br_bs_2_TSTG_low:d}}\caption{Band structure of the $\protect\IfStrEqCase{2}{{1}{\ket{\nu={}4} }
		{2}{\ket{\nu={}3, \mathrm{IVC}}}
		{3}{\ket{\nu={}3, \mathrm{VP}}}
		{4}{\ket{\nu={}2, \mathrm{K-IVC}}}
		{5}{\ket{\nu={}2, \mathrm{VP}}}
		{6}{\ket{\nu={}1, (\mathrm{K-IVC}+\mathrm{VP})}}
		{7}{\ket{\nu={}1, \mathrm{VP}}}
		{8}{\ket{\nu=0, \mathrm{K-IVC}}}
		{9}{\ket{\nu=0, \mathrm{VP}}}
	}
	[nada]
$ ground state candidate of TSTG for $\mathcal{E}=\SI{0}{\milli\electronvolt}$ at $T=\SI{5.6}{\kelvin}$. The total spectral function $\mathcal{A} \left( \omega \right)$ of the system is shown in (a) as a function of filling. (b)-(d) show the $\vec{k}$-resolved spectral function $\mathcal{A} \left( \omega, \vec{k}\right)$ of the correlated state at integer filling and for small hole and electron doping around it, respectively.}\label{app:fig:sym_br_bs_2_TSTG_low}\end{figure}\begin{figure}[!h]\includegraphics[width=\textwidth]{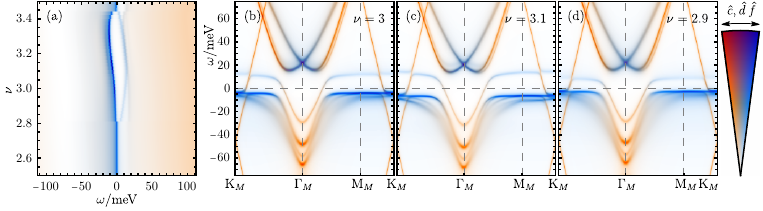}\subfloat{\label{app:fig:sym_br_bs_2_TSTGu_low:a}}\subfloat{\label{app:fig:sym_br_bs_2_TSTGu_low:b}}\subfloat{\label{app:fig:sym_br_bs_2_TSTGu_low:c}}\subfloat{\label{app:fig:sym_br_bs_2_TSTGu_low:d}}\caption{Band structure of the $\protect\IfStrEqCase{2}{{1}{\ket{\nu={}4} }
		{2}{\ket{\nu={}3, \mathrm{IVC}}}
		{3}{\ket{\nu={}3, \mathrm{VP}}}
		{4}{\ket{\nu={}2, \mathrm{K-IVC}}}
		{5}{\ket{\nu={}2, \mathrm{VP}}}
		{6}{\ket{\nu={}1, (\mathrm{K-IVC}+\mathrm{VP})}}
		{7}{\ket{\nu={}1, \mathrm{VP}}}
		{8}{\ket{\nu=0, \mathrm{K-IVC}}}
		{9}{\ket{\nu=0, \mathrm{VP}}}
	}
	[nada]
$ ground state candidate of TSTG for $\mathcal{E}=\SI{25}{\milli\electronvolt}$ at $T=\SI{5.6}{\kelvin}$. The total spectral function $\mathcal{A} \left( \omega \right)$ of the system is shown in (a) as a function of filling. (b)-(d) show the $\vec{k}$-resolved spectral function $\mathcal{A} \left( \omega, \vec{k}\right)$ of the correlated state at integer filling and for small hole and electron doping around it, respectively.}\label{app:fig:sym_br_bs_2_TSTGu_low}\end{figure}

\subsubsection{High temperature}\label{app:sec:results_corr_ins_2_high}
\begin{figure}[!h]\includegraphics[width=\textwidth]{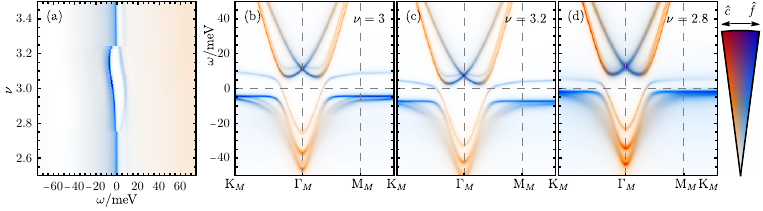}\subfloat{\label{app:fig:sym_br_bs_2_TBG_high:a}}\subfloat{\label{app:fig:sym_br_bs_2_TBG_high:b}}\subfloat{\label{app:fig:sym_br_bs_2_TBG_high:c}}\subfloat{\label{app:fig:sym_br_bs_2_TBG_high:d}}\caption{Band structure of the $\protect\IfStrEqCase{2}{{1}{\ket{\nu={}4} }
		{2}{\ket{\nu={}3, \mathrm{IVC}}}
		{3}{\ket{\nu={}3, \mathrm{VP}}}
		{4}{\ket{\nu={}2, \mathrm{K-IVC}}}
		{5}{\ket{\nu={}2, \mathrm{VP}}}
		{6}{\ket{\nu={}1, (\mathrm{K-IVC}+\mathrm{VP})}}
		{7}{\ket{\nu={}1, \mathrm{VP}}}
		{8}{\ket{\nu=0, \mathrm{K-IVC}}}
		{9}{\ket{\nu=0, \mathrm{VP}}}
	}
	[nada]
$ ground state candidate of TBG at $T=\SI{6}{\kelvin}$. The total spectral function $\mathcal{A} \left( \omega \right)$ of the system is shown in (a) as a function of filling. (b)-(d) show the $\vec{k}$-resolved spectral function $\mathcal{A} \left( \omega, \vec{k}\right)$ of the correlated state at integer filling and for small hole and electron doping around it, respectively.}\label{app:fig:sym_br_bs_2_TBG_high}\end{figure}\begin{figure}[!h]\includegraphics[width=\textwidth]{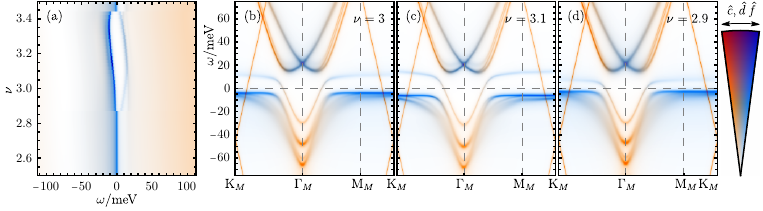}\subfloat{\label{app:fig:sym_br_bs_2_TSTG_high:a}}\subfloat{\label{app:fig:sym_br_bs_2_TSTG_high:b}}\subfloat{\label{app:fig:sym_br_bs_2_TSTG_high:c}}\subfloat{\label{app:fig:sym_br_bs_2_TSTG_high:d}}\caption{Band structure of the $\protect\IfStrEqCase{2}{{1}{\ket{\nu={}4} }
		{2}{\ket{\nu={}3, \mathrm{IVC}}}
		{3}{\ket{\nu={}3, \mathrm{VP}}}
		{4}{\ket{\nu={}2, \mathrm{K-IVC}}}
		{5}{\ket{\nu={}2, \mathrm{VP}}}
		{6}{\ket{\nu={}1, (\mathrm{K-IVC}+\mathrm{VP})}}
		{7}{\ket{\nu={}1, \mathrm{VP}}}
		{8}{\ket{\nu=0, \mathrm{K-IVC}}}
		{9}{\ket{\nu=0, \mathrm{VP}}}
	}
	[nada]
$ ground state candidate of TSTG for $\mathcal{E}=\SI{0}{\milli\electronvolt}$ at $T=\SI{8.4}{\kelvin}$. The total spectral function $\mathcal{A} \left( \omega \right)$ of the system is shown in (a) as a function of filling. (b)-(d) show the $\vec{k}$-resolved spectral function $\mathcal{A} \left( \omega, \vec{k}\right)$ of the correlated state at integer filling and for small hole and electron doping around it, respectively.}\label{app:fig:sym_br_bs_2_TSTG_high}\end{figure}\begin{figure}[!h]\includegraphics[width=\textwidth]{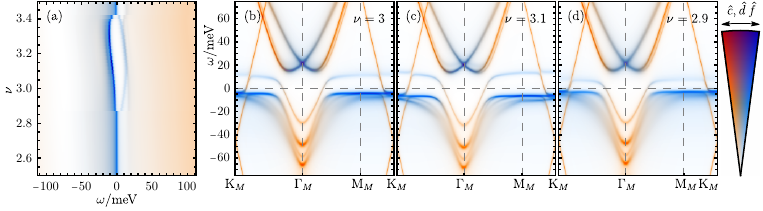}\subfloat{\label{app:fig:sym_br_bs_2_TSTGu_high:a}}\subfloat{\label{app:fig:sym_br_bs_2_TSTGu_high:b}}\subfloat{\label{app:fig:sym_br_bs_2_TSTGu_high:c}}\subfloat{\label{app:fig:sym_br_bs_2_TSTGu_high:d}}\caption{Band structure of the $\protect\IfStrEqCase{2}{{1}{\ket{\nu={}4} }
		{2}{\ket{\nu={}3, \mathrm{IVC}}}
		{3}{\ket{\nu={}3, \mathrm{VP}}}
		{4}{\ket{\nu={}2, \mathrm{K-IVC}}}
		{5}{\ket{\nu={}2, \mathrm{VP}}}
		{6}{\ket{\nu={}1, (\mathrm{K-IVC}+\mathrm{VP})}}
		{7}{\ket{\nu={}1, \mathrm{VP}}}
		{8}{\ket{\nu=0, \mathrm{K-IVC}}}
		{9}{\ket{\nu=0, \mathrm{VP}}}
	}
	[nada]
$ ground state candidate of TSTG for $\mathcal{E}=\SI{25}{\milli\electronvolt}$ at $T=\SI{8.4}{\kelvin}$. The total spectral function $\mathcal{A} \left( \omega \right)$ of the system is shown in (a) as a function of filling. (b)-(d) show the $\vec{k}$-resolved spectral function $\mathcal{A} \left( \omega, \vec{k}\right)$ of the correlated state at integer filling and for small hole and electron doping around it, respectively.}\label{app:fig:sym_br_bs_2_TSTGu_high}\end{figure}

\subsection{The \texorpdfstring{$\protect\IfStrEqCase{3}{{1}{\ket{\nu={}4} }
		{2}{\ket{\nu={}3, \mathrm{IVC}}}
		{3}{\ket{\nu={}3, \mathrm{VP}}}
		{4}{\ket{\nu={}2, \mathrm{K-IVC}}}
		{5}{\ket{\nu={}2, \mathrm{VP}}}
		{6}{\ket{\nu={}1, (\mathrm{K-IVC}+\mathrm{VP})}}
		{7}{\ket{\nu={}1, \mathrm{VP}}}
		{8}{\ket{\nu=0, \mathrm{K-IVC}}}
		{9}{\ket{\nu=0, \mathrm{VP}}}
	}
	[nada]
$}{nu=3 VP} correlated ground state candidate}
\label{app:sec:results_corr_ins:3}
\subsubsection{Low temperature}\label{app:sec:results_corr_ins_3_low}
\begin{figure}[!h]\includegraphics[width=\textwidth]{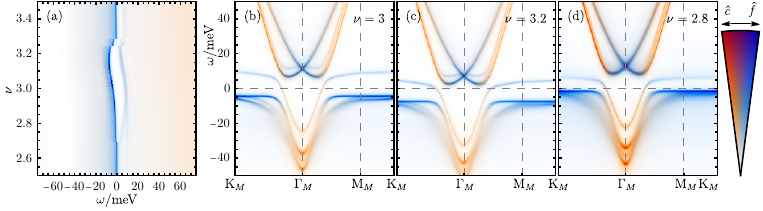}\subfloat{\label{app:fig:sym_br_bs_3_TBG_low:a}}\subfloat{\label{app:fig:sym_br_bs_3_TBG_low:b}}\subfloat{\label{app:fig:sym_br_bs_3_TBG_low:c}}\subfloat{\label{app:fig:sym_br_bs_3_TBG_low:d}}\caption{Band structure of the $\protect\IfStrEqCase{3}{{1}{\ket{\nu={}4} }
		{2}{\ket{\nu={}3, \mathrm{IVC}}}
		{3}{\ket{\nu={}3, \mathrm{VP}}}
		{4}{\ket{\nu={}2, \mathrm{K-IVC}}}
		{5}{\ket{\nu={}2, \mathrm{VP}}}
		{6}{\ket{\nu={}1, (\mathrm{K-IVC}+\mathrm{VP})}}
		{7}{\ket{\nu={}1, \mathrm{VP}}}
		{8}{\ket{\nu=0, \mathrm{K-IVC}}}
		{9}{\ket{\nu=0, \mathrm{VP}}}
	}
	[nada]
$ ground state candidate of TBG at $T=\SI{4}{\kelvin}$. The total spectral function $\mathcal{A} \left( \omega \right)$ of the system is shown in (a) as a function of filling. (b)-(d) show the $\vec{k}$-resolved spectral function $\mathcal{A} \left( \omega, \vec{k}\right)$ of the correlated state at integer filling and for small hole and electron doping around it, respectively.}\label{app:fig:sym_br_bs_3_TBG_low}\end{figure}\begin{figure}[!h]\includegraphics[width=\textwidth]{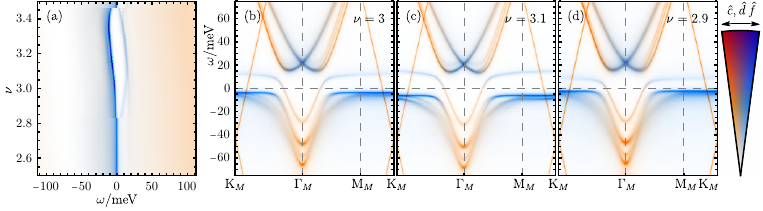}\subfloat{\label{app:fig:sym_br_bs_3_TSTG_low:a}}\subfloat{\label{app:fig:sym_br_bs_3_TSTG_low:b}}\subfloat{\label{app:fig:sym_br_bs_3_TSTG_low:c}}\subfloat{\label{app:fig:sym_br_bs_3_TSTG_low:d}}\caption{Band structure of the $\protect\IfStrEqCase{3}{{1}{\ket{\nu={}4} }
		{2}{\ket{\nu={}3, \mathrm{IVC}}}
		{3}{\ket{\nu={}3, \mathrm{VP}}}
		{4}{\ket{\nu={}2, \mathrm{K-IVC}}}
		{5}{\ket{\nu={}2, \mathrm{VP}}}
		{6}{\ket{\nu={}1, (\mathrm{K-IVC}+\mathrm{VP})}}
		{7}{\ket{\nu={}1, \mathrm{VP}}}
		{8}{\ket{\nu=0, \mathrm{K-IVC}}}
		{9}{\ket{\nu=0, \mathrm{VP}}}
	}
	[nada]
$ ground state candidate of TSTG for $\mathcal{E}=\SI{0}{\milli\electronvolt}$ at $T=\SI{5.6}{\kelvin}$. The total spectral function $\mathcal{A} \left( \omega \right)$ of the system is shown in (a) as a function of filling. (b)-(d) show the $\vec{k}$-resolved spectral function $\mathcal{A} \left( \omega, \vec{k}\right)$ of the correlated state at integer filling and for small hole and electron doping around it, respectively.}\label{app:fig:sym_br_bs_3_TSTG_low}\end{figure}\begin{figure}[!h]\includegraphics[width=\textwidth]{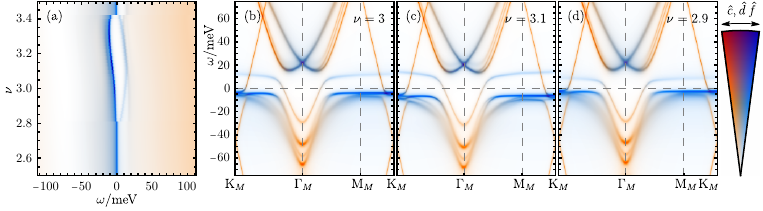}\subfloat{\label{app:fig:sym_br_bs_3_TSTGu_low:a}}\subfloat{\label{app:fig:sym_br_bs_3_TSTGu_low:b}}\subfloat{\label{app:fig:sym_br_bs_3_TSTGu_low:c}}\subfloat{\label{app:fig:sym_br_bs_3_TSTGu_low:d}}\caption{Band structure of the $\protect\IfStrEqCase{3}{{1}{\ket{\nu={}4} }
		{2}{\ket{\nu={}3, \mathrm{IVC}}}
		{3}{\ket{\nu={}3, \mathrm{VP}}}
		{4}{\ket{\nu={}2, \mathrm{K-IVC}}}
		{5}{\ket{\nu={}2, \mathrm{VP}}}
		{6}{\ket{\nu={}1, (\mathrm{K-IVC}+\mathrm{VP})}}
		{7}{\ket{\nu={}1, \mathrm{VP}}}
		{8}{\ket{\nu=0, \mathrm{K-IVC}}}
		{9}{\ket{\nu=0, \mathrm{VP}}}
	}
	[nada]
$ ground state candidate of TSTG for $\mathcal{E}=\SI{25}{\milli\electronvolt}$ at $T=\SI{5.6}{\kelvin}$. The total spectral function $\mathcal{A} \left( \omega \right)$ of the system is shown in (a) as a function of filling. (b)-(d) show the $\vec{k}$-resolved spectral function $\mathcal{A} \left( \omega, \vec{k}\right)$ of the correlated state at integer filling and for small hole and electron doping around it, respectively.}\label{app:fig:sym_br_bs_3_TSTGu_low}\end{figure}

\subsubsection{High temperature}\label{app:sec:results_corr_ins_3_high}
\begin{figure}[!h]\includegraphics[width=\textwidth]{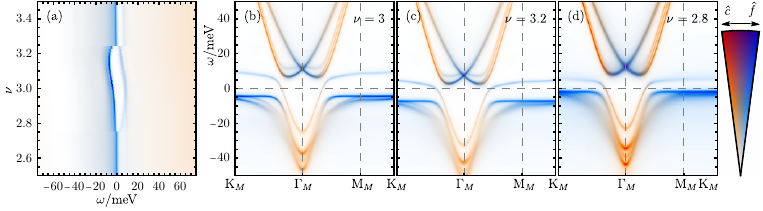}\subfloat{\label{app:fig:sym_br_bs_3_TBG_high:a}}\subfloat{\label{app:fig:sym_br_bs_3_TBG_high:b}}\subfloat{\label{app:fig:sym_br_bs_3_TBG_high:c}}\subfloat{\label{app:fig:sym_br_bs_3_TBG_high:d}}\caption{Band structure of the $\protect\IfStrEqCase{3}{{1}{\ket{\nu={}4} }
		{2}{\ket{\nu={}3, \mathrm{IVC}}}
		{3}{\ket{\nu={}3, \mathrm{VP}}}
		{4}{\ket{\nu={}2, \mathrm{K-IVC}}}
		{5}{\ket{\nu={}2, \mathrm{VP}}}
		{6}{\ket{\nu={}1, (\mathrm{K-IVC}+\mathrm{VP})}}
		{7}{\ket{\nu={}1, \mathrm{VP}}}
		{8}{\ket{\nu=0, \mathrm{K-IVC}}}
		{9}{\ket{\nu=0, \mathrm{VP}}}
	}
	[nada]
$ ground state candidate of TBG at $T=\SI{6}{\kelvin}$. The total spectral function $\mathcal{A} \left( \omega \right)$ of the system is shown in (a) as a function of filling. (b)-(d) show the $\vec{k}$-resolved spectral function $\mathcal{A} \left( \omega, \vec{k}\right)$ of the correlated state at integer filling and for small hole and electron doping around it, respectively.}\label{app:fig:sym_br_bs_3_TBG_high}\end{figure}\begin{figure}[!h]\includegraphics[width=\textwidth]{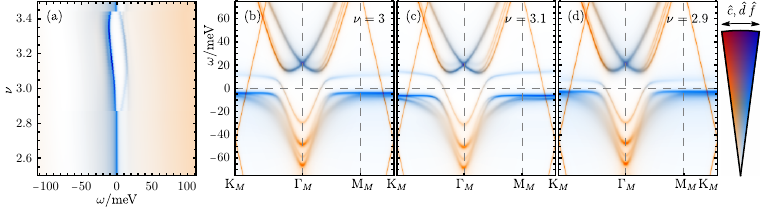}\subfloat{\label{app:fig:sym_br_bs_3_TSTG_high:a}}\subfloat{\label{app:fig:sym_br_bs_3_TSTG_high:b}}\subfloat{\label{app:fig:sym_br_bs_3_TSTG_high:c}}\subfloat{\label{app:fig:sym_br_bs_3_TSTG_high:d}}\caption{Band structure of the $\protect\IfStrEqCase{3}{{1}{\ket{\nu={}4} }
		{2}{\ket{\nu={}3, \mathrm{IVC}}}
		{3}{\ket{\nu={}3, \mathrm{VP}}}
		{4}{\ket{\nu={}2, \mathrm{K-IVC}}}
		{5}{\ket{\nu={}2, \mathrm{VP}}}
		{6}{\ket{\nu={}1, (\mathrm{K-IVC}+\mathrm{VP})}}
		{7}{\ket{\nu={}1, \mathrm{VP}}}
		{8}{\ket{\nu=0, \mathrm{K-IVC}}}
		{9}{\ket{\nu=0, \mathrm{VP}}}
	}
	[nada]
$ ground state candidate of TSTG for $\mathcal{E}=\SI{0}{\milli\electronvolt}$ at $T=\SI{8.4}{\kelvin}$. The total spectral function $\mathcal{A} \left( \omega \right)$ of the system is shown in (a) as a function of filling. (b)-(d) show the $\vec{k}$-resolved spectral function $\mathcal{A} \left( \omega, \vec{k}\right)$ of the correlated state at integer filling and for small hole and electron doping around it, respectively.}\label{app:fig:sym_br_bs_3_TSTG_high}\end{figure}\begin{figure}[!h]\includegraphics[width=\textwidth]{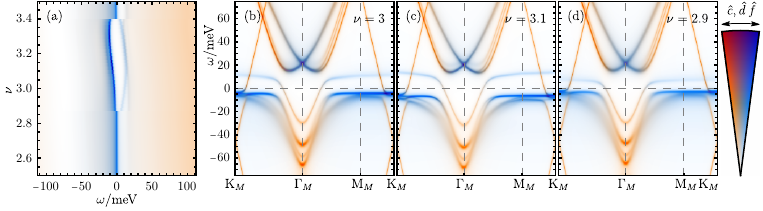}\subfloat{\label{app:fig:sym_br_bs_3_TSTGu_high:a}}\subfloat{\label{app:fig:sym_br_bs_3_TSTGu_high:b}}\subfloat{\label{app:fig:sym_br_bs_3_TSTGu_high:c}}\subfloat{\label{app:fig:sym_br_bs_3_TSTGu_high:d}}\caption{Band structure of the $\protect\IfStrEqCase{3}{{1}{\ket{\nu={}4} }
		{2}{\ket{\nu={}3, \mathrm{IVC}}}
		{3}{\ket{\nu={}3, \mathrm{VP}}}
		{4}{\ket{\nu={}2, \mathrm{K-IVC}}}
		{5}{\ket{\nu={}2, \mathrm{VP}}}
		{6}{\ket{\nu={}1, (\mathrm{K-IVC}+\mathrm{VP})}}
		{7}{\ket{\nu={}1, \mathrm{VP}}}
		{8}{\ket{\nu=0, \mathrm{K-IVC}}}
		{9}{\ket{\nu=0, \mathrm{VP}}}
	}
	[nada]
$ ground state candidate of TSTG for $\mathcal{E}=\SI{25}{\milli\electronvolt}$ at $T=\SI{8.4}{\kelvin}$. The total spectral function $\mathcal{A} \left( \omega \right)$ of the system is shown in (a) as a function of filling. (b)-(d) show the $\vec{k}$-resolved spectral function $\mathcal{A} \left( \omega, \vec{k}\right)$ of the correlated state at integer filling and for small hole and electron doping around it, respectively.}\label{app:fig:sym_br_bs_3_TSTGu_high}\end{figure}

\subsection{The \texorpdfstring{$\protect\IfStrEqCase{4}{{1}{\ket{\nu={}4} }
		{2}{\ket{\nu={}3, \mathrm{IVC}}}
		{3}{\ket{\nu={}3, \mathrm{VP}}}
		{4}{\ket{\nu={}2, \mathrm{K-IVC}}}
		{5}{\ket{\nu={}2, \mathrm{VP}}}
		{6}{\ket{\nu={}1, (\mathrm{K-IVC}+\mathrm{VP})}}
		{7}{\ket{\nu={}1, \mathrm{VP}}}
		{8}{\ket{\nu=0, \mathrm{K-IVC}}}
		{9}{\ket{\nu=0, \mathrm{VP}}}
	}
	[nada]
$}{nu=2 KIVC} correlated ground state candidate}
\label{app:sec:results_corr_ins:4}
\subsubsection{Low temperature}\label{app:sec:results_corr_ins_4_low}
\begin{figure}[!h]\includegraphics[width=\textwidth]{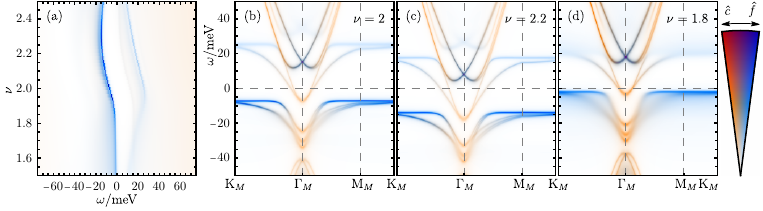}\subfloat{\label{app:fig:sym_br_bs_4_TBG_low:a}}\subfloat{\label{app:fig:sym_br_bs_4_TBG_low:b}}\subfloat{\label{app:fig:sym_br_bs_4_TBG_low:c}}\subfloat{\label{app:fig:sym_br_bs_4_TBG_low:d}}\caption{Band structure of the $\protect\IfStrEqCase{4}{{1}{\ket{\nu={}4} }
		{2}{\ket{\nu={}3, \mathrm{IVC}}}
		{3}{\ket{\nu={}3, \mathrm{VP}}}
		{4}{\ket{\nu={}2, \mathrm{K-IVC}}}
		{5}{\ket{\nu={}2, \mathrm{VP}}}
		{6}{\ket{\nu={}1, (\mathrm{K-IVC}+\mathrm{VP})}}
		{7}{\ket{\nu={}1, \mathrm{VP}}}
		{8}{\ket{\nu=0, \mathrm{K-IVC}}}
		{9}{\ket{\nu=0, \mathrm{VP}}}
	}
	[nada]
$ ground state candidate of TBG at $T=\SI{5}{\kelvin}$. The total spectral function $\mathcal{A} \left( \omega \right)$ of the system is shown in (a) as a function of filling. (b)-(d) show the $\vec{k}$-resolved spectral function $\mathcal{A} \left( \omega, \vec{k}\right)$ of the correlated state at integer filling and for small hole and electron doping around it, respectively.}\label{app:fig:sym_br_bs_4_TBG_low}\end{figure}\begin{figure}[!h]\includegraphics[width=\textwidth]{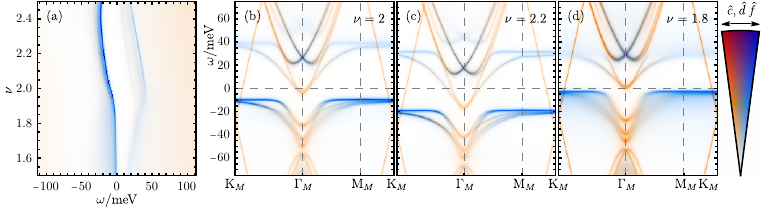}\subfloat{\label{app:fig:sym_br_bs_4_TSTG_low:a}}\subfloat{\label{app:fig:sym_br_bs_4_TSTG_low:b}}\subfloat{\label{app:fig:sym_br_bs_4_TSTG_low:c}}\subfloat{\label{app:fig:sym_br_bs_4_TSTG_low:d}}\caption{Band structure of the $\protect\IfStrEqCase{4}{{1}{\ket{\nu={}4} }
		{2}{\ket{\nu={}3, \mathrm{IVC}}}
		{3}{\ket{\nu={}3, \mathrm{VP}}}
		{4}{\ket{\nu={}2, \mathrm{K-IVC}}}
		{5}{\ket{\nu={}2, \mathrm{VP}}}
		{6}{\ket{\nu={}1, (\mathrm{K-IVC}+\mathrm{VP})}}
		{7}{\ket{\nu={}1, \mathrm{VP}}}
		{8}{\ket{\nu=0, \mathrm{K-IVC}}}
		{9}{\ket{\nu=0, \mathrm{VP}}}
	}
	[nada]
$ ground state candidate of TSTG for $\mathcal{E}=\SI{0}{\milli\electronvolt}$ at $T=\SI{7}{\kelvin}$. The total spectral function $\mathcal{A} \left( \omega \right)$ of the system is shown in (a) as a function of filling. (b)-(d) show the $\vec{k}$-resolved spectral function $\mathcal{A} \left( \omega, \vec{k}\right)$ of the correlated state at integer filling and for small hole and electron doping around it, respectively.}\label{app:fig:sym_br_bs_4_TSTG_low}\end{figure}\begin{figure}[!h]\includegraphics[width=\textwidth]{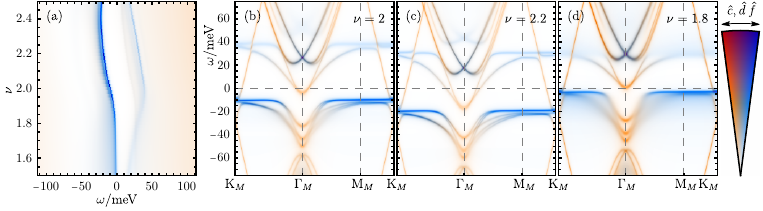}\subfloat{\label{app:fig:sym_br_bs_4_TSTGu_low:a}}\subfloat{\label{app:fig:sym_br_bs_4_TSTGu_low:b}}\subfloat{\label{app:fig:sym_br_bs_4_TSTGu_low:c}}\subfloat{\label{app:fig:sym_br_bs_4_TSTGu_low:d}}\caption{Band structure of the $\protect\IfStrEqCase{4}{{1}{\ket{\nu={}4} }
		{2}{\ket{\nu={}3, \mathrm{IVC}}}
		{3}{\ket{\nu={}3, \mathrm{VP}}}
		{4}{\ket{\nu={}2, \mathrm{K-IVC}}}
		{5}{\ket{\nu={}2, \mathrm{VP}}}
		{6}{\ket{\nu={}1, (\mathrm{K-IVC}+\mathrm{VP})}}
		{7}{\ket{\nu={}1, \mathrm{VP}}}
		{8}{\ket{\nu=0, \mathrm{K-IVC}}}
		{9}{\ket{\nu=0, \mathrm{VP}}}
	}
	[nada]
$ ground state candidate of TSTG for $\mathcal{E}=\SI{25}{\milli\electronvolt}$ at $T=\SI{7}{\kelvin}$. The total spectral function $\mathcal{A} \left( \omega \right)$ of the system is shown in (a) as a function of filling. (b)-(d) show the $\vec{k}$-resolved spectral function $\mathcal{A} \left( \omega, \vec{k}\right)$ of the correlated state at integer filling and for small hole and electron doping around it, respectively.}\label{app:fig:sym_br_bs_4_TSTGu_low}\end{figure}

\subsubsection{High temperature}\label{app:sec:results_corr_ins_4_high}
\begin{figure}[!h]\includegraphics[width=\textwidth]{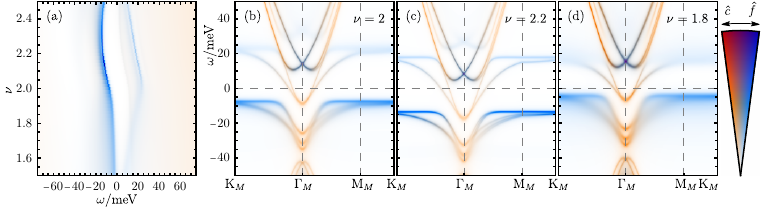}\subfloat{\label{app:fig:sym_br_bs_4_TBG_high:a}}\subfloat{\label{app:fig:sym_br_bs_4_TBG_high:b}}\subfloat{\label{app:fig:sym_br_bs_4_TBG_high:c}}\subfloat{\label{app:fig:sym_br_bs_4_TBG_high:d}}\caption{Band structure of the $\protect\IfStrEqCase{4}{{1}{\ket{\nu={}4} }
		{2}{\ket{\nu={}3, \mathrm{IVC}}}
		{3}{\ket{\nu={}3, \mathrm{VP}}}
		{4}{\ket{\nu={}2, \mathrm{K-IVC}}}
		{5}{\ket{\nu={}2, \mathrm{VP}}}
		{6}{\ket{\nu={}1, (\mathrm{K-IVC}+\mathrm{VP})}}
		{7}{\ket{\nu={}1, \mathrm{VP}}}
		{8}{\ket{\nu=0, \mathrm{K-IVC}}}
		{9}{\ket{\nu=0, \mathrm{VP}}}
	}
	[nada]
$ ground state candidate of TBG at $T=\SI{17}{\kelvin}$. The total spectral function $\mathcal{A} \left( \omega \right)$ of the system is shown in (a) as a function of filling. (b)-(d) show the $\vec{k}$-resolved spectral function $\mathcal{A} \left( \omega, \vec{k}\right)$ of the correlated state at integer filling and for small hole and electron doping around it, respectively.}\label{app:fig:sym_br_bs_4_TBG_high}\end{figure}\begin{figure}[!h]\includegraphics[width=\textwidth]{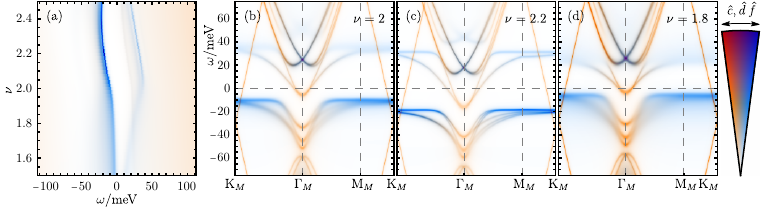}\subfloat{\label{app:fig:sym_br_bs_4_TSTG_high:a}}\subfloat{\label{app:fig:sym_br_bs_4_TSTG_high:b}}\subfloat{\label{app:fig:sym_br_bs_4_TSTG_high:c}}\subfloat{\label{app:fig:sym_br_bs_4_TSTG_high:d}}\caption{Band structure of the $\protect\IfStrEqCase{4}{{1}{\ket{\nu={}4} }
		{2}{\ket{\nu={}3, \mathrm{IVC}}}
		{3}{\ket{\nu={}3, \mathrm{VP}}}
		{4}{\ket{\nu={}2, \mathrm{K-IVC}}}
		{5}{\ket{\nu={}2, \mathrm{VP}}}
		{6}{\ket{\nu={}1, (\mathrm{K-IVC}+\mathrm{VP})}}
		{7}{\ket{\nu={}1, \mathrm{VP}}}
		{8}{\ket{\nu=0, \mathrm{K-IVC}}}
		{9}{\ket{\nu=0, \mathrm{VP}}}
	}
	[nada]
$ ground state candidate of TSTG for $\mathcal{E}=\SI{0}{\milli\electronvolt}$ at $T=\SI{23.8}{\kelvin}$. The total spectral function $\mathcal{A} \left( \omega \right)$ of the system is shown in (a) as a function of filling. (b)-(d) show the $\vec{k}$-resolved spectral function $\mathcal{A} \left( \omega, \vec{k}\right)$ of the correlated state at integer filling and for small hole and electron doping around it, respectively.}\label{app:fig:sym_br_bs_4_TSTG_high}\end{figure}\begin{figure}[!h]\includegraphics[width=\textwidth]{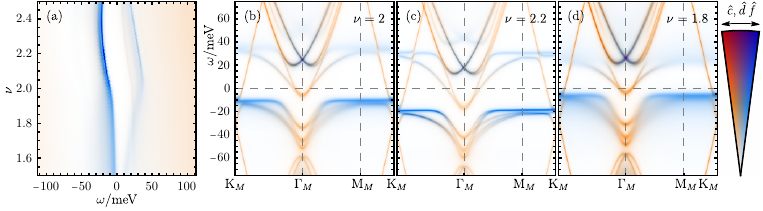}\subfloat{\label{app:fig:sym_br_bs_4_TSTGu_high:a}}\subfloat{\label{app:fig:sym_br_bs_4_TSTGu_high:b}}\subfloat{\label{app:fig:sym_br_bs_4_TSTGu_high:c}}\subfloat{\label{app:fig:sym_br_bs_4_TSTGu_high:d}}\caption{Band structure of the $\protect\IfStrEqCase{4}{{1}{\ket{\nu={}4} }
		{2}{\ket{\nu={}3, \mathrm{IVC}}}
		{3}{\ket{\nu={}3, \mathrm{VP}}}
		{4}{\ket{\nu={}2, \mathrm{K-IVC}}}
		{5}{\ket{\nu={}2, \mathrm{VP}}}
		{6}{\ket{\nu={}1, (\mathrm{K-IVC}+\mathrm{VP})}}
		{7}{\ket{\nu={}1, \mathrm{VP}}}
		{8}{\ket{\nu=0, \mathrm{K-IVC}}}
		{9}{\ket{\nu=0, \mathrm{VP}}}
	}
	[nada]
$ ground state candidate of TSTG for $\mathcal{E}=\SI{25}{\milli\electronvolt}$ at $T=\SI{23.8}{\kelvin}$. The total spectral function $\mathcal{A} \left( \omega \right)$ of the system is shown in (a) as a function of filling. (b)-(d) show the $\vec{k}$-resolved spectral function $\mathcal{A} \left( \omega, \vec{k}\right)$ of the correlated state at integer filling and for small hole and electron doping around it, respectively.}\label{app:fig:sym_br_bs_4_TSTGu_high}\end{figure}

\subsection{The \texorpdfstring{$\protect\IfStrEqCase{5}{{1}{\ket{\nu={}4} }
		{2}{\ket{\nu={}3, \mathrm{IVC}}}
		{3}{\ket{\nu={}3, \mathrm{VP}}}
		{4}{\ket{\nu={}2, \mathrm{K-IVC}}}
		{5}{\ket{\nu={}2, \mathrm{VP}}}
		{6}{\ket{\nu={}1, (\mathrm{K-IVC}+\mathrm{VP})}}
		{7}{\ket{\nu={}1, \mathrm{VP}}}
		{8}{\ket{\nu=0, \mathrm{K-IVC}}}
		{9}{\ket{\nu=0, \mathrm{VP}}}
	}
	[nada]
$}{nu=2 VP} correlated ground state candidate}
\label{app:sec:results_corr_ins:5}
\subsubsection{Low temperature}\label{app:sec:results_corr_ins_5_low}
\begin{figure}[!h]\includegraphics[width=\textwidth]{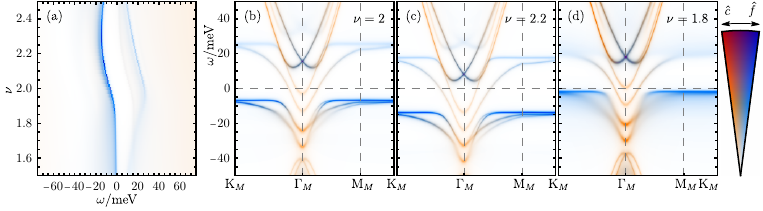}\subfloat{\label{app:fig:sym_br_bs_5_TBG_low:a}}\subfloat{\label{app:fig:sym_br_bs_5_TBG_low:b}}\subfloat{\label{app:fig:sym_br_bs_5_TBG_low:c}}\subfloat{\label{app:fig:sym_br_bs_5_TBG_low:d}}\caption{Band structure of the $\protect\IfStrEqCase{5}{{1}{\ket{\nu={}4} }
		{2}{\ket{\nu={}3, \mathrm{IVC}}}
		{3}{\ket{\nu={}3, \mathrm{VP}}}
		{4}{\ket{\nu={}2, \mathrm{K-IVC}}}
		{5}{\ket{\nu={}2, \mathrm{VP}}}
		{6}{\ket{\nu={}1, (\mathrm{K-IVC}+\mathrm{VP})}}
		{7}{\ket{\nu={}1, \mathrm{VP}}}
		{8}{\ket{\nu=0, \mathrm{K-IVC}}}
		{9}{\ket{\nu=0, \mathrm{VP}}}
	}
	[nada]
$ ground state candidate of TBG at $T=\SI{5}{\kelvin}$. The total spectral function $\mathcal{A} \left( \omega \right)$ of the system is shown in (a) as a function of filling. (b)-(d) show the $\vec{k}$-resolved spectral function $\mathcal{A} \left( \omega, \vec{k}\right)$ of the correlated state at integer filling and for small hole and electron doping around it, respectively.}\label{app:fig:sym_br_bs_5_TBG_low}\end{figure}\begin{figure}[!h]\includegraphics[width=\textwidth]{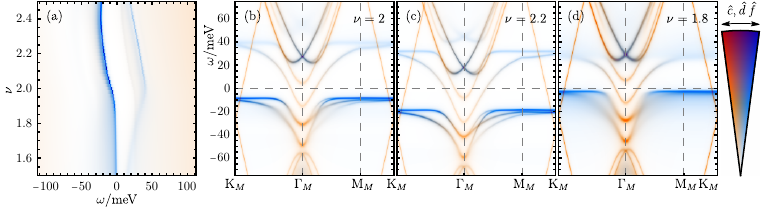}\subfloat{\label{app:fig:sym_br_bs_5_TSTG_low:a}}\subfloat{\label{app:fig:sym_br_bs_5_TSTG_low:b}}\subfloat{\label{app:fig:sym_br_bs_5_TSTG_low:c}}\subfloat{\label{app:fig:sym_br_bs_5_TSTG_low:d}}\caption{Band structure of the $\protect\IfStrEqCase{5}{{1}{\ket{\nu={}4} }
		{2}{\ket{\nu={}3, \mathrm{IVC}}}
		{3}{\ket{\nu={}3, \mathrm{VP}}}
		{4}{\ket{\nu={}2, \mathrm{K-IVC}}}
		{5}{\ket{\nu={}2, \mathrm{VP}}}
		{6}{\ket{\nu={}1, (\mathrm{K-IVC}+\mathrm{VP})}}
		{7}{\ket{\nu={}1, \mathrm{VP}}}
		{8}{\ket{\nu=0, \mathrm{K-IVC}}}
		{9}{\ket{\nu=0, \mathrm{VP}}}
	}
	[nada]
$ ground state candidate of TSTG for $\mathcal{E}=\SI{0}{\milli\electronvolt}$ at $T=\SI{7}{\kelvin}$. The total spectral function $\mathcal{A} \left( \omega \right)$ of the system is shown in (a) as a function of filling. (b)-(d) show the $\vec{k}$-resolved spectral function $\mathcal{A} \left( \omega, \vec{k}\right)$ of the correlated state at integer filling and for small hole and electron doping around it, respectively.}\label{app:fig:sym_br_bs_5_TSTG_low}\end{figure}\begin{figure}[!h]\includegraphics[width=\textwidth]{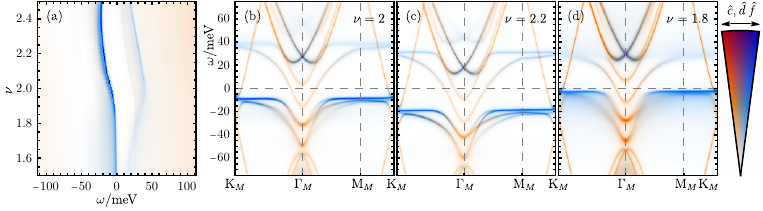}\subfloat{\label{app:fig:sym_br_bs_5_TSTGu_low:a}}\subfloat{\label{app:fig:sym_br_bs_5_TSTGu_low:b}}\subfloat{\label{app:fig:sym_br_bs_5_TSTGu_low:c}}\subfloat{\label{app:fig:sym_br_bs_5_TSTGu_low:d}}\caption{Band structure of the $\protect\IfStrEqCase{5}{{1}{\ket{\nu={}4} }
		{2}{\ket{\nu={}3, \mathrm{IVC}}}
		{3}{\ket{\nu={}3, \mathrm{VP}}}
		{4}{\ket{\nu={}2, \mathrm{K-IVC}}}
		{5}{\ket{\nu={}2, \mathrm{VP}}}
		{6}{\ket{\nu={}1, (\mathrm{K-IVC}+\mathrm{VP})}}
		{7}{\ket{\nu={}1, \mathrm{VP}}}
		{8}{\ket{\nu=0, \mathrm{K-IVC}}}
		{9}{\ket{\nu=0, \mathrm{VP}}}
	}
	[nada]
$ ground state candidate of TSTG for $\mathcal{E}=\SI{25}{\milli\electronvolt}$ at $T=\SI{7}{\kelvin}$. The total spectral function $\mathcal{A} \left( \omega \right)$ of the system is shown in (a) as a function of filling. (b)-(d) show the $\vec{k}$-resolved spectral function $\mathcal{A} \left( \omega, \vec{k}\right)$ of the correlated state at integer filling and for small hole and electron doping around it, respectively.}\label{app:fig:sym_br_bs_5_TSTGu_low}\end{figure}

\subsubsection{High temperature}\label{app:sec:results_corr_ins_5_high}
\begin{figure}[!h]\includegraphics[width=\textwidth]{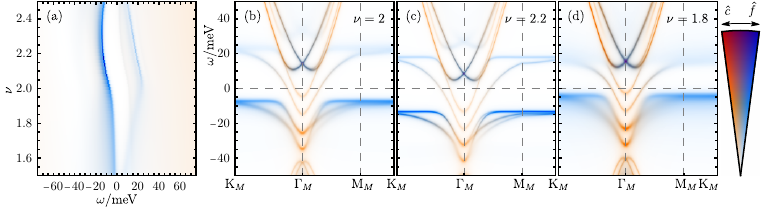}\subfloat{\label{app:fig:sym_br_bs_5_TBG_high:a}}\subfloat{\label{app:fig:sym_br_bs_5_TBG_high:b}}\subfloat{\label{app:fig:sym_br_bs_5_TBG_high:c}}\subfloat{\label{app:fig:sym_br_bs_5_TBG_high:d}}\caption{Band structure of the $\protect\IfStrEqCase{5}{{1}{\ket{\nu={}4} }
		{2}{\ket{\nu={}3, \mathrm{IVC}}}
		{3}{\ket{\nu={}3, \mathrm{VP}}}
		{4}{\ket{\nu={}2, \mathrm{K-IVC}}}
		{5}{\ket{\nu={}2, \mathrm{VP}}}
		{6}{\ket{\nu={}1, (\mathrm{K-IVC}+\mathrm{VP})}}
		{7}{\ket{\nu={}1, \mathrm{VP}}}
		{8}{\ket{\nu=0, \mathrm{K-IVC}}}
		{9}{\ket{\nu=0, \mathrm{VP}}}
	}
	[nada]
$ ground state candidate of TBG at $T=\SI{17}{\kelvin}$. The total spectral function $\mathcal{A} \left( \omega \right)$ of the system is shown in (a) as a function of filling. (b)-(d) show the $\vec{k}$-resolved spectral function $\mathcal{A} \left( \omega, \vec{k}\right)$ of the correlated state at integer filling and for small hole and electron doping around it, respectively.}\label{app:fig:sym_br_bs_5_TBG_high}\end{figure}\begin{figure}[!h]\includegraphics[width=\textwidth]{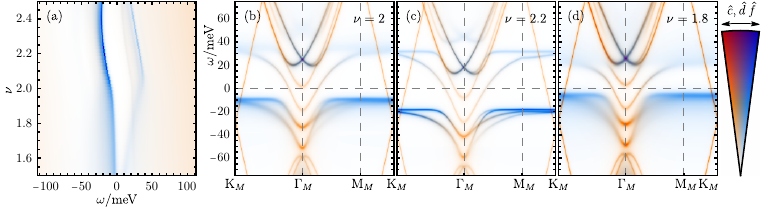}\subfloat{\label{app:fig:sym_br_bs_5_TSTG_high:a}}\subfloat{\label{app:fig:sym_br_bs_5_TSTG_high:b}}\subfloat{\label{app:fig:sym_br_bs_5_TSTG_high:c}}\subfloat{\label{app:fig:sym_br_bs_5_TSTG_high:d}}\caption{Band structure of the $\protect\IfStrEqCase{5}{{1}{\ket{\nu={}4} }
		{2}{\ket{\nu={}3, \mathrm{IVC}}}
		{3}{\ket{\nu={}3, \mathrm{VP}}}
		{4}{\ket{\nu={}2, \mathrm{K-IVC}}}
		{5}{\ket{\nu={}2, \mathrm{VP}}}
		{6}{\ket{\nu={}1, (\mathrm{K-IVC}+\mathrm{VP})}}
		{7}{\ket{\nu={}1, \mathrm{VP}}}
		{8}{\ket{\nu=0, \mathrm{K-IVC}}}
		{9}{\ket{\nu=0, \mathrm{VP}}}
	}
	[nada]
$ ground state candidate of TSTG for $\mathcal{E}=\SI{0}{\milli\electronvolt}$ at $T=\SI{23.8}{\kelvin}$. The total spectral function $\mathcal{A} \left( \omega \right)$ of the system is shown in (a) as a function of filling. (b)-(d) show the $\vec{k}$-resolved spectral function $\mathcal{A} \left( \omega, \vec{k}\right)$ of the correlated state at integer filling and for small hole and electron doping around it, respectively.}\label{app:fig:sym_br_bs_5_TSTG_high}\end{figure}\begin{figure}[!h]\includegraphics[width=\textwidth]{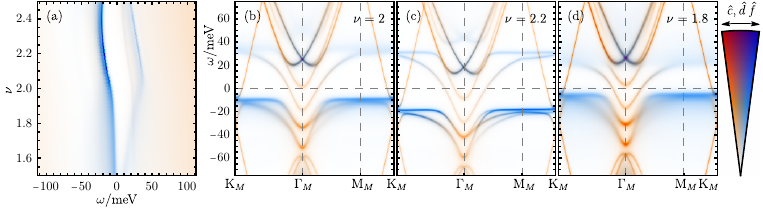}\subfloat{\label{app:fig:sym_br_bs_5_TSTGu_high:a}}\subfloat{\label{app:fig:sym_br_bs_5_TSTGu_high:b}}\subfloat{\label{app:fig:sym_br_bs_5_TSTGu_high:c}}\subfloat{\label{app:fig:sym_br_bs_5_TSTGu_high:d}}\caption{Band structure of the $\protect\IfStrEqCase{5}{{1}{\ket{\nu={}4} }
		{2}{\ket{\nu={}3, \mathrm{IVC}}}
		{3}{\ket{\nu={}3, \mathrm{VP}}}
		{4}{\ket{\nu={}2, \mathrm{K-IVC}}}
		{5}{\ket{\nu={}2, \mathrm{VP}}}
		{6}{\ket{\nu={}1, (\mathrm{K-IVC}+\mathrm{VP})}}
		{7}{\ket{\nu={}1, \mathrm{VP}}}
		{8}{\ket{\nu=0, \mathrm{K-IVC}}}
		{9}{\ket{\nu=0, \mathrm{VP}}}
	}
	[nada]
$ ground state candidate of TSTG for $\mathcal{E}=\SI{25}{\milli\electronvolt}$ at $T=\SI{23.8}{\kelvin}$. The total spectral function $\mathcal{A} \left( \omega \right)$ of the system is shown in (a) as a function of filling. (b)-(d) show the $\vec{k}$-resolved spectral function $\mathcal{A} \left( \omega, \vec{k}\right)$ of the correlated state at integer filling and for small hole and electron doping around it, respectively.}\label{app:fig:sym_br_bs_5_TSTGu_high}\end{figure}

\subsection{The \texorpdfstring{$\protect\IfStrEqCase{6}{{1}{\ket{\nu={}4} }
		{2}{\ket{\nu={}3, \mathrm{IVC}}}
		{3}{\ket{\nu={}3, \mathrm{VP}}}
		{4}{\ket{\nu={}2, \mathrm{K-IVC}}}
		{5}{\ket{\nu={}2, \mathrm{VP}}}
		{6}{\ket{\nu={}1, (\mathrm{K-IVC}+\mathrm{VP})}}
		{7}{\ket{\nu={}1, \mathrm{VP}}}
		{8}{\ket{\nu=0, \mathrm{K-IVC}}}
		{9}{\ket{\nu=0, \mathrm{VP}}}
	}
	[nada]
$}{nu=1 KIVC + VP} correlated ground state candidate}
\label{app:sec:results_corr_ins:6}
\subsubsection{Low temperature}\label{app:sec:results_corr_ins_6_low}
\begin{figure}[!h]\includegraphics[width=\textwidth]{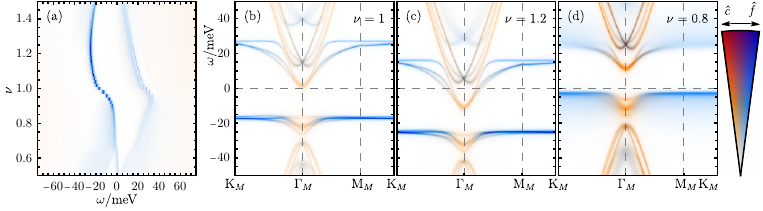}\subfloat{\label{app:fig:sym_br_bs_6_TBG_low:a}}\subfloat{\label{app:fig:sym_br_bs_6_TBG_low:b}}\subfloat{\label{app:fig:sym_br_bs_6_TBG_low:c}}\subfloat{\label{app:fig:sym_br_bs_6_TBG_low:d}}\caption{Band structure of the $\protect\IfStrEqCase{6}{{1}{\ket{\nu={}4} }
		{2}{\ket{\nu={}3, \mathrm{IVC}}}
		{3}{\ket{\nu={}3, \mathrm{VP}}}
		{4}{\ket{\nu={}2, \mathrm{K-IVC}}}
		{5}{\ket{\nu={}2, \mathrm{VP}}}
		{6}{\ket{\nu={}1, (\mathrm{K-IVC}+\mathrm{VP})}}
		{7}{\ket{\nu={}1, \mathrm{VP}}}
		{8}{\ket{\nu=0, \mathrm{K-IVC}}}
		{9}{\ket{\nu=0, \mathrm{VP}}}
	}
	[nada]
$ ground state candidate of TBG at $T=\SI{7}{\kelvin}$. The total spectral function $\mathcal{A} \left( \omega \right)$ of the system is shown in (a) as a function of filling. (b)-(d) show the $\vec{k}$-resolved spectral function $\mathcal{A} \left( \omega, \vec{k}\right)$ of the correlated state at integer filling and for small hole and electron doping around it, respectively.}\label{app:fig:sym_br_bs_6_TBG_low}\end{figure}\begin{figure}[!h]\includegraphics[width=\textwidth]{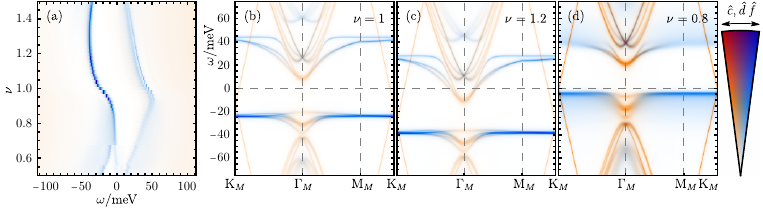}\subfloat{\label{app:fig:sym_br_bs_6_TSTG_low:a}}\subfloat{\label{app:fig:sym_br_bs_6_TSTG_low:b}}\subfloat{\label{app:fig:sym_br_bs_6_TSTG_low:c}}\subfloat{\label{app:fig:sym_br_bs_6_TSTG_low:d}}\caption{Band structure of the $\protect\IfStrEqCase{6}{{1}{\ket{\nu={}4} }
		{2}{\ket{\nu={}3, \mathrm{IVC}}}
		{3}{\ket{\nu={}3, \mathrm{VP}}}
		{4}{\ket{\nu={}2, \mathrm{K-IVC}}}
		{5}{\ket{\nu={}2, \mathrm{VP}}}
		{6}{\ket{\nu={}1, (\mathrm{K-IVC}+\mathrm{VP})}}
		{7}{\ket{\nu={}1, \mathrm{VP}}}
		{8}{\ket{\nu=0, \mathrm{K-IVC}}}
		{9}{\ket{\nu=0, \mathrm{VP}}}
	}
	[nada]
$ ground state candidate of TSTG for $\mathcal{E}=\SI{0}{\milli\electronvolt}$ at $T=\SI{9.8}{\kelvin}$. The total spectral function $\mathcal{A} \left( \omega \right)$ of the system is shown in (a) as a function of filling. (b)-(d) show the $\vec{k}$-resolved spectral function $\mathcal{A} \left( \omega, \vec{k}\right)$ of the correlated state at integer filling and for small hole and electron doping around it, respectively.}\label{app:fig:sym_br_bs_6_TSTG_low}\end{figure}\begin{figure}[!h]\includegraphics[width=\textwidth]{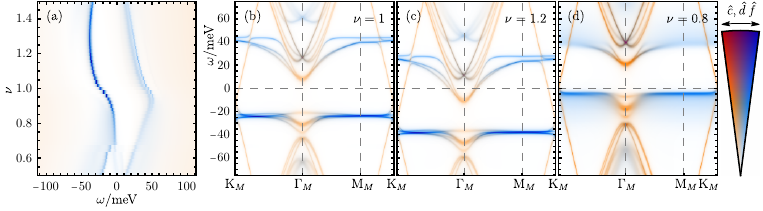}\subfloat{\label{app:fig:sym_br_bs_6_TSTGu_low:a}}\subfloat{\label{app:fig:sym_br_bs_6_TSTGu_low:b}}\subfloat{\label{app:fig:sym_br_bs_6_TSTGu_low:c}}\subfloat{\label{app:fig:sym_br_bs_6_TSTGu_low:d}}\caption{Band structure of the $\protect\IfStrEqCase{6}{{1}{\ket{\nu={}4} }
		{2}{\ket{\nu={}3, \mathrm{IVC}}}
		{3}{\ket{\nu={}3, \mathrm{VP}}}
		{4}{\ket{\nu={}2, \mathrm{K-IVC}}}
		{5}{\ket{\nu={}2, \mathrm{VP}}}
		{6}{\ket{\nu={}1, (\mathrm{K-IVC}+\mathrm{VP})}}
		{7}{\ket{\nu={}1, \mathrm{VP}}}
		{8}{\ket{\nu=0, \mathrm{K-IVC}}}
		{9}{\ket{\nu=0, \mathrm{VP}}}
	}
	[nada]
$ ground state candidate of TSTG for $\mathcal{E}=\SI{25}{\milli\electronvolt}$ at $T=\SI{9.8}{\kelvin}$. The total spectral function $\mathcal{A} \left( \omega \right)$ of the system is shown in (a) as a function of filling. (b)-(d) show the $\vec{k}$-resolved spectral function $\mathcal{A} \left( \omega, \vec{k}\right)$ of the correlated state at integer filling and for small hole and electron doping around it, respectively.}\label{app:fig:sym_br_bs_6_TSTGu_low}\end{figure}

\subsubsection{High temperature}\label{app:sec:results_corr_ins_6_high}
\begin{figure}[!h]\includegraphics[width=\textwidth]{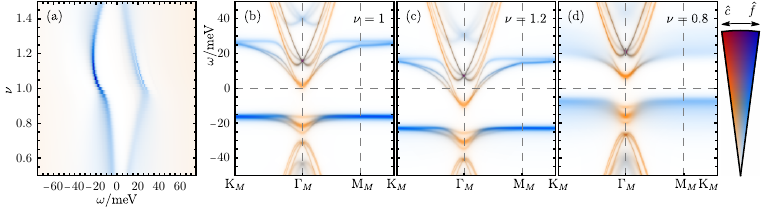}\subfloat{\label{app:fig:sym_br_bs_6_TBG_high:a}}\subfloat{\label{app:fig:sym_br_bs_6_TBG_high:b}}\subfloat{\label{app:fig:sym_br_bs_6_TBG_high:c}}\subfloat{\label{app:fig:sym_br_bs_6_TBG_high:d}}\caption{Band structure of the $\protect\IfStrEqCase{6}{{1}{\ket{\nu={}4} }
		{2}{\ket{\nu={}3, \mathrm{IVC}}}
		{3}{\ket{\nu={}3, \mathrm{VP}}}
		{4}{\ket{\nu={}2, \mathrm{K-IVC}}}
		{5}{\ket{\nu={}2, \mathrm{VP}}}
		{6}{\ket{\nu={}1, (\mathrm{K-IVC}+\mathrm{VP})}}
		{7}{\ket{\nu={}1, \mathrm{VP}}}
		{8}{\ket{\nu=0, \mathrm{K-IVC}}}
		{9}{\ket{\nu=0, \mathrm{VP}}}
	}
	[nada]
$ ground state candidate of TBG at $T=\SI{27}{\kelvin}$. The total spectral function $\mathcal{A} \left( \omega \right)$ of the system is shown in (a) as a function of filling. (b)-(d) show the $\vec{k}$-resolved spectral function $\mathcal{A} \left( \omega, \vec{k}\right)$ of the correlated state at integer filling and for small hole and electron doping around it, respectively.}\label{app:fig:sym_br_bs_6_TBG_high}\end{figure}\begin{figure}[!h]\includegraphics[width=\textwidth]{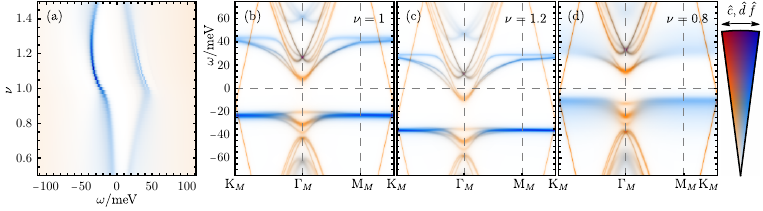}\subfloat{\label{app:fig:sym_br_bs_6_TSTG_high:a}}\subfloat{\label{app:fig:sym_br_bs_6_TSTG_high:b}}\subfloat{\label{app:fig:sym_br_bs_6_TSTG_high:c}}\subfloat{\label{app:fig:sym_br_bs_6_TSTG_high:d}}\caption{Band structure of the $\protect\IfStrEqCase{6}{{1}{\ket{\nu={}4} }
		{2}{\ket{\nu={}3, \mathrm{IVC}}}
		{3}{\ket{\nu={}3, \mathrm{VP}}}
		{4}{\ket{\nu={}2, \mathrm{K-IVC}}}
		{5}{\ket{\nu={}2, \mathrm{VP}}}
		{6}{\ket{\nu={}1, (\mathrm{K-IVC}+\mathrm{VP})}}
		{7}{\ket{\nu={}1, \mathrm{VP}}}
		{8}{\ket{\nu=0, \mathrm{K-IVC}}}
		{9}{\ket{\nu=0, \mathrm{VP}}}
	}
	[nada]
$ ground state candidate of TSTG for $\mathcal{E}=\SI{0}{\milli\electronvolt}$ at $T=\SI{37.8}{\kelvin}$. The total spectral function $\mathcal{A} \left( \omega \right)$ of the system is shown in (a) as a function of filling. (b)-(d) show the $\vec{k}$-resolved spectral function $\mathcal{A} \left( \omega, \vec{k}\right)$ of the correlated state at integer filling and for small hole and electron doping around it, respectively.}\label{app:fig:sym_br_bs_6_TSTG_high}\end{figure}\begin{figure}[!h]\includegraphics[width=\textwidth]{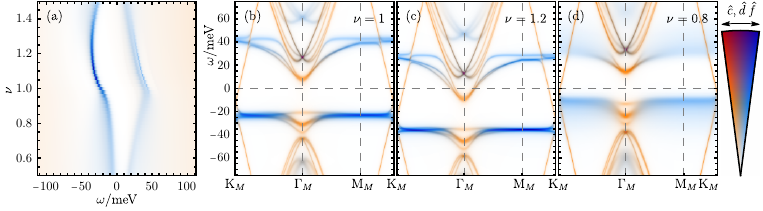}\subfloat{\label{app:fig:sym_br_bs_6_TSTGu_high:a}}\subfloat{\label{app:fig:sym_br_bs_6_TSTGu_high:b}}\subfloat{\label{app:fig:sym_br_bs_6_TSTGu_high:c}}\subfloat{\label{app:fig:sym_br_bs_6_TSTGu_high:d}}\caption{Band structure of the $\protect\IfStrEqCase{6}{{1}{\ket{\nu={}4} }
		{2}{\ket{\nu={}3, \mathrm{IVC}}}
		{3}{\ket{\nu={}3, \mathrm{VP}}}
		{4}{\ket{\nu={}2, \mathrm{K-IVC}}}
		{5}{\ket{\nu={}2, \mathrm{VP}}}
		{6}{\ket{\nu={}1, (\mathrm{K-IVC}+\mathrm{VP})}}
		{7}{\ket{\nu={}1, \mathrm{VP}}}
		{8}{\ket{\nu=0, \mathrm{K-IVC}}}
		{9}{\ket{\nu=0, \mathrm{VP}}}
	}
	[nada]
$ ground state candidate of TSTG for $\mathcal{E}=\SI{25}{\milli\electronvolt}$ at $T=\SI{37.8}{\kelvin}$. The total spectral function $\mathcal{A} \left( \omega \right)$ of the system is shown in (a) as a function of filling. (b)-(d) show the $\vec{k}$-resolved spectral function $\mathcal{A} \left( \omega, \vec{k}\right)$ of the correlated state at integer filling and for small hole and electron doping around it, respectively.}\label{app:fig:sym_br_bs_6_TSTGu_high}\end{figure}

\subsection{The \texorpdfstring{$\protect\IfStrEqCase{7}{{1}{\ket{\nu={}4} }
		{2}{\ket{\nu={}3, \mathrm{IVC}}}
		{3}{\ket{\nu={}3, \mathrm{VP}}}
		{4}{\ket{\nu={}2, \mathrm{K-IVC}}}
		{5}{\ket{\nu={}2, \mathrm{VP}}}
		{6}{\ket{\nu={}1, (\mathrm{K-IVC}+\mathrm{VP})}}
		{7}{\ket{\nu={}1, \mathrm{VP}}}
		{8}{\ket{\nu=0, \mathrm{K-IVC}}}
		{9}{\ket{\nu=0, \mathrm{VP}}}
	}
	[nada]
$}{nu=1 VP} correlated ground state candidate}
\label{app:sec:results_corr_ins:7}
\subsubsection{Low temperature}\label{app:sec:results_corr_ins_7_low}
\begin{figure}[!h]\includegraphics[width=\textwidth]{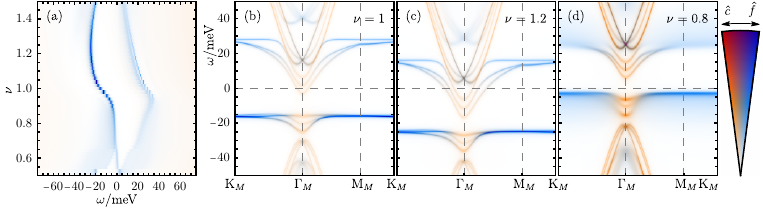}\subfloat{\label{app:fig:sym_br_bs_7_TBG_low:a}}\subfloat{\label{app:fig:sym_br_bs_7_TBG_low:b}}\subfloat{\label{app:fig:sym_br_bs_7_TBG_low:c}}\subfloat{\label{app:fig:sym_br_bs_7_TBG_low:d}}\caption{Band structure of the $\protect\IfStrEqCase{7}{{1}{\ket{\nu={}4} }
		{2}{\ket{\nu={}3, \mathrm{IVC}}}
		{3}{\ket{\nu={}3, \mathrm{VP}}}
		{4}{\ket{\nu={}2, \mathrm{K-IVC}}}
		{5}{\ket{\nu={}2, \mathrm{VP}}}
		{6}{\ket{\nu={}1, (\mathrm{K-IVC}+\mathrm{VP})}}
		{7}{\ket{\nu={}1, \mathrm{VP}}}
		{8}{\ket{\nu=0, \mathrm{K-IVC}}}
		{9}{\ket{\nu=0, \mathrm{VP}}}
	}
	[nada]
$ ground state candidate of TBG at $T=\SI{7}{\kelvin}$. The total spectral function $\mathcal{A} \left( \omega \right)$ of the system is shown in (a) as a function of filling. (b)-(d) show the $\vec{k}$-resolved spectral function $\mathcal{A} \left( \omega, \vec{k}\right)$ of the correlated state at integer filling and for small hole and electron doping around it, respectively.}\label{app:fig:sym_br_bs_7_TBG_low}\end{figure}\begin{figure}[!h]\includegraphics[width=\textwidth]{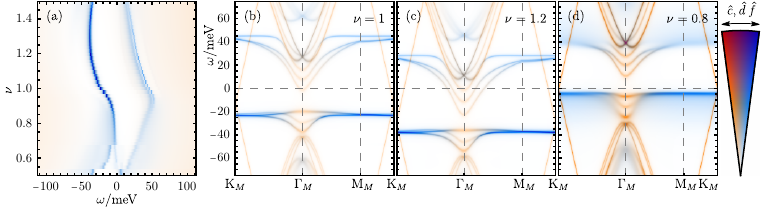}\subfloat{\label{app:fig:sym_br_bs_7_TSTG_low:a}}\subfloat{\label{app:fig:sym_br_bs_7_TSTG_low:b}}\subfloat{\label{app:fig:sym_br_bs_7_TSTG_low:c}}\subfloat{\label{app:fig:sym_br_bs_7_TSTG_low:d}}\caption{Band structure of the $\protect\IfStrEqCase{7}{{1}{\ket{\nu={}4} }
		{2}{\ket{\nu={}3, \mathrm{IVC}}}
		{3}{\ket{\nu={}3, \mathrm{VP}}}
		{4}{\ket{\nu={}2, \mathrm{K-IVC}}}
		{5}{\ket{\nu={}2, \mathrm{VP}}}
		{6}{\ket{\nu={}1, (\mathrm{K-IVC}+\mathrm{VP})}}
		{7}{\ket{\nu={}1, \mathrm{VP}}}
		{8}{\ket{\nu=0, \mathrm{K-IVC}}}
		{9}{\ket{\nu=0, \mathrm{VP}}}
	}
	[nada]
$ ground state candidate of TSTG for $\mathcal{E}=\SI{0}{\milli\electronvolt}$ at $T=\SI{9.8}{\kelvin}$. The total spectral function $\mathcal{A} \left( \omega \right)$ of the system is shown in (a) as a function of filling. (b)-(d) show the $\vec{k}$-resolved spectral function $\mathcal{A} \left( \omega, \vec{k}\right)$ of the correlated state at integer filling and for small hole and electron doping around it, respectively.}\label{app:fig:sym_br_bs_7_TSTG_low}\end{figure}\begin{figure}[!h]\includegraphics[width=\textwidth]{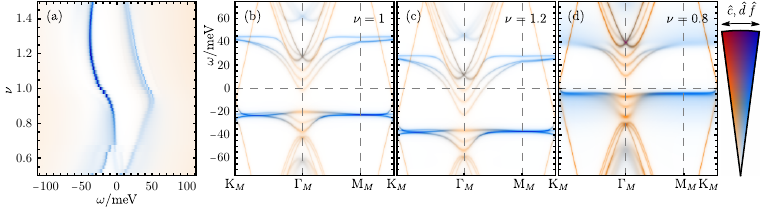}\subfloat{\label{app:fig:sym_br_bs_7_TSTGu_low:a}}\subfloat{\label{app:fig:sym_br_bs_7_TSTGu_low:b}}\subfloat{\label{app:fig:sym_br_bs_7_TSTGu_low:c}}\subfloat{\label{app:fig:sym_br_bs_7_TSTGu_low:d}}\caption{Band structure of the $\protect\IfStrEqCase{7}{{1}{\ket{\nu={}4} }
		{2}{\ket{\nu={}3, \mathrm{IVC}}}
		{3}{\ket{\nu={}3, \mathrm{VP}}}
		{4}{\ket{\nu={}2, \mathrm{K-IVC}}}
		{5}{\ket{\nu={}2, \mathrm{VP}}}
		{6}{\ket{\nu={}1, (\mathrm{K-IVC}+\mathrm{VP})}}
		{7}{\ket{\nu={}1, \mathrm{VP}}}
		{8}{\ket{\nu=0, \mathrm{K-IVC}}}
		{9}{\ket{\nu=0, \mathrm{VP}}}
	}
	[nada]
$ ground state candidate of TSTG for $\mathcal{E}=\SI{25}{\milli\electronvolt}$ at $T=\SI{9.8}{\kelvin}$. The total spectral function $\mathcal{A} \left( \omega \right)$ of the system is shown in (a) as a function of filling. (b)-(d) show the $\vec{k}$-resolved spectral function $\mathcal{A} \left( \omega, \vec{k}\right)$ of the correlated state at integer filling and for small hole and electron doping around it, respectively.}\label{app:fig:sym_br_bs_7_TSTGu_low}\end{figure}

\subsubsection{High temperature}\label{app:sec:results_corr_ins_7_high}
\begin{figure}[!h]\includegraphics[width=\textwidth]{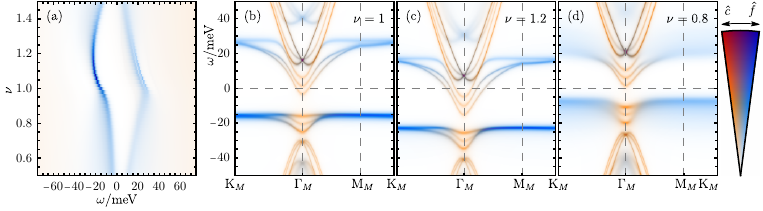}\subfloat{\label{app:fig:sym_br_bs_7_TBG_high:a}}\subfloat{\label{app:fig:sym_br_bs_7_TBG_high:b}}\subfloat{\label{app:fig:sym_br_bs_7_TBG_high:c}}\subfloat{\label{app:fig:sym_br_bs_7_TBG_high:d}}\caption{Band structure of the $\protect\IfStrEqCase{7}{{1}{\ket{\nu={}4} }
		{2}{\ket{\nu={}3, \mathrm{IVC}}}
		{3}{\ket{\nu={}3, \mathrm{VP}}}
		{4}{\ket{\nu={}2, \mathrm{K-IVC}}}
		{5}{\ket{\nu={}2, \mathrm{VP}}}
		{6}{\ket{\nu={}1, (\mathrm{K-IVC}+\mathrm{VP})}}
		{7}{\ket{\nu={}1, \mathrm{VP}}}
		{8}{\ket{\nu=0, \mathrm{K-IVC}}}
		{9}{\ket{\nu=0, \mathrm{VP}}}
	}
	[nada]
$ ground state candidate of TBG at $T=\SI{27}{\kelvin}$. The total spectral function $\mathcal{A} \left( \omega \right)$ of the system is shown in (a) as a function of filling. (b)-(d) show the $\vec{k}$-resolved spectral function $\mathcal{A} \left( \omega, \vec{k}\right)$ of the correlated state at integer filling and for small hole and electron doping around it, respectively.}\label{app:fig:sym_br_bs_7_TBG_high}\end{figure}\begin{figure}[!h]\includegraphics[width=\textwidth]{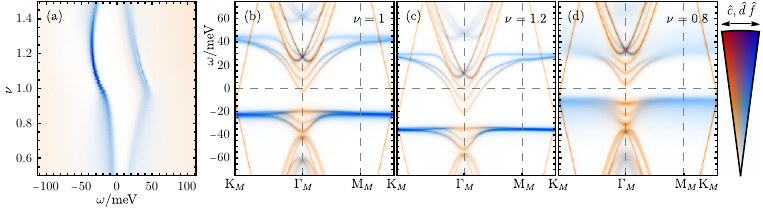}\subfloat{\label{app:fig:sym_br_bs_7_TSTG_high:a}}\subfloat{\label{app:fig:sym_br_bs_7_TSTG_high:b}}\subfloat{\label{app:fig:sym_br_bs_7_TSTG_high:c}}\subfloat{\label{app:fig:sym_br_bs_7_TSTG_high:d}}\caption{Band structure of the $\protect\IfStrEqCase{7}{{1}{\ket{\nu={}4} }
		{2}{\ket{\nu={}3, \mathrm{IVC}}}
		{3}{\ket{\nu={}3, \mathrm{VP}}}
		{4}{\ket{\nu={}2, \mathrm{K-IVC}}}
		{5}{\ket{\nu={}2, \mathrm{VP}}}
		{6}{\ket{\nu={}1, (\mathrm{K-IVC}+\mathrm{VP})}}
		{7}{\ket{\nu={}1, \mathrm{VP}}}
		{8}{\ket{\nu=0, \mathrm{K-IVC}}}
		{9}{\ket{\nu=0, \mathrm{VP}}}
	}
	[nada]
$ ground state candidate of TSTG for $\mathcal{E}=\SI{0}{\milli\electronvolt}$ at $T=\SI{37.8}{\kelvin}$. The total spectral function $\mathcal{A} \left( \omega \right)$ of the system is shown in (a) as a function of filling. (b)-(d) show the $\vec{k}$-resolved spectral function $\mathcal{A} \left( \omega, \vec{k}\right)$ of the correlated state at integer filling and for small hole and electron doping around it, respectively.}\label{app:fig:sym_br_bs_7_TSTG_high}\end{figure}\begin{figure}[!h]\includegraphics[width=\textwidth]{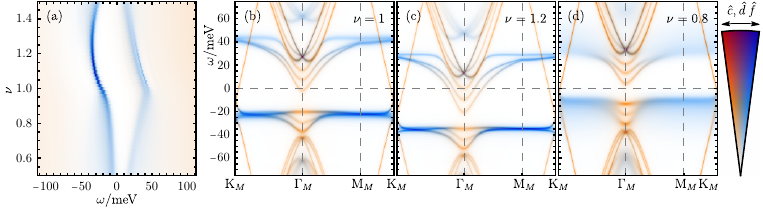}\subfloat{\label{app:fig:sym_br_bs_7_TSTGu_high:a}}\subfloat{\label{app:fig:sym_br_bs_7_TSTGu_high:b}}\subfloat{\label{app:fig:sym_br_bs_7_TSTGu_high:c}}\subfloat{\label{app:fig:sym_br_bs_7_TSTGu_high:d}}\caption{Band structure of the $\protect\IfStrEqCase{7}{{1}{\ket{\nu={}4} }
		{2}{\ket{\nu={}3, \mathrm{IVC}}}
		{3}{\ket{\nu={}3, \mathrm{VP}}}
		{4}{\ket{\nu={}2, \mathrm{K-IVC}}}
		{5}{\ket{\nu={}2, \mathrm{VP}}}
		{6}{\ket{\nu={}1, (\mathrm{K-IVC}+\mathrm{VP})}}
		{7}{\ket{\nu={}1, \mathrm{VP}}}
		{8}{\ket{\nu=0, \mathrm{K-IVC}}}
		{9}{\ket{\nu=0, \mathrm{VP}}}
	}
	[nada]
$ ground state candidate of TSTG for $\mathcal{E}=\SI{25}{\milli\electronvolt}$ at $T=\SI{37.8}{\kelvin}$. The total spectral function $\mathcal{A} \left( \omega \right)$ of the system is shown in (a) as a function of filling. (b)-(d) show the $\vec{k}$-resolved spectral function $\mathcal{A} \left( \omega, \vec{k}\right)$ of the correlated state at integer filling and for small hole and electron doping around it, respectively.}\label{app:fig:sym_br_bs_7_TSTGu_high}\end{figure}

\subsection{The \texorpdfstring{$\protect\IfStrEqCase{8}{{1}{\ket{\nu={}4} }
		{2}{\ket{\nu={}3, \mathrm{IVC}}}
		{3}{\ket{\nu={}3, \mathrm{VP}}}
		{4}{\ket{\nu={}2, \mathrm{K-IVC}}}
		{5}{\ket{\nu={}2, \mathrm{VP}}}
		{6}{\ket{\nu={}1, (\mathrm{K-IVC}+\mathrm{VP})}}
		{7}{\ket{\nu={}1, \mathrm{VP}}}
		{8}{\ket{\nu=0, \mathrm{K-IVC}}}
		{9}{\ket{\nu=0, \mathrm{VP}}}
	}
	[nada]
$}{nu=0 KIVC} correlated ground state candidate}
\label{app:sec:results_corr_ins:8}
\subsubsection{Low temperature}\label{app:sec:results_corr_ins_8_low}
\begin{figure}[!h]\includegraphics[width=\textwidth]{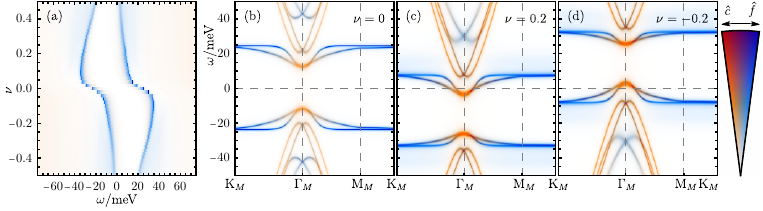}\subfloat{\label{app:fig:sym_br_bs_8_TBG_low:a}}\subfloat{\label{app:fig:sym_br_bs_8_TBG_low:b}}\subfloat{\label{app:fig:sym_br_bs_8_TBG_low:c}}\subfloat{\label{app:fig:sym_br_bs_8_TBG_low:d}}\caption{Band structure of the $\protect\IfStrEqCase{8}{{1}{\ket{\nu={}4} }
		{2}{\ket{\nu={}3, \mathrm{IVC}}}
		{3}{\ket{\nu={}3, \mathrm{VP}}}
		{4}{\ket{\nu={}2, \mathrm{K-IVC}}}
		{5}{\ket{\nu={}2, \mathrm{VP}}}
		{6}{\ket{\nu={}1, (\mathrm{K-IVC}+\mathrm{VP})}}
		{7}{\ket{\nu={}1, \mathrm{VP}}}
		{8}{\ket{\nu=0, \mathrm{K-IVC}}}
		{9}{\ket{\nu=0, \mathrm{VP}}}
	}
	[nada]
$ ground state candidate of TBG at $T=\SI{8}{\kelvin}$. The total spectral function $\mathcal{A} \left( \omega \right)$ of the system is shown in (a) as a function of filling. (b)-(d) show the $\vec{k}$-resolved spectral function $\mathcal{A} \left( \omega, \vec{k}\right)$ of the correlated state at integer filling and for small hole and electron doping around it, respectively.}\label{app:fig:sym_br_bs_8_TBG_low}\end{figure}\begin{figure}[!h]\includegraphics[width=\textwidth]{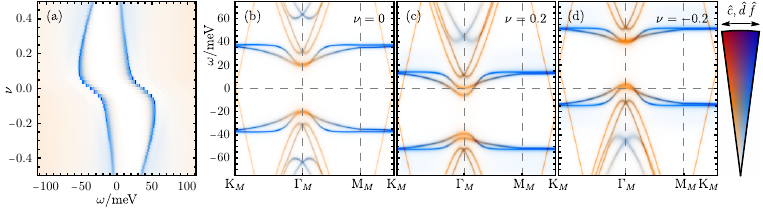}\subfloat{\label{app:fig:sym_br_bs_8_TSTG_low:a}}\subfloat{\label{app:fig:sym_br_bs_8_TSTG_low:b}}\subfloat{\label{app:fig:sym_br_bs_8_TSTG_low:c}}\subfloat{\label{app:fig:sym_br_bs_8_TSTG_low:d}}\caption{Band structure of the $\protect\IfStrEqCase{8}{{1}{\ket{\nu={}4} }
		{2}{\ket{\nu={}3, \mathrm{IVC}}}
		{3}{\ket{\nu={}3, \mathrm{VP}}}
		{4}{\ket{\nu={}2, \mathrm{K-IVC}}}
		{5}{\ket{\nu={}2, \mathrm{VP}}}
		{6}{\ket{\nu={}1, (\mathrm{K-IVC}+\mathrm{VP})}}
		{7}{\ket{\nu={}1, \mathrm{VP}}}
		{8}{\ket{\nu=0, \mathrm{K-IVC}}}
		{9}{\ket{\nu=0, \mathrm{VP}}}
	}
	[nada]
$ ground state candidate of TSTG for $\mathcal{E}=\SI{0}{\milli\electronvolt}$ at $T=\SI{11.2}{\kelvin}$. The total spectral function $\mathcal{A} \left( \omega \right)$ of the system is shown in (a) as a function of filling. (b)-(d) show the $\vec{k}$-resolved spectral function $\mathcal{A} \left( \omega, \vec{k}\right)$ of the correlated state at integer filling and for small hole and electron doping around it, respectively.}\label{app:fig:sym_br_bs_8_TSTG_low}\end{figure}\begin{figure}[!h]\includegraphics[width=\textwidth]{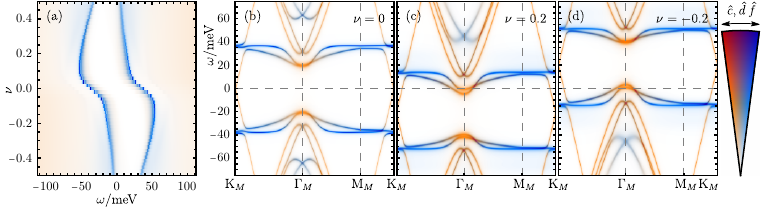}\subfloat{\label{app:fig:sym_br_bs_8_TSTGu_low:a}}\subfloat{\label{app:fig:sym_br_bs_8_TSTGu_low:b}}\subfloat{\label{app:fig:sym_br_bs_8_TSTGu_low:c}}\subfloat{\label{app:fig:sym_br_bs_8_TSTGu_low:d}}\caption{Band structure of the $\protect\IfStrEqCase{8}{{1}{\ket{\nu={}4} }
		{2}{\ket{\nu={}3, \mathrm{IVC}}}
		{3}{\ket{\nu={}3, \mathrm{VP}}}
		{4}{\ket{\nu={}2, \mathrm{K-IVC}}}
		{5}{\ket{\nu={}2, \mathrm{VP}}}
		{6}{\ket{\nu={}1, (\mathrm{K-IVC}+\mathrm{VP})}}
		{7}{\ket{\nu={}1, \mathrm{VP}}}
		{8}{\ket{\nu=0, \mathrm{K-IVC}}}
		{9}{\ket{\nu=0, \mathrm{VP}}}
	}
	[nada]
$ ground state candidate of TSTG for $\mathcal{E}=\SI{25}{\milli\electronvolt}$ at $T=\SI{11.2}{\kelvin}$. The total spectral function $\mathcal{A} \left( \omega \right)$ of the system is shown in (a) as a function of filling. (b)-(d) show the $\vec{k}$-resolved spectral function $\mathcal{A} \left( \omega, \vec{k}\right)$ of the correlated state at integer filling and for small hole and electron doping around it, respectively.}\label{app:fig:sym_br_bs_8_TSTGu_low}\end{figure}

\subsubsection{High temperature}\label{app:sec:results_corr_ins_8_high}
\begin{figure}[!h]\includegraphics[width=\textwidth]{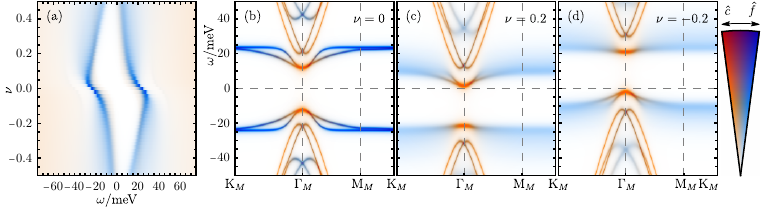}\subfloat{\label{app:fig:sym_br_bs_8_TBG_high:a}}\subfloat{\label{app:fig:sym_br_bs_8_TBG_high:b}}\subfloat{\label{app:fig:sym_br_bs_8_TBG_high:c}}\subfloat{\label{app:fig:sym_br_bs_8_TBG_high:d}}\caption{Band structure of the $\protect\IfStrEqCase{8}{{1}{\ket{\nu={}4} }
		{2}{\ket{\nu={}3, \mathrm{IVC}}}
		{3}{\ket{\nu={}3, \mathrm{VP}}}
		{4}{\ket{\nu={}2, \mathrm{K-IVC}}}
		{5}{\ket{\nu={}2, \mathrm{VP}}}
		{6}{\ket{\nu={}1, (\mathrm{K-IVC}+\mathrm{VP})}}
		{7}{\ket{\nu={}1, \mathrm{VP}}}
		{8}{\ket{\nu=0, \mathrm{K-IVC}}}
		{9}{\ket{\nu=0, \mathrm{VP}}}
	}
	[nada]
$ ground state candidate of TBG at $T=\SI{32}{\kelvin}$. The total spectral function $\mathcal{A} \left( \omega \right)$ of the system is shown in (a) as a function of filling. (b)-(d) show the $\vec{k}$-resolved spectral function $\mathcal{A} \left( \omega, \vec{k}\right)$ of the correlated state at integer filling and for small hole and electron doping around it, respectively.}\label{app:fig:sym_br_bs_8_TBG_high}\end{figure}\begin{figure}[!h]\includegraphics[width=\textwidth]{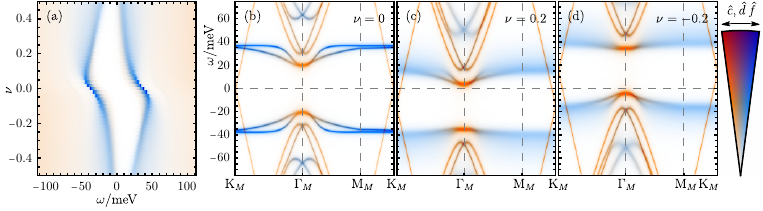}\subfloat{\label{app:fig:sym_br_bs_8_TSTG_high:a}}\subfloat{\label{app:fig:sym_br_bs_8_TSTG_high:b}}\subfloat{\label{app:fig:sym_br_bs_8_TSTG_high:c}}\subfloat{\label{app:fig:sym_br_bs_8_TSTG_high:d}}\caption{Band structure of the $\protect\IfStrEqCase{8}{{1}{\ket{\nu={}4} }
		{2}{\ket{\nu={}3, \mathrm{IVC}}}
		{3}{\ket{\nu={}3, \mathrm{VP}}}
		{4}{\ket{\nu={}2, \mathrm{K-IVC}}}
		{5}{\ket{\nu={}2, \mathrm{VP}}}
		{6}{\ket{\nu={}1, (\mathrm{K-IVC}+\mathrm{VP})}}
		{7}{\ket{\nu={}1, \mathrm{VP}}}
		{8}{\ket{\nu=0, \mathrm{K-IVC}}}
		{9}{\ket{\nu=0, \mathrm{VP}}}
	}
	[nada]
$ ground state candidate of TSTG for $\mathcal{E}=\SI{0}{\milli\electronvolt}$ at $T=\SI{44.8}{\kelvin}$. The total spectral function $\mathcal{A} \left( \omega \right)$ of the system is shown in (a) as a function of filling. (b)-(d) show the $\vec{k}$-resolved spectral function $\mathcal{A} \left( \omega, \vec{k}\right)$ of the correlated state at integer filling and for small hole and electron doping around it, respectively.}\label{app:fig:sym_br_bs_8_TSTG_high}\end{figure}\begin{figure}[!h]\includegraphics[width=\textwidth]{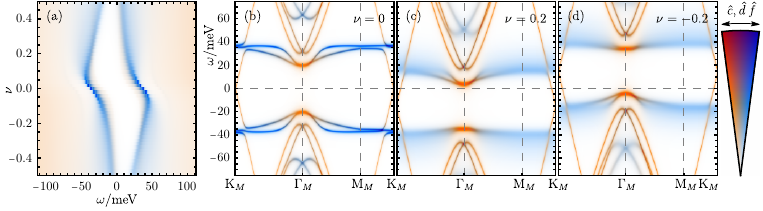}\subfloat{\label{app:fig:sym_br_bs_8_TSTGu_high:a}}\subfloat{\label{app:fig:sym_br_bs_8_TSTGu_high:b}}\subfloat{\label{app:fig:sym_br_bs_8_TSTGu_high:c}}\subfloat{\label{app:fig:sym_br_bs_8_TSTGu_high:d}}\caption{Band structure of the $\protect\IfStrEqCase{8}{{1}{\ket{\nu={}4} }
		{2}{\ket{\nu={}3, \mathrm{IVC}}}
		{3}{\ket{\nu={}3, \mathrm{VP}}}
		{4}{\ket{\nu={}2, \mathrm{K-IVC}}}
		{5}{\ket{\nu={}2, \mathrm{VP}}}
		{6}{\ket{\nu={}1, (\mathrm{K-IVC}+\mathrm{VP})}}
		{7}{\ket{\nu={}1, \mathrm{VP}}}
		{8}{\ket{\nu=0, \mathrm{K-IVC}}}
		{9}{\ket{\nu=0, \mathrm{VP}}}
	}
	[nada]
$ ground state candidate of TSTG for $\mathcal{E}=\SI{25}{\milli\electronvolt}$ at $T=\SI{44.8}{\kelvin}$. The total spectral function $\mathcal{A} \left( \omega \right)$ of the system is shown in (a) as a function of filling. (b)-(d) show the $\vec{k}$-resolved spectral function $\mathcal{A} \left( \omega, \vec{k}\right)$ of the correlated state at integer filling and for small hole and electron doping around it, respectively.}\label{app:fig:sym_br_bs_8_TSTGu_high}\end{figure}

\subsection{The \texorpdfstring{$\protect\IfStrEqCase{9}{{1}{\ket{\nu={}4} }
		{2}{\ket{\nu={}3, \mathrm{IVC}}}
		{3}{\ket{\nu={}3, \mathrm{VP}}}
		{4}{\ket{\nu={}2, \mathrm{K-IVC}}}
		{5}{\ket{\nu={}2, \mathrm{VP}}}
		{6}{\ket{\nu={}1, (\mathrm{K-IVC}+\mathrm{VP})}}
		{7}{\ket{\nu={}1, \mathrm{VP}}}
		{8}{\ket{\nu=0, \mathrm{K-IVC}}}
		{9}{\ket{\nu=0, \mathrm{VP}}}
	}
	[nada]
$}{nu=0 VP} correlated ground state candidate}
\label{app:sec:results_corr_ins:9}
\subsubsection{Low temperature}\label{app:sec:results_corr_ins_9_low}
\begin{figure}[!h]\includegraphics[width=\textwidth]{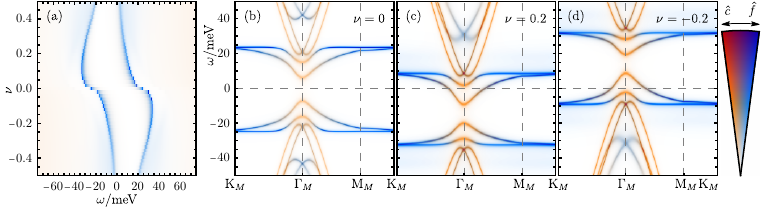}\subfloat{\label{app:fig:sym_br_bs_9_TBG_low:a}}\subfloat{\label{app:fig:sym_br_bs_9_TBG_low:b}}\subfloat{\label{app:fig:sym_br_bs_9_TBG_low:c}}\subfloat{\label{app:fig:sym_br_bs_9_TBG_low:d}}\caption{Band structure of the $\protect\IfStrEqCase{9}{{1}{\ket{\nu={}4} }
		{2}{\ket{\nu={}3, \mathrm{IVC}}}
		{3}{\ket{\nu={}3, \mathrm{VP}}}
		{4}{\ket{\nu={}2, \mathrm{K-IVC}}}
		{5}{\ket{\nu={}2, \mathrm{VP}}}
		{6}{\ket{\nu={}1, (\mathrm{K-IVC}+\mathrm{VP})}}
		{7}{\ket{\nu={}1, \mathrm{VP}}}
		{8}{\ket{\nu=0, \mathrm{K-IVC}}}
		{9}{\ket{\nu=0, \mathrm{VP}}}
	}
	[nada]
$ ground state candidate of TBG at $T=\SI{8}{\kelvin}$. The total spectral function $\mathcal{A} \left( \omega \right)$ of the system is shown in (a) as a function of filling. (b)-(d) show the $\vec{k}$-resolved spectral function $\mathcal{A} \left( \omega, \vec{k}\right)$ of the correlated state at integer filling and for small hole and electron doping around it, respectively.}\label{app:fig:sym_br_bs_9_TBG_low}\end{figure}\begin{figure}[!h]\includegraphics[width=\textwidth]{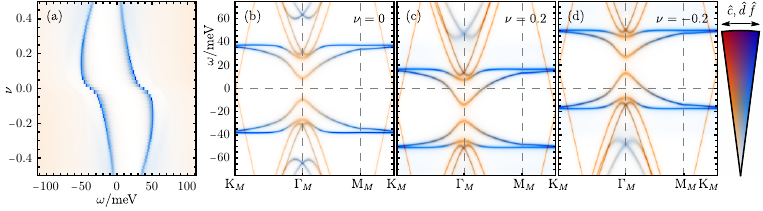}\subfloat{\label{app:fig:sym_br_bs_9_TSTG_low:a}}\subfloat{\label{app:fig:sym_br_bs_9_TSTG_low:b}}\subfloat{\label{app:fig:sym_br_bs_9_TSTG_low:c}}\subfloat{\label{app:fig:sym_br_bs_9_TSTG_low:d}}\caption{Band structure of the $\protect\IfStrEqCase{9}{{1}{\ket{\nu={}4} }
		{2}{\ket{\nu={}3, \mathrm{IVC}}}
		{3}{\ket{\nu={}3, \mathrm{VP}}}
		{4}{\ket{\nu={}2, \mathrm{K-IVC}}}
		{5}{\ket{\nu={}2, \mathrm{VP}}}
		{6}{\ket{\nu={}1, (\mathrm{K-IVC}+\mathrm{VP})}}
		{7}{\ket{\nu={}1, \mathrm{VP}}}
		{8}{\ket{\nu=0, \mathrm{K-IVC}}}
		{9}{\ket{\nu=0, \mathrm{VP}}}
	}
	[nada]
$ ground state candidate of TSTG for $\mathcal{E}=\SI{0}{\milli\electronvolt}$ at $T=\SI{11.2}{\kelvin}$. The total spectral function $\mathcal{A} \left( \omega \right)$ of the system is shown in (a) as a function of filling. (b)-(d) show the $\vec{k}$-resolved spectral function $\mathcal{A} \left( \omega, \vec{k}\right)$ of the correlated state at integer filling and for small hole and electron doping around it, respectively.}\label{app:fig:sym_br_bs_9_TSTG_low}\end{figure}\begin{figure}[!h]\includegraphics[width=\textwidth]{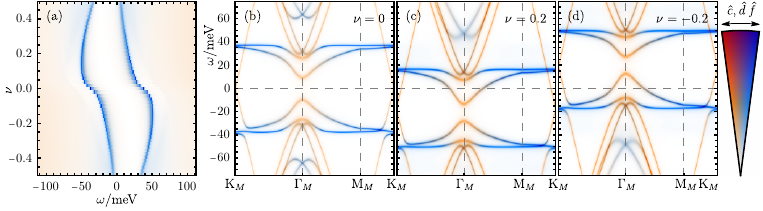}\subfloat{\label{app:fig:sym_br_bs_9_TSTGu_low:a}}\subfloat{\label{app:fig:sym_br_bs_9_TSTGu_low:b}}\subfloat{\label{app:fig:sym_br_bs_9_TSTGu_low:c}}\subfloat{\label{app:fig:sym_br_bs_9_TSTGu_low:d}}\caption{Band structure of the $\protect\IfStrEqCase{9}{{1}{\ket{\nu={}4} }
		{2}{\ket{\nu={}3, \mathrm{IVC}}}
		{3}{\ket{\nu={}3, \mathrm{VP}}}
		{4}{\ket{\nu={}2, \mathrm{K-IVC}}}
		{5}{\ket{\nu={}2, \mathrm{VP}}}
		{6}{\ket{\nu={}1, (\mathrm{K-IVC}+\mathrm{VP})}}
		{7}{\ket{\nu={}1, \mathrm{VP}}}
		{8}{\ket{\nu=0, \mathrm{K-IVC}}}
		{9}{\ket{\nu=0, \mathrm{VP}}}
	}
	[nada]
$ ground state candidate of TSTG for $\mathcal{E}=\SI{25}{\milli\electronvolt}$ at $T=\SI{11.2}{\kelvin}$. The total spectral function $\mathcal{A} \left( \omega \right)$ of the system is shown in (a) as a function of filling. (b)-(d) show the $\vec{k}$-resolved spectral function $\mathcal{A} \left( \omega, \vec{k}\right)$ of the correlated state at integer filling and for small hole and electron doping around it, respectively.}\label{app:fig:sym_br_bs_9_TSTGu_low}\end{figure}

\subsubsection{High temperature}\label{app:sec:results_corr_ins_9_high}
\begin{figure}[!h]\includegraphics[width=\textwidth]{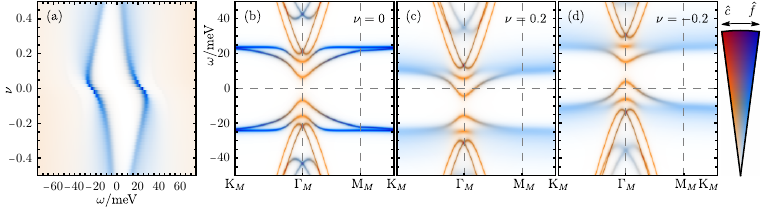}\subfloat{\label{app:fig:sym_br_bs_9_TBG_high:a}}\subfloat{\label{app:fig:sym_br_bs_9_TBG_high:b}}\subfloat{\label{app:fig:sym_br_bs_9_TBG_high:c}}\subfloat{\label{app:fig:sym_br_bs_9_TBG_high:d}}\caption{Band structure of the $\protect\IfStrEqCase{9}{{1}{\ket{\nu={}4} }
		{2}{\ket{\nu={}3, \mathrm{IVC}}}
		{3}{\ket{\nu={}3, \mathrm{VP}}}
		{4}{\ket{\nu={}2, \mathrm{K-IVC}}}
		{5}{\ket{\nu={}2, \mathrm{VP}}}
		{6}{\ket{\nu={}1, (\mathrm{K-IVC}+\mathrm{VP})}}
		{7}{\ket{\nu={}1, \mathrm{VP}}}
		{8}{\ket{\nu=0, \mathrm{K-IVC}}}
		{9}{\ket{\nu=0, \mathrm{VP}}}
	}
	[nada]
$ ground state candidate of TBG at $T=\SI{32}{\kelvin}$. The total spectral function $\mathcal{A} \left( \omega \right)$ of the system is shown in (a) as a function of filling. (b)-(d) show the $\vec{k}$-resolved spectral function $\mathcal{A} \left( \omega, \vec{k}\right)$ of the correlated state at integer filling and for small hole and electron doping around it, respectively.}\label{app:fig:sym_br_bs_9_TBG_high}\end{figure}\begin{figure}[!h]\includegraphics[width=\textwidth]{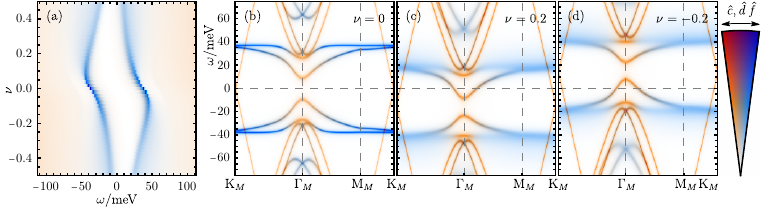}\subfloat{\label{app:fig:sym_br_bs_9_TSTG_high:a}}\subfloat{\label{app:fig:sym_br_bs_9_TSTG_high:b}}\subfloat{\label{app:fig:sym_br_bs_9_TSTG_high:c}}\subfloat{\label{app:fig:sym_br_bs_9_TSTG_high:d}}\caption{Band structure of the $\protect\IfStrEqCase{9}{{1}{\ket{\nu={}4} }
		{2}{\ket{\nu={}3, \mathrm{IVC}}}
		{3}{\ket{\nu={}3, \mathrm{VP}}}
		{4}{\ket{\nu={}2, \mathrm{K-IVC}}}
		{5}{\ket{\nu={}2, \mathrm{VP}}}
		{6}{\ket{\nu={}1, (\mathrm{K-IVC}+\mathrm{VP})}}
		{7}{\ket{\nu={}1, \mathrm{VP}}}
		{8}{\ket{\nu=0, \mathrm{K-IVC}}}
		{9}{\ket{\nu=0, \mathrm{VP}}}
	}
	[nada]
$ ground state candidate of TSTG for $\mathcal{E}=\SI{0}{\milli\electronvolt}$ at $T=\SI{44.8}{\kelvin}$. The total spectral function $\mathcal{A} \left( \omega \right)$ of the system is shown in (a) as a function of filling. (b)-(d) show the $\vec{k}$-resolved spectral function $\mathcal{A} \left( \omega, \vec{k}\right)$ of the correlated state at integer filling and for small hole and electron doping around it, respectively.}\label{app:fig:sym_br_bs_9_TSTG_high}\end{figure}\begin{figure}[!h]\includegraphics[width=\textwidth]{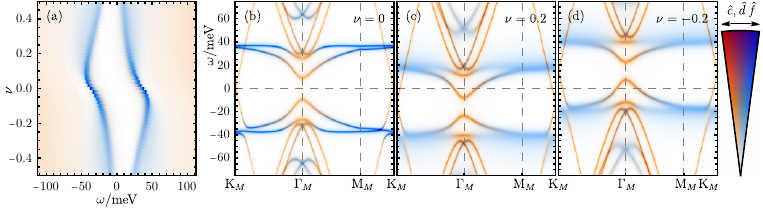}\subfloat{\label{app:fig:sym_br_bs_9_TSTGu_high:a}}\subfloat{\label{app:fig:sym_br_bs_9_TSTGu_high:b}}\subfloat{\label{app:fig:sym_br_bs_9_TSTGu_high:c}}\subfloat{\label{app:fig:sym_br_bs_9_TSTGu_high:d}}\caption{Band structure of the $\protect\IfStrEqCase{9}{{1}{\ket{\nu={}4} }
		{2}{\ket{\nu={}3, \mathrm{IVC}}}
		{3}{\ket{\nu={}3, \mathrm{VP}}}
		{4}{\ket{\nu={}2, \mathrm{K-IVC}}}
		{5}{\ket{\nu={}2, \mathrm{VP}}}
		{6}{\ket{\nu={}1, (\mathrm{K-IVC}+\mathrm{VP})}}
		{7}{\ket{\nu={}1, \mathrm{VP}}}
		{8}{\ket{\nu=0, \mathrm{K-IVC}}}
		{9}{\ket{\nu=0, \mathrm{VP}}}
	}
	[nada]
$ ground state candidate of TSTG for $\mathcal{E}=\SI{25}{\milli\electronvolt}$ at $T=\SI{44.8}{\kelvin}$. The total spectral function $\mathcal{A} \left( \omega \right)$ of the system is shown in (a) as a function of filling. (b)-(d) show the $\vec{k}$-resolved spectral function $\mathcal{A} \left( \omega, \vec{k}\right)$ of the correlated state at integer filling and for small hole and electron doping around it, respectively.}\label{app:fig:sym_br_bs_9_TSTGu_high}\end{figure}

\section{Numerical band structure results in the symmetric state}\label{app:sec:results_symmetry}

This \siSection{} presents comprehensive numerical results for the interacting spectral functions of TBG and TSTG in the symmetric phase. The calculations are carried out using the IPT method described in \cref{app:sec:se_symmetric} and elaborated in \cref{app:sec:se_symmetric_details}. A summary of the results is provided in \cref{app:tab:summary_symmetric}, with a detailed discussion following in \cref{app:sec:results_symmetry:results}. 

We begin by briefly highlighting the key features of the THF spectral function in the symmetric phase. We then introduce the Hubbard-I approximation~\cite{HUB63,HUB64} to describe the dynamical self-energy of the THF model in the symmetric state~\cite{HU25,LED25,LED25a}. This allows us to analytically interpret the numerically obtained spectral functions near integer fillings, particularly the emergence of gap-opening-like features.

{\renewcommand{\arraystretch}{1.2}
	\begin{table}[t]
		\centering
		\begin{tabular}{|l|c|c|}
			\hline
			Moir\'e system & $T / \si{\kelvin}$ & Figure \\
			\hline
			\multirow[c]{3}{*}{TBG} & $5$ & \cref{app:fig:sym_bs_tbg_low} \\
			\cline{2-3}
			 & $20$ & \cref{app:fig:sym_bs_tbg_med} \\
			\cline{2-3}
			 & $50$ & \cref{app:fig:sym_bs_tbg_high} \\
			\hline
			\multirow[c]{3}{*}{TSTG with $\mathcal{E} = \SI{0}{\milli\electronvolt}$} & $15$ & \cref{app:fig:sym_bs_tstg_noU_low} \\
			\cline{2-3}
			& $30$ & \cref{app:fig:sym_bs_tstg_noU_med} \\
			\cline{2-3}
			& $75$ & \cref{app:fig:sym_bs_tstg_noU_high}\\
			\hline
			\multirow[c]{3}{*}{TSTG with $\mathcal{E} = \SI{25}{\milli\electronvolt}$} & $7.5$ & \cref{app:fig:sym_bs_tstg_U_low}\\
			\cline{2-3}
			& $30$ & \cref{app:fig:sym_bs_tstg_U_med} \\
			\cline{2-3}
			& $75$ & \cref{app:fig:sym_bs_tstg_U_high} \\
			\hline
		\end{tabular}
		\caption{Overview of numerical results detailing the band structures of the TBG and TSTG in the symmetric states. We restrict to the positive integer-filled states from \cref{app:tab:model_states}. For each moir\'e system and temperature, we list the figure where the results are presented.}
		\label{app:tab:summary_symmetric}
\end{table}}

For each combination of moiré system and parameter set listed in \cref{app:tab:summary_symmetric}, we compute the interacting spectral function in the symmetric state using the IPT method. The corresponding results are displayed in a single figure for each case ({\it e.g.}{}, \cref{app:fig:sym_bs_tbg_med}). To facilitate direct comparison between different band structures, we adopt a consistent figure layout throughout.

In each figure, panel (a) shows the \emph{total} spectral function $\mathcal{A}(\omega)$ -- as defined in \cref{app:eqn:definition_total_spectral_function} -- plotted as a function of $\omega$ and the filling $\nu$ in the range $0 \leq \nu \leq 4.5$. Panels (b)-(j) display the $\vec{k}$-resolved spectral function along the high-symmetry lines of the moiré Brillouin zone at positive integer and half-integer fillings, as indicated in each panel. Both the $\vec{k}$-resolved spectral function and the total density of states are rendered using the same colormap as in \cref{app:sec:results_corr_ins}.

Consistent with \cref{app:sec:results_corr_ins}, we show results only for $\nu \geq 0$. The corresponding $\nu < 0$ results are obtained by applying the many-body charge conjugation symmetries of TBG and TSTG, as discussed in \cref{app:sec:hartree_fock:ground_states:ph_symmetry}.

\subsection{General discussion}\label{app:sec:results_symmetry:discussion}

\begin{figure}[!h]\includegraphics[width=\textwidth]{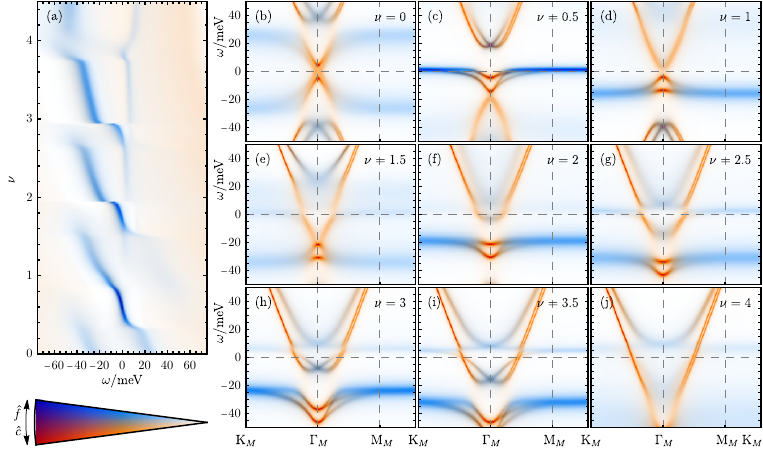}\caption{The symmetric phase of TBG within the THF model at $T = \SI{20}{\kelvin}$.}\label{app:fig:sym_bs_tbg_med}\end{figure}

\begin{figure}[!h]\includegraphics[width=\textwidth]{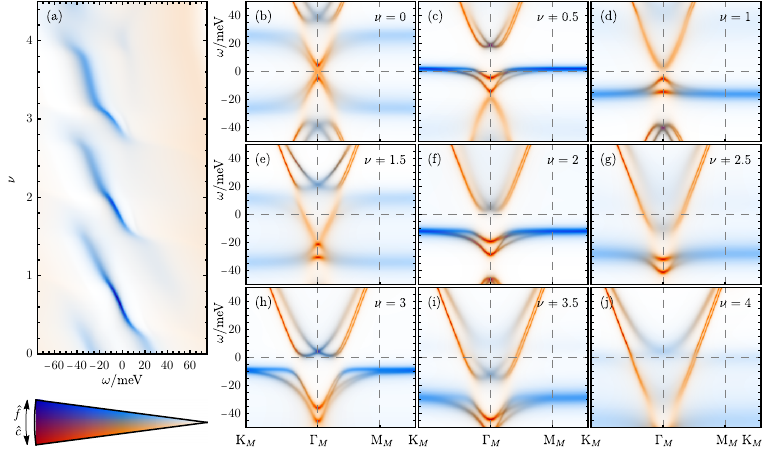}\caption{The symmetric phase of TBG within the THF model at $T = \SI{50}{\kelvin}$.}\label{app:fig:sym_bs_tbg_high}\end{figure}

We now discuss the key features of the spectral functions obtained in the symmetric-state calculations:
\begin{itemize}
	\item We begin with the high-temperature ($T = \SI{50}{\kelvin}$) TBG spectral function shown in \cref{app:fig:sym_bs_tbg_high}. At elevated temperatures, the dominant features are the Hubbard bands formed by the $f$-electrons. As the filling is varied, the energy of these bands exhibits a cascade of transitions. These transitions originate from changes between distinct $f$-electron occupancy states and can be understood in the zero-hybridization limit of the model~\cite{HU23i}.

	\item As the temperature is lowered to $T = \SI{20}{\kelvin}$, hybridization between the $f$- and $c$-electrons leads to the emergence of a Kondo resonance ({\it i.e.}{}, a zero-energy spectral peak) at specific fillings -- such as $\nu = 1.5$, $\nu = 2.5$, and $\nu \gtrsim 3$ -- as shown in \cref{app:fig:sym_bs_tbg_med}. In contrast, no Kondo resonance is observed near the integer fillings $\nu = 0$ and $\nu = 1$.
	
	\item As the temperature is further reduced to $T = \SI{5}{\kelvin}$, \cref{app:fig:sym_br_bs_1_TBG_low} shows that a Kondo resonance begins to develop at nearly all fillings. The corresponding zero-energy peak becomes noticeably sharper than at $T = \SI{20}{\kelvin}$.
	
	\item The spectral function of TSTG exhibits behavior similar to that of TBG. At high temperatures (\cref{app:fig:sym_br_bs_1_TSTG_high,app:fig:sym_br_bs_1_TSTGu_high}), we observe Hubbard bands and a cascade of transitions. Upon lowering the temperature, a Kondo resonance emerges, producing a sharp zero-energy peak at very low temperatures (\cref{app:fig:sym_br_bs_1_TSTG_low,app:fig:sym_br_bs_1_TSTGu_low}).
	
	\item The coexistence of these two features -- the Hubbard bands and the Kondo peak -- gives the spectral function a characteristic ``tail-feather'' appearance, as seen for example in \cref{app:fig:sym_br_bs_1_TBG_low}. This structure is consistent with observations from various spectroscopy experiments~\cite{WON20,CHO21,KIM22,NUC23,ZHA25a,XIA25}.
\end{itemize}

A comparison between the DMFT results obtained using the IPT impurity solver and those based on a QMC impurity solver~\cite{RAI23a,HU23} reveals excellent qualitative agreement, although some quantitative differences remain:
\begin{itemize}
	\item DMFT simulations with a QMC impurity solver suggest that Kondo physics is irrelevant at the integer fillings $\nu = 0, 1, 2$, down to temperatures as low as $T \sim \SI{2}{\kelvin}$~\cite{HU23}. While our IPT-based approach also shows a suppressed Kondo effect at integer fillings compared to non-integer ones, it does predict a slightly higher Kondo temperature: a small Kondo peak is still discernible around integer fillings even at $T = \SI{5}{\kelvin}$. At non-integer fillings, both our results and those from QMC-based DMFT consistently indicate a strong tendency toward Kondo physics~\cite{HU23}.
	
	\item Since the IPT method only \emph{approximately} satisfies the Luttinger theorem (as discussed in \cref{app:sec:se_symmetric_details:IPT:fix_mu_tilde}), the gap-opening-like features observed in the QMC-DMFT solution~\cite{RAI23a}, which occur almost \emph{exactly} at integer fillings, appear \emph{slightly} displaced in our IPT simulations -- by approximately $0.1$ in filling. This shift is visible in panel (a) of the figures in \cref{app:sec:results_symmetry:results}. 
\end{itemize}

In DMFT simulations employing either the IPT or QMC impurity solver, we find that the symmetric solution exhibits gap-opening behavior -- or, at the very least, a significant suppression of the numerically computed density of states at the Fermi level -- around non-zero integer fillings. To provide analytical insight into this feature, we introduce the Hubbard-I approximation for the dynamical self-energy of the symmetric phase in the following \cref{app:sec:results_symmetry:hubbard_I}, and analyze the spectrum of TBG within the THF model at integer fillings.

\subsection{Hubbard-I approximation}
\label{app:sec:results_symmetry:hubbard_I}

In this section, we describe the Hubbard-I approximation, as applied to the THF model. Within this approach, the interacting Green's function of a system is obtained using the atomic self-energy. While this approximation captures the formation of Hubbard bands, it does not account for the emergence of the Kondo resonance. However, as demonstrated in Ref.~\cite{HU23}, the Kondo effect becomes less relevant near integer fillings. This makes the Hubbard-I approximation particularly suitable for explaining the behavior of electrons in this regime.

Here, we use the Hubbard-I approximation to analyze the gap-opening behavior observed in the symmetric state near integer fillings. In \cref{app:sec:results_symmetry:hubbard_I:green_fun_hubb_I}, we present the Hubbard-I approximation in detail and derive the corresponding Green's function. We also show how this approach captures the emergence of gaps in the interacting spectrum at non-zero integer fillings.

\subsubsection{Mapping an interacting Green's function to a non-interacting Green's function}
\label{app:sec:results_symmetry:hubbard_I:mapp_green_fun}

Before presenting the results of the Hubbard-I approximation applied to the THF model, we first outline a general method for expressing an \emph{interacting} Green's function in terms of an equivalent non-interacting Green's function by introducing auxiliary fermionic degrees of freedom, as also discussed in Ref.~\cite{HU25}.

To illustrate this, we consider a generic multi-orbital system with electron operators $\hat{\gamma}^\dagger_{i}$, where the index $i$ may represent spin, momentum, orbital, or sublattice labels. The system is described by the following generic interacting Hamiltonian
\begin{equation}
	\hat{H} = \sum_{i,j} t_{ij} \hat{\gamma}^\dagger_{i}\hat{\gamma}_{j} + \sum_{i,j,l,m} U_{ijlm} \hat{\gamma}^\dagger_{i}\hat{\gamma}_{j}\hat{\gamma}^\dagger_{l}\hat{\gamma}_{m}
\end{equation}
where $t_{ij}$ denotes the hopping matrix elements, and $U_{ijlm}$ encodes the interaction terms.

The interacting Green's function of the system is defined as
\begin{align}
	\mathcal{G}_{i j} ( \tau ) &= -\left\langle \mathcal{T}_{\tau} \hat{\gamma}_{i} ( \tau ) \hat{\gamma}^\dagger_{j} ( 0 )  \right\rangle, \nonumber\\ 
	\mathcal{G}_{i j} \left(i \omega_n \right) &= \int_{0}^{\beta} \dd{\tau} e^{i \omega_n \tau} 	\mathcal{G}_{ij} ( \tau ) 
	= \left[ \left( i\omega_n \mathbb{1} - t - h^{\text{MF}} - \Sigma \left( i\omega_n \right) \right)^{-1} \right]_{ij},
	\label{app:eqn:hubbard_i_int_green}
\end{align}
where $h^{\text{MF}}_{ij}$ and $\Sigma_{ij} \left( i\omega_n \right)$ represent the static and dynamical self-energy contributions, respectively. The static part corresponds to the Hartree-Fock mean-field interaction Hamiltonian.

The dynamical self-energy itself admits a spectral representation~\cite{LUT61,PAV19}
\begin{equation}
	\label{app:eqn:self-energy_spectral_rep_hub_i}
	\Sigma_{ij} \left( i\omega_n \right) = \int_{-\infty} ^{\infty} \dd{\omega} \frac{ \rho^{\Sigma}_{ij} (\omega) }{i\omega_n - \omega },
\end{equation}
with $\rho^{\Sigma}_{ij}(\omega)$ denoting the spectral function of the self-energy, which is given explicitly by  
\begin{equation}
	\rho^{\Sigma}(\omega) = -\frac{1}{2\pi i }\left( \Sigma \left( \omega + i 0^{+} \right) -\Sigma^\dagger \left( \omega + i 0^{+} \right) \right)
	\label{app:eqn:hubbard_i_spec_self_energy}
\end{equation}

Similar to the spectral function of the Green's function defined in \cref{app:eqn:hubbard_i_int_green}, $\rho^{\Sigma}(\omega)$ is a Hermitian, positive semi-definite matrix. Hermiticity,
\begin{align}
	\rho^{\Sigma}(\omega) = \left( \rho^{\Sigma}(\omega) \right)^{\dagger},
\end{align}
follows directly from \cref{app:eqn:hubbard_i_spec_self_energy}. To see that $\rho^{\Sigma}(\omega)$ is also positive semi-definite, we note that from \cref{app:eqn:hubbard_i_int_green}, the self-energy just above the real axis can be written as
\begin{equation}
	\Sigma \left( \omega + i 0^{+} \right) = \omega + i 0^{+} - t - h^{\text{MF}} - \mathcal{G}^{-1} \left( \omega + i 0^{+} \right).
\end{equation}
This allows the spectral function of the dynamical self-energy to be rewritten as
\begin{align}
	\rho^{\Sigma}(\omega) &=  \frac{1}{2\pi i }\left( \mathcal{G}^{-1} \left( \omega + i 0^{+} \right) - \left[ \mathcal{G}^{\dagger} \left( \omega + i 0^{+} \right) \right]^{-1} \right) \nonumber \\
	&=  -\frac{1}{2\pi i } \mathcal{G}^{-1} \left( \omega + i 0^{+} \right) \left(  \mathcal{G} \left( \omega + i 0^{+} \right) - \mathcal{G}^\dagger \left( \omega + i 0^{+} \right)  \right) \left[ \mathcal{G}^{-1} \left( \omega + i 0^{+} \right) \right]^{\dagger} \nonumber \\ 
	&=  \mathcal{G}^{-1} \left( \omega + i 0^{+} \right) A (\omega) \left[ \mathcal{G}^{-1} \left( \omega + i 0^{+} \right) \right]^{\dagger}, \label{app:eqn:explaination_positive_semidefinite_self_energy}
\end{align}
which is manifestly a positive semi-definite matrix.

The fact that $\rho^{\Sigma}(\omega)$ is Hermitian and positive semi-definite allows it to be expressed as a sum of $\delta$-functions multiplied by positive semi-definite matrices. These matrices can, without loss of generality, be written as outer products of vectors
\begin{equation} 
	\label{app:eqn:self_energy_spectral_into_delta}
	\rho_{ij}^{\Sigma}(\omega) \approx \sum_{\alpha} v^{\alpha}_i \left( v^{\alpha}_j \right)^* \delta(\omega- \epsilon^\alpha)
\end{equation} 
where $v^{\alpha}_i$ are complex vectors and $\epsilon^\alpha$ are real frequencies. The parameters $v^{\alpha}_i$ and $\epsilon^\alpha$ are introduced to accurately describe the spectral function $\rho^{\Sigma}$. An arbitrarily good approximation can be achieved by including sufficiently many terms in the sum in \cref{app:eqn:self_energy_spectral_into_delta}. In the following, we assume that a sufficient number of terms have been included so that the approximation is effectively exact. Using the spectral representation of the self-energy from \cref{app:eqn:self-energy_spectral_rep_hub_i}, we then obtain
\begin{align}
	\label{app:eqn:hubbard_i_self_energy}
	\Sigma_{ij} \left( i\omega_n \right) = \sum_{\alpha} \frac{v^{\alpha}_i \left( v^{\alpha}_j \right)^*}{i\omega_n-\epsilon^\alpha }
\end{align}

\newcommand{\tAux}{\text{aux}}

We now prove that the interacting Green's function from \cref{app:eqn:hubbard_i_int_green} is equivalent to the Green's function of a suitably defined non-interacting system. Consider the following non-interacting Hamiltonian
\begin{equation}
	\label{app:eqn:aux_ham}
	\hat{H}^{\tAux} = \sum_{ij} \left( t_{ij} + h^{\text{MF}}_{ij} \right)\hat{\gamma}^\dagger_{i}\hat{\gamma}_{j} + \sum_{\alpha} \epsilon^\alpha \hat{a}^\dagger_{\alpha} \hat{a}_{\alpha} + \sum_{\alpha} \left( v_i^{\alpha} \hat{a}^\dagger_{\alpha} \hat{\gamma}_{i} + \text{h.c.} \right)
\end{equation} 
We proceed by computing the Green's function of the non-interacting Hamiltonian $\hat{H}^{\tAux}$. Defining the spinor $\hat{\Psi}^\dagger_{} = \begin{pmatrix} \hat{\gamma}^\dagger_{1}, \hat{\gamma}^\dagger_{2}, \dots, \hat{a}^\dagger_{1}, \hat{a}^\dagger_{2}, \dots \end{pmatrix}$, the full Green's function matrix is given by
\begin{equation} 
	\mathcal{G}^{\tAux}( \tau) = - \left\langle \mathcal{T}_{\tau} \hat{\Psi}_{}(\tau) \hat{\Psi}^\dagger_{} (0)  \right\rangle^{\tAux}, \qq{with} \mathcal{G}^{\tAux} \left( i \omega_n \right) = \begin{pmatrix}
		i \omega_n \mathbb{1} - t - h^{\text{MF}} & - V \\
		- V^{\dagger} & i \omega_n \mathbb{1} - \mathcal{E}		
	\end{pmatrix}^{-1},
\end{equation} 
where the hybridization matrix $V$ has elements $v_i^{\alpha}$ in the $i$-th row and $\alpha$-th column, while $\mathcal{E}$ is a diagonal matrix with entries $\epsilon^{\alpha}$. To extract the Green's function for the $\hat{\gamma}^\dagger_{i}$ fermions, we use the standard block matrix inversion identity for invertible matrices $A$ and $D$,
\begin{equation}
	\begin{pmatrix}
		A & B \\ 
		C & D 
	\end{pmatrix}^{-1} = \begin{pmatrix}
		( A - B D^{-1} C )^{-1} & -\left( A - B D^{-1} C \right)^{-1} B D^{-1} \\ 
		-( D - C A^{-1} B )^{-1} C A^{-1} & \left( D - C A^{-1} B \right)^{-1}
	\end{pmatrix}.
\end{equation}
Applying this identity to $\mathcal{G}^{\tAux} \left( i\omega_n \right)$ and extracting the upper-left block, we obtain the Green's function of the physical $\hat{\gamma}^\dagger_{i}$ electrons under the Hamiltonian $H^{\tAux}$
\begin{align} 
	\mathcal{G}^{\tAux,\gamma}_{ij} (\tau) &= -\left\langle \mathcal{T}_{\tau} \hat{\gamma}^\dagger_{i} (\tau) \hat{\gamma}_{j} (0) \right\rangle^{\tAux}, \nonumber \\
	\mathcal{G}^{\tAux,\gamma} \left( i \omega_n \right) &= \left[ i\omega_n \mathbb{1} - t - h^{\text{MF}} - V \left( i \omega_n \mathbb{1} - \mathcal{E} \right)^{-1} V^{\dagger} \right]^{-1},
\end{align} 
such that 
\begin{equation}
	\label{app:eqn:eqn:hubbard_i_self_energy_2}
	\left[ \left( \mathcal{G}^{\tAux,\gamma} \left( i \omega_n \right) \right)^{-1} \right]_{ij} = i \omega_n \delta_{ij} - t_{ij} - h^{\text{MF}}_{ij} - \sum_{\alpha} \frac{v^{\alpha}_i \left( v^{\alpha}_j \right)^*}{i\omega_n-\epsilon^\alpha },
\end{equation}
from where we can immediately conclude that
\begin{equation}
	\mathcal{G}^{\tAux,\gamma} \left( i\omega_n \right) = \mathcal{G} \left( i \omega_n \right).
\end{equation}

We have thus shown that the interacting Green's function of the original interacting Hamiltonian $H$ can be equivalently described by the Green's function of an appropriately constructed \emph{non-interacting} Hamiltonian $H^{\tAux}$. While this mapping does not provide a method for computing the interacting Green's function directly, it enables a more convenient framework for analyzing the band topology of the interacting system through the lens of an effective non-interacting model.

\subsubsection{The THF Green's function within the Hubbard-I approximation}\label{app:sec:results_symmetry:hubbard_I:green_fun_hubb_I}
\newcommand{\tHI}{\text{Hubbard-I}}

We now introduce the Green's function of the THF model within the Hubbard-I approximation. At integer fillings and in the low-temperature limit ($\beta U_1 \gg 1$), the self-energy of the $f$-electrons in the atomic limit was derived in \cref{app:eqn:atomic_self_energy_low_temp_integer}, and is reproduced here for convenience
\begin{equation}
	\label{app:eqn:atomic_self_energy_low_temp_integer_2}
	\tilde{\Sigma}^{\text{At}} \left( i \omega_n \right) \approx \frac{1}{N_f^2}\frac{ U_1^2 r\left(N_f-r \right)}{ i \omega_n - U_1 (r /N_f- \frac{1}{2})}, \qq{for} n = \frac{r}{N_f} \qq{with} r \in \mathbb{Z}, 0 \leq r \leq N_f, 
\end{equation}
where $r$ denotes the number of filled $f$-electron flavors (we assume an integer number of $f$-electron bands are filled, as in the zero-hybridization model of Ref.~\cite{HU23i}). In the Hubbard-I approximation, the \emph{interacting} Green's function is computed by approximating the dynamical self-energy of the system with its atomic-limit expression
\begin{equation}
	\label{app:eqn:hubbard_i_GF}
	\mathcal{G}^{\tHI} \left( i \omega_n, \vec{k} \right) = \left[\left( i \omega_n + \mu \right) \mathbb{1} - h^{\text{MF}} \left( \vec{k} \right) - \Sigma^{\text{At}} \left( i \omega_n \right)  
	\right]^{-1},
\end{equation}
where the atomic self-energy (in matrix form) is given by 
\begin{eqnarray}
	\Sigma^{\text{At}}_{i \eta s; i' \eta' s'} \left( i \omega_n \right) =\begin{cases}
		\tilde{\Sigma}^{\text{At}} \left( i \omega_n \right) \delta_{ii'} \delta_{\eta \eta'} \delta_{s s'} & \qq{if} i=5,6 \\
		0 & \qq{otherwise}
	\end{cases}. 
\end{eqnarray}

For TBG, the mean-field Hartree-Fock Hamiltonian $h^{\text{MF}}(\vec{k})$ can be approximated as
\begin{align}
	h^{\text{MF}}_{i \eta s,i' \eta' s'} \left( \vec{k} \right) &= \delta_{\eta,\eta'} \delta_{s,s'} \tilde{h}^{\text{MF},\eta }_{ii'} \left( \vec{k} \right) \label{app:eqn:hub_i_mf_ham_general}\\ 
	\tilde{h}^{\text{MF},\eta } \left( \vec{k} \right) &=
	\begin{pmatrix}
		\epsilon_{c,1} & 0  & v_\star \left( \eta k_x + ik_y \right)  & 0 & \gamma & v_\star^\prime \left( \eta k_x +ik_y \right) \\ 
		0 & \epsilon_{c,1} & 0 & v_\star \left( \eta k_x -ik_y \right) & v_\star^\prime \left( \eta k_x -ik_y \right) & \gamma \\ 
		v_\star \left( \eta k_x -ik_y \right) & 0 & \epsilon_{c,2} & M &  0 & 0 \\ 
		0 & v_\star\left( \eta k_x -ik_y \right) & M & \epsilon_{c,2} & 0 & 0 \\ 
		\gamma & v_\star^\prime \left( \eta k_x +ik_y \right) & 0 & 0 & \epsilon_f & 0 \\
		v_\star^\prime \left( \eta k_x -ik_y \right) & \gamma & 0 & 0 & 0 & \epsilon_f \nonumber
	\end{pmatrix} 
\end{align}
where 
\begin{align}
	\epsilon_{c,1} &= -\mu + \frac{V \left( \vec{0} \right)}{\Omega_0}\nu_c + W_1 \nu_f, \\
	\epsilon_{c,2} &=  -\mu + \frac{V\left( \vec{0} \right)}{\Omega_0}\nu_c + W_3 \nu_f, \\
	\epsilon_f &= -\mu + W_1 \left( \nu_{c,1}+\nu_{c,2} \right)+W_3 \left( \nu_{c,2}+\nu_{c,3} \right) + U_1 \left( r/N_f-\frac12 \right) \left( N_f - 1 \right).
\end{align}

In \cref{app:eqn:hub_i_mf_ham_general}, $\nu_{c,a}$ denotes the filling of $c$ electrons with orbital index $a$, while $\epsilon_{c,1}$, $\epsilon_{c,2}$, and $\epsilon_f$ denote the relative energies of the $c$- and $f$-electrons and include the chemical potential and the Hartree contributions stemming from the $H_V$, $H_W$, and $H_{U_1}$ terms of the THF interaction Hamiltonian. The corresponding Fock contributions -- which are ignored in \cref{app:eqn:hub_i_mf_ham_general} -- could, in principle, further renormalize the $f$–$c$ hybridization. However, since we are interested in the overall features of the spectrum of TBG in the symmetric state (such as gap openings), these effects are less important and have been ignored here, as have the other terms of the interacting Hamiltonian, namely $H_J$, $H_{\tilde{J}}$, $H_K$, and $H_{U_2}$. In what follows, we will restrict ourselves to integer total fillings $\nu \in \mathbb{Z}$ (with $\abs{\nu} \leq 3$) and approximate the filling of the $f$- and $c$-electrons by their zero-hybridization solution~\cite{HU23i}, whereby $\nu_c = 0$ and $\nu_f = \nu$.

We aim to describe the spectrum of the THF model of TBG under the Hubbard-I approximation by studying the Green's function from \cref{app:eqn:hubbard_i_GF}. We first employ the idea introduced in Ref.~\cite{HU25} and map the interacting THF Hamiltonian within the Hubbard-I approximation to an effective non-interacting one. To do so, we first note that the atomic self-energy from \cref{app:eqn:atomic_self_energy_low_temp_integer_2} admits the following spectral representation 
\begin{equation}
	\tilde{\Sigma}^{\text{At}} \left( i \omega_n \right)  = \int_{-\infty}^{\infty} 
	\frac{\rho^{\tilde{\Sigma}^{\text{At}}}(\omega)}{i\omega_n-\omega} \dd{\omega}, \qq{with} \rho^{\tilde{\Sigma}^{\text{At}}} ( \omega ) = \frac{U_1^2 r (N_f-r)}{N_f^2} \delta\left(\epsilon - U_1 \left(r/N_f-\frac{1}{2} \right) \right).
\end{equation}
We can then introduce the following effective non-interacting system: 
\begin{align}
	\hat{H}^{\tAux} = &\sum_{\substack{\vec{k},i, \eta, s \\ i', \eta', s'}} h^{\text{MF}}_{i \eta s;i' \eta' s'}\left( \vec{k} \right) \hat{\gamma}^\dagger_{\vec{k},i, \eta, s}  \hat{\gamma}_{\vec{k},i', \eta', s'}\nonumber\\ 
	& +\sum_{\vec{k},i, \eta, s} U_1 \left( r/N_f - \frac{1}{2} \right) \hat{a}^\dagger_{\vec{k},\alpha, \eta, s} \hat{a}_{\vec{k},\alpha, \eta, s}
	+ \sum_{\vec{k}, i, \eta, s} \frac{U_1 \sqrt{r \left( N_f-r \right)}}{N_f}\left( \hat{a}^\dagger_{\vec{k},\alpha, \eta, s} \hat{f}_{\vec{k},\alpha, \eta, s} + \text{h.c.} \right), 
	\label{app:eqn:hubbard_i_aux_ham}
\end{align}
where we have employed the unified notation from \cref{app:sec:hartree_fock:generic_not} for both $f$- and $c$-electrons. The operators $\hat{a}_{\vec{k},\alpha, \eta, s}$ denote the auxiliary fermions that couple to the $f$-electrons to reproduce the effect of the atomic self-energy, which is diagonal in the $f$-electron fermion flavors. Within the Hubbard-I approximation, the self-energy for each flavor contains a single pole. Consequently, we introduce one auxiliary fermion per flavor, coupled to the corresponding $f$-electron flavor. The Hamiltonian in \cref{app:eqn:hubbard_i_aux_ham} yields the same Green's function (for the $f$- and $c$-electrons) and the same single-particle spectrum as in \cref{app:eqn:hubbard_i_GF}.

We now discuss the single-particle spectrum of $\hat{H}^{\tAux}$. We begin by turning off the hybridization between the $f$- and $c$-electrons (by taking $\gamma = 0$ and $v_\star^\prime = 0$). In the zero-hybridization limit, the Hamiltonian $\hat{H}^{\tAux}$ decouples into two blocks: one for the $c$-electrons and one for the $f$- and $a$-fermions. The latter reads
\begin{align}
	\hat{H}^{\tAux,f,a} = &\sum_{\vec{k},\alpha, \eta, s} \epsilon_f \hat{\gamma}^\dagger_{\vec{k},\alpha, \eta, s}  \hat{\gamma}_{\vec{k},\alpha, \eta, s} \nonumber\\ 
	&+\sum_{\vec{k},i, \eta, s} U_1 \left( r/N_f - \frac{1}{2} \right) \hat{a}^\dagger_{\vec{k},\alpha, \eta, s} \hat{a}_{\vec{k},\alpha, \eta, s}
	+ \sum_{\vec{k}, i, \eta, s} \frac{U_1 \sqrt{r \left( N_f-r \right)}}{N_f}\left( \hat{a}^\dagger_{\vec{k},\alpha, \eta, s} \hat{f}_{\vec{k},\alpha, \eta, s} + \text{h.c.} \right),
\end{align}
with its dispersion given by
\begin{align}
	E_{f,1} =& \frac{\epsilon_f +U_1 \left( r/N_f - \frac12 \right)}{2} + \sqrt{
		\bigg(  \frac{\epsilon_f -U_1 \left( r/N_f - \frac12 \right)}{2}\bigg)^2  + \frac{U_1^2r(N_f-r)}{N_f^2}
	} \nonumber\\ 
	E_{f,2} =& \frac{\epsilon_f +U_1 \left( r/N_f - \frac12 \right)}{2} - \sqrt{
		\bigg(  \frac{\epsilon_f -U_1 \left( r/N_f - \frac12 \right)}{2}\bigg)^2  + \frac{U_1^2r(N_f-r)}{N_f^2}
	} 
\end{align}
These two eigenmodes represent the upper and lower Hubbard bands of the $f$-electrons. To further illustrate this point, we consider the spectrum at charge neutrality ($\nu = 0$), with $\nu_f = \nu_{c,a} = \nu_c = 0$, $r = N_f/2$, $\mu = 0$, and $\epsilon_f = 0$, for which we find $E_{f,1} = \frac{U}{2}$ and $E_{f,2} = -\frac{U}{2}$ -- which are exactly the energies of the Hubbard bands. The corresponding eigenfunctions are given by
\begin{align} 
	\hat{g}_{\vec{k},1,\alpha, \eta, s} = u \hat{f}_{\vec{k},\alpha, \eta, s} + v \hat{a}_{\vec{k},\alpha, \eta, s}, \quad 
	\hat{g}_{\vec{k},2,\alpha, \eta, s} = v \hat{f}_{\vec{k},\alpha, \eta, s} - u \hat{a}_{\vec{k},\alpha, \eta, s},
\end{align}
where 
\begin{align}
	u =& \frac{1}{\sqrt{2 E_{f,0} \left[E_{f,0} +  \frac{\epsilon_f -U_1 \left( r/N_f - \frac12 \right)}{2} \right]}
	}\left[ \frac{\epsilon_f -U_1 \left( r/N_f - \frac12 \right)}{2}+E_{f,0}
	\right], \nonumber \\
	v =& \frac{\frac{U_1\sqrt{r(N_f-r)}}{N_f}}{\sqrt{2E_{f,0} \left[E_{f,0} +  \frac{\epsilon_f -U_1 \left( r/N_f - \frac12 \right)}{2} \right]}
	}, \nonumber\\ 
	E_{f,0} &= \sqrt{ \left[ \frac{\epsilon_f -U_1 \left( r/N_f - \frac12 \right)}{2} \right]^2  + \frac{U_1^2r(N_f-r)}{N_f^2}
	}. 
\end{align}

In this new basis, $\hat{H}^{\tAux,f,a}$ can be written simply as
\begin{equation}
	\hat{H}^{\tAux,f,a} =  \sum_{\vec{k}, \alpha, \eta, s} \left(
	E_{f,1}\hat{g}^\dagger_{\vec{k},1,\alpha, \eta, s} \hat{g}_{\vec{k},1,\alpha, \eta, s} + E_{f,2} \hat{g}^\dagger_{\vec{k},2,\alpha, \eta, s} \hat{g}_{\vec{k},2,\alpha, \eta, s} \right) .
\end{equation}
Still in the zero-hybridization limit, the Hamiltonian corresponding to the $c$-electrons reads
\begin{align}
	\hat{H}^{\tAux,c} = &\sum_{\substack{ \vec{k},a,\eta,s \\ a',\eta,s}}
	\hat{c}^\dagger_{\vec{k},a,\eta,s} 
	\begin{pmatrix}
		\epsilon_{c,1} & 0  & v_\star \left( \eta k_x + ik_y \right)  & 0  \\ 
		0 & \epsilon_{c,1} & 0 & v_\star \left( \eta k_x -ik_y \right)  \\ 
		v_\star\left(\eta k_x -ik_y \right) & 0 & \epsilon_{c,2} & M  \\ 
		0 & v_\star\left( \eta k_x -ik_y \right) & M & \epsilon_{c,2} 
	\end{pmatrix}_{aa'}
	\hat{c}_{\vec{k}, a', \eta, s}. 
\end{align} 
To simplify the calculations, we also set $\epsilon_{c,1} = \epsilon_{c,2} \equiv \epsilon_c$, which leads to the following dispersion
\begin{align}
	E_{c,1} \left( \vec{k} \right) = \epsilon_c - \frac{M}{2} - \sqrt{ \frac{M^2}{4} 
		+ v_\star^2 \abs{\vec{k}}^2 } ,\quad 
	E_{c,2}\left( \vec{k} \right) = \epsilon_c - \frac{M}{2} + \sqrt{ \frac{M^2}{4} 
		+ v_\star^2 \abs{\vec{k}}^2 } \nonumber\\ 
	E_{c,3}\left( \vec{k} \right) = \epsilon_c + \frac{M}{2} - \sqrt{ \frac{M^2}{4} 
		+ v_\star^2 \abs{\vec{k}}^2 } ,\quad 
	E_{c,4}\left( \vec{k} \right) = \epsilon_c + \frac{M}{2} + \sqrt{ \frac{M^2}{4} 
		+ v_\star^2 \abs{\vec{k}}^2 }. \nonumber\\ 
\end{align}
We observe that, for each valley and spin, the $c$-electrons develop a node between $E_{c,2}(\vec{k})$ and $E_{c,3}(\vec{k})$ at the $\Gamma_M$ point, where $E_{c,2}(\vec{k} = \vec{0}) = E_{c,3}(\vec{k} = \vec{0})$.

We now analyze the effect of hybridization and focus on the evolution of the nodes formed by the $c$-electrons. We first note that, due to the formation of Hubbard bands, in the zero-hybridization limit the $f$- and $a$-electrons are gapped. As such, we can treat them as high-energy degrees of freedom that hybridize with the low-energy $c$-electrons. This allows us to treat the hybridization terms $\gamma$ and $v_\star^\prime$ perturbatively. To proceed, we first rewrite the $f$-$c$ hybridization term in the $g$-electron basis, which gives
\begin{align}
	\hat{H}^{\tAux,\text{hyb}} =&\sum_{\vec{k},\alpha, \eta, s} \left[ \gamma \left( u \hat{g}^\dagger_{\vec{k},1,\alpha, \eta, s}  + v \hat{g}^\dagger_{\vec{k},2,\alpha \eta s} \right) \hat{c}_{\vec{k},\alpha, \eta, s} + \text{h.c.} \right] \nonumber\\ 
	&+ \sum_{\vec{k}, \eta , s} \left[ v_\star^\prime \left( \eta k_x-ik_y \right) \left( u \hat{g}^\dagger_{\vec{k},1,2,\eta,s}	+ v \hat{g}^\dagger_{\vec{k},2,2,\eta,s} \right)
	\hat{c}_{\vec{k},1,\eta,s} + \text{h.c.} \right] \nonumber\\
	&+ \sum_{\vec{k}, \eta , s} \left[ v_\star^\prime \left( \eta k_x+ik_y \right) \left( u \hat{g}^\dagger_{\vec{k},1,1,\eta,s} + v \hat{g}^\dagger_{\vec{k},2,1,\eta,s} \right) \hat{c}_{\vec{k},2,\eta,s} + \text{h.c.} \right]. \label{app:sec:results_symmetry:hubbard_I:green_fun_hubb_I_aux_hyb1}
\end{align}
We then use second-order perturbation theory to integrate out the high-energy bands originating from the $g$-electrons. This yields the following low-energy effective Hamiltonian for the $c$-electrons
\begin{align}
	\hat{H}^{\tAux, \text{eff}} = &\sum_{\substack{ \vec{k},a,\eta,s \\ a',\eta,s}}
	\hat{c}^\dagger_{\vec{k},a,\eta,s} 
	\begin{pmatrix}
		\epsilon_{c,1} & 0  & v_\star \left( \eta k_x + ik_y \right)  & 0  \\ 
		0 & \epsilon_{c,1} & 0 & v_\star \left( \eta k_x -ik_y \right)  \\ 
		v_\star\left(\eta k_x -ik_y \right) & 0 & \epsilon_{c,2} & M  \\ 
		0 & v_\star\left( \eta k_x -ik_y \right) & M & \epsilon_{c,2} 
	\end{pmatrix}_{aa'}
	\hat{c}_{\vec{k}, a', \eta, s} \nonumber\\ 
	&+\sum_{\vec{k},a\eta s,a'\eta s}
	\hat{c}_{\vec{k},a,\eta,s}^\dag\bigg[ \frac{u^2}{E_{f,1}}+\frac{v^2}{E_{f,2}}\bigg] 
	\begin{pmatrix}
		-\left[ \gamma^2+ \left( v_\star^\prime \right)^2 \abs{\vec{k}}^2 \right]   &
		-\gamma v_\star^\prime \left( \eta k_x +ik_y \right)
		& 0 & 0  \\ 
		-\gamma v_\star^\prime \left( \eta k_x -ik_y \right) &    -\left[ \gamma^2+ \left( v_\star^\prime \right)^2 \abs{\vec{k}}^2 \right] & 0 & 0  \\ 
		0 & 0 & 0 & 0   \\ 
		0 & 0 & 0 & 0
	\end{pmatrix}_{aa'}
	\hat{c}_{\vec{k}, a', \eta, s}. \label{app:sec:results_symmetry:hubbard_I:green_fun_hubb_I_aux_hyb2}
\end{align} 
In the limit $v_\star^\prime = 0$, the resulting dispersion simplifies to
\begin{align}
	E_{c,1} \left( \vec{k} \right) &= \epsilon_c +\frac{ M+m}{2}  -\sqrt { \left( \frac{M-m}{2} \right)^2+v_\star^2 \abs{\vec{k}}^2 },\quad 
	E_{c,2} \left( \vec{k} \right) = \epsilon_c + \frac{ M+m}{2}  +\sqrt { \left( \frac{M-m}{2} \right)^2+v_\star^2 \abs{\vec{k}}^2 },\nonumber\\ 
	E_{c,3} \left( \vec{k} \right) &= \epsilon_c - \frac{M-m}{2}  -\sqrt { \left( \frac{M+m}{2} \right)^2+v_\star^2 \abs{\vec{k}}^2 },\quad  
	E_{c,4} \left( \vec{k} \right) = \epsilon_c - \frac{M-m}{2}  +\sqrt { \left( \frac{M+m}{2} \right)^2+v_\star^2 \abs{\vec{k}}^2 },
	\label{app:eqn:hubbard_i_band}
\end{align}
with 
\begin{align}
	m = - \gamma^2 \left( \frac{u^2}{E_{f,1}}+\frac{v^2}{E_{f,2}} \right).
\end{align}

Therefore, for each valley and spin, there are four bands near the Fermi energy and close to the $\Gamma_M$ point. We now discuss whether the system is gapped at integer fillings. In practice, near integer filling, for each valley and each spin, the lowest two ($c$-electron) bands described by \cref{app:eqn:hubbard_i_band} will be filled. Exactly at the $\Gamma_M$ point, we find
\begin{align}
	E_{c,1} \left( \vec{0} \right) =\epsilon_c + \frac{m+M -|M-m|}{2} ,\quad E_{c,2} \left( \vec{0} \right) =\epsilon_c + \frac{m+M +|M-m|}{2},\nonumber\\ 
	E_{c,3} \left( \vec{0} \right) =\epsilon_c + \frac{m-M -|M+m|}{2} ,\quad E_{c,4} \left( \vec{0} \right) =\epsilon_c + \frac{m-M +|M+m|}{2}.
\end{align}

Depending on the value of $M$, three distinct scenarios can occur:
\begin{itemize}
	\item Case I: $|m|<M$: 
	\begin{align}
		&E_{c,1} \left( \vec{0} \right) =\epsilon_c + m,\quad E_{c,2} \left( \vec{0} \right) =\epsilon_c + M\nonumber\\ 
		&E_{c,3} \left( \vec{0} \right) =\epsilon_c -M ,\quad E_{c,4} \left( \vec{0} \right) =\epsilon_c + m
	\end{align}
	    In this case, there is a node formed by the middle two bands $E_{c,1}(\vec{k})$ and $E_{c,4}(\vec{k})$, and the system (with the two lowest bands fully filled) is gapless. 
	\item Case II: $m>M$:
	\begin{align}
		&E_{c,1} \left( \vec{0} \right) =\epsilon_c + M,\quad E_{c,2} \left( \vec{0} \right) =\epsilon_c + m,\nonumber\\ 
		&E_{c,3} \left( \vec{0} \right) =\epsilon_c -M ,\quad E_{c,4} \left( \vec{0} \right) =\epsilon_c + m,
	\end{align}
    Here, the upper two bands $E_{c,2}(\vec{k})$ and $E_{c,4}(\vec{k})$ form a node. In this case, the system (with the two lowest bands filled) acquires a gap near the $\Gamma_M$ point.
	\item Case III: $m<-M$:
	\begin{align}
		&E_{c,1} \left( \vec{0} \right) =\epsilon_c + m,\quad E_{c,2} \left( \vec{0} \right) =\epsilon_c + M,\nonumber\\ 
		&E_{c,3} \left( \vec{0} \right) =\epsilon_c +m ,\quad E_{c,4} \left( \vec{0} \right) =\epsilon_c -M,
	\end{align}
	In this situation, a node forms between the lower two bands $E_{c,1}(\vec{k})$ and $E_{c,3}(\vec{k})$. As in the previous case, the system (with its two lowest bands filled) becomes gapped near the $\Gamma_M$ point.
\end{itemize}

It is worth noting that at charge neutrality, particle-hole symmetry enforces $m = 0$, and the system is gapless with nodes. At non-zero integer fillings, we are generally in the regime $|M| < |m|$, and the system develops a finite gap.

\clearpage
\subsection{Results}\label{app:sec:results_symmetry:results}
\begin{figure}[!h]\includegraphics[width=\textwidth]{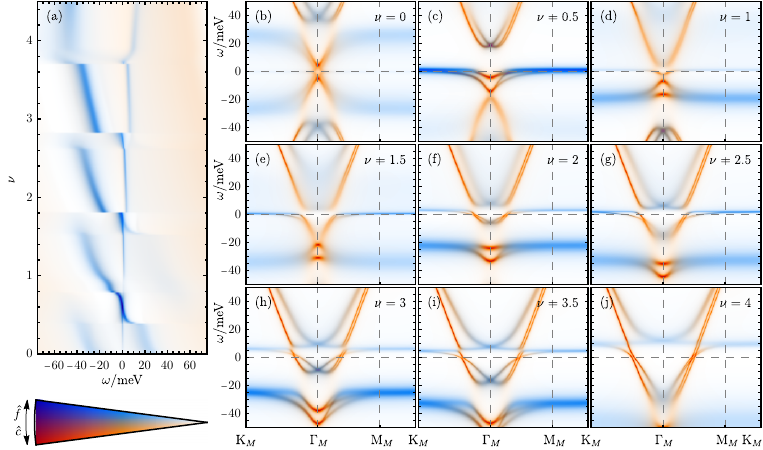}\caption{The symmetric phase of TBG within the THF model at $T = \SI{5}{\kelvin}$.}\label{app:fig:sym_bs_tbg_low}\end{figure}\enlargethispage{20mm}
\begin{figure}[!h]\includegraphics[width=\textwidth]{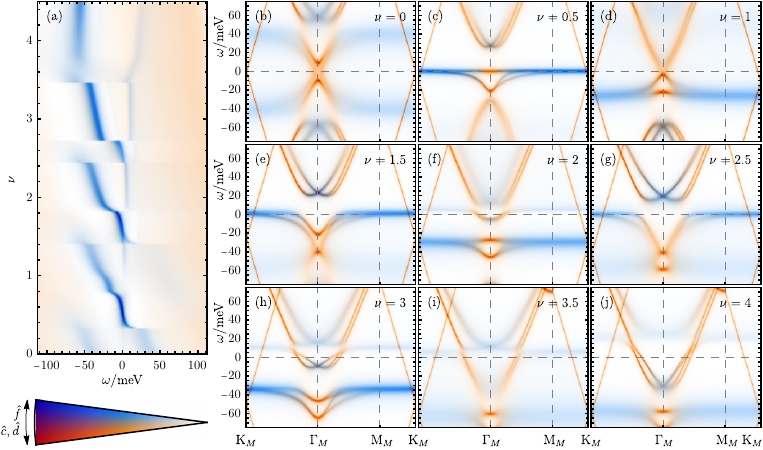}\caption{The symmetric phase of TSTG within the THF model for $\mathcal{E}=\SI{0}{\milli\electronvolt}$ at $T = \SI{15}{\kelvin}$.}\label{app:fig:sym_bs_tstg_noU_low}\end{figure}
\begin{figure}[!h]\includegraphics[width=\textwidth]{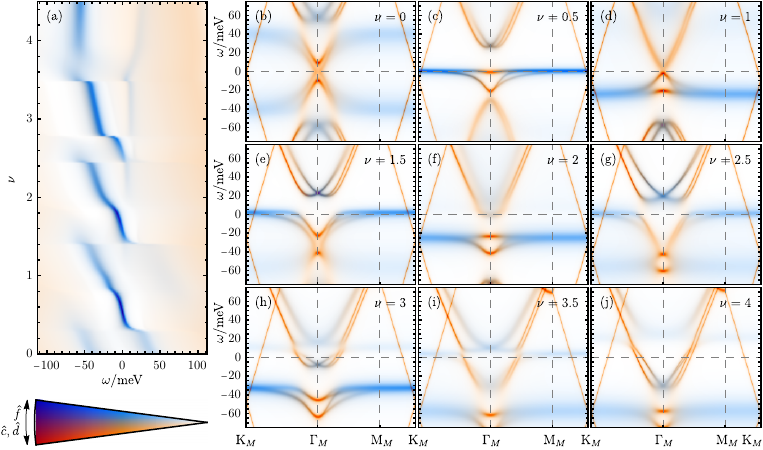}\caption{The symmetric phase of TSTG within the THF model for $\mathcal{E}=\SI{0}{\milli\electronvolt}$ at $T = \SI{30}{\kelvin}$.}\label{app:fig:sym_bs_tstg_noU_med}\end{figure}
\begin{figure}[!h]\includegraphics[width=\textwidth]{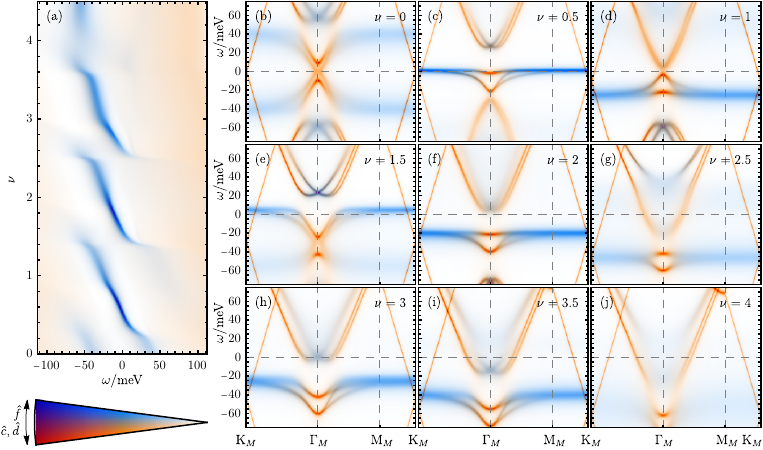}\caption{The symmetric phase of TSTG within the THF model for $\mathcal{E}=\SI{0}{\milli\electronvolt}$ at $T = \SI{75}{\kelvin}$.}\label{app:fig:sym_bs_tstg_noU_high}\end{figure}
\begin{figure}[!h]\includegraphics[width=\textwidth]{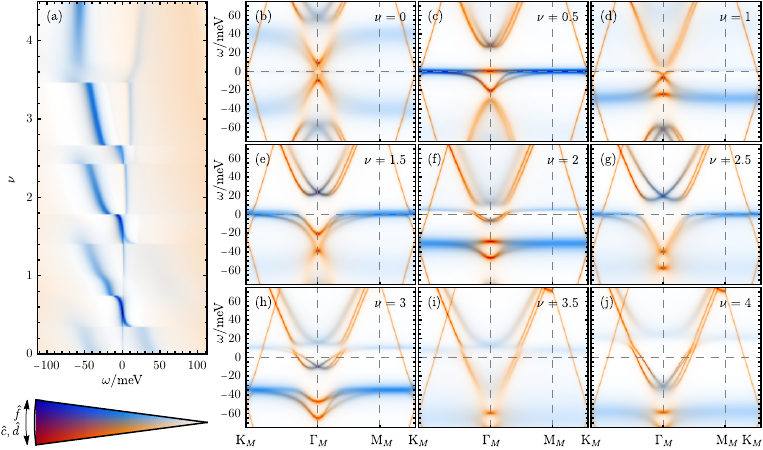}\caption{The symmetric phase of TSTG within the THF model for $\mathcal{E}=\SI{25}{\milli\electronvolt}$ at $T = \SI{7.5}{\kelvin}$.}\label{app:fig:sym_bs_tstg_U_low}\end{figure}
\begin{figure}[!h]\includegraphics[width=\textwidth]{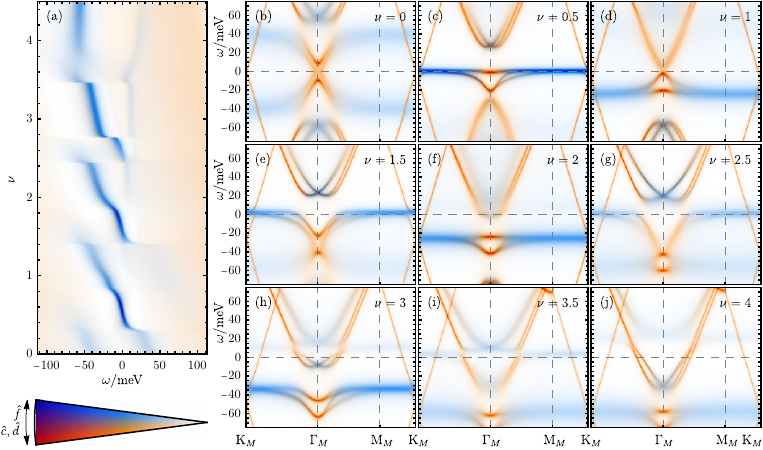}\caption{The symmetric phase of TSTG within the THF model for $\mathcal{E}=\SI{25}{\milli\electronvolt}$ at $T = \SI{30}{\kelvin}$.}\label{app:fig:sym_bs_tstg_U_med}\end{figure}
\begin{figure}[!h]\includegraphics[width=\textwidth]{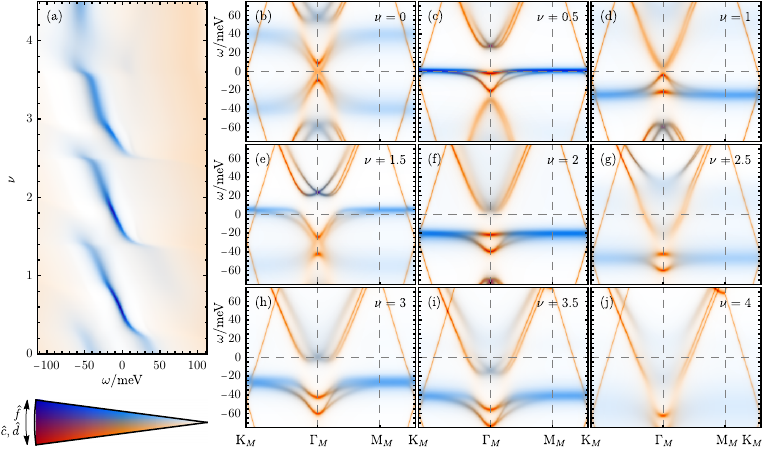}\caption{The symmetric phase of TSTG within the THF model for $\mathcal{E}=\SI{25}{\milli\electronvolt}$ at $T = \SI{75}{\kelvin}$.}\label{app:fig:sym_bs_tstg_U_high}\end{figure}
\FloatBarrier

\end{document}